\documentclass[11pt,twoside,a4paper]{cernrep} 
\usepackage{rep_common}

\usepackage{amsmath}
\usepackage{amssymb}
\usepackage{anysize}
\usepackage{bold-extra}
\usepackage{color}
\usepackage{enumerate}
\usepackage{fancyhdr}
\usepackage{graphicx}
\usepackage[linktocpage,colorlinks,urlcolor=blue]{hyperref}
\usepackage{multirow}
\usepackage[final]{pdfpages}
\usepackage{rotating}
\usepackage{subfig}
\usepackage{textcomp}
\usepackage{url}
\usepackage{verbatim}
\usepackage{wrapfig}
\usepackage{amsfonts}
\usepackage{comment}
\usepackage{epigraph}
\usepackage[utf8]{inputenc}
\usepackage{slashed}
\usepackage{bbm}
\usepackage{cancel}
\usepackage{epsf, array, color}
\usepackage{a4}
\usepackage{epsfig}
\usepackage{latexsym}
\usepackage{bm}
\usepackage{braket}
\usepackage[T1]{fontenc} 
\usepackage[utf8]{inputenc}
\usepackage{placeins}
\usepackage[capitalize, noabbrev]{cleveref}
\usepackage[utf8]{inputenc}
\usepackage{fullpage}
\usepackage{amsmath}
\usepackage{amssymb}
\usepackage{xspace}

\usepackage{amsmath} 
\usepackage{amsthm} 
\usepackage{amssymb}	
\usepackage{bm}	
\usepackage{graphicx} 
\usepackage{multirow}
\usepackage{booktabs}
\usepackage{makecell}
\usepackage[export]{adjustbox}
\usepackage{multirow}
\usepackage{fullpage}
\usepackage{amsmath,amssymb,mathtools}
\usepackage{enumerate}
\usepackage{booktabs}
\usepackage{multirow}
\usepackage{graphicx}
\usepackage{xspace}
\usepackage{multicol}
\usepackage{lineno}

\newcommand*{\ttH}{\ensuremath{t\bar{t}H}\xspace}

\newcommand*{\ttW}{\ensuremath{t\bar{t}W}\xspace}
\newcommand*{\ttZ}{\ensuremath{t\bar{t}Z}\xspace}

\DeclareOldFontCommand{\rm}{\normalfont\rmfamily}{\mathrm}
\DeclareOldFontCommand{\sf}{\normalfont\sffamily}{\mathsf}
\DeclareOldFontCommand{\tt}{\normalfont\ttfamily}{\mathtt}
\DeclareOldFontCommand{\bf}{\normalfont\bfseries}{\mathbf}
\DeclareOldFontCommand{\it}{\normalfont\itshape}{\mathit}
\DeclareOldFontCommand{\sl}{\normalfont\slshape}{\@nomath\sl}
\DeclareOldFontCommand{\sc}{\normalfont\scshape}{\@nomath\sc}

\newcommand{\main}{.}
\def\biblio{}

\def\bibfiles{\main/bib/chapter,\main/electroweak/electroweak,\main/top/top,\main/qcd/qcd,\main/frwd_physics/frwd_physics,\main/eft/eft,\main/theory_tools/theory_tools,\main/experiments/experiments,\main/hl-lhc_he-lhc_notes}
\providecommand{\biblio}{\nocite{article-minimal}\bibliographystyle{report}\clearpage\bibliography{\bibfiles}}  

\begin{document}


\title{{\normalfont\bfseries\boldmath\huge
\begin{center}
Standard Model Physics\\
at the HL-LHC and HE-LHC\\
\begin{normalsize} 
  \href{http://lpcc.web.cern.ch/hlhe-lhc-physics-workshop}{Report from Working Group 1 on the Physics of the HL-LHC, and Perspectives at the HE-LHC}
\end{normalsize}
\end{center}\vspace*{0.2cm}
}}

\author{Editors: \\P. Azzi$^{1}$, S. Farry$^{2}$, P. Nason$^{3,4}$, A. Tricoli$^{5}$, D. Zeppenfeld$^{6}$
\\ \vspace*{4mm} 
Contributors: 
\\ R. Abdul Khalek$^{7,8}$, J. Alimena$^{9}$, N. Andari$^{10}$, L. Aperio Bella$^{11}$, A.J. Armbruster$^{11}$
% ...
\vspace*{1cm} 
}
\institute{$^{1}$~INFN Sezione di Padova, Padova, Italy \\$^{2}$~Oliver Lodge Laboratory, University of Liverpool, Liverpool, United Kingdom \\$^{3}$~INFN Sezione di Milano-Bicocca, Milano, Italy \\$^{4}$~Universit{\`a} degli Studi di Milano-Bicocca, Dipartimento di Fisica ''G.Occhialini'', Piazza della Scienza 3, 20126 Milan, Italy \\$^{5}$~Brookhaven National Laboratory (BNL), USA \\$^{6}$~Karlsruher Institut f{\"u}r Technologie (KIT), Institut f{\"u}r Theoretische Physik (TP), Wolfgang-Gaede-Str. 1, 76131 Karlsruhe, Germany \\$^{7}$~Department of Physics and Astronomy, VU University, NL-1081 HV Amsterdam, Netherlands \\$^{8}$~Nikhef National Institute for Subatomic Physics, Amsterdam, Netherlands \\$^{9}$~The Ohio State University, Columbus, USA \\$^{10}$~IRFU, CEA, Universit{\'e} Paris-Saclay, France \\$^{11}$~European Laboratory for Particle Physics, CERN, Geneva, Switzerland \\$^{12}$~Institut for Theoretical Physics, Eberhard Karls University T{\"u}bingen, Auf der Morgenstelle 14, D-72076 T{\"u}bingen, Germany \\$^{13}$~Rudolf Peierls Centre, Beecroft Building, Parks Road, Oxford, OX1 3PU, United Kingdom \\$^{14}$~Institute of Physics, University of Belgrade, Belgrade, Serbia \\$^{15}$~Purdue University, West Lafayette, USA \\$^{16}$~The University of Kansas, Lawrence, USA \\$^{17}$~Imperial College, London, United Kingdom \\$^{18}$~Universit{\`a} di Padova, Padova, Italy \\$^{19}$~Vrije Universiteit Brussel, Brussel, Belgium \\$^{20}$~Universit{\'e} de Strasbourg, CNRS, IPHC UMR 7178, Strasbourg, France \\$^{21}$~University of Rochester, Rochester, NY, USA \\$^{22}$~Skobeltsyn Institute of Nuclear Physics, Lomonosov Moscow State University, Moscow, Russia \\$^{23}$~Universidad de Buenos Aires, Buenos Aires, Argentina \\$^{24}$~University of Florida, Gainesville, USA \\$^{25}$~Theory Division, IFIC, University of Valencia-CSIC, E-46980 Paterna, Valencia, Spain \\$^{26}$~Department of Physics and Astronomy, University College London, London, United Kingdom \\$^{27}$~Fermi National Accelerator Laboratory, P.O. Box 500 Batavia, 60510 USA \\$^{28}$~Department of Physics and State Key Laboratory of Nuclear Physics and Technology, Peking University, Beijing 100871, China \\$^{29}$~ Collaborative Innovation Center of Quantum Matter, Beijing 100871, China \\$^{30}$~ Center for High Energy Physics, Peking University, Beijing 100871, China \\$^{31}$~Department of Physics and Astronomy, University of Sussex, Brighton, United Kingdom \\$^{32}$~Institute for Particle Physics Phenomenology, University of Durham, Durham, UK \\$^{33}$~High Energy Physics Institute of Tbilisi State University, Tbilisi, Georgia \\$^{34}$~LLR, Ecole polytechnique, CNRS/IN2P3, Universit{\'e} Paris-Saclay, Palaiseau, France \\$^{35}$~Key Laboratory of Quark and Lepton Physics (MoE) and Institute of Particle Physics, Central China Normal University, China \\$^{36}$~Department of Mathematics, Statistics, and Physics, Wichita State University, Wichita KS, USA \\$^{37}$~INFN Sezione di Roma Tre, Via della Vasca Navale 84, I-00146 Roma, Italy \\$^{38}$~INFN, Laboratori Nazionali di Frascati, Frascati, Italy \\$^{39}$~ Istituto Nazionale di Fisica Nucleare (INFN) Sezione di Torino , Italy \\$^{40}$~ Universit{\`a} di Torino, Torino, Italy \\$^{41}$~ Universita' di Milano, Dipartimento di Fisica, Milano, Italy \\$^{42}$~RWTH Aachen University, III. Physikalisches Institut A, Aachen, Germany \\$^{43}$~University of Louisville, Louisville, Kentucky, USA \\$^{44}$~Narodowe Centrum Bada{\'n} J{\k{a}}drowych (NCBJ), Ho{\.z}a 69, 00-681 Warszawa, Poland \\$^{45}$~Department of Physics and Astronomy, Michigan State University, East Lansing, Michigan, USA \\$^{46}$~Department of Physics, Yale University, New Haven CT, USA \\$^{47}$~Universit{\"a}t W{\"u}rzburg, Institut f{\"u}r Theoretische Physik und Astrophysik, Emil-Hilb-Weg 22, 97074 W{\"u}rzburg, Germany \\$^{48}$~Laboratoire de Physique Nucl{\'e}aire et de Hautes Energies (LPNHE), Sorbonne Universit{\'e}, Paris-Diderot Sorbonne Paris Cit{\'e}, CNRS/IN2P3, France \\$^{49}$~LAPP, Univ. Grenoble Alpes, Univ. Savoie Mont Blanc, CNRS/IN2P3, Annecy, France \\$^{50}$~Department of Physics and Astronomy, University of Pennsylvania, Philadelphia, Pennsylvania, USA \\$^{51}$~Deutsches Elektronen-Synchrotron, Hamburg, Germany \\$^{52}$~School of Physics, Institute for Research in Fundamental Sciences (IPM), P.O. Box 19395-5531, Tehran, Iran \\$^{53}$~Laborat{\'o}rio de Instrumenta{\c{c}}{\~a}o e F{\'\i}sica Experimental de Part{\'\i}culas - LIP, Lisboa, Portugal \\$^{54}$~INFN, Sezione di Milano, Via Celoria 16, I-20133 Milan, Italy \\$^{55}$~Physics Department, New York City College of Technology, CUNY, 300 Jay Street, Brooklyn, NY 11201 USA \\$^{56}$~Boston University, Department of Physics, Boston, Massachusetts, USA \\$^{57}$~Science Department, Borough of Manhattan Community College, New York, NY 10007, USA \\$^{58}$~INFN, Sezione Roma I, Roma, Italy \\$^{59}$~Technische Universit{\"a}t M{\"u}nchen, M{\"u}nchen, Germany \\$^{60}$~Laboratoire de Physique Th{\'e}orique et Hautes {\'E}nergies (LPTHE), Boite 126, T13-14 4{\`e}me {\'e}tage, 4 place Jussieu, F-75252 Paris CEDEX 05, France \\$^{61}$~Institut Universitaire de France, 103 boulevard Saint-Michel, 75005 Paris, France \\$^{62}$~Laborat{\'o}rio de Instrumenta{\c{c}}{\~a}o e F{\'\i}sica Experimental de Part{\'\i}culas - LIP, Coimbria, Portugal \\$^{63}$~School of Physics and Astronomy, Shanghai Jiao Tong University, Shanghai, China \\$^{64}$~Physik-Institut, Universitaet Zurich, 8057 Zurich, Switzerland \\$^{65}$~Department of Physics, ETH Zurich, CH-8093 Zurich, Switzerland \\$^{66}$~University of Split, Split, Croatia \\$^{67}$~Universit{\`a} di Roma Tor Vergata, Dipartimento di Fisica, Roma, Italy \\$^{68}$~Laboratoire de Physique Theorique et Hautes Energies, UMR 7589, Sorbonne Universit{\'e} et CNRS, Bo{\^\i}te 126, T13-14 4{\`e}me {\'e}tage, 4 place Jussieu, F-75252 Paris CEDEX 05, France \\$^{69}$~University of Hamburg, Hamburg, Germany \\$^{70}$~Denys Wilkinson Building, University of Oxford, Oxford, OX1 3RH, United Kingdom \\$^{71}$~Department of Physics, University of Warwick, Coventry, United Kingdom \\$^{72}$~Max Planck Institute for Physics, Foehringer Ring 6, 80805 Munich, Germany \\$^{73}$~University of Wisconsin - Madison, Madison, USA \\$^{74}$~University of Vienna, Vienna, Austria \\$^{75}$~Bethel University, St. Paul, USA \\$^{76}$~Brown University, Providence, USA \\$^{77}$~Tata Institute of Fundamental Research, Mumbai, India \\$^{78}$~School of Physics, University of the Witwatersrand, South Africa \\$^{79}$~PRISMA Cluster of Excellence and Institute for Physics (THEP), Johannes Gutenberg University Mainz, D-55099 Mainz, Germany \\$^{80}$~Department of Physics, University of Science and Technology of Mazandaran, P.O.Box 48518-78195, Behshahr, Iran \\$^{81}$~Department of Physics, University of Tehran, North Karegar Ave., Tehran 14395-547, Iran \\$^{82}$~Cornell University, Physics Department, Ithaca, NY, USA \\$^{83}$~Petersburg Nuclear Physics Institute (PNPI), Gatchina, Russia \\$^{84}$~Institut f{\"u}r Hochenergiephysik, Wien, Austria \\$^{85}$~University of Cape Town, South Africa \\$^{86}$~Department of Physics, University of Toronto, Toronto, Canada \\$^{87}$~Department of Physics, University of Michigan, Ann Arbor, Michigan, USA \\$^{88}$~DESY, Hamburg and Zeuthen, Germany \\$^{89}$~Lawrence Berkeley National Laboratory, 1 Cyclotron Road, Berkeley CA 94720 USA \\$^{90}$~Durham University, Institute for Particle Physics Phenomenology, Ogden Centre for Fundamental Physics, South Road, Durham DH1 3LE, United Kingdom \\$^{91}$~College of Nuclear Science and Technology,Beijing Normal University,100875 Beijing, China, Beijing Radiation Center, Beijing 100875, China \\$^{92}$~University of Sheffield, Sheffield, United Kingdom \\$^{93}$~Institute for High Energy Physics of National Research Centre 'Kurchatov Institute', Protvino, Russia \\$^{94}$~Henryk Niewodniczanski Institute of Nuclear Physics  Polish Academy of Sciences, Krak{\'o}w, Poland \\$^{95}$~Research Institute for Nuclear Problems of Byelorussian State University, Minsk, Belarus \\$^{96}$~Institute of Particle and Nuclear Studies, High Energy Accelerator Research Organization (KEK), Tsukuba 305-0801, Japan \\$^{97}$~Cavendish Laboratory, University of Cambridge, Cambridge, United Kingdom \\$^{98}$~Czech Technical University in Prague, Prague, Czech Republic \\$^{99}$~ Sapienza Universita' di Roma, Dipartimento di Fisica \\$^{100}$~Waseda University, Waseda Research Institute for Science and engineering (WISE), Tokyo, Japan \\$^{101}$~Physics Department, University of Minho, 4710 - 057, Braga, Portugal \\$^{102}$~University of California, Berkeley, California, USA \\$^{103}$~Universidad de Oviedo, Oviedo, Spain \\$^{104}$~CFTP, Instituto Superior T{\'e}cnico, Universidade de Lisboa, Avenida Rovisco Pais 1, 1049 Lisboa, Portugal \\$^{105}$~LAPTh, 9 Chemin de Bellevue, F-74941 Annecy Cedex, France \\$^{106}$~Physics Department, Florida State University, Tallahassee, Florida, USA \\$^{107}$~Rudjer Boskovic Institute, Bijenicka cesta 54, P.O. Box 180, 10002 Zagreb, Croatia \\$^{108}$~Department of Astronomy and Theoretical Physics, Lund University, Lund, Sweden \\$^{109}$~Kirchhoff-Institut f{\"u}r Physik, Heidelberg, Germany \\$^{110}$~University of Sydney, Sydney, Australia \\$^{111}$~International Center for Elementary Particle Physics and Department of Physics, The University of Tokyo, Japan \\$^{112}$~Physics Department, Universidad Autonoma de Madrid, Spain \\$^{113}$~University of Science and Technology of China, Hefei, China \\$^{114}$~Institute of High Energy Physics, Chinese Academy of Sciences, Beijing, China \\$^{115}$~University of California, San Diego, La Jolla, USA
}

\begin{titlepage}

\vspace*{-1.8cm}

\noindent
\begin{tabular*}{\linewidth}{lc@{\extracolsep{\fill}}r@{\extracolsep{0pt}}}
\vspace*{-1.2cm}\mbox{\!\!\!\includegraphics[width=.14\textwidth]{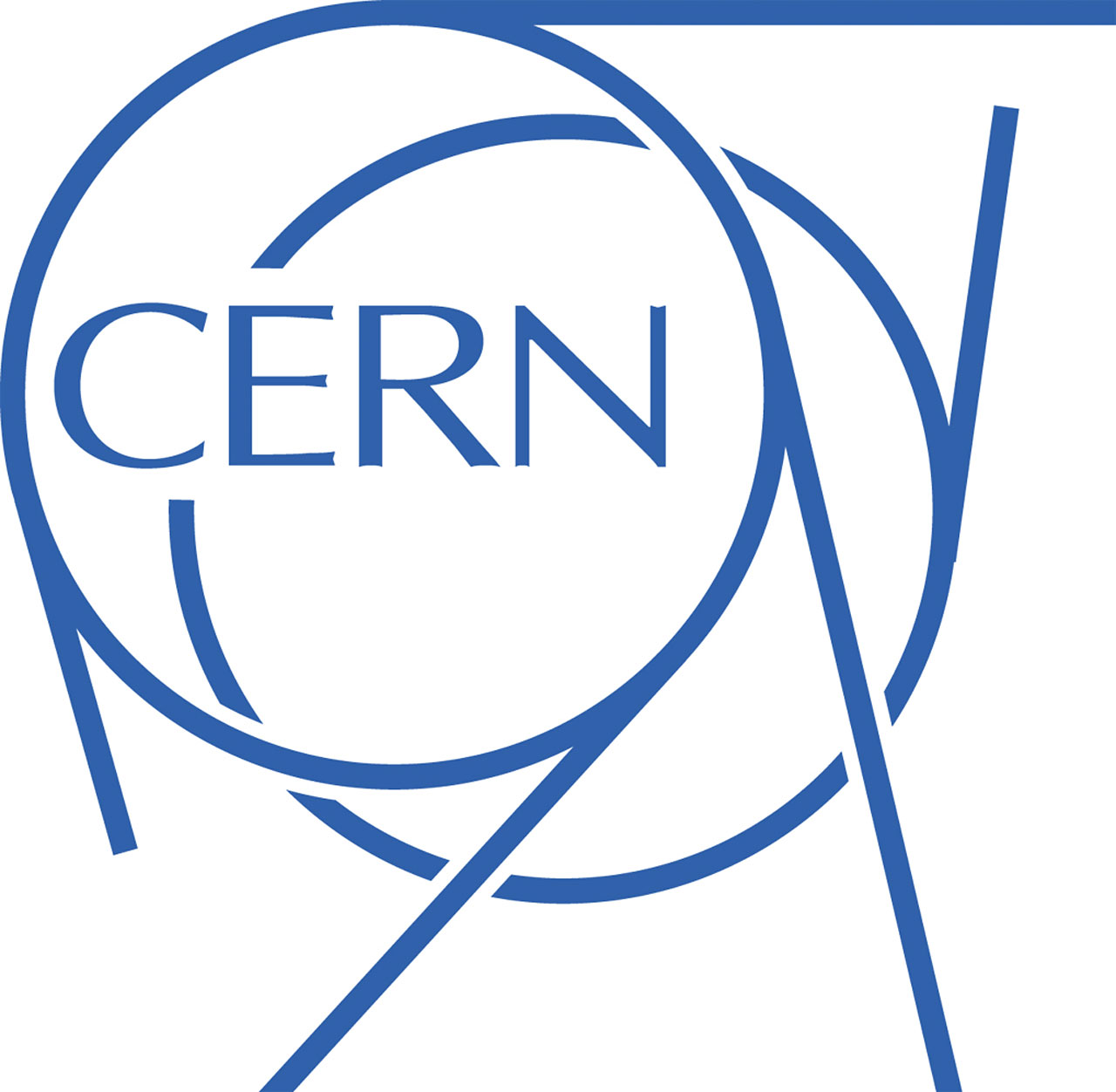}} & & \\
 & & CERN-LPCC-2018-03 \\  
 & & \today \\ 
 & & \\
\hline
\end{tabular*}

\vspace*{0.3cm}
\maketitle
\vspace{\fill}
%

\begin{abstract}
The successful operation of the Large Hadron Collider (LHC) and the excellent performance of the ATLAS, CMS, LHCb and ALICE detectors in Run-1 and Run-2  with $pp$ collisions at center-of-mass energies of 7, 8 and 13 TeV as well as the giant leap in precision calculations and modeling of fundamental interactions at hadron colliders have allowed an extraordinary breadth of physics studies including precision measurements of a variety physics processes. The LHC results have so far confirmed the validity of the Standard Model of particle physics up to unprecedented energy scales and with great precision in the sectors of strong and electroweak interactions as well as flavour physics, for instance in top quark physics. The upgrade of the LHC to a High Luminosity phase (HL-LHC) at 14 TeV center-of-mass energy with 3 ab$^{-1}$ of integrated luminosity will probe the Standard Model with even greater precision and will extend the sensitivity to possible anomalies in the Standard Model, thanks to a ten-fold larger data set, upgraded detectors and expected improvements in the theoretical understanding. This document summarises the physics reach of the HL-LHC in the realm of strong and electroweak interactions and top quark physics, and provides a glimpse of the potential of a possible further upgrade of the LHC to a 27 TeV $pp$ collider, the High-Energy LHC (HE-LHC), assumed to accumulate an integrated luminosity of 15 ab$^{-1}$. 

\end{abstract}
\vspace*{2.0cm}
\vspace{\fill}

\end{titlepage}

\setcounter{tocdepth}{3}
\tableofcontents
\newpage

\documentclass[../report.tex]{subfiles}
\providecommand{\main}{..}
\IfEq{\jobname}{\currfilebase}{\AtEndDocument{\biblio}}{}


\section{Introduction}

The Large Hadron Collider (LHC) is one of largest scientific instruments ever built. To extend its discovery potential, the LHC will undergo a major upgrade in the 2020s, the High-Luminosity LHC (HL-LHC). The HL-LHC will collide protons against protons at 14 TeV centre-of-mass energy with an instantaneous luminosity a factor of five greater than the LHC and will accumulate ten times more data, resulting in an integrated luminosity of 3 ab$^{-1}$. 

The LHC results have so far confirmed the validity of the Standard Model of particle physics up to unprecedented energy scales and with great precision in the sectors of strong and electroweak interactions, Higgs boson as well as flavour physics including top quark properties. The HL-LHC program, thanks to a ten-fold larger data set, upgraded detectors and expected improvements in the theoretical understanding, will extend the sensitivity to new physics in direct and indirect searches for processes with low production cross sections 
and harder signatures. 
In addition, a considerable improvement is expected in precise measurements of properties of the Higgs boson, e.g. couplings measurements at the percent level, and of Standard Model (SM) production processes. Several of these measurements will be limited by the uncertainties on the knowledge of the partonic inner structure of the proton, i.e. Parton Denstity Functions (PDFs). Global PDF fits of several HL-LHC measurements will allow a significant improvement in PDF uncertainties and, in turn, in measurements of SM parameters, e.g. the weak mixing angle and the W boson mass. Anomalies in precision measurements in the SM sector can become significant when experimental measurements and theoretical predictions reach the percent level of precision, and when probing unprecedented energy scales in the multi-TeV regime. These anomalies could give insights to new physics effects from higher energy scales. 

Additional studies on the potential of a possible further upgrade of the LHC to a 27 TeV $pp$ collider, the High-Energy LHC (HE-LHC), assumed to accumulate an integrated luminosity of 15 ab$^{-1}$, have also been carried out.  

A year long Workshop organized at CERN in 2017-2018 brought together experimentalists from the ATLAS, CMS, LHCb, and ALICE Collaborations and theorists to study the expected physics reach of the HL-LHC project and its possible upgrade to the HE-LHC. Studies of the Workshop in the sectors of electroweak and strong interactions as well as top physics were carried out within the Working Group 1 (WG1) and the results are summarized in this report that constitutes a chapter of the HL/HE-LHC Yellow Report volume to be submitted to the European Strategy Group.

The report first introduces the theoretical tools used for the following theoretical projections and their expected future improvements as well as the experimental performance assumed in the following experimental analyses. Dedicated sections summarize the results of the studies in the areas of electroweak processes, strong interactions, top physics including effective coupling interpretations, and proposes studies of forward physics that are possible with new forward detectors. The sections focus on physics projections for the HL-LHC and the expected improvements in measurement precision or kinematic reach compared to LHC. In some cases the studies are extended to HE-LHC highlighting the larger statistics and energy reach of HE-LHC compared to HL-LHC.
In the following sections the authors of the theoretical contributions are listed in footnotes to the section titles. Where the authors are not explicitly indicated, they are the experimental LHC Collaborations.

\newpage

\section{Theoretical tools}


\subsection[High Order QCD calculations]{High Order QCD calculations\footnote{Contributed by G. Zanderighi.}}


In order to exploit the full potential of the High-Luminosity LHC
physics program, the high precision of experimental data must be
compared to theoretical predictions that have the same
accuracy. Precision calculations in QCD are typically classified into
fixed-order expansions in the coupling constant $\alpha_s$, and into
predictions that resum large logarithms to all orders in
$\alpha_s$.  The latter are usually also subdivided into
numerical parton-shower approaches and analytic resummed
calculations. In recent years, a lot of work has been  devoted also to
matching and merging fixed-order and resummed calculations, so as to
have an improved accuracy in all regions of phase space.

The technical ingredients required for a fixed-order
calculation to higher orders are the computation of real, virtual or,
from two loop on, mixed real-virtual amplitudes, the calculations
of the required master integrals and a procedure to regularize
intermediate soft and collinear divergences.
The first non-trivial contribution is of next-to-leading order
(NLO).
%
Here, the basis of master integrals required to compute any process
at one-loop in QCD had been known for a long time, and is now available in
public codes~\cite{Ellis:2007qk,vanHameren:2010cp}. In addition, two general
subtraction methods (FKS~\cite{Frixione:1995ms} and
CS~\cite{Catani:1996vz}), well suited for automation, were developed.
%
The tensor reduction of virtual amplitudes (i.e.
the reduction of virtual amplitude into a combination
of master integrals) proved to be the most difficult problem,
since the most straightforward approaches
yielded too complex results for generic processes.
Around ten to fifteen years ago, a number of breakthrough
ideas~\cite{Britto:2004nc,Ossola:2006us,Ellis:2008ir,Giele:2008ve,Ellis:2011cr,Cascioli:2011va}
led to algorithms for tensor reduction that can be automatized
efficiently.
%
With all ingredients in place, 
a number of tools to compute NLO cross sections for generic LHC
processes in an automated way were developed.
These tools are today heavily used at the LHC and will be
indispensable for future phenomenology. The most widely used tools
include
\textsc{Gosam}~\cite{Cullen:2014yla},
\textsc{MadLoop}~\cite{Alwall:2014hca}, or 
\textsc{OpenLoops}~\cite{Cascioli:2011va}.
%
It is interesting to note that, in the early days of NLO calculations,
also slicing approaches were suggested to handle intermediate
divergences~(see~e.g.~\cite{Eynck:2001en}).  They were however soon
abandoned in favour of subtraction approaches.

While NLO tools are certainly more appropriate then leading-order (LO) generators to
accurately predict LHC distributions, already with Run-2 data it is
clear that an even better perturbative accuracy is required to match
the precision of data.
One of the first explicit demonstrations of this fact was given by the $WW$
cross section~\cite{ATLAS:2012mec,Chatrchyan:2013yaa,Chatrchyan:2013oev},
that raised interest because of discrepancies in the extrapolated
total cross section between theory and data both at 7~TeV and 8~TeV, and
both at ATLAS and CMS. The discrepancy could be resolved thanks to the
inclusion of next-to-next-to-leading (NNLO) corrections and thanks to the observation that the
extrapolation from the fiducial to the inclusive cross section had a
larger uncertainty than the estimated one. 
This example highlights the importance of quoting also fiducial
cross sections, prior to any Monte Carlo based extrapolation, and of
including NNLO corrections when comparing to high-precision data.

Current years are seeing an incredibly fast progress in
the calculation of NNLO cross sections (for recent short reviews see~e.g. Ref.~\cite{Heinrich:2017una,Dittmaier:2017vus}). The current
status is that all non-loop induced $2\to2$ SM processes are known at
NNLO, including dijet production~\cite{Currie:2017eqf} that has the
most complicated subprocess and singularity structure.
This breakthrough was possible thanks to the development of new
methods to compute two-loop integrals. One idea that was exploited to
a great extent is the fact that polylogarithmic integrals can be
calculated by means of differential
equations~\cite{Kotikov:1991hm,Kotikov:1991pm,Henn:2013pwa,Papadopoulos:2014lla}.
Currently, the processes that are more difficult to compute are those that
involve internal masses, since they lead not only to polylogarithms
but also to elliptic integrals. Examples include loop-induced processes
like gluon-fusion Higgs or di-Higgs production with full top-mass
dependence, or gluon induced di-boson production. 

With the High-Luminosity run of the LHC, it will be possible to explore the Higgs
transverse momentum spectrum up to almost 1 TeV, where the large-$m_t$
approximation is
well-known to fail. Recently, two-loop NLO results for the Higgs
transverse momentum spectrum became
available~\cite{Jones:2018hbb,Lindert:2018iug}, but genuine NNLO
predictions for these loop-induced processes are still out of reach.

The calculations of multi-scale two-loop amplitudes with massive
internal particles relevant for Higgs-, top- and vector-boson
production, and in particular the mathematical structures beyond
multiple polylogarithms that appear in these amplitudes, is a very
active area of research 
today~\cite{Adams:2014vja,Henn:2014lfa,Caola:2014lpa,Gehrmann:2014bfa,Papadopoulos:2014hla,Papadopoulos:2015jft,Remiddi:2017har,Bonciani:2016qxi,Adams:2018kez,Adams:2017tga,Adams:2016xah,Ablinger:2017bjx,vonManteuffel:2017hms,Broedel:2018iwv}.
The developments of yet new ideas and computational methods are
eagerly needed. 
Approaches for the full numerical calculation of master integrals also exist (see
e.g. Ref.~\cite{Becker:2012bi,Czakon:2013goa,Bogner:2017vim,Borowka:2017idc}
and references therein), requiring however considerable computing power
as the complexity increases.

As far as the problem of canceling divergences, quite a number of
different approaches are being pursued now. They can be broadly
divided into subtractions methods (antenna
subtraction~\cite{GehrmannDeRidder:2005cm}, sector-improved residue
subtraction~\cite{Heinrich:2002rc,Czakon:2010td,Boughezal:2011jf,Czakon:2014oma},
nested subtraction~\cite{Caola:2017dug}, colourful
subtraction~\cite{DelDuca:2016csb}, projection to
Born~\cite{Cacciari:2015jma}) or slicing methods
($q_T$-subtraction~\cite{Catani:2009sm},
$N$-jettiness~\cite{Boughezal:2015dva,Gaunt:2015pea}).
These methods are being scrutinized, compared, and refined, and while it
is not clear yet which method will prevail,
it seems realistic to assume that, by the beginning of
the High-Luminosity phase, the issue of handling intermediate
divergences in NNLO calculations will be considered solved.
An ambitious goal is in fact to have $2\to3$ NNLO results by the
beginning of the High-Luminosity phase. A milestone would be certainly
to have NNLO prediction for $ttH$ production.
Motivated by the success at one-loop, a lot of effort is devoted to
extending generalized unitarity and the OPP methods beyond one
loop (see e.g. Ref.~\cite{Badger:2012dp}).
Currently, $2 \to 3$ processes are a very active subject of study, with
initial results of $3$-jet amplitudes starting to
appear~\cite{Badger:2013gxa,Badger:2015lda,Gehrmann:2015bfy,Badger:2016ozq,Dunbar:2016gjb,Badger:2017jhb,Abreu:2017hqn}.

Beyond NNLO, two calculations of LHC processes exist today at N$^3$LO
for inclusive Higgs production in the large $m_t$
approximation~\cite{Anastasiou:2016cez,Mistlberger:2018etf} and for
vector-boson-fusion (VBF) Higgs production in the structure function
approximation~\cite{Dreyer:2016oyx}. The complexity of these
calculations suggest that it will be very hard to extend this level of
accuracy to more complicated processes, since the technology they use
explicitly exploits the simplicity of these two
processes, and cannot be easily extended to more complex
ones.

Besides fixed-order, also resummed calculations have seen a leap in
recent years. The accuracy with which particular observables can be
resummed analytically reaches N$^3$LL
(see e.g. Ref,~\cite{Bizon:2017rah,Chen:2018pzu,Bizon:2018foh}), which means three towers
of logarithmic terms down compared to the leading logarithms that
arise when only soft and collinear gluons are correctly accounted
for.
These results are properly matched to fixed order NNLO calculations.

Resummed calculations rely either on methods based upon coherent
branching~\cite{Collins:1984kg,Catani:1992ua} or upon Soft Collinear
Effective Theory (SCET)~\cite{Bauer:2001yt}. So far, the two
approaches have also been considered as complementary, in fact both
methods proceed by performing a systematic expansion of the
contributions to the cross section. Recent work highlights the
connection between the two methods~\cite{Bauer:2018svx}.

While the logarithmic accuracy of resummed calculations is impressive,
the formal accuracy of parton showers is much less advanced. Unlike
resummed calculations, that are targeted to a well defined cross
section or distribution, Monte Carlo generators make predictions for several kind
of observables at the same time, and, at present, a
rigorous way to qualify their accuracy is missing. First studies in this direction can
be found in~\cite{Dasgupta:2018nvj}. Nevertheless, attempts to improve
some aspects of the shower algorithms are the focus of recent work.
Different approaches are taken: one
can incorporate the spin-color interference into
showers~\cite{Nagy:2015hwa}, include higher-order splitting functions and
$1\to 3$ splitting kernels into
showers~\cite{Jadach:2016zgk,Hoche:2017iem} or consider different
shower evolution variables~\cite{Hoche:2015sya,Fischer:2016vfv}.
It seems likely that by the start of the High-Luminosity program we
will have a much better theoretical control on the parton shower
evolution and the uncertainty associated to it.

In the same way as the progress in NLO went hand in hand with the
development of matching procedures of NLO and parton shower, a number
of approaches have been suggested recently to match NNLO calculations
and parton
showers~\cite{Alioli:2013hqa,Hamilton:2013fea,Hoeche:2014aia}. The
bottleneck in these approaches is currently the fact that they rely on
a reweighing procedure that is differential in the Born phase
space. Such a reweighing is possible for relatively simple processes
but becomes numerically unfeasible for more complicated ones. It
seems reasonable to expect that in the next years better NNLOPS
approaches will be developed that do not rely on any reweighing to
the NNLO. This would make it possible to have NNLO
predictions matched to parton shower (PS), also called NNLOPS, to more generic processes for which an NNLO calculation is
available, as is currently the case at NLO.

\subsection[Electroweak corrections]{Electroweak corrections\footnote{Contributed by D. Pagani, M. Zaro and M. Sch\"onherr.}}\label{sec:theory_tools_ew}


\newcommand{\mgamc}{MadGraph5\_aMC@NLO}
\newcommand{\Sherpa}{\textsc{Sherpa}\xspace}
\newcommand{\Recola}{\textsc{Recola}\xspace}
\newcommand{\OpenLoops}{\textsc{OpenLoops}\xspace}
\newcommand{\GoSam}{\textsc{GoSam}\xspace}
\newcommand{\MadDip}{\textsc{MadDipole}\xspace}
\newcommand{\Herwig}{\textsc{Herwig}\xspace}
\newcommand{\Pythia}{\textsc{Pythia}\xspace}

\subsubsection*{Existing tools}
In the last few years,
the automation of electroweak (EW) NLO corrections has witnessed an impressive progress, for what concerns both
one-loop and real-emission contributions (and their combination), by collaborations such as
\Recola~\cite{Actis:2016mpe,Actis:2012qn} with \Sherpa~\cite{Gleisberg:2008ta,
Schonherr:2017qcj}, \OpenLoops~\cite{Cascioli:2011va} with \Sherpa, 
\GoSam~\cite{Cullen:2011ac,Cullen:2014yla} with either 
\MadDip~\cite{Frederix:2008hu,Gehrmann:2010ry} or \Sherpa, 
and \mgamc~\cite{Alwall:2014hca, Frederix:2018nkq}. For most of these codes tuned 
comparisons have also been published~\cite{Badger:2016bpw, Bendavid:2018nar}, displaying
excellent agreement among them. Although the capabilities and reach in process complexity can differ from one 
computer program to another, recent results obtained with these
tools~\cite{Denner:2014ina,Denner:2014wka,Denner:2015yca,Kallweit:2014xda,
Frixione:2014qaa,Chiesa:2015mya,Kallweit:2015dum,Frixione:2015zaa,
Biedermann:2016guo,Biedermann:2016yvs,Denner:2016jyo,Biedermann:2016yds,
Biedermann:2016lvg,Denner:2016wet,Frederix:2016ost,Pagani:2016caq,
Biedermann:2017yoi,Kallweit:2017khh,Chiesa:2017gqx,Biedermann:2017bss,Biedermann:2017oae,
Czakon:2017wor,Greiner:2017mft,Frederix:2017wme,Gutschow:2018tuk,Schonherr:2018jva} clearly 
demonstrate how automation has made it possible to tackle problems whose complexity 
is too great to justify their solutions through traditional approaches.\\

Stemming from these advances, newer applications have become possible, one of these 
is the computation of the so-called ``complete-NLO'' corrections. In general, a given scattering processes can proceed
through $n$ different coupling combinations at LO (for example, $t\bar t$ or dijet production receives contributions
at order $\alpha_s^2$, $\alpha_s \alpha$ and $\alpha^2$); typically only the term with the largest power of 
$\alpha_s$ is retained, owing to the fact that $\alpha_s \gg \alpha$. This structure generates a similar 
one at N$^p$LO, with $n+p$ contributions, and the term ``complete-NLO'' means the (simultaneous) computation
of all the terms entering at LO and NLO. Among the computer programs cited above, some have been employed for
the computation of the complete-NLO corrections. In most of the cases the impact of the various contributions
closely follows the pattern one would expect from the coupling powers, as it is the case for dijet 
production~\cite{Frederix:2016ost}, top-pair~\cite{Czakon:2017wor} possibly with one extra 
jet~\cite{Gutschow:2018tuk}. However, there exist processes for which the coupling hierarchy is violated, 
or even flipped. Examples are same-sign $W$ production with two jets~\cite{Biedermann:2017bss}, top-pair production
in association with a $W$ boson and four-top production~\cite{Frederix:2017wme}.

\subsubsection*{Corrections beyond NLO}

Similarly to the NLO case, also NNLO corrections can be organized in powers of $\alpha$ and $\alpha_s$. At the moment, $ \mathcal{O}(\alpha_s^2)$ NNLO QCD calculations have been performed for many production processes at the LHC. Conversely, complete NNLO mixed QCD-EW calculations of $\mathcal{O}(\alpha_s \alpha)$ have not been performed for any process yet. These calculations are essential in order to pin down the theoretical uncertainties for processes that at the HL- and HE-LHC will be measured with very high precision. For this reason a great effort has been already invested for achieving this result and great progress can be expected in the next years. We recall the calculations that have been performed for Drell-Yan production~\cite{Dittmaier:2014qza, Dittmaier:2015rxo} in the resonance region via the pole   approximation. For this kind of calculations two-loop amplitudes~\cite{Kotikov:2007vr,Kilgore:2011pa,Bonciani:2011zz,Bonciani:2016ypc, vonManteuffel:2017myy} as well as regularized double-real emissions~\cite{Bonciani:2016wya} are necessary ingredients. Similarly, NNLO mixed QCD-EW corrections to gluon-gluon-fusion ($gg$F) Higgs production, which are induced by three-loop diagrams, have been estimated in ref.~\cite{Anastasiou:2008tj}.  Further recent calculations \cite{Bonetti:2018ukf,Anastasiou:2018adr} support those results and, in particular, support the fact that they can be correctly approximated via the so-called multiplicative approach. In short: NNLO mixed QCD-EW $\sim$  NLO QCD $\times$ NLO EW. 
   
     The aforementioned multiplicative approach is in general a very good approximation when the bulk of QCD and EW corrections at NLO is dominated by soft effects and Sudakov logarithms, respectively. Given the current lack of exact  NNLO mixed QCD-EW calculations, this approximation is already being used for estimating these corrections and/or missing higher orders uncertainties of different processes. First (N)NNLO QCD calculations including NLO EW corrections via the multiplicative approach have already appeared \cite{Anastasiou:2016cez, Czakon:2017wor, Lindert:2017olm} 
and are already necessary for a correct interpretation of current data; this level of accuracy will be mandatory for more processes at HL and HE-LHC.

Besides NNLO mixed QCD-EW corrections of $\mathcal{O}(\alpha_s \alpha)$, non-negligible contributions can emerge also from large $\mathcal{O}(\alpha^n)$ corrections with $n>1$. These typically involve final-state radiation (FSR) from massless/light particles and Sudakov logarithms. Both effects can be resummed, (at LL) via shower simulations (see the following sections on
matching with QED showers and with EW showers), or analytically. In the case of Sudakov logarithms, general methods for their calculation \cite{Denner:2000jv, Denner:2001gw} and techniques for resumming them \cite{Chiu:2007yn,Chiu:2008vv} are already known since quite some time. Based on the study already performed for 100 TeV proton--proton collisions \cite{Mangano:2016jyj}, at the HE-LHC, the resummation of Sudakov effects may be relevant in the tail of distributions.

\subsubsection*{Matching with QED shower}
\label{Qedmatch}

Fixed order computations need to be matched to parton showers, which 
compute a fully differential numerical resummation and implement 
the evolution of both QCD and EW particles from the hard scale to 
low scales, connecting it to the non-perturbative hadronization stage 
to arrive at fully differential particle level that can be subjected 
to detector level data.
This matching has been fully automated for NLO QCD calculations. 
At NLO EW accuracy only selected process specific solutions exist 
\cite{Bernaciak:2012hj,Barze:2012tt,Barze:2013fru,Muck:2016pko,Granata:2017iod}. 
As all parton showers incorporate a joint QCD+QED parton evolution,
general matching procedures, which are still lacking at the moment, will become 
available in the near future. 
This will enable precise particle level predictions that can be subjected 
to detector simulations for highly realistic and detailed studies.

Additionally, first solutions exist to incorporate approximate electroweak 
corrections in multijet merged calculations 
\cite{Chiesa:2013yma,Kallweit:2015dum,Gutschow:2018tuk}. 
In these approximations, the universal nature of EW corrections in 
the high energy limit, where they are dominated by Sudakov-type 
logarithms of virtual origin, is exploited. 
Thus, these methods will form the cornerstone of precise particle-level
predictions at large transverse momenta, which are at the 
basis of the increased reach of both the HL-- and HE--LHC 
new physics search program.

\subsubsection*{Weak showers}
\label{EWshower}
All parton showers publicly available in the major Monte-Carlo 
event generators \Herwig, \Pythia and \Sherpa contain both QCD and 
QED splitting functions to numerically resum the respective 
logarithms at (N)LL accuracy. First steps towards parton showers 
incorporating also weak effects in their splitting functions have 
been taken recently \cite{Christiansen:2014kba,Krauss:2014yaa}. 
The now complete electroweak splitting functions suffer from their 
strong dependence on the helicity of the propagating parton. 
These parton showers, however, operate in the spin-averaged 
approximation, neglecting all spin-correlations. 
The current effort to understand the full spin dependence of the 
electroweak part of the evolution of partons 
\cite{Bauer:2017isx,Chen:2016wkt} in analytic resummations is 
complemented by efforts to keep the full colour and spin structure, 
including non-diagonal parts of the (now matrix-valued) evolution 
equations, in the parton shower community. 
In time for the High Luminosity Upgrade fully spin-dependent parton 
evolution will then be incorporated in fully differential parton 
shower resummations that can then produce accurate predictions for 
the emission probabilities of secondary weakly interacting particles 
and gauge bosons.

\subsection[Monte Carlo generators]{Monte Carlo generators\footnote{Contributed by F. Maltoni, M. Sch\"onherr and P. Nason.}}

The complexity of the final states, together with the complexity of the
detectors that analyse them, are such that a full simulation of an
event, yielding a realistic multi-particle final state distribution,
is an indispensable theoretical tool for the physics of high-energy
hadron colliders. Driven by the needs of the Tevatron and LHC,
the physics of Monte Carlo (MC) generators has seen steady progress from its
inception to the present, and is, at the moment, a field in active
development. The current LHC physics program, as well as the requirements
for its HL-LHC and eventually its HE-LHC phases,
has
evidenced several areas of development that need to be addressed by theorists.
These are mainly driven by the quest for higher precision and accuracy,
but also by practical issues, such as the need for generating
very large samples for the most abundant LHC processes, and for
the efficient handling of the variations of the input parameters
needed in order to study uncertainties.

Much progress in this field takes place within the main collaborations
that maintain the widely used general purpose Monte Carlo generators,
i.e.~\mbox{\textsc{Herwig}}\xspace~\cite{Corcella:2000bw,Bahr:2008pv,Bellm:2015jjp}, ~\mbox{\textsc{Pythia}}\xspace~\cite{Sjostrand:2006za,Sjostrand:2014zea}
and ~\mbox{\textsc{Sherpa}}\xspace~\cite{Gleisberg:2008ta}, but there is also a large theoretical
community that works on more specialised aspects of Monte Carlo
generators,
such as formal/theoretical advances to improve the resummation accuracy, and
to improve the fixed-order accuracy in
the generation of the primary event and of the hardest radiations accompanying it.

In spite of the several
challenges ahead of us, considering the evolution of the field in the last twenty years, it can be anticipated that
considerable progress will be made from now up to the beginning
 (around 2025) and in the following ten-fifteen years 
of the high luminosity program. This progress will take place in
particularly favourable conditions, as the running of the LHC and the data accumulated will provide continuous feedback to the theoretical work in the field.

It can be can anticipated major developments in the following directions:
precision for inclusive observables, logarithmic accuracy,
technical improvements for fast and efficient generation of events,
and improvements in the modeling of hadronization and underlying event.

\subsubsection*{Precision for inclusive observables}
In this context, let us generically refer to ``precision" as a measure of the accuracy of the result as well as of the size of the left-over uncertainties that can be achieved in the computation
of inclusive quantities, i.e. those that can be computed directly
in fixed-order calculations. Fixed-order calculation have always been, and
are now, ahead of the precision that Monte Carlo generators can provide
for inclusive observables. Since their wide use started, and up to about
twenty years ago, shower MC's had typically leading order precision
for inclusive observables, while the state of the art for
fixed order computations was at the Next-to-Leading-Order (NLO) level. Thanks to the introduction
of general methods for interfacing shower Monte Carlo to fixed-order
NLO calculations, like {\sc aMC@NLO} \cite{Frixione:2002ik},
\mbox{\textsc{POWHEG}}\xspace \cite{Nason:2004rx},
and more recently the \mbox{\textsc{Krk-NLO}}\xspace method \cite{Jadach:2015mza},
the state of the art for shower MC's precision has reached the NLO level.
On the other hand,  progress in fixed-order computations, including the evaluation of  two-loop amplitudes and the development of several subtraction methods, allowed NNLO calculations to become available for a rather large set of processes. It is therefore natural to wonder whether
\emph{general methods for interfacing Shower generators to NNLO calculation
will be available at the start of the High Luminosity program.}
NNLO-PS methods have already appeared for relatively simple processes,
typically in the production of massive colourless final
states~\cite{Re:2018vac,Hamilton:2013fea,Hoche:2014dla,Alioli:2015toa,Alioli:2013hqa}. 
However, the methods used so far do not
seem to have the generality needed to handle processes of increasing complexity,
and it is very likely that new theoretical breakthroughs will be needed.

Achieving NNLO accuracy for a given final state, for example for Higgs production,
implies also the NLO accuracy for the the same final state in association
with a jet, i.e. the $HJ$ process in the Higgs example. In practical applications,
the less ambitious goal of having NLO accuracy for inclusive
result, and also achieve NLO accuracy
for the final states that also include associated production of jets,
thus achieving an extension of the  CKKW~\cite{Catani:2001cc} method to NLO order,
can be extremely useful.

The availability of automated NLO corrections for arbitrary processes including a
relatively large number of associated jets has paved the way to important developments in this direction. Several proposals
to merge samples with different jet multiplicity computed at the NLO, usually called ``NLO-PS matrix-element merging'',
have been put forward. These are the \mbox{\textsc{FxFx}}\xspace  method~\cite{Frederix:2012ps},
implemented in the {\sc aMC@NLO} framework; the \mbox{\textsc{UNLOPS}}\xspace method~\cite{Lonnblad:2012ix},
implemented in \mbox{\textsc{Pythia}}\xspace and the \mbox{\textsc{MePS\@NLO}}\xspace method~\cite{Hoeche:2012yf},
implemented in \mbox{\textsc{Sherpa}}\xspace. All methods introduce a separation scale that defines
the jet multiplicity
for a given event, and allows to generate inclusive samples out of non-overlapping samples
with different jet multiplicity.  Whether
these procedures really achieve NLO accuracy for observables involving
different jet multiplicity also when generic (i.e. different from those
used at the generation level) separation scales are chosen, is a delicate
question, which is still a matter of debate. Alternative merging
procedures, that consider more carefully the problems that may
arise at the boundary of the merging regions and also aim at improving the
resummation accuracy , have been proposed
in the \mbox{\textsc{GENEVA}}\xspace approach~\cite{Alioli:2012fc}, and presently applied to
Drell-Yan production~\cite{Alioli:2015toa,Alioli:2016wqt}.
The goal of achieving NLO accuracy for different jet multiplicity
has also been achieved without the use of merging with the so called \mbox{\textsc{MiNLO}}\xspace procedure~\cite{Hamilton:2012rf,Frederix:2015fyz}. 

While NLO-PS generators for
standard QCD processes can be obtained with a fairly high level
of automation, there are processes that require particular attention,
typically the loop induced ones. An example of one such process is Higgs-pair
production, that has been implemented first in {\sc aMC@NLO}~\cite{Maltoni:2014eza} using an approximation for the yet unknown two-loop contributions and then in \mbox{\textsc{POWHEG}}\xspace and {\sc aMC@NLO}~\cite{Heinrich:2017kxx}
as soon as the results of the two-loop computation has become available.~\cite{Borowka:2016ehy,Borowka:2016ypz}. There are several other $gg$ loop-induced processes for which a full NLO+PS implementation is still missing which, thanks to the quick developments in computation of two-loop amplitudes, are expected to become available in the coming years.

Another important direction where there has been considerable progress recently is the automation of the  computation of EW corrections~\cite{Actis:2012qn,Kallweit:2014xda,Schonherr:2017qcj,Frederix:2018nkq}
to the point that fixed-order NLO QCD and EW corrections are readily
available for virtually all processes of interest.
Details can be found in Section \ref{sec:theory_tools_ew}.
An general interface of these calculations to shower
generators  that correctly account for QED radiation for these computations, however, is not yet available.
The problem in this case is
the consistent handling of photon radiation, that can arise
both from the shower and from the fixed-order calculation. These pose
new problems compared to the production of coloured partons, where
the presence of individual partons cannot be required in the final state, and thus
showers develop inclusively generating jets from partons.
Photons, on the other hand, can be explicitly detected in the
final state, and an NLO+PS scheme should take care of handling
both shower generated photons and those originating in the NLO
calculation in a consistent way, in order to give a reliable description
of both collinear photons embedded in jets and highly energetic
isolated ones. A scheme for achieving this in the Drell-Yan case
has been presented in Ref.~\cite{Barze:2013fru,Barze:2012tt} in
the context of the \mbox{\textsc{POWHEG}}\xspace method. A scheme using fragmentation functions has been introduced in Ref.~\cite{Frederix:2018nkq}.

Finally, it is to be noted that the progress achieved recently  to account for intermediate resonant states in the NLO+PS context~\cite{Jezo:2015aia,Jezo:2016ujg,Frederix:2016rdc}
will likely
be essential in the framework of electro-weak corrections. In this case, weak vector bosons are part of the electroweak corrections and their presence entail a correct handling of their decays also in presence of extra QED radiation. It is expected that interfacing complete NLO EW-QCD calculations with a shower approach (QED+QCD) will become standard procedure by the beginning of HL-LHC.

\subsubsection*{Accuracy in resummation}
As current state of the art, shower generators rely upon
the first order Altarelli-Parisi splitting kernels, together
with some appropriate scheme to handle soft emissions,
either by angular ordering in parton shower cascades
or using dipole shower algorithms.
Several studies have appeared recently aiming at improving
parton showers by increasing the accuracy of specific
ingredients, either by developing novel shower schemes that
remain within the standard parton or dipole branching, such as
\mbox{\textsc{Dire}}\xspace~\cite{Hoche:2015sya} and {\textsc Vincia} \cite{Giele:2007di,Fischer:2016vfv};
by going beyond the typical probabilistic cascade of the shower
algorithms and handling directly the quantum density matrix~\cite{Nagy:2017dxh};
and by incorporating higher order splitting functions
\cite{Jadach:2015mza,Jadach:2010aa,Hoche:2017hno,Hoche:2017iem,Dulat:2018vuy}.

While fixed order improvements in shower MC
generators have the clear goal of reaching the same
fixed order accuracy as the corresponding computations
for inclusive observables,  it is less straightforward to quantify
how improvements in the shower algorithms impact the precision
of the description of observables
that require resummation. In a recent study~\cite{Dasgupta:2018nvj},
some criteria were proposed in order to address this problem. 
In particular, two criteria were examined:
the first refers to the ability of a shower algorithm
to correctly reproduce the singularity structure of
$n$-parton matrix elements, while the second measures
the level of accuracy of a shower algorithm in the computation
of a general class of observables that require resummation.
It was found that there are regions where commonly used
shower algorithms fail to reproduce the correct singularity
structure of the matrix elements, and that this affects
the logarithmic resummation accuracy of the shower already 
in the leading term, yet at subleading number of colours, and in
the next-to-leading term at leading colour.

Thus, the current trend of research moves along parallel directions,
not only by seeking improvements in the shower algorithms in particular
areas, but also by critical examination of the shower formalism
in an attempt to qualify their accuracy in a more solid way.

\subsubsection*{Technical improvements}

The pressing requirements of the LHC physics program have already had
an impact in driving technical improvements in Monte Carlo
generators. In particular, the need to study uncertainties,
corresponding to a large set of combination of parameter variations
when generating a sample, often leading to several hundreds
variations, has led to the development of procedures to implement the
variation of parameters by reweighting the same event, rather than
generating independent samples. Besides the obvious simplification of
having to deal with a single event sample, this has also the advantage
that the effects of variations of the input parameters are affected by
smaller theoretical errors, since they all apply to the same generated
event. A method for reweigthing the full shower development was
presented in Ref.~\cite{Bellm:2016voq} and implemented in \mbox{\textsc{Herwig}}\xspace
in Ref.~\cite{Bellm:2016rhh}. A similar method was presented
in~\cite{Mrenna:2016sih} for \mbox{\textsc{Pythia}}\xspace, and in Ref. \cite{Bothmann:2016nao} for
\mbox{\textsc{Sherpa}}\xspace.  Reweighting techniques to evaluate uncertainties as well as for other applications are  available in {\sc MadGraph5\_aMC@NLO} \cite{Alwall:2014hca,Mattelaer:2016gcx} and in \mbox{\textsc{POWHEG}}\xspace.

For certain common Standard Model processes, a large statistics is often
required, and is especially needed to populate the kinematic tails
at large transverse momenta. The most advanced generators usually
suffer from poor performance, especially in such areas of the phase space, and thus the need for more accurate tools must be balanced with the practical needs for large samples.
These problems will need to be addressed on a case by case basis,
depending upon the process that is been considered, and the
specific purpose that a generator for that process should serve.
The presence of negative weights, for example, should be minimised for
generators that must produce large samples to be fed through
detector simulators. The sampling of suppressed tails of phase
space, on the other hand, may be easily increased by suitable
bias functions. It is also apparent that attention should be given to
whether new computer architectures may be advantageously explored
for Monte Carlo generators, such as MPIs and GPU architectures,
and that new software techniques
making use of Boosted Decision Trees or Deep Neural Networks
may provide advantages over traditional techniques of Monte Carlo
integration and phase space generation~\cite{Bendavid:2017zhk}.

\subsubsection*{Hadronization and underlying event}

A recent fascinating direction in parton shower MC's is towards
establishing a unified picture in the description of multi-parton
dynamics in pp, pA, and AA collisions~\cite{Sjostrand:2018xcd}.
Traditionally, pp collisions have been described through the picture
of double-, single- and non-diffractive interactions of partons in a
vacuum in pp collisions. AA collisions, on the other hand,
are typically described in terms of the dynamics of a quark
gluon plasma, with a formalism more related to hydrodynamics than
particle physics.
A series of observations in
high-multiplicity pp events at the LHC, however, have exposed
remarkable similarities and features in common with those observed in
pA and AA collisions, at least with respect to flavour composition and
flow. The question therefore arises whether a new state of matter, the
quark gluon plasma, is actually formed in high-multiplicity pp events
and how this could be tested quantitatively. Efforts and new ideas
have recently emerged towards having a unified MC description of such
events. This has started with a simple stacking of (soft and hard) pp
events \cite{Bellm:2018sjt}. A recent proposal,
Angantyr\cite{Bierlich:2018xfw}, has been inspired by the old Fritiof
model~\cite{Andersson:1986gw}
and the notion of wounded nucleons. While more elaborated than a
stacking approach, it does not yet feature a description of collective
effects.  In the coming years, progress will be achieved by first
identifying the experimental features that are genuine signatures of
the formation of a quark gluon plasma, and those which could be
associated to other effects. Alternative explanations would likely
also be of a collective character, yet without requiring a phase
transition.

The intense ongoing theoretical and experimental work in this
framework is likely to lead to new breakthrough in the modeling of the
hadronization phase and the underlying event before the beginning of
the HL-LHC running.

In the description of the underlying event in $pp$ collisions, a key role is
played by multi-parton interactions (MPI, see Sec.~\ref{sec:UEMPI}).
There has been recent progress in the theoretical understanding
of double parton scattering that has been summarised in Sec.~\ref{sec:DPS}.
There it is also shown that at the HL-LHC it may be possible to find evidence
of correlations in double parton interactions. This opens the possibility
of constructing improved models of MPI in MC generators, to be eventually
refined in the first few years of running of the HL-LHC.


\subsection[PDF calculations and tools]{PDF calculations and tools%
\footnote{Contributed by L. Harland-Lang, J. Gao and J. Rojo.}}

\def\smallfrac#1#2{\hbox{$\frac{#1}{#2}$}}
\newcommand{\be}{\begin{equation}}
\newcommand{\ee}{\end{equation}}
\newcommand{\bea}{\begin{eqnarray}}
\newcommand{\eea}{\end{eqnarray}}
\newcommand{\bi}{\begin{itemize}}
\newcommand{\ei}{\end{itemize}}
\newcommand{\ben}{\begin{enumerate}}
\newcommand{\een}{\end{enumerate}}
\newcommand{\la}{\left\langle}
\newcommand{\ra}{\right\rangle}
\newcommand{\lc}{\left[}
\newcommand{\rc}{\right]}
\newcommand{\lp}{\left(}
\newcommand{\rp}{\right)}
\newcommand{\as}{\alpha_s}
\newcommand{\aq}{\alpha_s\left( Q^2 \right)}
\newcommand{\amz}{\alpha_s\left( M_Z^2 \right)}
\newcommand{\aqq}{\alpha_s \left( Q^2_0 \right)}
\newcommand{\aqz}{\alpha_s \left( Q^2_0 \right)}
\newcommand{\Ord}{\mathcal{O}}
\newcommand{\MSbar}{\overline{\text{MS}}}
\def\toinf#1{\mathrel{\mathop{\sim}\limits_{\scriptscriptstyle
{#1\rightarrow\infty }}}}
\def\tozero#1{\mathrel{\mathop{\sim}\limits_{\scriptscriptstyle
{#1\rightarrow0 }}}}
\def\toone#1{\mathrel{\mathop{\sim}\limits_{\scriptscriptstyle
{#1\rightarrow1 }}}}
\def\frac#1#2{{{#1}\over {#2}}}
\def\gsim{\gtrsim}
\def\lsim{\lesssim}    
\newcommand{\mrexp}{\mathrm{exp}}
\newcommand{\dat}{\mathrm{dat}}
\newcommand{\one}{\mathrm{(1)}}
\newcommand{\two}{\mathrm{(2)}}
\newcommand{\art}{\mathrm{art}} 
\newcommand{\rep}{\mathrm{rep}}
\newcommand{\net}{\mathrm{net}}
\newcommand{\stopp}{\mathrm{stop}}
\newcommand{\sys}{\mathrm{sys}}
\newcommand{\stat}{\mathrm{stat}}
\newcommand{\diag}{\mathrm{diag}}
\newcommand{\pdf}{\mathrm{pdf}}
\newcommand{\tot}{\mathrm{tot}}
\newcommand{\minn}{\mathrm{min}}
\newcommand{\mut}{\mathrm{mut}}
\newcommand{\partt}{\mathrm{part}}
\newcommand{\dof}{\mathrm{dof}}
\newcommand{\NS}{\mathrm{NS}}
\newcommand{\cov}{\mathrm{cov}}
\newcommand{\gen}{\mathrm{gen}}
\newcommand{\cut}{\mathrm{cut}}
\newcommand{\parr}{\mathrm{par}}
\newcommand{\val}{\mathrm{val}}
\newcommand{\tr}{\mathrm{tr}}
\newcommand{\checkk}{\mathrm{check}}
\newcommand{\reff}{\mathrm{ref}}
\newcommand{\Mll}{M_{ll}}
\newcommand{\extra}{\mathrm{extra}}
\newcommand{\draft}[1]{}
\newcommand{\muf}{\mu_\text{F}}
\newcommand{\mur}{\mu_\text{R}}

\def\beq{\begin{equation}}  
\def\eeq{\end{equation}}

\def\({\left(}
\def\){\right)}
\def\[{\left[}
\def\]{\right]}
\let\originalleft\left
\let\originalright\right
\renewcommand{\left}{\mathopen{}\mathclose\bgroup\originalleft}
\renewcommand{\right}{\aftergroup\egroup\originalright}



\newcommand{\tmop}[1]{\ensuremath{\operatorname{#1}}}
\newcommand{\tmtextit}[1]{{\itshape{#1}}}
\newcommand{\tmtextrm}[1]{{\rmfamily{#1}}}
\newcommand{\tmtexttt}[1]{{\ttfamily{#1}}}


At the HL-LHC, a precise knowledge of the quark and gluon
structure of the proton will be essential for many analyses.
These include the profiling of the Higgs boson sector~\cite{deFlorian:2016spz},
direct searches for new heavy BSM states~\cite{Beenakker:2015rna}, indirect BSM searches
by e.g. means of the SMEFT~\cite{Alioli:2017jdo}, and
the measurement of fundamental SM parameters such as the $W$ boson mass~\cite{Aaboud:2017svj},
the Weinberg mixing angle~\cite{Aaltonen:2018dxj} or the strong coupling
constant~\cite{Ball:2018iqk} and its running.

This section gives a brief review the PDF tools that will be used
in this Report for the studies of the SM chapter. Those aspects of modern PDF fits
that are more relevant for studies at the HL-LHC will be also highlighted.
The end of this section will provide some
 perspectives about the role of PDFs
  at the HE-LHC.
It must be stressed that this document is not intended to be a review of
recent developments on PDFs, and the reader is referred
to~\cite{Gao:2017yyd,Rojo:2015acz,Forte:2013wc} and reference therein,
for further details in this sense.

The studies presented in this Report will be based mostly
on the PDF4LHC15 set~\cite{Butterworth:2015oua},
constructed from the statistical
combination and subsequent
reduction~\cite{Gao:2013bia,Carrazza:2015hva,Carrazza:2015aoa} of the
CT14~\cite{Dulat:2015mca}, MMHT14~\cite{Harland-Lang:2014zoa},
and NNPDF3.0~\cite{Ball:2014uwa}
global analyses.
The PDF4LHC15 set is interfaced to matrix-elements calculators and Monte Carlo
shower programs by means of the {\sc LHAPDF6} package~\cite{Buckley:2014ana}.

\begin{figure}
  \begin{center}
    \includegraphics[scale=0.42]{\main/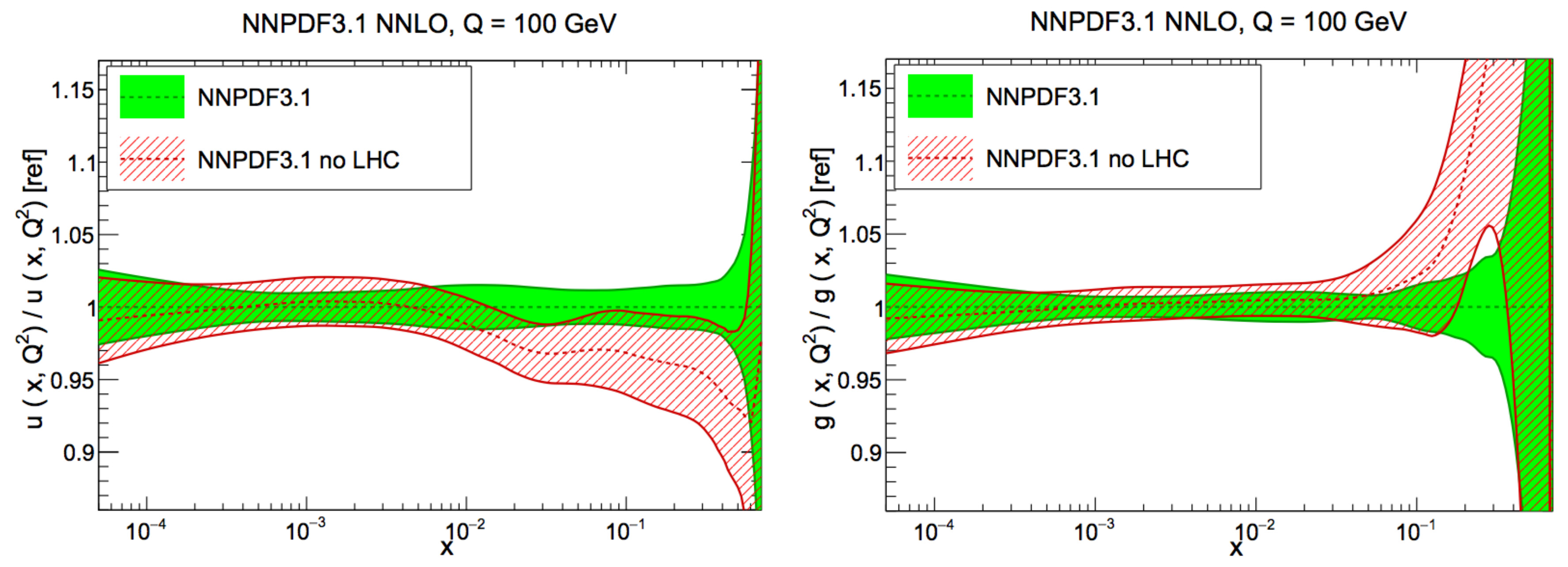}
    \caption{\small Comparison of the NNPDF3.1NNLO fits with and without
      LHC data, normalized to the central value of the former
      at $Q=100$ GeV.
      The up quark (left) and the gluon (right plot) are shown.
      The bands indicate the 68\% confidence level PDF uncertainty.
     \label{fig:impact_LHC_data} }
  \end{center}
\end{figure}

\subsubsection*{Quantifying the impact of LHC measurements.}
In recent years, one of the main developments in global PDF fits has been the increasingly significant role played by LHC processes in providing stringent PDF constraints. The combination of high precision LHC data with state-of-the art NNLO theory calculations for such hadronic processes as top-quark pair production~\cite{Czakon:2016olj},
the transverse momentum spectrum of $Z$ bosons~\cite{Boughezal:2017nla}, 
direct photon production~\cite{Campbell:2018wfu}, and inclusive jet production~\cite{Currie:2016bfm} is having an important impact on precision PDF fits.
To illustrate this, Fig.~\ref{fig:impact_LHC_data}
compares the recent NNPDF3.1 fit~\cite{Ball:2017nwa}
with and without the LHC data at $Q=100$ GeV for the up quark
and gluon PDFs.
The marked impact of the LHC data for $x\gsim 0.005$ can be
observed both for central values and for the PDF
uncertainties.
It is of particular note that only Run-1 data has been included
in these fits. Thus, it is clear
that the addition of data from Run-2 and -3 first and then 
from the HL-LHC, for which the precision and reach will be greatly increased, 
should lead to further improvements in the determination of the proton structure.
A subsequent section of this report will quantify the impact of HL-LHC
measurements, demonstrating that a significant reduction can be expected and providing a public PDF set including the expected
constraints from the final HL-LHC dataset.

\subsubsection*{Fast interfaces to (N)NLO calculations}
To avoid the direct evaluation of the lengthy (N)NLO hadronic cross sections during the fit itself, a method of fast interfaces is generally applied, whereby the CPU time intensive part of the higher--order calculation is pre--computed once using a complete interpolation basis for the input PDFs. For a number of years, the {\sc APPLgrid}~\cite{Carli:2010rw}  and {\sc Fastnlo}~\cite{Wobisch:2011ij} tools have been available for a range of NLO processes.
The former is interfaced to the {\sc MCFM}~\cite{Boughezal:2016wmq}
and {\sc NLOjet++}~\cite{Nagy:2001fj} programs. More recently, the {\sc aMCfast} interface~\cite{amcfast} to
{\sc MadGraph5\_aMC@NLO}~\cite{Alwall:2014hca} has also been developed. Results within the {\sc Fastnlo}  framework for differential top quark production at NNLO are already available~\cite{Czakon:2016dgf,Czakon:2017dip}, while work is ongoing within the {\sc APPLfast} project to extend the {\sc FastNLO} and {\sc APPLgrid} technology to NNLO. This will be interfaced by default to the {\sc NNLOJET} program~\cite{Currie:2016bfm}, but will be reusable for other theory codes. Thus, for future PDF fits, relevant to HL and HE-LHC running, fast interface implementations of NNLO theory calculations are expected to be the standard.

\subsubsection*{Theoretical uncertainties}
Given the high precision expected for HL-LHC data, it will be crucial to
include all sources of experimental, methodological,
and theoretical uncertainties associated with PDFs in order
to ensure robust predictions.
An important issue in this context is to estimate the
theoretical uncertainties in PDFs due to missing
higher orders (MHOU) in the perturbative expansion for the
theory prediction~\cite{Bagnaschi:2014wea}, which
are so far ignored in all global fits.
There is by now some evidence that MHOUs can be comparable,
if not larger, than the nominal PDF uncertainties based
on the propagation of experimental and methodological
uncertainties.
In this context, HL-LHC projections should ideally be based
on PDFs that consistently account for MHOUs in addition to
other sources of uncertainties.

\begin{figure}
  \begin{center}
    \includegraphics[scale=0.4]{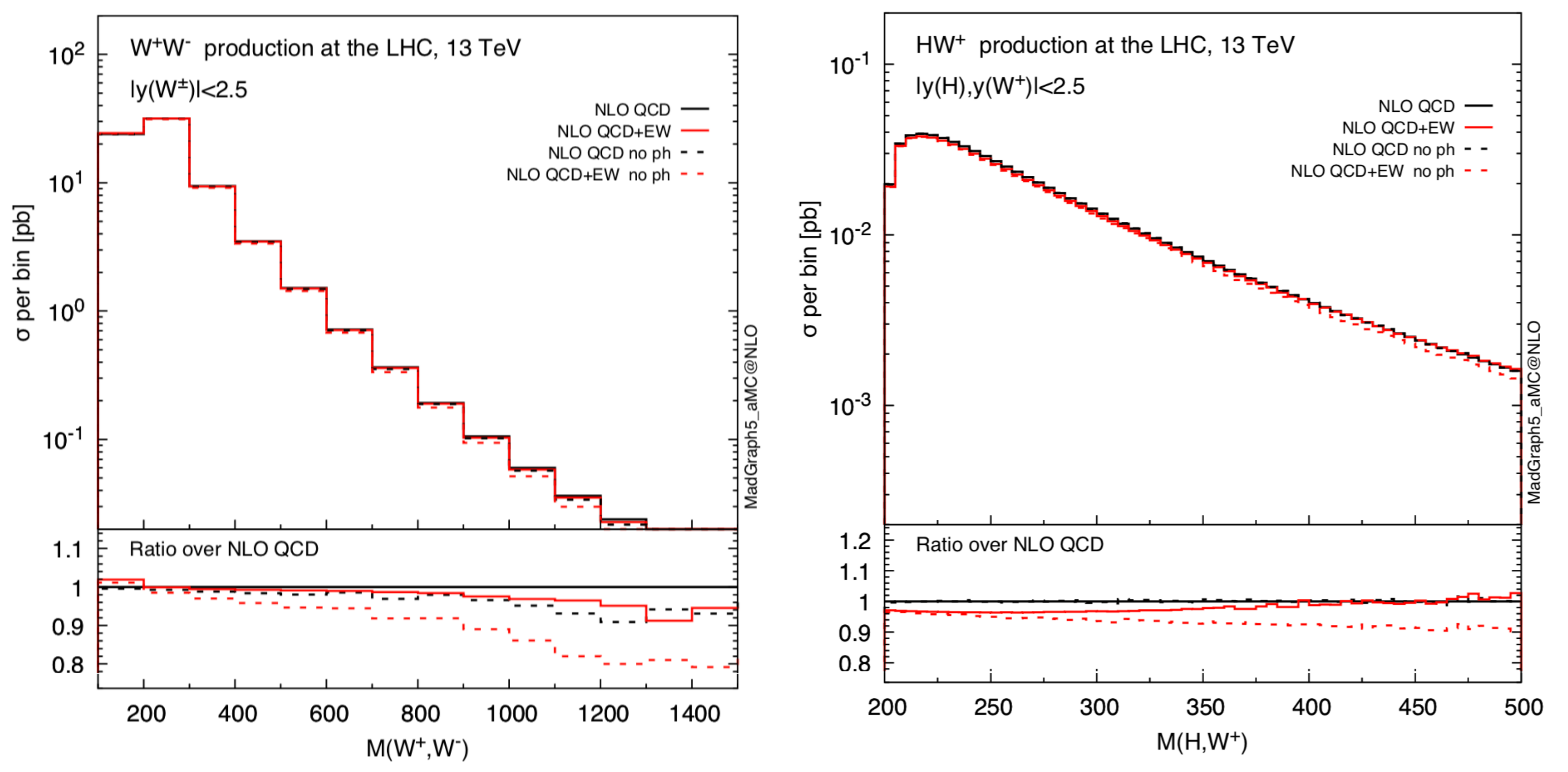}
    \caption{
      Photon-initiated contributions partially cancel 
      the NLO EW corrections in the TeV region, as shown
      for the case of $W^+W^-$ production (left)
      and $hW^+$ production (right plot) at 13 TeV.
     \label{fig:theory_PI_PDFs} }
  \end{center}
\end{figure}

To keep such uncertainties to a minimum, global PDF fits will need to include higher-order
perturbative corrections either at fixed-order or at all-orders
using some form of resummation.
In the former case, encouraging recent progress with N${}^3$LO
splitting functions~\cite{Moch:2017uml} suggest that an (approximate)
N${}^3$LO fit might be within the reach of the HL-LHC era, to match the precision
of partonic cross-sections for processes such as Higgs production in gluon
fusion~\cite{Anastasiou:2016cez,Dulat:2017prg}.
In the latter case, one can use threshold (BFKL) resummation~\cite{Bonvini:2015ira,Ball:2017otu}
to reduce theoretical uncertainties at the large-$x$ (small-$x$)
kinematic regions.
Indeed, several state-of-the-art predictions for LHC processes
include threshold resummation, such as for example
top quark pair production~\cite{Czakon:2018nun}.

\subsubsection*{Electroweak effects and photon-initiated contributions}
The enhanced coverage of the TeV region  at the HL-LHC requires not
only higher-order QCD corrections to be accounted for, but also electroweak ones,
which can be enhanced due to Sudakov logarithms~\cite{Mishra:2013una}.
In the context of PDF studies, there are two main considerations
to take into account.
First of all, exploiting the constraints from the HL-LHC measurements
for PDF fits will require systematically accounting for
NLO EW corrections.
Secondly, PDFs with QED effects and thus with photon-initiated
contributions should become the baseline.
It has now been demonstrated~\cite{Manohar:2016nzj,Manohar:2017eqh} (see Ref.~\cite{Bertone:2017bme} for a recent implementation within a global fit) that the photon PDF can be determined
with percent--level uncertainties and carry up to $\simeq 0.5\%$ of
the proton's momentum.
For certain processes, in the TeV region the photon-initiated
contributions can have a comparable size but opposite sign to the NLO
virtual EW corrections, and therefore it is crucial to include
both consistently.
This is illustrated in Fig.~\ref{fig:theory_PI_PDFs} in the
specific cases of  $W^+W^-$ and $hW^+$
production at 13 TeV.
A more detailed discussion of EW corrections
for HL-LHC studies is presented later in the report.

\subsubsection*{Perspectives at the High Energy LHC}
At a centre-of-mass energy of $\sqrt{s}=27$ GeV, a number
of novel phenomena are expected to arise, due to the increased
phase space available.
Much of this has already been discussed in the context
of the Future Circular Collider (FCC) studies
at $\sqrt{s}=100$ TeV~\cite{Mangano:2016jyj,Rojo:2016kwu}.
To begin with, as
illustrated in Fig.~\ref{fig:HELHC},
when going to higher energies one becomes
more sensitive to the small-$x$ region, even for electroweak-scale
observables, implying that BFKL resummation effects could become relevant.

Indeed, for $M_X\simeq 100$ GeV the NNPDF3.1sx results~\cite{Ball:2017otu} at NNLO and at NNLO+NLL$x$ for the $gg$ luminosities are found to differ at the $\simeq 5\%$ level
at the HE--LHC. In Ref.~\cite{Bonvini:2018iwt} a detailed study of SM Higgs boson production via gluon fusion has been performed, consistently including BFKL resummation in the PDFs (see Ref.~\cite{Ball:2017otu}) and coefficient functions. The role of the former is found to be dominant, and while the impact is mild at the LHC, for the HE--LHC a larger increase is seen relative to the N${}^3$LO result with fixed--order NNLO PDFs, that lies outside the fixed--order PDF uncertainty bands, see Fig.~\ref{fig:HELHC} (right). This highlights the important role such effects will play at high energies and precision.

\begin{figure}
  \begin{center}
    \includegraphics[scale=0.55]{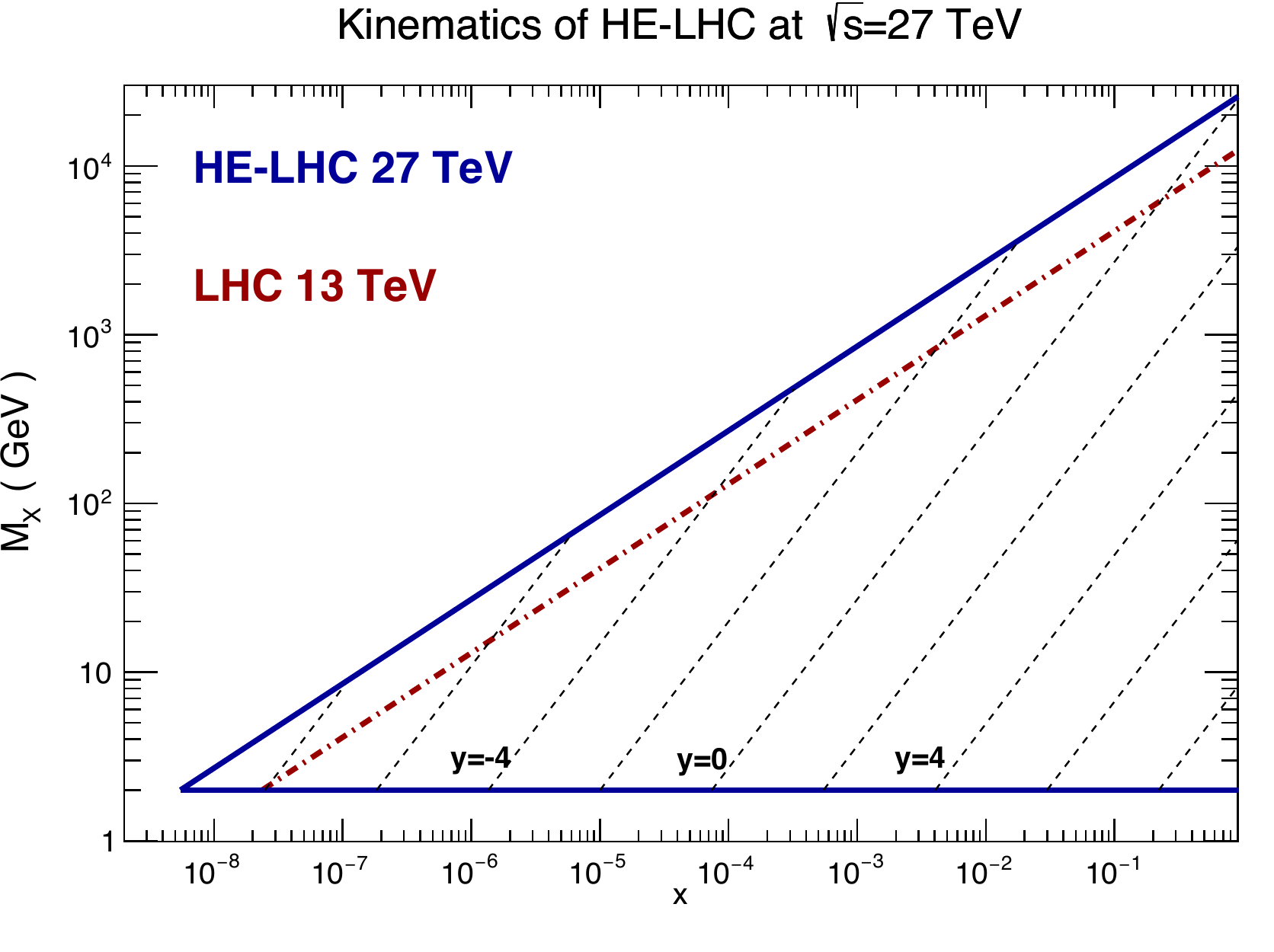}\qquad
    \includegraphics[scale=1.2]{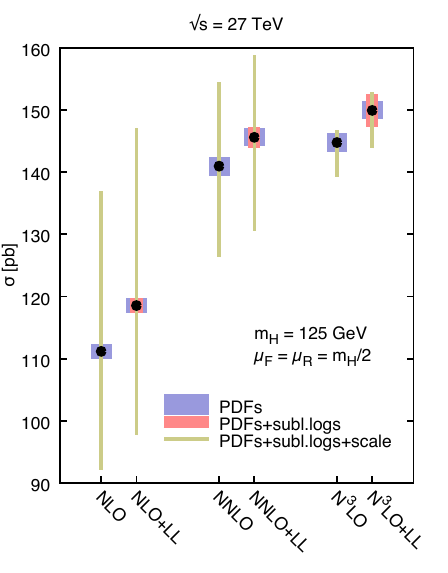}
    \caption{\small Left: kinematic coverage of the HE-LHC at
      $\sqrt{s}=27$ TeV compared to 13 TeV.
      Right: the Higgs cross section at the HE-LHC,
      for different orders and with/without (LL)
      low--$x$ resummation, and with uncertainty bands from PDF, subleading logarithms, and scale variations~\cite{Bonvini:2018iwt}.
     \label{fig:HELHC} }
  \end{center}
\end{figure}

Another effect that might become relevant at the HE-LHC are
the electroweak PDFs~\cite{Fornal:2018znf,Bauer:2017isx}
from the resummation of large
collinear logarithms of the masses of the $W$ and $Z$ bosons, which become
effectively massless at high energies. Related to this is the top quark PDF, which can be (and is) straightforwardly generated within the standard PDF framework. When included with a suitably matched flavour scheme, this may provide a more accurate description of processes involving top quarks~\cite{Han:2014nja,Bertone:2017djs}.
In addition, at $\sqrt{s}=27$ TeV, knowledge of the small-$x$ PDFs will
be also required for the modeling of soft and semi-hard
QCD dynamics in Monte Carlo event generators~\cite{Skands:2014pea,Mangano:2016jyj}.
In turn, an improved understanding of the PDFs in the ultra-low-$x$
regime will have implications in high-energy astrophysics,
for processes such as cosmic ray detection and
for signal and background event rates in neutrino telescopes~\cite{Gauld:2016kpd}.

\subsection[Effective Field Theory calculations and tools]{Effective Field Theory calculations and tools\footnote{Contributed by E. Vryonidou.}}

\subsubsection*{State of the art}
The success of the Standard Model Effective Theory (SMEFT) programme
at the LHC relies on the availability of public tools for calculations in this framework. Among the most important of these are Monte Carlo (MC) tools for providing realistic predictions for collider processes both for phenomenological studies and experimental analyses. 
In this respect, significant efforts are being made to implement the effects of dimension-6 operators in MC event generators. 
Concerning Leading Order (LO) predictions, recent progress includes \mbox{\textsc{SMEFTsim}}\xspace, a complete implementation of the dimension-6 operators in the Warsaw basis~\cite{Brivio:2017btx}, an alternative implementation of the Warsaw basis in the $R_{\xi}$ gauge~\cite{Dedes:2017zog}, \mbox{\textsc{dim6top}}\xspace, an implementation of top quark operators under various flavour assumptions~\cite{AguilarSaavedra:2018nen} and the Higgs Effective Lagrangian (HEL)~\cite{Alloul:2013naa} implementation of SILH basis operators. Complementary to SMEFT implementations, there also exist several models of anomalous couplings such as the Higgs Characterisation~\cite{Maltoni:2013sma,Demartin:2014fia,Demartin:2015uha} and BSM Characterisation models~\cite{Falkowski:2015wza}. These models are all made available in the Universal  \mbox{\textsc{FeynRules}}\xspace Output (UFO) format that can be imported into general purpose Monte Carlo tools, such as {\sc MadGraph5\_aMC@NLO} or \mbox{\textsc{Sherpa}}\xspace, to generate events and interface them to parton shower generators (PS). A powerful aspect of this workflow is that, once implemented, the model is generic enough to enable event generation for any desired process.

Implementations of particular processes in the presence of dimension-6 operators exist also in other frameworks. An example is the weak production of Higgs in association with a vector boson in \mbox{\textsc{POWHEG}}\xspace based on the NLO computation of~\cite{Mimasu:2015nqa}, the implementation of Higgs pair production in the EFT in \mbox{\textsc{Hpair}}\xspace (including approximate NLO corrections)~\cite{Grober:2017gut} and in \mbox{\textsc{Herwig}}\xspace ~\cite{Goertz:2014qta,Bellm:2015jjp}. Two well-known tools for calculating cross sections for Higgs production via gluon fusion including higher order QCD corrections, \mbox{\textsc{HiGlu}}\xspace~\cite{Spira:1995mt,Spira:1996if} and \mbox{\textsc{SusHi}}\xspace~\cite{Harlander:2016hcx}, can also include the effects of modified top and bottom quark Yukawas and the dimension-5 Higgs-gluon-gluon operator. The latter code also permits event generation at NLOQCD+PS accuracy via \mbox{\textsc{aMCSusHi}}\xspace~\cite{Mantler:2015vba} including modified top and bottom quark Yukawa couplings. 
For a variety of processes with electroweak and Higgs bosons in the final state (VBF H, W and Z production, weak boson pair production, vector-boson-scattering processes, triboson production) the \mbox{\textsc{VBFNLO}}\xspace program~\cite{Arnold:2008rz,Arnold:2011wj} provides NLO QCD corrections together with implementations of dimension-6 operators and, in the case of VBS and triboson production, dimension-8 operators.

There are also EFT-specific tools providing a number of useful interfaces and calculations. 
\\\mbox{\textsc{eHDECAY}}\xspace~\cite{Djouadi:1997yw,Contino:2014aaa} is a package for the calculation of Higgs boson branching fractions including SMEFT effects parametrised by SILH basis operators. The freedom of basis choice in the SMEFT implies that arbitrarily many equivalent descriptions of the model can be formulated. This has important consequences for the development of EFT tools given that any numerical implementation of EFT effects requires choosing a specific basis.  A SMEFT basis translation tool, \mbox{\textsc{Rosetta}}\xspace~\cite{Falkowski:2015wza}, can be used to numerically transform points in parameter space from one basis to another. It adopts the SLHA convention for model parameter specification and provides an interface to Monte Carlo event generation tools through the aforementioned BSMC model. Furthermore, additional interfaces exist to other programs such as \mbox{\textsc{eHDECAY}}\xspace, internal routines testing compatibility of Higgs signal-strength and EW precision measurements as well as providing predictions for di-Higgs production cross sections in the SMEFT. {\textsc Rosetta} provides SMEFT basis-independent access to these functionalities. A related tool is {\textsc DEFT}~\cite{Gripaios:2018zrz}, a python code that can check if a set of operators forms a basis, generate a basis and change between bases. A similar implementation based on \mbox{\textsc{FeynRules}}\xspace is \mbox{\textsc{AllYourBases}}\xspace 
, that performs the reduction of an arbitrary dimension-6 operator into the Warsaw basis operator set. Efforts are also underway to establish a common format for the Wilson coefficients~\cite{Aebischer:2017ugx}, which will allow interfacing various programs computing the matching and running of the operators such as \mbox{\textsc{Dim6Tools}}\xspace~\cite{Celis:2017hod}  and \mbox{\textsc{Wilson}}\xspace~\cite{Aebischer:2018bkb}.
A public fitting framework that can be used to obtain constraints on the EFT is \mbox{\textsc{HEPfit}}\xspace, which is based on the Bayesian Analysis Toolkit, and includes Higgs and electroweak precision observables.

\subsubsection*{Future Developments}
There is significant progress in computing NLO QCD corrections for the EFT, in both the top and Higgs sector~\cite{Degrande:2016dqg,Mimasu:2015nqa,Alioli:2018ljm,Franzosi:2015osa,Zhang:2016omx,Bylund:2016phk,Maltoni:2016yxb,Degrande:2018fog,deBeurs:2018pvs}. This progress, now on a process-by-process basis, will eventually lead to a full automation of QCD corrections for the SMEFT. As experimental measurements become increasingly systematics dominated, the importance of higher order calculations grows.
The complete implementation of dimension-6 operators at NLO, including some flavour symmetry assumptions, is in preparation. This implementation will enable the computation of NLO-QCD corrections to any tree-level process, bringing the Monte Carlo automation to the same level as the Standard Model. 

Another direction in which progress is expected over the coming years is the computation of weak corrections in the SMEFT. A small sample of computations has been done, e.g. weak corrections to Higgs production and decay due to top quark loops~\cite{Vryonidou:2018eyv} and due to modified trilinear Higgs coupling~\cite{Degrassi:2016wml,Bizon:2016wgr,DiVita:2017eyz} as well as Higgs and Z-boson decays~\cite{Hartmann:2015aia,Hartmann:2015oia,Hartmann:2016pil,Dawson:2018pyl,Dedes:2018seb,Dawson:2018liq}. Due to the behaviour of the Sudakov logarithms, weak corrections are typically important for high transverse momentum regions. Therefore at HE/HL-LHC their impact is expected to be enhanced. It can be expected that the recent progress on a process-by-process basis will eventually lead to the automation of the computation of weak loops in the EFT, as in the Standard Model. 

Finally progress is expected in linking tools which compute the running and mixing of the operators with Monte Carlo tools. This will allow the automatic computation of cross-sections and differential distributions taking into account the mixing and running of the operator coefficients.

\newpage


\section{Experimental environment at HL-LHC}
\label{sec:exp}

\subsection{Analysis methods, particle reconstruction and identification}

\newcommand{\delphes}[1]{\textsc{Delphes#1}}
\newcommand{\contributors}[1]{{\bf Contribution from: #1}}
\newcommand{\pt}{p_{T}}
\newcommand{\pythia}[1]{\textsc{Pythia#1}}
\newcommand{\geant}[1]{\textsc{Geant#1}}

Different approaches have been used by the experiments and in theoretical prospect studies, hereafter named projections, to assess the sensitivity in searching for new physics at the HL-LHC and HE-LHC.
For some of the projections, a mix of the approaches described below is used, in order to deliver the most realistic result.
The total integrated luminosity for the HL-LHC dataset is assumed to be $3000$~fb$^{-1}$ at a centre-of-mass energy of $14$~TeV. For HE-LHC studies the dataset is assumed to be $15$~ab$^{-1}$ at a centre-of-mass of $27$~TeV.
The effect of systematic uncertainties is taken into account based on the studies performed for the existing analyses and using common guidelines for projecting the expected improvements that are foreseen thanks to the large dataset and upgraded detectors, as described in Section~\ref{sec:methods:syst}.

{\bf Detailed-simulations} are used to assess the performance of reconstructed objects in the upgraded detectors and HL-LHC conditions, as described in Sections~\ref{sec:methods:perf},\ref{sec:methods:perf_LHCb}.
For some of the projections, such simulations are directly interfaced to different event generators, parton showering (PS) and hadronisation generators. Monte Carlo (MC) generated events are used for Standard Model (SM) and beyond-the-Standard-Model (BSM) processes, and are employed in the various projections to estimate the expected contributions of each process.

{\bf Extrapolations} of existing results rely on the existent statistical frameworks to estimate the expected sensitivity for the HL-LHC dataset.
The increased centre-of-mass energy and the performance of the upgraded detectors are taken into account for most of the extrapolations using scale factors on the individual processes contributing to the signal regions. Such scale factors are derived from the expected cross sections and from detailed simulation studies.

{\bf Fast-simulations} are employed for some of the projections in order to produce a large number of Monte Carlo events and estimate their reconstruction efficiency for the upgraded detectors. The upgraded CMS detector performance is taken into account encoding the expected performance of the upgraded detector in \delphes~\cite{deFavereau:2013fsa}, including the effects of pile-up interactions. Theoretical contributions use \delphes~\cite{deFavereau:2013fsa} with the commonly accepted HL-LHC card corresponding to the upgraded ATLAS and CMS detectors.

{\bf Parametric-simulations} are used for some of the projections to allow a full re-optimization of the analysis selections that profit from the larger available datasets.
Particle-level definitions are used for electrons, photons, muons, taus, jets and missing transverse momentum. These are constructed from stable particles of the MC event record with a lifetime larger than $0.3 \times 10^{-10}$~s within the observable pseudorapidity range. Jets are reconstructed using the anti-$k_{\rm T}$ algorithm~\cite{Cacciari:2008gp} implemented in the Fastjet~\cite{fastjet} library, with a radius parameter of 0.4. All stable final-state particles are used to reconstruct the jets, except the neutrinos, leptons and photons associated to $W$ or $Z$ boson or $\tau$ lepton decays. The effects of an upgraded ATLAS detector are taken into account by applying energy smearing, efficiencies and fake rates to generator level quantities, following parameterisations based on detector performance studies with the detailed simulations. The effect of the high pileup at the HL-LHC is incorporated by overlaying pileup jets onto the hard-scatter events. Jets from pileup are randomly selected as jets to be considered for analysis with $\sim 2\%$ efficiency, based on studies of pile-up jet rejection and current experience.

\subsubsection{ATLAS and CMS performance}
\label{sec:methods:perf}

The expected performance of the upgraded ATLAS and CMS detectors has been studied in detail in the context of the Technical Design Reports
and subsequent studies; the assumptions used for this report and a more detailed description are available in Ref.~\cite{ATLAS_PERF_Note,Collaboration:2650976}. For CMS, the object performance in the central region assumes a barrel calorimeter aging corresponding to an integrated luminosity of $1000$~fb$^{-1}$.

The triggering system for both experiments will be replaced and its impact on the triggering abilities of each experiment assessed;
new capabilities will be added, and, despite the more challenging conditions, most of the trigger thresholds for common objects are expected
to either remain similar to the current ones or to even decrease~\cite{ATLAS_TDAQ_TDR,CMSL1interim}.

The inner detector is expected to be completely replaced by both experiments, notably extending its coverage to $|\eta|<4.0$.
The performance for reconstructing charged particles has been studied in detail in Ref.~\cite{ATLAS_Pixel_TDR,ATLAS_Strip_TDR,CMS_Tracker_TDR}.

Electrons and photons are reconstructed from energy deposits in the electromagnetic calorimeter and information from the inner tracker\cite{ATLAS_LAr_TDR,CMS_Barrel_TDR,CMS_HGCAL_TDR,CMS_MTD_TP}.
Several identification working points have been studied and are employed by the projection studies as most appropriate.

Muons are reconstructed combining muon spectrometer and inner tracker information~\cite{ATLAS_Muon_TDR,CMS_Muon_TDR}.

Jets are reconstructed by clustering energy deposits in the electromagnetic and hadronic calorimeters\cite{ATLAS_Tile_TDR,ATLAS_LAr_TDR,CMS_Barrel_TDR} using the anti-$k_{\rm T}$ algorithm\cite{Cacciari:2008gp}.
B-jets are identified via $b$-tagging algorithms. B-tagging is performed if the jet is within the tracker acceptance ($|\eta|<4.0$).
Multivariate techniques are employed in order to identify $b-$jets and $c-$jets, and were fully re-optimized for the upgraded detectors~\cite{ATLAS_Pixel_TDR,CMS_Tracker_TDR}.
An 70\% $b-$jet efficiency working point is used, unless otherwise noted.

High $p_{\rm T}$ boosted jets are reconstructed using large-radius anti-$k_{\rm T}$ jets with a distance parameter of 0.8. Various jet substructure variables are employed to identify boosted $W$/$Z$/Higgs boson and top quark jets with good discrimination against generic QCD jets. 

Missing transverse energy is reconstructed following similar algorithms as employed in the current data taking.
Its performance has been evaluated for standard processes, such as top pair production~\cite{ATLAS_Pixel_TDR,Contardo:2020886}.

The addition of new precise-timing detectors and its effect on object reconstruction has also been studied in Ref.~\cite{ATLAS_TP_HGTD,CMS_MTD_TP}, although its results are only taken into account in a small subset of the projections in this report.

\subsubsection{LHCb performance}
\label{sec:methods:perf_LHCb}
The LHCb upgrades are shifted with respect to those of ATLAS and CMS. A first upgrade will happen at the end of Run-2 of the LHC, to run at a luminosity five times larger  ($2\times 10^{33}\text{cm}^{-2}\text{s}^{-1}$) in LHC Run-3 compared to those in Runs-1 and-2, while maintaining or improving the current detector performance. This first upgrade (named \mbox{Upgrade~I}) will be followed by by the so-called \mbox{Upgrade~II} (planned at the end of Run-4) to run at a luminosity of $\sim 2\times 10^{34}\text{cm}^{-2}\text{s}^{-1}$.

The LHCb MC simulation used in this document mainly relies on the \pythia~8 generator~\cite{Sjostrand:2007gs} with a specific LHCb configuration~\cite{LHCb-PROC-2010-056}, using the CTEQ6 leading-order set of parton density functions~\cite{cteq6}. The interaction of the generated particles with the detector, and its response, are implemented using the \geant{} toolkit~\cite{Allison:2006ve,Agostinelli:2002hh}, as described in Ref.~\cite{LHCb-PROC-2011-006}. 

The reconstruction of jets is done using a particle flow algorithm, with the output of this clustered using
the anti-$k_{\rm T}$ algorithm as implemented in \mbox{\textsc{Fastjet}}\xspace, with a distance parameter of
0.5. Requirements are placed on the candidate jet in order to reduce the background
formed by particles which are either incorrectly reconstructed or produced in additional pp interactions in the same event.

Concerning the increased pile-up, different assumptions are made, but in general the effect is assumed to be similar to the one in Run-2.

\subsection{Treatment of systematic uncertainties}
\label{sec:methods:syst}
It is a significant challenge to predict the expected systematic uncertainties of physics results at the end of HL-LHC running.
It is reasonable to anticipate improvements to techniques of determining systematic uncertainties over an additional decade of data-taking.
To estimate the expected performance, experts in the various physics objects and detector systems from ATLAS and CMS have looked at current limitations to
systematic uncertainties in detail to determine which contributions are limited by statistics and where there are more fundamental limitations.
Predictions were made taking into account the increased integrated luminosity and expected potential gains in technique.
These recommendations were then harmonized between the experiments to take advantage of a wider array of expert opinions and to allow the experiments to make sensitivity predictions on equal footing~\cite{ATLAS_PERF_Note,Collaboration:2650976}. For theorists' contributions, a simplified approach is often adopted, loosely inspired by the improvements predicted by experiments. 

General guide-lining principles were defined in assessing the expected systematic uncertainties.
Theoretical uncertainties are assumed to be reduced by a factor of two with respect to the current knowledge, thanks to both
higher-order calculation as well as reduced PDF uncertainties~\cite{Khalek:2018mdn}.
All the uncertainties related to the limited number of simulated events are neglected, under the assumption that sufficiently large simulation samples will be available by the time the HL-LHC becomes operational. For all scenarios, the intrinsic statistical uncertainty in the measurement is reduced by a factor $1/\sqrt{\text{L}}$, where $\text{L}$ is the projection integrated luminosity divided by that of the reference Run-2 analysis.
Systematics driven by intrinsic detector limitations are left unchanged, or revised according to detailed simulation studies of the upgraded detector.
Uncertainties on methods are kept at the same value as in the latest public results available, assuming that the harsher HL-LHC conditions will be compensated by method improvements.

The uncertainty in the integrated luminosity of the data sample is expected to be reduced down to 1\% by a better understanding of the calibration methods and
their stability employed in its determination, and making use of the new capabilities of the upgraded detectors.

In addition to the above scenario (often referred to as ``YR18 systematics uncertainties'' scenario), results are often
compared to the case where the current level of understanding of systematic uncertainties is assumed (``Run-2 systematic uncertainties'')
or to the case of statistical-only uncertainties.


\subsection{Precision Luminosity}

\subsubsection*{Motivation}

Measurements of production cross sections provide fundamental tests of theoretical predictions. Ultimate precision both of the experimental measurements and the theoretical predictions is required in order to determine fundamental parameters of the Standard Model and to constrain or discover beyond-the-Standard-Model phenomena. 
At the LHC, the precision of cross section measurements is limited by the uncertainty of the integrated luminosity, 
currently about 2\%. The impact of all other experimental uncertainties combined 
is smaller than $\sim1$\% (2--3\%) for Drell-Yan ($t\bar{t}$) cross section measurements, respectively~\cite{Aad:2016naf,Sirunyan:2017uhy}. 
For the HL-LHC~\cite{Apollinari:2284929}, significant improvements of the luminosity measurement are being planned.
A target uncertainty of 1\% has been set, and this is also assumed for many of the results presented in this report.
Such improvement is expected to be achieved by combination of improved luminosity detector instrumentation, 
currently in the design phase, and refined analysis techniques, rapidly developing 
during the analysis of Run-2 data. In the following, we provide a short description of the general plan
towards the 1\% target for the integrated luminosity at the HL-LHC.

\subsubsection*{Van der Meer Scans}

At hadron colliders, the precision of theoretical predictions for inclusive cross sections, e.g.\,for Z/$\gamma^*$ production, 
is limited by the knowledge of the parton density functions (PDFs) in the proton, and the uncertainty is of the order of 3--5\%~\cite{Ogul:2017zjd}. 
A more precise, and purely experimental method to determine the luminosity is based on the Van der Meer (VdM) scan technique~\cite{vanderMeer:296752}.
In VdM scans, beam axes are moved in the transverse planes, $x$ and $y$, across each other such that the beam overlap integral can be determined. 
From the measured overlap integral, and the beam currents, the instantaneous luminosity during the VdM scan is determined~\cite{Grafstrom:2015foa}.

In practice, VdM scan data are typically recorded with a small number of low pile-up bunches well separated in time, with special interaction-region optics optimised for the measurement of the luminous-region parameters~\cite{Grafstrom:2015foa,Bartosik:1590405,CMS-PAS-LUM-17-004}, and with the bunch intensity lowered to about 3/4 of that during physics runs so as to reduce beam-beam biases while retaining adequate statistics in the luminometers. To transfer the luminosity information from VdM scans to high pileup operation, rate measurements are performed during the VdM scan, in several detectors. The absolute scale, i.e.\,the relation between the measured rate in a given detector and the luminosity measurement is a detector-specific calibration constant, usually referred to as visible cross section $\sigma_{\rm vis}$, relating the measured event rate ${\rm d}N/{\rm d}t$ to the instantaneous luminosity through the relation ${\rm d}N/{\rm d}t = L \cdot \sigma_{\rm vis}$. The integrated luminosity for a complete data taking period, e.g.\,a full year of data taking is then obtained by continuous rate measurements throughout the year. The integrated normalized rate measurement then corresponds to the integrated luminosity.

\subsubsection*{Systematic Uncertainties}

The uncertainty in the integrated luminosity consists of three components~\cite{CMS-PAS-LUM-17-004,Aaboud:2016hhf}: the absolute-scale uncertainty, i.e. that on the measured visible cross-sections extracted from the VdM-scan analysis; the calibration-transfer uncertainty, which affects the extrapolation of the visible cross-section from the low pile-up, low luminosity VdM regime to the high pile-up, high luminosity physics regime; and the stability uncertainty, that arises from possible time-dependencies and degradations of the detector response affecting the rate measurement over time. Improved analysis techniques, better detectors and extended data takings dedicated to precision luminosity measurements are required to reduce the current uncertainty towards the 1\% goal.

\subsubsubsection*{Absolute Scale Uncertainty}

Dominant uncertainties in the luminosity scale arise from the modeling of, and the potential non-linear correlations between, the horizontal and vertical beam profiles; from inconsistencies between equivalent visible cross-section measurements carried out during the same calibration session or using different luminometers; from the absolute displacement scale of the beams during the scans; and from beam-orbit stability. In Run-2, these and other uncertainties have been reduced using refined methods and dedicated additional data have been recorded for such specific purposes. Improvements of the uncertainty can be achieved by combination of different complementary approaches, of results obtained using different detectors, and of datasets obtained from different VdM scans. 

An alternative technique, complementary to VdM scans, was established by the LHCb experiment~\cite{Aaij:2014ida}. The shape of a single beam is measured as the distribution of beam-gas interactions. For this purpose a gas is injected into the interaction region during the VdM fill. The combination of VdM-scan and beam-gas imaging measurements leads to further reduction of the uncertainty, at least for LHCb, thanks to the exquisite performance of the VELO vertex detector.

\subsubsubsection*{Calibration-transfer Uncertainty}
In the HL-LHC area, the VdM calibration will typically be carried out under similar conditions as in Run-2, i.e.\,at a pile-up level of about 0.5 interactions per bunch crossing, and with a luminosity of a few Hz/$\mu$barn. In contrast, the physics running during the HL-LHC, will be characterized by pile-up parameters of up to 200 interactions per bunch crossing, and by average instantaneous luminosities of 50 Hz/nb, two to three times the peak instantaneous luminosity achieved so far. This will lead to an increase of the uncertainties associated with non-linearities in luminometer response. Most luminosity detectors for HL-LHC are still being designed. Drawing on Run-1 and Run-2 experience with precision luminosity measurements, the design of the future detectors aims to  reduce the associated systematic uncertainties. HL-LHC detectors are required to behave linearly over several orders of magnitude in their track, energy or hit rate measurements, with residual non-linearities that are reproducible and monitorable. Special runs with scans at intermediate instantaneous luminosity can be used to pin down nonlinear behaviour further.

\subsubsubsection*{Long-term stability and consistency of luminosity measurements}
In the past, one obvious way to determine stability and linearity effects has been to devise and compare the luminosity measurements by several detectors, using different technologies, with uncorrelated systematics. Since 2016, experiments started to exploit so-called emittance scans. These are short VdM scans (duration of minutes) performed at standard physics optics and currents, regularly at the beginning and at the end of fills~\cite{Hostettler:2016puf,em_scans,LHC_Emit_Scans,DubnaConf_CMSEmit_Scans}. While the emittance scans are not primarily designed for the precision determination of $\sigma_{\rm vis}$, trends over time, or as a function of instantaneous luminosity, can be used to determine stability effects, such as aging, independently for each given detector. The combination of emittance scans and of rate comparisons between redundant and independent detector systems has been successfully used to discover and control drifts and trends throughout Run-2, the longest LHC data-taking period so far, during which 150 fb$^{-1}$ worth of data were recorded. As a result, the uncertainty in the integrated luminosity in recent years remained at around 2-2.5\% even though the pile-up extrapolation range and the duration of the integration periods increased significantly.

\subsubsubsection*{Recent Ideas}

Additional methods are being discussed among luminosity experts of the LHC experiments and machine. One method recently developed is to use the rate measurement of ${Z}\rightarrow \mu\mu$ production
~\cite{Salfeld_Nebgen_2018}.
This is a high-rate physics process with in-situ calibration capabilities. Luminosity and Z boson rate are experimentally related through the following formula: $\sigma_{Z} = N_{Z} / (L \times \epsilon_{{Z}\rightarrow \mu\mu})$ where $N_{Z}$  is the number of reconstructed Z bosons, $L$ the integrated luminosity, and $\epsilon_{{Z}\rightarrow\mu\mu}$ the ${Z}\rightarrow\mu\mu$ event reconstruction efficiency. If $\epsilon_{{ Z}\rightarrow\mu\mu}$ and $L$ are known, then the fiducial Z boson production cross section $\sigma_{Z}$ can directly be determined from the measured event rate. To minimize the uncertainties associated with luminometer non-linearities and long-term stability, the fiducial Z boson cross section is measured from data recorded during an extended proton--proton production run at low pileup. This run should be close in time to one or two extended VdM scans. The efficiency $\epsilon_{{Z}\rightarrow\mu\mu}$ can be determined in situ, using the tag-and-probe method on the same event sample
~\cite{Salfeld_Nebgen_2018}.
Once the cross section is measured at sub-percent level precision, the continuous rate measurement can be used to transfer the calibration to the high pileup dataset. The integrated luminosity will be given by the total number of produced $Z$ bosons, corrected by the time-integrated muon identification efficiency with an uncertainty consisting of the absolute scale uncertainty from the VdM scan (or, in LHCb, beam-gas imaging scan), and a remaining uncertainty in the pileup dependency of the muon identification efficiency.

\subsubsection*{Conclusions Towards HL-LHC}

The aim for HL-LHC is to measure luminosity with substantially improved precision.
This aim can be achieved by combination of three ingredients:
\begin{enumerate}
\item High precision luminosity detectors are needed to provide high-granularity bunch-by-bunch luminosity measurements, with very good linearity and stability.
\item Advanced, multiple and redundant VdM scans and refined VdM analysis techniques can lead to substantial improvements.
\item Novel techniques, such as the measurement of fiducial Z boson production rates exploiting in-situ efficiency determination, provide handles for advancement of the integrated luminosity uncertainty towards the 1\% target.
\end{enumerate}

In order to achieve these goals during HL-LHC, a suite of tests and proof-of-concept measurements is being developed which should be carried out already during Run-3.

\clearpage

\newpage

\section{Electroweak processes}

The study of electroweak processes is a central topic of SM tests. Given the small electroweak couplings, high luminosity provides a crucial handle for gaining precision in these measurements, in particular for complex final states with relatively small cross sections. Prospects for those measurements and for their theoretical description are considered in the following for vector boson fusion (VBF) and vector boson scattering (VBS) processes, for di-boson and tri-boson production, and for single weak boson production processes, which promise unprecedented precision on W-mass and weak mixing angle measurements.

\subsection[Vector boson fusion]{Vector boson fusion\footnote{Contribution by F.~Campanario, T.~Chen, J.~M.~Cruz-Martinez, T.~Figy, A.~Karlberg, S.~Pl\"atzer and M.~Sj\"odahl.}}
\label{sec:vbf}







\newcommand{\NNLO}{\mathrm{NNLO}}
\newcommand{\QCD}{\mathrm{QCD}}
\newcommand{\alphas}{\alpha_{\mathrm{s}}}
\newcommand{\qt}{q_{\mathrm{T}}}
\newcommand{\kt}{k_{\mathrm{T}}}
\newcommand{\pt}{p_{\mathrm{T}}}
\newcommand{\as}{\alpha_s}
\newcommand{\muF}{\mu_F}
\newcommand{\muR}{\mu_R}
\newcommand{\diff}{\mathrm{d}}
\newcommand{\GeV}{\text{GeV}\xspace}
\newcommand{\TeV}{\text{TeV}\xspace}
\providecommand\DY{\mathrm{DY}}
\providecommand\DIS{\mathrm{DIS}}
\providecommand\VBF{\mathrm{VBF}}
\providecommand\NLL{\mathrm{NLL}}
\providecommand\VH{\mathrm{VH}}
\providecommand\WH{\mathrm{WH}}
\providecommand\ZH{\mathrm{ZH}}
\providecommand\ELWK{\mathrm{EW}}
\providecommand\HAWK{{\sc HAWK}}
\providecommand\MCFM{{\sc MCFM}}
\providecommand\VBFNLO{{\sc VBFNLO}}
\providecommand\VHNNLO{{\sc VH@NNLO}}
\providecommand\vhnnlo{{\sc VHNNLO}}
\providecommand\POWHEG{{\sc POWHEG}}
\providecommand\POWHEGBOX{{\sc POWHEG BOX}}
\providecommand\HERWIG{{\sc HERWIG}}
\providecommand\PYTHIA{{\sc PYTHIA}}
\providecommand\HDECAY{{\sc HDECAY}}
\providecommand{\kT}{\ensuremath{k\sb{\scriptstyle\mathrm{T}}}}
\providecommand\NNLOPS{{\sc NNLOPS}}
\providecommand\HVNNLO{{\sc HVNNLO}} 
\providecommand{\HVNNLOPS}{{\sc HVNNLOPS}} 
\providecommand{\HWJMINLOPS}{{\sc HWJ-MiNLO (Pythia8+hadr)}}
\providecommand{\HWNNLOPS}{{\sc HW-NNLOPS (Pythia8+hadr)}}
\providecommand{\HWNNLOPSshort}{{\sc HW-NNLOPS}}
\providecommand\DYNNLOPS{{\sc DYNNLOPS}}
\providecommand\MINLO{{\sc MiNLO}}
\providecommand\HWJMINLO{{\sc HWJ-MiNLO}}
\providecommand\FASTJET{{\sc FastJet}}
\providecommand\HNNLOPS{{\sc HNNLOPS}}
\providecommand\HW{{\sc HW}}
\providecommand\PhiHW{\Phi_{\scriptscriptstyle HW^*}} 
\providecommand\PhiHWsimp{\Phi_{\scriptscriptstyle HW}} 
\providecommand\thetacs{\theta^*}
\providecommand\phics{\phi^*}

This sub-section discusses the prospects for vector boson
fusion Higgs production at the HL-LHC and the HE-LHC, respectively. A particular focus is to investigate how hard and how forward the two tag jets
are expected to be at $27 ~\TeV$. The efficiency of VBF
cuts will be discussed, and fiducial cross sections and differential
distributions for a set of typical analysis cuts will be determined.
Finally, the quality of the VBF approximation will be considered, 
in particular when extra jet activity in addition to the two tag jets is required.

The relevant parameters used for the calculations in this chapter are reported here. More details can be found in LHC Higgs Cross Section Working Group report~\cite{deFlorian:2016spz}. The gauge boson masses and widths are set to
\begin{align}
  m_W = 80.385~\GeV, \qquad \Gamma_W = 2.085\GeV.
\end{align}
\begin{align}
  m_Z = 91.1876~\GeV, \qquad \Gamma_Z = 2.4952~\GeV.
\end{align}
and the Fermi constant is
\begin{align}
  G_F = 1.16637\cdot 10^{-5}~\GeV^{-2}.
\end{align}
The Higgs is described in the narrow width approximation with mass $m_H = 125~\GeV$.
The parton distribution function PDF4LHC15\_nnlo\_100\_pdfas is used and the central renormalization and factorization scale is set to $\mu_0 = m_W$, unless otherwise specified.

\subsubsection*{Detector requirements}
\label{sec:detector}
\begin{figure}
    \centering\includegraphics[width=0.9\textwidth]{\main/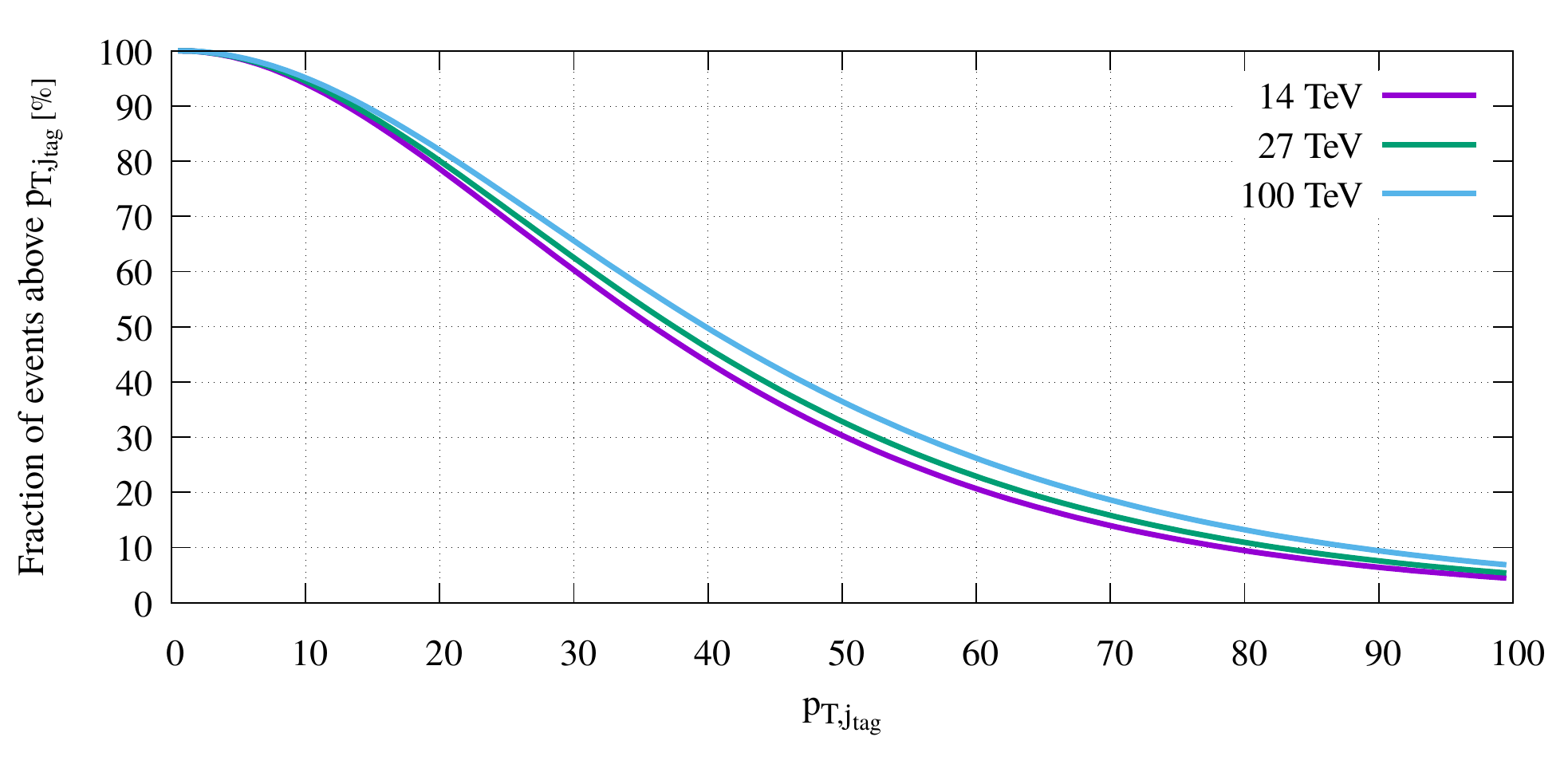}\hfill
    \caption{Fraction of the total VBF cross section surviving a $\pt$ cut on the two hardest jets of $p_{\mathrm{T,j_{tag}}}$ for three different collider energies. The results shown here are computed at LO.}
    \label{fig:higgsDifEnergy}
\end{figure}
\begin{figure}
    \centering\includegraphics[width=0.9\textwidth]{\main/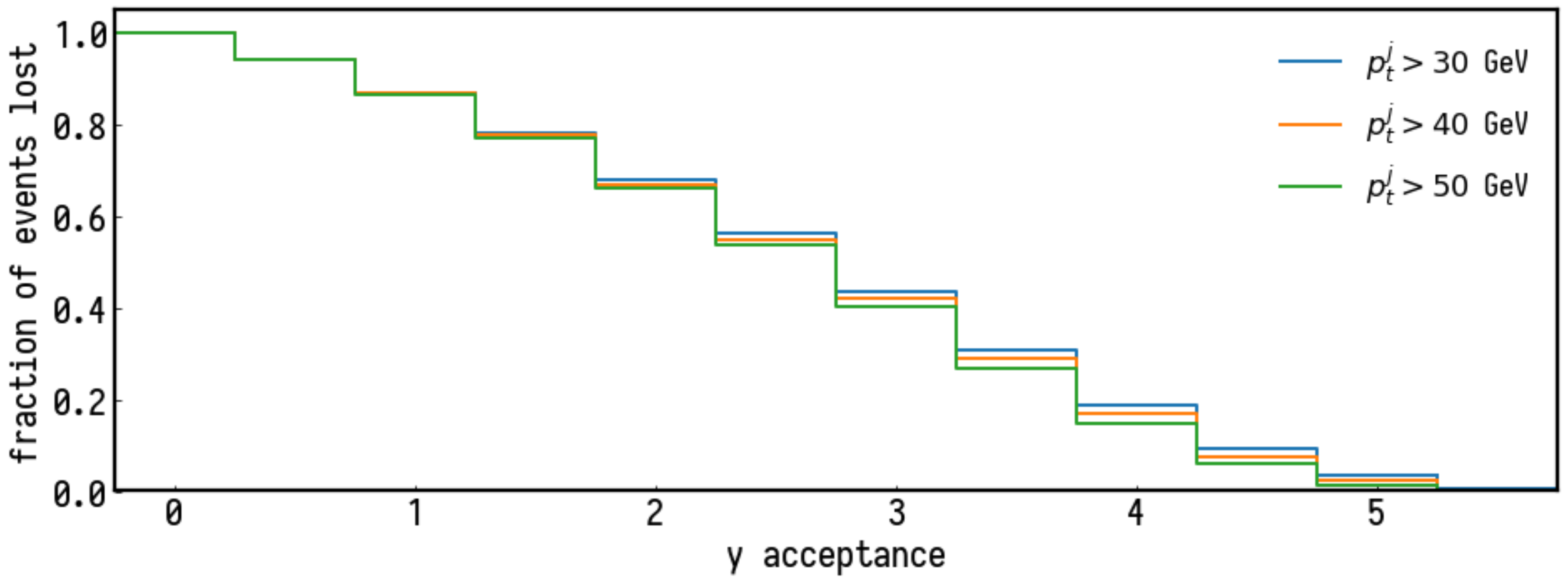}\hfill
    \caption{Fraction of events lost as a function of the rapidity acceptance of the detector at a collider energy of $\sqrt{s} = 27~\TeV$. Results shown for three different tag jet transverse momentum cuts. The results shown here are computed at LO.}
    \label{fig:higgs27acceptance}
\end{figure}

VBF production is characterized by two hard
and forward jets accompanying the two bosons. The requirement of two such jets can significantly
reduce the QCD induced background along with the electroweak
production stemming from s-channel processes. The transverse hardness
of the VBF jets is fundamentally set by the mass scale of the virtual
vector bosons. It is therefore expected that the jet spectrum is not
very sensitive to the collider centre-of-mass energy, and in
particular that the jets do not get appreciably harder when increasing
the energy.

Figure~\ref{fig:higgsDifEnergy} shows the fraction of total VBF cross sections
that survives the cut on the transverse momentum of the two tag jets
for the three collider energies $14$, $27$, and $100~\TeV$. As can be
seen, the cross section drops rapidly as the $p_{\rm T}$-cut is increased. In
particular, at $27~\TeV$, roughly $60\%$ survive for
$p_{\rm{T},tag}>30~\GeV$, which diminishes to $30\%$ of the total VBF cross section for $p_{\rm{T},tag}>50~\GeV$. It will
therefore be of great importance to the VBF program to be able to keep
the jet definition not too hard.

Given that the two tag jets tend to be forward in the detector volume,
it is of interest to study how many jets are lost above a certain rapidity threshold. Figure~\ref{fig:higgs27acceptance} shows the
fraction of events with
$\max{\left|y_{j_1}\right|,\left|y_{j_2}\right|}$ above some threshold
at $\sqrt{s}=27~\TeV$ for various jet $p_{\rm T}$ definitions. As can be
seen from the plot, about $20\%$ of the cross section has
$\max{\left|y_{j_1}\right|,\left|y_{j_2}\right|}>4$. For comparison,
this number is $\sim 5\%$ at $14~\TeV$. Additionally one finds that these losses increase to $\sim 30\%$ when imposing the dedicated VBF cuts for 27~TeV defined below. Hence, in order to maximize the potential of
VBF analyses at the HE-LHC it will be 
highly desirable that the detectors have a
rapidity reach beyond $4.0$.

\subsubsection*{HL-LHC}
\label{sec:HLLHC}
For VBF production with a centre of mass energy of $\sqrt{s} =14~\TeV$, VBF cuts as in Ref. ~\cite{deFlorian:2016spz} are used, with two anti-$k_{\rm T}$ jets with $R=0.4$ and
\begin{align}
    p_{\mathrm{T}}^{j} > 20~\GeV, \qquad |y_j| < 5.0, \qquad |y_{j_1}-y_{j_2}| > 3.0, \qquad M_{jj} > 130~\GeV.
  \label{eq:VBFcuts14}
\end{align}
The requirement on the rapidity separation and invariant mass significantly reduces background contributions to the process $pp\rightarrow Hjj$. 

Table~\ref{tab:vbf_XSfiducial} reports the fiducial VBF cross
section under the above cuts. The cross section includes NNLO-QCD
corrections in the DIS approximation and NLO-EW corrections including
photon induced contributions. Shown separately is the $s$-channel
contribution which is not included in the total number. The NNLO-QCD
corrections have been computed with {\mbox{\textsc{proVBFH-1.1.0}}\xspace} ~\cite{Cacciari:2015jma,Dreyer:2016oyx,Alioli:2010xd,Nason:2009ai,Jager:2014vna} and the electroweak
contributions with
{\mbox{\textsc{HAWK-2.0}}\xspace}~\cite{Ciccolini:2007ec,Ciccolini:2007jr,Denner:2011id,Denner:2014cla}.
\begin{table}
  \caption{Fiducial VBF cross sections including QCD and EW
    corrections and their uncertainties for collider energy
    $\sqrt{s}=14~\TeV$ and for a Higgs-boson mass $m_H=125~\GeV$. The
    QCD corrections have been updated compared to those reported in
    Ref.~\cite{deFlorian:2016spz}.}
  \label{tab:vbf_XSfiducial}
  \begin{center}%
    \begin{small}%
      \tabcolsep5pt
      \begin{tabular}{|ccc|cc|c|c|}%
        \hline
        $\sigma^{\VBF}$[fb] & $\Delta_{\mathrm{scale}}$[\%] & 
        $\Delta_{\mathrm{PDF\oplus\alphas}}$[\%] &
        $\sigma_{\NNLO \QCD}^{\DIS}$[fb] & $\delta_{\ELWK}$[\%] & $\sigma_{\gamma}$[fb] & $\sigma_{\mbox{\scriptsize $s$-channel}}$[fb]
        \\
        \hline\hline
        $2259$ &$^{+1.5}_{-1.3}$ &$\pm 2.1/\pm 0.4/\pm2.1$ & $2401$ & $-6.9$ & $23.6$ & $32.9$
        \\
        \hline
      \end{tabular}%
    \end{small}%
  \end{center}%
\end{table}

\begin{figure}
    \includegraphics[width=0.46\textwidth]{\main/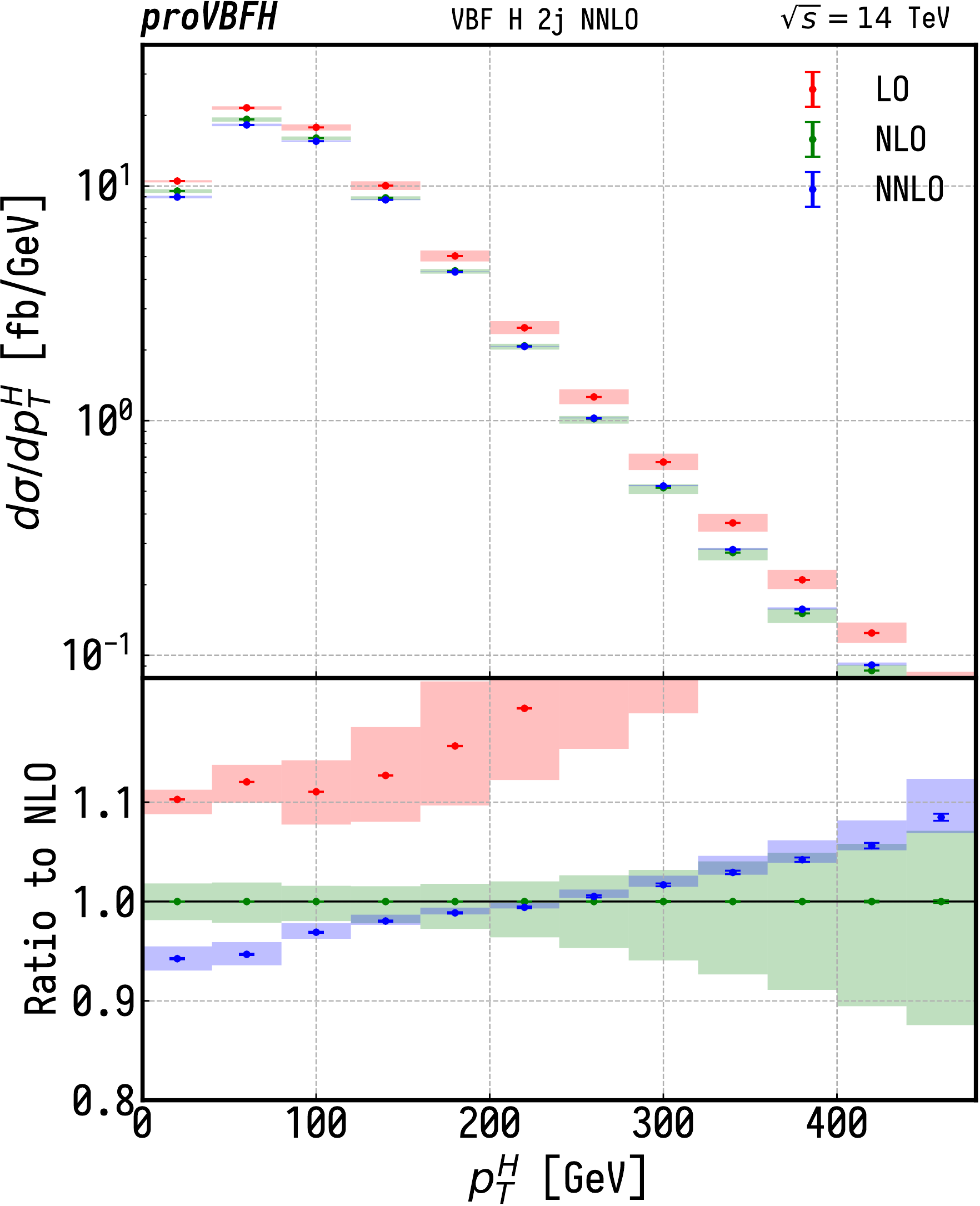}\hfill
    \includegraphics[width=0.46\textwidth]{\main/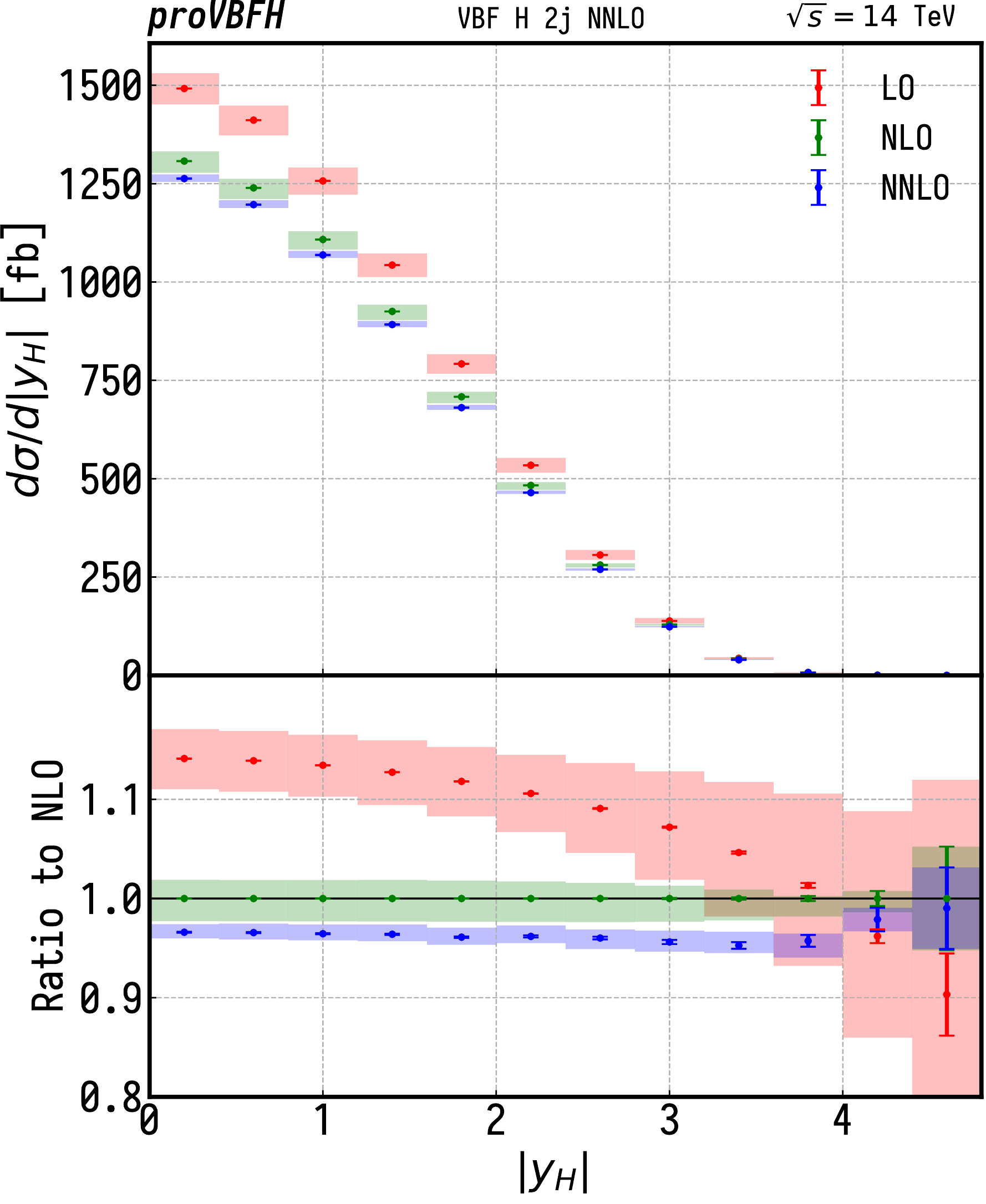}
    \caption{Transverse momentum and rapidity of the Higgs boson after the cuts of eq.~\eqref{eq:VBFcuts14} and at a collider energy $\sqrt{s} = 14~\TeV$.}
    \label{fig:higgs14}
\end{figure}

\begin{figure}
    \includegraphics[width=0.46\textwidth]{\main/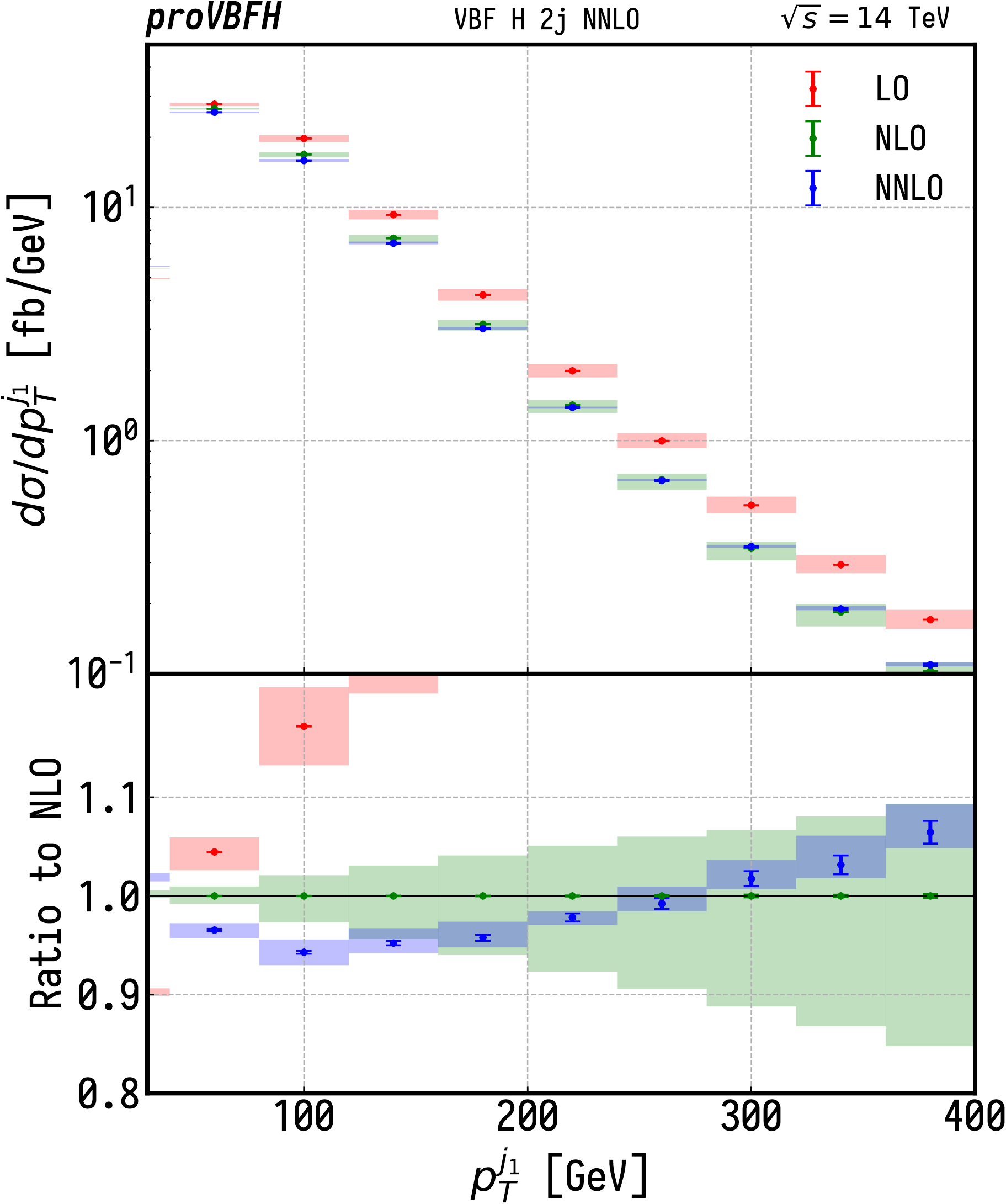}\hfill
    \includegraphics[width=0.46\textwidth]{\main/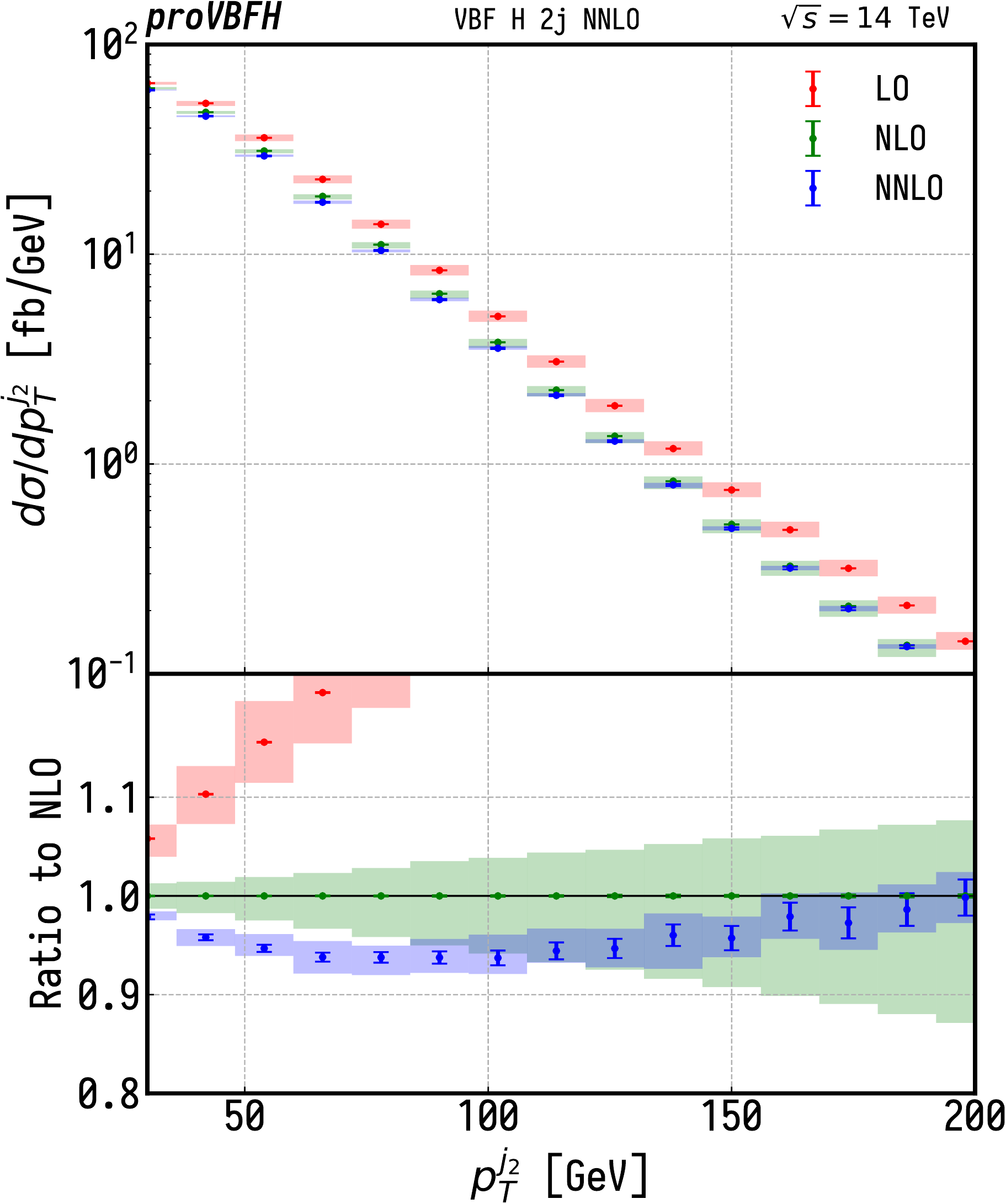}\hfill
    \includegraphics[width=0.46\textwidth]{\main/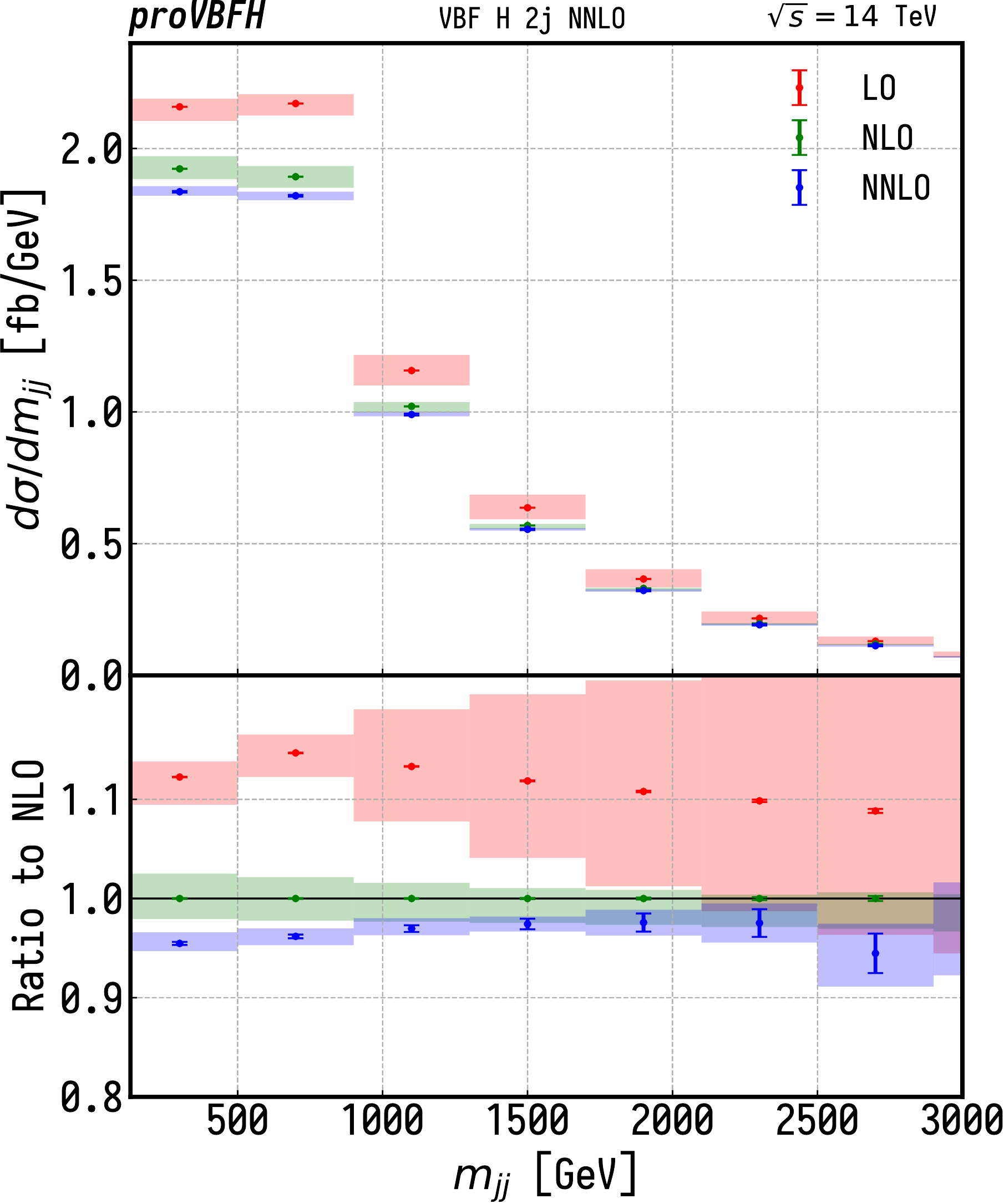}\hfill
    \includegraphics[width=0.46\textwidth]{\main/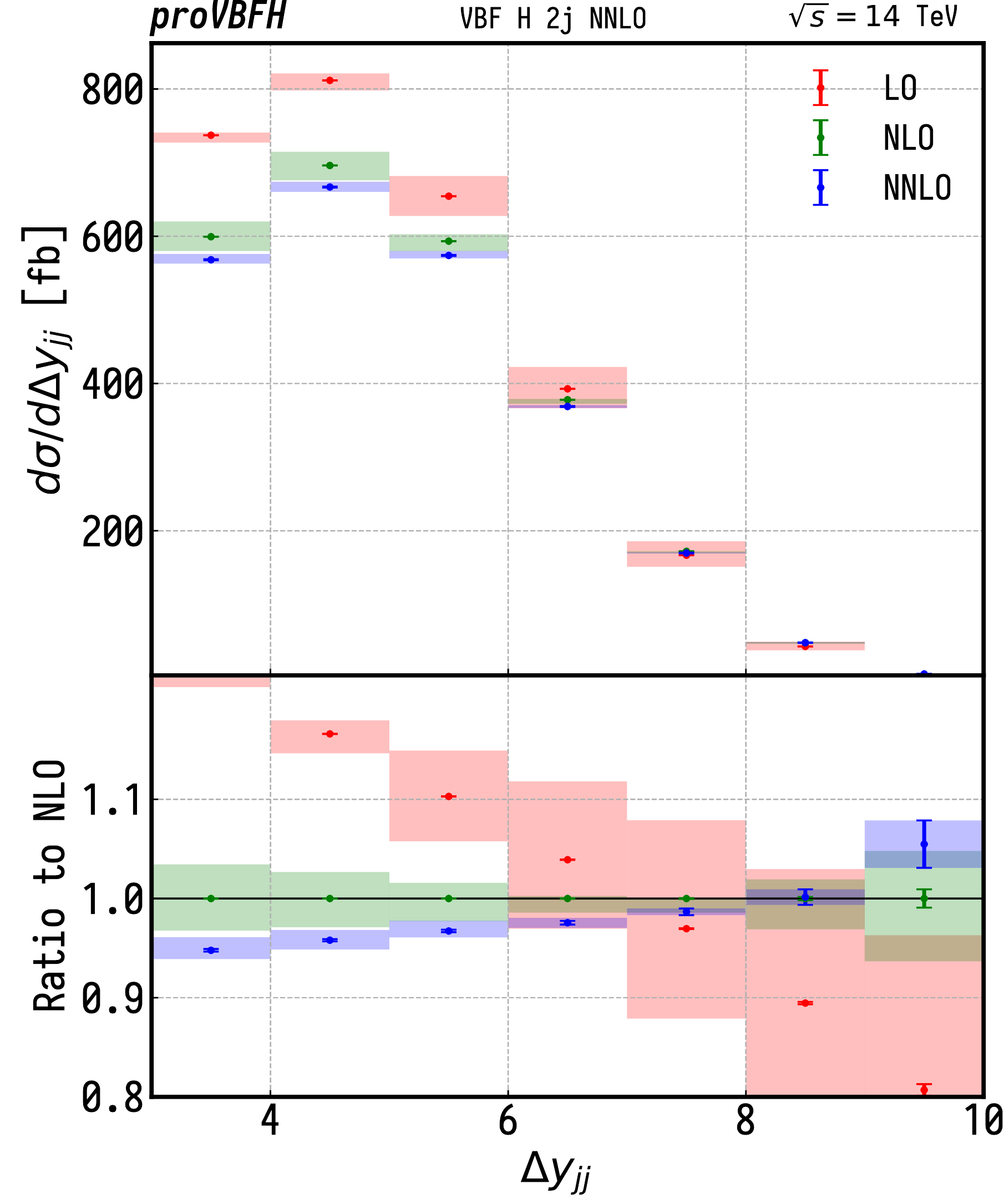}
    \caption{In the top row the transverse momentum the two hardest jet after the cuts of eq.~\eqref{eq:VBFcuts14} and at collider energy $\sqrt{s} = 14~\TeV$. In the bottom row the invariant mass and absolute rapidity gap between the two hardest jets.}
    \label{fig:1jet14}
\end{figure}

\subsubsection*{HE-LHC}
\label{sec:helhc}
For fiducial cross sections at a centre-of-mass energy of $\sqrt{s}=27~\TeV$, all physical parameters are kept unchanged with respect to the previous sections. The contributions of the gluon fusion (ggF) and VBF channels to Hjj production are compared, and results are presented for 
the effects of the NLO and NNLO QCD corrections to VBF Hjj production as computed in {\mbox{\textsc{NNLOJET}}\xspace}
~\cite{CRUZMARTINEZ2018672} with a redefined set of VBF cuts for the new energy choice. 


For the comparison of VBF to the ggF background, any kind of VBF cut is omitted, requiring only two jets with
\begin{align}
    p_{\mathrm{T}}^{j} > 30~\GeV, && \qquad |y_j| < 5.0,
    \label{eq:inclusivecut}
\end{align}
defined using the anti-$k_{\rm T}$ algorithm~\cite{Cacciari:2005hq} with $R=0.4$.
The total cross section for ggF and VBF is shown in Table~\ref{table:ggfVsVBF27}. Both the ggF and VBF contributions are computed with the parton-level Monte Carlo {\mbox{\textsc{NNLOJET}}\xspace} which includes ggF Higgs production in the heavy top limit (HTL)~\cite{Chen:2014gva, Chen:2016zka, Chen:2018pzu, Bizon:2018foh, Cieri:2018oms} among other processes~\cite{Gehrmann:2018szu, Ridder:2015dxa, Currie:2016ytq, Currie:2016bfm, Gehrmann-DeRidder:2017mvr,Currie:2017eqf, Currie:2018fgr, NIEHUES2019243, Currie:2018xkj}. The comparison of Table~\ref{table:ggfVsVBF27} is done at NLO QCD since Higgs plus two jets in gluon fusion is only available at this accuracy level.

In order to define a set of cuts which enhance the VBF contribution, the invariant mass ($m_{jj}$) and the spacial distribution (through the rapidity gap between both jets, $\Delta y_{jj}$) of the dijet system formed by the two leading jets is considered. The VBF production mode dominates over ggF in the large rapidity separation region ($\Delta y_{jj} > 4.5$) as well as for moderate and high values of the dijet invariant mass ($m_{jj} > 700~\GeV$).

\begin{table}[bh]
    \centering
\caption{Comparison between Higgs production by gluon fusion and vector boson fusion for a centre-of-mass energy $\sqrt{s} = 27~\TeV$, at NLO QCD. Errors correspond to Monte Carlo statistics.} \label{table:ggfVsVBF27}
    \begin{tabular}{ |c | c | c| }
        \hline Production mode & Total cross section (fb) & \% of Total    \\ \hline\hline
        ggF (HTL) & 21984 $\pm$ 10 & 75.32 $\pm$ 0.04     \\ 
        VBF & 7203 $\pm$ 2 & 24.68 $\pm$ 0.01     \\  
        \hline
    \end{tabular}
   
\end{table}

\begin{figure}
    \includegraphics[width=0.43\textwidth, height=0.4\textheight]{\main/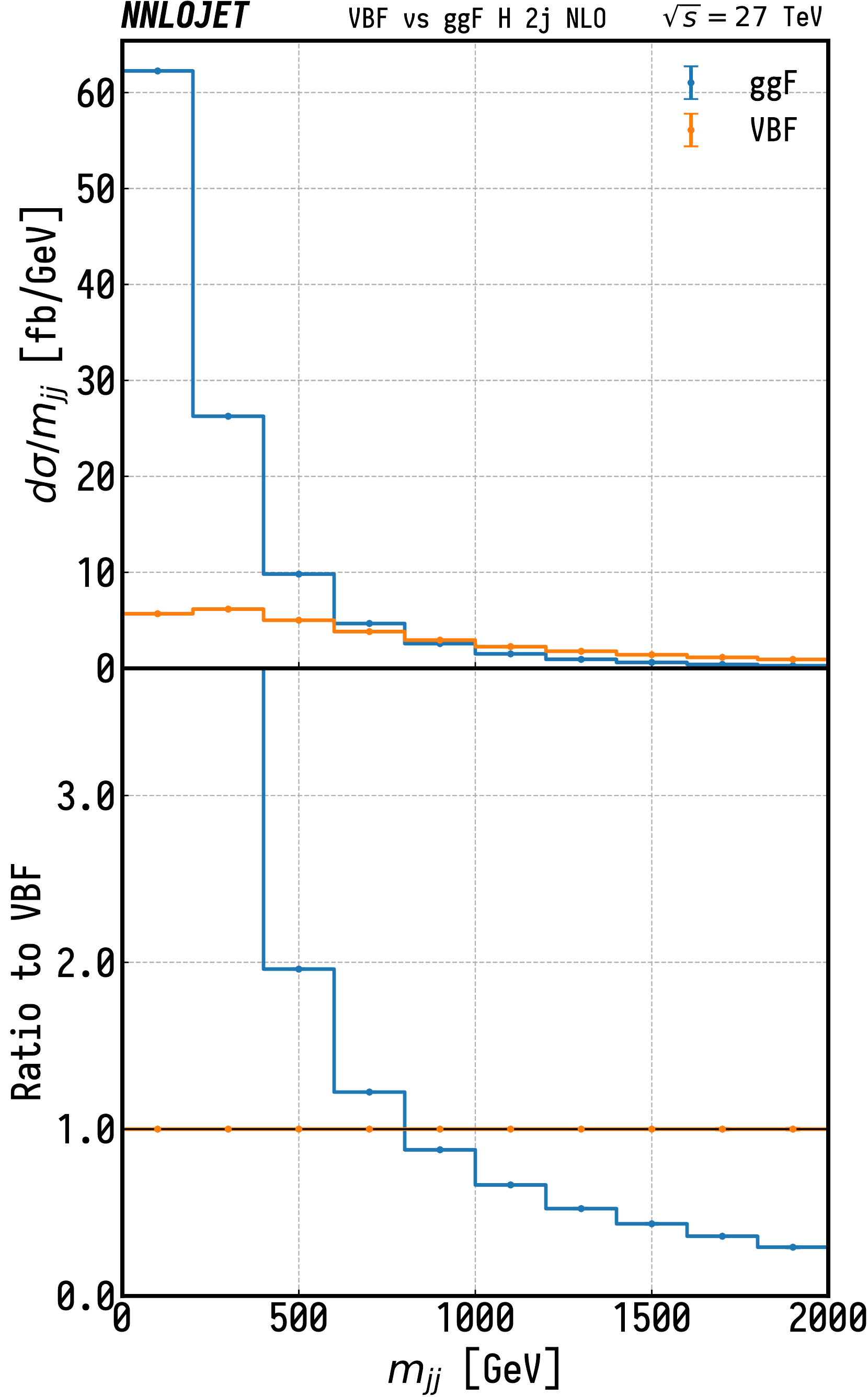}\hfill
    \includegraphics[width=0.43\textwidth, height=0.4\textheight]{\main/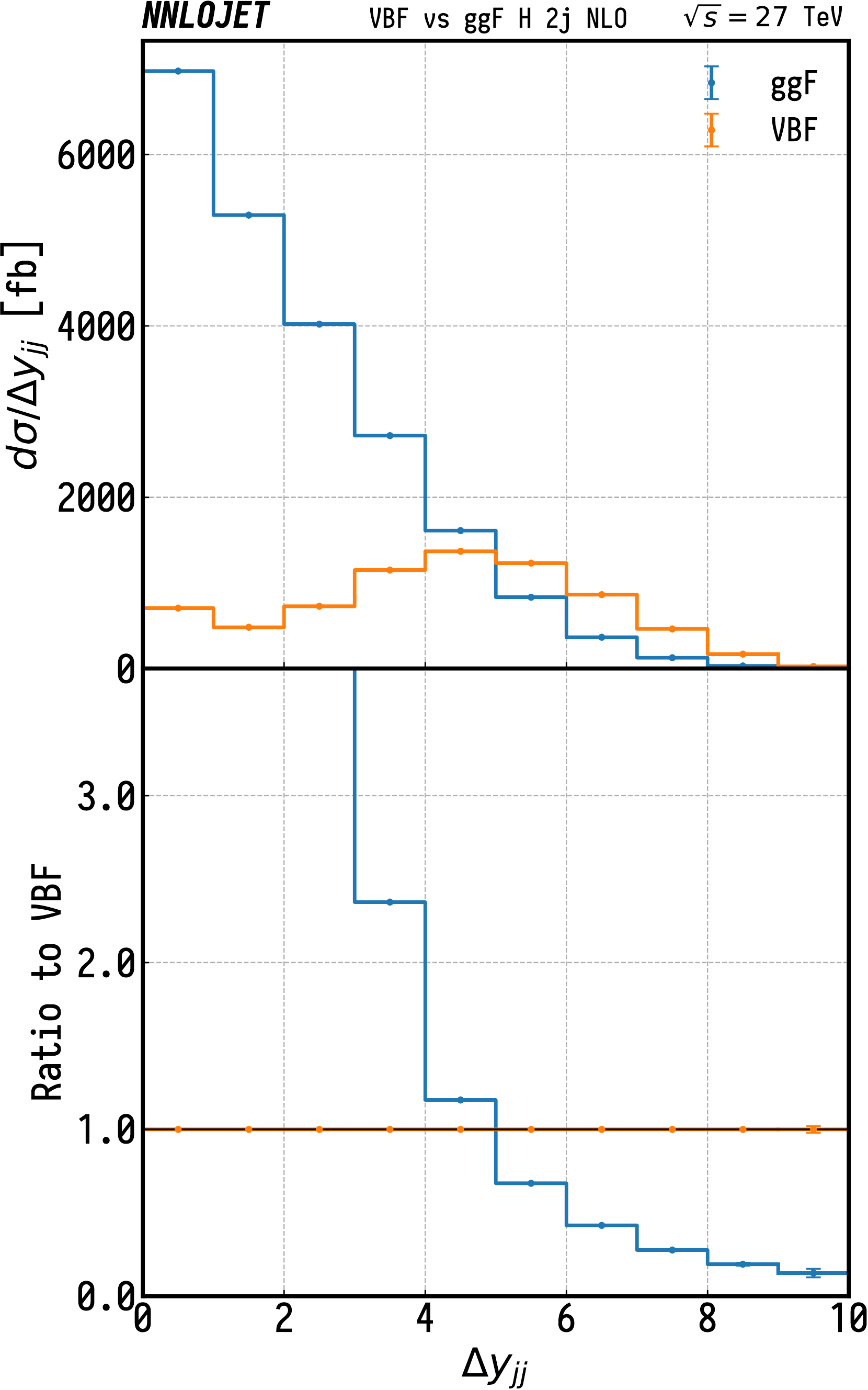}
    \caption{Differential distributions for the invariant mass (left) and spatial distribution (right) of the dijet system. At lower values of $m_{jj}$ and $\Delta y_{jj}$ one observes a strong dominance of the ggF channel. For larger values of both observables, however, the VBF channel gains importance.}
    \label{fig:ggfVsVBF27}
\end{figure}

Fiducial cross sections for VBF at $\sqrt{s} = 27~\TeV$ are defined with a set of tight VBF cuts,
\begin{align}
    \Delta y_{jj} > 4.5, && m_{jj} > 600~\GeV, \label{eq:tight}
\end{align}
requiring the two leading jets to be found in opposite rapidity hemispheres with a maximum rapidity of $|y_{j}| < 5.0$.
\begin{table}
  \caption{Fiducial VBF cross sections including QCD and EW
    corrections and their uncertainties for collider energy
$\sqrt{s}=27~\TeV$ ($m_H=125~\GeV$). 
For completeness the $s$-channel contribution (corresponding to $pp \to HV \to q\bar{q}$) is also included.}
  \label{tab:vbf_XSfiducial_27}
  \begin{center}%
      \tabcolsep5pt
      \begin{tabular}{|cc|cc|c|cc|}%
        \hline
        $\sigma^{\VBF}$[fb] & $\Delta_{\mathrm{scale}}$[\%] & 
        $\sigma_{\NNLO \QCD}^{\DIS}$[fb] & $\delta_{\ELWK}$[\%] & $\sigma_{\gamma}$[fb] & $\sigma_{\mbox{\scriptsize $s$-channel}}$[fb] &$\pt^{j}$ cut [GeV] 
        \\
        \hline\hline
        $2805$ &$^{+1.05}_{-0.02}$  & $3059$ & $-9.6$  & $39.8$ & $5.9$ & $30$ \\
        $2087$ &$^{+1.13}_{-1.05}$  & $2283$ & $-10.0$ & $32.3$ & $4.4$ & $40$ \\
        $1442$ &$^{+1.43}_{-1.61}$  & $1586$ & $-10.5$ & $22.3$ & $3.0$ & $50$ 
        \\
        \hline
      \end{tabular}%
  \end{center}%
\end{table}
In Table~\ref{tab:vbf_XSfiducial_27} the fiducial cross section is computed for three choices of the cut on the transverse momentum of the two leading jets: $\pt^{j} > \{30, 40, 50\}~\GeV$ while differential distributions for $\pt^{j} > 30~\GeV$ are shown in Figs.~\ref{fig:higgs27} and~\ref{fig:dijet27}. The Hjj contribution in the VBF approximation as well as plots in this section are calculated at NNLO QCD accuracy with {\mbox{\textsc{NNLOJET}}\xspace}, electroweak corrections and the $s$-channel contribution shown in Table~\ref{tab:vbf_XSfiducial_27} are again computed with {\mbox{\textsc{HAWK-2.0}}\xspace}.
Shaded boxes in all plots represent scale variations with $\muR = \muF = \{0.5, 2\}\mu_{0}$ with the central scale $\mu_{0} = m_{W}$ and error bars represent statistical uncertainties from the Monte Carlo integration.
In Fig.~\ref{fig:higgs27} the transverse momentum and rapidity distribution of the Higgs boson is shown. The kinematical variables for the system formed by the two leading jets are shown in Fig.~\ref{fig:dijet27}.

\begin{figure}
    \includegraphics[width=0.46\textwidth]{\main/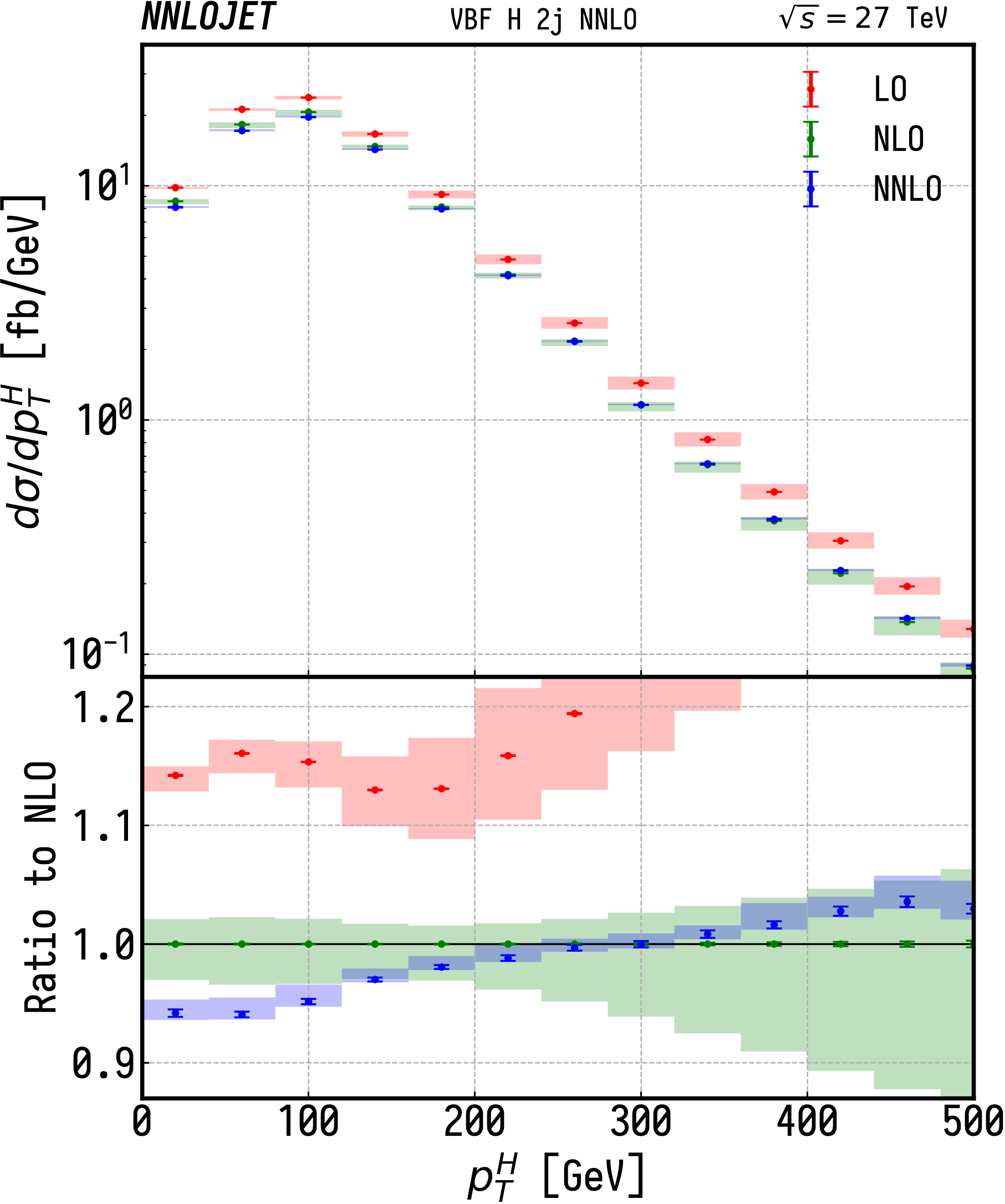}\hfill
    \includegraphics[width=0.46\textwidth]{\main/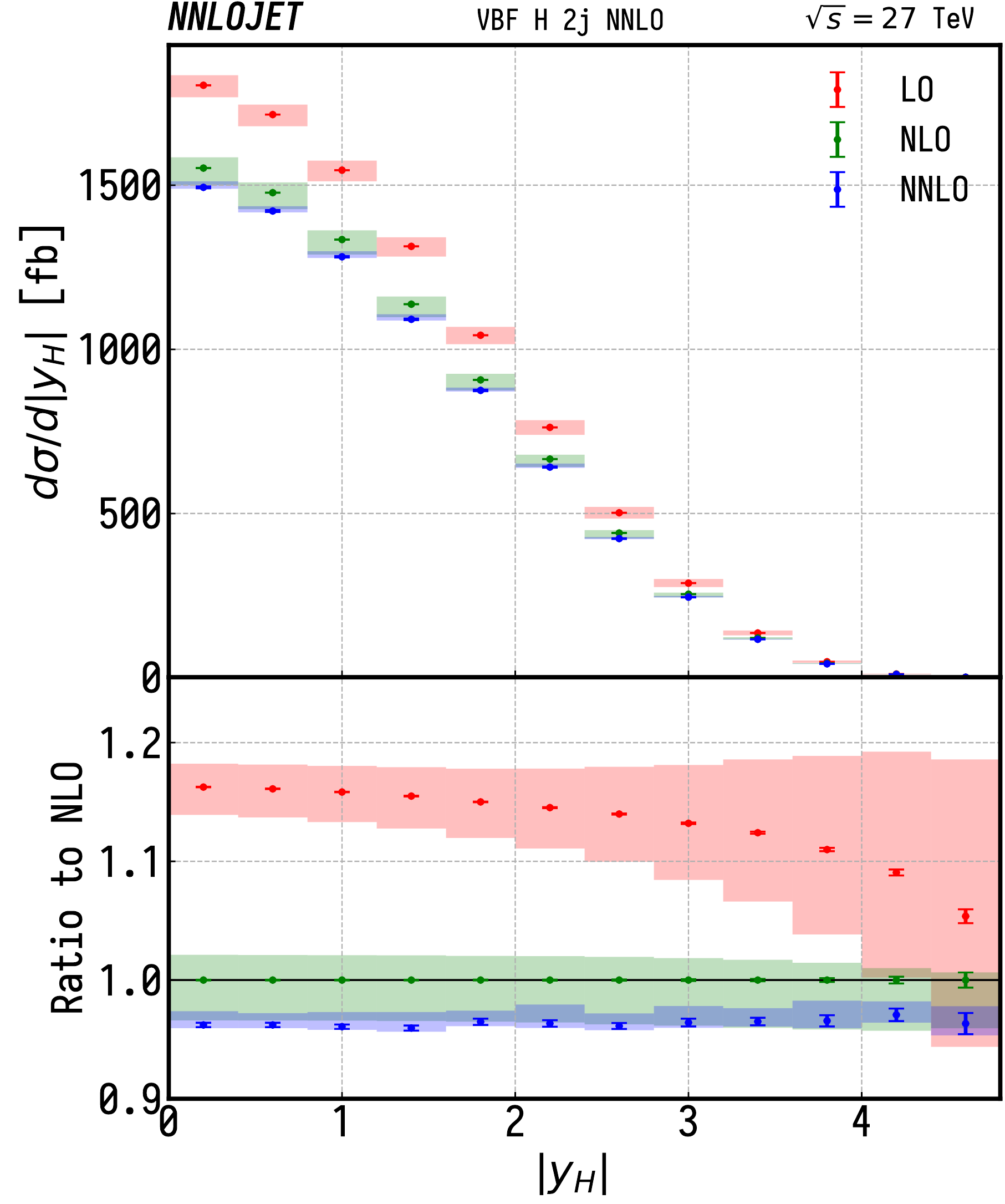}
    \caption{Kinematical variables for the Higgs boson at $\sqrt{s} = 27~\TeV$ for tight VBF cuts. The NLO corrections are of more than -10~\% across the whole considered range. The NNLO corrections, much smaller than NLO, show good convergence of the perturbative series. The NNLO corrections changes sign for high transverse momentum (left). For the rapidity distribution (right) they remain stable across the entire range of the observable.}
    \label{fig:higgs27}
\end{figure}
\begin{figure}
    \includegraphics[width=0.46\textwidth]{\main/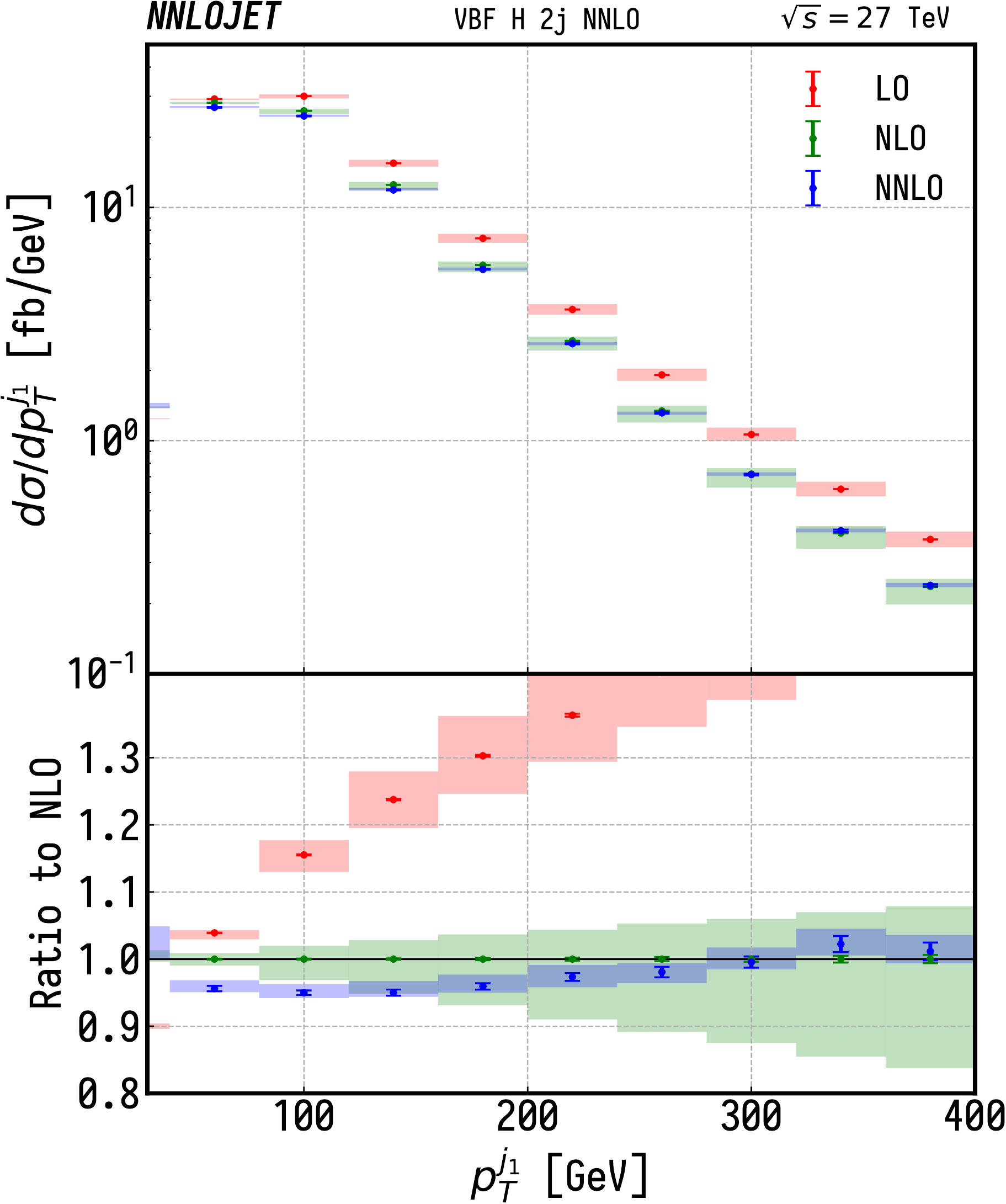}\hfill
    \includegraphics[width=0.46\textwidth]{\main/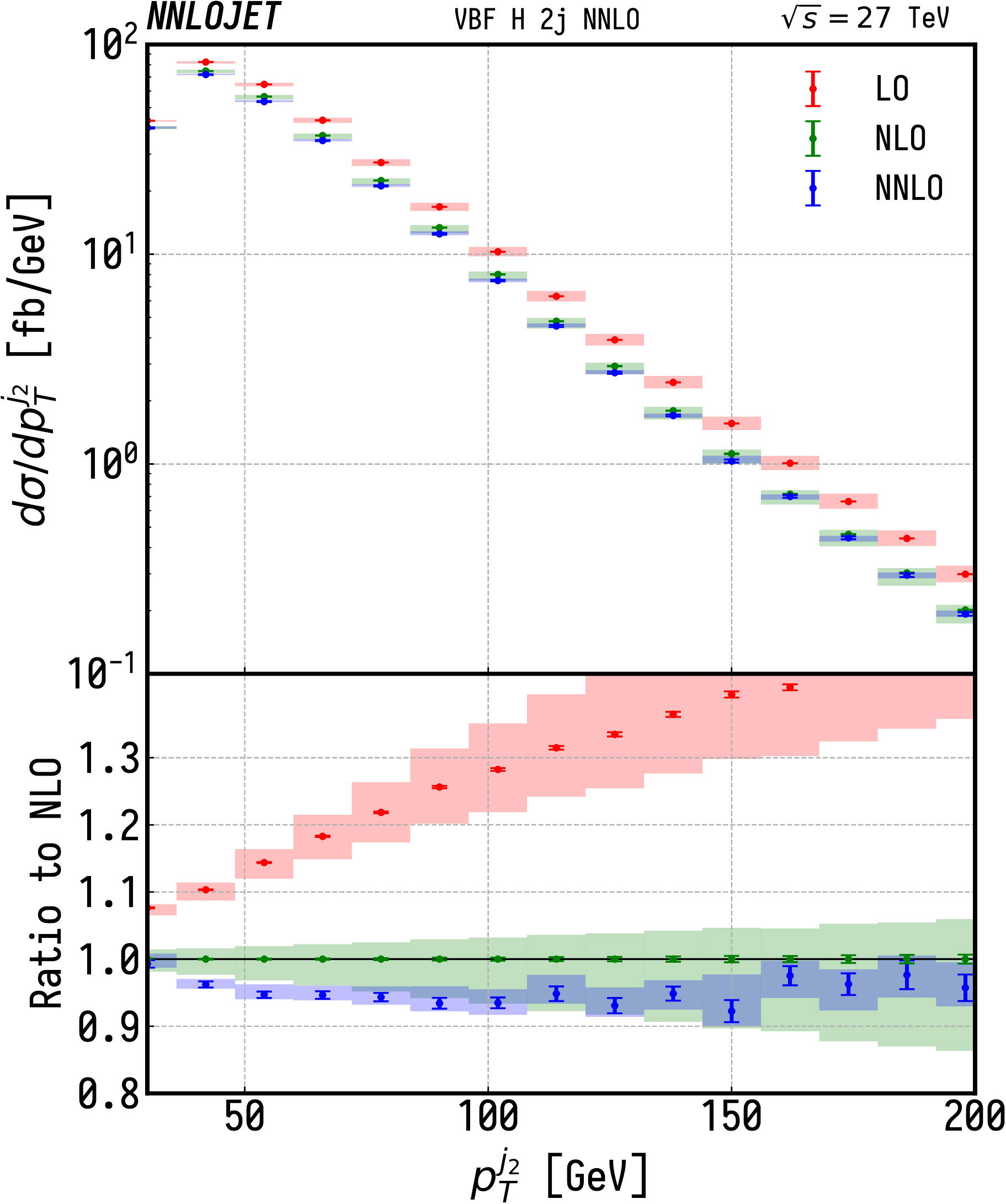}\hfill
    \includegraphics[width=0.46\textwidth]{\main/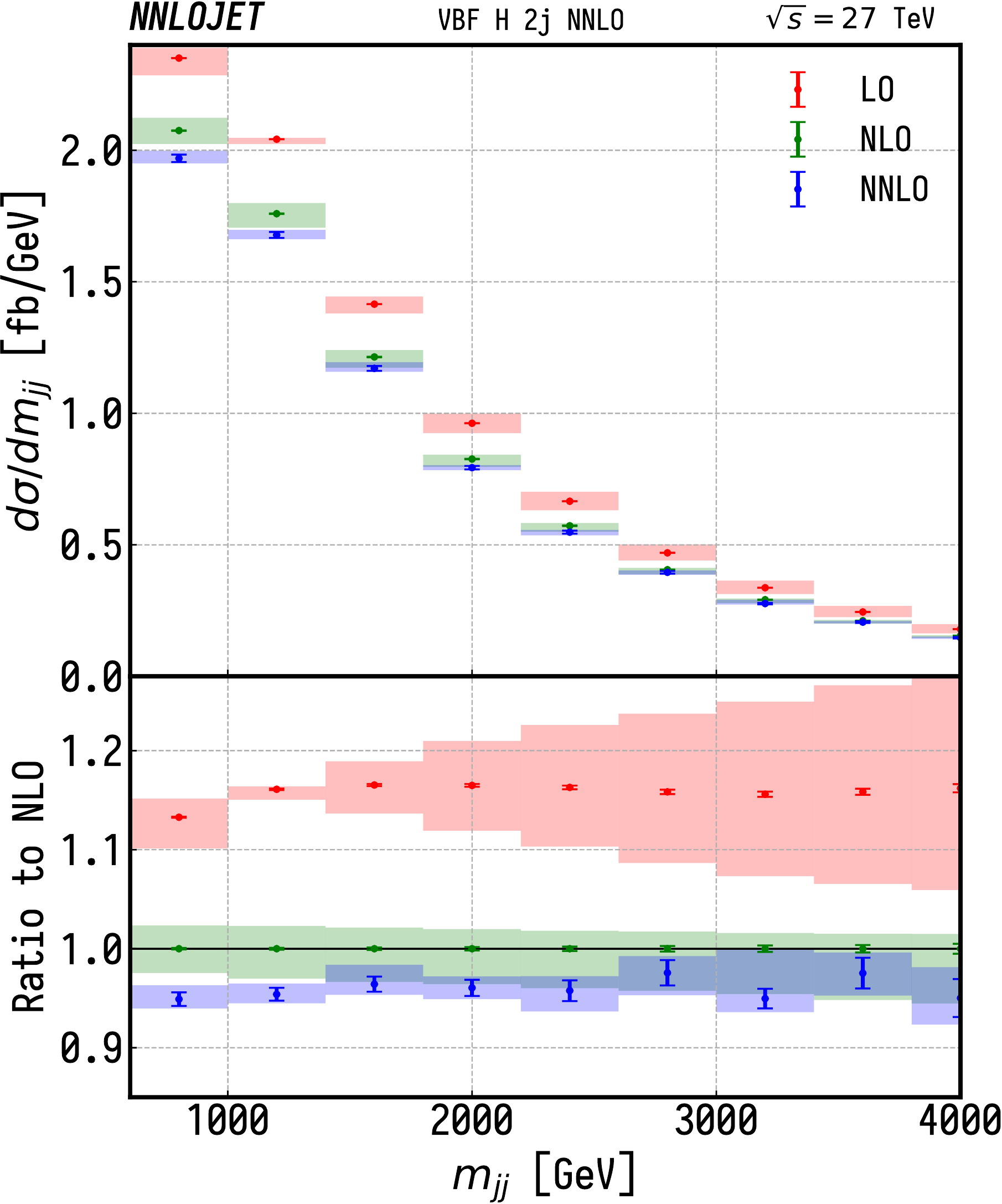}\hfill
    \includegraphics[width=0.46\textwidth]{\main/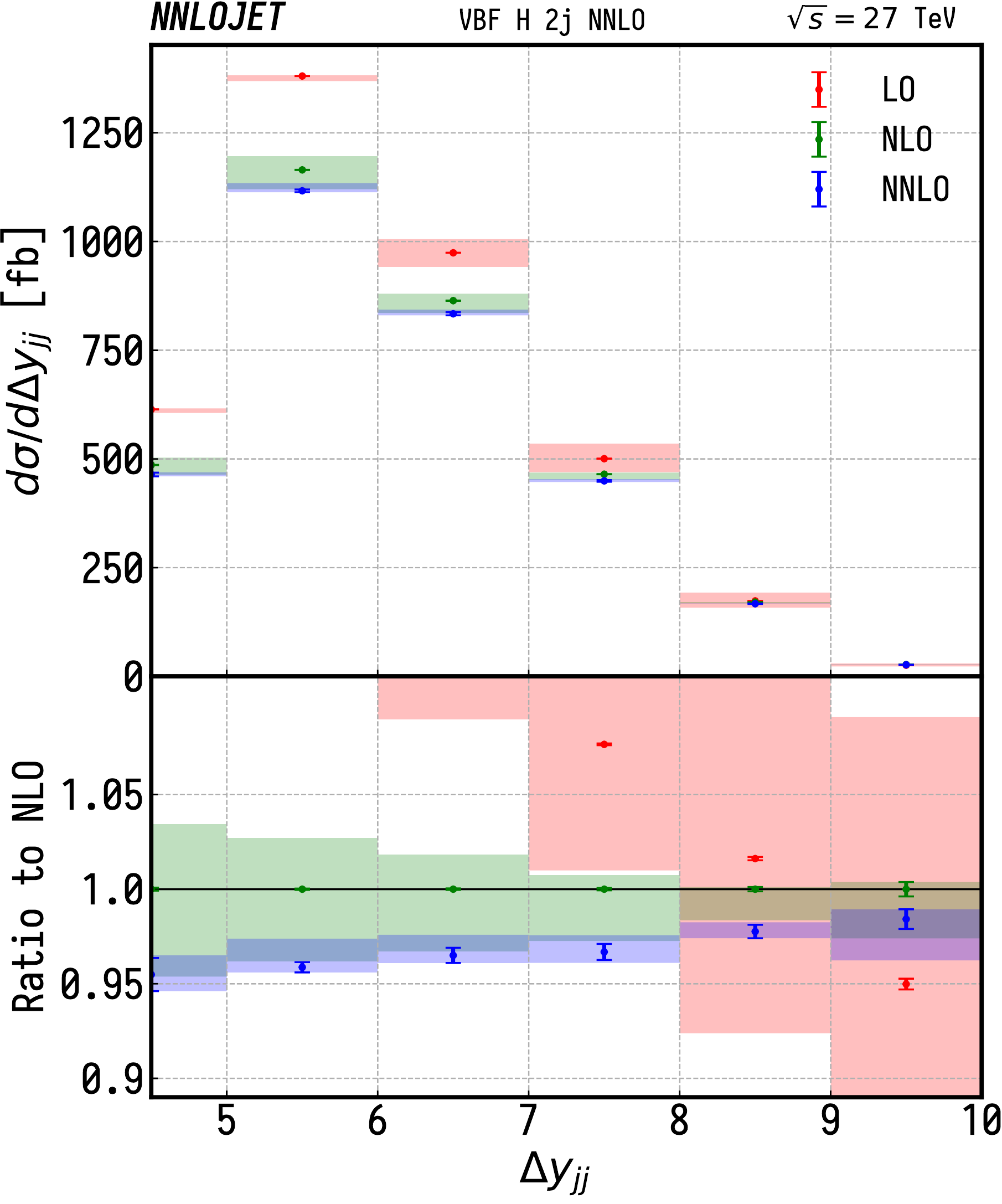}
    \caption{The top row shows the transverse momentum of the two leading jets ordered in rapidity at $\sqrt{s}=27~\TeV$ for tight VBF cuts. The bottom row depicts the kinematical variables for the dijet system they form. Note that NNLO corrections noticeably reduce the scale uncertainties for both observables over the entire range considered. NLO corrections are big for moderate and high transverse momentum with a scale uncertainty that grows with the transverse momentum. This behaviours is softened by the NNLO corrections. }
    \label{fig:dijet27}
\end{figure}

\begin{figure} 
		\includegraphics[width=0.5\textwidth]{\main/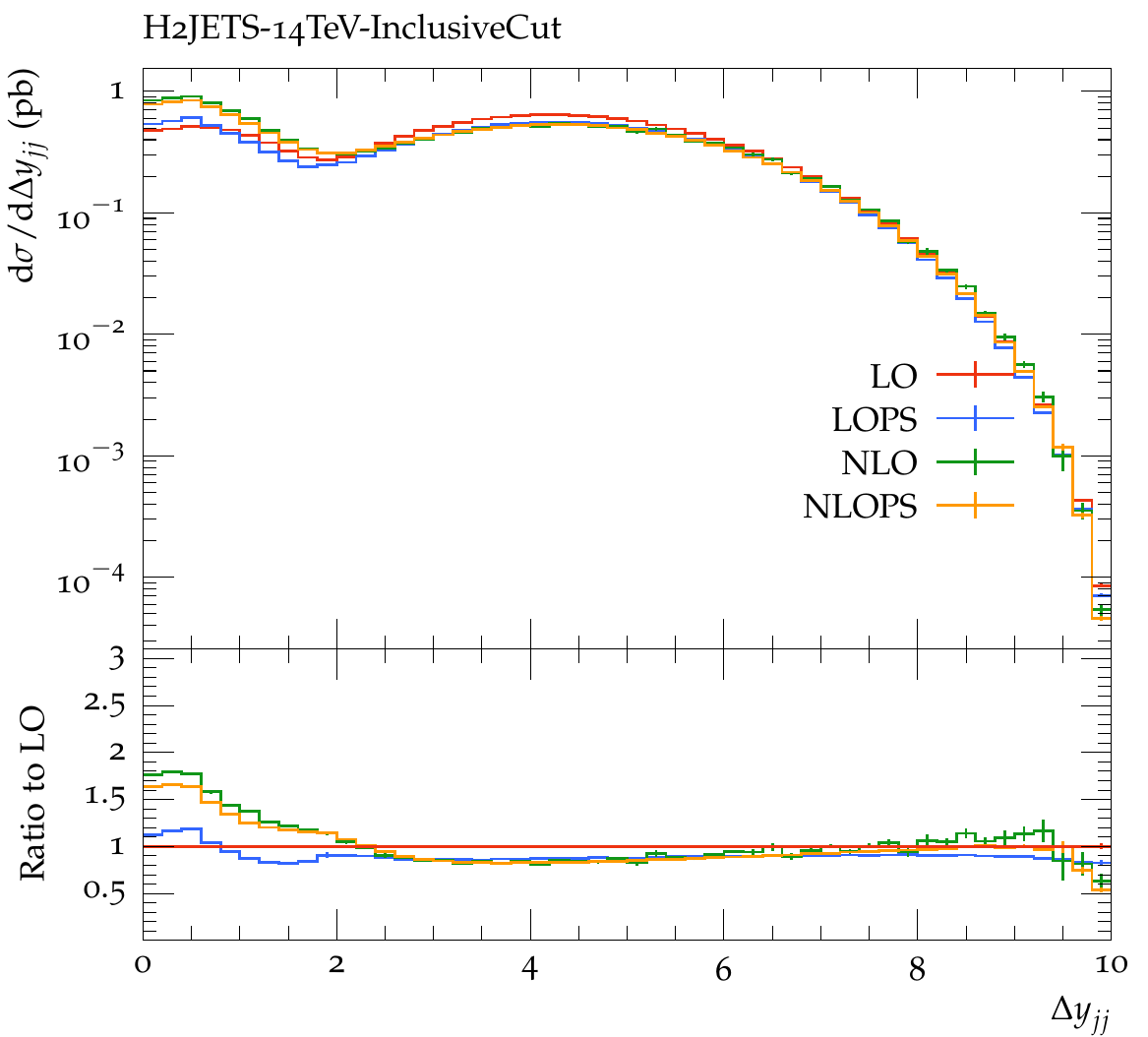} 
		\includegraphics[width=0.5\textwidth]{\main/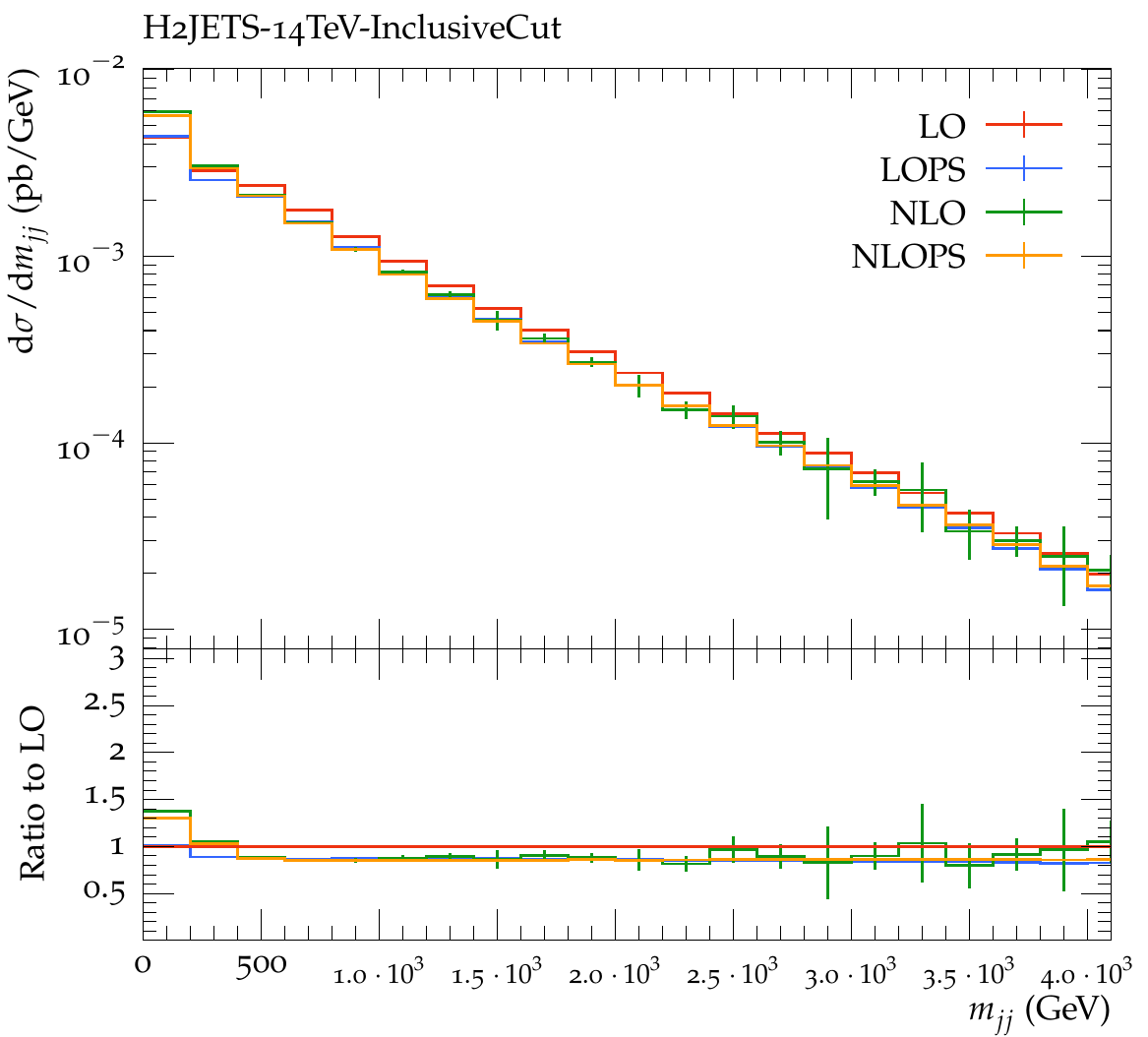} 
		\includegraphics[width=0.5\textwidth]{\main/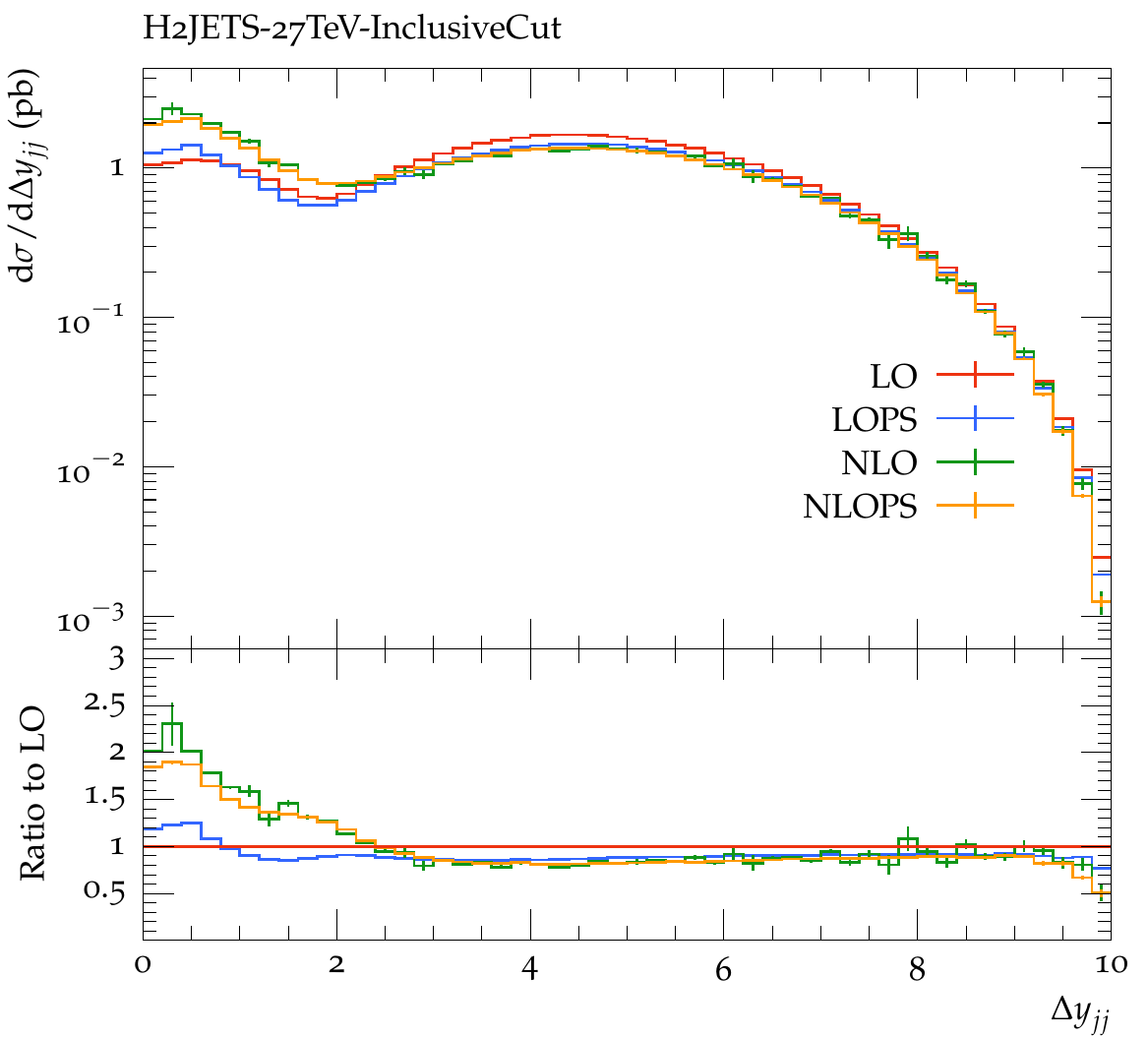} 
		\includegraphics[width=0.5\textwidth]{\main/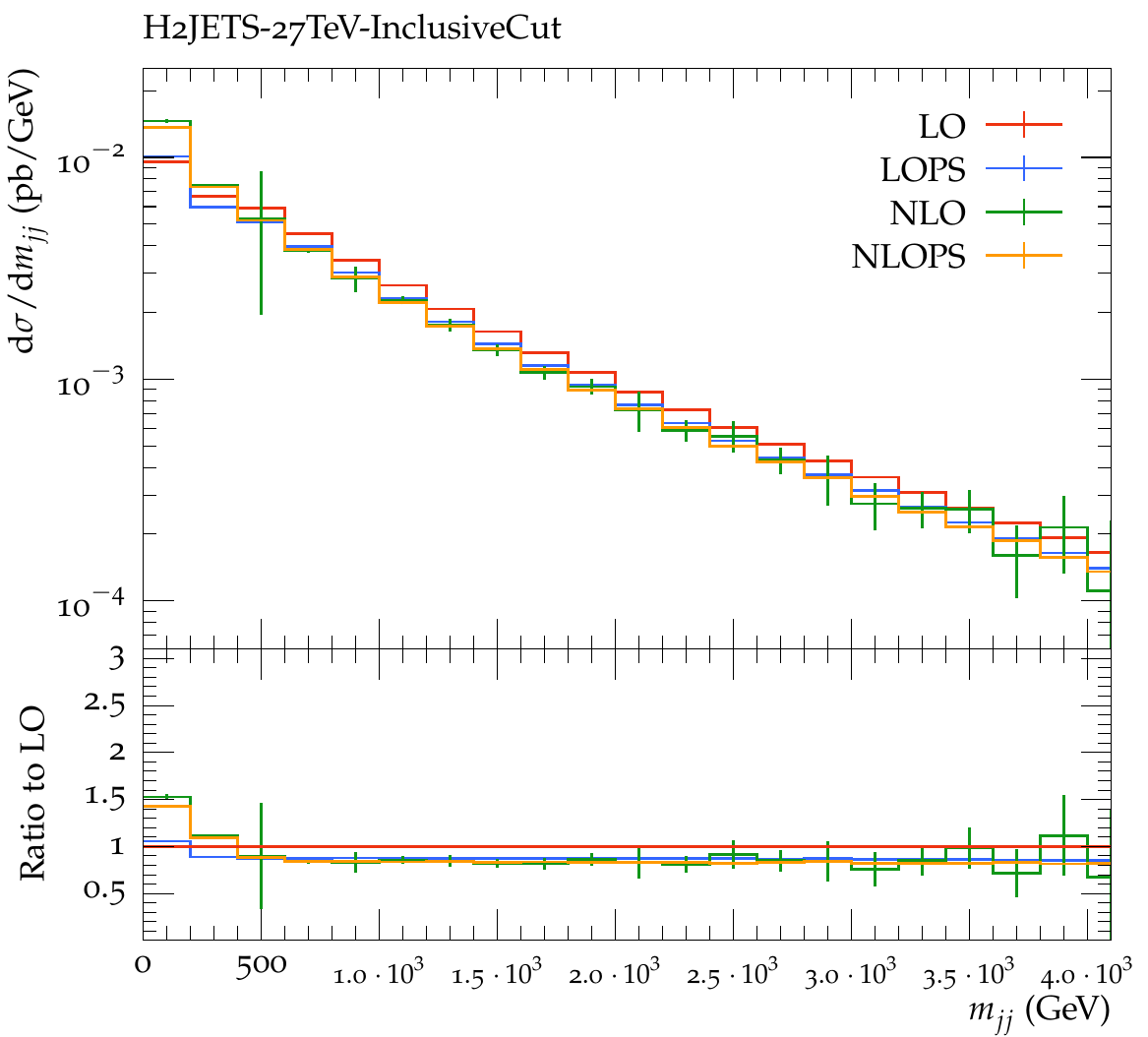} 
		\caption{Differential distributions of $\Delta y_{jj}$ and $m_{jj}$ at $\sqrt{s}=$ 14 TeV (top row) and $\sqrt{s}=$ 27 TeV (bottom row). {\mbox{\textsc{HJets++}}\xspace} matrix elements and inclusive cuts are used in the $H+2$ jets calculations. }
	\label{fig:h2jets_hjets} 
\end{figure}

\subsubsection*{Comparison of {\mbox{\textsc{HJets++}}\xspace} and {\mbox{\textsc{VBFNLO}}\xspace} for Higgs boson production}
The {\mbox{\textsc{HJets++~1.1}}\xspace} module implements~\cite{Campanario:2013fsa,Campanario:2013nca,Campanario:2014aia,PhysRevD.98.033003} electroweak Higgs boson plus two and three jet production. 
The one-loop integrals are computed using the techniques discussed in Ref.~\cite{Campanario:2011cs} and the colour algebra is performed using 
{\mbox{\textsc{ColorFull}}\xspace} ~\cite{Sjodahl:2014opa}.
For the VBF approximation, the matrix elements encoded in {\mbox{\textsc{VBFNLO}}\xspace} version $3.0$ beta $5$~\cite{Baglio:2014uba,Arnold:2011wj,Arnold:2008rz,Figy:2007kv} are used, with {\mbox{\textsc{Herwig 7}}\xspace} as the event generator~\cite{Platzer:2011bc,Bellm:2015jjp,Bellm:2017bvx,Bahr:2008pv}.
Jet reconstruction is performed on final state partons using the anti-$k_{\rm T}$ algorithm~\cite{Cacciari:2005hq} in the \FASTJET ~library~\cite{Cacciari:2011ma}. Simulated events are analyzed via {\mbox{\textsc{Rivet}}\xspace}~\cite{Buckley:2010ar}. 


For comparison plots of Higgs plus two jet calculations,  collider energies of $\sqrt{s}=14$ TeV and $\sqrt{s}=27$ TeV are considered. 
Two kinematic variables, namely the invariant mass, $m_{jj}$, and the spatial distribution, $\Delta y _{jj}$, of the two tag jets are chosen to present their differential distributions.  
Parton distribution functions PDF4\_LHC15\_nlo\_100 are used, while all other input parameters are the same as given at the beginning of Section \ref{sec:vbf}. 
Differential distributions for leading order, leading order plus parton shower, next-to-leading order, and next-to-leading order plus parton shower are shown in Fig.~\ref{fig:h2jets_hjets}, with the inclusive cuts defined in eq. (\ref{eq:inclusivecut}).
Comparison plots between two different matrix elements, {\mbox{\textsc{HJets++}}\xspace} and {\mbox{\textsc{VBFNLO}}\xspace} are shown in Fig.~\ref{fig:h2jets_14TeV}. 
{\mbox{\textsc{VBFNLO}}\xspace} uses the VBF approximation throughout, i.e. s-channel contributions such as $pp\to VH\to jjH$ production need to be added as separate processes. The comparison between {\mbox{\textsc{HJets++}}\xspace} and {\mbox{\textsc{VBFNLO}}\xspace} thus also serves to highlight the phase space regions where the VBF approximation is warranted.
 
The tight VBF cuts applied for $\sqrt{s}=14$ TeV are defined as
 \begin{align}
 p_{\mathrm{T}}^{j} > 30~\GeV, \qquad |y_j| < 5.0, \qquad |y_{j_1}-y_{j_2}| > 3.0, \qquad M_{jj} > 130~\GeV.
 \label{eq:30GEVVBFcuts14}
 \end{align}
 For $\sqrt{s}=27$ TeV comparison plots, the tight VBF cuts defined in eq. (\ref{eq:tight}) are used. 
 The {\mbox{\textsc{VBFNLO}}\xspace} calculation is consistent with the {\mbox{\textsc{HJets++}}\xspace}  calculation after applying the tight VBF cut.

\begin{figure}
	
		\includegraphics[width=0.5\textwidth]{\main/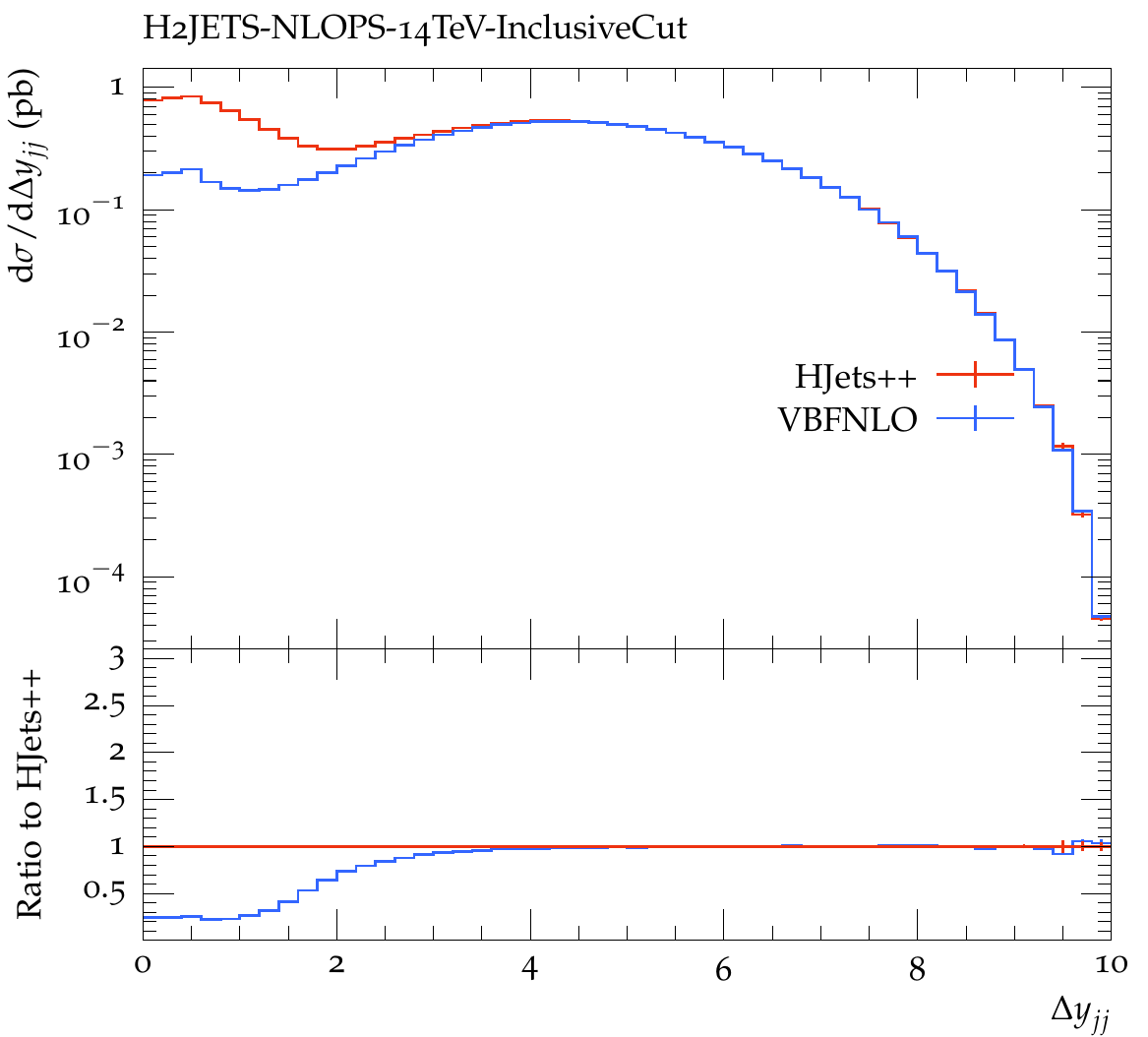} 
		\includegraphics[width=0.5\textwidth]{\main/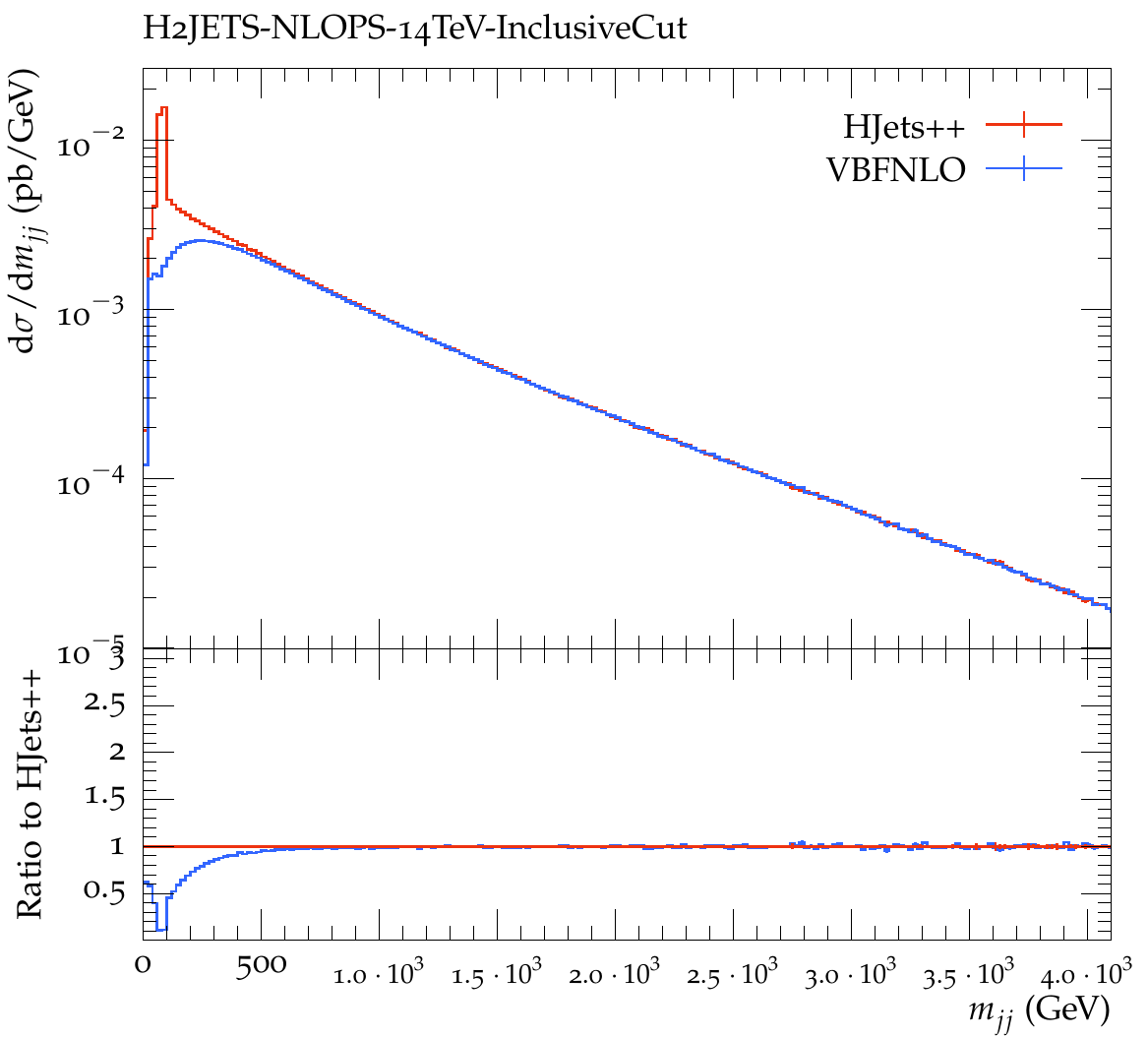} 
		\includegraphics[width=0.5\textwidth]{\main/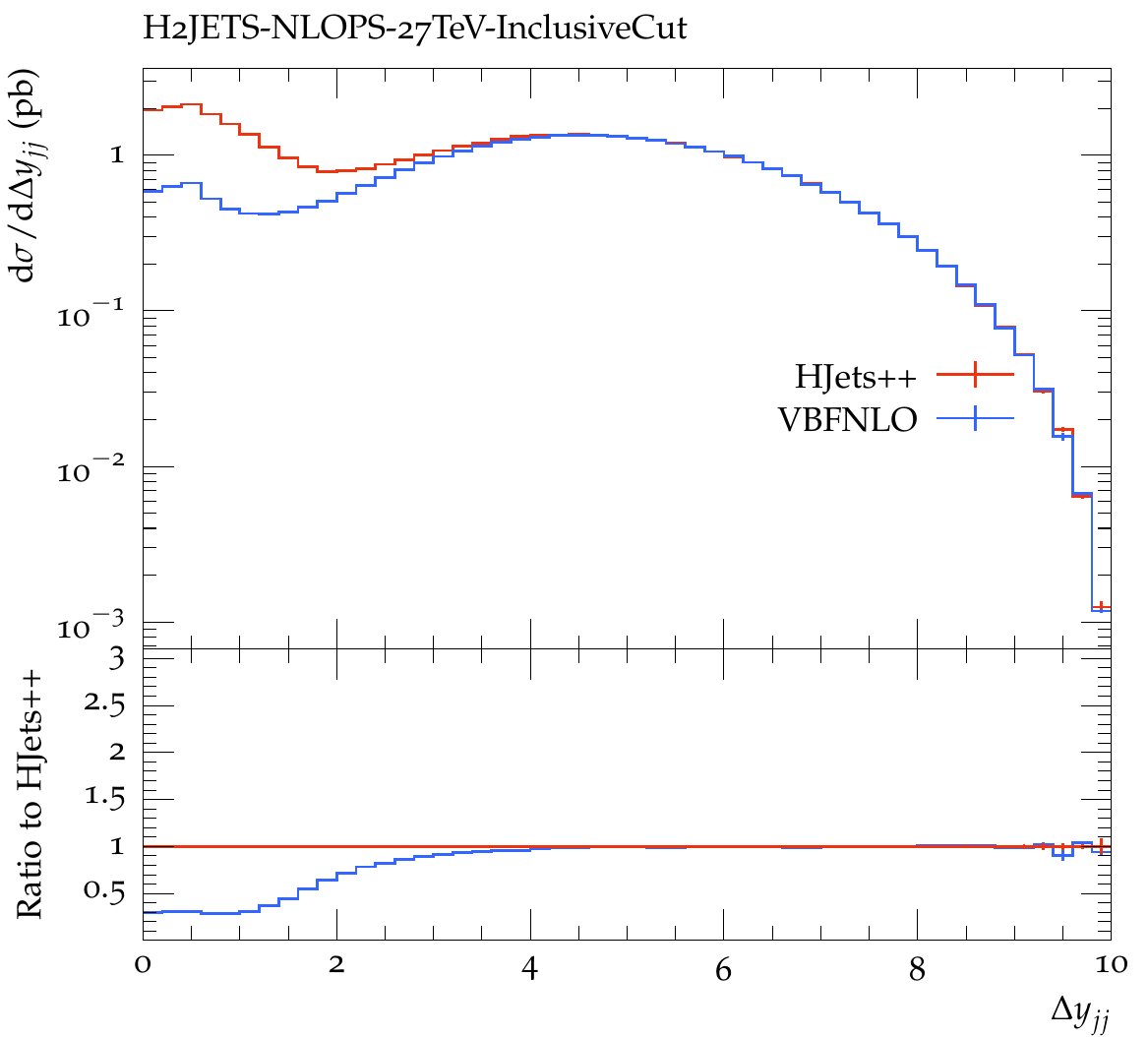} 
		\includegraphics[width=0.5\textwidth]{\main/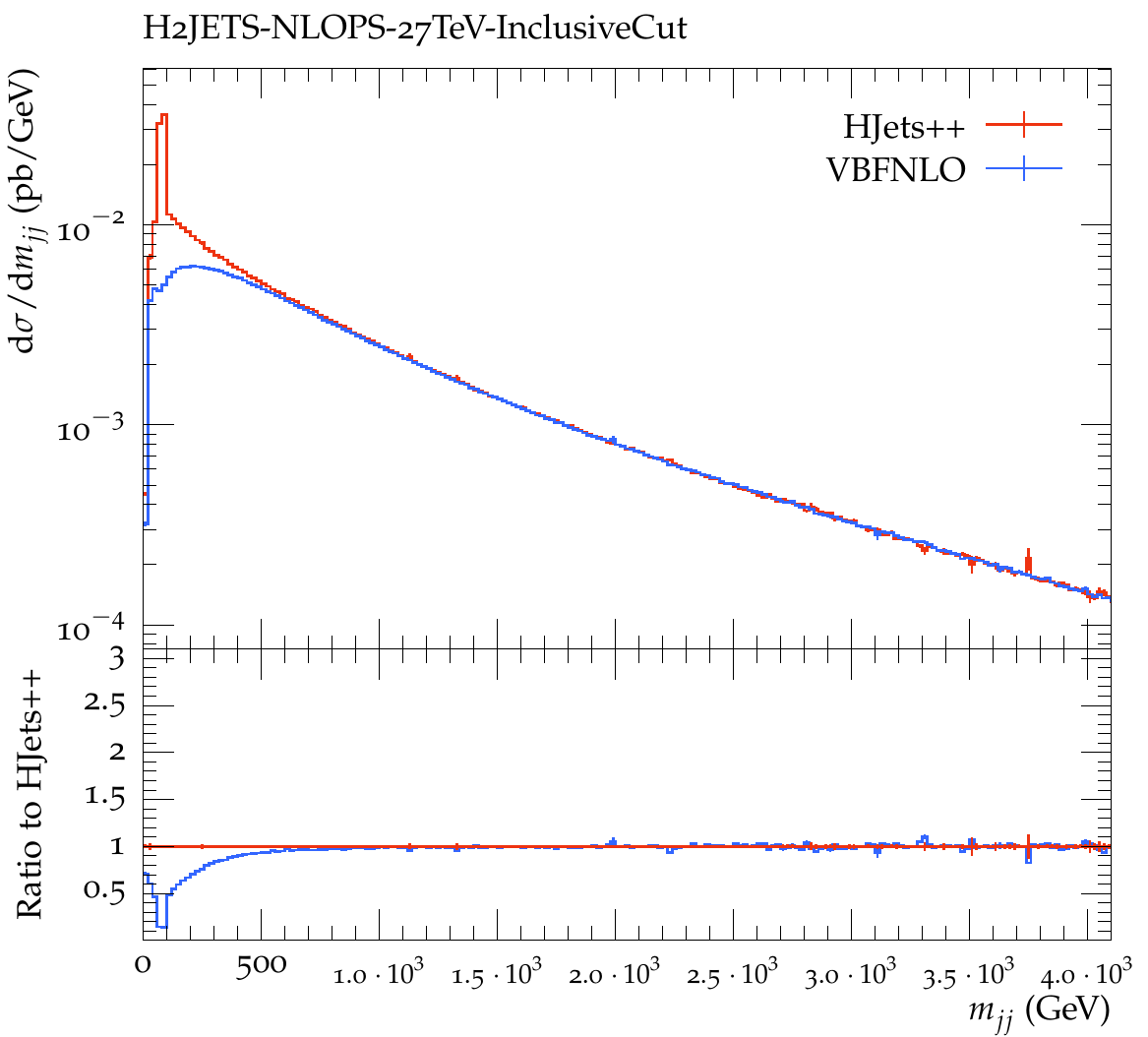}  
   
		\caption{The distributions of kinematic variables in $H+2$ jets at $\sqrt{s}=14~\TeV$ (top row) and $\sqrt{s}=27~\TeV$ (bottom row). Comparisons are between the {\mbox{\textsc{HJets++}}\xspace} matrix elements and the {\mbox{\textsc{VBFNLO}}\xspace} matrix elements at NLO plus parton shower. Plots indicate that both {\mbox{\textsc{HJets++}}\xspace} and {\mbox{\textsc{VBFNLO}}\xspace} calculations agree once the tight VBF cuts are applied. }
	\label{fig:h2jets_14TeV} 
\end{figure}

Fig.~\ref{fig:h3jet-27tev} shows differential distributions of kinematics variables for the NLO full and approximate results at $\sqrt{s}=14~\TeV$ and
$\sqrt{s}=27~\TeV$.
The comparison of the full and approximate calculations are shown in the second and third rows of Fig.~\ref{fig:h3jet-27tev} for tight VBF cuts for
the transverse momentum of the third jet $p^{j_3}_{\rm T}$ and the centrality of the third jet $y_{j_{3}}^{\star}=(y_{j_{3}}-\frac{1}{2}(y_{j_{1}}+y_{j_{2}}))/|y_{j_{1}}-y_{j_{2}}|$. 
For the $\sqrt{s}=27~\TeV$ tight VBF cuts ($\Delta y_{jj}>4.5$, $m_{jj}>600~\GeV$, and $y_{j_1} \cdot y_{j_2}<0$), one observes excellent agreement between the full and approximate calculation.
For the $\sqrt{s}=14~\TeV$ tight VBF cuts ($\Delta y_{jj}>3.0$, $m_{jj}>130~\GeV$, and $y_{j_1} \cdot y_{j_2}<0$), the full and approximate calculations still do not converge.  However, for $\Delta y_{jj}>4.0$ or $m_{jj}>600~\GeV$ the full and approximate calculations would compare quite well.

%
%

\begin{figure}
    \includegraphics[width=0.5\textwidth]{\main/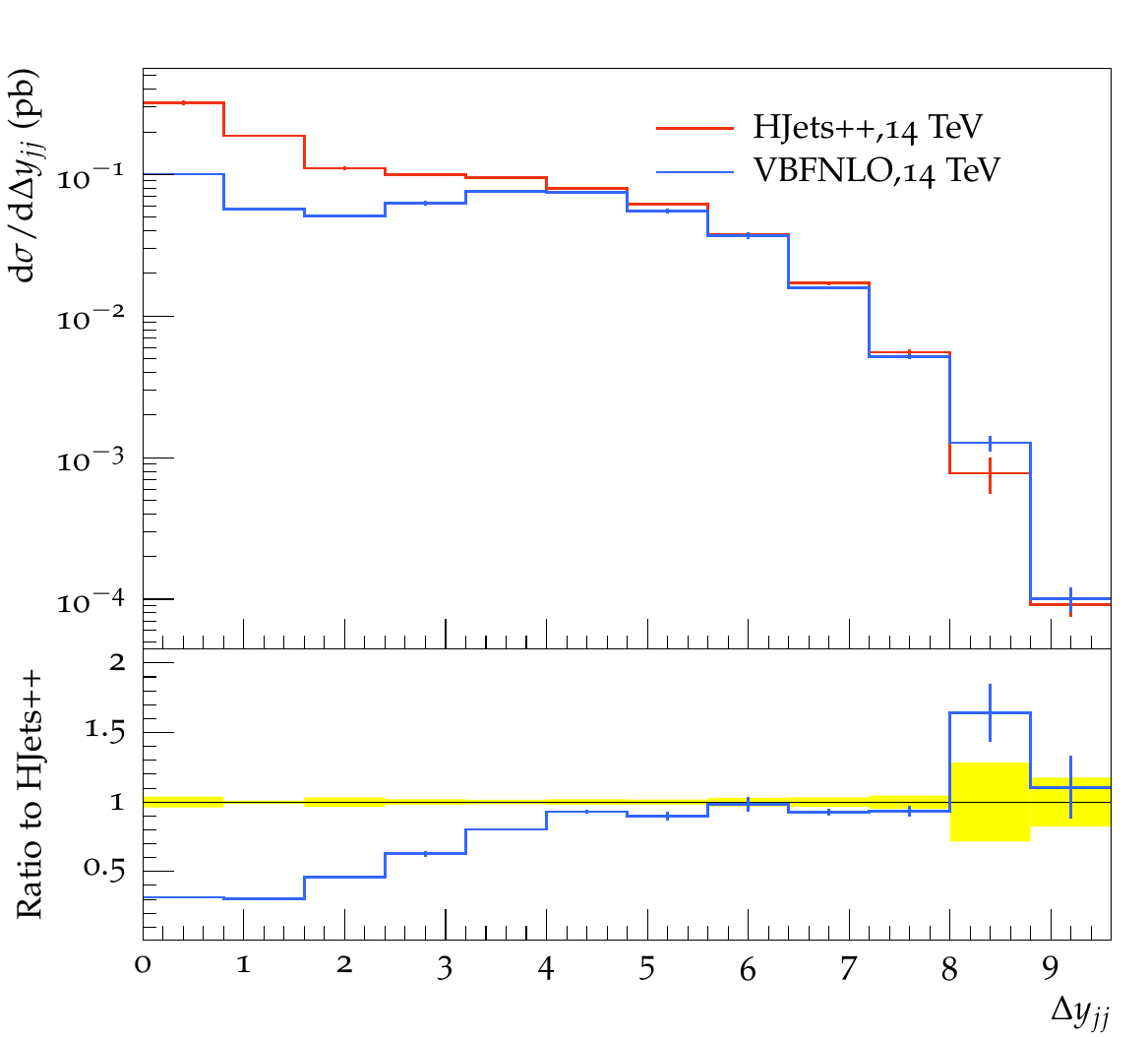}\hfill
    \includegraphics[width=0.5\textwidth]{\main/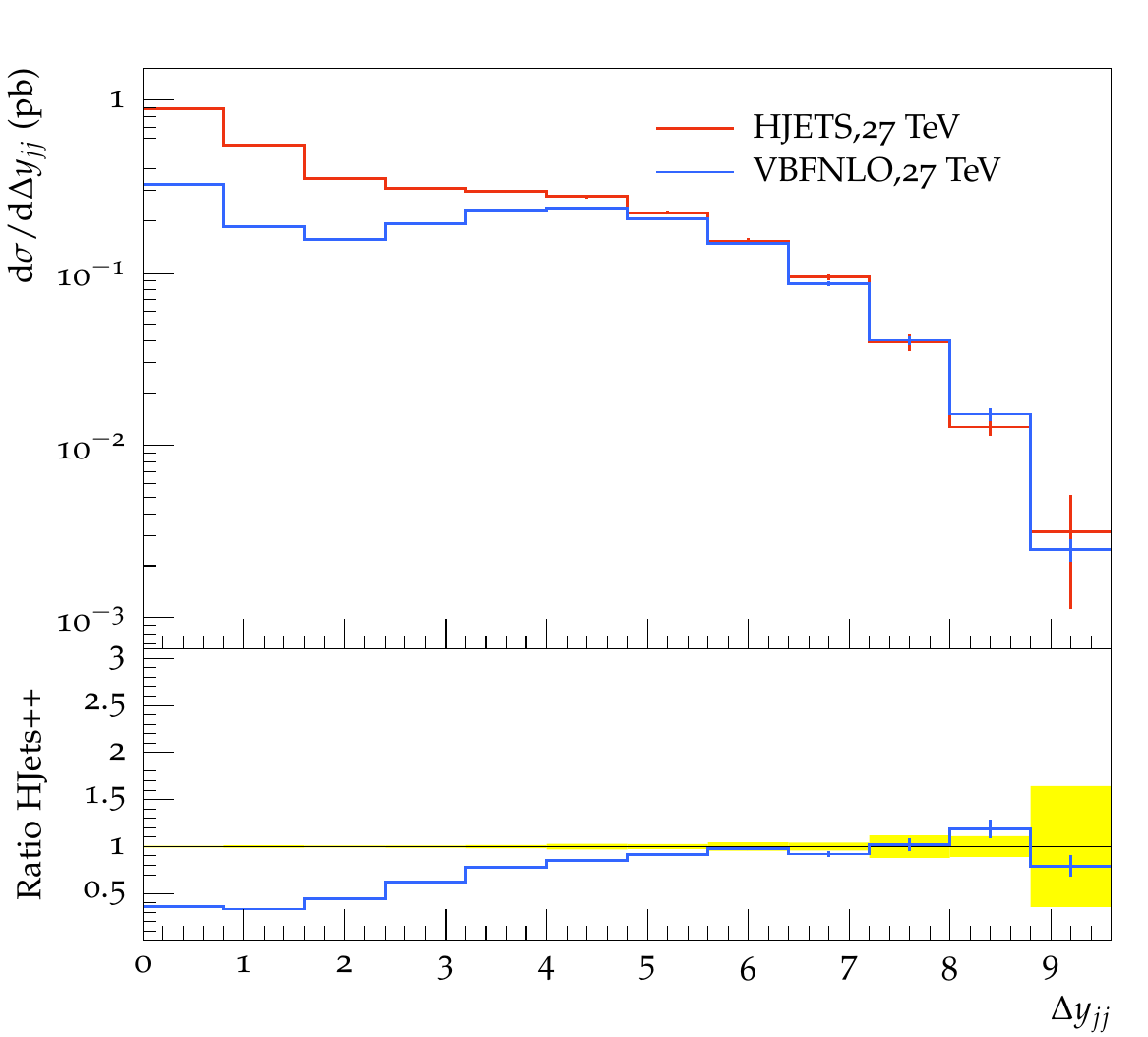} \
    \includegraphics[width=0.5\textwidth]{\main/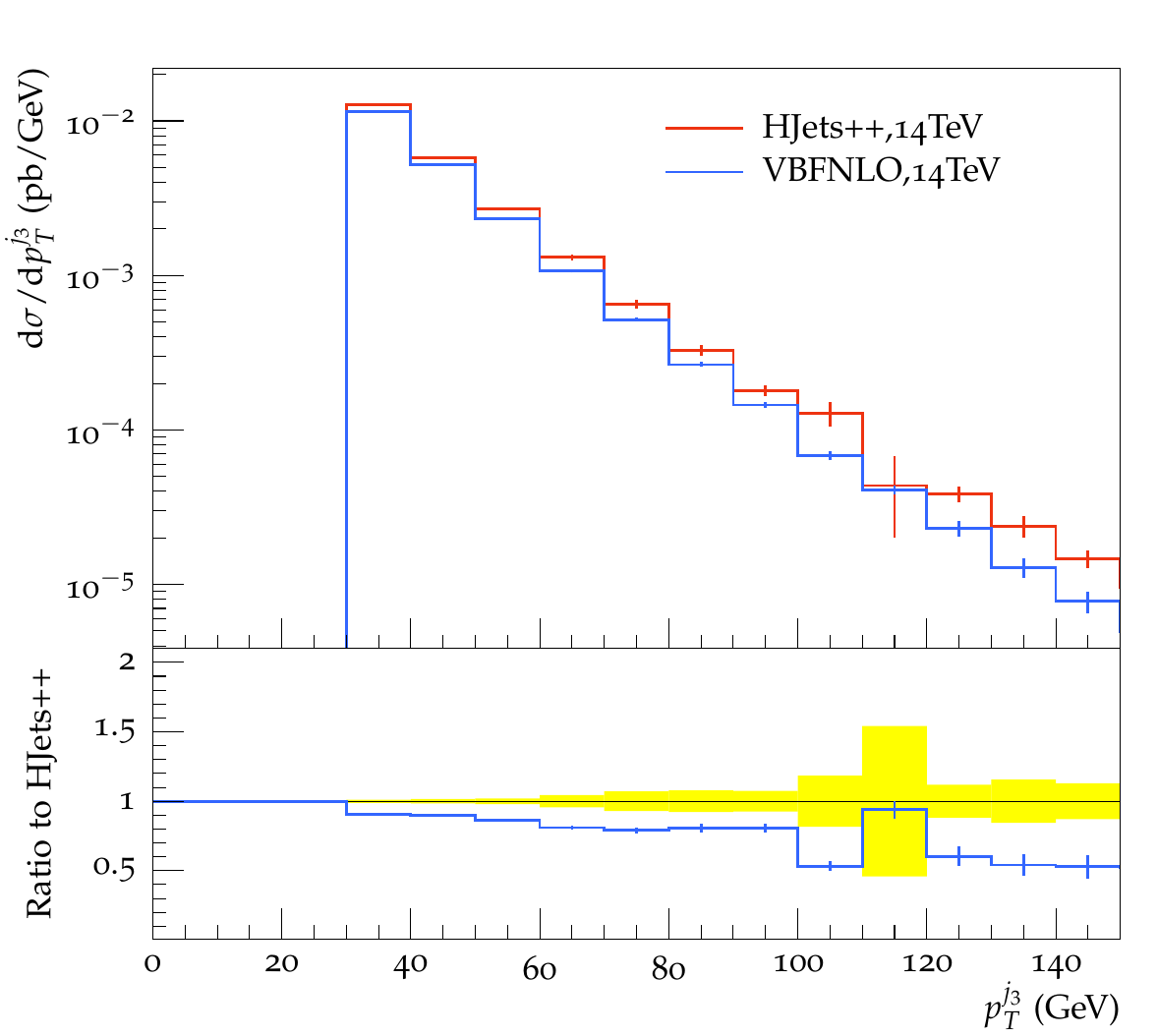}\hfill
    \includegraphics[width=0.5\textwidth]{\main/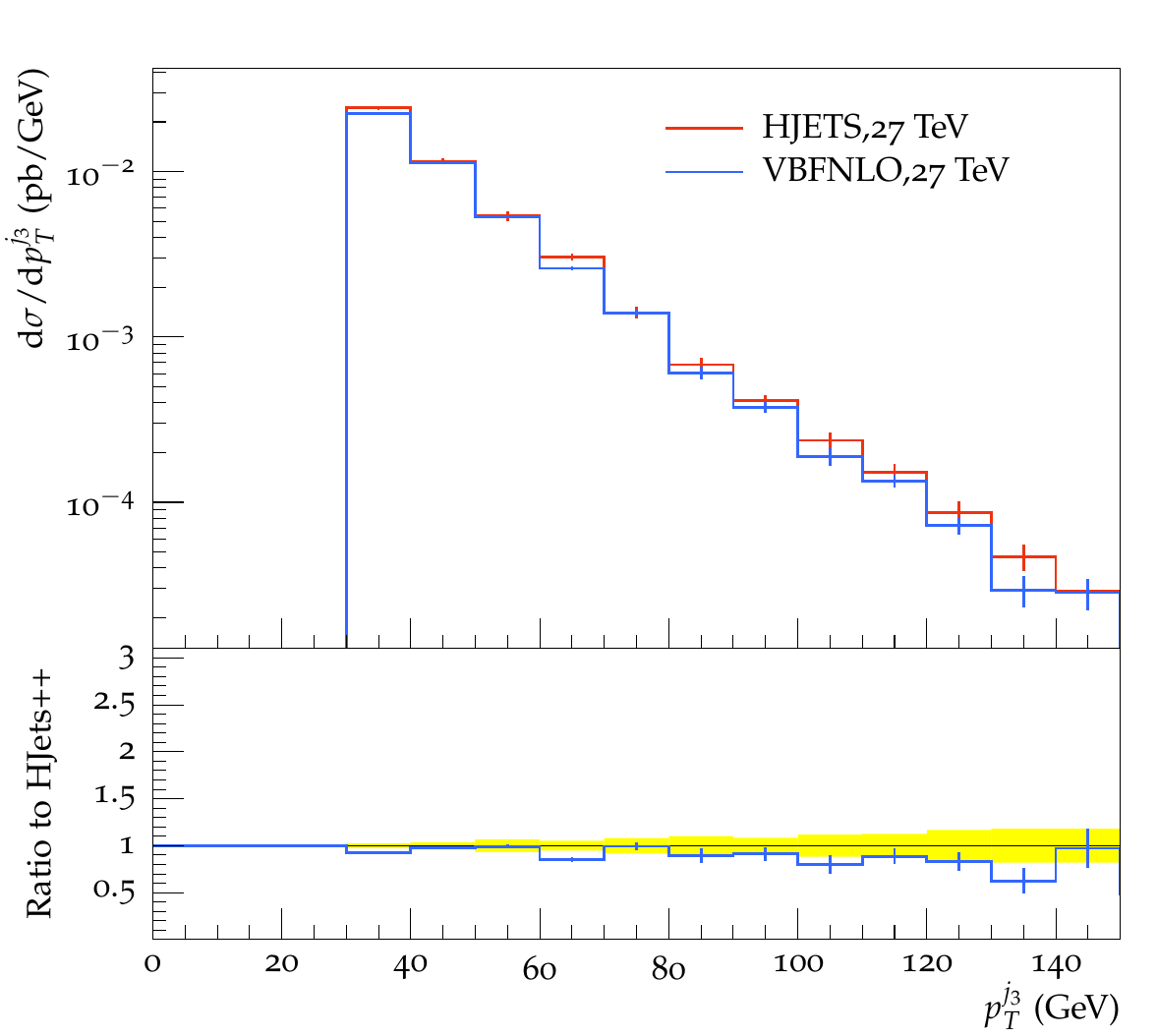} \
    \includegraphics[width=0.5\textwidth]{\main/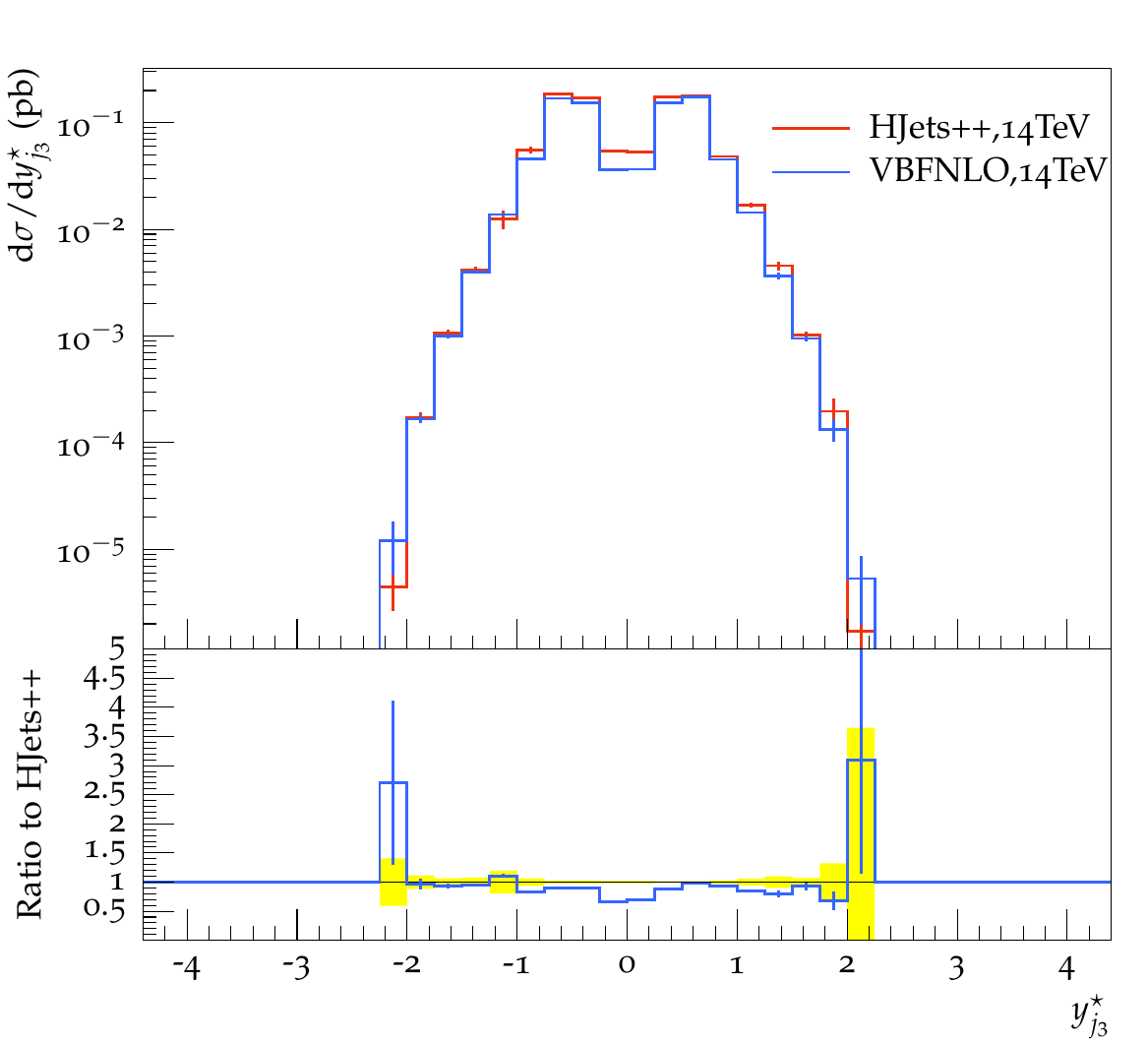} \hfill
    \includegraphics[width=0.5\textwidth]{\main/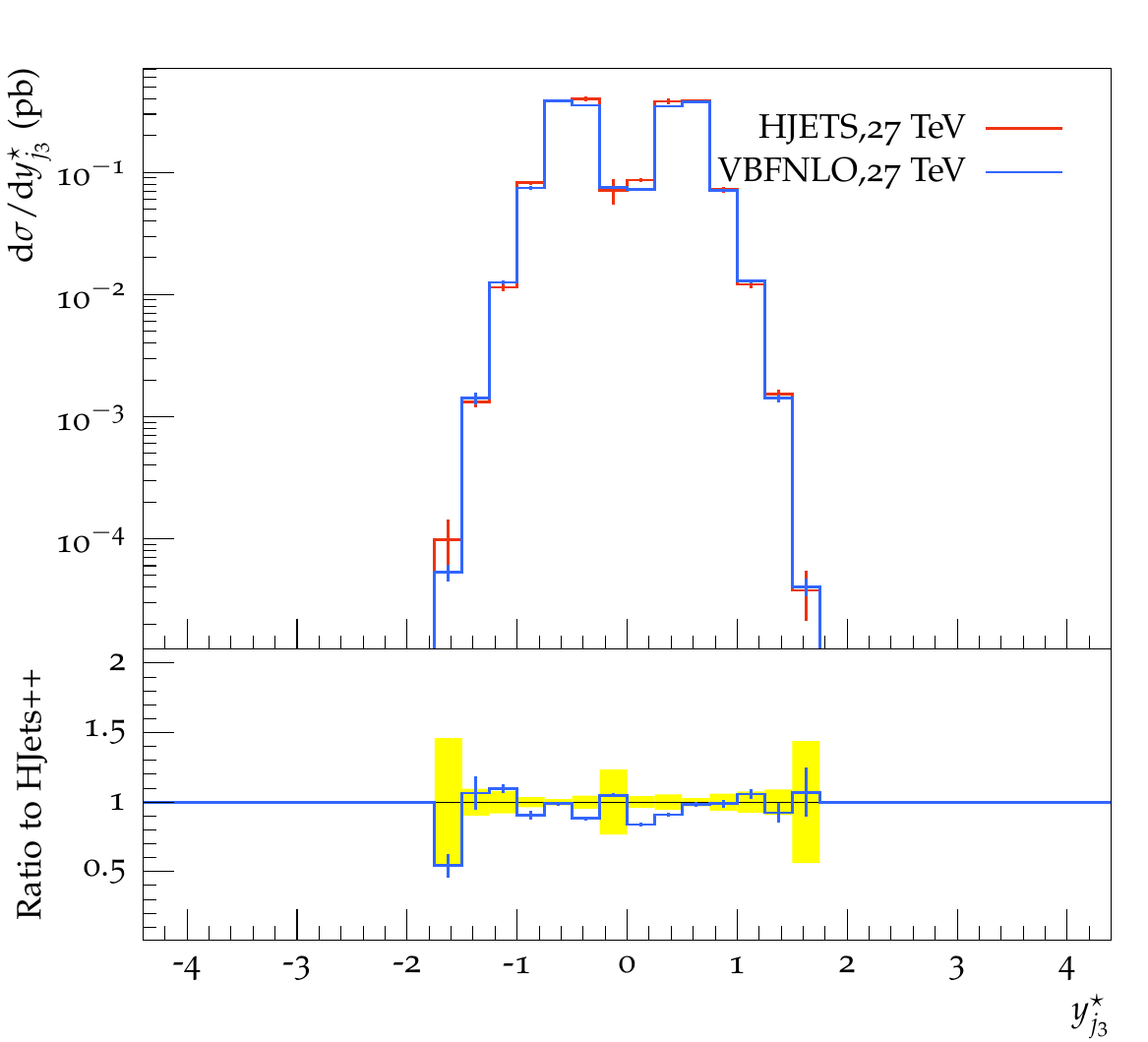}
    \caption{ Kinematics distributions for $H+3$ jet production at NLO for the full result ({\mbox{\textsc{HJets++}}\xspace}) 
    and the approximate result ({\mbox{\textsc{VBFNLO}}\xspace}) for $\sqrt{s}=14~\TeV$ (first column ) and $\sqrt{s}=27~\TeV$ (second column).  The kinematic distribution $\Delta y_{jj}$ (top row) is shown 
    for inclusive selection cuts. The kinematic distributions for $p_{\rm T}^{j_3}$ and $y_{j_3}^{\star}$ are shown for VBF tight selection cuts.  } 
	\label{fig:h3jet-27tev}
\end{figure}

\FloatBarrier

\subsection{Vector boson scattering processes}

\renewcommand\xspace{}

\providecommand{\alphas}{\alpha_{\mathrm s}}




\providecommand{\Pj}{\ensuremath{\text{j}}\xspace}
\providecommand{\Pt}{\ensuremath{\text{t}}\xspace}
                                    
\providecommand{\Mt}{\ensuremath{m_\Pt}\xspace}
\providecommand{\MWOS}{\ensuremath{M_\text{W}^{\text{OS}}}\xspace}
\providecommand{\MZOS}{\ensuremath{M_{\text{Z}}^{\text{OS}}}\xspace}
\providecommand{\Gt}{\ensuremath{\Gamma_\Pt}\xspace}
\providecommand{\GH}{\ensuremath{\Gamma_{\PH}}\xspace}
\providecommand{\GZOS}{\ensuremath{\Gamma_{\text{Z}}^\text{OS}}\xspace}
\providecommand{\GWOS}{\ensuremath{\Gamma_{\text{W}}^\text{OS}}\xspace}

\providecommand{\ptsub}[1]{\ensuremath{p_{\text{T},#1}}\xspace}

\providecommand{\MVOS}{\ensuremath{M_{V}^\text{OS}}\xspace}%
\providecommand{\GVOS}{\ensuremath{\Gamma_{V}^\text{OS}}\xspace}%

\providecommand{\fbinv} {\mbox{\ensuremath{\,\text{fb}^\text{$-$1}}}\xspace} 
\providecommand{\TeV}{\ensuremath{\,\text{Te\hspace{-.08em}V}}\xspace}
\providecommand{\GeV}{\ensuremath{\,\text{Ge\hspace{-.08em}V}}\xspace}

\providecommand{\zepp}{\ensuremath{\left|\eta_{3\ell} - \frac{1}{2}(\eta_{j_{1}} + \eta_{j_{2}})\right|}}
\providecommand{\WZEW}{\ensuremath {\mathrm{EW-}WZjj} }
\providecommand{\WZQCD}{\ensuremath {\mathrm{QCD-}WZ} }
\providecommand{\ZZEW}{\ensuremath {\mathrm{EW-}ZZ} }
\providecommand{\ZZQCD}{\ensuremath {\mathrm{EW-}ZZ} }
\providecommand{\Pt}{\ensuremath {\mathrm {p_{T}}}}
\providecommand{\PT}{\ensuremath {\mathrm {p_{T} }}}
\providecommand{\Ptj}{\ensuremath {\mathrm {p_{T}^{jet}}}}
\providecommand{\Ptl}{\ensuremath {\mathrm {p_{T}^{l}}}}
\providecommand{\Etmiss}{\ensuremath {\mathrm {E_{\rm{T}}^{miss}}}}
\providecommand{\Eta}{\ensuremath {\mathrm {\eta}}}
\providecommand{\Etaj}{\ensuremath{\mathrm {\eta^{jet}}}}
\providecommand{\Etal}{\ensuremath{\mathrm {\eta^{l}}}} 
\providecommand{\MZ}{\ensuremath{\mathrm {M_{Z}}}}
\providecommand{\MW}{\ensuremath{\mathrm {M_{W}}}}
\providecommand{\Mjj}{\ensuremath{\mathrm {M_{jj}}}}
\providecommand{\WZtau}{\ensuremath{\mathrm WZ\rightarrow\tau X}}
\providecommand{\sovsqrtb}{\ensuremath{\mathrm S/\sqrt{B}}}
\providecommand{\sovsqrtspb}{\ensuremath{\mathrm S/\sqrt{S+B}}}
\providecommand{\sovsqrtspbA}{\ensuremath{\mathrm S/\sqrt{S+B+\Delta B(5\%)}}}
\providecommand{\sovsqrtspbB}{\ensuremath{\mathrm S/\sqrt{S+B+\Delta B(10\%)}}}
\providecommand{\cthsZ}{\ensuremath{\mathrm \cos\theta^{*}_{Z}}}
\providecommand{\cthsW}{\ensuremath{\mathrm \cos\theta^{*}_{W}}}
\providecommand{\fb}{\ensuremath{\mathrm {fb^{-1}}}}
\newcommand{\ifb}{fb$^{-1}$}
\newcommand{\ZZjj}{\ensuremath{\PZ\PZ\mathrm{jj}}\xspace} 
\newcommand{\SHERPA}{{\mbox{\textsc{Sherpa}}\xspace}} 
\newcommand{\MADGRAPH}{\mbox{\textsc{MadGraph}}\xspace}
\newcommand{\MCATNLO}{\textsc{mc@nlo}} 
\newcommand{\MCFM}{\textsc{mcfm}} 
\newcommand{\llll}{\ensuremath{4\ell}~} 
\newcommand{\mZOne}{\ensuremath{m_{\mathrm{Z_1}}}}
\newcommand{\mZPDG}{\ensuremath{m_{\mathrm{PDG}}}~}
\newcommand{\mZTwo}{\ensuremath{m_{\mathrm{Z_2}}}}
\newcommand{\mjj}{\ensuremath{m_{jj}}}
\newcommand{\detajj}{\ensuremath{\vert \Delta\eta_{jj} \vert}}
%



The study of the scattering of two massive vector bosons $V=W,Z$ (vector boson scattering, VBS) provides a key opportunity to probe the nature of the electroweak symmetry breaking (EWSB) mechanism as well as physics beyond the Standard Model (SM)~\cite{Bagger:1993zf,Bagger:1995mk}. It is still unknown whether the discovered Higgs boson~\cite{Chatrchyan:2012xdj} preserves unitarity of the longitudinal $VV$ scattering amplitude at all energies, or if other new physics processes are involved~\cite{Veltman:1976rt,Lee:1977yc,Lee:1977eg,P5:2014pwa,Aleksan:1628377}. In the VBS topology, two incoming quarks radiate bosons which interact, yielding a final state of two jets from the outgoing quarks, and two massive bosons which decay into fermions. This final state can be the result of $VVjj$ electroweak (EW) production with and without a scattering topology, or of processes involving the strong interaction.

\subsubsection{Measurements of $W^{\pm}W^{\pm}$ scattering and extraction of the longitudinal scattering component}

\newcommand*{\eee}{\ensuremath{eee}\xspace}
\newcommand*{\mmm}{\ensuremath{\mu\mu\mu}\xspace}
\newcommand*{\eem}{\ensuremath{ee\mu}\xspace}
\newcommand*{\mme}{\ensuremath{\mu\mu e}\xspace}
\newcommand*{\pt}{\ensuremath{p_{\rm T}}\xspace}
\newcommand*{\ETMISS}{\ensuremath{E_{\rm{T}}^{miss}}\xspace}
\newcommand*{\zepl}{\ensuremath{Z_{3\ell}}\xspace}
\newcommand*{\zgamma}{\ensuremath{Z/\gamma}\xspace}
\newcommand*{\TTbar}{\ensuremath{t\bar{t}}\xspace}
\newcommand*{\PYTHIA}{\ensuremath{\sc Pythia}\xspace}
\newcommand*{\POWHEG}{\ensuremath{\sc POWHEG}\xspace}

\newcommand{\cutoff}[1]{}  
\def\ssWW{\ensuremath{W^{\pm}W^{\pm}jj}\xspace}
\def\ssWWLL{\ensuremath{W_{\rm{L}}^{\pm}W_{\rm{L}}^{\pm}jj}\xspace}
\def\pT{\ensuremath{p_{\mathrm{T}}}\xspace}
\def\WWemu{\ensuremath{WW \to e \mu}\xspace}
\def\metrel{\ensuremath{E_{\mathrm{T,~Rel}}^{\mathrm{miss}}}\xspace}
\def\ptmiss{\ensuremath{p_{\mathrm{T}}^{\mathrm{miss}}}\xspace}
\def\abseta{\ensuremath{|\eta|}\xspace}
\def\Zsig{\ensuremath{Z_{\sigma}}\xspace}
\def\precision{\ensuremath{\frac{\Delta \mu}{\mu}}\xspace}
\def\RpT{\ensuremath{R_{p_{\rm T}}}\xspace}
\def \sumpt{\ensuremath{\Sigma p_{\rm T}}\xspace}
\def \sumet{\ensuremath{\Sigma E_{\rm T}}\xspace}
\def\deltaR{\ensuremath{\Delta R}\xspace}
\def \ee{\ensuremath{ee}\xspace}
\def \emu{\ensuremath{e\mu}\xspace}
\def \me{\ensuremath{\mu e}\xspace}
\def \mm{\ensuremath{\mu\mu}\xspace}
\def \eep{\ensuremath{e^{\pm}e^{\pm}}\xspace}
\def \emp{\ensuremath{e^{\pm}\mu^{\pm}}\xspace}
\def \mep{\ensuremath{\mu^{\pm} e^{\pm}}\xspace}
\def \mmp{\ensuremath{\mu^{\pm}\mu^{\pm}}\xspace}
\def \nsigopt{2431\xspace}
\def \nbkgopt{1460\xspace}
\def \xsecopt{\ensuremath{16.94~\pm~0.36~(\mathrm{stat})~\pm~0.53~(\mathrm{theory})~\pm~0.78~(\mathrm{sys})}\xspace}
\def \LLsig{\ensuremath{1.8\sigma}\xspace}
\def \LLxsec{\ensuremath{16.94~\pm~5.79~(\mathrm{stat})~\pm~0.68~(\mathrm{theory})~\pm~7.92~(\mathrm{sys})}\xspace}
\def \xsecprecision{\ensuremath{6\%}\xspace}

\def\nnpdf{\textsc{NNPDF3.0}}
\def\mgamcatnlo{\mbox{\textsc{Madgraph5\_aMC@NLO}}\xspace}
\def\pythia{\textsc{PYTHIA}}
\def\pythiaEight{\mbox{\textsc{Pythia v8}}\xspace}


 With the largest cross section ratio of electroweak to strong production~\cite{Accomando:2005hz,Zhu:2010cz}, events with $W^{\pm}W^{\pm}$ plus two jets (\ssWW) provide one of the best opportunities to study the scattering of two vector bosons. ATLAS and CMS have both observed the EW process at 13 TeV with significances of 6.9 $\sigma$ and 5.5 $\sigma$, respectively~\cite{ATLAS-CONF-2018-030,Sirunyan:2017ret}.
\\
\\
This section describes the prospects for the study of \ssWW\ at $\sqrt{s} = 14$ TeV at the HL-LHC, with the HL-LHC upgraded ATLAS and CMS detectors~\cite{ATL-PHYS-PUB-2018-052,CMS-PAS-FTR-18-005}. Results are presented for a range of integrated luminosities ${\cal L}$, from 300\fbinv through 8000\fbinv, where the first value corresponds to one year of data taking, and the latter to 10 years of combined data sets collected by the ATLAS and CMS experiments in the most optimistic scenario.
\\
\\
In both ATLAS and CMS analyses, the signal (VBS and non-VBS EW) and background (QCD) \ssWW\ events are simulated at leading order using \mgamcatnlo\ ~\cite{Alwall:2014hca,Frixione:2002ik} with the \nnpdf\ set~\cite{Ball:2012cx,Ball:2014uwa}, interfaced with \pythiaEight~\cite{Sjostrand:2014zea} for parton showering, hadronization and underlying event modelling. The information about the polarization of the individual $W$ bosons in the signal process is extracted by generating a separate set of events using the {\mbox{\textsc{DECAY}}\xspace} package of \MADGRAPH (v1.5.14). The other backgrounds -- top ($t\bar{t}$ + jets, single-top), Drell-Yan, diboson ($W\gamma$, $W^\pm W^\pm$ and $ W Z$) and triboson ($ W W\gamma$, $ W Z\gamma$, $ W W W$, $ W W Z$, $ W Z Z$,  $ Z Z Z$) -- are generated with either \mgamcatnlo, {\mbox{\textsc{POWHEG}}\xspace}~\cite{Campbell:2014kua}, or \pythiaEight. The analyses use generated events obtained either using a fully simulated description of the HL-LHC CMS detector, implemented using the {\mbox{\textsc{GEANT4}}\xspace} package~\cite{Agostinelli:2002hh} (CMS) or using a parameterised description of the detector response~\cite{ATLAS_PERF_Note} (ATLAS). Additional details for each analysis are provided in the relevant reports from CMS~\cite{CMS-PAS-FTR-18-005} and ATLAS~\cite{ATL-PHYS-PUB-2018-052}.
\\
\begin{table}[!htbp]\renewcommand{\arraystretch}{1.2}
  \begin{center}
    \caption{ATLAS and CMS event selection criteria for \ssWW\ candidate events, with $\ell = e,\mu$ and $j$ as the leading(sub-leading) lepton or jet.} 
  \label{tab:selection2}
  \begin{tabular}{|l|c|c|}
    \hline
Selection requirement                            &   ATLAS Selection  & CMS Selection  \\
\hline\hline
Signal lepton \pT     	&   \pT$>$ 28(25) \GeV  & \pT$>$ 20 \GeV\ \\
Signal lepton $\eta$     	&  $\abseta \leq 4.0$ & $\abseta \leq 3.0$\\
Tag jet \pT     		&   \pT$>$ 90(45) \GeV\  & \pT$>$ 50 \GeV\ \\
Tag jet $\eta$     		&  $\abseta \leq 4.5	$ & $\abseta \leq 4.7 $ 	\\
\hline
Dilepton mass	&  $m_{\ell \ell}>$ 28 \GeV & $m_{\ell \ell}>$ 20 \GeV 	 	\\
$Z_{ee}$ veto     &  $|m_{ee} - m_{Z}| >$ 10 \GeV & $|m_{ee} - m_{Z}| >$ 15 \GeV		\\
\ETMISS     	& \ETMISS$>$~40 \GeV & \ETMISS$>$~40 \GeV		\\
Number of $b$-tagged jets    	&  0 & 0		\\
Jet selection     	&  Anti-$k_{\rm T}$ \cite{antikt-jet} jets with $\Delta R_{\ell, j} >$ 0.3 &  Anti-$k_{\rm T}$ PUPPI~\cite{Bertolini:2014bba} jets with $\Delta R_{\ell, j} >$ 0.4 \\
Preselected lepton veto     	&  \pt $>$ 7(6)\GeV\ 	& \pt $>$ 10 \GeV\	\\
Dijet rapidity separation    	&  $\Delta \eta_{j,j} >$ 2.5 		& $\Delta \eta_{j,j} >$ 2.5 		\\
Dijet mass    	 &  $m_{jj}>$ 520 \GeV &  $m_{jj}>$ 500 \GeV 		\\
Lepton centrality    	& $\zeta$ > -0.5 & $Z_{\rm MAX}$ < 0.75 \\
\hline
  \end{tabular}
  \end{center}
\end{table}
\\
The experimental signature of the \ssWW\ scattering process consists of exactly two isolated leptons (electrons or muons) with the same electric charge, two jets well-separated in rapidity, and moderate \ETMISS. The event selection requirements for the two experiments are listed in Table~\ref{tab:selection2}. A minimum requirement on the dilepton mass reduces the contamination from low-mass Drell-Yan processes, with an additional restriction excluding the $Z$ mass in the dielectron channel where the likelihood of charge misidentification is higher. A requirement on \ETMISS further reduces the background from charge misidentified events, and events containing any $b$-tagged jets\footnote{The $b$-tagging of jets in CMS is performed with the Deep Combined Secondary Vertex discriminator based on a deep neural network~\cite{PhysRevD.94.112002}. } are vetoed to suppress background contribution from ${t\bar t}$ production. A veto on additional preselected leptons significantly reduces background from $WZ$ events. The two leading jets are required to have a large invariant mass, and large angular separation, to satisfy the expected VBS topology. Since leptons in the EW \ssWW\ process are expected to be located in the central region defined by the forward-backward jets, non-VBS background can be suppressed with a requirement on the centrality of the two leptons. CMS uses the Zeppenfeld variable ~\cite{Rainwater:1996ud}, defined for a given lepton with pseudorapidity $\eta_\ell$ as
$$Z_\ell = \frac {[\eta_\ell - 0.5(\eta_1 +\eta_2)]}{|(\eta_1 -\eta_2)|} ,$$
where $\eta _1, \eta_2$ refer to the pseudorapidities of the leading and subleading jets. The maximum value of this variable, $Z_{\rm MAX}$, for any of the leptons is required to be less than 0.75. ATLAS uses a requirement on the function $\zeta$, where $\zeta = \min [\min (\eta_{\ell1}, \eta_{\ell2} )-\min(\eta_{j1},\eta_{j2}), \max(\eta_{j1},\eta_{j2})-\max(\eta_{\ell1},\eta_{\ell2}) ]$
\cutoff{The event should also contain two jets, reconstructed using the anti-$k_{\rm T}$ Pileup Per Particle Identification (PUPPI)~\cite{Bertolini:2014bba} algorithm, with distance parameter of 0.4, with $p_{\rm T} > 50$ \GeV and $|\eta| < 4.7$. The jets are not considered if they overlap with any of the isolated lepton or photon candidates, within $\Delta R = 0.4$.
In order to suppress background contribution from ${\rm t\bar t}$ production, events are rejected if there is a $b$-tagged jet in the event with $|\eta| < 2.4$. To further suppress background events should have missing transverse momentum above 40\GeV, and pass VBS selection,  M$_{jj} \ge 500$ GeV and  $|\Delta \eta_{jj}| \ge 2.5$. }
\\
\\
The event selections are optimized to maximize signal acceptance (CMS) or minimize fake background (ATLAS). ATLAS uses tight electron requirements, which have a lower efficiency (around 50\%~\cite{ATLAS_PERF_Note}). 
\\
\\
The expected event yields are summarized in Table~\ref{tab:yields} for CMS, and Table~\ref{tab:neventsoptimised} for ATLAS. The $m_{jj}$ distributions after the full event selection for ${\cal L}= 3000\fbinv$ are presented in Fig.~\ref{fig:mjj} .  The main background contributions in the final signal region are due to inclusive ${t \bar t}$  and ${WZ}$ productions, where the third lepton in the event was not reconstructed within the detector acceptance. ATLAS explicitly models the background contributions from jets faking electrons and lepton charge misidentification, which also contribute significantly in the signal region, while CMS includes the fake contribution under $t\bar{t}$ and does not consider the charge-misidentified or triboson backgrounds in this study, since their contributions were found to be negligible. The integrated number of signal and background events as a function of the dilepton invariant mass is shown in Figure~\ref{fig:dilepton} for the ATLAS selection. 
\\
\begin{table}[htbp]
\caption{CMS expected yields for signal and background contributions for ${\cal L} = $ 3000\fbinv.}
\label{tab:yields}
\begin{center}
\begin{tabular}{ | l | c |}
\hline
Process & Expected yield, ${\cal L} = 3000\fbinv$ \\
\hline\hline
$W^\pm W^\pm$ (QCD) & 196  \\
${t \bar t}$   &  5515   \\
${WZ}$      & 1421  \\
${W} \gamma$    & 406 \\
\hline
Total Background & 7538 \\
Signal $W^\pm W^\pm$ (EW) & 5368  \\
\hline
\end{tabular}
\end{center}
\end{table}

\begin{table}[!htbp]\renewcommand{\arraystretch}{1.2}
  \begin{center}
   \caption{The ATLAS expected signal and background event yields after the optimised full event selection for a corresponding integrated luminosity of $\cal{L}$=3000 fb$^{-1}$. Events tagged as either "charge misidentification" or "jets faking leptons" are summed for all background samples and combined into a single entry each in the table. Remaining events are listed separately per process. Both QCD and EW production of $WZ$ processes are included in the diboson background.} 
  \label{tab:neventsoptimised}
   \begin{tabular}{|l|c|cccc|}
    \hline
Process & All channels 
& \mmp & \eep 
& \mep & \emp  \\
\hline\hline
\ssWW (QCD) & 168.7 & 74.6 & 19.7 & 32.2 & 42.2\\
Charge Misidentification & 200 & 0.0 & 11 & 30 & 160\\
Jets faking electrons & 460 & 0.0 & 130 & 260 & 70\\
$WZ+ZZ$ & 1286 & 322 & 289 & 271 & 404\\
Tribosons & 76 & 30.1 & 9.6 & 15.1 & 21.6\\
Other non-prompt & 120 & 29 & 16.6 & 50 & 19\\
\hline
Total Background & 2310 & 455 & 480 & 660 & 710\\
\hline
Signal \ssWW (EW) & 2958 & 1228 & 380 & 589 & 761\\
\hline
  \end{tabular}
  \end{center}
\end{table} 

The uncertainty of the expected cross section measurement as a function 
of integrated luminosity is measured by fitting the $m_{jj}$ 
distribution, using a binned  maximum likelihood approach with all 
systematic uncertainties in the form of nuisance parameters with log-
normal distributions. The correlations among different sources of 
uncertainties are  taken into account while different final states  are 
considered as independent channels in the fit. CMS considers three 
channels categorised by lepton flavour ($\Pe\Pe$, $\Pe\Pgm$ and $\Pgm\Pgm$), while ATLAS uses eight channels by lepton flavour and charge ($e^+ e^+$, $e^- e^-$, $e^+\mu^+$, $e^-\mu^-$, $\mu^+ e^+$, $\mu^- e^-$, $\mu^+\mu^+$, $\mu^-\mu^-$).
\\
\\
The experimental uncertainties, statistical and systematic, in the CMS analysis contribute to a total uncertainty on the signal strength of 3.2$\%$ for 3000\fbinv. Including a theoretical uncertainty of 3$\%$ and an uncertainty on the luminosity of 1$\%$, the total uncertainty reaches a value of 4.5$\%$ for 3000\fbinv.
For the ATLAS analysis experimental systematics on the trigger, leptons, jets, and flavour tagging are taken from the 13 TeV analysis unchanged, while for the baseline estimation, rate uncertainties on the backgrounds are halved. An "optimistic" set of uncertainties is also presented, where the uncertainties on the non-data-driven backgrounds are aggressively reduced. The total uncertainty is presented in Fig.~\ref{fig:lumiproj} as a function of the integrated luminosity. The values of ${\cal L}$ exceeding 3000\fbinv are an estimation of a combination of the measurements from CMS and ATLAS, effectively doubling the total integrated luminosity.
\\
 \begin{figure}[htbp]
  \begin{center}
\includegraphics[width=0.4\textwidth]{\main/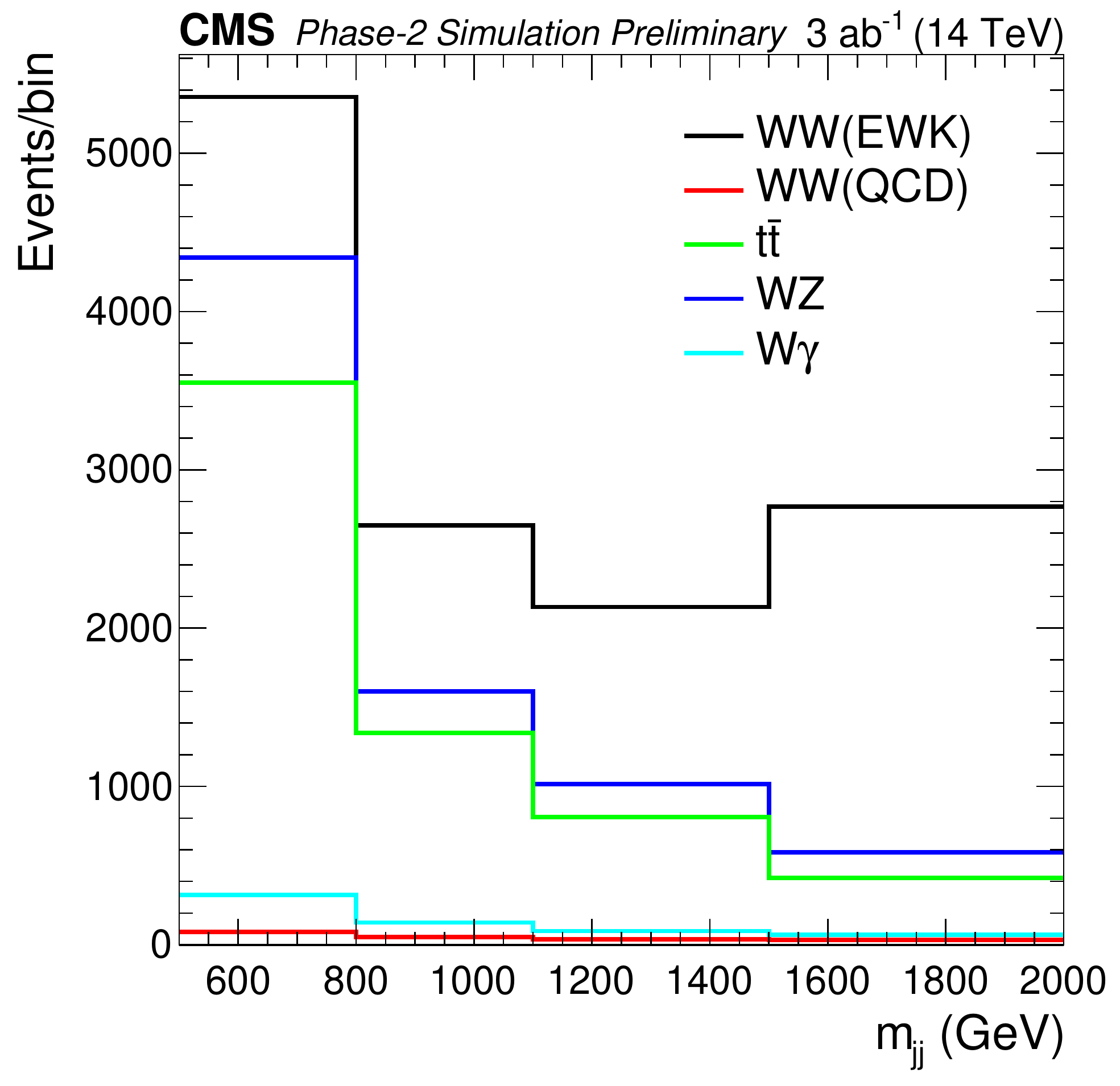} 
    \includegraphics[width=0.55\textwidth]{\main/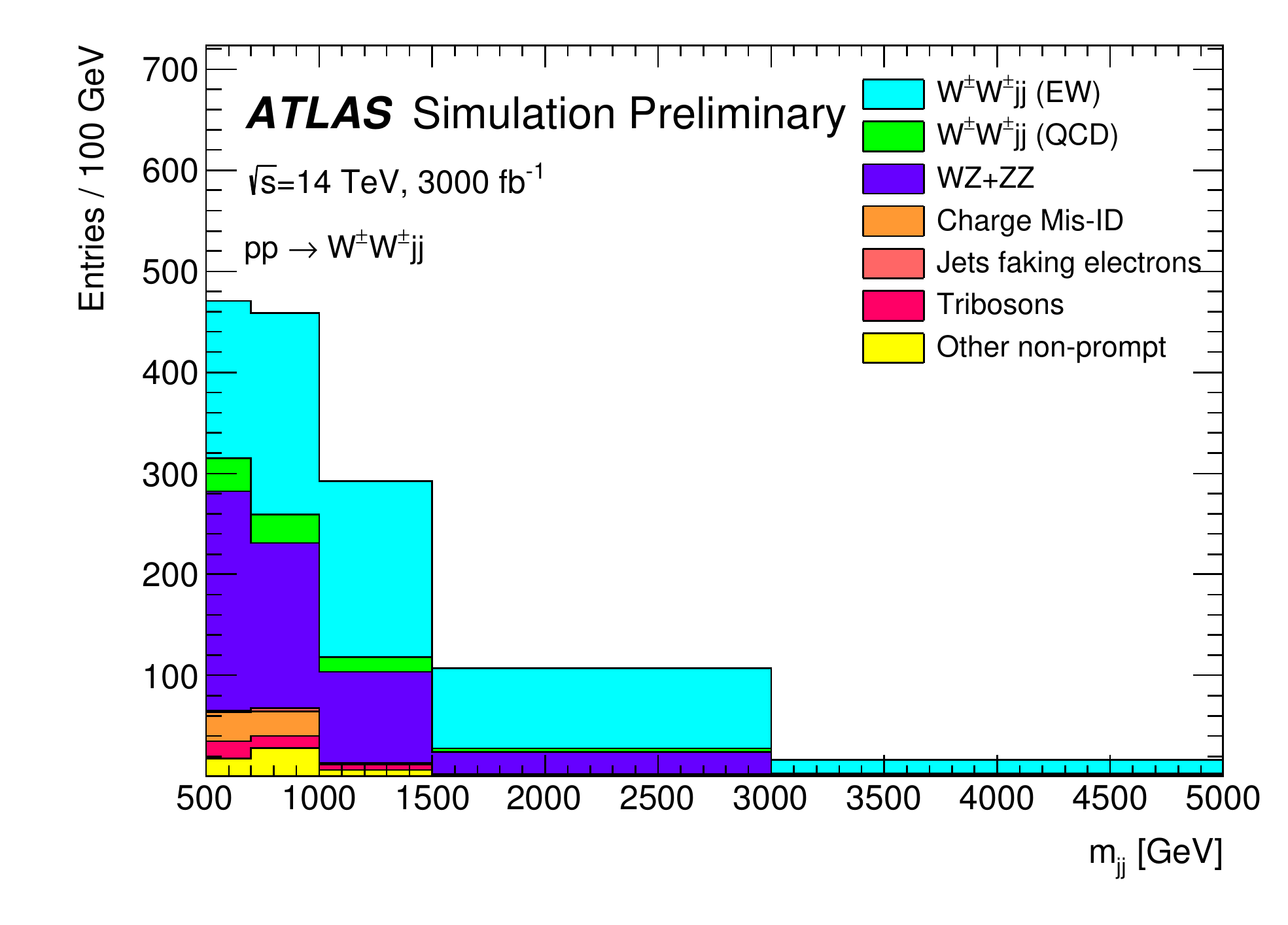}
    \caption{The distribution of the invariant mass of the two leading jets after the selection requirements for an integrated luminosity of 3000\fbinv, for CMS (left) and ATLAS (right).  }
    \label{fig:mjj}
  \end{center}
\end{figure}

\begin{figure}[htbp]
  \begin{center}
    \includegraphics[width=0.4\textwidth]{\main/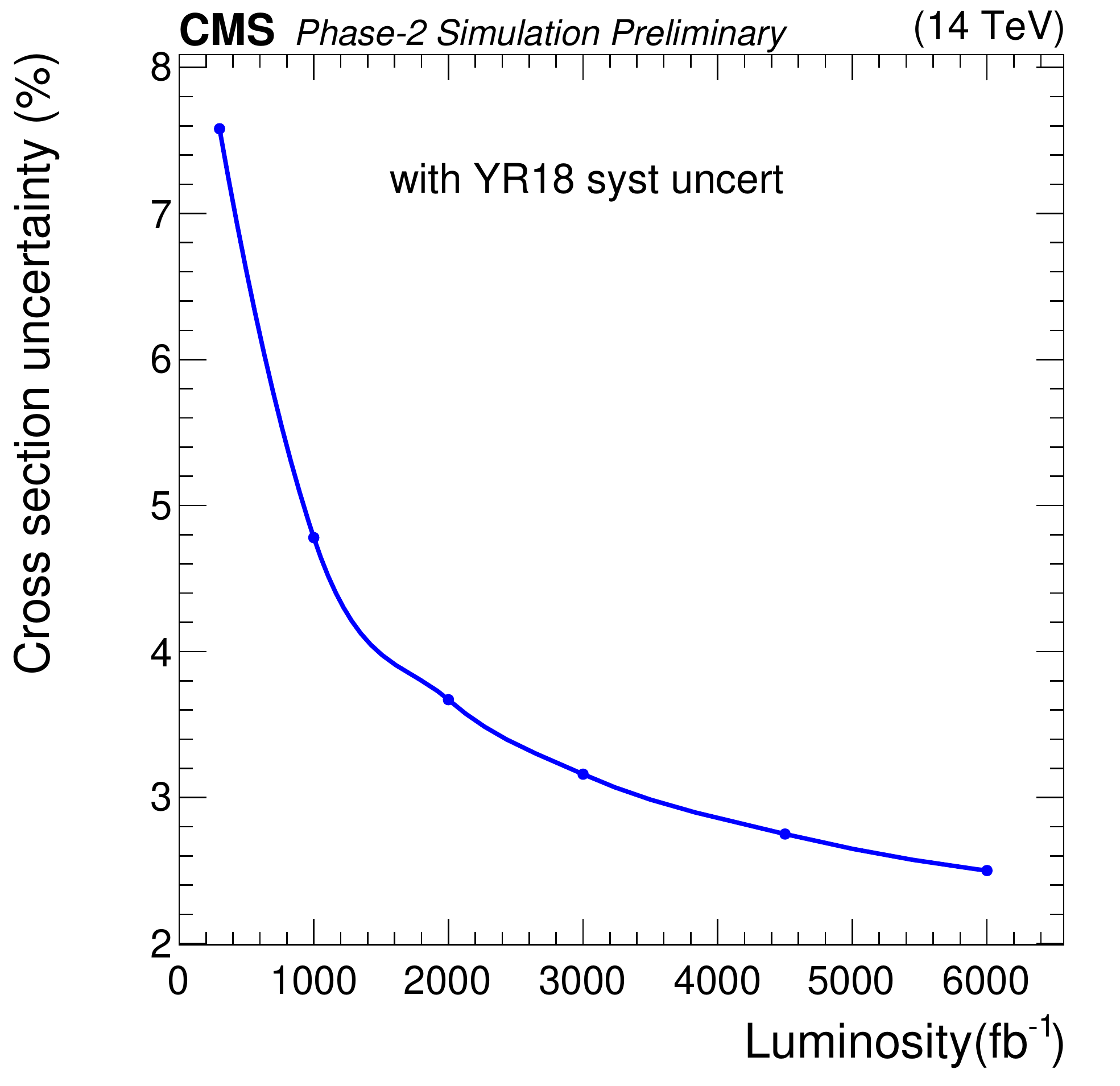}
    \includegraphics[width=0.55\textwidth]{\main/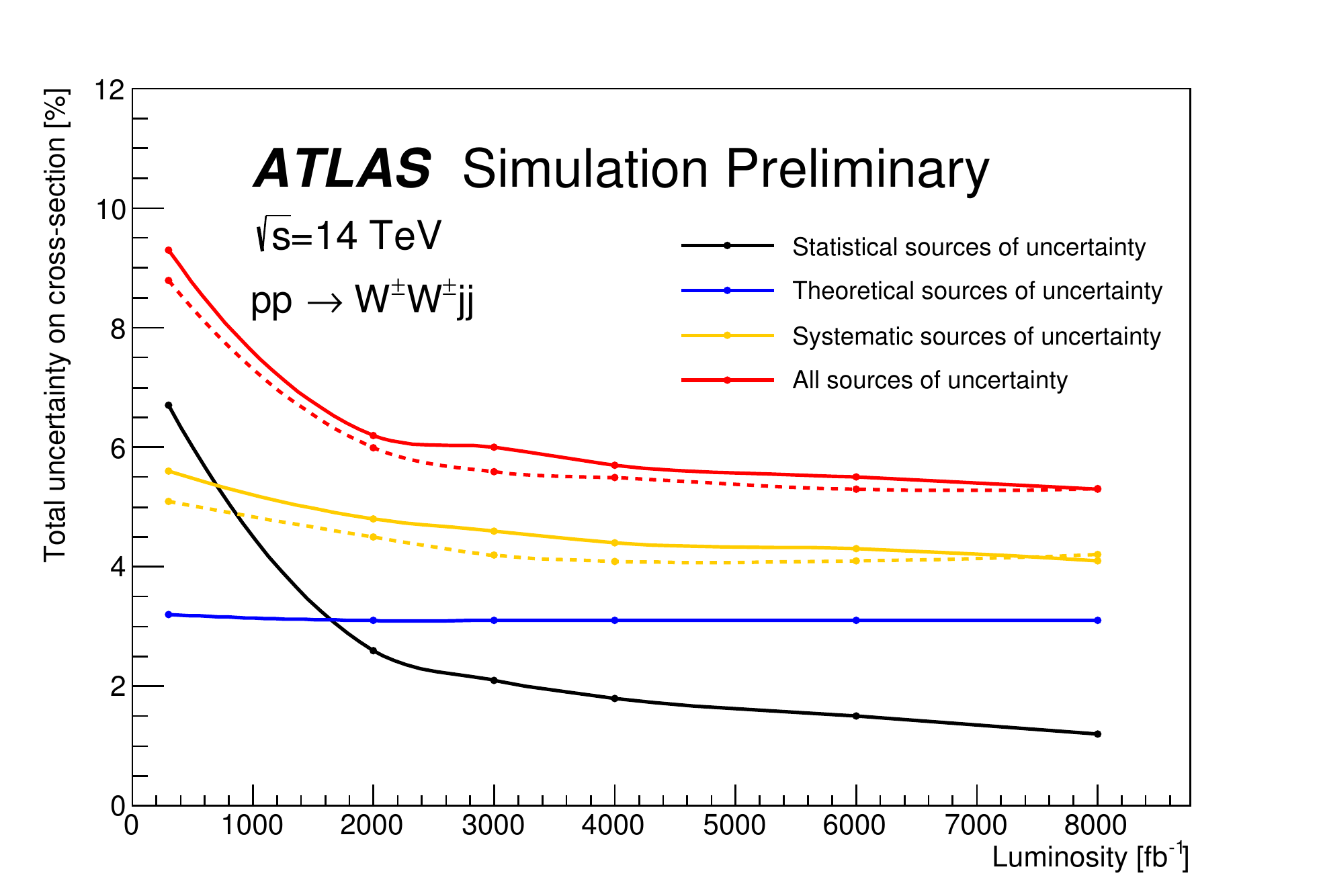}
    \caption{The estimated uncertainty of the EW $ W^\pm W^\pm$  cross section measurement as a function of the integrated luminosity, for CMS (left), only statistical and experimental systematic uncertainties are considered, and ATLAS (right). }
    \label{fig:lumiproj}
  \end{center}
\end{figure}

\cutoff{\subsection{Measurement of the longitudinally polarized $ W^\pm W^\pm$ scattering}}

The total \ssWW\;   VBS cross section can be decomposed into the polarized components based on the decays of the individual $ W$ bosons. Either or both can be longitudinally (L) or transversely (T) polarized, giving rise to final states of LL, TT as well as the mixed state LT (with TL combination implied). The LL component, \ssWWLL, is expected to be only about 6-7\% of the total VBS cross section
for  jet $p_{\rm T} > 50\GeV$. 
The  difference in azimuthal angle between the two leading jets, $\Delta\phi_{jj}$, has the potential for discriminating the LL component of the VBS scattering from TT and LT contributions. Since the signal-to-background separation for the EW \ssWW\  process improves with increasing $m_{jj}$ as shown in  Fig.~\ref{fig:mjj} (left), the $\Delta\phi_{jj}$ distributions are studied in two ranges of $m_{jj}$: for 500-1100\GeV\ and above 1100\GeV. Figure~\ref{fig:LL} shows the combination of signal and background yields as a function of $\Delta\phi_{jj}$ for high $m_{jj}$ regions. Using a simultaneous fit 
to two mass regions\footnote{The low $m_{jj}$ region serves to constrain the $t\bar{t}$/fake background.}, the significance for the observation of the LL process is estimated as a function of integrated luminosity. The significance is found to be up to 2.7 standard deviations for ${\cal L} = $ 3000\fbinv. The gradual improvement of signal significance as a function of integrated luminosity is shown in Fig.~\ref{fig:significance}~right. A combination of ATLAS and CMS results, using fully simulated ATLAS events and improved electron efficiency, is expected to reach an expected 
significance of 3 standard deviations with 2000~fb$^{-1}$ per experiment. In addition, recent studies~\cite{Searcy:2015apa} have shown that advances in machine learning can also improve the prospects for the measurement of the \ssWWLL\ process.

\begin{figure}[htbp]
  \begin{center}
    \includegraphics[width=0.55\textwidth]{\main/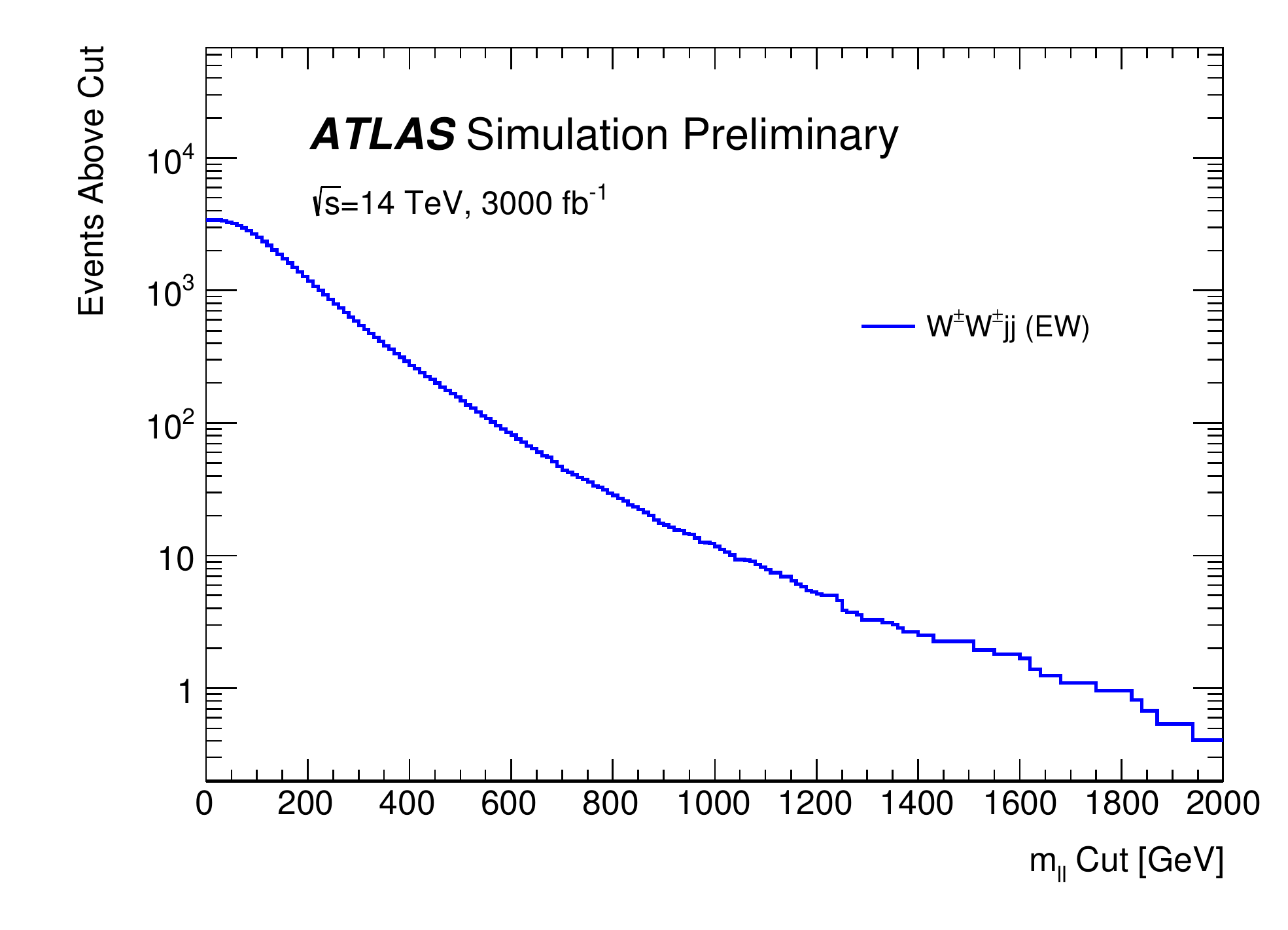}
    \caption{Integrated number of events as a function of dilepton invariant mass for events passing all selection criteria of the ATLAS signal region.}
    \label{fig:dilepton}
  \end{center}
\end{figure}

\begin{figure}[htbp]
  \begin{center}
    \includegraphics[width=0.4\textwidth]{\main/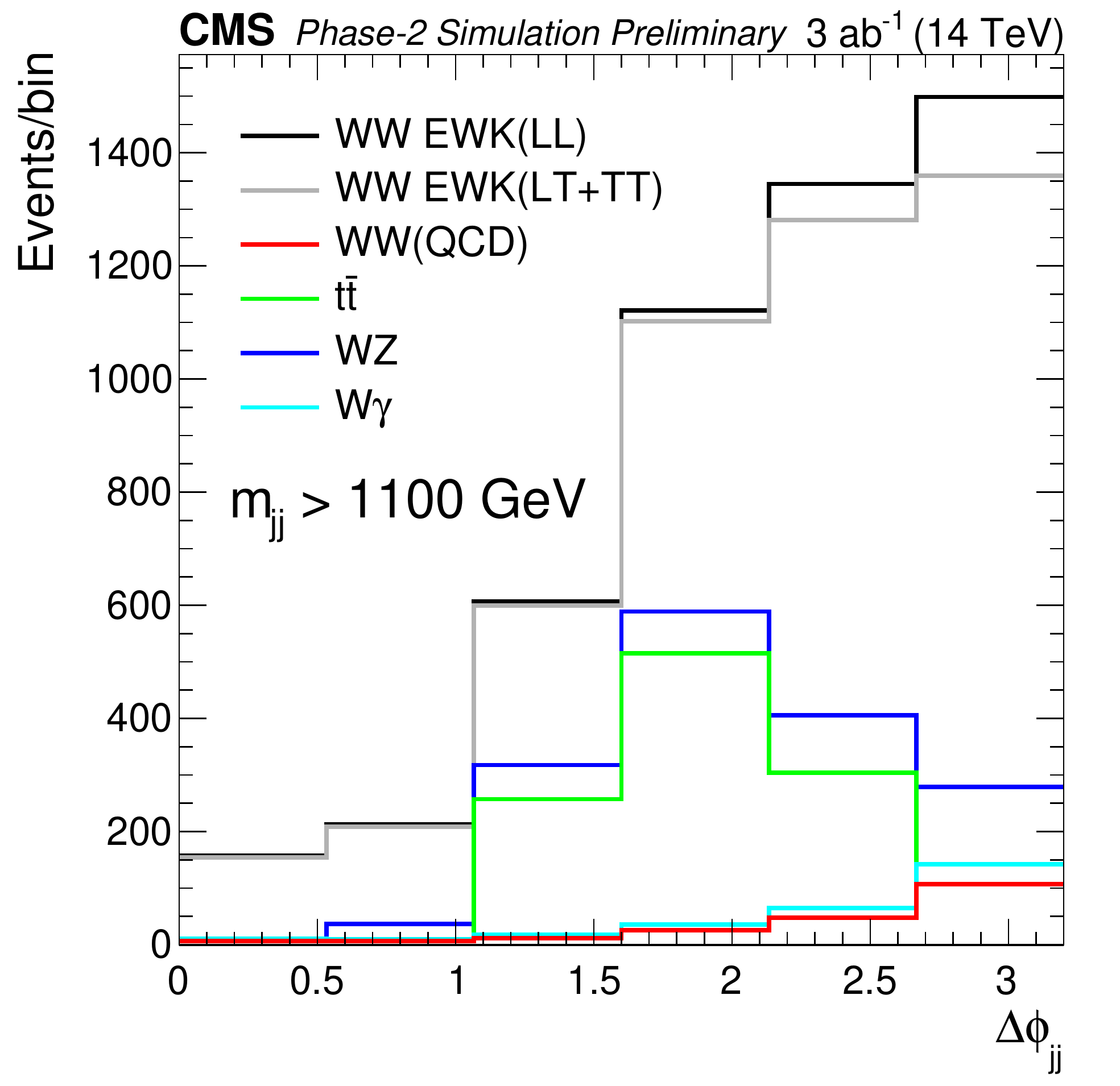}
    \includegraphics[width=0.55\textwidth]{\main/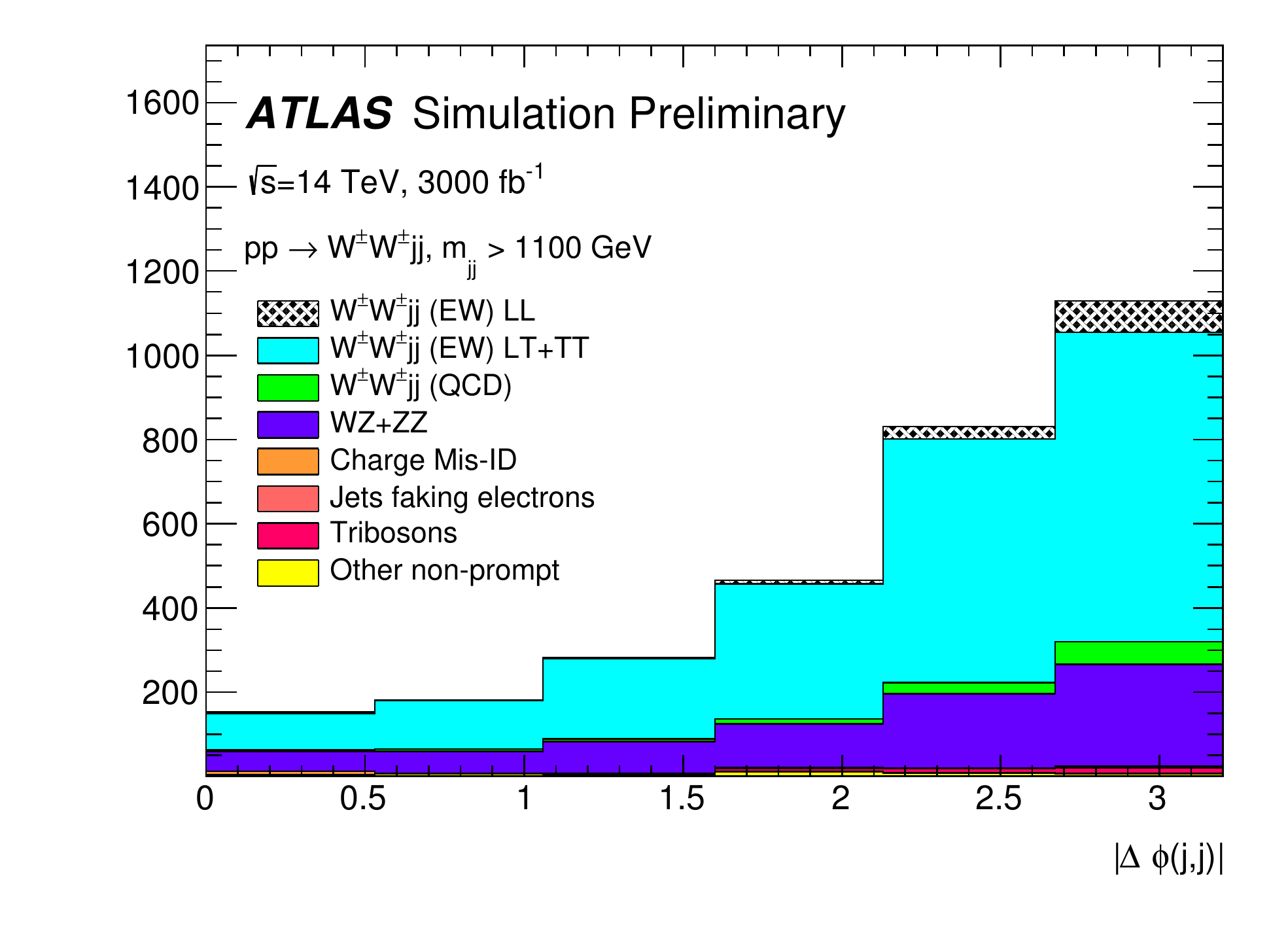}
    \caption{Distribution of the azimuthal angle difference between two leading jets for dijet invariant mass above 1100 GeV.}
    \label{fig:LL}
  \end{center}
\end{figure}

\begin{figure}[htbp]
  \begin{center}
    \includegraphics[width=0.4\textwidth]{\main/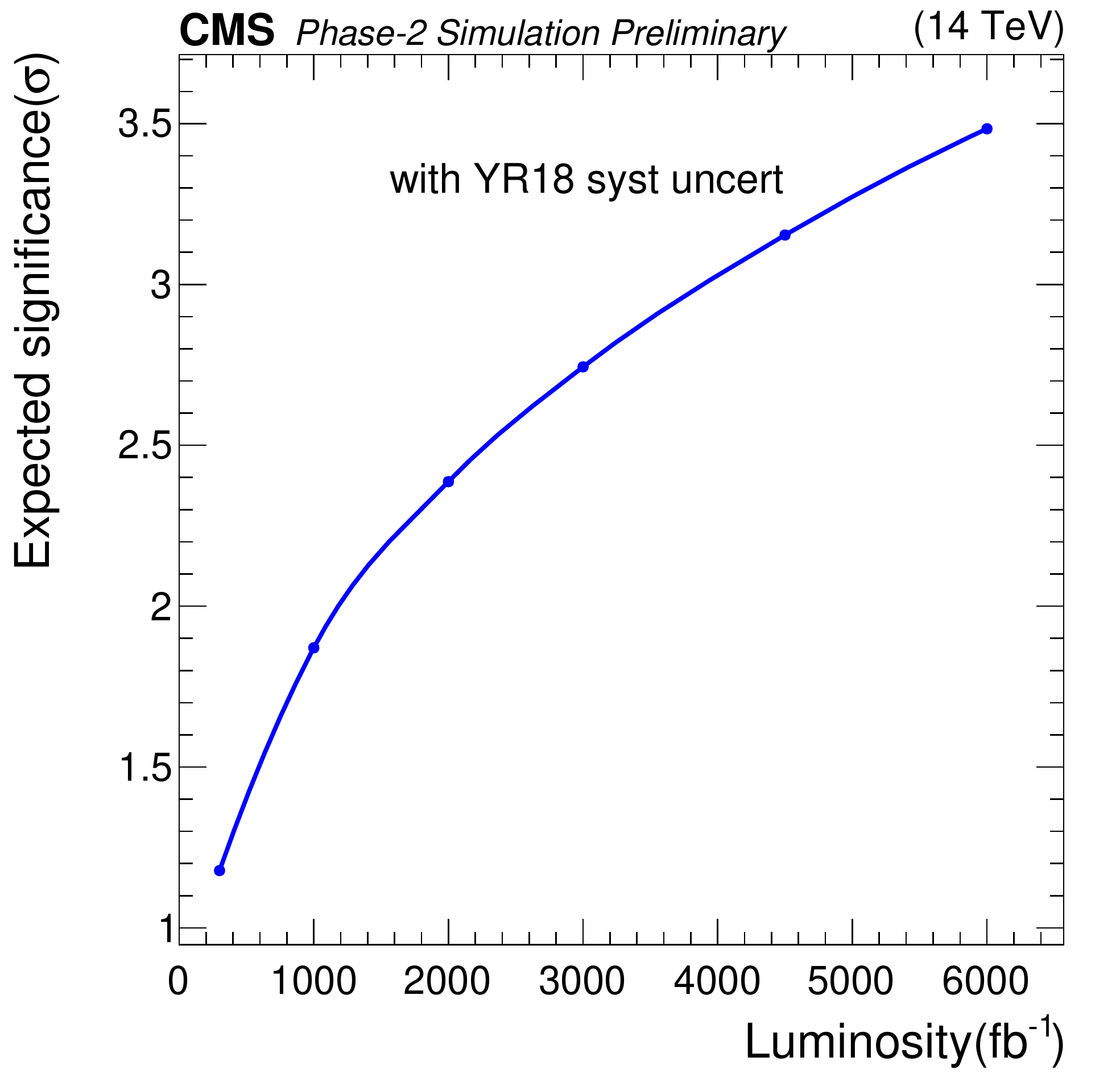}
    \includegraphics[width=0.55\textwidth]{\main/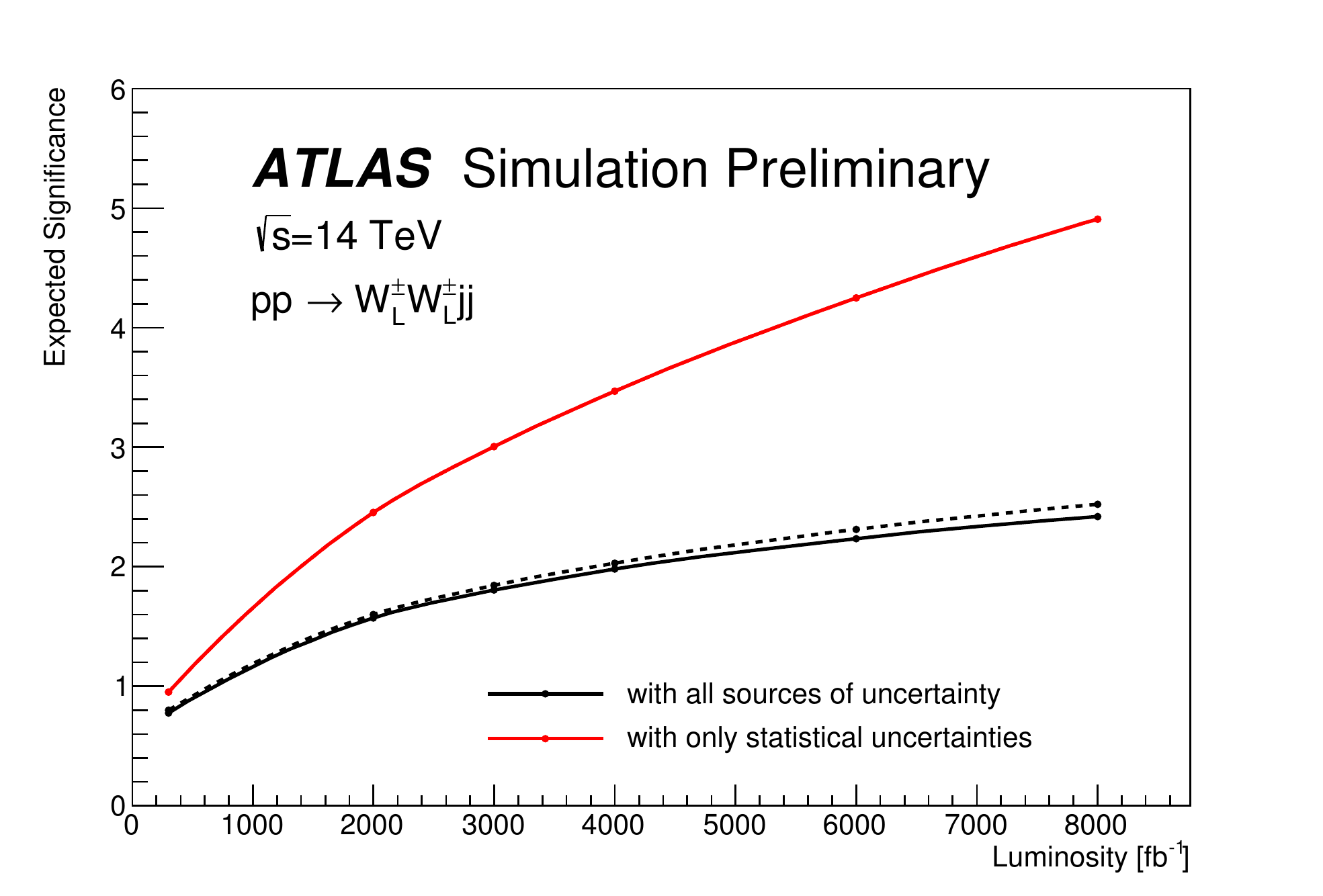}
    \caption{Significance of the observation of the scattering of a pair of longitudinally polarized $W$ bosons as a function of the integrated luminosity at CMS (left) and ATLAS (right).}
    \label{fig:significance}
  \end{center}
\end{figure}

\subsubsection[High Order corrections in VBS W$^\pm$W$^\pm$ production]{High Order corrections in VBS W$^\pm$W$^\pm$ production\footnote{Contribution by A.~Denner and M.~Pellen.}}\label{sec:theory-HO-ssWW}

The expected experimental precision in the measurement of VBS processes 
offers great opportunities to probe the electroweak (EW) sector and
its associated symmetry breaking mechanism
(see Refs.~\cite{Mangano:2016jyj,Goncalves:2017gzy,Jager:2017owh} for
$100\TeV$-collider studies).
Therefore, it is of prime importance to make precise theoretical
predictions available for the future operation of the LHC.
In this contribution, predictions for NLO EW corrections are provided
for the LHC running in its high-luminosity and high-energy configurations.
The HL set-up corresponds to a centre-of-mass energy of $14\TeV$ while
the HE one refers to $27\TeV$.
For both centre-of-mass energies the same type of event selections has
been used.
These predictions represent important benchmarks as they indicate the
expected rates when accounting for NLO EW corrections.
The NLO EW corrections have been shown to be very large for VBS
processes \cite{Biedermann:2016yds} and even the dominating NLO
contribution for same-sign $WW$ scattering \cite{Biedermann:2017bss}.
Nonetheless, the inclusion of NLO QCD corrections is necessary as they
can significantly distort the shape of jet-related observables
\cite{Jager:2006zc,Jager:2006cp,Bozzi:2007ur,Jager:2009xx,Jager:2011ms,Denner:2012dz,Rauch:2016pai,Biedermann:2017bss,Ballestrero:2018anz,Jager:2018cyo}.
In addition, they drastically reduce theoretical uncertainties.
The QCD corrections for all VBS signatures can be obtained from public
programs such as {\mbox{\textsc{MadGraph5\_aMC@NLO}}\xspace}~\cite{Alwall:2014hca}, {\mbox{\textsc{Powheg}}\xspace}~\cite{Nason:2004rx,Frixione:2007vw,Alioli:2010xd}, {\mbox{\textsc{Sherpa}}\xspace}~\cite{Gleisberg:2008ta,Gleisberg:2003xi}, or {\mbox{\textsc{VBFNLO}}\xspace}~\cite{Arnold:2008rz,Arnold:2011wj,Baglio:2014uba}.

In this study, the NLO EW corrections have been obtained from {\mbox{\textsc{MoCaNLO+Recola}}\xspace}
\cite{Bendavid:2018nar,Actis:2016mpe,Actis:2016mpe} based on a
full NLO computation \cite{Biedermann:2017bss} for the same-sign WW
signature.
While the exact value of the corrections is expected to be different
for other signatures, their magnitudes and nature should be similar.


The hadronic scattering processes are simulated at the LHC with a centre-of-mass energies $\sqrt s = 14 \TeV$ and $\sqrt s = 27 \TeV$.
    The NNNPDF~3.1 LUXQED parton distribution
    functions~(PDFs)~\cite{Bertone:2017bme} with five massless
    flavours,\footnote{For the process considered, no bottom
      (anti-)quarks appear in the initial or final state at LO and
      NLO, as they would lead to top quarks rather than light jets in the final state.} 
    NLO-QCD evolution, and a strong coupling constant $\alphas( \MZ ) = 0.118$ are employed.\footnote{The corresponding identifier {\tt lhaid} in the program LHAPDF6~\cite{Buckley:2014ana} is 324900.}
    Initial-state collinear singularities are factorised according to
    the ${\overline{\rm MS}}$ scheme, consistently with the conventions in the NNPDF set.

 The other input parameters have been chosen as in Ref.~\cite{Ballestrero:2018anz}.
    For the massive particles, the following masses and decay widths are used:
    \begin{alignat}{2}
                      \Mt   &=  173.21\GeV,       & \quad \quad \quad \Gt &= 0 \GeV,  \nonumber \\
                    \MZOS &=  91.1876\GeV,      & \quad \quad \quad \GZOS &= 2.4952\GeV,  \nonumber \\
                    \MWOS &=  80.385\GeV,       & \GWOS &= 2.085\GeV,  \nonumber \\
                    M_{\rm H} &=  125.0\GeV,       &  \GH   &=  4.07 \times 10^{-3}\GeV.
    \end{alignat}
    The measured on-shell (OS) values for the masses and widths of the W and Z bosons are converted into pole values for the gauge bosons ($V= W, Z$) according to Ref.~\cite{Bardin:1988xt},
    \begin{equation}
    \begin{split}
            M_V &= \MVOS/\sqrt{1+(\GVOS/\MVOS)^2}\,, \\
       \Gamma_V &= \GVOS/\sqrt{1+(\GVOS/\MVOS)^2}.
    \end{split}
    \end{equation}
    The EW coupling is fixed in the $G_\mu$ scheme \cite{Denner:2000bj} according to 
    \begin{equation}
    \alpha =  \frac{\sqrt{2}}{\pi} G_{\mu} M_{\rm W}^2 \left(1-\frac{M_{\rm W}^2}{M_{\rm Z}^2} \right),
    \end{equation}
    with
    \begin{equation}
        G_{\mu}    = 1.16637\times 10^{-5}\GeV^{-2},
    \end{equation}
    and $M_V^2$ corresponds to the real part of the squared pole mass.
    The complex-mass scheme~\cite{Denner:1999gp,Denner:2005fg,Denner:2006ic} is used throughout to treat unstable intermediate particles in a gauge-invariant manner.

    The central value of the renormalisation and factorisation scales is set to 
    \begin{equation}
    \label{eq:defscale}
     \mu_{\rm R} = \mu_{\rm F} = \sqrt{p_{\rm T, j_1}\, p_{\rm T, j_2}} .
    \end{equation}
    The transverse momenta are those of the two hardest jets.
    This choice of scale has been shown to provide stable NLO-QCD predictions \cite{Denner:2012dz}.

    Following experimental measurements \cite{Aad:2014zda,Aaboud:2016ffv,Khachatryan:2014sta,Zhu:2010cz} and prospect studies \cite{ATL-PHYS-PUB-2017-023}, the event selection used in the present study is:

    \begin{itemize}
        \item The two same-sign charged leptons are required to fulfill
          cuts on transverse momentum, rapidity, separation in the
          rapidity--azimuthal-angle, and the lepton-pair invariant mass, 
            \begin{align}
            \label{cut:1}
             \ptsub{\Pl} >  20\GeV,\qquad |y_{\Pl}| < 4.0, \qquad \Delta R_{\Pl\Pl}> 0.3, \qquad m_{\Pl\Pl}>20\GeV.
            \end{align}
        \item The total missing transverse momentum, computed from the vectorial sum of the transverse momenta of the two neutrinos, is required to be
            \begin{align}
            \label{cut:2}
              p_{\rm T, miss} >  40\GeV\,.
            \end{align}
        \item QCD partons (light quarks and gluons) are clustered  using the anti-$k_{\rm T}$ algorithm~\cite{Cacciari:2008gp} with jet-resolution parameter $R=0.4$.
        Cuts on the jets' transverse momenta and rapidities are imposed,  
            \begin{align}
            \label{cut:3}
             \ptsub{j} >  30\GeV, \qquad |y_j | < 4.0. 
            \end{align}
            VBS cuts are applied to the two jets with largest transverse momentum, specifically on the 
             in\-vari\-ant mass of the di-jet system, as well as on the rapidity separation of the two jets and their separation from leptons,
            \begin{align}
            \label{cut:4}
             m_{j  j } >  500\GeV,\qquad |\Delta y_{j  j }| > 2.5, \qquad \Delta R_{j \Pl} > 0.3 .
            \end{align}
        \item Finally, the centrality of the leptons is enforced according to Ref.~\cite{ATL-PHYS-PUB-2017-023}:
            \begin{align}
            \label{cut:5}
             \zeta = \text{min}\left[\text{min}\left(y_{\Pl_1},y_{\Pl_2}\right) - \text{min}\left(y_{j _1},y_{j _2}\right), \text{max}\left(y_{j _1},y_{j _2}\right) - \text{max}\left(y_{\Pl_1},y_{\Pl_2}\right)\right]> 0 .
            \end{align}
            
        \item For EW corrections, real photons and charged fermions are clustered using the anti-$k_{\rm T}$ algorithm with
            radius parameter $R=0.1$. In this case, leptons and quarks are understood as {\it dressed fermions}.
    \end{itemize}


In the following discussion of SM predictions for the HL- and HE-LHC both QCD and EW corrections have been combined.
For VBS processes EW corrections are particularly large and therefore of prime importance.
The leading contributions originate from the exchange of massive gauge bosons in the virtual corrections.
They tend to grow large and negative in the high-energy limit owing to
so-called Sudakov double logarithms.
As shown in Ref.~\cite{Biedermann:2016yds}, large EW corrections are an intrinsic feature of VBS at the LHC.
While this study is based on the same-sign $ W$ channel, it
has been further confirmed recently by the computation of large EW corrections to the $ W Z$ channel \cite{SchwanTalk,Denner:2019tmn}.

Given their size and the foreseen experimental
precision, these corrections are actually measurable.
Because they involve interactions of the EW sector, their measurement would constitute a further test of the SM.
On the left hand-side of Fig.~\ref{fig:ew}, the distribution in the invariant mass of the two leading jets is shown at LO and NLO EW for the process $pp \to \mu^+ \nu_{\mu} \Pe^+ \nu_{\Pe} j j $ at $14\TeV$.
The yellow band describes the expected statistical uncertainty for a
HL LHC collecting 3000\fbinv. 
On the right hand-side for Fig.~\ref{fig:ew}, a similar plot for the absolute rapidity of the jet pair is shown.
It is thus clear that with the expected luminosity, one is not only
sensitive to the VBS process but also to its EW corrections.
\begin{figure}
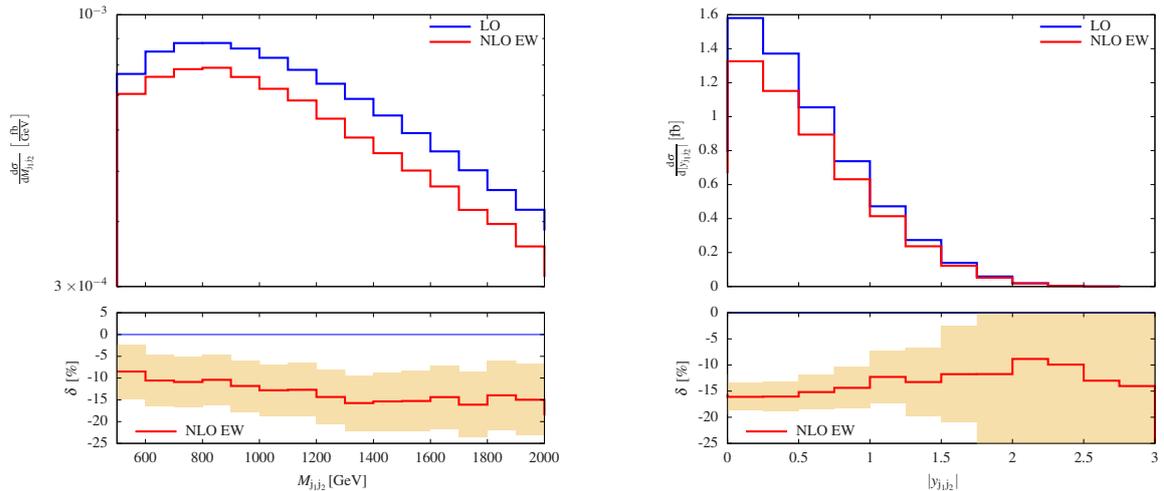

\includegraphics[width=.5\textwidth]{{{\main/electroweak/img/vbs_zz/histogram_invariant_mass_mjj12_rauch}}}
\includegraphics[width=.5\textwidth]{{{\main/electroweak/img/vbs_zz/histogram_rapidity_j1j2}}}
\caption{
Differential distributions in the invariant mass of the two jets (left) and their rapidity (right)
in $pp\to\mu^+\nu_\mu\Pe^+\nu_{\Pe}j j $ at $14\TeV$ including NLO EW corrections (upper panel) and relative NLO EW corrections (lower panel).
The yellow band describes the expected statistical
uncertainty for a high-luminosity LHC collecting 3000\fbinv and
represents a relative variation of $\pm 1/\sqrt{N_{\rm obs}}$ where $N_{\rm obs}$ is the number of observed events in each bin.
}
\label{fig:ew}
\end{figure}

\begin{figure}
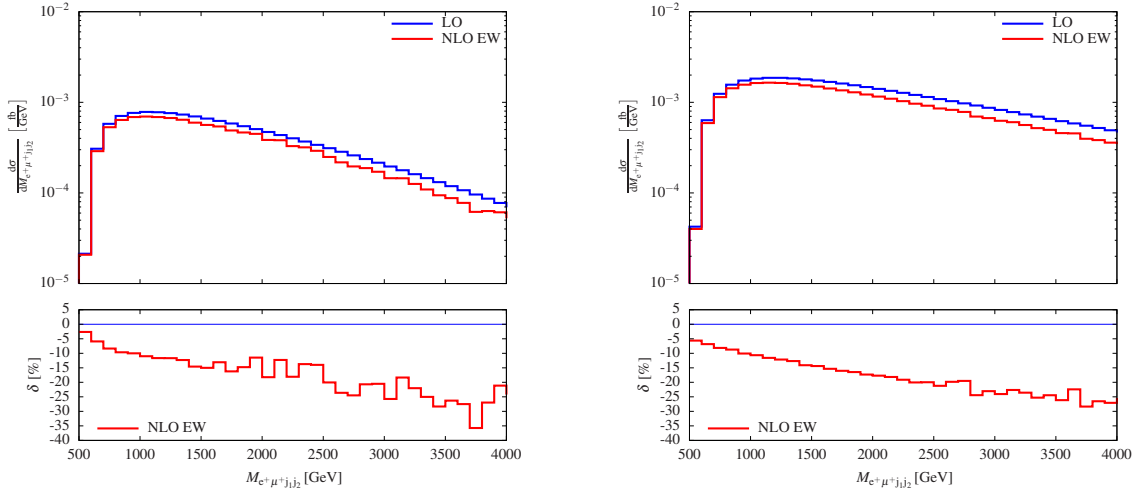

\includegraphics[width=.5\textwidth]{{{\main/electroweak/img/vbs_zz/histogram_invariant_mass_all_HL}}}
\includegraphics[width=.5\textwidth]{{{\main/electroweak/img/vbs_zz/histogram_invariant_mass_all_HE}}}
\caption{
Differential distribution in the invariant mass of the visible system ($\Pe^+\mu^+j j $) in $pp\to\mu^+\nu_\mu\Pe^+\nu_{\Pe}j j $ at $14\TeV$ (left) and $27\TeV$ (right) including NLO EW corrections (upper panel) and relative NLO EW corrections (lower panel).
}
\label{fig:ew2}
\end{figure}

In Fig.~\ref{fig:ew2}, the distributions in the invariant mass of the visible system ($\Pe^+\mu^+j j $) at both $14\TeV$ (left) and $27\TeV$ (right) are shown.
As expected, the corrections are larger for higher centre-of-mass energy due to the higher representative scale of the process.
In the tail of the distribution where new physics could play an important role, the corrections are particularly large and reach about $25\%$ for the $27\TeV$ set-up.
Note that in the present predictions, the real radiation of massive gauge bosons is not taken into account. This effect has been estimated to be of the order of few percent for 
the HL set-up when considering the total cross section.
While this effect is for now negligible, for the HL and HE mode of the 
LHC, it will become relevant in the same way as the use of VBS 
approximations in theoretical predictions~\cite{Ballestrero:2018anz}.
These observations are further confirmed via the cross sections for the two centre-of-mass energies at LO (using full matrix element) and NLO EW given in Table \ref{tab:EWxsec}.
At $27\TeV$ the EW corrections are a few percent larger than at $14\TeV$ ($-18.9\%$ against $-15.1\%$, respectively).
Note that the jump in energy from $14\TeV$ to $27\TeV$ is accompanied by an increase by more than a factor 3 in the cross section at LO.

\begin{table}
\begin{center}
\caption{Cross sections at LO ($\mathcal{O}\left(\alpha^6 \right)$) and NLO EW ($\mathcal{O}\left(\alpha^7 \right)$) for $pp \to \mu^+ \nu_\mu \Pe^+ \nu_{\Pe} j j $ at both $14\TeV$ and $27\TeV$ at the LHC.
The relative EW corrections are given in percent, and the digits in parentheses indicate the integration error.}
\label{tab:EWxsec}
\begin{tabular}
{|c|ccc|}
\hline
  & $\sigma^{\rm LO}$~[fb] &  $\sigma^{\rm NLO}_{\rm EW}$~[fb] & $\delta_{\rm EW}~[\%]$
\\
\hline\hline
$14\TeV$ & $\phantom{1}1.4282(2)$ & $\phantom{1}1.213(5)$& $-15.1$ \\
$27\TeV$ & $\phantom{1}4.7848(5)$ & $\phantom{1}3.881(7)$& $-18.9$ 
\\
\hline
\end{tabular}
\end{center}
\end{table}


\subsubsection{Measurements of $WZ$ scattering at the HL-HLC}

Prospects are presented for measuring the $WZ$ electroweak production in fully leptonic final state at the HL-LHC. This  work includes studies of the polarised $WZ$ production: measurements of the vector bosons in longitudinally polarized states are of particular importance, since they give  direct access to the nature of the electroweak symmetry breaking via the exchange of a Higgs bosons in the t-channel as shown in Fig.~\ref{Fig:Feymann}.  
Another relevant aspect of $WZ$ production lies in the probe of the non-abelian structure of the Standard Model via sensitive tests to triple and quartic gauge couplings, a topic which is partially addressed in the next subsection.
\begin{figure}[htbp]
    \centering
    \includegraphics{\main/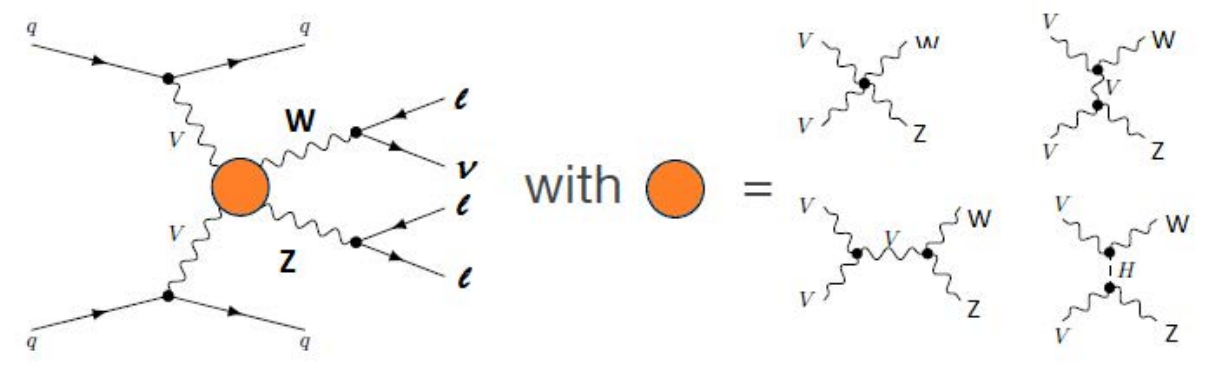}
    \caption{Feymann diagrams contributing to VBS $WZ$ production.}
    \label{Fig:Feymann}
\end{figure}
Measurements of the electroweak production using 36~\fbinv of the proton-proton collisions at 13 TeV were reported by both the ATLAS \cite{Aaboud:2018ddq} and CMS \cite{CMS-PAS-SMP-18-001} collaborations. The existing results are strongly limited by the statistical uncertainties of the data samples, therefore the integrated luminosity expected at the end of the HL-LHC operation is mandatory to fully exploit the physics behind VBS in $WZ$ production via  measurement of  differential distributions and the polarization of the final state bosons. 

In proton-proton collisions, the VBS process results from the interaction of two bosons radiated by the initial quarks  leading to a final state with two centrally produced  bosons and two forward jets.
The main irreducible background is represented by events in which the same final state is mediated by strong interactions (\WZQCD ) and where the two bosons are not the direct result of a scattering process. Other backgrounds consist of different di-boson final states ($ZZ$, $Z\gamma$), tri-bosons and $tV$ or $t\bar tV$ production, where $V$ is a $Z$ or a $W$ boson. The amount of the non-prompt backgrounds, where one or more lepton candidates are coming from jets misidentified as leptons, ultimately depends on the detector geometry, reconstruction technique and  event selection requirements.

The signal selection requires events with three isolated leptons with \PT\ > 15 GeV with $|\eta| < 4$ for ATLAS and $|\eta|<2.8\ (3.0)$ for muons (electrons) for CMS.
In addition, at least one lepton should pass the single lepton trigger (ATLAS). In order to suppress the background from $ZZ$ processes, events containing  four or more lepton candidates are discarded. At least one of the three lepton candidates is required to have \PT\ > 25 GeV. The event must have at least one pair of leptons of the same flavor and opposite charge, with an invariant mass that is consistent with the nominal $Z$ boson mass at $M_Z =~$91.188 GeV within 10 GeV for ATLAS and 15 GeV for CMS. This pair is considered as a $Z$ boson candidate. The third lepton is assigned to the $W$ boson and its \PT\ is required to be greater than 20 GeV. Finally, $E_{T}^{miss}$ (CMS) or the transverse mass of the $W$ candidate computed using $E_{T}^{miss}$ and the \PT\ of the third lepton (ATLAS) is required to be above 30 GeV.  The VBS signature is characterized by the presence of two forward jets. Jets are reconstructed with the anti-$k_{\rm T}$ algorithm with distance parameter 0.4. 
For ATLAS, the event is selected if it contains two jets in opposite hemispheres  with \Ptj\ greater than 30 GeV and |\Etaj | < 3.8. For CMS, the event is selected if it contains two jets with \Ptj\ > 50 GeV and |\Etaj | < 4.7. In addition, the pseudorapidity separation between jets, $\Delta \eta_{jj}$, is requested to be greater than 2.5. Finally, the dijet mass $m_{jj}$ is required to be greater than 500 GeV. The full list of selection requirements is summarized in Table~\ref{tab:WZEWKselections}.

\begin{table}[htbp]
  \begin{center}
\caption{Summary of event selection requirements.}
\label{tab:WZEWKselections}
    \begin{tabular}{|c|c|c|}
  \hline
      Variables & ATLAS & CMS \\
  \hline\hline
     $\PT(\ell) $   [GeV]       &  $ > 15 $     &  $ > 15 $  \\ 
     $\PT(\ell_{lead})$          &  $ > 25 $     &    --      \\
     	     $\PT(\ell_{ Z,1}),\ \PT(\ell_{ Z,2}) $ [GeV]  &   &  $>25$, $> 15$ \\  
    $\PT(\ell_{ W})           $ [GeV]            & $ > 20 $     & $> 20$      \\
    $|\eta(\Pgm)|   $                         & $< 4.0$ & $< 2.8$     \\
    $|\eta(\Pe)|     $                              & $< 4.0$ & $< 3.0$     \\
    $|m_{ Z}-m^{\mathrm{PDG}}_{ Z}|$ [GeV]   & $< 10$ & $ < 15 $    \\
    $m_{3\ell}                $ [GeV]                   & -- & $> 100$     \\
    $m_{\ell\ell}             $ [GeV]                   & -- & $> 4$       \\
     $E_{\rm{T}}^{miss}$   [GeV]                        & --           & $> 30$      \\
     $M_{\rm{T}}^{W}$ [GeV]         & $ > 30$  &  --  \\  
     \hline
    $n_{\mathrm{j}}           $                         & $\ge$  2 & $\ge 2$     \\
$|\eta(j)|     $                           & $< 3.8$ & $< 4.7$     \\
    $\Ptj                   $ [GeV]                  & $ > 30$   & $ > 50$     \\
  $\Delta R(j, \ell)$                                   & -- & $ > 0.4$    \\

  $\PT(b)                   $ [GeV]             &   --      & $ > 30$     \\
    $n_{b-\rm{jet}}       $                         & -- & $= 0$       \\
\hline 
      $m_{jj}                   $                           & $ > 500$  & $> 500$     \\
    $\Delta\eta_{jj}$        & Opp. hemis. & $> 2.5$     \\
    $\zepp$                           & --                  & $< 2.5$     \\

\hline
  \end{tabular}
  \end{center}
\end{table}

\begin{table}[htbp]
  \centering
\caption{ Expected signal and background yields corresponding to the event selection listed in Table~\ref{tab:WZEWKselections} for 3000\fbinv. Background contributions are grouped differently for ATLAS and CMS.}
    \label{tab:yields3000fbWZEWK}
  \begin{tabular}{ |c|c|c| }
    \hline
        Process  & ATLAS & CMS  \\
    \hline\hline
        \WZEW\  & 3889         & 2757  \\
\hline
        \WZQCD\ & 29754  & 3486 \\
        $t\bar{t}V$ & 3145 & -- \\
        $tZ$ & 2221 & -- \\
        tV/VVV  & --        & 1374  \\
       Non prompt & -- & 1192 \\
       $ZZ$ & 1970 & -- \\
           VV    & -- & 398 \\
       Z$\gamma$ & --  & 296  \\
    \hline
  \end{tabular}
\end{table}

Distinct approaches are used by ATLAS and CMS, respectively  based on simulation at 14~\TeV\ and on extrapolation from Run-2 results. 
ATLAS uses Monte Carlo samples generated with a fast simulation based on the parameteriation of the performance of the HL-LHC detector and where jets from pileup (PU) interactions corresponding to <$\mu$> = 200 are added to the event record; a loose event selection and a conservative background hypothesis is used. The signal events are generated at LO with {\mbox{\textsc{Sherpa}}\xspace} 2.2.2\cite{Gleisberg:2008ta} and the \WZQCD\ background is simulated at NLO with {\mbox{\textsc{Sherpa}}\xspace} 2.2.0: in Ref.~\cite{Aaboud:2018ddq}, it was shown that the \WZQCD\ background predictions might be overestimated by 40\% in certain regions of the phase-space. And with a \Ptj cut as low as 30 GeV, an $| \eta^{jet} |$ cut less than 3.8, corresponding to the HL-LHC tracker acceptance, was found necessary to maintain the contamination of PU jets in signal (resp. \WZQCD ) events from 18\% (resp. 69\%) to 2\% (resp. 11\%). 
  
The CMS projection is based on MC samples  with full simulation of the CMS detector at 13~\TeV\ and data driven background estimates, see Ref.~\cite{CMS-PAS-FTR-18-038}. The cross sections of samples are scaled for this projection from 13 to 14\TeV\ using SM predictions, for the data-driven backgrounds the scaling is done using appropriate mixture of simulated events. The performance of the CMS detector at the HL-LHC at pileup 200 is simulated using {\mbox{\textsc{Delphes}}\xspace}. It is proven that lepton and PUPPI~\cite{Bertolini:2014bba} jet reconstruction allow to keep the same or better level of reconstruction efficiency and background rejection as in existing data; no additional corrections
are applied in  the projection. An additional scaling factor is applied to account for the increased pseudorapidity coverage of the HL-LHC CMS detector.
The ATLAS and CMS signal and background yields are summarized in Table~\ref{tab:yields3000fbWZEWK} for the total integrated luminosity of 3000\fbinv. 
\begin{figure}[htbp]
\begin{center}
\includegraphics[width=0.4\textwidth]{\main/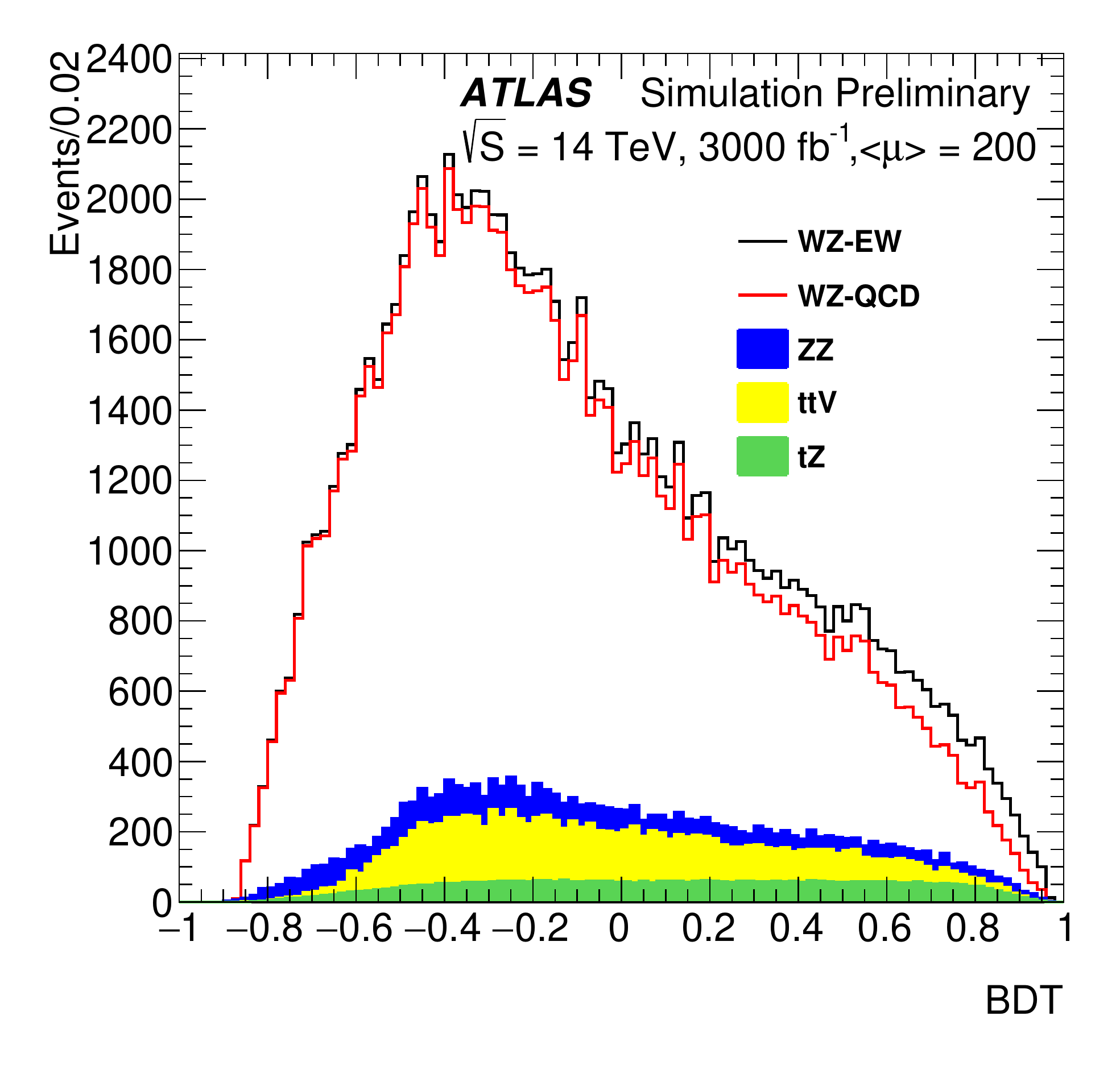}
\includegraphics[width=0.49\textwidth]{\main/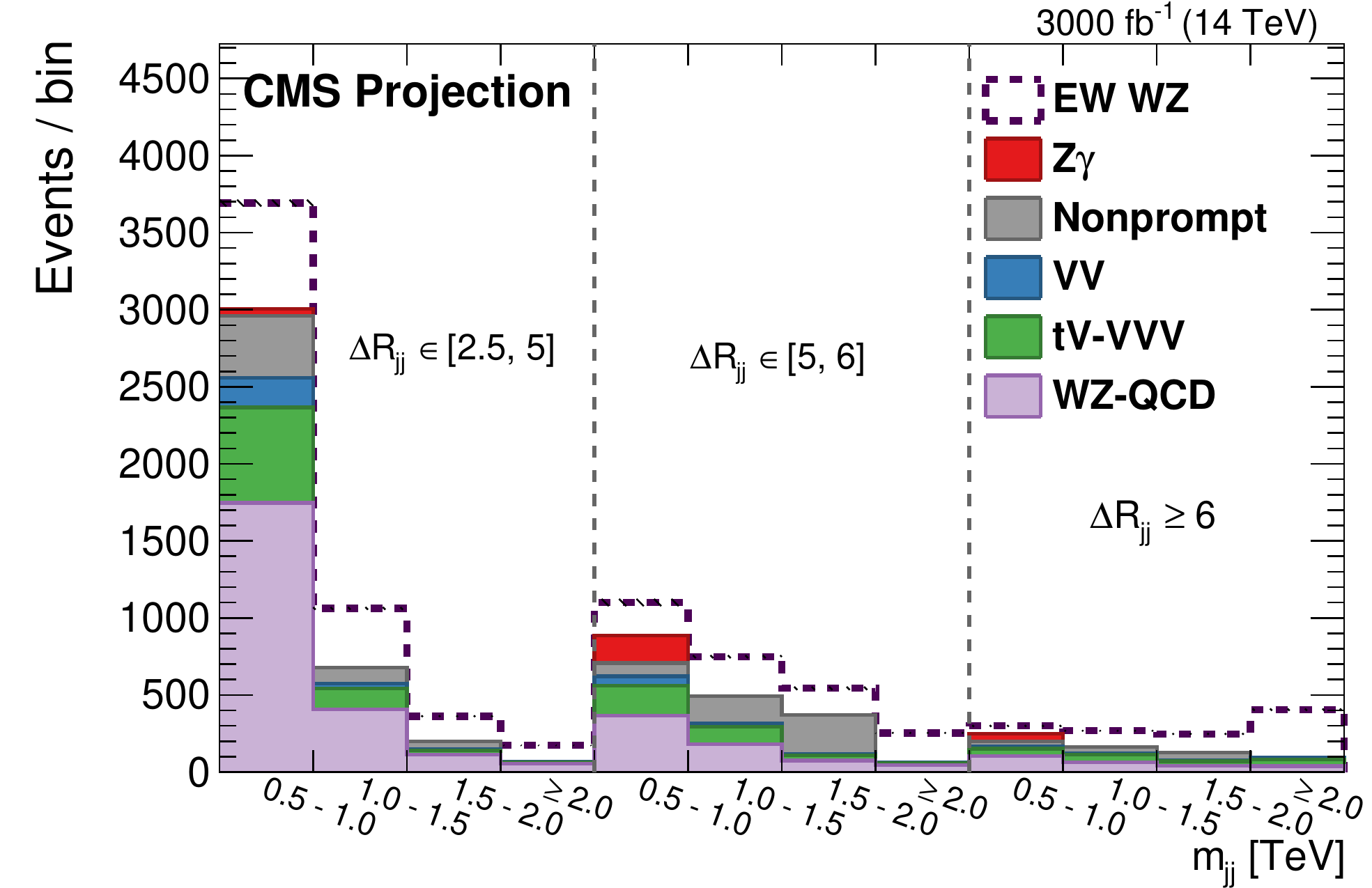}
\caption{Example of BDT distribution for 3000\fbinv (left). The $m_{jj}$ distributions in bins of $\Delta R_{jj}$ for 3000\fbinv (right). \label{Fig:mjj}}
\end{center}
\end{figure}

To extract the electroweak signal, ATLAS uses nominally a final $m_{jj}$ cut optimised at 600 GeV or a multivariate analysis (BDT) based on 25 variables that are shown to best separate the signal and background events. The shape of the BDT output is shown in Fig.~\ref{Fig:mjj} left. In the CMS case, a 2D distribution of dijet invariant mass in bins of dijet angular separation is used, as shown in Fig.~\ref{Fig:mjj} right. The measurement of the \WZEW\ production cross section results from a maximum likelihood fit of this distribution performed simultaneously for four different lepton combinations in the final states, each combination being considered as independent decay channel. The systematic uncertainties are represented by nuisance parameters in the fit and are allowed to vary according to their probability density functions. The correlations across bins, between different sources of uncertainty  and decay channels are taken into account. The background contributions are allowed to vary within the estimated uncertainties.

\begin{figure}[htbp]
\begin{center}
\includegraphics[width=0.45\textwidth]{\main/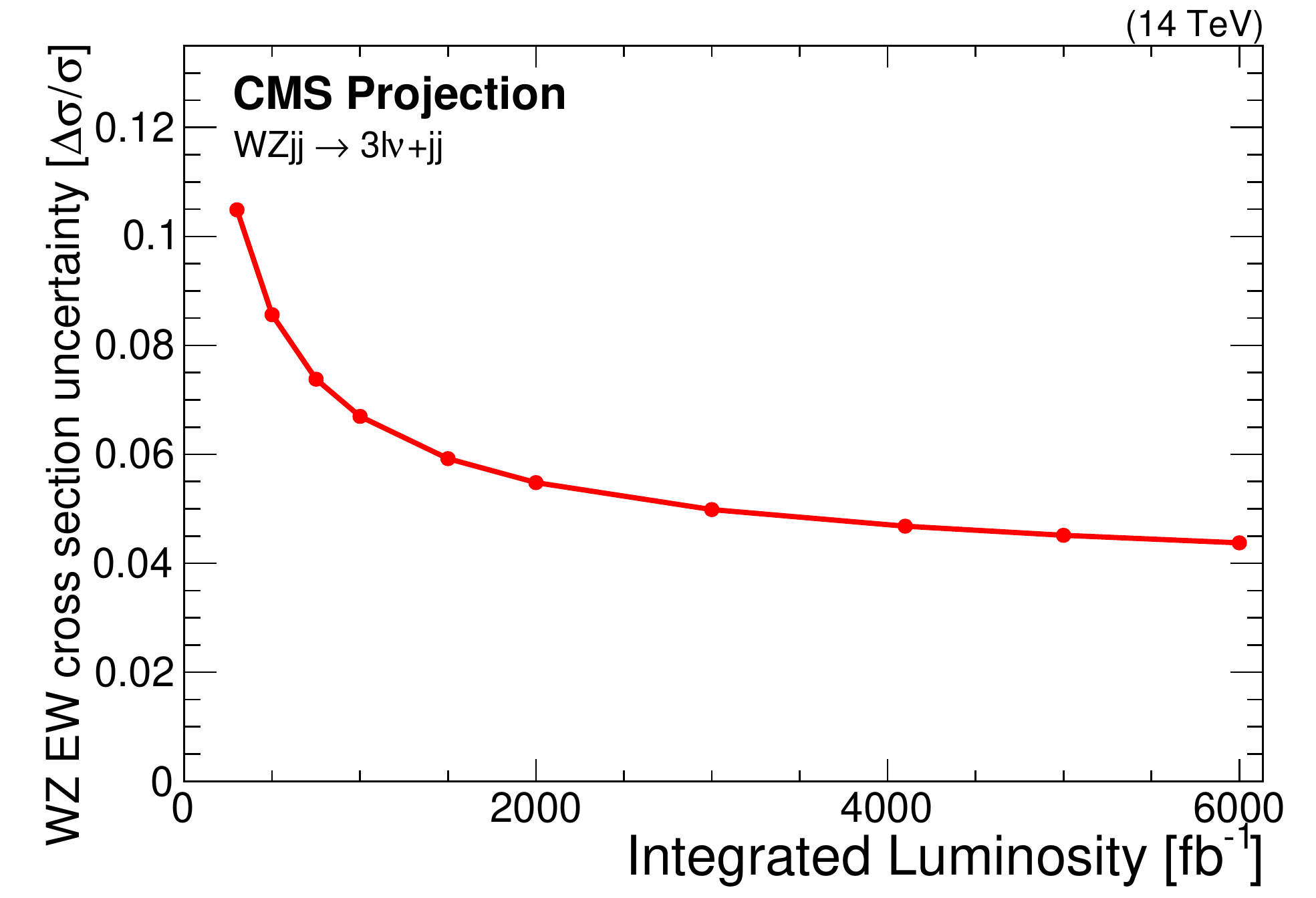}
\includegraphics[width=0.45\textwidth]{\main/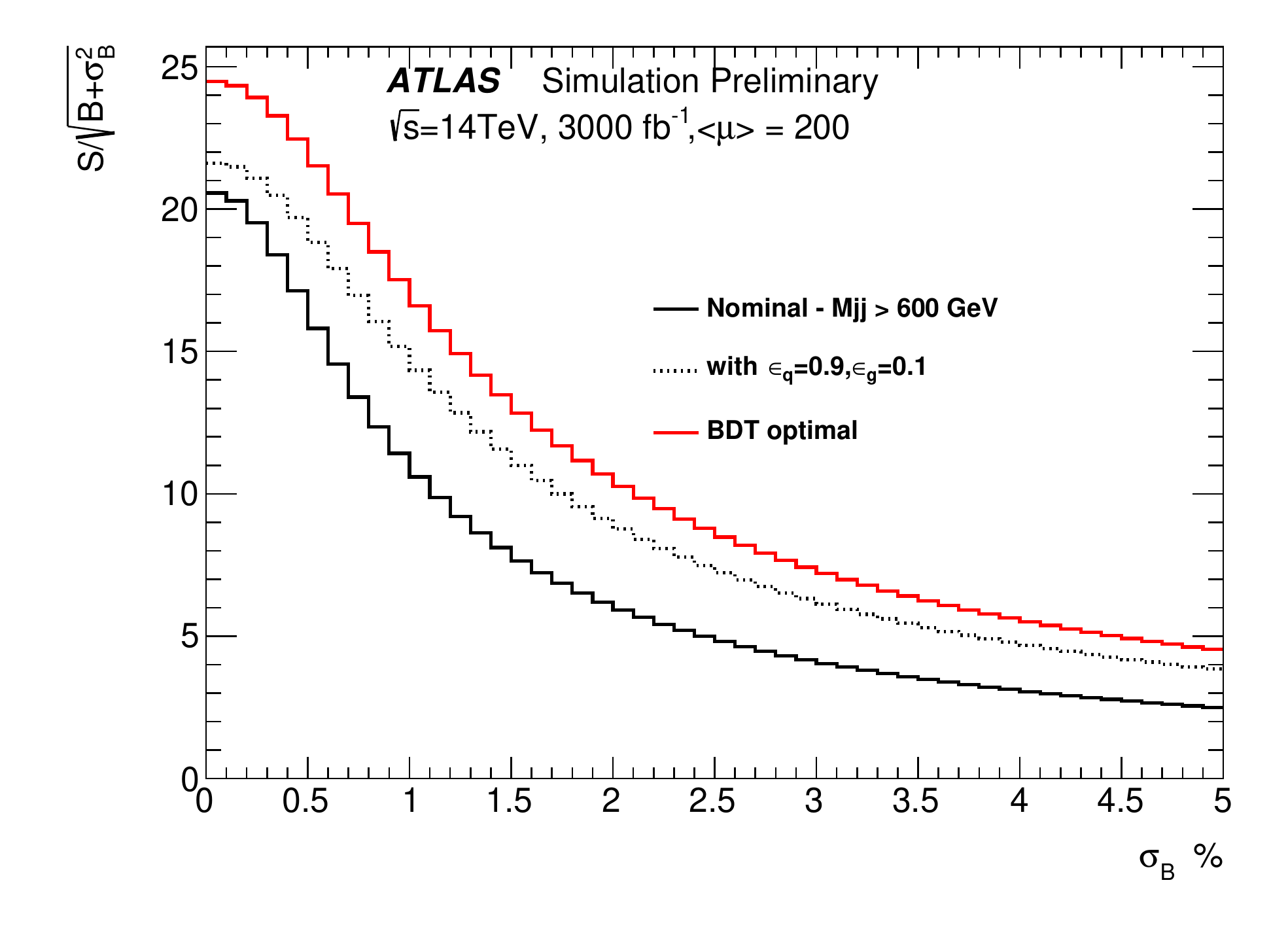}
\caption{Expected uncertainty on the cross section measurement as a function of the integrated luminosity for the CMS projection (left). Signal significance versus the total background uncertainty for the ATLAS simulation (right), presented for the nominal selection, along with two alternative selections meant to mitigate the \WZQCD\ background. \label{Fig:syst}} 
\end{center}
\end{figure}

The experimental systematic error will be dominated by the jet energy related uncertainties, and amounts to a maximum of 5\%. The non-prompt background uncertainty may also be significant depending on the final state. Depending on the level of \WZQCD\ background, the theoretical error affecting its modeling will eventually dominate. However it is expected that the impact of these uncertainties can be controlled to less that 5\% using refined and diverse control regions allowed by the large statistics at HL-LHC. The total uncertainty of the electroweak cross section measurement as a function of luminosity is shown in Fig.~\ref{Fig:syst} left for the CMS projection, while the signal significance as a function of the projected total uncertainty on background is presented in Fig.~\ref{Fig:syst} right for the ATLAS simulation as it is arguable whether the theoretical uncertainty can be precisely predicted at this stage.

The polarisation of the final state bosons can be measured inclusively for each boson in two different final state configurations, $ZW^{+}$ and $ZW^{-}$ or combined in a doubly longitudinally polarised final state. 
The \cthsZ\ (~\cthsW ~), where $\theta_{Z}^{*}$ represents the angle of the lepton with the $Z$ ($W$) direction in the $WZ$ rest frame, is the most sensitive differential distribution to the polarisation of the $Z$ ($W$) boson. An example of the \cthsZ\ distribution is shown in Fig.~\ref{Fig:ATL_polar} left for the \WZEW\ signal and the sum of backgrounds for $Z(W^{+})$ final state; the distribution is fitted with three parameters: the longitudinal polarised fraction F0, the left-handed minus right-handed contributions and the number of \WZEW\ events using three polarisation templates plus the two background contributions. The result of the fit is shown in Fig.~\ref{Fig:ATL_polar} left, where the fraction of \WZEW\ events where the $Z$-boson is longitudinally, left or right polarised are represented, while the log-likelihood profile corresponding to F0 is presented in Fig.~\ref{Fig:ATL_polar} right. The significance to measure F0, computed as $\sqrt{-2log(\lambda(F0=0))}$, is estimated to be between 1.5 and 2.5 $\sigma$ for $Z(W^{+})$ and 0.7 and 1.5 $\sigma$ for $W^{-}$ depending on the final selection that affects the signal purity, and systematic assumptions on the total background normalisation. 

\begin{figure}[htbp]
\begin{center}
\includegraphics[width=7cm]{\main/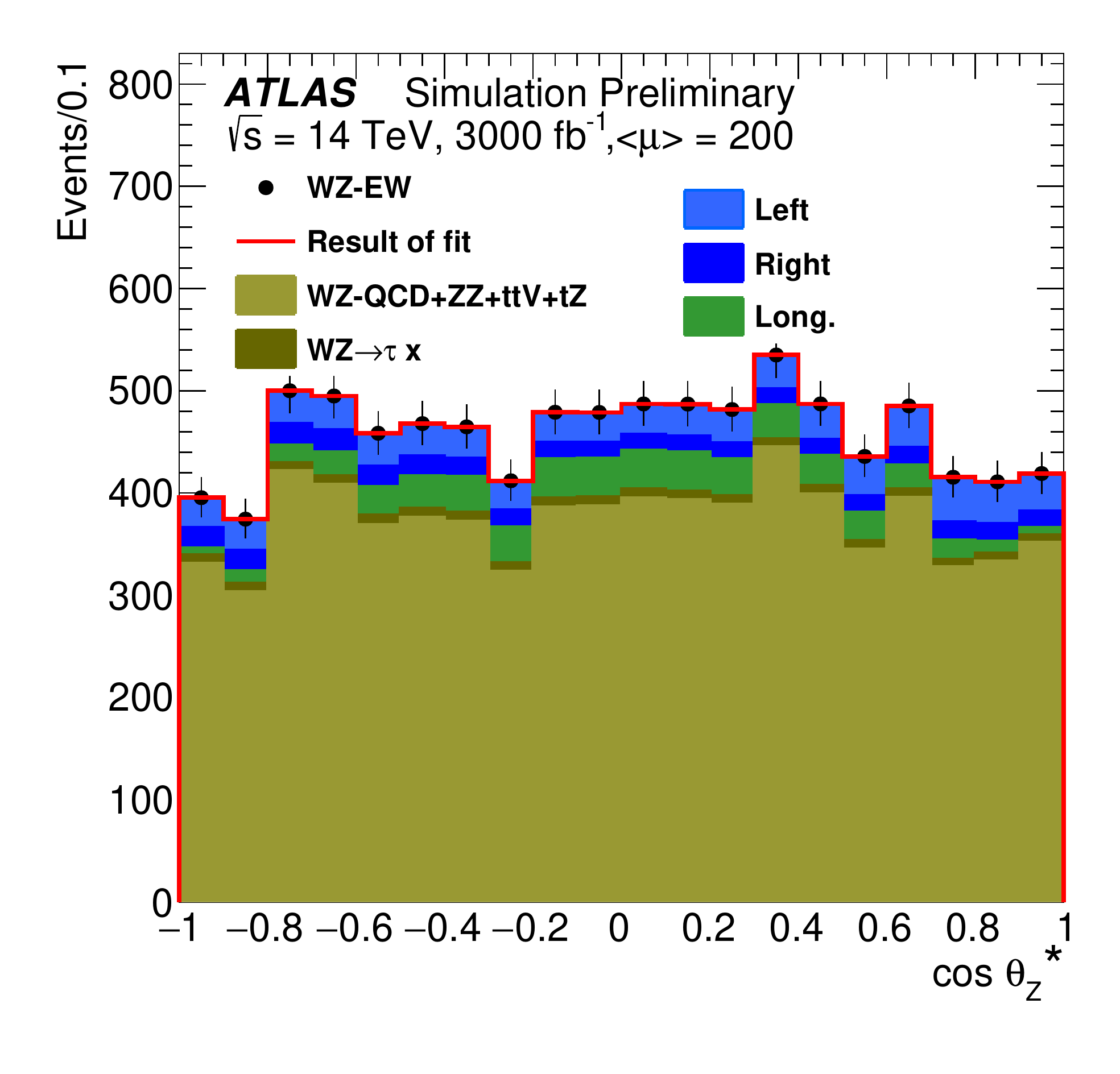}
\includegraphics[width=7cm]{\main/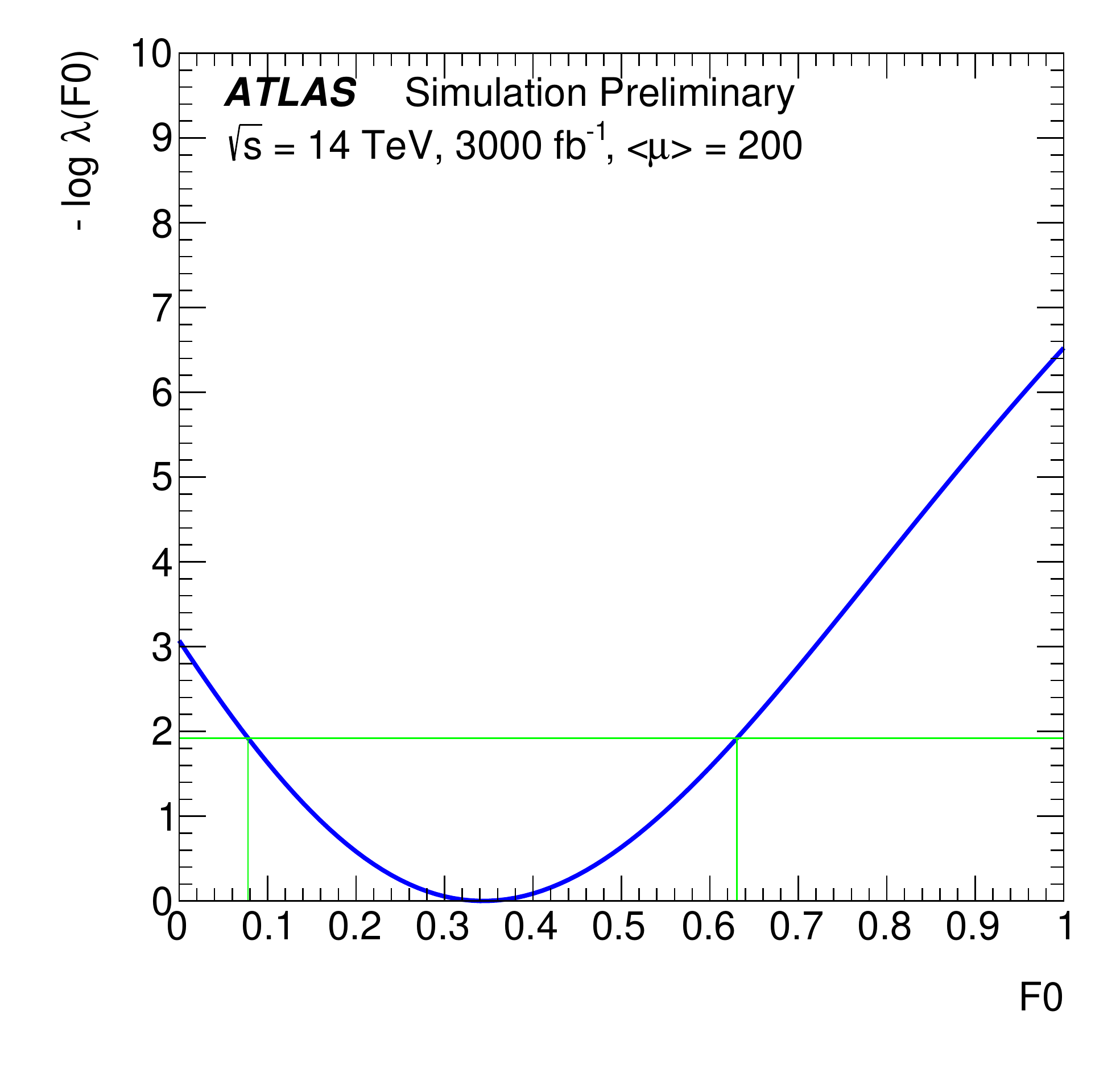}
\vspace{-0.5cm}
\caption{Distribution of \cthsZ\ for 3000\fbinv and result of the fit of the long, polarised contribution (F0), left- and right-handed contributions on top of the sum of backgrounds and of the $WZ \rightarrow x\tau$ background, both contributions taken into account with a normalisation error of 2.5\% (left). Shape of the log-likelihood profile for the F0 parameter around its minimum (right) \label{Fig:ATL_polar}.}
\end{center}
\end{figure}

To measure the doubly longitudinal (LL) process, an approach based on the jets kinematics similar to this for the total \WZEW\ cross section is used by CMS. 
The LL fraction is expected to be of the order of 5\% of the total $\WZEW$ production (\cite{Alwall:2011uj}) and its unrolled 2D distribution is shown in Fig.~\ref{fig:LLSignificance} left for 3000\fbinv. It can be observed that the LL contribution is increasing from~2-3\% to~7-8\% for high angular separation between jets and for high invariant mass of the dijet system. In the fit, the LL fraction is considered as signal, while the rest of the \WZEW\ process is considered as an additional background. The systematic uncertainties of the LL and non-LL fractions are considered as fully correlated within the total electroweak cross section. The significance of the LL observation as a function of integrated luminosity is shown in Fig.~\ref{fig:LLSignificance} right: the red curve presents the significance if only statistical uncertainties of the measurement are taken into account and the black line presents the results including the systematics as discussed above. 

\begin{figure}[htbp]
  \centering
   \includegraphics[width=0.45\textwidth]{\main/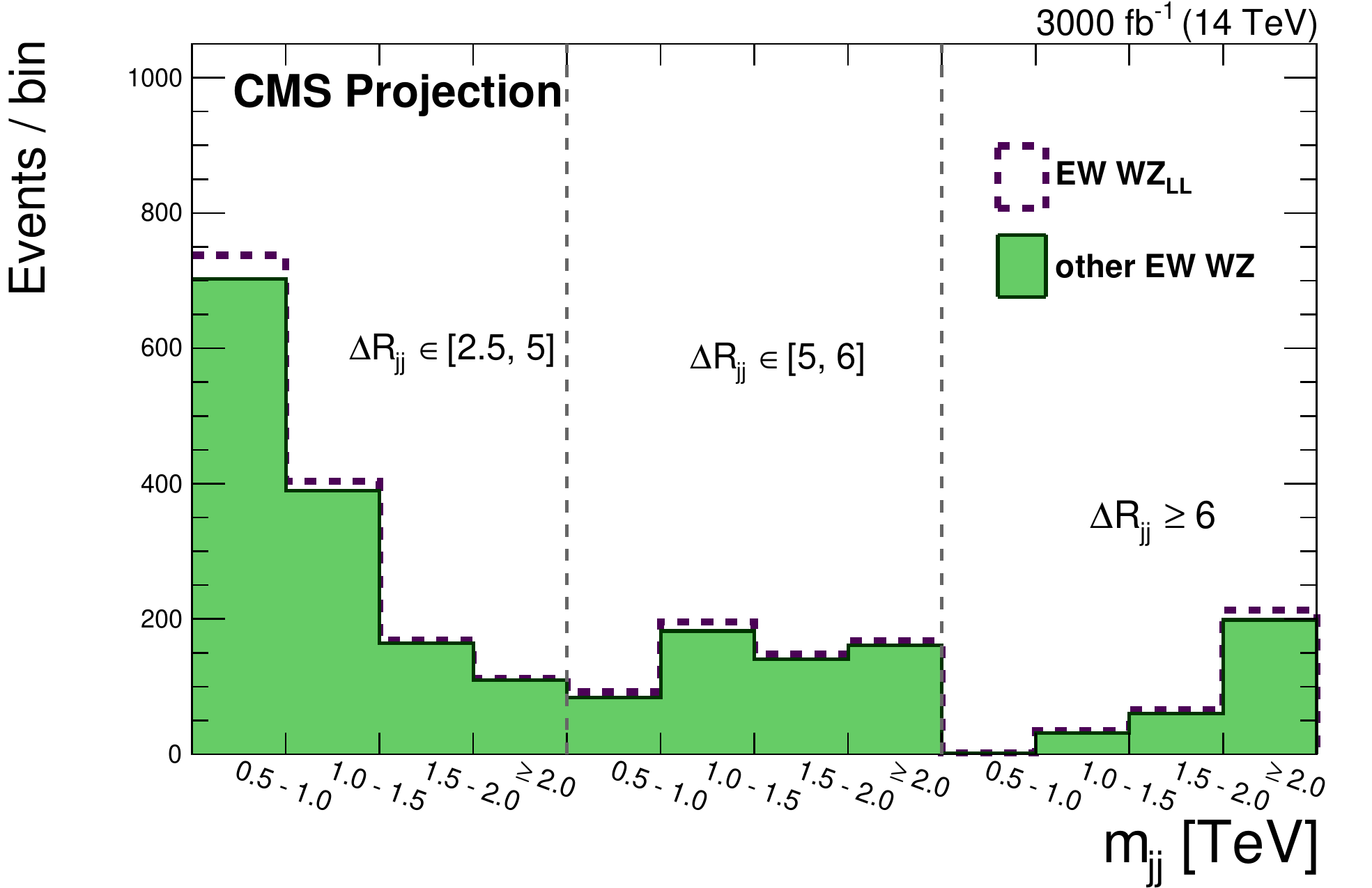}
   \includegraphics[width=0.45\textwidth]{\main/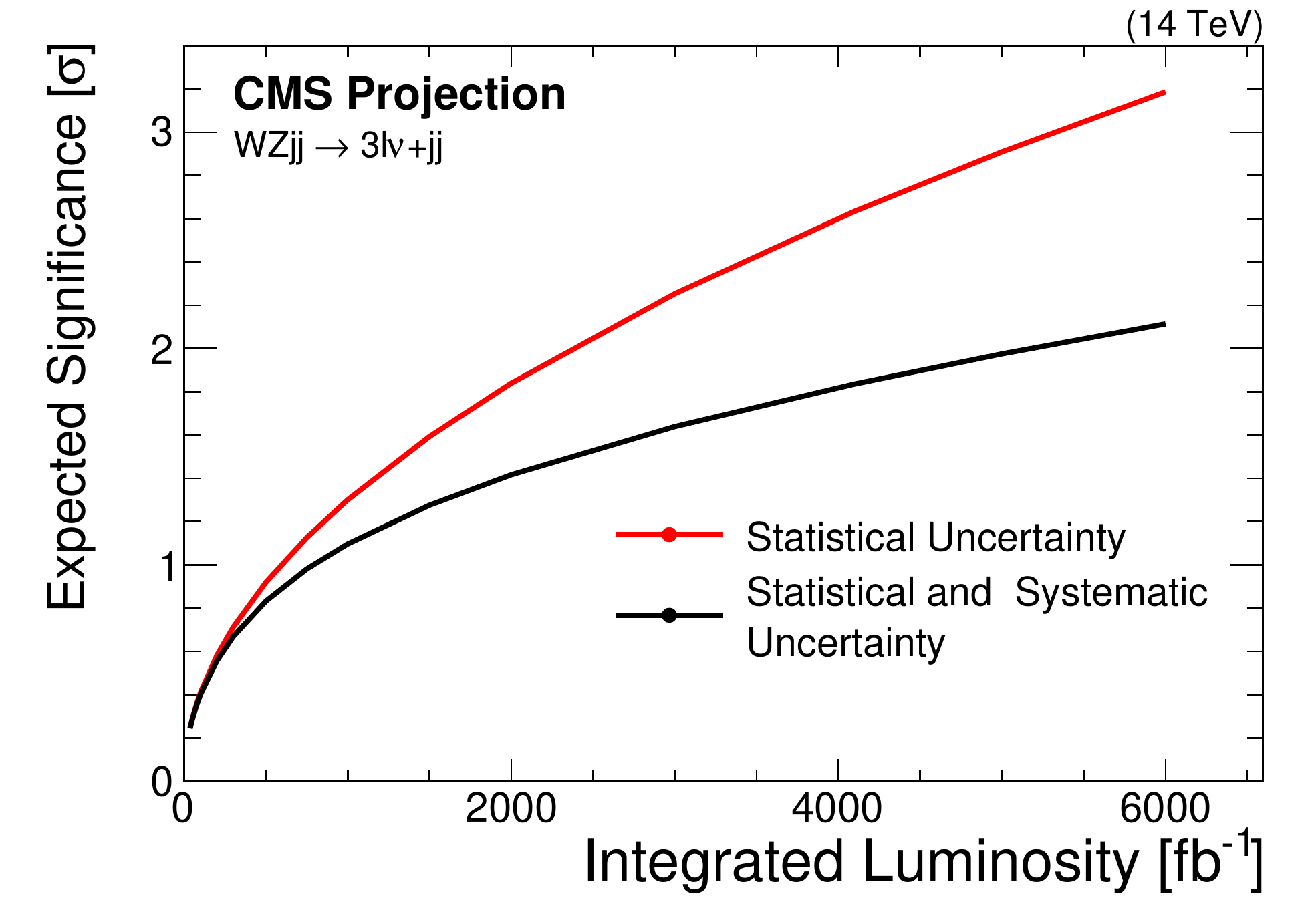}
  \caption{Unrolled 2D ($\Delta R_{jj};m_{jj}$) LL and non-LL distribution for 3000\fbinv (left). Significance of the LL observation with and without systematic error (right).}
 \label{fig:LLSignificance}
\end{figure}

The results presented in this section confirm that the \WZEW\ cross section can be measured with accuracy at the HL-LHC if the jets from pileup collisions in the events are well separated from the jets produced in the hard interactions. Increased pseudorapidity coverage of the detectors should improve precision of such measurement. The single polarized cross sections can also be measured and the double polarized measurement requires more sophisticated methods, including development of multivariate discriminants for better separation of the signal from background. Systematic uncertainties also start to play a significant role at the HL-LHC, in particular those affecting the theoretical prediction.

\subsubsection[Prospects for quartic gauge coupling measurements in VBS]{Prospects for quartic gauge coupling measurements in VBS\footnote{Contribution by H.~Sch\"afer-Siebert and D.~Zeppenfeld.}}
%

Due to the strong gauge theory cancellations between the different Feynman graphs present in VBS (Fig.~\ref{Fig:Feymann}) the various VBS processes provide excellent probes for the structure of gauge boson interactions, in particular for the quartic gauge couplings. Deviations from SM predictions can conveniently be parameterised by an effective Lagrangian, ${\cal L}_{\rm EFT} = \sum_i f_i/\Lambda^{d_i-4} \, {\cal O}_i^{(d_i)}$, where the operators ${\cal O}_i^{(d_i)}$ of energy dimension $d_i$ are built with the covariant derivative of the SM Higgs doublet field, $D_\mu\Phi$, and the $SU(2)_L$ and $U(1)_Y$ field strength tensors $\widehat{W}_{\mu\nu}$ and $\widehat{B}_{\mu\nu}$ (normalized according to  $[D_\mu,D_\nu]=\widehat{W}_{\mu\nu}+\widehat{B}_{\mu \nu}$). At the dimension six level, all allowed operators in ${\cal L}_{\rm EFT}$ also contribute to trilinear couplings of electroweak gauge bosons or to $hVV$ couplings, which are better measured in $q\bar q\rightarrow VV$ processes or in Higgs boson decay. Thus, operators of energy dimension eight, which do not give rise to anomalous trilinear couplings, are used for a parameterisation of anomalous quartic gauge couplings (aQGC), which is sufficiently general for the present purpose. In the following, the operator basis of Ref.~\cite{Eboli:2006wa,Eboli:2016kko} with {\mbox{\textsc{VBFNLO}}\xspace}  normalization~\cite{Rauch:2016pai,Baak:2013fwa,Perez:2018kav} is used to assess the sensitivity of VBS $W^\pm W^\pm jj$ and $WZjj$ production to aQGC, with the subset of operators
\begin{subequations}
  \label{eq:dim8-longitudinal}
  \begin{alignat}{2}
    {\cal O}_{S_0} &= \left [ \left ( D_\mu \Phi \right)^\dagger
      D_\nu \Phi \right ] &&\times
    \left [ \left ( D^\mu \Phi \right)^\dagger
      D^\nu \Phi \right ] \, ,\\
    {\cal O}_{S_1} &= \left [ \left ( D_\mu \Phi \right)^\dagger
      D^\mu \Phi  \right ] &&\times
    \left [ \left ( D_\nu \Phi \right)^\dagger
      D^\nu \Phi \right ] \, \\
    {\cal O}_{T_0} &=   \hbox{Tr}\left [ {\widehat{W}}_{\mu\nu} {\widehat{W}}^{\mu\nu} \right ]
    &&\times   \hbox{Tr}\left [ {\widehat{W}}_{\alpha\beta} {\widehat{W}}^{\alpha\beta} \right ]  
    \, ,\\
    {\cal O}_{T_1} &=   \hbox{Tr}\left [ {\widehat{W}}_{\alpha\nu} {\widehat{W}}^{\mu\beta} \right ] 
    &&\times   \hbox{Tr}\left [ {\widehat{W}}_{\mu\beta} {\widehat{W}}^{\alpha\nu} \right ]  \\
  {\cal O}_{M_0} &=   \hbox{Tr}\left [ {\widehat{W}}_{\mu\nu} {\widehat{W}}^{\mu\nu} \right ]
  &&\times  \left [ \left ( D_\beta \Phi \right)^\dagger
    D^\beta \Phi \right ]  
  \, ,\\
  {\cal O}_{M_1} &=   \hbox{Tr}\left [ {\widehat{W}}_{\mu\nu} {\widehat{W}}^{\nu\beta} \right ]
  &&\times  \left [ \left ( D_\beta \Phi \right)^\dagger
    D^\mu \Phi \right ]  
  \,  .
\end{alignat}
\end{subequations}
in ${\cal L}_{\rm EFT}= \sum_i \frac{f_i}{\Lambda^4} \, {\cal O}_i$. At high invariant masses, $\sqrt s$, of the $VV\rightarrow VV$ subprocess, the tree level insertions of the dimension eight operators lead to matrix elements which grow like $s^2$ and violate unitarity within the accessible energy range of the LHC. This unphysical behaviour is avoided below by using the unitarization scheme of Ref.~\cite{Perez:2018kav}, dubbed $T_u$-model, which is a variant of K-matrix unitarization, producing close to maximal absolute values of the partial wave amplitudes at high energies.

In the presence of aQGC which signify strong interactions in the bosonic sector, VBS cross sections are enhanced at high $VV$ invariant masses, which feeds into observables correlated to $m_{VV}$ such as the integrated dilepton invariant mass distribution for \ssWW\ events shown in Fig.~\ref{fig:dilepton} or the integrated $WZ$ transverse mass distribution shown in Fig.~\ref{fig:MT-WZjj-cuts}. The $m_{\rm{T}}(WZ)$-distribution is obtained from the ATLAS $WZjj$ analysis (see Table~\ref{tab:WZEWKselections}) with the additional cuts $m_{jj}> 600$~GeV, $\Delta\eta_{jj}>3.0$ on the invariant mass and the rapidity separation of the tagging jets, and $|\eta_\mu| <2.7$ on muon rapidity. Also shown in Fig.~\ref{fig:MT-WZjj-cuts} are $m_{\rm{T}}(WZ)$-distributions for $f_{M_0}/\Lambda^4=3.8$~TeV$^{-4}$ within the pure EFT and including the unitarization of the $T_u$-model for the VBS $WZjj$ signal. Detector effects are included by assuming the same efficiencies in each $m_{\rm{T}}(WZ)$ bin as for the SM EW  signal. The processes contributing to the background distribution in Fig.~\ref{fig:MT-WZjj-cuts} are listed in Table~\ref{tab:yields3000fbWZEWK}. The aQGC leads to an excess of events at very high $m_{\rm{T}}(WZ)$. Assuming that no significant excess is observed in the high energy tail, one finds the expected 95\% CL bounds on aQGC listed in Table~\ref{table:aQGC}. Also shown in the Table are bounds expected from \ssWW\ production, based on the dilepton invariant mass distribution of Fig.~\ref{fig:dilepton}. The expected bounds for the HE-LHC are obtained in a similar fashion, assuming the same signal to background ratio as at 14 TeV for the SM case, and generating VBS \ssWW\ and $WZjj$ events with {\mbox{\textsc{VBFNLO}}\xspace} at LO QCD.

The above procedure provides conservative estimates for the sensitivity to aQGC in VBS: The experimental VBS analyses focused on the significance of the various SM VBS signals and did not try to optimize sensitivity to deviations at highest $VV$ invariant masses, as would be favorable for aQGC measurements. Taking into account weak boson rapidity and transverse momentum distributions and correlations, the sensitivity to aQGC could be improved somewhat. On the other hand, dedicated analyses including Sudakov suppression at high invariant mass, as discussed in Section~\ref{sec:theory-HO-ssWW}, which is expected to slightly decrease sensitivity to aQGC, have not been performed yet in the above setting.

\begin{figure}
\centering
\includegraphics[scale=0.6]{\main/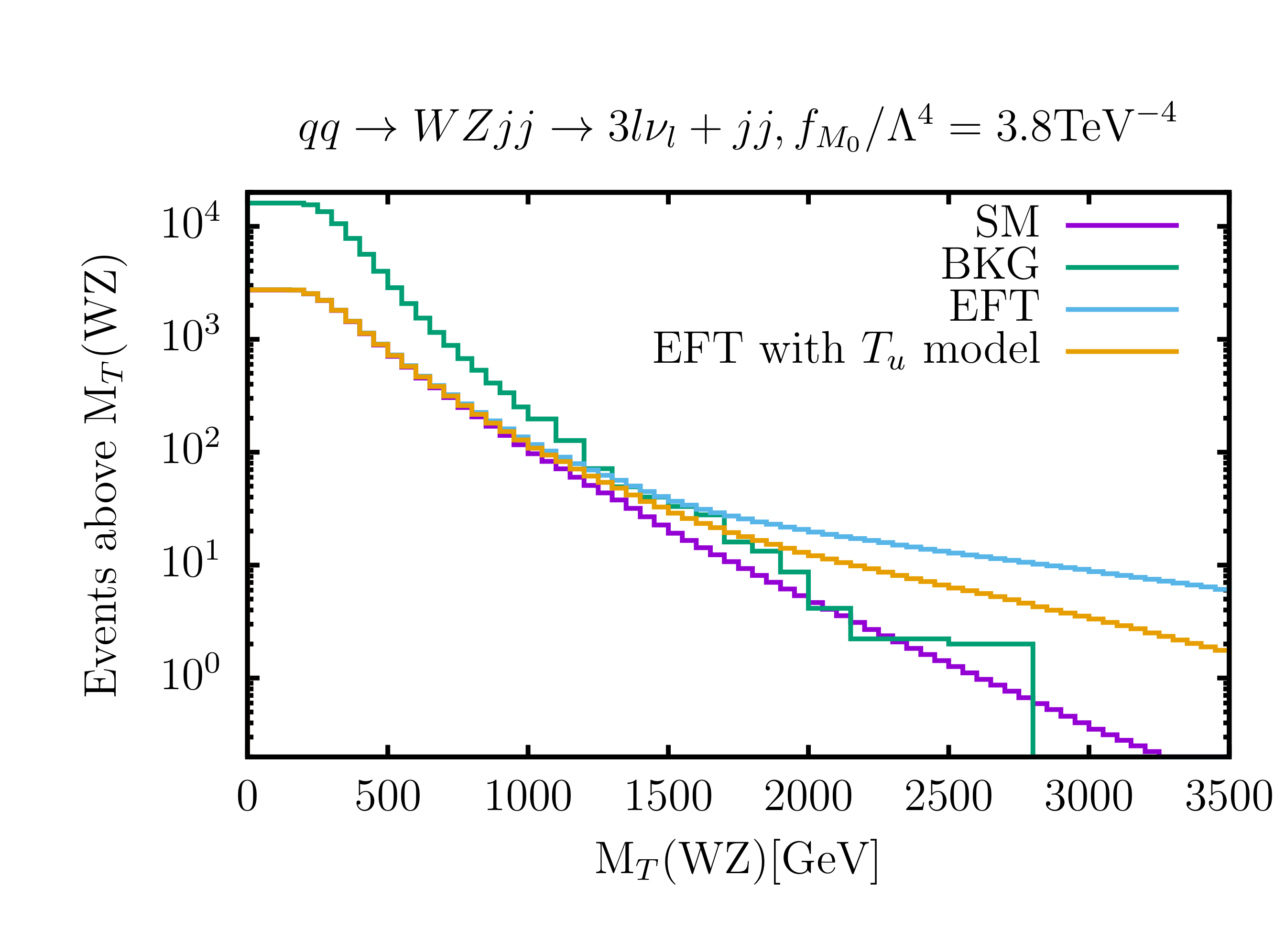}
\caption{Integrated $WZ$ transverse mass distribution for $f_{M_0}/\Lambda^4=3.8$~TeV$^{-4}$ within the pure EFT, the unitarization of the $T_u$-model as well as the SM VBS signal and the background predictions based on the ATLAS $WZjj$ analysis.}
\label{fig:MT-WZjj-cuts}
\end{figure}

\begin{table}
\caption{Expected bounds (in $\mathrm{TeV}^{-4}$) on the coefficients of dimension-8 operators, assuming no significant excess in the integrated $m_{\rm{T}}(WZ)$ ($WZjj$) or $m_{ll}$ ($W^{\pm}W^{\pm}jj$) distributions at high mass.}
\label{table:aQGC}
\centering
\begin{tabular}{|c|c|c|c|c|}
\hline
{}&\multicolumn{2}{|c|}{$14 \:\mathrm{TeV}$} &\multicolumn{2}{c|}{$27 \:\mathrm{TeV}$}\\
{}&$WZjj$&$W^{\pm}W^{\pm}jj$&$WZjj$&$W^{\pm}W^{\pm}jj$\\
\hline\hline
$f_{S_0}/\Lambda^4$&[-8,8] &[-6,6]&[-1.5,1.5]&[-1.5,1.5]\\
$f_{S_1}/\Lambda^4$&[-18,18] &[-16,16]&[-3,3]&[-2.5,2.5]\\
$f_{T_0}/\Lambda^4$&[-0.76,0.76]&[-0.6,0.6]&[-0.04,0.04]&[-0.027,0.027]\\
$f_{T_1}/\Lambda^4$&[-0.50,0.50]&[-0.4,0.4]&[-0.03,0.03]&[-0.016,0.016]\\
$f_{M_0}/\Lambda^4$&[-3.8,3.8]&[-4.0,4.0]&[-0.5,0.5]&[-0.28,0.28]\\
$f_{M_1}/\Lambda^4$&[-5.0,5.0]&[-12,12]&[-0.8,0.8]&[-0.90,0.90]\\
\hline
\end{tabular}

\end{table}

\subsubsection{Measurements of $ZZ$ scattering}

This section presents the studies performed for VBS in the $ZZ$ fully leptonic decay channel for HL-LHC and HE-LHC. 
Despite the very low cross section times branching fraction, the reconstruction of all final state leptons allows to precisely measure the angular distributions of the $Z$ decays to optimally separate the longitudinal from the dominating transverse polarizations. In addition, a precise measurement of the hard scattering centre-of-mass energy is possible from the reconstructed four-leptons invariant mass. Last but not least, the reducible background in this channel is very small, making it an ideal case for high statistics measurement since the impact of associated experimental systematics uncertainties is expected to be very small.

The ATLAS analysis is performed with simulated events at generator level
at 14 TeV, where the detector effects of lepton and jet reconstruction and identification were estimated by corrections, assuming a mean number of interactions per bunch crossing of 200. The CMS analysis is based on the experimental investigation of VBS in the $ZZ$ channel using 36\,\ifb of data collected in 2016~\cite{Sirunyan:2278242} which showed an observed significance of 2.7 standard deviations. This analysis is projected to HL-LHC conditions\cite{CMS-PAS-FTR-18-014} by scaling the expected yields for the signal and  background processes, taking into account the increase in luminosity and scattering energy as well as the changes in acceptance and selection efficiencies between the LHC Phase-1 (13 TeV) and the HL-LHC (14 TeV) configurations. The {\mbox{\textsc{Delphes}}\xspace} simulation~\cite{deFavereau:2013fsa} is then used to assess the sensitivity to VBS ${Z_{\rm {L}}Z_{\rm {L}}}$. The HL-LHC result is further projected to the HE-LHC configuration.

Several Monte Carlo event generators were used to simulate the signal and background contributions. In the ATLAS analysis, both the EW-$ZZjj$ and QCD-$ZZjj$ processes with the $ZZ \rightarrow \llll$ decays are modeled using \SHERPA~v2.2.2~\cite{Gleisberg:2008ta} with the NNPDF3.0NNLO~\cite{ball2015parton} parton distribution functions (PDFs) set. The signal sample is generated with two jets at Matrix Element (ME) level. The background process is modeled with next-to-leading order (NLO) QCD accuracy for events with up to one outgoing parton and with leading order (LO) accuracy for the case with two and three partons, in a phase space of $m_{\ell\ell} > $ 4~GeV and at least two leptons with $\pT > $ 5~GeV. Other backgrounds have minor contributions to the \llll channel and therefore are not included. The CMS analysis uses \mgamcatnlo~v2.3.3~\cite{MGatNLO} to simulate the EW-$ZZjj$ signal and QCD-$ZZjj$ background samples with zero, one, and two outgoing partons at Born level at NLO. The different jet multiplicities are merged using the FxFx scheme~\cite{Frederix:2012ps} with a merging scale of 30~GeV, and leptonic $ Z$ boson decays were simulated using \textsc{MadSpin}~\cite{Artoisenet:2012st}. The gluon loop-induced production of two Z bosons ($ggZZ$) is simulated at LO with \MCFM~v.7.0.1~\cite{MCFM}, and checked with a dedicated simulation of the loop-induced $\mathrm{gg} \to ZZjj$  process using \mgamcatnlo. The NNPDF3.0 PDF set is also used. The interference between EW-$ZZjj$ and QCD-$ZZjj$ processes is found to be small and is neglected in both analyses. Simulated samples with polarization information on the outgoing Z bosons are generated using \mgamcatnlo~v1.5.14
and the \textsc{decay} package from this version. 

The selections are based on Run-2 analyses and have been modified according to the expected changes for the detectors at HL-LHC. The foreseen forward lepton coverage is up to  $\vert \eta \vert = 4.0$ for both electrons and muons in ATLAS, while it is is up to
$\vert \eta \vert = 3.0 (2.8)$ for electrons (muons) in the CMS upgrade, with an option for an extension of up to $\vert \eta \vert = 4.0$ for electrons.
Candidate events should contain two pairs of oppositely charged isolated leptons (electrons or muons), consistent with the decays of two on-shell Z bosons. The VBS topology is ensured by requiring at least two jets with large invariant mass and $\eta$ separation in the cut based analysis, whereas an inclusive selection is used when the signal extraction is performed with a multivariate discriminant (BDT). Table~\ref{tab:selections} summarizes the details of the selection criteria used by the ATLAS and CMS collaborations.
\begin{table}[!h]
\begin{center}
\caption{Event selections used in ATLAS and CMS analyses. For the leptons $\eta$ and $\pT$ in CMS the first number refers to electrons and the second, in parenthesis, to muons.}
 \label{tab:selections}
 \begin{tabular}{| l | c | c |}
   \hline 
   & ATLAS  & CMS \\
   \hline\hline
     lepton $\eta$ &  $\vert \eta \vert < 4.0$ & $\vert \eta \vert < 3.0 (2.8)$ ($\vert \eta \vert < 4.0 (2.8),~\mathrm{extended~option})$\\
     lepton $\pT$  &  $\pT > 20,20,10,7$ GeV & $\pT > 20,12(10),10,7(5)$ GeV \\
     N leptons     & exactly 4 &  $\geq 4$ \\ 
     Z mass        & $60 < m_{ll} < 120 $ GeV & $60 < m_{ll} < 120 $ GeV \\
     Z$_{1}$ definition & $m_{ll}$ closest to PDG~\cite{Patrignani:2016xqp} value & $\pT$-leading Z \\
     jet $\eta$    & $\vert \eta \vert < 4.5$  & $\vert \eta \vert < 4.7$ \\
     jet $\pT$     &  $\pT > 30(70)$ GeV for $\vert \eta \vert < 3.8(>3.8)$ & $\pT > 30$ GeV \\
     N jets        & $\geq 2$, with $\eta^{j_1} \times \eta^{j_2} < 0$ & $\geq 2$ \\ 
     VBS cuts      & $m_{jj} > 600$ GeV and $\vert \Delta\eta_{jj} \vert > 2$ & $m_{jj} > 100$ GeV, signal extraction from BDT \\
   \hline
   \end{tabular}
   \end{center}
\end{table}

The distributions of the $ZZ$ invariant mass ($m_{ZZ}$) and the azimuthal angular difference between the two Z bosons ($|\Delta\phi(ZZ)|$)
are shown in Fig.~\ref{fig:ewk_qcd_normalized}, after the ATLAS event selection.
\begin{figure}[!b]
\centering
\includegraphics[width=0.48\textwidth]{\main/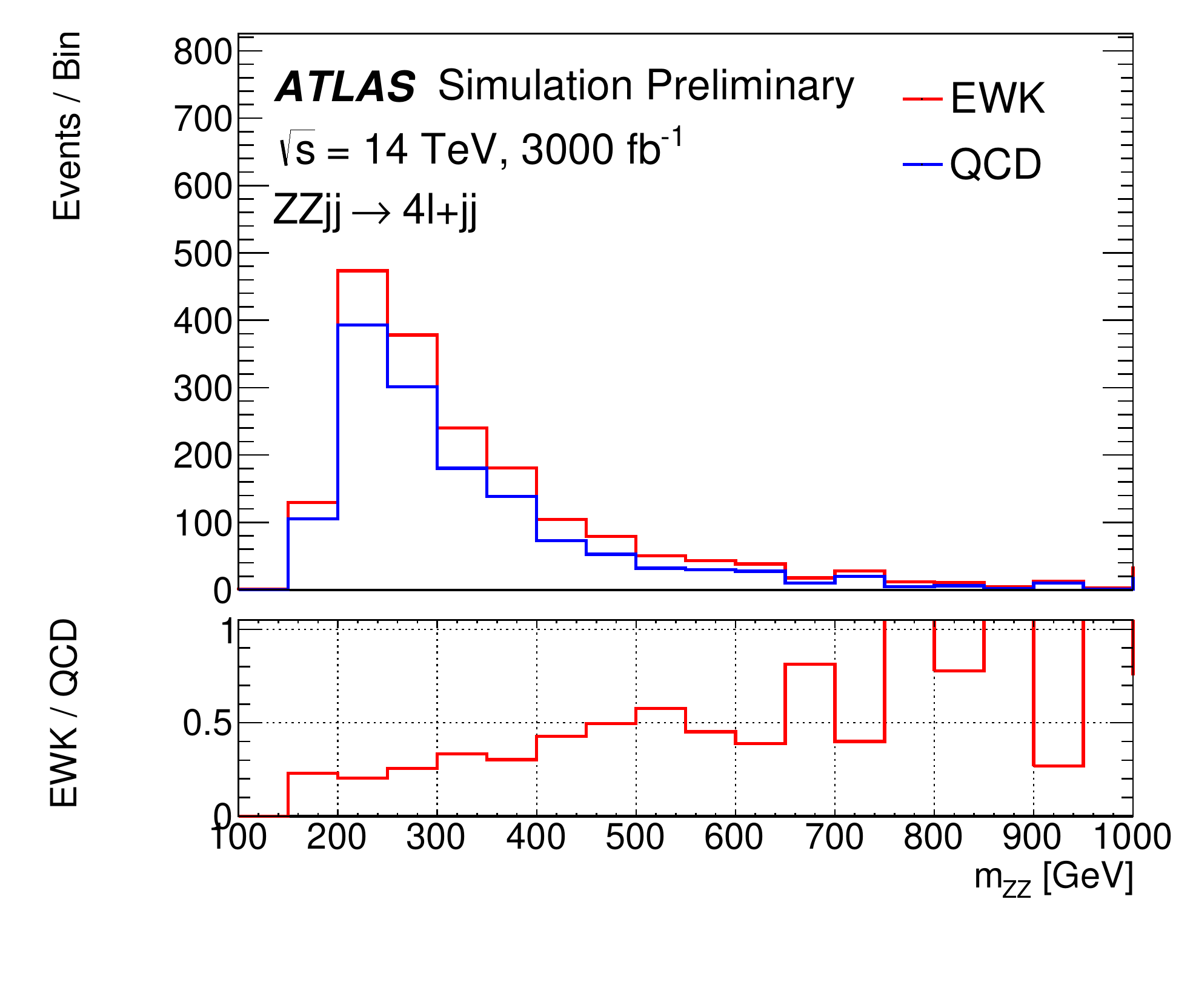}
\includegraphics[width=0.48\textwidth]{\main/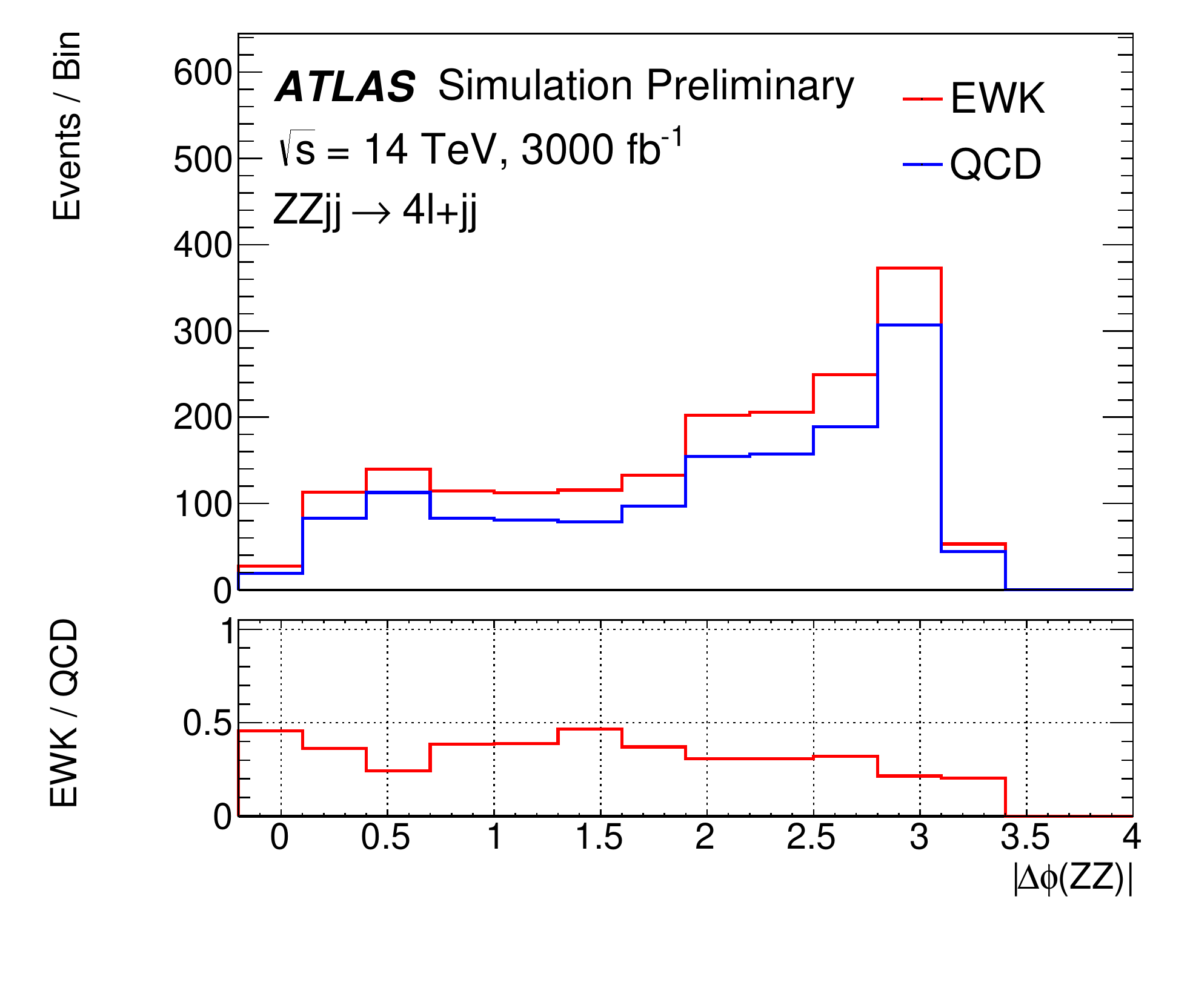}
\caption{ 
Detector level distributions of $m_{ZZ}$~and  $|\Delta\phi(ZZ)|$ for the EW and QCD $ZZjj$ processes after the cut-based event selection, normalized to 3000~\ifb{}. }
\label{fig:ewk_qcd_normalized}
\end{figure}
The numbers of selected signal and background events are quoted in Table~\ref{tab:yield}, normalized to 3000~\ifb{} of integrated luminosity. In addition to the baseline selection, two alternative selections are also studied to compare different detector scenarios at the HL-LHC. Uncertainties in the table refer to expected data statistical uncertainty at 14~TeV with 3000~\ifb{}.
\begin{table}[!h]
\begin{center}
\caption{Comparison of event yields for the signal ($N_{\mathrm{EW}-ZZjj}$) and background ($N_{\mathrm{QCD}-ZZjj}$) processes, and expected significance of EW-$ZZjj$ processes, normalized to 3000~\ifb{} data at 14~TeV, for baseline and alternative selections.}
\label{tab:yield}
  \begin{tabular}{|l|c|c|c|}
    \hline
    Selection & $N_{\mathrm{EW}-ZZjj}$ & $N_{\mathrm{QCD}-ZZjj}$ & $N_{\mathrm{EW}-ZZjj}$ / $\sqrt{N_{\mathrm{QCD}-ZZjj}}$   \\
    \hline\hline
    Baseline                                 & 432 $\pm$ 21 & 1402 $\pm$ 37   & 11.5 $\pm$ 0.6 \\
    Leptons with $|\eta|<$ 2.7               & 373 $\pm$ 19 & 1058 $\pm$ 33   & 11.5 $\pm$ 0.6 \\
    PU jet suppression only in $|\eta|<$ 2.4 & 536 $\pm$ 23 & 15470 $\pm$ 124 &  4.3 $\pm$ 0.2  \\
    \hline
  \end{tabular}
  \end{center}
\end{table}
The benefit of the extension for the rejection of PU jets is clear. The extended tracking coverage improves the lepton detection efficiency and increases the number of signal events, providing larger event yield for differential cross section measurements and for the longitudinal scattering. However, the overall significance of observing the EW-$ZZjj$ process does not improve as much, due to larger increase of the QCD-$ZZjj$ background contribution. 
This is due to the $ZZ$ system being more centrally produced in EW processes than in QCD processes. These results, however, do not include the gluon-induced contribution, for which the $ZZ$ system is found to be more centrally produced than for the leading quark-induced contribution. Moreover, in the case of the longitudinal scattering, the $\eta$ distribution of longitudinally polarized Z bosons is peaked in the forward region, therefore extended coverage is beneficial in this case as will be shown in the following.

The dominant systematics for \llll channel are from theoretical modeling of the QCD-$ZZjj$ background processes. The ATLAS analysis considers different sizes of systematic uncertainty in the background modeling of 5, 10 and 30\%{}. The 30\%{} uncertainty is a conservative estimation from direct calculation by comparing different choices of PDF sets and QCD renormalization and factorization scales, following recommendation from PDF4LHC~\cite{Butterworth:2015oua}. The 5\%{} one is an optimistic estimation where enough data events from QCD enriched control region at the HL-LHC could be used to provide constraints on the theoretical modeling of QCD-$ZZjj$ processes. For the experimental sources, the jet uncertainties have been checked following the studies in Ref.~\cite{ATL-PHYS-PUB-2016-026} and the effect is within fluctuation of the simulated events, which is at the 5\%{} level.
Thus a 5\%{} uncertainty is used as a conservative estimate of the experimental uncertainties. In this analysis these uncertainties are treated as uncorrelated and summed up quadratically. The CMS analysis considers two scenarios for the systematic uncertainties. The first scenario ('Run-2 scenario') consists in using the same systematic uncertainties as those used for the Run-2 analysis, apart from the uncertainty in the gluon-induced background contribution for which a 10\% uncertainty is considered. In the second scenario ('YR18 scenario'), improved systematic uncertainties are assumed to be obtained from the more data and better understanding of the detector. In this scenario, the theory systematic uncertainties (PDF and QCD scales) are halved with respect to the Run-2 scenario. In this analysis the systematic uncertainties are considered as nuisances in the fit and profiled.

Figure~\ref{fig:sig_scan_3000} (left) shows the result of a scan over different \mjj~cuts in addition to the ATLAS baseline selection, for an integrated luminosity of 3000~\ifb. The expected significance of EW-$ZZjj$ production processes is calculated as 
$\mathrm{Significance} = S / \sqrt{\sigma(B)_{stat.}^2 + \sigma(B)_{syst.}^2}$,
where S denotes the number of signal events after the selection,
and $\sigma(B)_{stat.}$ and $\sigma(B)_{syst.}$ refer to the statistical and systematic uncertainties in background yield. The statistical uncertainty is estimated from expected data yield at 14~TeV with 3000~\ifb.

\begin{figure}[!htbp]
\centering
    \resizebox{7cm}{!}{\includegraphics[width=0.4\textwidth, height=0.4\textwidth,trim=0 0.6cm 0 -0.6cm]{\main/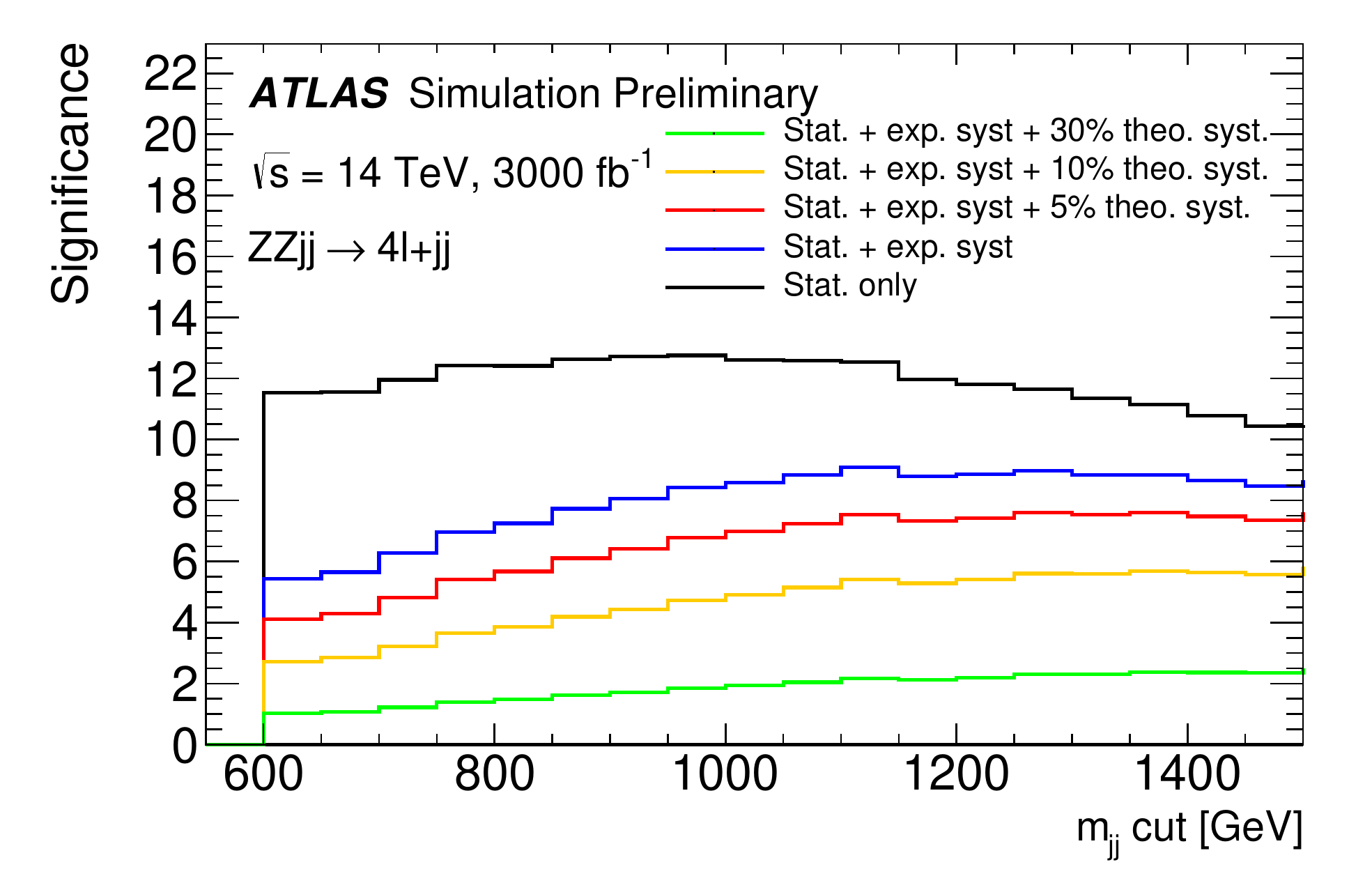}}
    \resizebox{7cm}{!}{\includegraphics[width=0.4\textwidth, height=0.398\textwidth]{\main/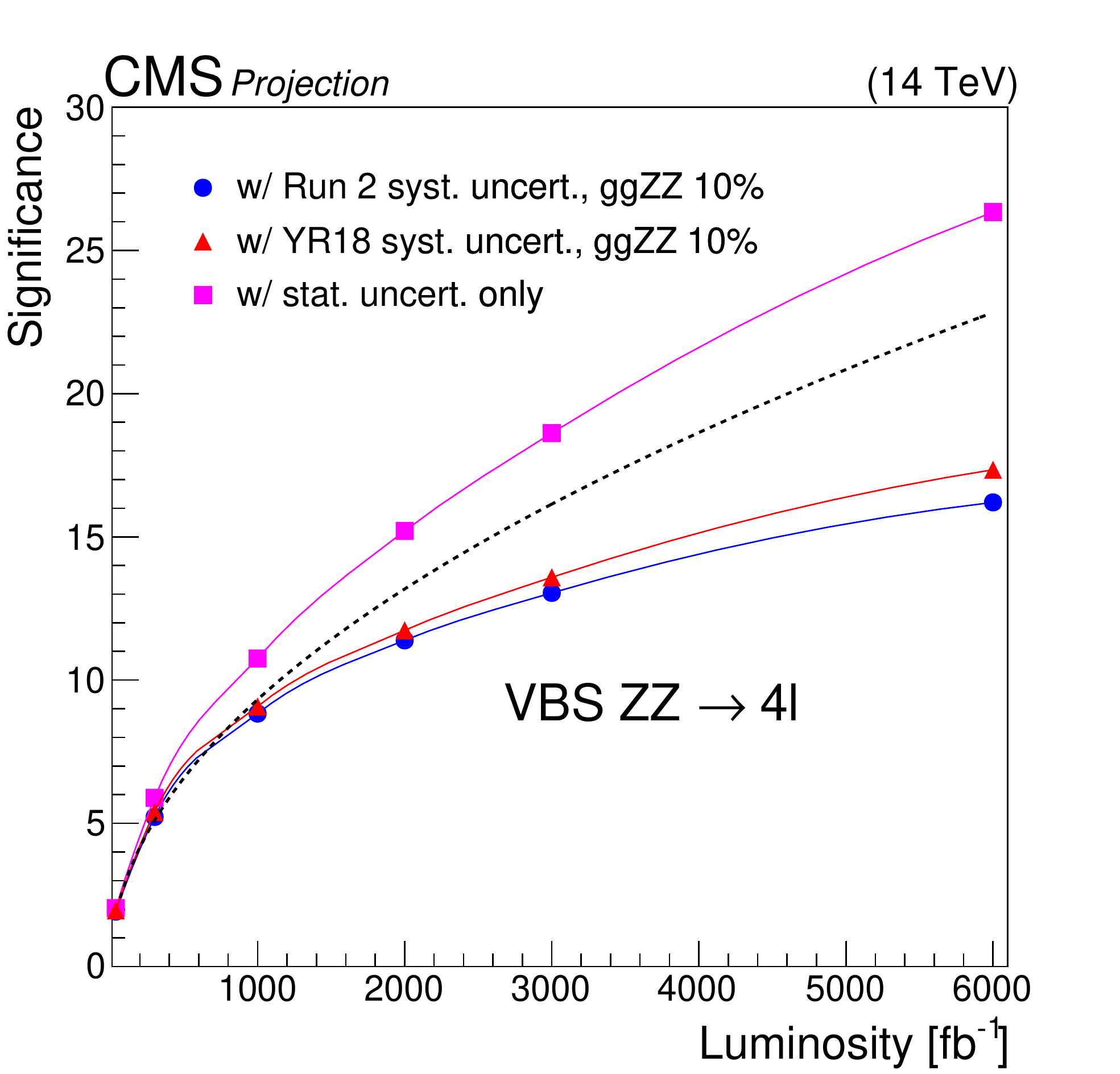}}
\caption{
Expected significance of EW-$ZZjj$ processes as a function of different \mjj~cuts for 3000~\ifb, for different sizes of theoretical uncertainties in the QCD-$ZZjj$ background modeling (left).  Projected significance in the multivariate analysis as a function of the integrated luminosity for the two considered scenario and a 10\% uncertainty in the loop-induced $ggZZ$ background yield, as well as with only the statistical uncertainties included (right).}
\label{fig:sig_scan_3000}
\end{figure}

The CMS analysis employs a multivariate discriminant based on a boosted decision tree (BDT) to extract the EW-$ZZjj$ signal from the QCD-$ZZjj$ background processes. Seven observables are used in the BDT, including $m_\textrm{jj}$, $|\Delta\eta_\textrm{jj}|$, $m_{{\mathrm{Z}\mathrm{Z}}}$, as well as the Zeppenfeld variables~\cite{Rainwater:1996ud} $\eta^{*}_{{Z_{1,2}}}=\eta_{Z_{1,2}} - (\eta_{\mathrm{jet_{1}}} + \eta_{\mathrm{jet_{2}}})/2$ of the two $Z$ bosons, and the ratio between the $p_{\mathrm T}$ of the tagging jet system and the scalar sum of $p_{\mathrm T}$ of the tagging jets ($R(p_{\mathrm T})^{\textrm{jets}}$). The BDT also exploits the event balance $R(p_{\mathrm T} )^{\textrm{hard}}$, defined as the transverse component of the vector sum of the Z bosons and tagging jets momenta, normalized to the scalar $p_{\mathrm T}$ sum of the same objects \cite{Khachatryan:2014dea}. The modeling of all these observables was checked with Run-2 data in a background-enriched region~\cite{Sirunyan:2278242}. A maximum likelihood fit of the BDT distributions for signal and backgrounds is used to extract the signal strength. The shape and normalization of each distribution are allowed to vary within their respective uncertainties. Figure~\ref{fig:sig_scan_3000} (right) shows the projected significance for a 10\% uncertainty in the loop-induced $ggZZ$ background yield, as a function of the integrated luminosity and for the two scenarios described above, as well as for a scenario with only the statistical uncertainty included. The dashed line shows the projected significance as obtained scaling the 2016 result with statistical uncertainty only by the luminosity ratio. The impact of a multivariate analysis is clear for such small signal. The expected significance is 13.0$\sigma$ (13.6$\sigma$) for the Run-2 (YR18) systematic scenario, with a 10\% uncertainty in the loop-induced $ggZZ$ background yield and an integrated luminosity of 3000~\ifb. 

A fiducial phase space is defined at generator level with the same kinematic selections as listed in Table~\ref{tab:selections}, and is used to study the expected precision of the cross section measurements. Table~\ref{tab:xs} shows the expected cross section measurement in this phase space for 3000~\ifb{}, with the statistical only case, and the cases with different sizes of theoretical uncertainties. The statistical uncertainty is at 10\%{} level and the integrated cross section measurement becomes dominated by experimental and modeling uncertainty in the QCD-$ZZjj$ background. For the possible extension of the HL-LHC run to 4000~\ifb{}, the statistical uncertainty will be further reduced to 8\%{} level.

\begin{table}[htbp]
  \small
  \centering
  \caption{
  Summary of expected cross section measurements for different theoretical uncertainties. The statistical uncertainty is estimated from expected data yield at 14~TeV with 3000~\ifb. Different uncertainties are summed up quadratically.}
\label{tab:xs}
  \begin{tabular}{|c|c|c|c|c|c|c|}
    \hline
     & Cross section [fb] & Stat. only & Plus exp. & Plus $5\%$ theo. & Plus $10\%$ theo. & Plus $30\%$ theo. \\
    \hline\hline
    EW-$ZZjj$ & 0.21 & $\pm0.02$ & $\pm0.04$ & $\pm0.05$ & $\pm 0.08$ & $\pm 0.21$ \\
    \hline
  \end{tabular}
\end{table}

The projected measurement uncertainty from the CMS analysis is 9.8\% (8.8\%) for the Run-2 (YR18) scenario and for a 10\% uncertainty in the loop-induced $ggZZ$ background yield, for an integrated luminosity of 3000~\ifb and a coverage of up to $\vert \eta \vert = 3$ for electrons. Extending the coverage up to $\vert \eta \vert = 4$ for electrons, the expected measurement uncertainty becomes 9.5\% and 8.5\%, respectively. In these estimates it is assumed that a fiducial cross section close to the detector volume is used, such that the measurement is to first order insensitive to theoretical uncertainties in the signal cross section.

In addition, the expected differential cross section measurements of the EW-$ZZjj$ processes at 14~TeV have been studied in the defined phase space,
as a function of \mjj, and $m_{ZZ}$,
as shown in Fig.~\ref{fig:xs_mjj_mzz}.
The expected differential cross section measurements are calculated bin by bin as 
\begin{equation}
  \sigma = \frac{N_{\mathrm{pseudo-data}} - N_{\mathrm{QCD}-ZZjj}}{L*C_{\mathrm{EW}-ZZjj}}, ~
  C_{\mathrm{EW}-ZZjj} = \frac{N_{\mathrm{EW}-ZZjj}^{det.}}{N_{\mathrm{EW}-ZZjj}^{part.}},
\end{equation}
where $N_{\mathrm{pseudo-data}}$ is the expected number of data events with 3000~\ifb{} luminosity,
and $N_{\mathrm{QCD}-ZZjj}$ and $N_{\mathrm{EW}-ZZjj}$ are the number of predicted events from QCD-$ZZjj$ and EW-$ZZjj$ processes, respectively.
The $C_{\mathrm{EW}-ZZjj}$ factor refers to the detector efficiency for EW-$ZZjj$ processes,
calculated as number of selected signal events at detector level ($N_{\mathrm{EW}-ZZjj}^{det.}$),
divided by number of selected events at particle level in the fiducial phase space ($N_{\mathrm{EW}-ZZjj}^{part.}$). Both the statistical only case (statistical uncertainty is estimated from expected data yield at 14~TeV with 3000~\ifb) and the ones with different sizes of theoretical uncertainties on the background modeling are shown in Fig.~\ref{fig:xs_mjj_mzz}.

\begin{figure}[!htbp]
\centering
\includegraphics[width=0.48\textwidth]{\main/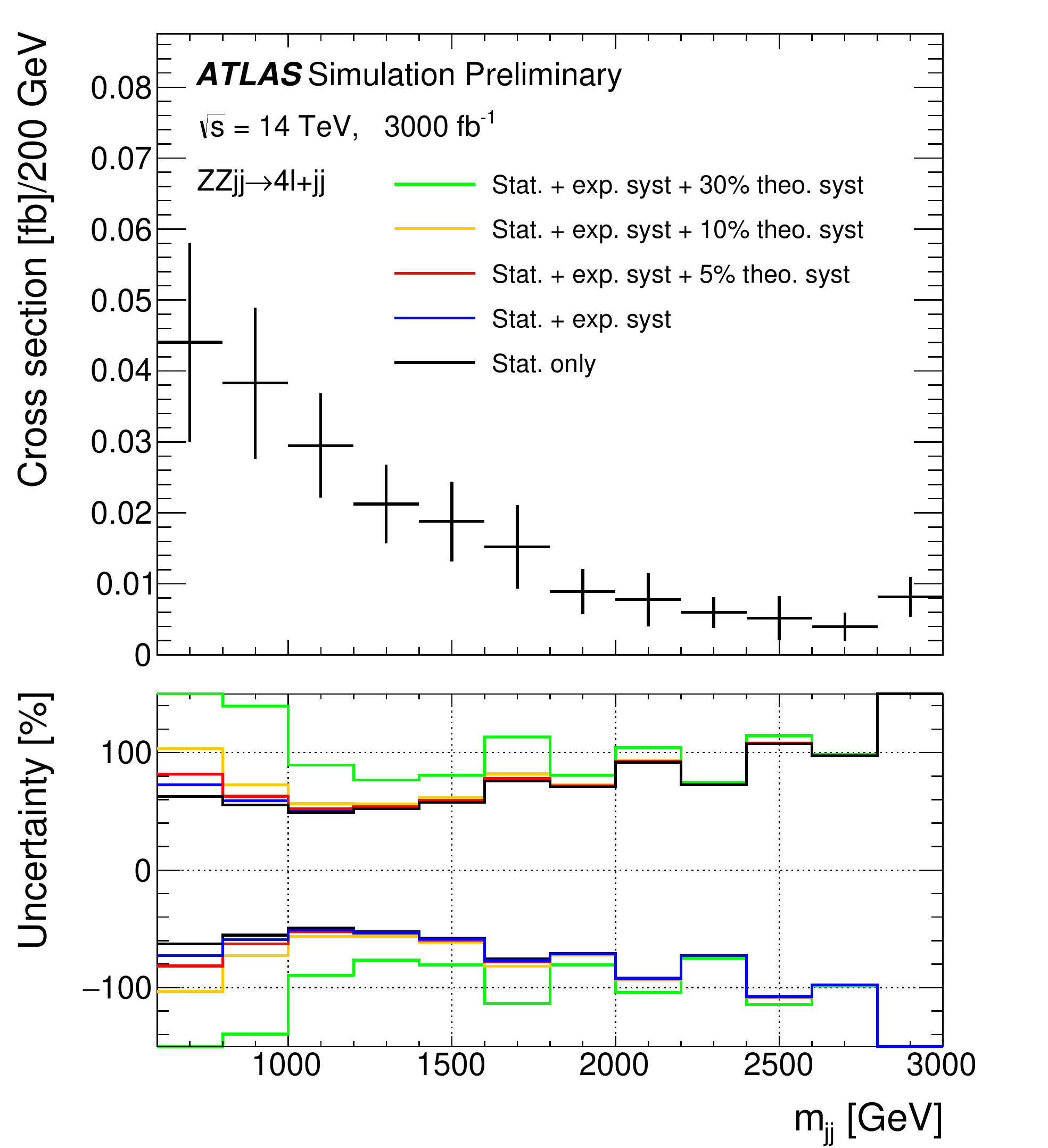}
\includegraphics[width=0.48\textwidth]{\main/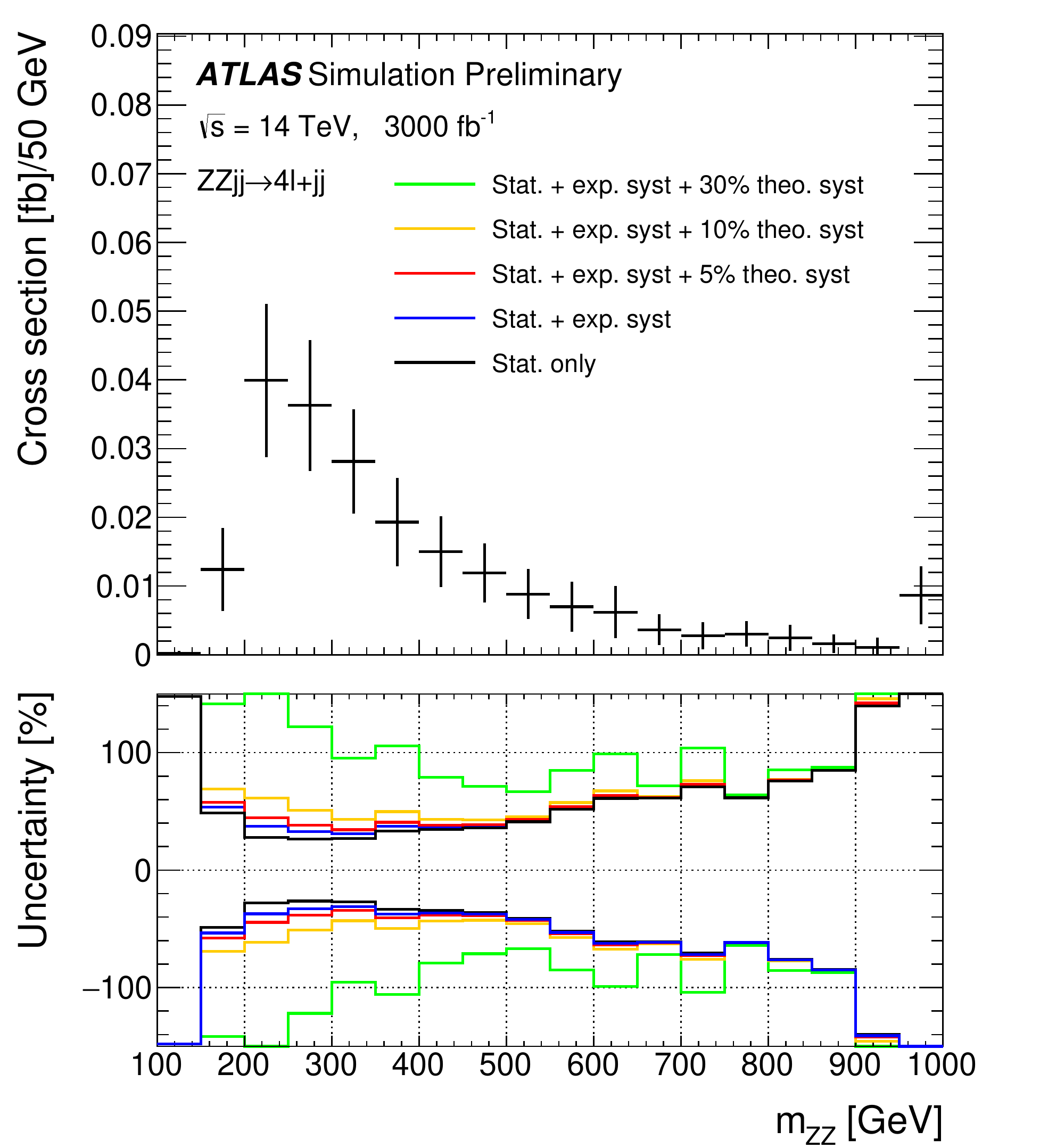}
\caption{
Expected differential cross sections at 14~TeV for the EW-$ZZjj$ processes as a function of \mjj (left) and $m_{ZZ}$ (right). Results are shown with different sizes of systematic uncertainties.
}
\label{fig:xs_mjj_mzz}
\end{figure}

The decay angle $\cos \theta^{*}$ of the lepton direction in the $Z$ decay rest frame with respect to the $Z$ momentum direction in the laboratory frame is the most distinctive feature of longitudinal $Z$ bosons (${Z_{\rm {L}}}$).  The Z boson $\pT$ and $\eta$ distributions also carry information on ${Z_{\rm {L}}Z_{\rm {L}}}$ production, in particular longitudinal $Z$ bosons are produced with a lower $\pT$ and more forward, compared to transverse polarizations ($Z_{\rm{T}}$). The distributions of $\cos \theta^{*}$, $\pT$ and $\eta$ of both $Z$ bosons, together with the distributions of all observables used to separate VBS processes from QCD backgrounds and described above are employed as input to a BDT to separate the VBS $Z_{\rm {L}} Z_{\rm {L}}$ signal from all backgrounds. The BDT is trained separately to discriminate the VBS $Z_{\rm {L}} Z_{\rm {L}}$ signal from the QCD backgrounds (QCD BDT) and to discriminate the VBS $Z_{\rm {L}} Z_{\rm {L}}$ signal from the VBS background (VBS BDT). Cut values are defined on the QCD BDT and on the VBS BDT output values, which maximizes the overall significance estimator $S/\surd{B}$ for the selected events. The corresponding signal efficiency is 14.1\% and the VBS, leading QCD-$ZZjj$ and loop-induced $ggZZ$ background efficiencies are 1.6\%, 0.03\% and 0.05\%, respectively. It is assumed that the VBS $Z_{\rm {L}} Z_{\rm {L}}$ fraction, defined as VBS $Z_{\rm {L}} Z_{\rm {L}}$ / VBS~($Z_{\rm {L}} Z_{\rm {L}} + Z_{\rm {L}}Z_{\rm {T}} + Z_{\rm {T}}Z_{\rm {T}}$) will be measured, rather than the absolute VBS $Z_{\rm {L}}\ {Z}_{\rm{L}}$ cross section. In such ratio measurement, the systematic uncertainties from luminosity, and selection efficiency, as well as theoretical uncertainties on the VBS and VBS background cross section cancel out, such that only the uncertainties in the QCD backgrounds yields are considered.

Figure~\ref{fig:res_significance_vs_lumi_LL_fullBDT} shows the expected significance for the VBS $Z_{\rm {L}} Z_{\rm {L}}$ fraction as a function of the integrated luminosity and for the two scenarios described above and a 10\% uncertainty in the loop-induced $ggZZ$ background yield, as well as for a scenario with only the statistical uncertainty included. A significance of 1.4$\sigma$ is reached for 3000~\ifb. As expected from the ratio measurement, the effect of systematic uncertainties is very small. Results are also shown for an integrated luminosity of 6000~\ifb, which would approximately correspond to combining ATLAS and CMS after 3000~\ifb.
\begin{figure}[!htb]
\begin{center}
\resizebox{7cm}{!}{\includegraphics{\main/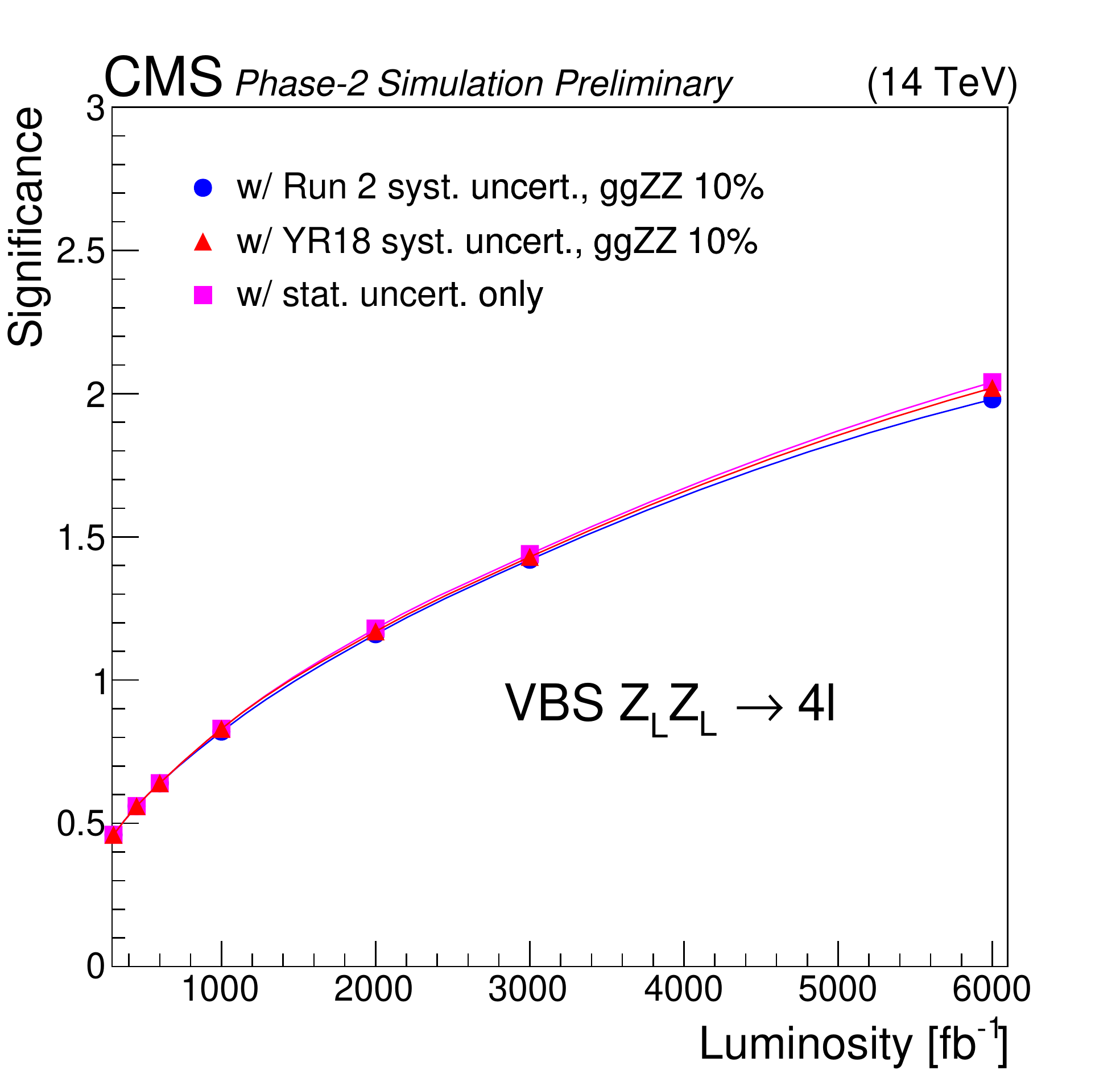}}
\caption{Expected significance for the VBS ${Z_{\rm {L}} Z_{\rm {L}}}$ fraction as a function of the integrated luminosity and for systematic uncertainties according to the Run-2 and YR18 scenario, as well as with only the statistical uncertainties included.}
\label{fig:res_significance_vs_lumi_LL_fullBDT}
\end{center}
\end{figure}
Table~\ref{tab:acceptance} presents the expected significance and relative uncertainty in the VBS ${Z_{\rm {L}} Z_{\rm {L}}}$ fraction for various $\eta$ coverage configurations. The foreseen coverage extension of up to $\vert \eta \vert = 3~(2.8)$ for electrons (muons) leads to a $\sim 13\%$ improvement for the significance  and precision on the VBS ${Z_{\rm {L}} Z_{\rm {L}}}$ fraction. An extension of up to $\vert \eta \vert = 4$ for electrons would allow to further improve by $\sim 4\%$ both significance and cross section measurement uncertainty.

\begin{table}[!h]
\begin{center}
\caption{Significance and measurement uncertainty for the VBS ${Z_{\rm {L}} Z_{\rm {L}}}$ fraction for different acceptance configurations at HL-LHC. In the quoted $\eta$ coverages, the first number corresponds to electrons while the number in parentheses corresponds to muons.}
\label{tab:acceptance}
   \begin{tabular}{|c|c|c|}
   \hline
	$\eta$ coverage        		&  significance  &  VBS ${Z_{\rm {L}} Z_{\rm {L}}}$ fraction uncertainty (\%)\\
   \hline\hline
	$\vert \eta \vert < 2.5~(2.4)$   &  1.22$\sigma$           &  88     	                  \\
	$\vert \eta \vert < 3.0~(2.8)$   &  1.38$\sigma$           &  78     	                  \\
	$\vert \eta \vert < 4.0~(2.8)$   &  1.43$\sigma$           &  75     	                  \\
  \hline
\end{tabular}
\end{center}
\end{table}

Finally, a simple scaling of the signal and background cross sections is performed to assess the sensitivity to the VBS ${Z_{\rm {L}} Z_{\rm {L}}}$ fraction at HE-LHC. An integrated luminosity of 15 ab$^{-1}$ is considered, together with a c.o.m energy of 27 TeV. The cross section ratios $\sigma_{27\TeV}$ / $\sigma_{14\TeV}$ are evaluated at LO with \MADGRAPH (v5.4.2) ~\cite{Alwall:2011uj} for the EW signal and the leading QCD-$ZZjj$ background, and with \MCFM (v.7.0.1) ~\cite{MCFM} for the $ggZZ$ loop-induced background. Table~\ref{tab:helhc} shows the expected significance and relative uncertainty for the VBS ${Z_{\rm {L}} Z_{\rm {L}}}$ fraction at HE-LHC, compared to HL-LHC. The HE-LHC machine would allow to bring the sensitivity (uncertainty) for the measurement of the VBS ${Z_{\rm {L}} Z_{\rm {L}}}$ fraction to the level of $\sim 5\sigma$ ($\sim 20\%$).

\begin{table}[!h]
\begin{center}
\caption{Expected significance and measurement uncertainty for the VBS ${Z_{\rm {L}}Z_{\rm {L}}}$ fraction at HL-LHC and HE-LHC with and without systematic uncertainties included.}
   \label{tab:helhc}
   \begin{tabular}{|l|c|c|c|c|}
   \hline
      & \multicolumn{2}{c|}{significance} & \multicolumn{2}{c|}{precision (\%)} \\ 
      & w/ syst. uncert. &  w/o syst. uncert. & w/ syst. uncert. & w/o syst. uncert.) \\
   \hline\hline
	HL-LHC   &  1.4$\sigma$    &  1.4$\sigma$   &  75\%  &   75\%                \\
	HE-LHC   &  5.2$\sigma$    &  5.7$\sigma$   &  20\%  &   19\%                \\
  \hline
   
   \end{tabular}
   \end{center}
\end{table}

\subsubsection{The production of $WW$ / $WZ$ via vector boson scattering with semi-leptonic final states}





The existing Run-2 VBS measurements and the above analyses have focused on channels involving the fully leptonic boson decays, or decay modes involving photons. The semileptonic channels can however offer some interesting advantages: the $V\to q\bar q$ branching fractions are much larger than the leptonic ones and the use of jet substructure techniques with large-radius jet reconstruction allows to  reconstruct and identify the $V$-boson produced in the high-$p_{\mathrm{T}}$ region, which is the most sensitive to new physics effects. This section presents the sensitivity of the ATLAS experiment to VBS in the $V(qq)W(\ell\nu)$ final state, assuming an integrated luminosity of 300 or 3000 fb$^{-1}$  of $pp$ collisions at $\sqrt{s}$= 14 TeV.

This analyses uses generator-level samples of the main signal and background processes, combined with the parameterisations of the detector performance (muon and jet reconstruction and selection efficiencies and momentum resolutions) expected at the HL-LHC from fully simulated samples. The parametrized detector resolutions are used to smear the generator-level particle transverse momenta,
while the parametrized efficiencies are used to reweigh the selected events.
All generated samples were produced at $\sqrt{s}$= 14 TeV and normalized to luminosities of 300 or 3000 fb$^{-1}$ when the results are presented.

The electroweak (EW) $VVjj$\ production is modeled using \mgamcatnlo~v2.3.3~\cite{Alwall:2014hca}, plus \textsc{PYTHIA}8~\cite{Sjostrand:2007gs} for fragmentation. 
The main background sources are $W$ bosons produced in association with jets ($W$+jets),  with significant contributions from top-quark production (both $t\bar{t}$ pair and single-top), non-resonant vector-boson pair production ($ZZ$, $WZ$ and $WW$) and $Z$ bosons produced in association with jets ($Z$+jets). Background originating from multi-jet processes are expected to be negligible due to the event selection requirements. Details about the samples generation can be found in Ref.~\cite{ATL-PHYS-PUB-2018-022}.

To increase the purity of considered events, several requirements are placed on the constituents of an event. Events are required to have exactly one lepton. Generator-level electrons or muons are required to be isolated and pass the tight identification criteria~\cite{ATL-PHYS-PUB-2016-026} and to have $p_{\mathrm{T}} >27$~GeV.
Events are required to contain a hadronically-decaying $W/Z$ candidate, reconstructed either from two small-$R$ 
jets, defined as the resolved channel, or from one large-$R$ jet, designated the boosted channel. 
Small-$R$ jets are defined using the anti-$k_{\rm T}$ algorithm~\cite{antikt_algorithm} with a radius parameter of R = 0.4.  The identification of jets originating from $b$-quarks is done by finding jets with generator-level $b$-hadron within a cone of $\Delta R <$ 0.4  around the jet direction. Similarly, the anti-$k_{\rm T}$ algorithm with a radius parameter of R = 1.0 is used to reconstruct large-$R$ jets. The large-$R$ jets are trimmed using the standard ATLAS trimming parameters~\cite{fatjet_trimming}. It is assumed that the performance of a future $W/Z$-boson tagger at the HL-LHC conditions will have similar, if not better, performance as existing boson taggers. To simulate the effect of Run-2 $W/Z$-boson tagging performance~\cite{Aad:2015rpa,BosonTagPreRec}  events which contain a large-$R$ jet are scaled by the expected boson tagging efficiency for the $V \to qq$ with kinematics corresponding to the large-$R$ jet, calculated from fully-simulated 13~TeV Monte-Carlo (MC) samples. The missing transverse energy $E_{\mathrm{T}}^{miss}$ is required to be greater than 60 GeV, which suppresses the expected multijet background to a negligible level.
By constraining the $E_{\mathrm{T}}^{miss}$ + lepton system to be consistent with the $W$ mass, the $z$ component of the neutrino ($\nu$) momentum can be reconstructed by solving a quadratic equation. 

Experimentally, VBS is characterized by the presence of a pair of vector bosons and two forward jets with a large
separation in pseudorapidity and a large dijet invariant mass. Therefore the VBS search is required to have 2 additional forward VBS-topology tagging jets in the event in addition to jets associated with the boson decay, similar to the resonant VBF search. The VBS tagging jets are required to be non-$b$-tagged, be in the opposite hemispheres, $\eta(j_1^{\mathrm{tag}}) \cdot \eta(j_2^{\mathrm{tag}}) < 0$, and to have the highest dijet invariant mass among all pairs of jets remaining in the event after the $V\to jj$ jet selection.   After the tagging jet pair are selected, it is required that both tagging jets should have $p_{\mathrm{T}}>$30~GeV, and that the invariant mass of the two tagging jets system is greater than 400~GeV . In the merged selection, events are required to have at least one large-$R$ jet with $p_{\mathrm{T}}(J) > 200$~GeV and $|\eta(J)| < 2$. From those candidate large-$R$ jets, the one with the smallest ${|m(J) - m(W/Z)|}$ is selected as the signal large-$R$ jet. Mass window cuts and boson tagging efficiencies are applied as described above. To suppress backgrounds with top quarks, an event is rejected if any of the reconstructed jets outside the large R jet, is identified as containing a $b$-quark.
If events fail the merged VBS selection, the resolved selection is then applied. Signal jets are chosen as the pair with $m(jj)$ closest to the $W/Z$ mass. The signal jet pairs are then required to have $|m(jj)-m(W/Z)|<15$ GeV.
To suppress backgrounds with top quarks, an event is rejected if any of the reconstructed jets is identified as containing a $b$-quark.

To optimize the signal sensitivity, Boosted Decision Trees (BDT) for the resolved and merged searches were trained on the background and signal MC samples in the respective regions.
Four variables are included in the merged BDT: the invariant mass of the $l\nu J$ system, the lepton $\eta$, the second tag jet $p_{\rm T}$ and the boson centrality $\zeta_V$. 
The boson centrality is defined as $\zeta_V=min(\Delta\eta_+,\Delta\eta_{-})$  where $\Delta\eta_{+}=max(\eta(j_1^{\mathrm{tag}}),\eta(j_2^{\mathrm{tag}}))-max(\eta(\ell\nu),\eta(J))$ and $\Delta\eta_{-}=min(\eta(\ell\nu),\eta(J))-min(\eta(j_1^{\mathrm{tag}}),\eta(j_2^{\mathrm{tag}}))$.
In the resolved BDT, eight variables were used: the invariant mass of the $WVjj$ system , the lepton $\eta$, the  $p_{\rm T}$ of both VBS-tagging jets and sub-leading signal jet, the boson centrality defined similarly to above, the $\Delta\eta$ between signal jets, and the $\Delta R$ between the lepton and neutrino candidate.
These variables were chosen as they are the minimal subset of variables with the greatest separation between the signal and background, that provide significant improvement when added during the training. 
The BDT were trained using a gradient descent BDT algorithm, maximizing the Gini index, in the TMVA package~\cite{Hocker:2007ht}.
The BDT are chosen as the discriminants and their distributions are used in the final fit for the VBS search shown in Figure~\ref{fig:VBSHLHC}.

If an event fails either a mass-window cut or a $b$-veto but passes all other events then the event is categorized as a $W$ or top control region.  These regions are used to constrain the normalization and shape systematics of the background.

The results are extracted by performing a simultaneous binned maximum-likelihood fit to the BDT distributions in the signal regions and the $W$+jets and $t \bar{t}$ control regions. A test statistic based on the profile likelihood ratio~\cite{Cowan:2010js} is used to test hypothesized values of the signal cross section. The likelihood is defined as the product of the Poisson likelihoods for all signal and control regions for a given production mechanism category and channel.  Systematic uncertainties are taken into account as constrained nuisance parameters with Gaussian or log-normal distributions. The main background modelling systematics, namely the $W$+jets and $t\bar{t}$ shape uncertainties, are constrained by the corresponding control regions and are treated as uncorrelated among the resolved and merged signal regions.  

The expected significance for the SM VBS process is 5.7$\sigma$ at 300 fb$^{-1}$ as shown in Fig.~\ref{fig:VBSHLHC}. The expected cross section uncertainties are 18\% at  300 fb$^{-1}$ and 6.5\% at 3000 fb$^{-1}$. The effects of unfolding were not considered for the cross section estimates. If control regions are not used to constrain the systematics the expected significance is reduced to 3.6$\sigma$ at 300 fb$^{-1}$. Likewise the cross section uncertainty are increased to 28\% at 300 fb$^{-1}$ and 10\% at 3000 fb$^{-1}$ when control regions are ignored.

\begin{figure}
\centering
\includegraphics[width=0.45\textwidth]{\main/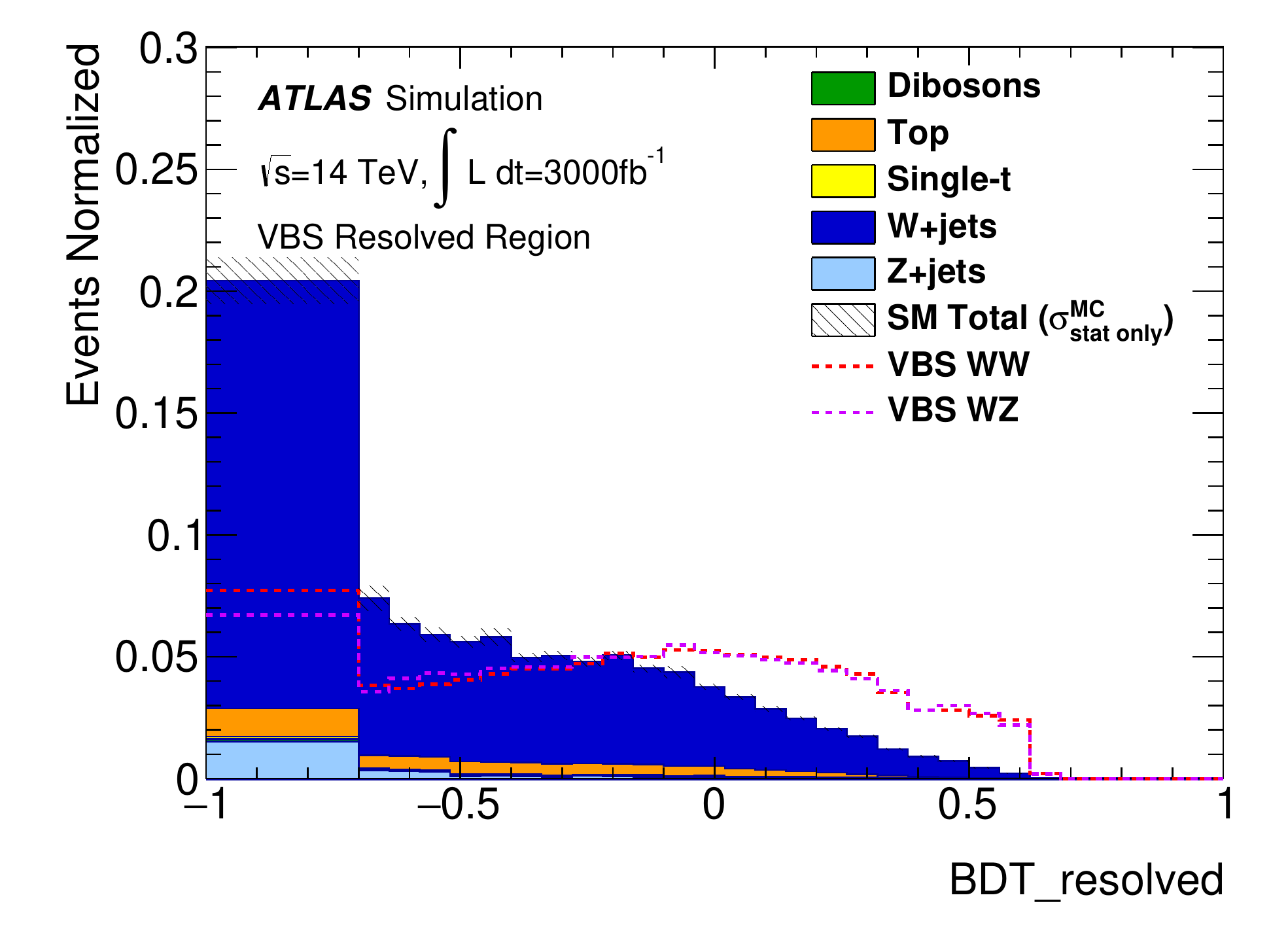}
\includegraphics[width=0.45\textwidth]{\main/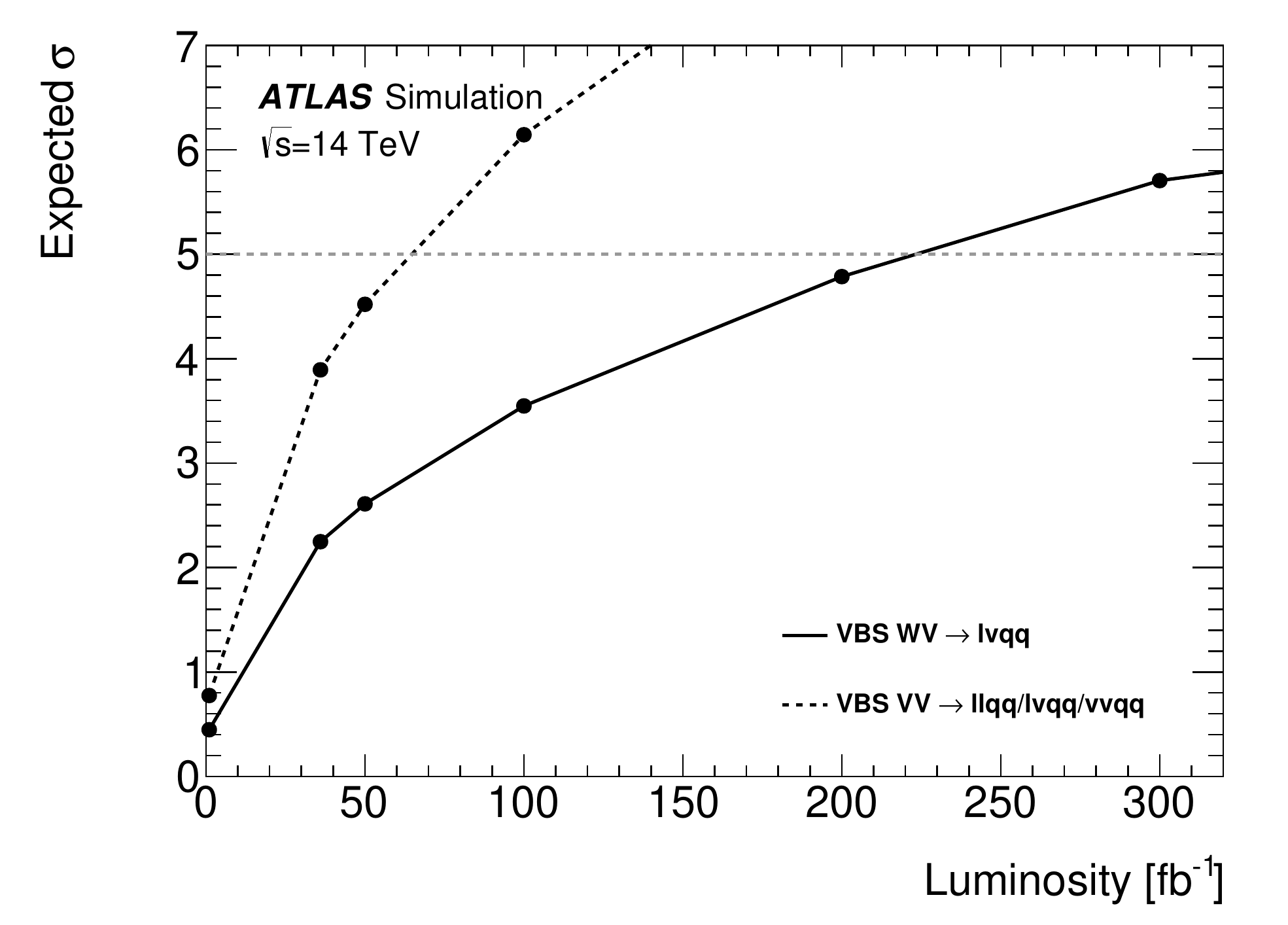}
\caption{Final signal and background distributions for the VBS search in the respective resolved  signal region for the normalized BDT response. Background distributions are separated into production type. VBS signals in $WW$ and $WZ$ mode are overlaid as dashed curves where appropriate. Both background and signal BDT distributions are normalized to unity (left). Expected signal significance as a function of integrated luminosity up to 300 fb$^{-1}$. The solid black curve is the significance from the $\ell \nu qq $channel, while the black dashed curve shows the expected significance from all semi-leptonic channels assuming equal sensitivity (right). 
 }
\label{fig:VBSHLHC}
\end{figure}

\subsubsubsection{Electroweak $WW$ / $WZ$ production analysis at HE-LHC}

The prospect analysis at HE-LHC \cite{Cavaliere:2018zcf}  mimics the analysis at HL-LHC but the {\mbox{\textsc{Delphes}}\xspace} simulation is used ~\cite{Delphes}. 
VBS signal samples are produced in the same manner as the HL-LHC analysis. The major backgrounds $W$+jets and $t\bar{t}$ production are simulated with \textsc{Madgraph} and \textsc{aMC@NLO} respectively, interfaced with {\mbox{\textsc{PYTHIA}}\xspace}. $Z$+jets, single top and diboson contribution are not simulated and are expected to contribute at most 10\% to the total background.

The unprecedented energy of $pp$ collisions at the HE-LHC will significantly improve sensitivity to new multi-TeV particles over LHC and HL-LHC. However, the experimental environment is expected to be challenging at the HE-LHC, primarily due to a significant increase of the number of $pp$ collisions in a same and nearby bunch crossings (pile-up). The HE-LHC is planned to be operated at a centre-of-mass energy of 27~TeV with 800 pile-up collisions at the peak luminosity. Such extreme pile-up conditions are expected to be particularly challenging for identifying hadronically decaying $W$/$Z$ boson as the extra contribution of particles produced from pile-up collisions into jets could degrade the performance of $W$/$Z$ boson tagger significantly. It is therefore important to assess the performance of pile-up mitigation technique at the HE-LHC in order to have a reliable estimate of the search sensitivity.

The study presented here focuses on the performance of pile-up mitigation techniques and $W$/$Z$ boson tagging. The VBS signal events are produced with the overlay of minimum-bias $pp$ interactions generated using {\mbox{\textsc{PYTHIA}}\xspace}~8. The minimum-bias interactions are overlaid onto hard scattering event using Poisson probability distribution with the mean number of interactions ($\mu_{\text{pileup}}$) varied from 0 to 100, 200, 400 and 800. Furthermore, the minimum-bias interactions are distributed randomly in $z$ and timing using Gaussian profiles of $\sigma_z=5.3$~cm and $\sigma_t=160$~ps, respectively ($z$=0 at the detector centre and $t$=0 for hard scattering event). The overlaid VBS signal events are processed through {\mbox{\textsc{Delphes}}\xspace} with two pile-up mitigation techniques: the Pile-up Per Particle Identification (PUPPI) algorithm~\cite{Bertolini:2014bba} used in CMS and the trimming procedure used in ATLAS. The trimming parameters of the $p_{\rm T}$ fraction cut and the sub-jet reclustering radius are chosen to be the same as those used in ATLAS. For the PUPPI algorithm the standard {\mbox{\textsc{Delphes}}\xspace} implementation is used. 

Figure~\ref{fig:PUPPI_trimmed_jet} shows the leading large-$R$ jet mass ($m_J$) for the PUPPI-only jets and the PUPPI+trimmed jets, both required to have $p_{\rm T}>200$~GeV. The $m_J$ distribution get shifted towards lower values with the trimming applied, enhancing the peak around $m_W$ . The residual pile-up effect is still visible as a shift towards larger values with increasing $\mu_{\text{pileup}}$, but the overall signal yield after the mass-window and $D_2$ requirements (e.g, $D_2<1.5$) is largely stable. This indicates that an impact to the $W$/$Z$-boson tagging performance from expected pile-up collisions at the HE-LHC can be mitigated to the level where the tagging performance is similar to what is expected at Run-2 or the HL-LHC. Therefore, the study presented in the rest of this note is based on the $W$/$Z$-boson tagging performance at Run-2.

\begin{figure}
\includegraphics[width=0.45\textwidth]{\main/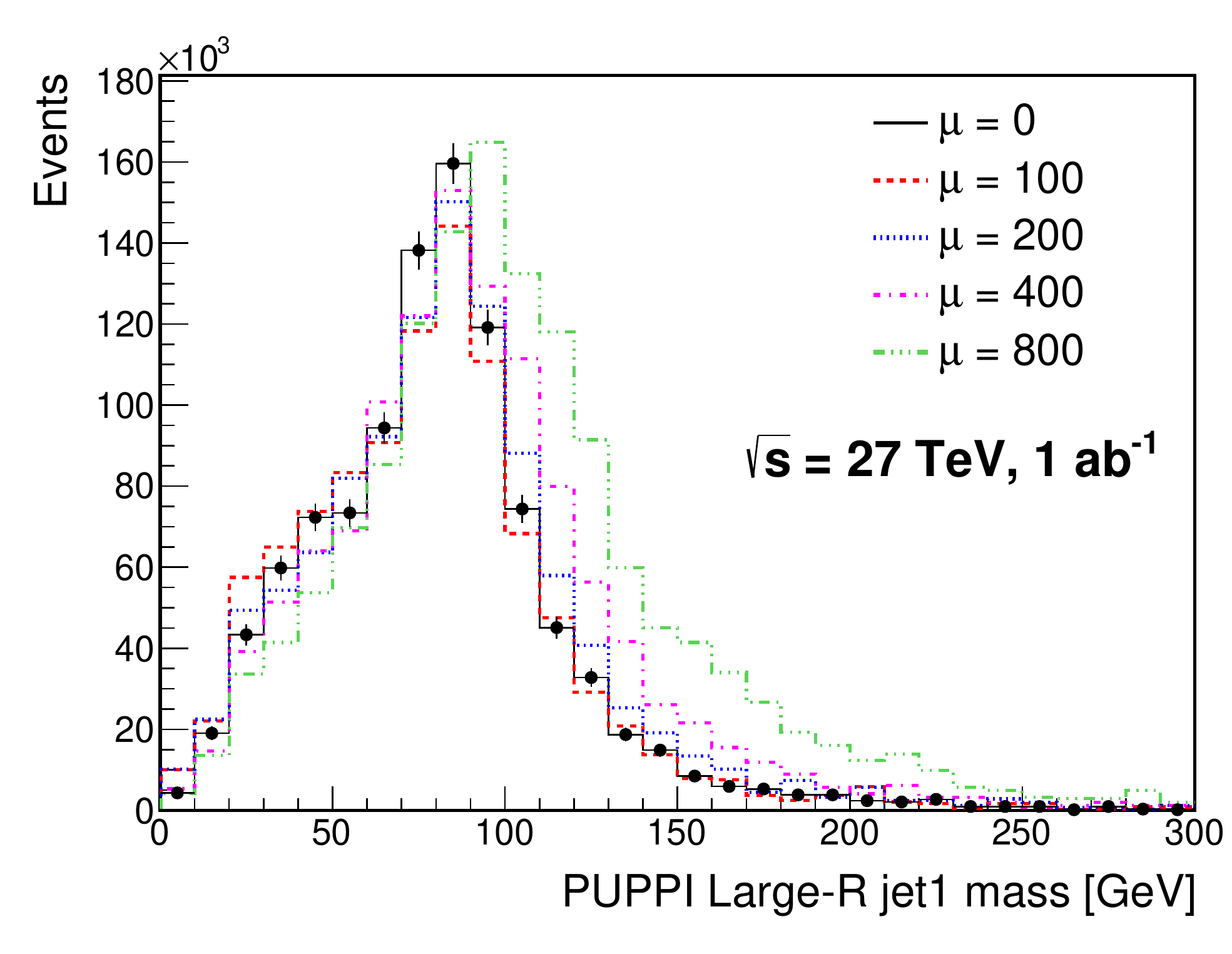}
\includegraphics[width=0.45\textwidth]{\main/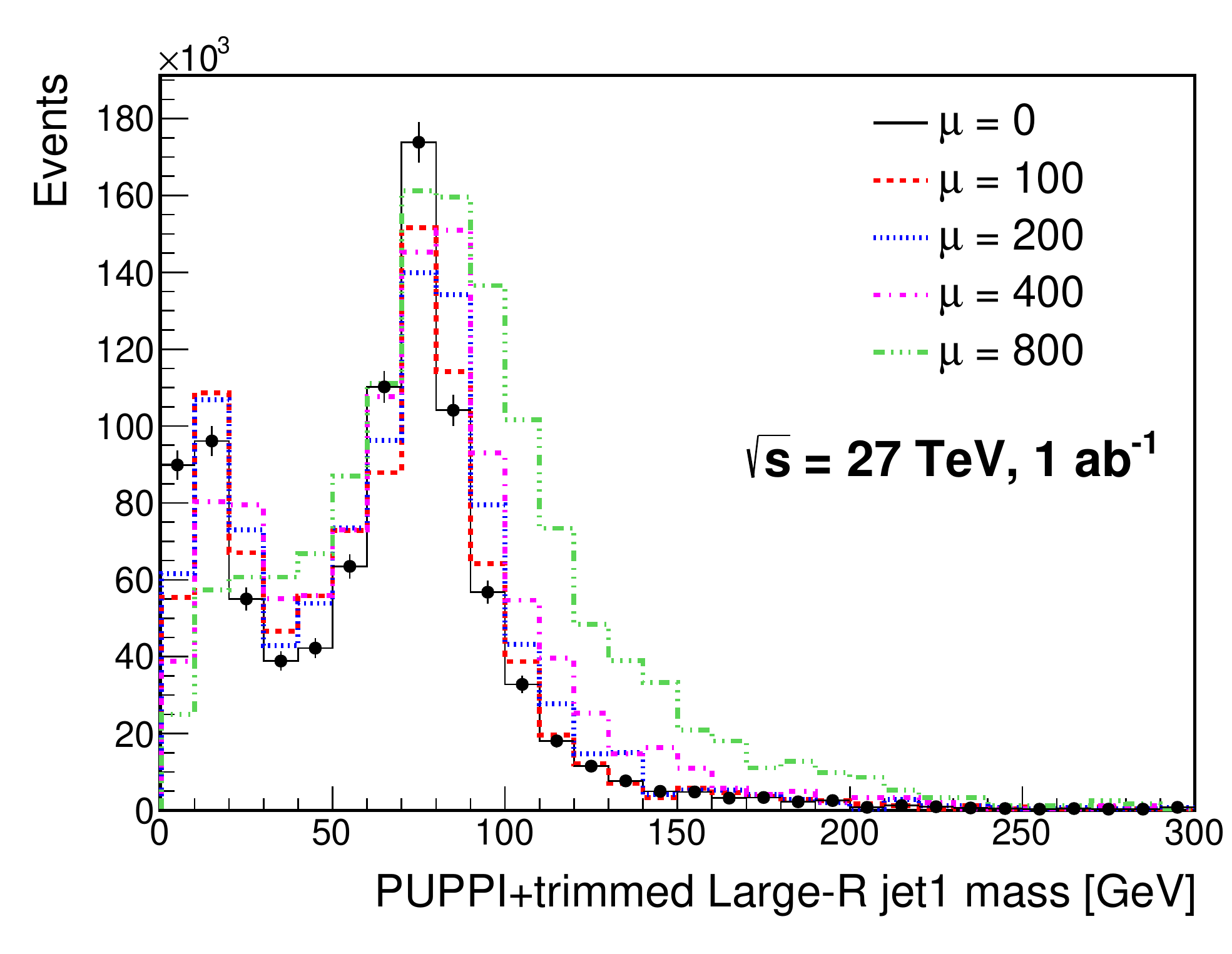}
\caption{Leading large-$R$ jet mass (left)  after applying the PUPPI algorithm at an integrated luminosity of 1~ab$^{-1}$ at $\sqrt{s}=27$~TeV with  five different pile-up overlay conditions of $\mu_{\text{pileup}}=0$, 100, 200, 400 and 800. The right plots shows the same
distribution but after additionally requiring that the jets are trimmed with the conditions described in the text.}
\label{fig:PUPPI_trimmed_jet}
\end{figure}

The sensitivity  to the VBS signal at 27 TeV is extracted in the same manner as the HL-LHC analysis. The event selection is similar and
a BDT is built using the same variables both in the resolved and boosted channel. For more details about the BDT and the setup used please refer to citation.  
Figure~\ref{fig:VBS-HE-LHC} shows the expected cross section uncertainty as function of integrated luminosity at 27 TeV compared to the one obtained at 14 TeV. The results are very consistent and show that given the same luminosity the same uncertainty can be reached at 27 TeV.
Prospects are also presented for the extraction of the longitudinal component of the $WW$ scattering. 
For the extraction of the longitudinal component in VBS processes, the electroweak $WWjj$ samples are generated with the {\mbox{\textsc{DECAY}}\xspace} program to identify the polarization state of the produced $V$ bosons.  The generated events are then classified according to the polarization state: both $V$ bosons are longitudinally (LL)  or transversely (TT) polarized, or in the mixed state (LT). Each event is showered using {\mbox{\textsc{PYTHIA}}\xspace} and 
then processed through the {\mbox{\textsc{Delphes}}\xspace} simulation.

In this case a BDT is built training the signal samples ($WW$ $LL$) against the sum of the backgrounds which include the TT and LT component of the  electroweak $WWjj$ samples. The observed significance expected with this simple setup is shown in the right figure of Fig.~\ref{fig:VBS-HE-LHC}. One line shows the results obtained by fitting a single variable, the total invariant mass of the system and the other one shows the expected significance using the BDT. The third line shows the expected significance assuming the combination of all three semi-leptonic channels with the same sensitivity. 
It is expected to reach 5$\sigma$ sensitivities with 3000 fb$^{-1}$ combining all the semileptonic channels.

\begin{figure}
\includegraphics[width=0.45\textwidth]{\main/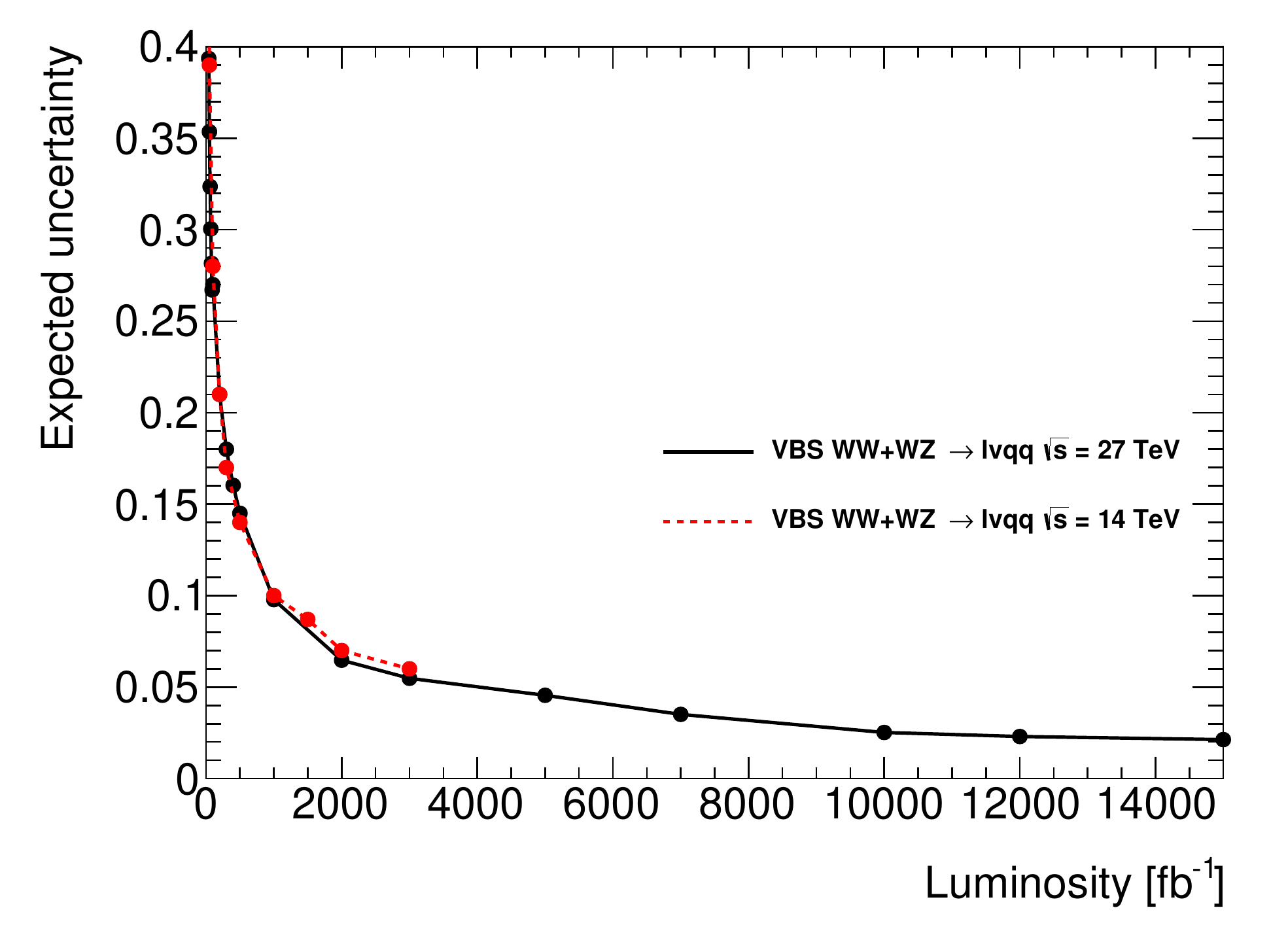}
\includegraphics[width=0.45\textwidth]{\main/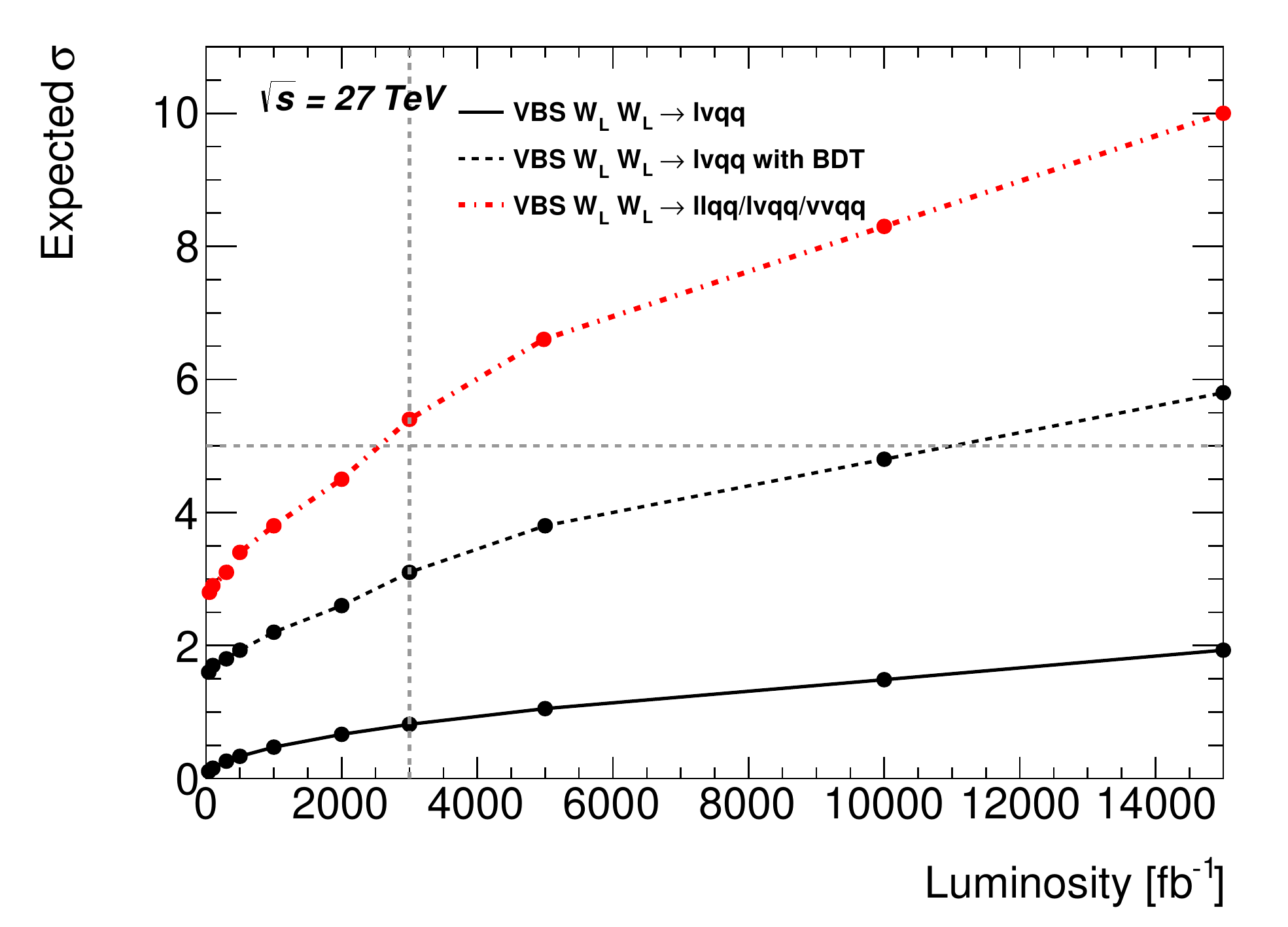}
\caption{The expected cross section uncertainty as function of integrated luminosity at 27 TeV  compared to the one obtained at 14 TeV (left). Right: Observed significance as a function of the luminosity and expected uncertainty for the EW $W_LW_L$ signal assuming a 10\% fraction predicted by {\mbox{\textsc{Madgraph}}\xspace} (right). One line shows the results obtained by fitting a single variable, the total invariant mass of the system and the other one shows the expected significance using the BDT. The third line shows the expected significance assuming the combination of all three semi-leptonic channels with the same sensitivity.}
\label{fig:VBS-HE-LHC}
\end{figure}

\subsection{Tri-boson production}




\makeatletter
\makeatother


\label{sec:triboson_intro}


The production of multiple heavy gauge bosons $V$ (= $W^{\pm}, Z$) opens up a multitude of potential decay channels categorised according to the number of charged leptons in the final state. The sensitivity prospect studies have been performed related to the production of $W^{\pm} W^{\pm} W^{\mp}$,  $W^{\pm} W^{\mp} Z$ or $W^{\pm} Z Z$ followed by the fully leptonic or semi-hadronic\footnote{In case of semi-hadronic channels we assume that one of the vector bosons decays hadronically while the other two decay leptonically.} decays: $W^{\pm} W^{\pm} W^{\mp} \to \ell^{\pm}\nu \ell^{\pm}\nu\ell^{\mp}\nu$, $W^{\pm} W^{\pm} W^{\mp} \to \ell^{\pm}\nu \ell^{\pm}\nu jj$,  $W^{\pm} W^{\mp} Z \to \ell^{\pm}\nu \ell^{\pm}\nu \ell^+\ell^-$,  $W^{\pm} W^{\mp} Z \to \ell^{\pm}\nu jj \ell^+\ell^-$, $W^{\pm} Z Z \to \ell^{\pm}\nu \ell^{+}\ell^{-} \ell^{+}\ell^{-}$, $W^{\pm} Z Z \to \ell^{\pm}\nu \ell^{+}\ell^{-} \nu\nu$,  $W^{\pm} Z Z \to jj \ell^{+}\ell^{-}  \ell^{+}\ell^{-} $ and $W^{\pm} Z Z \to \ell^{\pm}\nu  \ell^{+}\ell^{-} jj$, with $\ell = e$ or $\mu$. 
Prospect studies have been performed, using a cut-based analysis, corresponding to an integrated luminosity of 3000 fb$^{-1}$~and 4000 fb$^{-1}$~of proton--proton collisions at a centre-of-mass energy of $\sqrt{s}$ = 14 TeV, expected to be collected by  the ATLAS detector at the HL-LHC~\cite{Apollinari:2284929}.  In this section we summarize only results that are expected to provide the best sensitivity according to the full prospect studies documented in \cite{ATL-PHYS-PUB-2018-030}.

Monte Carlo (MC) simulated event samples are used to predict the background from SM processes and to
model the multi-boson signal production. The effects of an upgraded ATLAS detector are taken into account by applying
energy smearing, efficiencies and fake rates to generator level quantities, following parameterisations
based on detector performance studies with full simulation and HL-LHC conditions. 
The most relevant MC samples have equivalent luminosities (at 14 TeV) of at least 3000 fb$^{-1}$.
Several MC generators are used to model the production of signal and dominant SM background processes relevant for the analysis. 

For the generation of triboson signal events, matrix elements for all combinations of $pp \to VV$  ($V = W^\pm, Z$) have been generated using  \textsc{Sherpa} v2.2.2 \cite{Gleisberg:2008ta} with up to two additional partons in the final state, including full next-to-leading-order calculations (NLO) \cite{Lazopoulos:2008de,Garzelli:2012bn,Campbell:2012dh} accuracy for the inclusive process.  All diagrams with three electroweak couplings are taken into account, including diagrams involving Higgs propagators. However, since these samples use factorised decays with on-shell vector bosons, the resonant contribution from those diagrams can not be reached from the 125 GeV Higgs.  In order to account for the contribution coming from these diagrams the corresponding production of $VH$  ($V = W, Z$) bosons is added to the signal.  
Electroweak NLO corrections to the signal production cross sections are not considered in this analysis.
The diboson processes are generated with  \textsc{Sherpa} event generator following the approach described in \cite{ATL-PHYS-PUB-2017-005}. 
For the simulation of the top quark pair and the production of $VH$  ($V = W, Z$) bosons \textsc{Powheg}\cite{Nason:2004rx,Frixione:2007vw,Alioli:2010xd}+\textsc{Pythia}\cite{Sjostrand:2006za} was used as described in \cite{ATL-PHYS-PUB-2018-009}, 
while for the $t\bar{t}$ + $V$ ($V = W, Z, H$) \textsc{MadGraph5}\_a\textsc{MC@NLO} \cite{Alwall:2014hca} interfaced to \textsc{Pythia} was used as in \cite{ATL-PHYS-PUB-2016-005}. 

The expected multi-boson yields are normalised to the \textsc{Sherpa} predictions, while the $t\bar{t}$ + $V$ ($V = W, Z, H$) yields are normalized to NLO. 
The top quark pair-production contribution is normalised to approximate NNLO+NNLL accuracy~\cite{Czakon:2013goa,Czakon:2011xx}.

\subsubsection*{Experimental signatures}
\label{sec:signatures}

The experimental signature of the triboson processes considered in these studies consists of at least three charged leptons,  moderate $E_{\mathrm T}^{\mathrm{miss}}$ originating from the leptonic decay of $W$ bosons, and jets in case one of the vector bosons decays hadronically. 
The event selection starts from the one used in the published analysis in Ref.~ \cite{Aaboud:2016ftt}, but considers tighter selection criteria in terms of transverse momentum of the selected objects and missing transverse momentum of the event, in order to suppress higher pile-up contributions expected at the HL-LHC.    
The selection requirements used to define the signal regions are obtained from an optimization to maximize the sensitivity to $W^{\pm} W^{\pm} W^{\mp}$, $W^{\pm} W^{\mp} Z$ and $W^{\pm} Z Z$ processes and to reduce the contributions from SM background processes.  In the case of $W^{\pm} W^{\pm} W^{\mp} \to \ell^{\pm}\nu \ell^{\pm}\nu\ell^{\mp}\nu$ channel, three separate signal regions are defined based on the number of same-flavour opposite-sign (SFOS) lepton pairs in the event: 0SFOS ($e^{\pm} e^{\pm} \mu^{\mp}$, $\mu^{\pm} \mu^{\pm} e^{\mp}$), 1SFOS ($e^{\pm} e^{\mp} \mu^{\pm}$, $e^{\pm} e^{\mp} \mu^{\mp}$, $\mu^{\pm} \mu^{\mp} e^{\pm}$, $\mu^{\pm} \mu^{\mp} e^{\mp}$) and 2SFOS ($e^{\pm} e^{\pm} e^{\mp}$, $\mu^{\pm} \mu^{\pm} \mu^{\mp}$). Similarly, in $W^{\pm} W^{\mp} Z \to \ell^{\pm}\nu \ell^{\pm}\nu \ell^+\ell^-$ channel, two signal regions are defined based on the selection of SFOS or different-flavour opposite-sign (DFOS) lepton-pair events: SFOS ($e^{\pm} e^{\mp} \mu^{\mp} \mu^{\pm}$, $e^{\pm} e^{\mp} e^{\pm} e^{\mp}$, $\mu^{\mp} \mu^{\pm} \mu^{\mp} \mu^{\pm}$) and DFOS ($e^{\pm} e^{\mp} \mu^{\mp} e^{\pm}$, $\mu^{\mp} \mu^{\pm} \mu^{\mp} e^{\pm}$). 
To select $W^{\pm} W^{\pm} W^{\mp} \to \ell^{\pm}\nu \ell^{\pm}\nu jj$ candidates, events are required to have exactly two leptons with the same electric charge, and at least two jets. 
Three different final states are considered based on the lepton flavour, namely $e^{\pm}e^{\pm}$,  $e^{\pm}\mu^{\pm}$ and $\mu^{\pm}\mu^{\pm}$. In the case of $W^{\pm} Z Z$ process, separate set of selection criteria are defined in order to select events in which vector bosons undergo either fully leptonic of semi-hadronic decay. In all channels, events are rejected if they have identified $b$-jets. This selection requirement suppresses background involving top quarks, with marginal impact on the signal efficiency. 
Full description to the optimized selection criteria, estimated systematic uncertainties and expected signal and background event yields for all channels considered in the study are available in Ref.~\cite{ATL-PHYS-PUB-2018-030}. Three channels, 0SFOS $W^{\pm} W^{\pm} W^{\mp} \to3\ell$ $3 \nu$, DFOS $W^{\pm} W^{\mp} Z \to4\ell$ $2 \nu$ and $W^{\pm} Z Z \to5\ell$ $1 \nu$, for which we give details in the following, are estimated to provide best sensitivities. 
\cref{tab:WWW_1,tab:WWZ_1,tab:WZZ_1} show the kinematic selection criteria used to select signal events in these channels.

\begin{table}[htbp]
	\centering
 \caption{Event selection criteria for $W^{\pm} W^{\pm} W^{\mp}\to3\ell$ $3 \nu$ candidate events. 	} 
	\label{tab:WWW_1}	
	\footnotesize
	\begin{tabular}{|l|c|}
		\hline
		{$W^{\pm} W^{\pm} W^{\mp} \to \ell^{\pm}\nu \ell^{\pm}\nu\ell^{\mp}\nu$} & 0SFOS events: $e^{\pm} e^{\pm} \mu^{\mp}$, $\mu^{\pm} \mu^{\pm} e^{\mp}$ \\
		\hline
		\hline
		Preselection  & \multicolumn{1}{c|}{Exactly 3 charged {\it tight} leptons with {$p_{\mathrm T}>$ 30 GeV and $|\eta| <$ 4}} \\
		SFOS dilepton mass &$m_{\ell\ell}^{\mathrm{SFOS}} >$ 20 GeV \\
		Angle between the trilepton system and $\overrightarrow{E}_{\mathrm T}^{\mathrm miss}$ & \multicolumn{1}{c|}{$|\varphi^{3\ell} - \varphi^{\overrightarrow{E}_{\mathrm T}^{\mathrm miss}}| >$ 2.5} \\
		$Z$ boson veto & $|m_{\mathrm{ee}}-m_{\mathrm Z}|>$ 15 GeV  \\
		Jet veto & \multicolumn{1}{c|}{At most one jet with {$p_{\mathrm T}>$ 30 GeV} and $|\eta|<$ 2.5} \\
		$b$-jet veto & \multicolumn{1}{c|}{No identified $b$-jets with {$p_{\mathrm T}>$ 30 GeV} } \\
		\hline
	\end{tabular}
\end{table}

\begin{table}[htbp]
	\centering
    	\caption{Event selection criteria for $W^{\pm} W^{\mp} Z\to4\ell$ $2 \nu$ candidate events. The four-lepton mass $m_{4\ell}$ is calculated as invariant mass of the four-lepton system.  	} 	
	\label{tab:WWZ_1}	
	\footnotesize
	\begin{tabular}{|l|c|}
		\hline
		{$W^{\pm} W^{\mp} Z \to \ell^{\pm}\nu \ell^{\pm}\nu \ell^+\ell^-$}  & DFOS events: $e^{\pm} e^{\mp} \mu^{\mp} e^{\pm}$, $\mu^{\mp} \mu^{\pm} \mu^{\mp} e^{\pm}$ \\
		\hline
		\hline
Preselection  & \multicolumn{1}{c|}{Exactly 4 charged \textit{loose} (3$^{rd}$ and 4$^{th}$ \textit{tight}) leptons} \\
		 & \multicolumn{1}{c|}{with $p_{\mathrm T}(1,2)>$ 30 GeV, $p_{\mathrm T}(3,4)>$ 25 GeV and $|\eta| <$ 4} \\
SFOS dilepton mass & \multicolumn{1}{c|}{ $ |m_{\ell\ell}^{\mathrm{SFOS}} - 91$ GeV$ | < $ 15 GeV} \\
DFOS dilepton mass &  $m^{\mathrm{DFOS}}_{\ell\ell}> 40$ GeV \\
Four-lepton mass &  $m_{4\ell}>$ 250 GeV \\
$b$-jet veto & \multicolumn{1}{c|}{No identified $b$-jets with $p_{\mathrm T}>$ 30 GeV } \\
\hline
	\end{tabular}
\end{table}

\begin{table}[htbp]
	\centering
\caption{Event selection criteria for $W^{\pm} Z Z \to 5\ell$ $1 \nu$ candidate events. Two-lepton pairs of the same flavour and opposite charge have to satisfy same-flavour dilepton mass selection requirement. The transverse mass is calculated from the $E_{\mathrm T}^{\mathrm{miss}}$  and the lepton that does not pass dilepton mass requirement. }	
\label{tab:WZZ_1}
	\footnotesize
	\begin{tabular}{|l|c|}
		\hline
		 $W^{\pm} Z Z \to \ell^{\pm}\nu \ell^{+}\ell^{-} \ell^{+}\ell^{-}$ & $5\ell 1 \nu$ \\
\hline\hline
		Preselection  & Exactly 5 charged {\it loose} (4$^{rd}$ and 5$^{th}$ \textit{tight}) leptons with    \\
		& $p_{\mathrm T}(1,2,3)>$ 30 GeV, $p_{\mathrm T}(4,5)>$ 25 GeV and $|\eta|<$ 4   \\
SFOS dilepton mass & \multicolumn{1}{c|}{$ |m_{\ell\ell}^{\mathrm{SFOS}} - 91$ GeV$ | < $ 15 GeV} \\
Transverse mass & $m_{\mathrm T}>$ 40 GeV  \\
$b$-jet veto & \multicolumn{1}{c|}{No identified $b$-jets with $p_{\mathrm T}>$ 30 GeV} \\
\hline
\end{tabular}
\end{table}

\subsubsection*{Results}
\label{sec:result}

The SM processes that mimic the multi-boson signal signatures by producing at least three prompt leptons or two prompt leptons with the same electric charge, can be grouped into the following  categories:
\begin{itemize}
\item The $WZ$ and $ZZ$ processes, referred to as ``diboson background'';
\item The $WWW, WWZ, WZZ, ZZZ$ processes, excluding the signal process under study, referred to as ``triboson background'';
\item The $VH$ and $t\bar{t} H$ processes,  excluding the processes which are added to the signal,  referred to as ``Higgs+X background'';
\item The production of four top quarks, top quark associated with $WZ$ bosons or $t\bar{t}$ associated with $W, Z, WZ$ or $W^{\pm}W^{\mp}$ bosons, referred to as ``top background'';
\item Processes that have non-prompt leptons (electrons) originating from misidentified jets (referred to as ``fake-lepton background'');
\item Processes that produce prompt charged leptons, but the charge of one lepton is misidentified (referred to as ``charge-flip background''). 
\end{itemize}

The contributions from the $WW$ and $t\bar{t}$ processes are accounted for in the fake-lepton and charge-flip backgrounds. 
The diboson, triboson, Higgs+X and top background sources are estimated using simulated events, with the dominant irreducible background in most of the channels originating from the diboson processes.  
In some channels the contribution of the fake-lepton background, which is derived by applying the pre-defined ($p_{\mathrm{T}}$, $\eta$)-dependent likelihood as described in \cref{sec:exp}, becomes significant. The charge-flip background has been investigated and found to be negligible in all considered processes.

In $W^{\pm} W^{\pm} W^{\mp}\to3\ell$ $3 \nu$ channel, the background is dominated by the irreducible diboson background and fake-lepton contribution. The contribution of signal events containing Higgs decays are at the level of 40\%. In $W^{\pm} W^{\mp} Z\to4\ell$ $2 \nu$ channel with two leptons being of different flavour, this requirement suppresses a large fraction of the diboson background. Contribution of Higgs decays is quite smaller with respect to the one in $W^{\pm} W^{\pm} W^{\mp}\to3\ell$ $3 \nu$ due to smaller lepton $p_{\mathrm T}$ and invariant mass requirement $m^{\mathrm{DFOS}}_{\ell\ell}> 40$ GeV. 
In the $W^{\pm}ZZ$ channel, the most promising signal region is the one with five charged leptons. In this case, the fake-lepton contribution becomes significant. The background is dominated by rare top production of $t\bar{t}ZW$.

\cref{fig:triboson_plots} shows relevant distributions in the three channels: the $m_{\mathrm T}^{3\ell}$ distribution for the $W^{\pm} W^{\pm} W^{\mp}\to3\ell$ $3 \nu$ channel,  the distribution of transverse momenta of the two-lepton system $p_{\mathrm T}^{\ell\ell}$ in $W^{\pm} W^{\mp} Z\to4\ell$ 2$\nu$ channel and the distribution of two lepton invariant mass $p_{\mathrm T}^{\ell\ell}$ selected to give the mass closest to the mass of the $Z$ boson in $W^{\pm} Z Z\to5\ell$ $1 \nu$ channel.

\begin{figure}[htbp]
\centering
	\includegraphics[width=0.45\textwidth]{\main/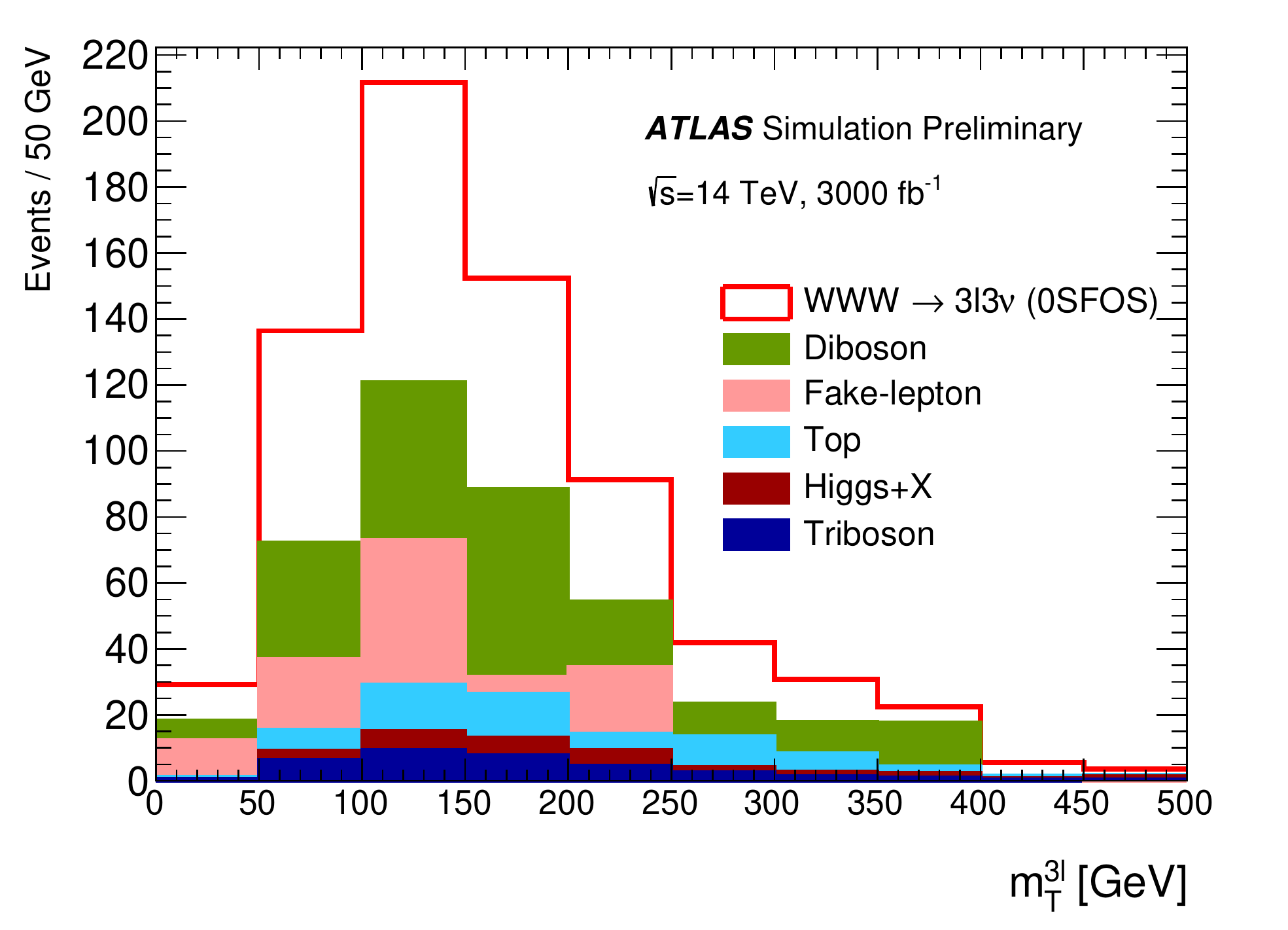}
	\includegraphics[width=0.45\textwidth]{\main/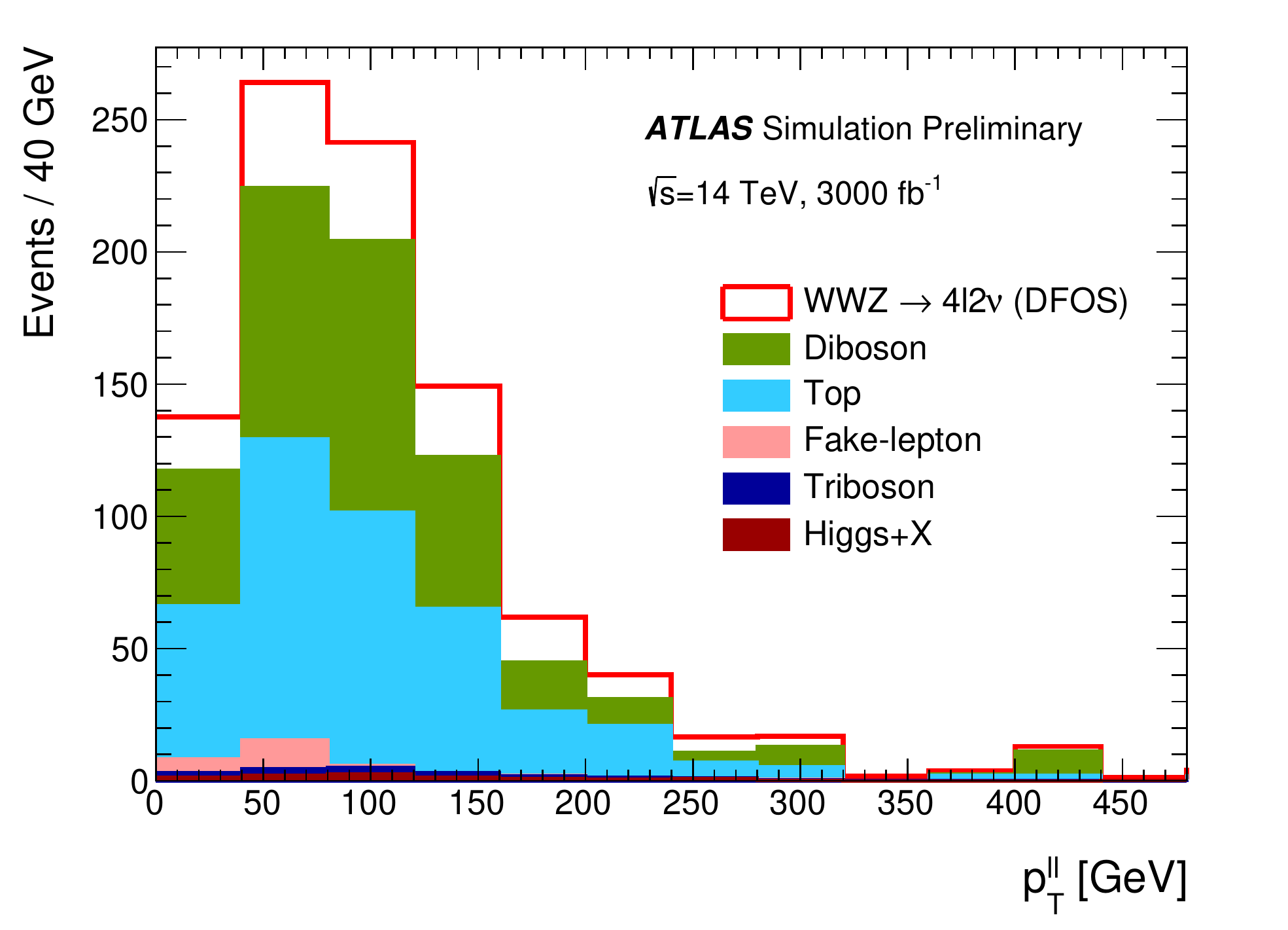}
	\includegraphics[width=0.45\textwidth]{\main/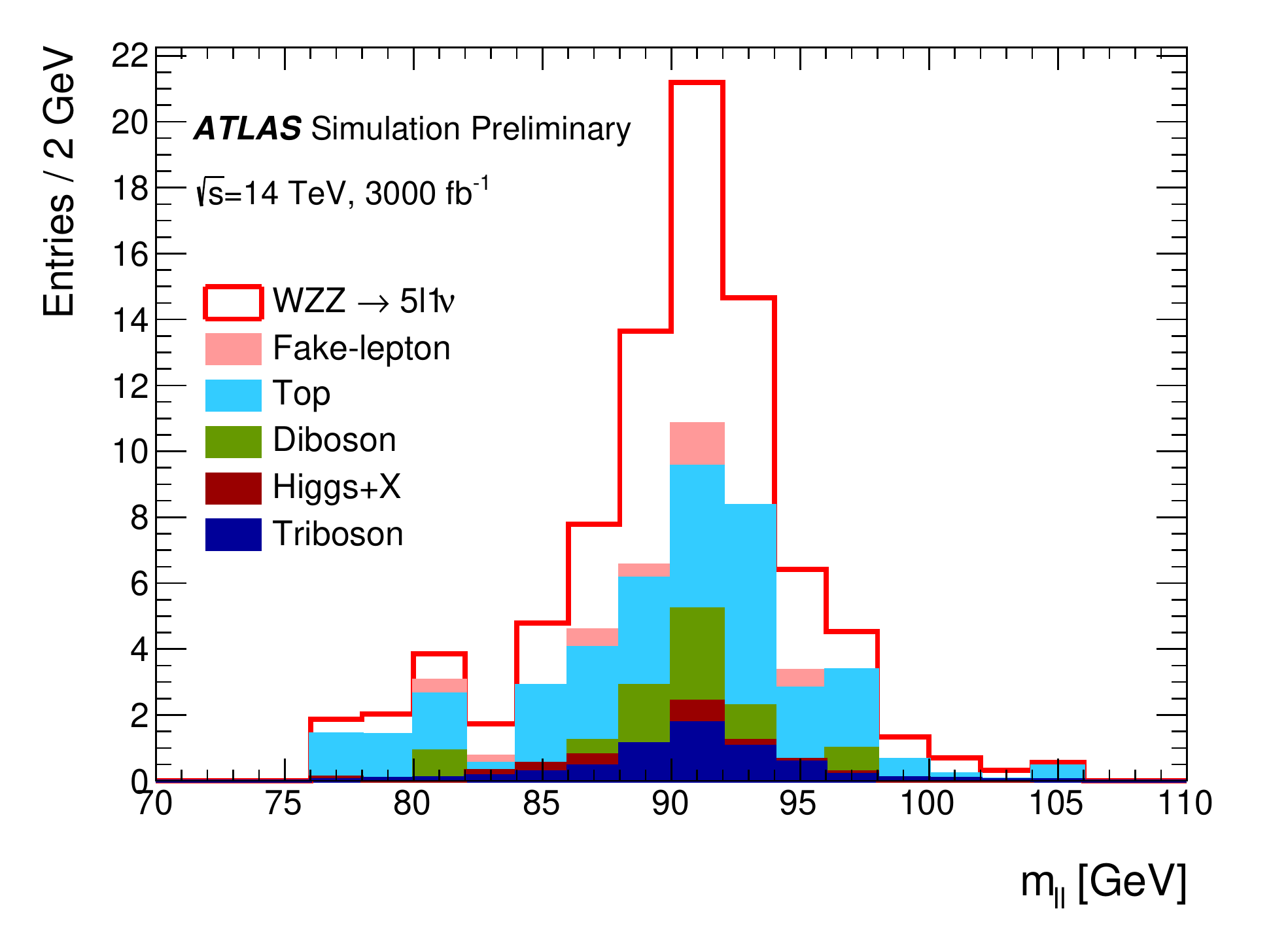}	
	\caption{The distribution of $m_{\mathrm T}^{3\ell}$ for the $W^{\pm} W^{\pm} W^{\mp}\to3\ell$ $3 \nu$ channel (top left),  the distribution of transverse momenta of the two-lepton system $p_{\mathrm T}^{\ell\ell}$ in $W^{\pm} W^{\mp} Z\to4\ell$ $2 \nu$ channel (top right) and the distribution of two lepton invariant mass $p_{\mathrm T}^{\ell\ell}$ selected to give the mass closest to the mass of the $Z$ boson in $W^{\pm} Z Z\to5\ell$ $1 \nu$ channel (bottom) as expected from the signal and background processes at 3000 fb$^{-1}$ after applying the selection criteria from \cref{tab:WWW_1,tab:WWZ_1,tab:WZZ_1}.}
	\label{fig:triboson_plots}
\end{figure}

Systematic uncertainties in the signal and background predictions arise from the uncertainties in the measurement of the integrated luminosity, from the experimental modelling of the signal acceptance and detection efficiency, and from the background normalisation. With the much larger integrated luminosity and a sophisticated understanding of the detector performance and backgrounds at the HL-LHC, we expect experimental uncertainties related to the lepton reconstruction and identification efficiencies as well as lepton energy/momentum resolution and scale modelling of 1\%, to the $E_{\mathrm T}^{\mathrm{miss}}$  modelling of 1\%, to the jet energy scale and resolution of 1.5\% and 5\% in the fully leptonic and leptons+jets channels, respectively, to the luminosity measurement of 1\% and to the expected pileup of 1\% \cite{ATLAS_PERF_Note}.  Based on the extrapolations of current ATLAS measurements and assuming a reduction of the uncertainty at the level of 15--80\%, depending on the process and the origin of the systematics, the following systematic uncertainties on the cross section normalisation for each of the background processes are assumed: 4\% on $\sigma_{\mathrm{diboson}}$, 30\% on $\sigma_{\mathrm{triboson}}$, 3\% on $\sigma_{t\bar{t}}$, 20\% on $\sigma_{t\bar{t}H}$, 6\% on $\sigma_{t\bar{t}Z}$, and 11\% on $\sigma_{t\bar{t}W}$. The uncertainty on the level of the fake-lepton background is estimated to be 10\%.  Taking these assumptions into account, we estimate the total systematic uncertainty on the background of 9\% for $W^{\pm} W^{\pm} W^{\mp}\to3\ell3\nu$ and $W^{\pm} Z Z\to 5\ell 1\nu$ channels and  6\% in  $W^{\pm} W^{\mp} Z\to4\ell 2\nu$ channel. 
Assuming that the number of signal events follows a Poissonian distribution and taking into account an estimated systematic uncertainty on the background,  
the signal significance $Z_{\sigma}$ and the estimated precision on the signal strength measurement, $\frac{\Delta \mu}{\mu}$ are calculated using the asymptotic formula from Ref.~\cite{Cowan:2010js}. Only experimental uncertainties are taken into account for the signal. Uncertainties related to the limited number of MC events are neglected.
The total number of signal and background events expected after applying the full set of selection requirements from \cref{tab:WWW_1,tab:WWZ_1,tab:WZZ_1} in three selected channels, the corresponding  signal significance and the expected precision on the signal strength measurement, for an integrated luminosity of 3000 fb$^{-1}$ are shown in \cref{tab:triboson_yields}.

\begin{table}[htbp]
	\centering
\caption{Expected number of signal and background events, the expected signal significance $Z_{\sigma}$ and the estimated precision on the signal strength measurement, $\frac{\Delta \mu}{\mu}$ in $W^{\pm} W^{\pm} W^{\mp}\to3\ell3\nu$, $W^{\pm} W^{\mp} Z\to4\ell2\nu$ and $W^{\pm} Z Z\to5\ell1\nu$ channels after applying the selection criteria from \cref{tab:WWW_1,tab:WWZ_1,tab:WZZ_1}.}
		\label{tab:triboson_yields}
         \footnotesize
	\begin{tabular}{|l|c|c|c|}
		\hline
		 & $W^{\pm} W^{\pm} W^{\mp} \to3\ell 3\nu$ & $W^{\pm} W^{\mp} Z \to4\ell 2\nu$ & $W^{\pm} Z Z \to 5\ell 1\nu$  \\
		\hline
		\hline
		Signal                           &  312 & 168  & 19 \\		
		\hline
		Diboson                    & 208  & 357 &   4.0 \\
		Triboson                    & 37   &   11 &   3.0 \\
		Higgs+X                    & 25   &   10 &  0.3 \\
		Top                            & 60   & 390 &  15 \\
		fake-lepton                & 97 &   16 &  3.0 \\ \hline
  	        Total:                         & 427  & 784 &  25 \\
		\hline\hline
		Significance $Z_{\sigma}$  & 6.7  &  3.0 & 3.0 \\ 
		Significance $Z_{\sigma}$ (4000 fb$^{-1}$)   & 7.0  &  3.1 & 3.4 \\ \hline		
		Precision $\frac{\Delta \mu}{\mu}$  & 11\% & 27\% & 36\% \\
                 Precision $\frac{\Delta \mu}{\mu}$ (4000 fb$^{-1}$) & 10\% & 25\% & 31\% \\
		\hline
	\end{tabular}
\end{table}

The HL-LHC offers a large improvement to multi-boson production, where this simple cut-and-count approach provides sensitivities larger than 3$\sigma$ in the three channels considered in this analysis. It should be noted that more mature analysis techniques such as MVA, would likely improve these results further. However, high level of background control, mainly diboson background as well as instrumental background arising from fake-leptons, will be needed in order to maintain desired level of precision.
 \FloatBarrier


\subsection{Precision electroweak measurements}


\newcommand\TwoFigBottom{-2}
\providecommand{\abinv} {\mbox{\ensuremath{\,\text{ab}^\text{$-$1}}}\xspace} 
\newcommand*{\met}{\ensuremath{E_{\text{T}}^{\text{miss}}}\xspace}


\subsubsection[NNLO predictions for Z-boson pair production]{NNLO predictions for Z-boson pair production\footnote{Contribution by G.~Heinrich, S.~Jahn, S.~Jones, M.~Kerner and J.~Pires.}}

The results presented in this section are produced using the program described in
Ref.~\cite{Heinrich:2017bvg} with the NNPDF3.0~\cite{Ball:2014uwa} set of parton distribution
functions. The parton densities
and $\alpha_s$ are evaluated at each corresponding order (i.e. ($n$+1)-loop
$\alpha_s$ is used at N$^{n}$LO, with $n=0,1,2$) and $N_f=5$
massless quark flavours are considered. 
For the renormalisation ($\mu_R$) and factorisation ($\mu_F$) scales
two choices are investigated:
$\mu_R=\mu_F=m_{Z}$ and the dynamic scale $\mu_R=\mu_F=m_{ZZ}/2$. 
The $G_\mu$ EW scheme is used where the EW input parameters have been set to $G_F=1.16639\times 10^{-5}$, $m_W=80.399$ GeV and
$m_Z=91.1876$ GeV. The top quark and Higgs boson masses that are included in the real-virtual one-loop contributions and in the loop-induced $gg$ channel have been set to $m_t=173.2$\,GeV and
$m_H=125$\,GeV, respectively. The one-loop contributions are
calculated with the program {\sc  GoSam}~\cite{Cullen:2011ac,Cullen:2014yla}.
For the NNLO real radiation the $N$-jettiness subtraction scheme~\cite{Stewart:2009yx,Boughezal:2015dva,Boughezal:2015aha,Gaunt:2015pea} is employed. The process dependent hard function has been extracted
from the two-loop amplitude computed in Ref.~\cite{Gehrmann:2015ora} and cross-checked with an in-house calculation.
The top quark contributions in the double virtual two-loop diagrams are not included in the results below.
Table~\ref{tab:Xsec} shows cross section results for the central scale
$\mu_R=\mu_F=m_{Z}$, including 7-point scale variations.
In Table~\ref{tab:Xsec_dynamicscale} results for the
dynamic scale $\mu_R=\mu_F=m_{ZZ}/2$ are given.

\begin{table}[htb]
\centering
\caption{Inclusive cross section for $ZZ$ production at the LHC for
  $\sqrt{s}=$14\,TeV and $\sqrt{s}=$27\,TeV at LO, NLO and NNLO with
  $\mu_R=\mu_F=m_{Z}$.
The uncertainties  are obtained
by varying the renormalisation and factorisation scales in the range $m_{Z}/2<\mu_R,\mu_F<2m_{Z}$ with the constraint 
$0.5<\mu_F/\mu_R<2$.}
\label{tab:Xsec}
\begin{tabular}{|c|l|l|l|c|}
\hline
  & $\sigma_{LO}$ [pb] & $\sigma_{NLO}$ [pb] & $\sigma_{NNLO}$ [pb] &$gg\to ZZ$ [pb]\\ \hline\hline
 & & & &\\  
14\,TeV & $10.80^{+5.7\%}_{-6.7\%}$    &  $15.55^{+3.0\%}_{-2.4\%}$   &    $18.50^{+3.0\%}_{-3.2\%} $ & $1.56^{+25\%}_{-18\%}$\\ 
 & & & &\\  
27\,TeV &  $23.59^{+10.0\%}_{-10.9 \%}$   & $35.59^{+3.2\%}_{-4.2\%}$   &    $ 44.52^{+3.7\%}_{-4.1\%} $ & $4.81^{+25\%}_{-18\%}$\\ 
 & & & &\\  
\hline
\end{tabular}
\end{table}

\begin{table}[htb]
\centering
\caption{Inclusive cross section for $ZZ$ production at the LHC for
  $\sqrt{s}=$14\,TeV and $\sqrt{s}=$27\,TeV at LO, NLO and NNLO with
  the dynamic scale choice 
  $\mu_R=\mu_F=m_{ZZ}/2$.
The uncertainties  are obtained
by varying the renormalisation and factorisation scales in the range $m_{ZZ}/4<\mu_R,\mu_F<m_{ZZ}$ with the constraint 
$0.5<\mu_F/\mu_R<2$.}
\label{tab:Xsec_dynamicscale}
\begin{tabular}{|c|l|l|l|c|}
\hline
  & $\sigma_{LO}$ [pb] & $\sigma_{NLO}$ [pb] & $\sigma_{NNLO}$ [pb] &$gg\to ZZ$ [pb]\\ \hline\hline
 & & & &\\  
14\,TeV & $11.03^{+5.2\%}_{-6.1\%}$ & $15.38^{+2.5\%}_{-2.0\%}$&$18.20^{+3.3\%}_{-2.3\%}$&$1.41^{+23\%}_{-18\%}$\\ 
 & & & &\\  
27\,TeV & $24.68^{+9.0\%}_{-9.8\%}$ &$35.43^{+2.6\%}_{-3.7\%}$&$43.71^{+3.3\%}_{-3.2\%}$&$4.41^{+23\%}_{-17\%}$\\
 & & & &\\  
\hline
\end{tabular}

\end{table}

\begin{figure}[htb]
\centering
\includegraphics[width=0.45\textwidth]{\main/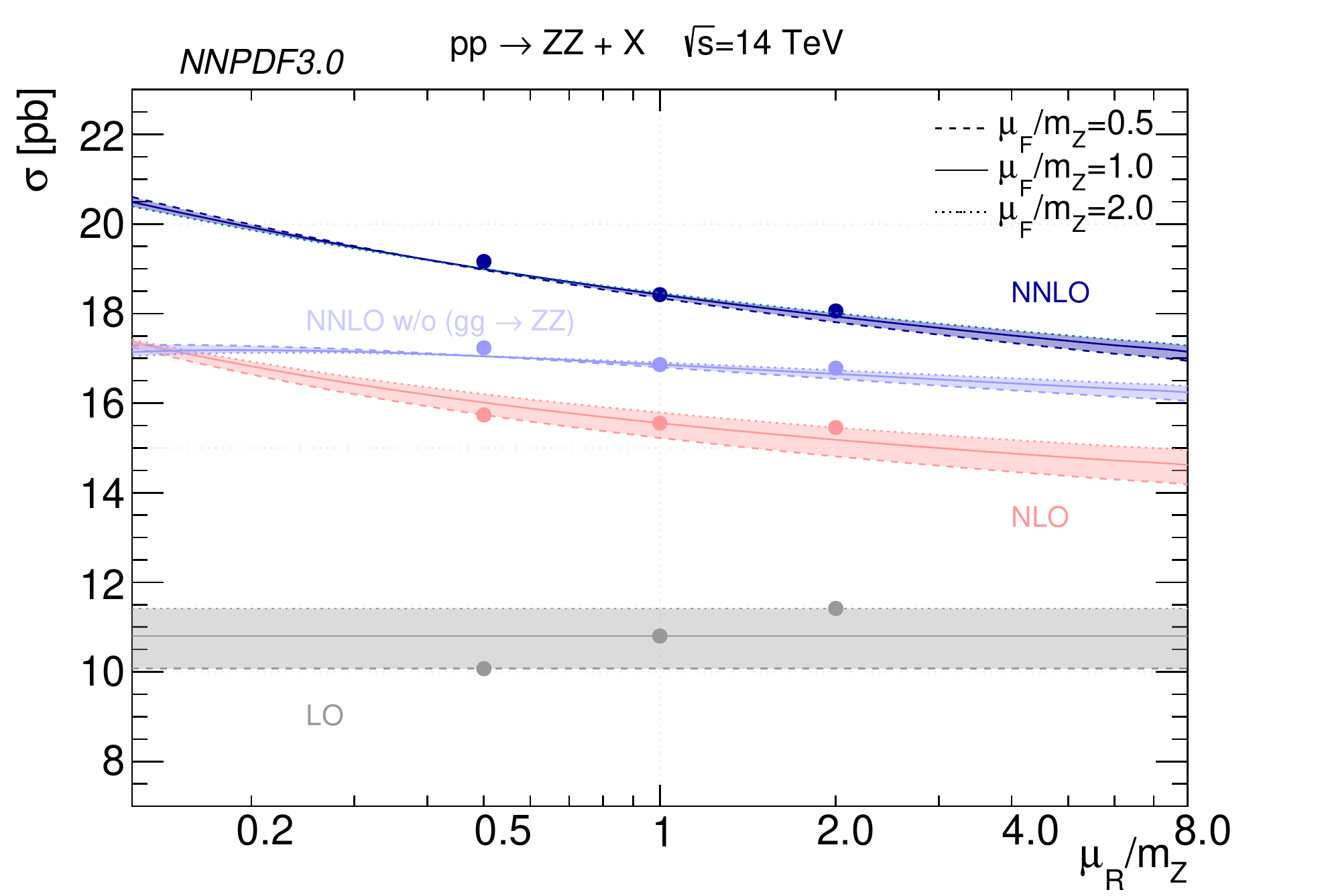}
\label{fig:scalevar14TeV}
\includegraphics[width=0.45\textwidth]{\main/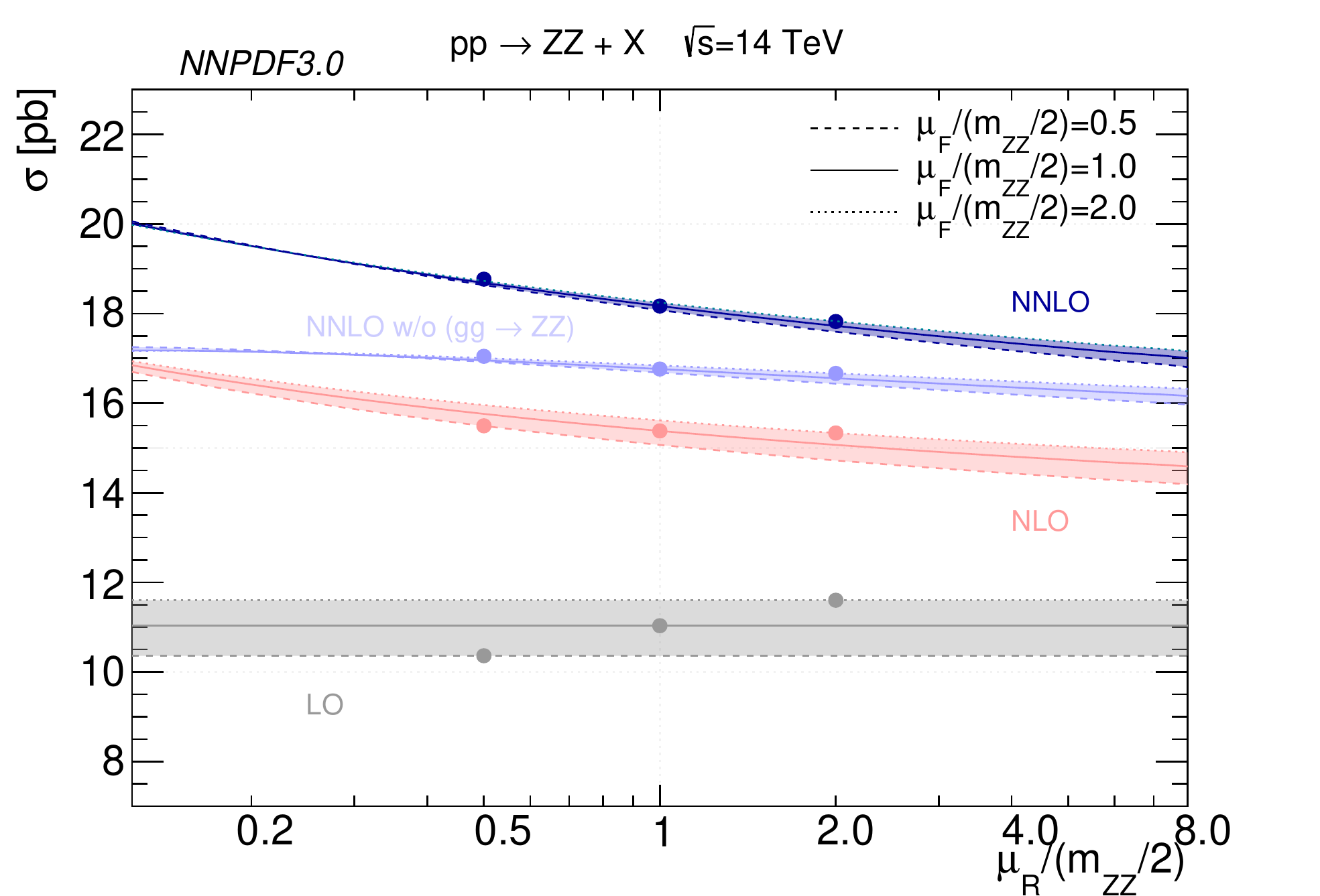}
\label{fig:scalevar_dynamic-14TeV}
\caption{Renormalisation and factorisation scale dependence of the
  $ZZ$ cross section for $\sqrt{s}=14$\,TeV   at LO, NLO and NNLO for
  the fixed central scale choice $\mu_R=\mu_F=m_Z$ (left) and for the
  dynamic central scale choice $\mu_R=\mu_F=m_{ZZ}/2$ (right). 
The NNLO result without the gluon fusion contributions is shown in
light blue.
The thickness of the bands show the variation with the factorisation
scale, while the slope shows the renormalisation scale dependence. The
scale uncertainties are the envelope of scale variations by a factor of two up and down with the constraint 
$0.5<\mu_F/\mu_R<2$, i.e. 7-point scale variations.} 
\label{fig:scalevariations-14TeV}
\end{figure}
\begin{figure}[htb]
\centering
\includegraphics[width=0.45\textwidth]{\main/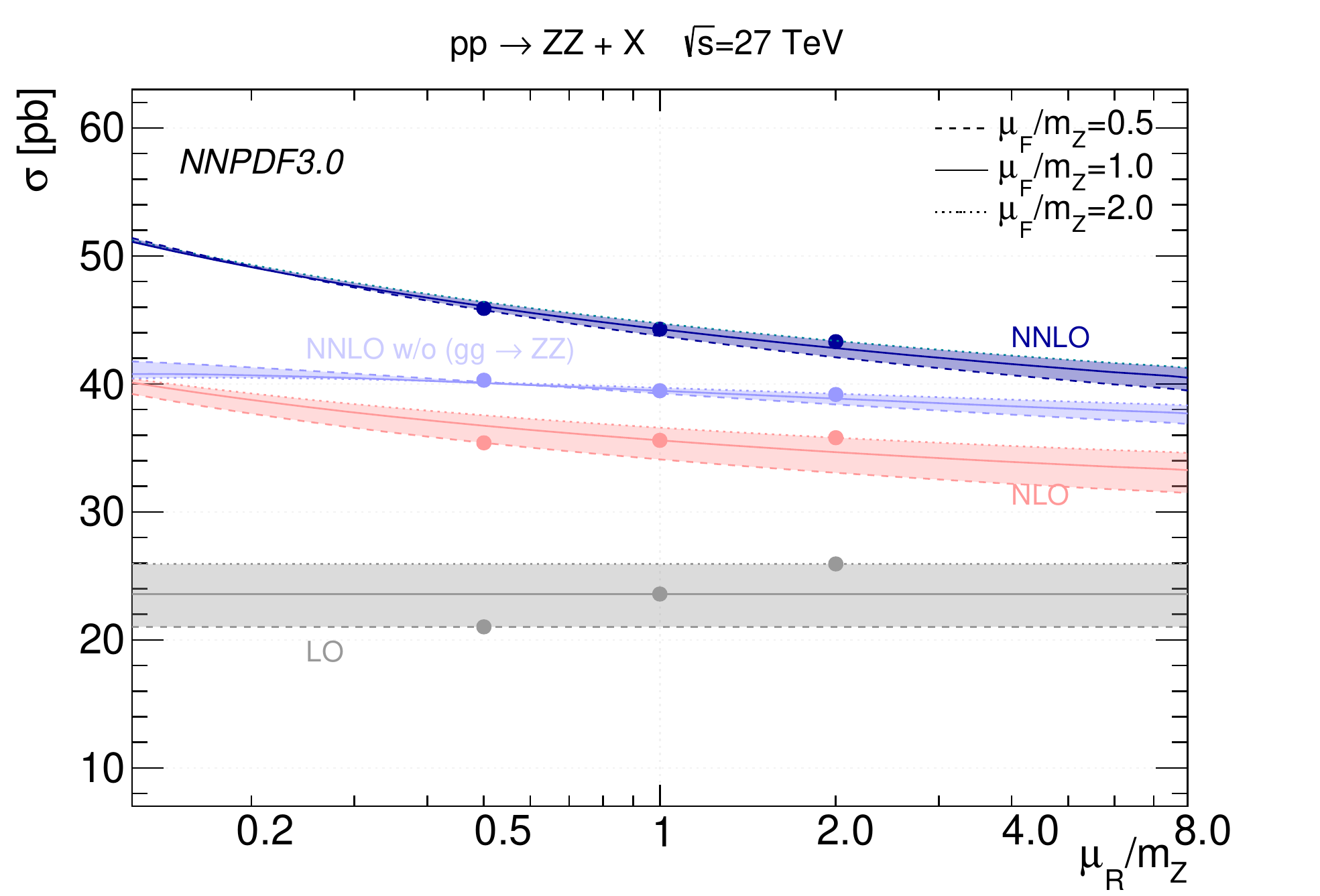}
\label{fig:scalevar27TeV}
\includegraphics[width=0.45\textwidth]{\main/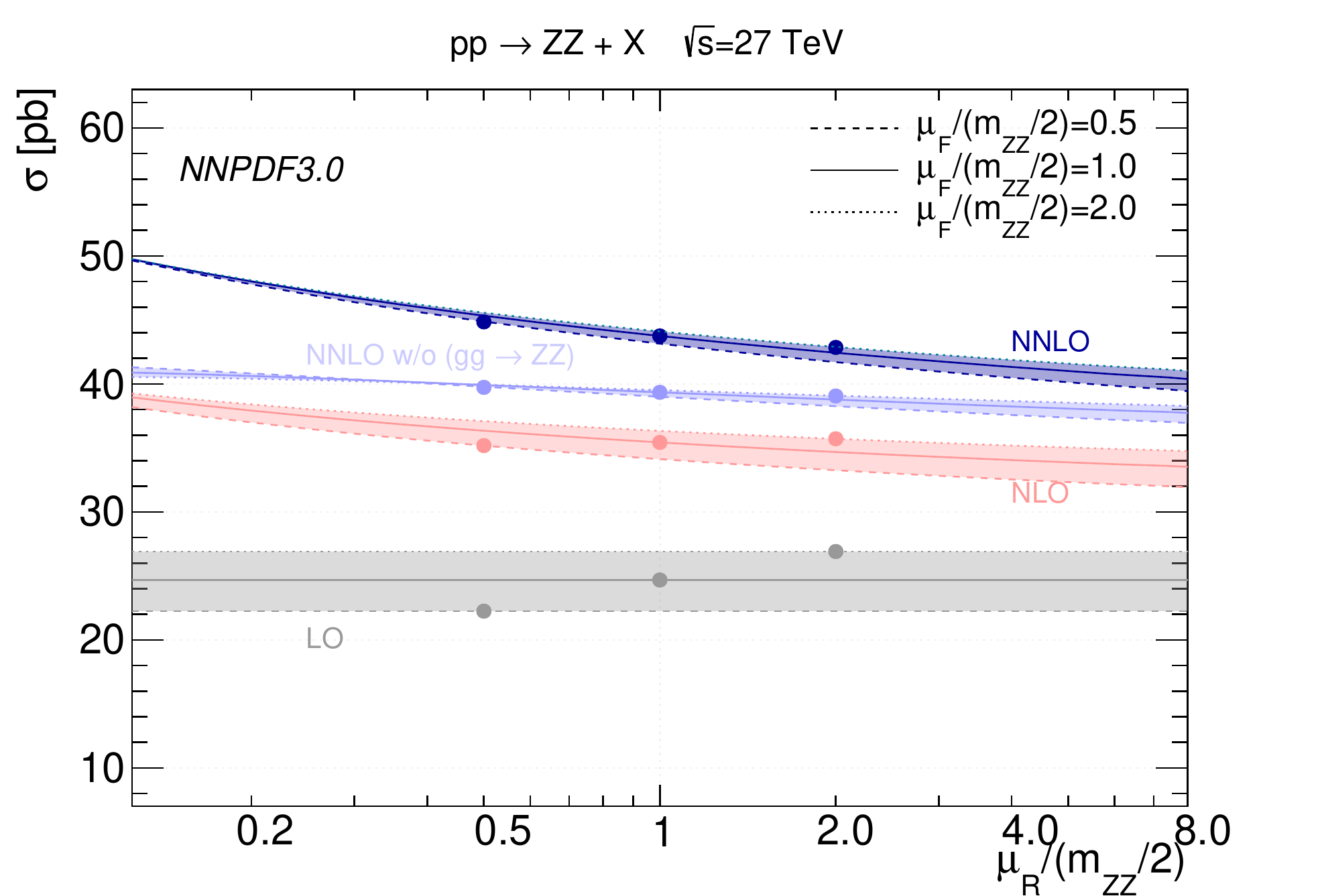}
\label{fig:scalevar_dynamic-27TeV}
\caption{Renormalisation and factorisation scale dependence of the
  $ZZ$ cross section for $\sqrt{s}=27$\,TeV   at LO, NLO and NNLO for
  the fixed central scale choice $\mu_R=\mu_F=m_Z$ (left) and for the
  dynamic central scale choice $\mu_R=\mu_F=m_{ZZ}/2$ (right). 
The NNLO result without the gluon fusion contributions is shown in
light blue, and the bands are produced in the same way as in Fig.~\ref{fig:scalevariations-14TeV}.} 
\label{fig:scalevariations-27TeV}
\end{figure}
Figures~\ref{fig:scalevariations-14TeV} and~\ref{fig:scalevariations-27TeV} show largely non-overlapping
scale uncertainty bands between NLO and NNLO, both for a fixed central scale choice $\mu=m_Z$ as well as for a dynamic central scale choice $\mu=m_{ZZ}/2$.
This demonstrates that for this process, the scale variations are insufficient to estimate
missing higher order terms in the perturbative expansion. This is mostly due to the fact that at NNLO, the loop-induced
gluon fusion channel $gg \to ZZ$ opens up, and due to the large gluon
flux it represents a numerically significant contribution, about 8\%  at $\sqrt{s}=14$\,TeV and 11\% at $\sqrt{s}=27$\,TeV of the
total NNLO cross section, for both central scale choices.
Further studies of the gluon channel can be seen in Refs.~\cite{Alioli:2016xab,Grazzini:2018owa}.
Since this new channel contributes for the first time at NNLO its contribution cannot be captured by the scale variations
of the NLO cross section. 
Therefore, with increasing perturbative order, 
a systematic reduction of the factorisation scale dependence of the cross section is observed
(indicated by the thickness of the scale uncertainty band), while there is no
significant reduction of the renormalisation scale dependence. 
To show that this effect can be attributed to the gluon fusion channel
opening up at NNLO, the NNLO result excluding this
channel is also shown in Figs.~\ref{fig:scalevariations-14TeV} and~\ref{fig:scalevariations-27TeV}.

\subsubsection[Gauge-boson pair production with M{\small ATRIX}]{Gauge-boson pair production with M{\small ATRIX}\footnote{Contribution by S.~Kallweit, M.~Grazzini and M.~Wiesemann.}}


%
%
\def\nn{\nonumber}
\def\sfrac#1#2{{\textstyle{#1\over #2}}}
\def\lsim{\mbox{\raisebox{-.6ex}{~$\stackrel{<}{\sim}$~}}}
\def\gsim{\mbox{\raisebox{-.6ex}{~$\stackrel{>}{\sim}$~}}}
\newcommand{\La}{\mathcal{L}}

\newcommand{\fbinv} {\mbox{\ensuremath{\,\text{fb}^\text{$-$1}}}\xspace}

%



\newcommand{\zz}{\ensuremath{ZZ}}                                                                                                                                                 
\newcommand{\ww}{\ensuremath{W^+W^-}}
\newcommand{\wz}{\ensuremath{W^\pm Z}}
\newcommand\as{\alpha_{\mathrm{S}}} 
\def\citere#1{\mbox{Ref.~\cite{#1}}}
\def\citeres#1{\mbox{Refs.~\cite{#1}}}
\newcommand\powheg{{\sc Powheg}}
\newcommand\Matrix{{\sc Matrix}}
\newcommand\Munich{{\sc Munich}}
\newcommand\OpenLoops{{\sc OpenLoops}}
\newcommand\Collier{{\sc Collier}}
\newcommand\GINAC{{\sc Ginac}}
\newcommand{\CutTools}{{\sc CutTools}}
\newcommand{\OneLOop}{{\sc OneLOop}}
\newcommand{\qt}{\ensuremath{q_T}}

\newcommand{\nameobservable}[1]{{#1}}
\newcommand{\captiontext}{}
\newcommand{\GeV}{\,\mathrm{GeV}}



NNLO QCD predictions for \ww{}, \wz{} and \zz{} production in proton--proton collisions are presented in this section.
Two LHC upgrade scenarios are considered,
namely the HL-LHC running at $\sqrt{s}=14\,{\rm TeV}$ with an
assumed integrated luminosity of 3\abinv, 
and the HE-LHC at  $\sqrt{s}=27\,{\rm TeV}$ with 15~\abinv. 
More precisely, the following inclusive hard-scattering processes are considered
\begin{align*}
  & pp\to \ell^+\nu_\ell \,\ell^{\prime -}  {\bar \nu}_{\ell^\prime}+X\,,\\
  & pp \to \ell\nu_\ell \,\ell^{\prime +} \ell^{\prime -}+X\,,\\
  & pp \to \ell^+ \ell^- \ell^{\prime +} \ell^{\prime -}+X\, ,
\end{align*}
where all off-shell effects and interference contributions are fully accounted for. 

All results are obtained with the public parton-level NNLO framework {\sc Matrix}. This program, and earlier versions of it,
have been used to compute state-of-the-art QCD predictions for gauge-boson pair production processes \cite{Grazzini:2013bna,Grazzini:2015nwa,Cascioli:2014yka,Grazzini:2015hta,Gehrmann:2014fva,Grazzini:2016ctr,Grazzini:2016swo,Grazzini:2017ckn,Kallweit:2018nyv}.\footnote{It was
  also used in the NNLL+NNLO computation for \ww{} and \zz{} production of \citere{Grazzini:2015wpa}, and in the NNLOPS computation for \ww{} production of Ref.~\cite{Re:2018vac}.}
All tree-level and one-loop amplitudes are evaluated with 
\OpenLoops{}\footnote{\OpenLoops{} which relies on the fast and stable
tensor reduction of \Collier{}~\cite{Denner:2014gla,Denner:2016kdg},
supported by a rescue system based on quad-precision
\CutTools\cite{Ossola:2007ax} with \OneLOop\cite{vanHameren:2010cp}
to deal with exceptional phase-space
points.}~\cite{Cascioli:2011va,Buccioni:2017yxi}.
At two-loop level the $q\bar{q}\to VV'$ amplitudes of \citere{Gehrmann:2015ora} are used.

The complex mass scheme~\cite{Denner:2005fg} is applied throughout, 
i.e.\ complex $W$- and $Z$-boson masses are used and the EW mixing angle is defined as $\cos\theta_W^2=(m_W^2-i\Gamma_W\,m_W)/(m_Z^2-i\Gamma_Z\,m_Z)$.
For the input of the weak parameters the $G_\mu$ scheme is employed with $\alpha=\sqrt{2}\,G_\mu |(m_W^2-i\Gamma_W\,m_W)\sin^2\theta_W|/\pi$\,.
The following parameters are set, $G_F = 1.16639\times 10^{-5}$\,GeV$^{-2}$, $m_W=80.399$\,GeV, $\Gamma_W=2.1054$\,GeV, $m_Z = 91.1876$\,GeV, $\Gamma_Z=2.4952$\,GeV, $m_H = 125$\,GeV and $\Gamma_H = 0.00407$\,GeV. Furthermore,  a diagonal CKM matrix is used.

The number of light quarks is chosen differently for the processes under consideration:
all \ww{} results are obtained by applying the four-flavour scheme (4FS) with massive top and bottom quarks in order to consistently
remove top-quark contamination by omitting the (separately IR finite) partonic processes with real bottom-quark emissions.
In the 4FS, the on-shell bottom mass $m_b = 4.92$\,GeV is used.
For all other processes the five-flavour scheme (5FS) is applied with a vanishing bottom mass $m_b = 0$.
The top quark is treated as massive and unstable throughout,
and $m_t$ is set to $173.2$\,GeV as well as $\Gamma_t = 1.44262$\,GeV.\footnote{Massive
top-quark contributions are neglected in the virtual two-loop corrections, but are kept everywhere else in the computations.}

The MMHT2014~\cite{Harland-Lang:2014zoa} sets of parton distribution functions~(PDFs) are used with $n_f=4$ or $n_f=5$ active quark flavours,
consistently with the flavour scheme under consideration.                             
N$^n$LO ($n=0,1,2$) predictions are obtained by using PDFs at the same perturbative order and
the evolution of $\as$ at $(n+1)$-loop order, as provided by the corresponding PDF set.
To be precise, in the 5FS MMHT2014lo68cl, MMHT2014nlo68cl, and
  MMHTnnlo68cl at LO, NLO, and NNLO are used.
  In the 4FS MSTW2008lo68cl\_nf4, MMHT2014nlo68cl\_nf4, and
 MMHT2014nnlo68cl\_nf4 at LO, NLO, and NNLO are used.

The central predictions are obtained by setting the factorization and renormalization 
scales to $\mu_F=\mu_R=\mu_0\equiv E_{\mathrm{T,}V_1}+E_{\mathrm{T,}V_2}$,
with $E_{\mathrm{T,}V_i}= \sqrt{M^2_{V_i}+p^2_{\mathrm{T,}V_i}}$, where $M_{V_i}$ is the invariant mass 
and $p_{{\rm T,}V_i}$ the transverse momentum of the respective vector boson.
Uncertainties from missing higher-order
contributions are estimated in the usual way by independently varying $\mu_F$ and $\mu_R$ in the range $0.5\mu_0<\mu_F,\mu_R<2\mu_0$ with the constraint $0.5<\mu_F/\mu_R<2$.

\begin{table}[t]
\begin{center}
\caption{\label{tab:dibosonXS} Inclusive cross sections for \ww{}, \wz{} and \zz{} production  where the leptonic decays of the bosons are included.}
\resizebox{\columnwidth}{!}{
\begin{tabular}{|c c | c c c c c|}
\hline
\multicolumn{2}{|c|}{$\sigma$ [fb]}
& LO
& NLO
& NLO$^\prime$$+gg$
& NNLO
& \multirow{2}{*}{\Large $\frac{\sigma_{\rm NNLO}({\rm 27\,TeV})}{\sigma_{\rm NNLO}({\rm 14\,TeV})}$} 
\\
\multicolumn{2}{|c|}{\footnotesize (correction)}
& 
& {\footnotesize (NLO/LO$-$1)}
& {\footnotesize (NLO$^\prime$$+gg$/NLO$-$1)}
& {\footnotesize (NNLO/NLO$-$1)} & 
\\
\hline\hline
\multirow{ 4}{*}{\vspace{-0.25cm}\ww{}}
& 
\multirow{ 2}{*}{$\sqrt{s}=14$\,TeV}
& $897.27(9)^{     +4.3 \%}_{     -5.3 \%}$
& $1303.3(1)^{     +2.7 \%}_{     -2.2 \%}$ 
& $1386.1(2)^{     +3.7 \%}_{     -2.9 \%}$ 
& $1485.(1)^{     +2.4 \%}_{     -2.2 \%}$ 
& \multirow{ 4}{*}{\vspace{-0.25cm} 2.33}
\\ 
&
&
& {\footnotesize ($+45.3\%$)} 
& {\footnotesize ($+6.4 \%$)}
& {\footnotesize ($+13.9 \%$)}
& 
\\[0.25cm]
& \multirow{ 2}{*}{$\sqrt{s}=27$\,TeV}
& $2091.5(2)^{     +7.6 \%}_{     -8.6 \%}$
& $2988.4(3)^{     +2.8 \%}_{     -2.9 \%}$
& $3213.0(4)^{     +4.1 \%}_{     -3.2 \%}$
& $3457.(4)^{     +2.8 \%}_{     -2.4 \%}$ 
&
\\
&
&
& {\footnotesize ($+42.9\%$)}
& {\footnotesize ($+7.0\%$)}
& {\footnotesize ($+15.6\%$)}
&
\\
\midrule
\multirow{ 4}{*}{\vspace{-0.25cm}$W^+Z${}}
& \multirow{ 2}{*}{$\sqrt{s}=14$\,TeV}
& $60.322(6)^{     +3.4 \%}_{     -4.3 \%}$
& $106.15(1)^{     +3.6 \%}_{     -3.0 \%}$ 
& \multirow{ 2}{*}{---}
& $120.5(1)^{     +2.0 \%}_{     -1.9 \%}$ 
& \multirow{ 4}{*}{\vspace{-0.25cm} 2.35}
\\
&
&
& {\footnotesize ($+76.0\%$)} 
& 
& {\footnotesize ($+13.5\%$)}
& 
\\[0.25cm]
& \multirow{ 2}{*}{$\sqrt{s}=27$\,TeV}
& $136.66(1)^{     +6.8 \%}_{     -7.8 \%}$
& $248.51(2)^{     +4.0 \%}_{     -3.3 \%}$
& \multirow{ 2}{*}{---}
& $283.4(3)^{     +2.1 \%}_{     -2.1 \%}$
&
\\
&
&
& {\footnotesize ($+81.8\%$)}
&
& {\footnotesize ($+14.0\%$)}
&
\\
\midrule
\multirow{ 4}{*}{\vspace{-0.25cm}$W^-Z${}}
& \multirow{ 2}{*}{$\sqrt{s}=14$\,TeV}
& $39.182(4)^{     +3.7 \%}_{     -4.7 \%}$
& $68.430(7)^{     +3.7 \%}_{     -3.0 \%}$ 
& \multirow{ 2}{*}{---}
& $77.63(7)^{     +1.9 \%}_{     -1.9 \%}$
& \multirow{ 4}{*}{\vspace{-0.25cm} 2.57}
\\
&
&
& {\footnotesize ($+74.6\%$)}
& 
& {\footnotesize ($+13.4\%$)}
&
\\[0.25cm]
& \multirow{ 2}{*}{$\sqrt{s}=27$\,TeV}
& $96.70(1)^{     +7.2 \%}_{     -8.2 \%}$
& $175.44(2)^{     +4.0 \%}_{     -3.3 \%}$
& \multirow{ 2}{*}{---}
& $199.7(2)^{     +2.0 \%}_{     -2.0 \%}$
&
\\
&
&
& {\footnotesize ($+81.4\%$)}
& 
& {\footnotesize ($+13.8\%$)}
&
\\
\midrule
\multirow{ 4}{*}{\vspace{-0.25cm}\zz{}}
& \multirow{ 2}{*}{$\sqrt{s}=14$\,TeV}
& $24.500(2)^{     +4.3 \%}_{     -5.3 \%}$
& $34.201(3)^{     +2.0 \%}_{     -1.8 \%}$ 
& $37.531(4)^{     +3.3 \%}_{     -2.6 \%}$
& $39.64(4)^{     +2.4 \%}_{     -2.1 \%}$ 
& \multirow{ 4}{*}{\vspace{-0.25cm} 2.40}
\\
&
&
& {\footnotesize ($+39.6\%$)}
& {\footnotesize ($+9.7\%$)}
& {\footnotesize ($+15.9 \%$)}
&
\\[0.25cm]
& \multirow{ 2}{*}{$\sqrt{s}=27$\,TeV}
& $58.622(6)^{     +7.9 \%}_{     -8.9 \%}$
& $79.757(8)^{     +2.2 \%}_{     -3.0 \%}$ 
& $89.89(1)^{     +3.7 \%}_{     -3.0 \%}$ 
& $95.20(9)^{     +2.9 \%}_{     -2.4 \%}$
&
\\
&
& 
& {\footnotesize ($+36.1\%$)}
& {\footnotesize ($+12.7\%$)}
& {\footnotesize ($+19.4\%$)}
&
\\
\bottomrule

\end{tabular}}
\end{center}
\renewcommand{\baselinestretch}{1.0}
\end{table}

\begin{table}[t]
\begin{center}
\renewcommand{\baselinestretch}{1.0}
\caption{\label{tab:dibosonXSpT100GeV} Cross sections with a $p_{\mathrm{T, min}}=100\,{\rm GeV}$ cut on the transverse momentum
of the charged leptons and the missing energy for \ww{}, \wz{} and \zz{} production.}
\resizebox{\columnwidth}{!}{%
\begin{tabular}{|c c | c c c c c|}
\hline
\multicolumn{2}{|c|}{$\sigma$ [fb]}
& LO
& NLO
& NLO$^\prime$$+gg$
& NNLO
& \multirow{2}{*}{\Large $\frac{\sigma_{\rm NNLO}({\rm 27\,TeV})}{\sigma_{\rm NNLO}({\rm 14\,TeV})}$}
\\
\multicolumn{2}{|c|}{\footnotesize (correction)}
& 
& {\footnotesize (NLO/LO$-$1)}
& {\footnotesize (NLO$^\prime$$+gg$/NLO$-$1)}
& {\footnotesize (NNLO/NLO$-$1)}
&
\\
\hline\hline
\multirow{ 4}{*}{\vspace{-0.25cm}\ww{}}
& 
\multirow{ 2}{*}{$\sqrt{s}=14$\,TeV}
& $     0.920(1)^{     +2.7 \%}_{     -2.7 \%}$
& $     2.827(5)^{     +9.7 \%}_{     -8.0 \%}$
& $     2.793(7)^{     +9.9 \%}_{     -8.1 \%}$
& $      3.51(1)^{     +5.2 \%}_{     -5.0 \%}$
& \multirow{ 4}{*}{\vspace{-0.25cm} 3.93}
\\
&
&
& {\footnotesize ($+207.1 \%$)} 
& {\footnotesize ($  -1.2 \%$)}
& {\footnotesize ($ +24.3 \%$)}
&
\\[0.25cm]
& \multirow{ 2}{*}{$\sqrt{s}=27$\,TeV}
& $     2.847(3)^{    +0.08 \%}_{     -0.5 \%}$
& $     10.83(2)^{     +8.2 \%}_{     -6.9 \%}$
& $     10.66(2)^{     +8.4 \%}_{     -7.1 \%}$
& $     13.80(4)^{     +5.3 \%}_{     -4.8 \%}$
&
\\
&
&
& {\footnotesize ($+280.5 \%$)}
& {\footnotesize ($  -1.6 \%$)}
& {\footnotesize ($ +27.3 \%$)}
&
\\
\midrule
\multirow{ 4}{*}{\vspace{-0.25cm}$W^+Z${}}
& \multirow{ 2}{*}{$\sqrt{s}=14$\,TeV}
& $   0.06524(8)^{     +3.3 \%}_{     -3.2 \%}$
& $    0.1273(3)^{     +7.1 \%}_{     -5.8 \%}$
& \multirow{ 2}{*}{---}
& $    0.1485(9)^{     +3.4 \%}_{     -3.3 \%}$ 
& \multirow{ 4}{*}{\vspace{-0.25cm} 3.82}
\\
&
&
& {\footnotesize ($ +95.2 \%$)} 
& 
& {\footnotesize ($ +16.6 \%$)}
&
\\[0.25cm]
& \multirow{ 2}{*}{$\sqrt{s}=27$\,TeV}
& $    0.1919(2)^{     +0.1 \%}_{     -0.5 \%}$
& $    0.4642(8)^{     +7.0 \%}_{     -5.8 \%}$
& \multirow{ 2}{*}{---}
& $     0.568(3)^{     +3.8 \%}_{     -3.6 \%}$
&
\\
&
&
& {\footnotesize ($+141.9 \%$)}
&
& {\footnotesize ($ +22.5 \%$)}
&
\\
\midrule
\multirow{ 4}{*}{\vspace{-0.25cm}$W^-Z${}}
& \multirow{ 2}{*}{$\sqrt{s}=14$\,TeV}
& $   0.03289(4)^{     +3.1 \%}_{     -3.1 \%}$
& $    0.0641(2)^{     +7.5 \%}_{     -6.0 \%}$
& \multirow{ 2}{*}{---}
& $    0.0767(5)^{     +3.4 \%}_{     -3.5 \%}$
& \multirow{ 4}{*}{\vspace{-0.25cm} 4.34}
\\
& 
& 
& {\footnotesize ($ +94.9 \%$)}
& 
& {\footnotesize ($ +19.7 \%$)}
&
\\[0.25cm]
& \multirow{ 2}{*}{$\sqrt{s}=27$\,TeV}
& $    0.1121(1)^{      +0. \%}_{     -0.3 \%}$
& $    0.2719(5)^{     +7.2 \%}_{     -5.9 \%}$
& \multirow{ 2}{*}{---}
& $     0.333(2)^{     +3.7 \%}_{     -3.5 \%}$
&
\\
&
&
& {\footnotesize ($+142.7 \%$)}
& 
& {\footnotesize ($ +22.5 \%$)}
&
\\
\midrule
\multirow{ 4}{*}{\vspace{-0.25cm}\zz{}}
& \multirow{ 2}{*}{$\sqrt{s}=14$\,TeV}
& $   0.02108(3)^{     +3.1 \%}_{     -3.1 \%}$
& $    0.0318(1)^{     +3.8 \%}_{     -3.2 \%}$
& $    0.0342(1)^{     +5.4 \%}_{     -4.3 \%}$
& $    0.0371(3)^{     +3.6 \%}_{     -3.0 \%}$
& \multirow{ 4}{*}{\vspace{-0.25cm} 3.70}
\\
&
&
& {\footnotesize ($ +50.6 \%$)}
& {\footnotesize ($  +7.7 \%$)}
& {\footnotesize ($ +16.9 \%$)}
&
\\[0.25cm]
& \multirow{ 2}{*}{$\sqrt{s}=27$\,TeV}
& $    0.0675(1)^{      +0. \%}_{     -0.2 \%}$
& $    0.1100(3)^{     +3.5 \%}_{     -2.8 \%}$
& $    0.1235(3)^{     +5.4 \%}_{     -4.3 \%}$
& $    0.1371(7)^{     +4.3 \%}_{     -3.5 \%}$
&
\\
&
& 
& {\footnotesize ($ +62.9 \%$)}
& {\footnotesize ($ +12.3 \%$)}
& {\footnotesize ($ +24.7 \%$)}
&
\\
\hline
\end{tabular}}
\end{center}
\end{table}

In Table~\ref{tab:dibosonXS} cross sections are presented for \ww{}, \wz{} and \zz{} production,
inclusive over the phase space of the final-state leptons, for $pp$ collisions at $\sqrt{s}=14$\,TeV and $\sqrt{s}=27$\,TeV.
Throughout, only a basic selection cut on $Z$ bosons is applied, by requiring the invariant masses
of all opposite-sign same-flavour lepton pairs to be within a $Z$-mass window of $66\GeV<m_{\mathrm{\ell^-\ell^+}}<116\GeV$, 
which is necessary to avoid divergencies induced by soft intermediate photons. The gain in the inclusive cross section at $\sqrt{s}=27$\,TeV is 
roughly a factor of $2.5$ for all processes under consideration, see last
column of Table~\ref{tab:dibosonXS}. 
The importance of QCD corrections is seen:
Higher-order contributions are huge, especially for \wz{} production.
The NLO corrections
range from about $+36\%$ to $+82\%$ depending on process and collider energy, while 
NNLO QCD corrections are still sizeable and induce a further increase of the cross sections of $13\%$ to $20\%$.
The cross-section ratio for $W^+Z$/$W^-Z$ production is about $1.55$ at NNLO for $\sqrt{s}=14$\,TeV, changes to
$1.42$ for $\sqrt{s}=27$\,TeV, and is essentially independent on the perturbative order.

It should be stressed that QCD radiative corrections may change quite significantly 
as soon as fiducial cuts on the leptonic final state are applied, or when 
kinematical distributions are considered. The corrections for the 
inclusive cross sections in Table~\ref{tab:dibosonXS} should therefore be understood as illustrative, and the use of inclusive $K$-factors
to obtain NNLO predictions from lower order results with different sets of cuts should be avoided in general.

It is interesting to quantify the size of the loop-induced gluon fusion contribution of the charge-neutral processes, 
which is part of the NNLO QCD corrections.
By NLO$^\prime$+$gg$ its sum is denoted with the NLO cross section computed with NNLO PDFs.
The NLO$^\prime$+$gg$ result for \ww{} production is $6.4\%$ ($7.0\%$) larger than the NLO result at $\sqrt{s}=14\,(27)$\,TeV, while 
their difference is even $9.7\%$ ($12.7\%$)  for \zz{} production.
These numbers amount to roughly half of the full NNLO correction of the \ww{}
process, and even about two-thirds for \zz{} production.
However, one has to bear in mind that under typical fiducial selection requirements on 
the leptons and missing transverse energy, the impact of the loop-induced 
contribution decreases significantly, especially for \ww{} production.
Furthermore, its relative contribution is strongly suppressed as far as 
the tails of the kinematical distributions are concerned, due to the large-$x$ suppression
of the gluon density.

To illustrate how strongly the radiative corrections may depend on the fiducial cuts, in Table~\ref{tab:dibosonXSpT100GeV} cross sections are shown
with a minimum $p_{\mathrm{T, min}}=100$\,GeV cut on the transverse momentum
of the charged leptons and the missing energy. More precisely, depending on the process the following cuts have been applied, as shown in Table \ref{tab:dibosoncuts}.

\begin{table}
\begin{center}
\renewcommand\arraystretch{1.5}
\caption{Selection cuts applied in the analysis for the different processes.}
\label{tab:dibosoncuts}
\begin{tabular}{|l|c|c|c|}
  \hline
& \ww{} & \wz{} & \zz{} \\
  \hline\hline
{lepton cuts} &
  $p_{\mathrm{T,\ell_{1/2}}}>p_{\mathrm{T, min}}$ &
  $p_{\mathrm{T,\ell_{1/2/3}}}>p_{\mathrm{T, min}}$ &
  $p_{\mathrm{T,\ell_{1/2/3/4}}}>p_{\mathrm{T, min}}$ \\
{neutrino cuts} &
  $p_{\mathrm{T,miss}}>p_{\mathrm{T, min}}$ &
  $p_{\mathrm{T,miss}}>p_{\mathrm{T, min}}$ &
  ---
  \\
  \hline
\end{tabular}
\end{center}
\end{table}

As can be read from Tables~\ref{tab:dibosonXS} and~\ref{tab:dibosonXSpT100GeV}, radiative corrections at NLO can be enormous for some processes with $p_{\mathrm{T, min}}=100$\,GeV,
ranging from $+51\%$ to even $+281\%$. Also the NNLO corrections are significantly increased with respect to the inclusive case,
and can be as large as $+27\%$. It is also apparent that the importance of the loop-induced gluon fusion contribution
is significantly reduced. For \ww{} production, due to the applied $p_{\mathrm{T,miss}}$ cut the NLO$^\prime$+$gg$ contribution is even smaller than the NLO cross section
by $-1.2\%$ ($-1.6\%$) at $\sqrt{s}=14\,(27)$\,TeV (i.e.\ the positive impact of the $gg$ channel is smaller than the negative effect
from using NNLO PDFs instead of NLO PDFs in the NLO$^\prime$+$gg$ prediction). For \zz{} production, it is still sizeable with $7.7\%$ ($12.3\%$), but its 
relative contribution at $\mathcal{O}(\as^2)$ has decreased from roughly two-thirds in the inclusive case to less than half
of the NNLO corrections for $p_{\mathrm{T, min}}=100$\,GeV. Furthermore, compared to the inclusive results
 an even more substantial increase of the cross sections is observed from $\sqrt{s}=14$\,TeV to $\sqrt{s}=27$\,TeV of roughly a factor of four.
This can be understood by the fact, that the additional energy enlarges the available phase-space, especially at high momentum transfer.

From the results in Table~\ref{tab:dibosonXS} and \ref{tab:dibosonXSpT100GeV} 
it is clear that the perturbative uncertainties at NLO cannot account for the additional loop-induced 
gluon fusion contribution that appears at NNLO. Besides that, also the genuine NNLO corrections to the quark--antiquark production mechanism
cannot be anticipated from NLO scale variations, which in turn means that the NLO uncertainties are underestimated.
The inclusion of NNLO corrections is therefore crucial. At this order all partonic channels 
are included for the first time, and scale variations can be used to obtain an estimate of the actual size of missing 
higher-order terms. However, the NLO corrections to the loop-induced gluon fusion contribution are relevant and should be included when possible,
especially at $\sqrt{s}=27$\,TeV where gluons with smaller $x$ are probed.
In particular in tails of high-energy observables, the inclusion of
NLO EW corrections and their interplay with QCD corrections will also need to be investigated.
Nevertheless NNLO QCD results are presented in the following, but the above-mentioned extensions
will become available well before the start of the HL-LHC.

The differential results in diboson processes in light of the HL and HE upgrades of 
the LHC are now discussed. Since the importance of highest-order predictions is evident from the previous discussion, only NNLO QCD accurate results are presented here. The cumulative cross section with a minimum $p_{\mathrm{T, min}}$ cut, as introduced above is considered first.
In order to analyse the number of expected events as a function of $p_{\mathrm{T, min}}$, the  
cross sections have been translated into event numbers by assuming an integrated luminosity of $3$\,ab$^{-1}$ at $14\,$TeV and of $15$\,ab$^{-1}$ at $27$\,TeV.

Figure~\ref{fig:ptmin_diboson} shows the expected number of events as a function of $p_{\mathrm{T, min}}$. 
Since the transverse momentum of all leptonic final states are restricted simultaneously, the reach in the tails may appear smaller 
than expected, and would be significantly larger if a cut on the transverse momentum of only the leading lepton
or the missing energy were to be considered.
However, the toy scenario considered is well suited to compare the three diboson production processes, and to 
quantify the relative gain of the additional energy and luminosity.

The curves in Fig.~\ref{fig:ptmin_diboson} show all production processes under consideration:
\begin{figure}
\begin{center}
\includegraphics[trim = 10mm 0mm 0mm 0mm, width=.5\textheight]{\main/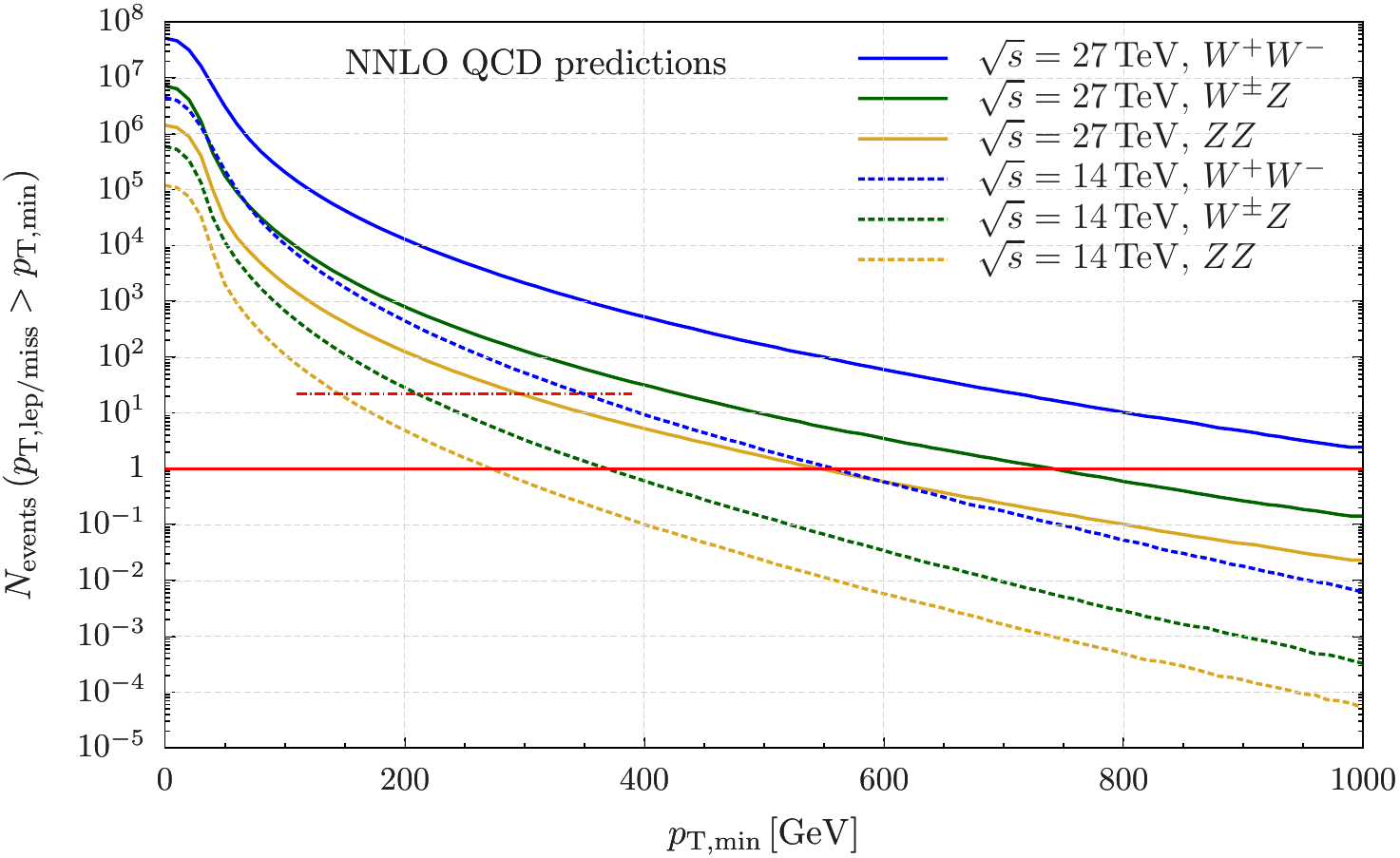}
\caption[]{{
\label{fig:ptmin_diboson}
Cumulative number of events as a function of $ p_{\mathrm{T, min}}$ for the following production processes: \ww{} (blue), \wz{} (green), and \zz{} (orange); at $14$\,TeV (dashed) and $27$\,TeV (solid).}}
\end{center}
\end{figure}
\ww{} (blue), \wz{} (green), and \zz{} (orange); at $14$\,TeV (dashed) and $27$\,TeV (solid). The horizontal red line shows the one-event threshold, below which no events are expected anymore.
The following features are evident in the plot: At $\sqrt{s}=14$\,TeV events up to $p_{\mathrm{T, min}}$ values of roughly
$550$\,GeV, $370$\,GeV, and $270$\,GeV are expected for \ww{}, \wz{}, and \zz{} production, respectively.
At $\sqrt{s}=27$\,TeV these values read $>${}$1000$\,GeV, $740$\,GeV, and $550$\,GeV.
To put these numbers into perspective, a dash-dotted red line for the present status at the end of Run-2 is added, which 
represents the one-event threshold for $150\,{\rm fb^{-1}}$ at $13\,{\rm TeV}$ 
($14\,{\rm TeV}\to13\,{\rm TeV}$ conversion approximated by a constant cross-section correction factor of $0.9$). 
Its intersection points with the $\sqrt{s}=14$\,TeV curves
  indicates the current reach of the LHC, which is roughly up to
  $350$\,GeV, $210$\,GeV, and $140$\,GeV for \ww{}, \wz{}, and \zz{} production, respectively.
The improved reach in the tails at $27$\,TeV is not only related to the larger inclusive cross section and higher luminosity, but also the enlarged phase-space available with higher energies plays an important role: Whereas the solid curves fall only by $7-8$ orders of magnitude in the range of $0\,{\rm GeV} \le p_{\mathrm{T, min}}\le 1000$\,GeV,
the dashed $14$\,TeV curves fall by more than $9$ orders of magnitude in the same region. This also explains why the $14$\,TeV \ww{} result, which has a much larger 
inclusive cross section, crosses the red one-event line at almost the same point as the $27$\,TeV \zz{} result.

In Fig.~\ref{fig:mcut_diboson} the reach of the three vector-boson 
pair production processes is considered for future LHC upgrades 
in the invariant-mass distributions of all produced charged leptons.
\begin{figure}
\begin{center}
\includegraphics[trim = 10mm 0mm 0mm 0mm, width=.5\textheight]{\main/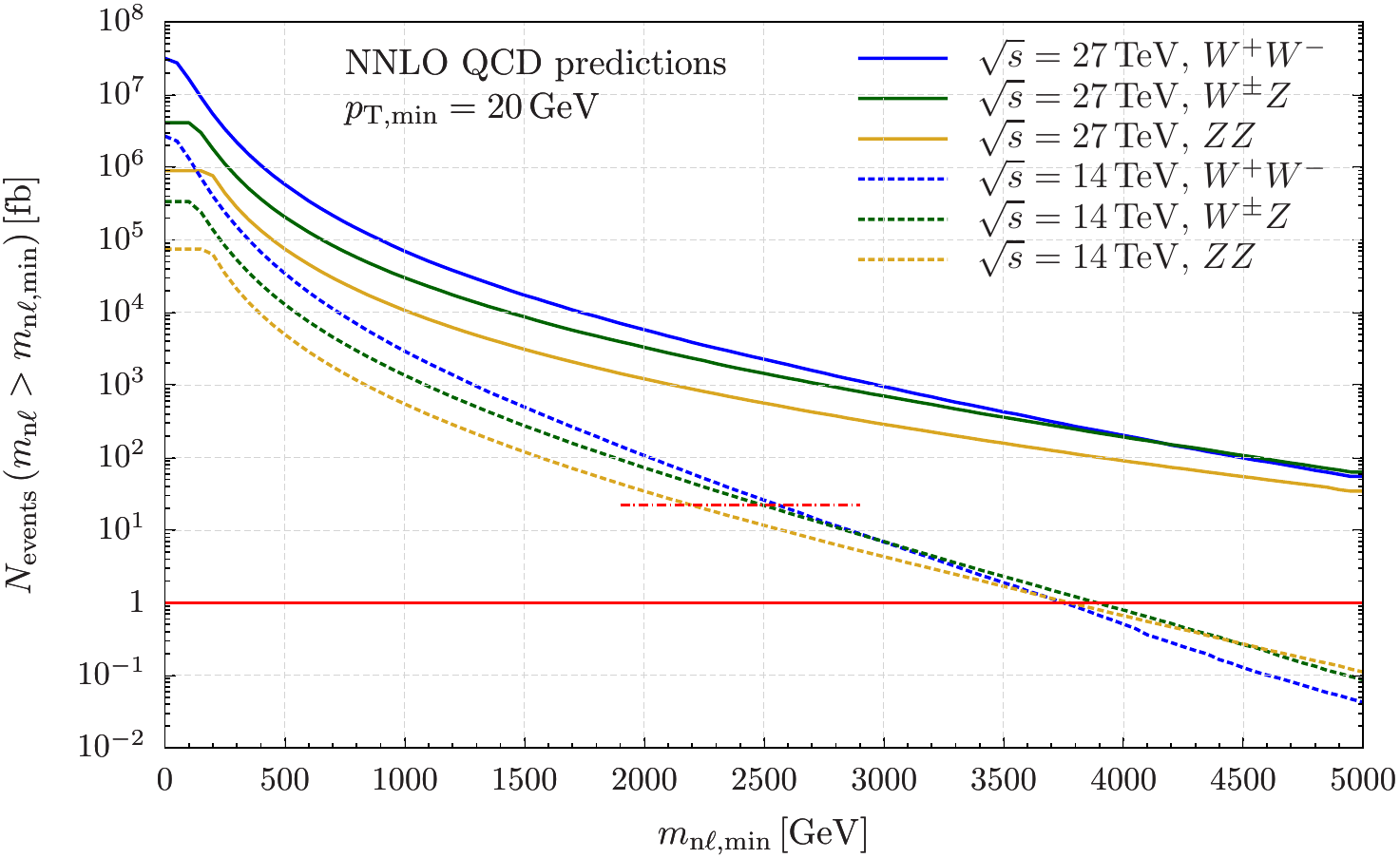}
\caption[]{{
\label{fig:mcut_diboson}
Cumulative number of events as a function of $m_{n\ell,{\rm min}}$ for the following production processes: \ww{} (blue), \wz{} (green), and \zz{} (orange); at $14$\,TeV (dashed) and $27$\,TeV (solid).}}
\end{center}
\end{figure}
A scenario is chosen where $p_{\mathrm{T, min}}$, defined as before in the 
three processes, is $20$\,GeV in order to have at least a rough definition of the fiducial phase-space.
The expected number of events, assuming the same integrated luminosities
as stated above, is shown for $\sqrt{s}=14$\,TeV (dashed) and $\sqrt{s}=27$\,TeV (solid) 
with a lower cut $m_{n\ell}>m_{n\ell,{\rm min}}$, where $n$ is the number of leptons
in the respective process, i.e., for \ww{} production it is the 
distribution in $m_{2\ell}$ (blue), for \wz{} it is the one in $m_{3\ell}$ (green),
and for \zz{} in $m_{4\ell}$ (orange). The significant reach in energy for both 
the HL run of the LHC and a potential HE upgrade is evident, where ``reach" refers to the point where the curves cross the red horizontal one-event threshold.
A resonance in the tails of the invariant masses of 
two leptons (plus missing transverse momentum) or of four leptons is
indeed a realistic signature predicted by many BSM theories. While with the current
Run-2 data (red, dash-dotted line crossing the $14$\,TeV results) searches can hardly pass the two TeV frontier, future LHC upgrades 
will probe mass scales of a few TeV at $14$\,TeV with $3$\,ab$^{-1}$, 
or potentially even up to ten TeV at $27$\,TeV with $15$\,ab$^{-1}$.
It is also apparent that despite $\sigma_{\ww{}}\gg \sigma_{\wz{}}\gg \sigma_{\zz{}}$ holding inclusively, the point where the three lines fall below one event is much 
closer. This is simply caused by the fact that the phase space of the 
four-lepton system in \zz{} production 
is larger than the one of the three-lepton system in \wz{} production, where
some energy is taken by the additional neutrino. An analogous interpretation applies
to \ww{} production. Furthermore, also here the significantly 
enlarged phase space induced by the increase in energy at $27$\,TeV is evident:
The $27$\,TeV results drop by roughly $4-5$ orders of magnitude in the displayed 
range, while the $14$\,TeV ones drop by more than $6$ orders.

The study is continued by analysing the importance of the additional 
fiducial phase space that becomes available with detector upgrades to 
enlarge the accessible rapidity range of charged leptons. Since very similar results were found for \ww{}, \wz{} and \zz{} 
production in that respect,  in Fig.~\ref{fig:rapidity_diboson} the rapidity efficiency of the four-lepton signature for \zz{} production only is shown.
\begin{figure}
\begin{center}
\includegraphics[trim = 10mm 0mm 0mm 0mm, width=.5\textheight]{\main/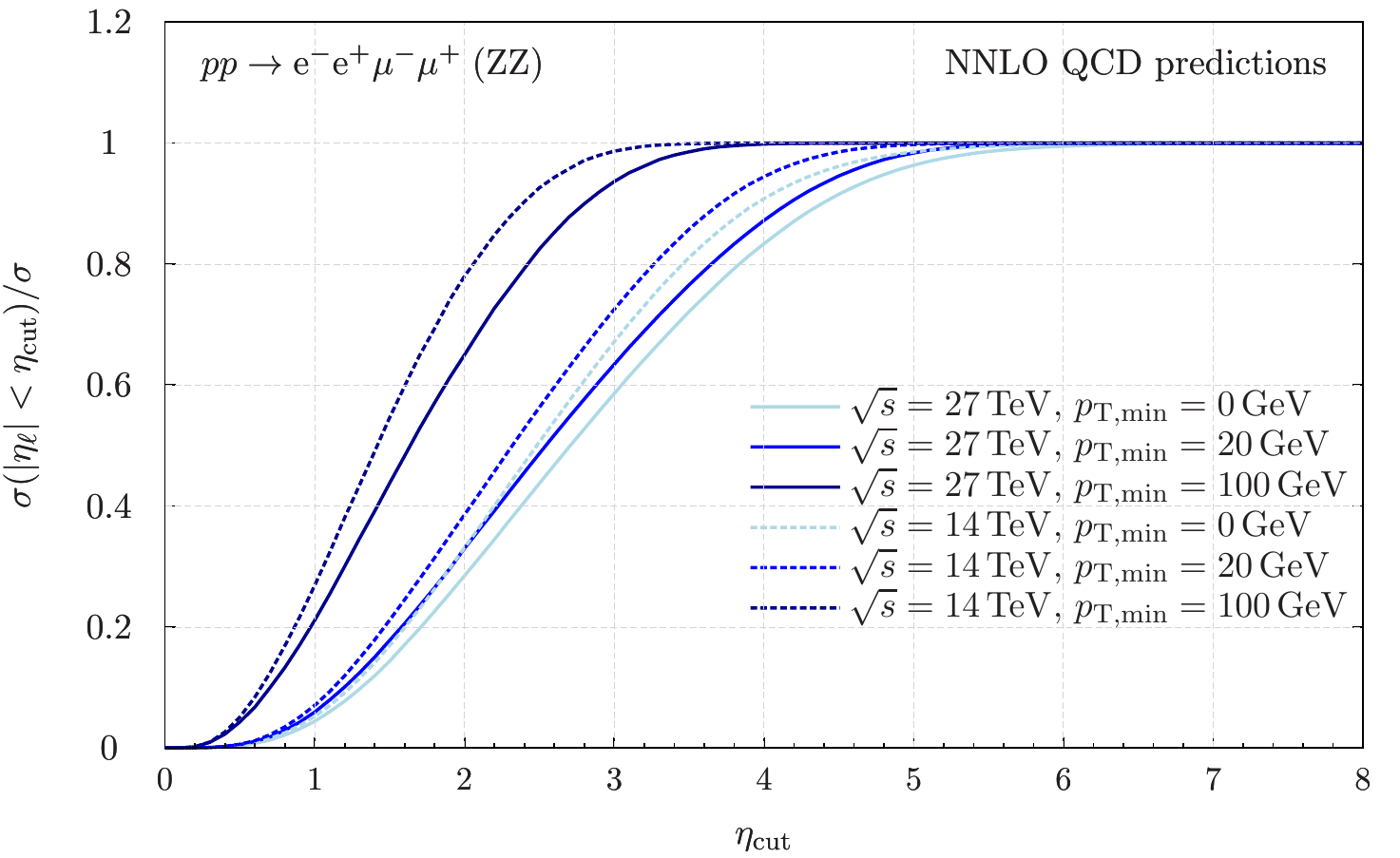}
\caption[]{{
\label{fig:rapidity_diboson}
Rapidity efficiency of the charged leptons.}}
\end{center}
\end{figure}
The rapidity efficiency is defined as the ratio of the cross section with an absolute-rapidity cut $\eta_{\rm cut}$ on all four charged leptons, divided by the 
inclusive cross section. As for $\eta_{\rm cut}\rightarrow \infty$ no cut is applied, the ratio tends to unity for large $\eta_{\rm cut}$ values.
The efficiency as a function of $\eta_{\rm cut}$ is studied for three $p_{\mathrm{T, min}}$ scenarios: inclusive (light blue), $p_{\mathrm{T, min}} = 20$\,GeV (blue), and $p_{\mathrm{T, min}} = 100$\,GeV (dark blue); at $14$\,TeV (dashed) and $27$\,TeV (solid). It is directly observed that the efficiency 
decreases with the machine energy. In other words, a small rapidity threshold at $27$\,TeV results in a much larger (relative) reduction
of the cross section than at $14$\,TeV. This is because the additional energy induces more forward (and boosted) leptons, and it shows that 
detector upgrades that enlarge the measurable rapidity range become even more important at the HE LHC.
Requiring minimum transverse-momentum cuts, on the other hand, has the effect
of increasing the rapidity efficiency, which is particularly striking for 
$p_{\mathrm{T, min}} = 100$\,GeV. The reason for this is simple: Leptons 
with high transverse momentum are predominantly produced at central rapidities.

The scenario with $p_{\mathrm{T, min}} = 20$\,GeV
provides the most realistic fiducial setup, which is actually not much 
different from the fully inclusive case, and is discussed here.
Typical rapidity cuts on charged leptons with the current LHC detectors are 
of the order of $\eta_\ell = 2.5$. Future detector upgrades for the HL
phase of the LHC can be expected to reach rapidities at the level of $\eta_\ell = 4$.
At $14\,(27)$\,TeV this would allow us to improve measurements of fiducial cross 
from a $<$$60\%$ ($\sim 50\%$) efficiency 
for $\eta_{\rm cut} = 2.5$ to a $>$$90\%$ ($\lesssim 90\%$) efficiency for $\eta_{\rm cut} = 4$.
This implies that the available inclusive cross section will be hardly reduced 
by fiducial rapidity requirements anymore once the detectors have been upgraded. 
This statement holds even more when considering scenarios with boosted leptons: For
$p_{\mathrm{T, min}} = 100$\,GeV the efficiency is practically $100\%$ for 
$\eta_{\rm cut} = 4$.

\subsubsection[Projections for measurements of anomalous 3-gauge boson couplings]{Projections for measurements of anomalous 3-gauge boson couplings\footnote{Contribution by J.~Baglio, S.~Dawson and I.~ M.~Lewis.}}


The sensitivity of the production of $W^+W^-$ pairs to anomalous gauge boson and anomalous fermion couplings at future LHC upgrades is now discussed. The $SU(2)\times U(1)$ structure of the electroweak sector of the Standard Model  determines the $W^+W^-V$ interactions ($V=\gamma, Z $). 
The amplitudes for the production of $W^+W^-$ pairs involve subtle cancellations between 
contributions that grow with energy, so  the pair production of gauge bosons is extremely 
sensitive to new physics interactions. Assuming C and P conservation, the most general Lorentz 
invariant $3-$gauge boson couplings can be written as in Ref.~\cite{Gaemers:1978hg,Hagiwara:1986vm}
\begin{eqnarray}
 \La_{V}&=&
-ig_{WWV}\biggl\{(1+\delta g_1^V)\left(W^+_{\mu\nu}W^{-\mu}V^\nu-W_{\mu\nu}^-W^{+\mu}V^\nu\right)+(1+\delta\kappa^V)W^+_\mu
            W^-_\nu V^{\mu\nu})\nonumber \\ &&+\frac{\lambda^V}{M^2_W}W^+_{\rho\mu}{W^{-\mu}}_\nu V^{\nu\rho}\biggr\},
\label{eq:lagdef}
\end{eqnarray}  
where $V=\gamma, Z$, $g_{WW\gamma}=e$, $g_{WWZ}=g \cos\theta_W$, 
 $s_W^{} \equiv \sin\theta_W^{}$ , $c_W^{} \equiv
\cos\theta_W^{}$, and in the SM, $\delta g_1^V = \delta\kappa_{}^V
= \lambda_{}^V = 0$.  Because of gauge invariance,
  this form can  be translated into the language of effective
field theory, where $\delta g_1^V,~\delta\kappa^V,~\lambda^V\sim {v^2\over\Lambda^2}$, with $\Lambda$ the scale of
BSM physics, $\Lambda \gg v$.

The effective couplings of fermions to gauge fields are parameterised as,
\begin{eqnarray}
  \La&=&{g\over c_W} Z_\mu\biggl[g_L^{Zq}+\delta g_{L}^{Zq}\biggr]
  {\overline q}_L\gamma_\mu q_L\
 +g_ZZ_\mu\biggl[g_R^{Zq}+\delta g_{R}^{Zq}\biggr]
  {\overline q}_R\gamma_\mu q_R\nonumber \\
  &&+{g\over \sqrt{2}}\biggl\{W_\mu\biggl[(1+\delta g_{L}^W){\overline q}_L\gamma_\mu q_L^\prime
  +\delta g_R^W
  {\overline q}_R\gamma_\mu q_R^\prime\biggr] +h.c.\biggr\}\, ,
  \label{eq:dgdef}
  \end{eqnarray}
 where  $Q_q$ is the electric  charge of the quarks, and $q$ denotes up-type or down-type quarks. 
 The anomalous fermion couplings also scale as $\delta g_{L,R}^{Zq}, \delta g_{L,R}^W\sim{v^2\over\Lambda^2}$.
 The SM quark couplings are
$g_R^{Zq}=-s_W^2 Q_q$ and $g_L^{Zq}=T_3^q -s_W^2 Q_q$ with
 $T_3^q=\pm \displaystyle \frac{1}{2}$.  $SU(2)$ invariance relates
the coefficients,
$
\delta g_L^W=\delta g_L^{Zf}-\delta g_L^{Zf'}, \,
\delta g_1^Z= \delta \kappa_{}^Z+{s_W^2\over c_W^2}\delta \kappa_{}^\gamma$ and
$
\lambda_{}^\gamma = \lambda_{}^Z$,
where $f$ denotes up-type quarks and $f'$  down-type quarks.

The anomalous 3-gauge boson and fermion couplings have been implemented
into the {\mbox{\textsc{POWHEG BOX}}\xspace} framework\cite{Melia:2011tj, Nason:2013ydw, Baglio:2018bkm} for $W^+W^-$ production and samples of events are generated with $pp\rightarrow W^+W^-\rightarrow \mu^\pm e^\mp\nu\nu$.
Fits to 8~TeV data\cite{Baglio:2017bfe,Alves:2018nof} illustrate the importance of including  both anomalous fermion and 3-gauge boson couplings.
The sensitivity to anomalous couplings results almost entirely 
from contributions quadratic in the anomalous couplings and  the effects of anomalous 3-gauge boson and 
fermion couplings are numerically similar.

To probe the sensitivity to anomalous couplings, events are generated using the cuts
\begin{equation}
p_{\rm T}^{l}>30~{\rm GeV},\,\mid \eta^l\mid<2.5,\,m_{ll}>10 ~{\rm GeV}, \met >20~{\rm GeV}\,.
\end{equation}
These cuts are similar to those applied in the ATLAS\cite{Aad:2016wpd} and CMS\cite{Khachatryan:2015sga} extractions of anomalous coupling limits using the 8~TeV data. 
A hypothetical future systematic uncertainty of 
$\delta_{sys}=16\%$ is postulated and a cut on the $p_{\rm T}$ of the leading lepton applied such that the systematic error is smaller than the statistical error, $\delta_{stat}={1\over \sqrt{L\sigma(p_{l,T}^{lead}>p_{\rm T}^{cut})}}>\delta_{sys}$,where $L$ is the integrated luminosity.
The integrated cross section above a $p_{\rm T}^{cut}$ is evaluated, assuming a $50~\%$ efficiency and the cuts set as
\begin{eqnarray}
27~{\rm TeV}~{\text{with}}~15\abinv: &p_{\rm T}^{cut}=750~{\rm GeV},\qquad 
14~{\rm TeV}~{\text{with}}~3\abinv: p_{\rm T}^{cut}=1350~{\rm GeV}\, .
\end{eqnarray}
The calculations are performed at NLO QCD, using CT14qed-inc-proton PDFs, and the renormalisation/factorisation scales are taken to be to be $M_{WW}/2$. It is assumed the 
$Wl\nu$ couplings in the decays are SM-like.

The results of the scans are shown in Figs.~\ref{fig:atfig14} and \ref{fig:atfig27}; the 
allowed regions are within the ellipses. 
A significant improvement going from 14~TeV to 27~TeV is seen, while
the improvement from reducing the systematic error, $\delta_{sys}=0.16\rightarrow 0.04$, is marginal. The fermion couplings are allowed to vary around $0$, assuming the $2\sigma$ errors from fits to LEP
data.
As can be seen, by including the anomalous fermion couplings, the sensitivity of 
the scan is significantly reduced~\cite{Baglio:2017bfe, Grojean:2018dqj, Baglio:2018bkm}. 
This effect is quite pronounced at $27$~TeV and implies that global fits to both anomalous 
fermion and $3$ gauge boson couplings are necessary.

 \begin{figure}
  \centering
\subfloat{\includegraphics[width=0.32\textwidth]{\main/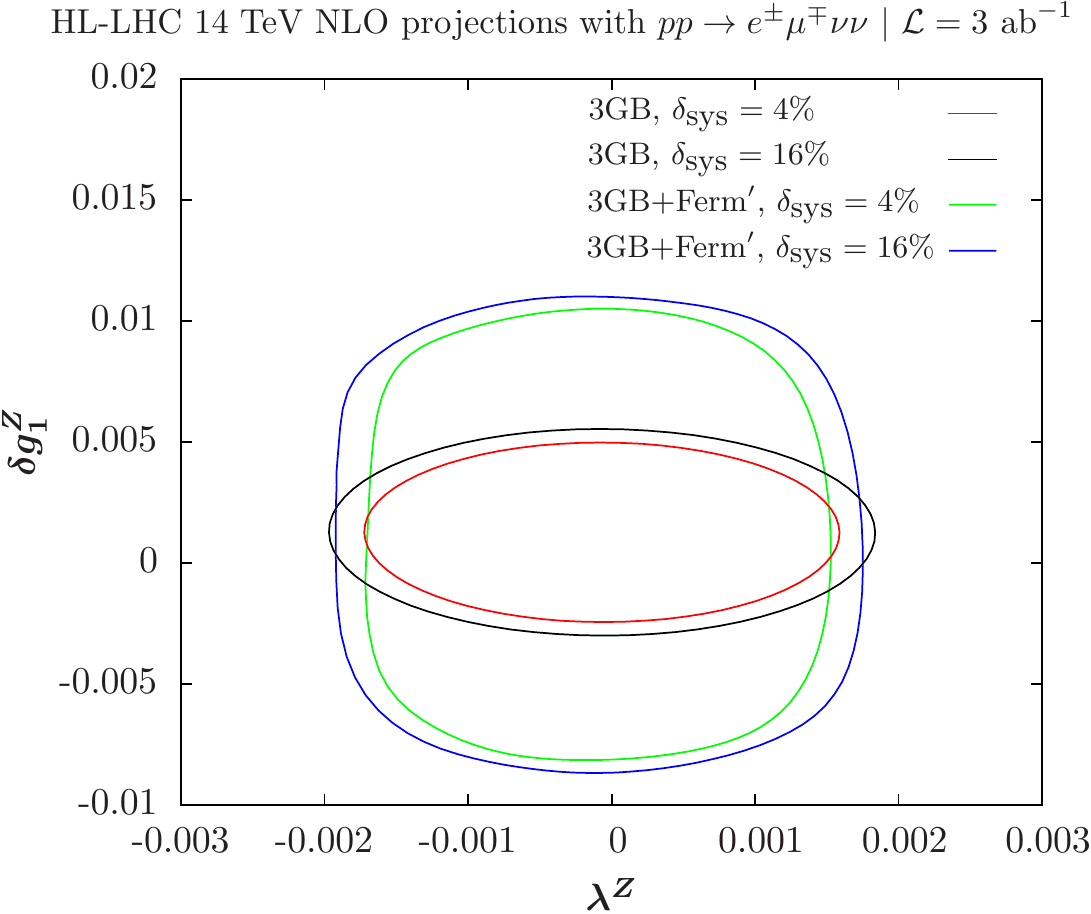}}
\subfloat{\includegraphics[width=0.32\textwidth]{\main/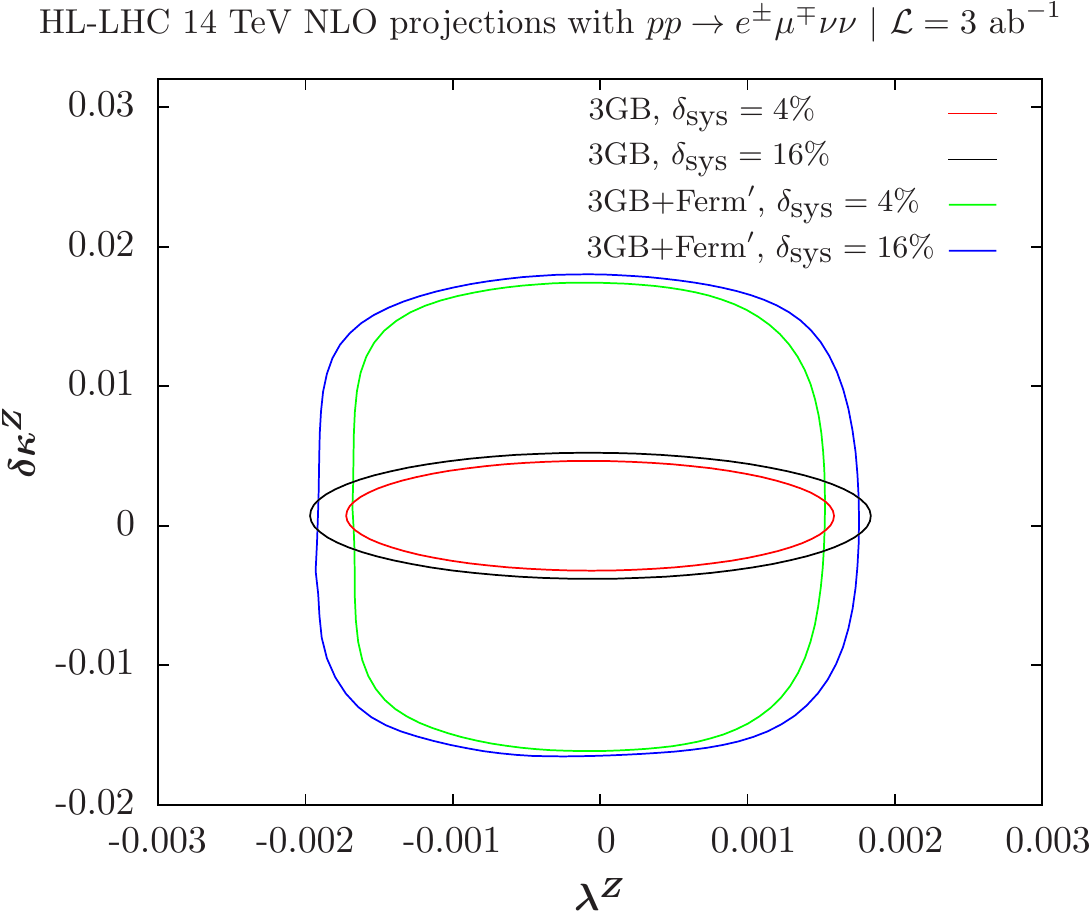}}
\subfloat{\includegraphics[width=0.32\textwidth]{\main/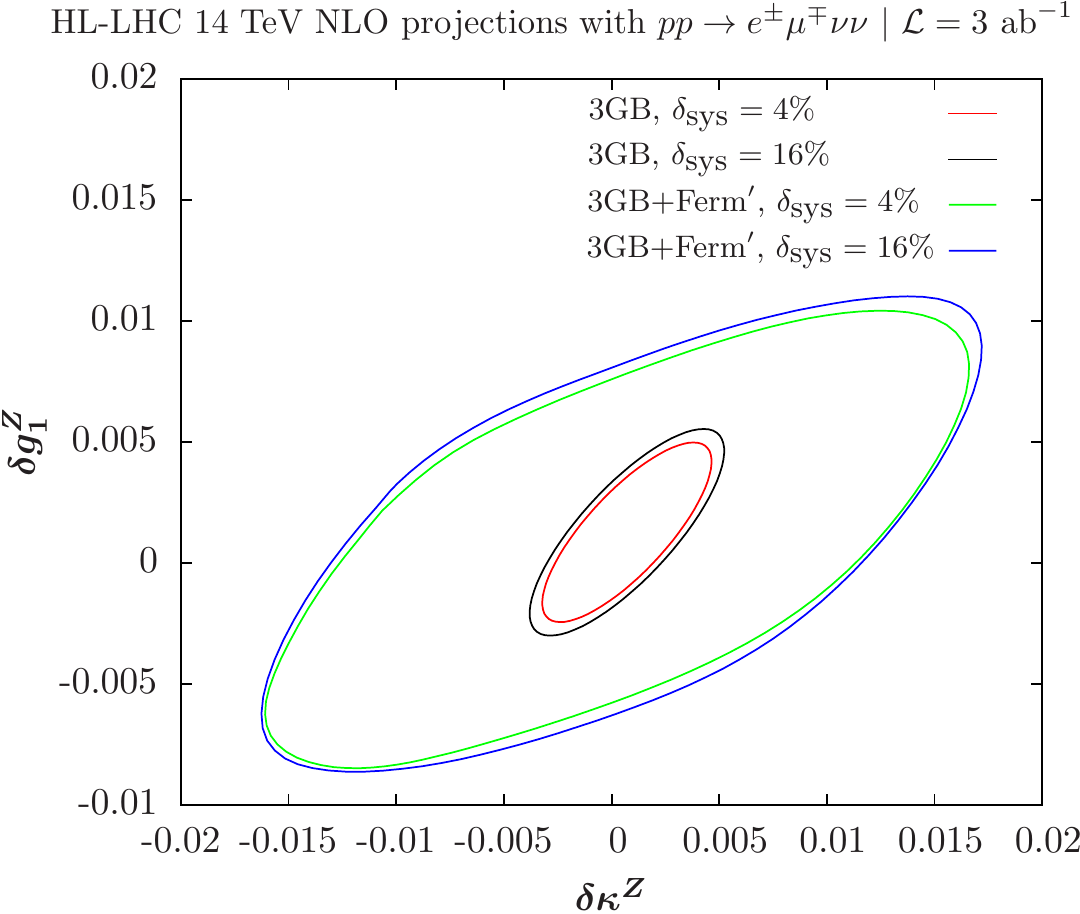}}
 \caption{Projections for $14$ TeV with 3\abinv.
 $p_{{\rm T},cut}=750$~GeV, corresponding to $\delta_{stat}=16\%$  with $\delta_{sys}=4\%$ and $\delta_{sys}=16\%$. The curves labelled 3GB have SM $Z$-fermion couplings, while the curves labelled 3GB +Ferm' allow the $Z$-fermion couplings to vary around a central value of $0$.}
   \label{fig:atfig14}
\end{figure}

\begin{figure}
  \centering
\subfloat{\includegraphics[width=0.32\textwidth]{\main/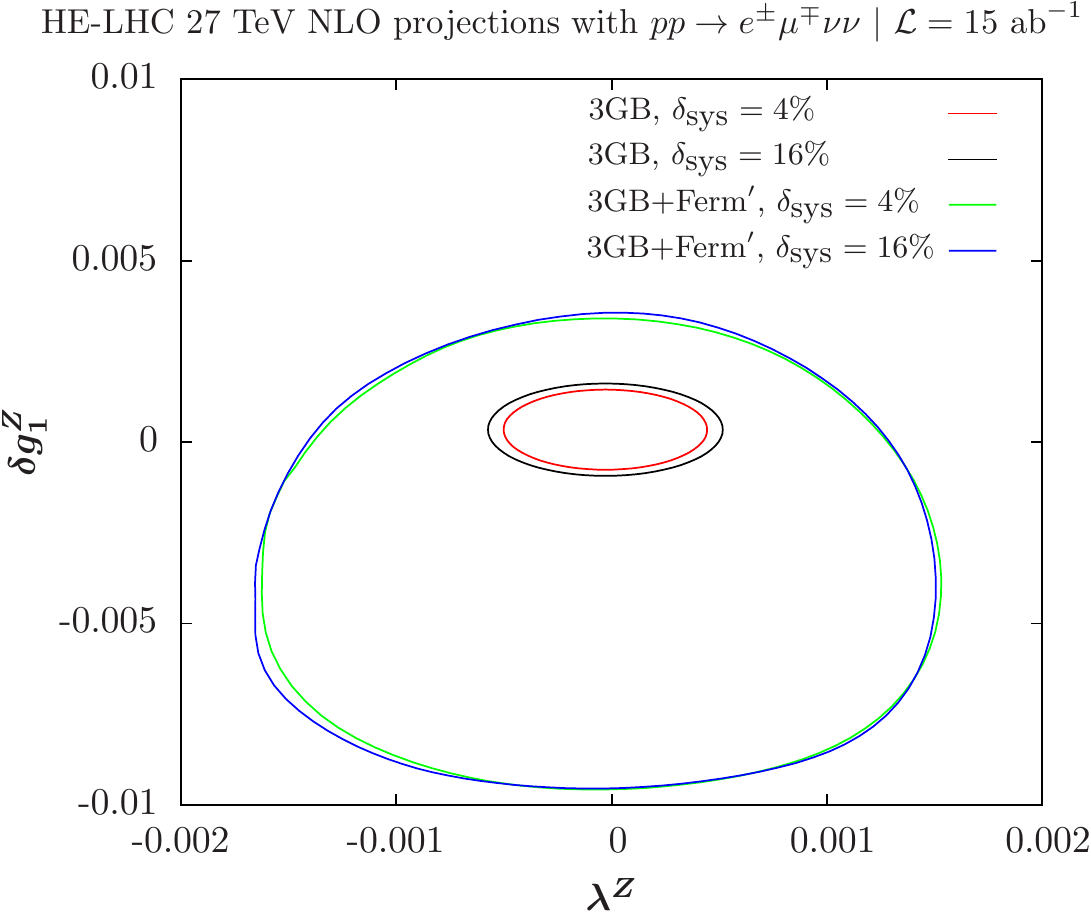}}
\subfloat{\includegraphics[width=0.32\textwidth]{\main/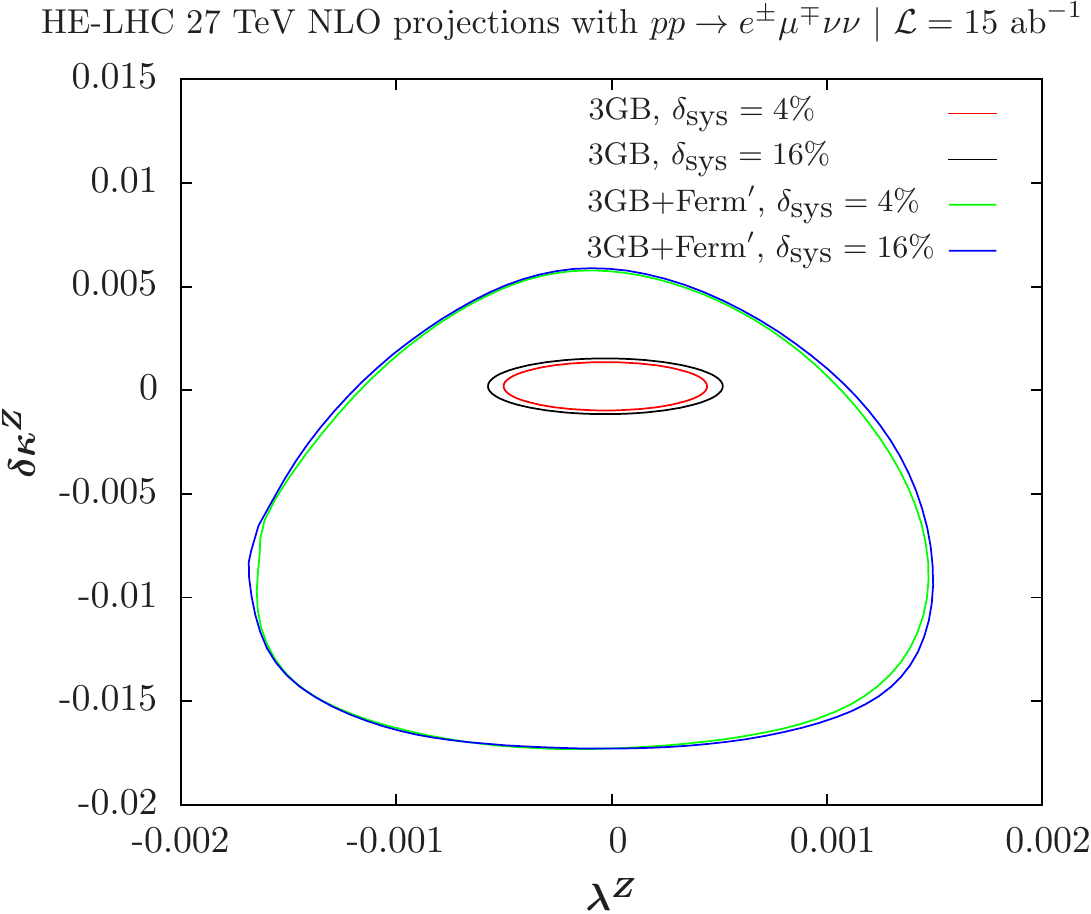}}
\subfloat{\includegraphics[width=0.32\textwidth]{\main/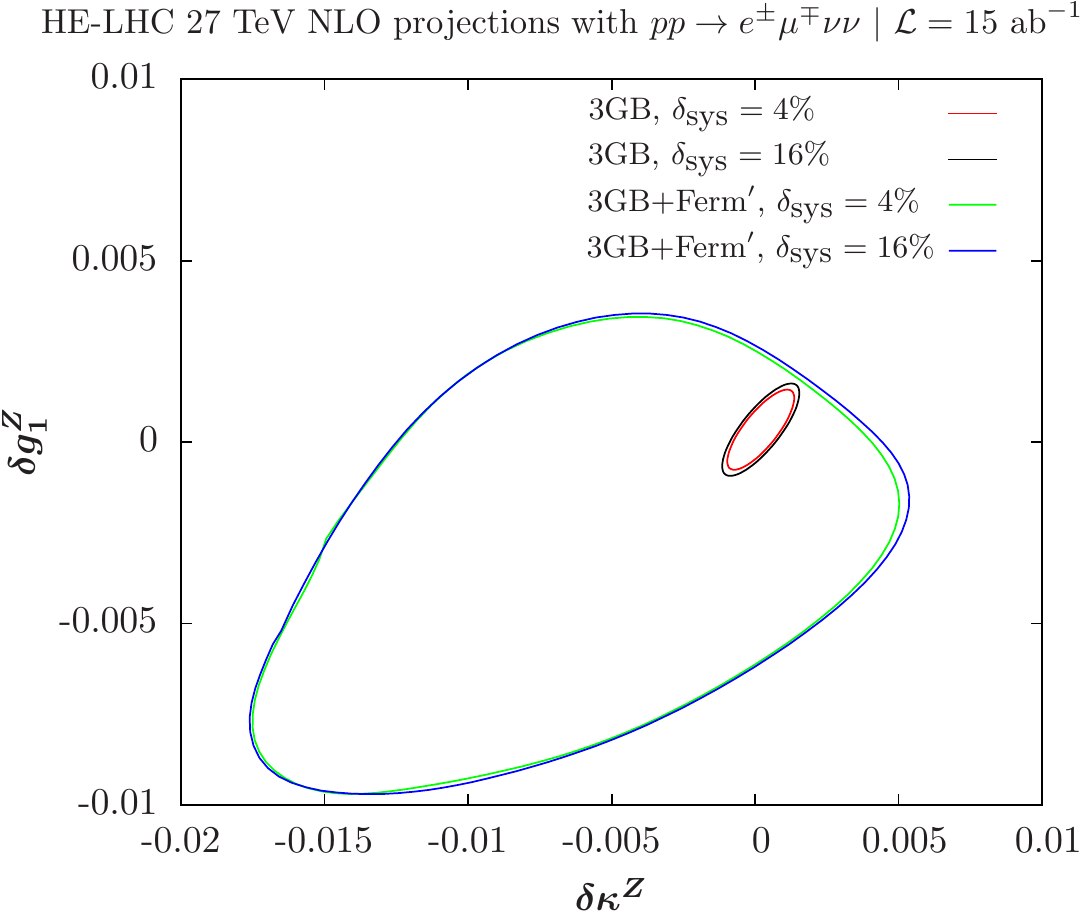}}
 \caption{Projections for $27$ TeV with 15\abinv.
 $p_{{\rm T},cut}=1350$~GeV, corresponding to $\delta_{stat}=16\%$  with $\delta_{sys}=4\%$ and $\delta_{sys}=16\%$. The curves labelled 3GB have SM $Z$-fermion couplings, while the curves labelled 3GB +Ferm' allow the $Z$-fermion couplings to vary around a central value of $0$. }
   \label{fig:atfig27}
\end{figure}

\subsubsection{Prospects for the measurement of the $W$-boson mass}
\label{sec:wmassprospects}



\newcommand{\avg}[1]{\ensuremath{\left< #1 \right>}}
\newcommand*{\pTl}{\ensuremath{\pT^{\ell}}\xspace}
\newcommand*{\vecpTl}{\ensuremath{\vecpT^{\ell}}\xspace}
\newcommand*{\pTel}{\ensuremath{\pT^{e}}\xspace}
\newcommand*{\pTmu}{\ensuremath{\pT^{\mu}}\xspace}
\newcommand*{\phil}{\ensuremath{\phi_{\ell}}\xspace}
\newcommand*{\phimiss}{\ensuremath{\phi_\text{miss}}\xspace}
\newcommand*{\mT}{\ensuremath{m_\text{T}}\xspace}
\newcommand*{\uT}{\ensuremath{u_\text{T}}\xspace}
\newcommand*{\pTW}{\ensuremath{\pT^{W}}\xspace}
\newcommand*{\pTZ}{\ensuremath{\pT^{Z}}\xspace}
\newcommand*{\pTV}{\ensuremath{\pT^{V}}\xspace}
\newcommand*{\vecpT}{\ensuremath{\vec{p}_{\text{T}}}\xspace}
\newcommand*{\vecmet}{\ensuremath{\vec{E}_{\text{T}}^{\text{miss}}}\xspace}
\newcommand*{\ipb}{\mbox{pb$^{-1}$}}
\newcommand*{\ifb}{\mbox{fb$^{-1}$}}

\setlength\parindent{0pt}                                                                                                                                                                                              

Special low pile-up proton-proton collision data at the HL-LHC (and HE-LHC) will be of large interest for $W$ boson physics. At $\sqrt{s}=14$~TeV and for an instantaneous luminosity of ${\cal L}\sim 5\times 10^{32}$~$\text{cm}^{-2} \text{s}^{-1}$, corresponding to two collisions per bunch crossing on average, about 2$\times 10^6$ $W$ boson events can be collected in one week. Such a sample provides a statistical sensitivity at the permille level for cross section measurements, at the percent level for measurements of the $W$ boson transverse momentum distribution, and of about 10~MeV for a measurement of $m_W$. The increased acceptance provided by the new inner detector in ATLAS, the ITk~\cite{ATLAS_Strip_TDR}, extends the coverage in pseudorapidity from $|\eta|<2.5$ to $|\eta|<4$ and allows further constraints on the parton density functions (PDFs) from cross section measurements, reducing the corresponding uncertainties in the measurement of $m_W$. An energy increase at the HE-LHC to $\sqrt{s}=27$~TeV~\cite{Zimmermann:2017bbr} could play a similar role. 
A first quantitative study of the potential improvement in the $W$-boson mass using low pile-up data at the HL-LHC and HE-LHC is discussed in~\cite{ATLAS:mWPubNote} considering only statistical and PDF uncertainties. Experimental systematic uncertainties can be maintained at a level similar to the statistical uncertainty, since they are largely dominated by the statistics of the low pile-up samples. Other theoretical uncertainties in the modelling of the $W$-boson production, like the description of the boson transverse momentum distribution, will also be constrained by measurements using these data.\\

Leptonic $W$ boson decays are characterised by an energetic, isolated electron or muon, and significant missing transverse momentum reflecting the decay neutrino. The hadronic recoil, \uT, is defined
from the vector sum of the transverse momenta of all reconstructed particles in the event excluding the charged lepton, and provides a measure of the $W$ boson transverse momentum. The lepton transverse momentum, \pTl, the missing transverse momentum, \met, and the hadronic recoil are related through $\vecmet = -(\vecpTl + \vec{\uT})$. The \pTl and \met distributions have sharp peaks at $\pTl\sim\met\sim m_W/2$. The transverse mass \mT, defined as $\mT=\sqrt{2 \pTl \met \cos(\phil-\phimiss)}$, peaks at $\mT\sim m_W$.

Events are generated at $\sqrt{s}=14$ and 27~TeV using the \textsc{W\_ew\_bmnnp} process~\cite{Barze:2012tt} of the {\mbox{\textsc{POWHEG}}\xspace} v1 event generator~\cite{Alioli:2010xd}, with electroweak corrections
switched off. The CT10 PDF set~\cite{Lai:2010vv} is used, and parton shower effects are included using the {\mbox{\textsc{PYTHIA}}\xspace} v8 event generator~\cite{Sjostrand:2014zea} with parameters set according to
the AZNLO tune~\cite{STDM-2012-23}. Final-state QED corrections are applied using \textsc{Photos}~\cite{Davidson:2010ew}. The energy resolutions of the lepton and hadronic recoil are parameterised as a function of the truth-related observables in order to emulate detector effects. These parameterised resolutions are checked against simulated distributions at the reconstructed level, and they agree at the level of a few percent. 

Events are selected by applying the following cuts to the object kinematics, after resolution corrections: $\pTl>25$~GeV, $\met>25$~GeV, $\mT>50$~GeV and $\uT<15$~GeV; $|\eta_{\ell}|<2.4$ or $2.4<|\eta_{\ell}|<4$. The first set of cuts selects the range of the kinematic peaks of the $W$ boson decay products, restricting to the region of small \pTW to maximise the sensitivity of the distributions to $m_W$. Two pseudorapidity ranges are considered, corresponding to the central region accessible with the current ATLAS detector, and to the forward region accessible in the electron channel with the ITk.

The Monte Carlo samples are produced using the CT10 PDF set, $m_W^\text{ref}=80.399$~GeV, and the corresponding Standard Model prediction for $\Gamma_W$. Kinematic distributions for the different values of $m_W$ are obtained by applying an event weight to the reference samples based on the ratio of the Breit--Wigner densities corresponding to $m_W$ and $m_W^\text{ref}$, for a given value of the final state invariant mass. A similar event weight, calculated internally by {\mbox{\textsc{POWHEG}}\xspace} and corresponding to the ratio of the event cross sections predicted by CT10 and several alternate PDFs, is used to obtain final state
distributions corresponding to the CT14~\cite{Dulat:2015mca}, MMHT2014~\cite{Harland-Lang:2014zoa}, HL-LHC~\cite{Khalek:2018mdn} and LHeC~\cite{Klein:1564929} PDF sets and their associated uncertainties. Compared to current sets such as CT14 and MMHT2014, the HL-LHC set incorporates the expected constraints from present and future LHC data; it starts from the PDF4LHC convention~\cite{Butterworth:2015oua} and comes in three scenarios corresponding to more or less optimistic projections of the experimental uncertainties. The LHeC PDF set represents the impact of a proposed future high-energy, high-luminosity $ep$ scattering experiment~\cite{AbelleiraFernandez:2012cc} on the uncertainties in the proton structure, using the theoretically best understood process for this purpose.

The shift in the measured value of $m_W$ resulting from a change in the assumed PDF set is estimated as follows. Considering a set of template distributions obtained for different values of $m_W$ and
a given reference PDF set, and ``pseudo-data'' distributions obtained for $m_W=m_W^\text{ref}$ and an alternate set $i$ (representing, for example, uncertainty variations with respect to the reference
set), the preferred value of $m_W$ for this set is determined by minimising the $\chi^2$ between the pseudo-data and the templates. The preferred value is denoted $m_W^i$, and the corresponding variation is defined as $\delta m_W^i = m_W^i -m_W^\text{ref}$. The statistical uncertainty on the measurement is estimated from the half width of the $\chi^2$ function one unit above the minimum.

The present study considers measurements of $m_W$ in separate categories, corresponding to $W^+$ and $W^-$ events; five pseudorapidity bins, $|\eta_{\ell}|<0.6$, $0.6<|\eta_{\ell}|<1.2$, $1.2<|\eta_{\ell}|<1.8$,
$1.8<|\eta_{\ell}|<2.4$, and $2.4<|\eta_{\ell}|<4$; \pTl and \mT distribution fits; and two centre-of-mass energies ($\sqrt{s}=14$ and $27$~TeV). For each category $\alpha$ and for the PDF sets considered here,
the Hessian uncertainty corresponding to a given set is estimated as $\delta m_{W\alpha}^+ = \left[\sum_i \left({\delta m_{W\alpha}^i}\right)^2\right]^{1/2}$, if $\delta m_{W\alpha}^i>0$, and as $\delta m_{W\alpha}^- = \left[\sum_i \left({\delta m_{W\alpha}^i}\right)^2\right]^{1/2}$, if $\delta m_{W\alpha}^i<0$, where $i$ runs over the uncertainty sets, and $\delta m_{W\alpha}^i$ is calculated with respect to the reference PDF set. For CT10 and CT14, the uncertainties are divided by a factor 1.645 to
match the 68\% CL. Only symmetrised uncertainties, $\delta m_{W\alpha} = (\delta m_{W\alpha}^+ + \delta m_{W\alpha}^-)/2$, are considered for simplicity. The correlation of PDF uncertainties between different measurement categories is
calculated as $\rho_{\alpha\beta} = \frac{\sum_i \delta m_{W\alpha}^i \delta m_{W\beta}^i}{\delta m_{W\alpha}\delta m_{W\beta}}$. 

PDF variations generate correlated variations in the \pTW and \pTZ distributions, while the latter are strongly constrained by experimental data~\cite{STDM-2012-23,STDM-2014-12}. These
constraints were used in the ATLAS measurement of $m_W$~\cite{Aaboud:2017svj}, bringing significant reduction in the PDF uncertainties. The uncertainties estimated here are thus conservative from
this perspective, and partly account for uncertainties in the \pTW distribution.

The overall measurement precision is evaluated by combining the results obtained in the different categories using the BLUE prescription~\cite{Valassi:2003mu}. Only statistical and PDF uncertainties
are considered. The former are assigned assuming an integrated luminosity of 200~\ipb, and normalising the samples to the expected cross-sections. 
The expected measurement uncertainties, together with their statistical and PDF components, are summarised in Fig.~\ref{fig:ct10unc} (a) for CT10. The numbers quoted for $0<|\eta_{\ell}|<2.4$ correspond to the combination of the four pseudorapidity bins in this range. Moderate or negative PDF uncertainty correlations, leading to reduced combined uncertainties, are observed between categories of different $W$-boson charges, and between central and forward pseudorapidities, at given $\sqrt{s}$. On the other hand, PDF uncertainty correlations tend to be large and positive between $\sqrt{s}=14$ and 27~TeV, for a given boson charge and lepton pseudorapidity range. With 200~\ipb of data collected at each energy, a total uncertainty of about 10~MeV is obtained. 

Table~\ref{tab:pdfcomp} and Fig.~\ref{fig:ct10unc} (b) compare the uncertainties obtained for different PDF sets. The CT10 and CT14 sets display similar uncertainty correlations, leading to similar
improvements under combination of categories, and yielding comparable final PDF uncertainties. The MMHT2014 uncertainties are about 30\% lower. The three projected HL-LHC PDF sets give very similar
uncertainties; the most conservative one is shown here. Compared to CT10 and CT14, a reduction in PDF uncertainty of about a factor of two is obtained. The LHeC projection results from a QCD fit to 1 ab$^{-1}$ of $ep$ scattering pseudodata, with $E_e=60$~GeV and $E_p=7$~TeV. Such a sample could be collected in about five years, synchronously with the HL-LHC operation. In this configuration, the neutral- and charged-current DIS samples are sufficient to disentangle the first and second generation parton densities without ambiguity, and reduce the PDF uncertainty below 2~MeV, a factor 5--6 compared to present knowledge. Also in this case the $m_W$ measurement will benefit from the large $W$ boson samples collected at the LHC, and from the anti-correlation between central and forward categories. In this context, PDF uncertainties would still be sub-leading with 1 \ifb\ of low pile-up data. \\ 

\begin{figure}
  \begin{center}
    \subfloat[]{\includegraphics[width=0.45\textwidth]{\main/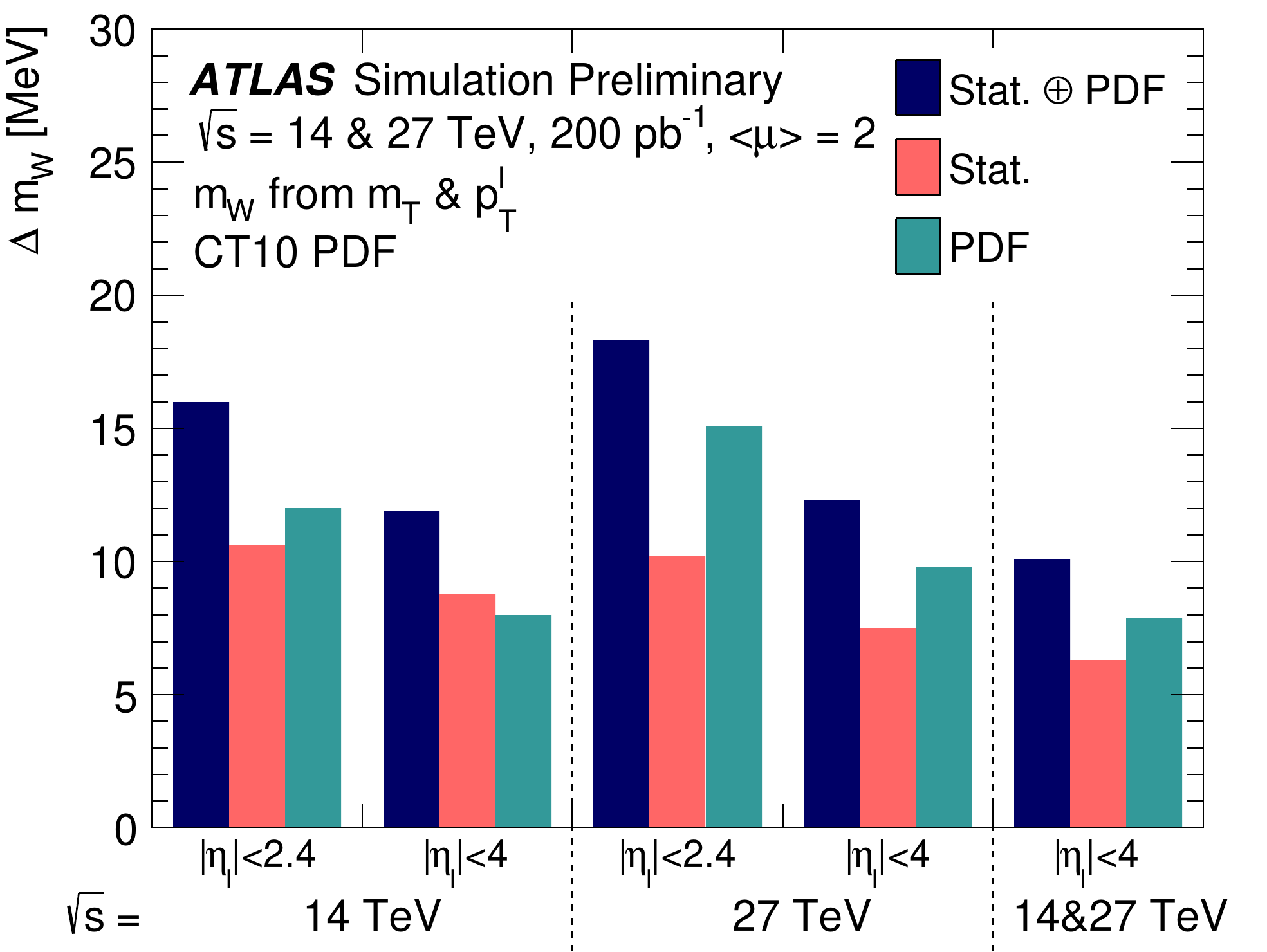}}
    \subfloat[]{\includegraphics[width=0.45\textwidth]{\main/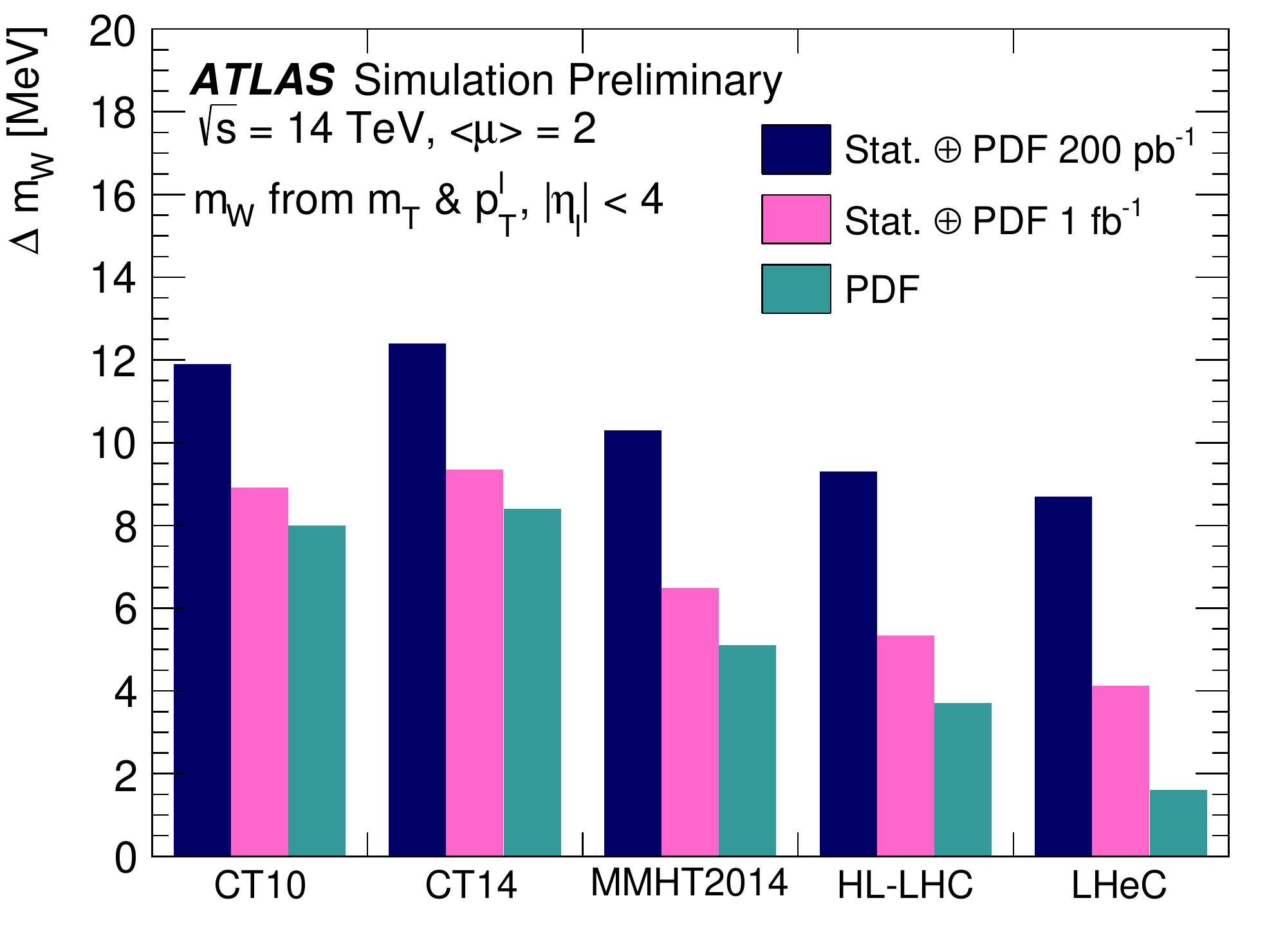}}
  \end{center}
  \caption{Measurement uncertainty for combined fits to the \pTl and \mT distributions (a) in different lepton acceptance regions and for different centre-of-mass energies, using the CT10 PDF set and for 200~\ipb collected at each energy and (b) for different PDF sets in $|\eta_{\ell}|<4$, for 200~\ipb and 1~\ifb collected at $\sqrt{s}=14$~TeV. The numbers quoted for $0<|\eta_{\ell}|<2.4$ correspond to the combination of the four pseudorapidity bins in this range. \label{fig:ct10unc}}
\end{figure}

\begin{table}
  \begin{center}
  \caption{Measurement uncertainty for different lepton acceptance regions, centre-of-mass energies and PDF sets, combined fits to the \pTl and \mT distributions, and for 200~\ipb collected at each energy. The numbers quoted for $0<|\eta_{\ell}|<2.4$ correspond to the combination of the four pseudorapidity bins in this range. In each case, the first number corresponds to the sum of statistical and PDF uncertainties, and the numbers between parentheses are the statistical and PDF components, respectively.\label{tab:pdfcomp}}
  
  \scalebox{0.9}{
    \begin{tabular}{|c|l|ccc|}
\hline
      $\sqrt{s}$ [TeV] & Lepton acceptance  & \multicolumn{3}{c|}{Uncertainty in $m_W$ [MeV]}    \\
&  & CT10       & CT14      & MMHT2014 \\
      \hline\hline
      14    & $|\eta_{\ell}|<2.4$   & 16.0 (10.6 $\oplus$ 12.0) & 17.3 (11.4 $\oplus$ 13.0) & 15.4 (10.7 $\oplus$ 11.1) \\
      14    & $|\eta_{\ell}|<4$     & 11.9 (8.8 $\oplus$ 8.0) & 12.4 (9.2 $\oplus$ 8.4) & 10.3 (9.0 $\oplus$ 5.1) \\
      27    & $|\eta_{\ell}|<2.4$   & 18.3 (10.2 $\oplus$ 15.1) & 18.8 (10.5 $\oplus$ 15.5) & 16.5 (9.4 $\oplus$ 13.5) \\
      27    & $|\eta_{\ell}|<4$     & 12.3 (7.5 $\oplus$ 9.8) & 12.7 (8.2 $\oplus$ 9.7) & 11.4 (7.9 $\oplus$ 8.3)\\
      14+27 & $|\eta_{\ell}|<4$     & 10.1 (6.3 $\oplus$ 7.9) & 10.1 (6.9 $\oplus$ 7.4) & 8.6 (6.5 $\oplus$ 5.5) \\
    \hline
    \end{tabular}}
  \end{center}
 
  \begin{center}
    \scalebox{0.9}{
    \begin{tabular}{|c|l|cc|}
    \hline
      $\sqrt{s}$ [TeV] & Lepton acceptance & \multicolumn{2}{c|}{Uncertainty in $m_W$ [MeV]}    \\
&  & HL-LHC & LHeC \\
\hline\hline  
  14    & $|\eta_{\ell}|<2.4$   & 11.5 (10.0 $\oplus$ 5.8 ) & 10.2 (9.9 $\oplus$ 2.2) \\
      14    & $|\eta_{\ell}|<4$     & 9.3 (8.6 $\oplus$ 3.7)    & 8.7 (8.5 $\oplus$ 1.6) \\
     \hline
    \end{tabular}}
  \end{center}
\end{table}

\subsubsection{Prospects for the measurement of the effective weak mixing angle}
\label{sec:s2tw}
\newcommand{\afb}{\ensuremath{\mathrm{A_{FB}}}\xspace}
\newcommand{\zmumu}{\ensuremath{\cPZ/\gamma^*\to \MM\xspace}}
\newcommand{\zee}{\ensuremath{\cPZ/\gamma^*\to \EE\xspace}}
\newcommand{\ztautau}{\ensuremath{\cPZ/\gamma^*\to \TT \xspace}}
\newcommand{\zll}{\ensuremath{\cPZ/\gamma^*\to \ell^+\ell^-\xspace}}
\newcommand{\csh}{\ensuremath{\cos\theta^{*}}\xspace}
\newcommand{\sineff}{\ensuremath{\sin^2\theta^{\text{lept}}_{\text{eff}}}\xspace}
\newcommand{\pt}{\ensuremath{p_{\mathrm{T}}}\xspace}
\newcommand{\llll}{\ensuremath{\ell\ell\xspace}}

At leading order dilepton pairs are produced through the annihilation of a quark and antiquark via the exchange of a  $Z$ boson or a virtual photon: $\Pq\bar{\Pq}\rightarrow Z/\gamma^*\rightarrow \ell^+\ell^-$. The definition of the forward-backward asymmetry, $\mathrm{A_{FB}}$, is based on the angle $\theta^*$ of the lepton ($\ell^-$) in the Collins-Soper~\cite{CSFrame,Chaichian:1981va} frame of the dilepton system:
\begin{equation} \mathrm{A_{FB}}=\frac{\sigma_\text{F}-\sigma_\text{B}}{\sigma_\text{F}+\sigma_\text{B}},
\end{equation}
where $\sigma_\text{F}$ and $\sigma_\text{B}$ are the cross sections in the forward ($\csh>0$) and backward ($\csh<0$) hemispheres, respectively. In this frame the $\theta^*$ is the angle of the $\ell^-$ direction with respect to the axis that bisects the angle between the direction of the quark and opposite direction of the anti-quark. In $pp$ collisions the direction of the quark is assumed to be in the boost direction of the dilepton pair. Here, \csh is calculated using  laboratory-frame quantities as follows:
\begin{equation} \csh=\frac{2(p_1^+p_2^- - p_1^-p_2^+)}{\sqrt{M^2(M^2+P_\text{T}^2)}}\times\frac{P_\text{z}}{|P_\text{z}|},
\end{equation}
where $M$, $P_\text{T}$, and $P_\text{z}$ are the mass, transverse momentum, and longitudinal momentum, respectively, of the dilepton system, and $p_1(p_2)$ are defined in terms of energy, $e_1 (e_2)$, and longitudinal momentum, $p_{\text{z},1}(p_{\text{z},2})$, of the negatively (positively) charged lepton as $p_\text{i}^\pm=(e_i\pm p_{\text{z},i})/\sqrt{2}$~\cite{CSFrame}. 

A non-zero $\mathrm{A_{FB}}$ in dilepton events arises from the vector and axial-vector couplings of electroweak bosons to fermions. At tree level, the vector $v_\text{f}$ and axial-vector $a_\text{f}$ couplings of $\PZ$ bosons to fermions ($f$) are:
 \begin{align}
	v_\text{f}&= T_3^\text{f}-2Q_\text{f}\sin^2\theta_\text{W}, \\ 
	a_\text{f}&= T_3^\text{f},
    \end{align}
where $T_3^\text{f}$ and $Q_\text{f}$ are the third component of the weak isospin and the charge of the fermion, respectively, and $\sin^2\theta_\text{W}$ is the weak mixing angle, which is related to the masses of the $W$ and $Z$ bosons by the relation $\sin^2\theta_\text{W} =1-M_\text{W}^2/M_\text{Z}^2$. Electroweak radiative corrections affect these leading-order relations. 
An effective weak mixing angle, 
$\sin^2\theta_{\text{eff}}^\text{f}$,  
is defined based on the relation between these couplings: 
$v_\text{f}/a_\text{f}=1-4|Q_\text{f}|\sin^2\theta_{\text{eff}}^\text{f}$, 
with 
$\sin^2\theta_{\text{eff}}^\text{f}=\kappa_{\text{f}} \sin^2\theta_\text{W}$, 
where flavour-dependent $\kappa_{\text{f}}$ is determined by electroweak corrections. Consequently, precise measurements of $\mathrm{A_{FB}}$ can be used to extract the effective leptonic  weak mixing angle (\sineff).

The most precise  previous measurements of \sineff were performed by the LEP and SLD experiments~\cite{ALEPH:2005ab}. There is, however, a known tension of about 3 standard deviations between the two most precise measurements. Measurements of \sineff have also been performed by the LHC and Tevatron experiments~\cite{Chatrchyan:2011ya,Aad:2015uau,Aaij:2015lka,Aaltonen:2014loa,Aaltonen:2016nuy,Abazov:2014jti}.

In measurements of $\mathrm{A_{FB}}$ (or associated angular variables) in leptonic decays of $Z$ bosons at a $pp$ collider, the assignment of the $z$-axis is crucial. At low rapidities, there is a two-fold ambiguity in the direction of the initial state quark and anti-quark; the colliding quark is equally likely to be in either proton and the parton level asymmetry is diluted. However, at higher rapidities, the $Z$ boson tends to be produced in the direction of travel of the quark, since the (valence) quark tends to be at higher Bjorken-$x$ than the anti-quark. This means that the dilution between parton level and proton level quantities is significantly smaller at larger rapidities, illustrated in Fig.~\ref{fig:s2twlhcb1}, and a larger forward-backward asymmetry is induced. Consequently, the forward acceptance of LHCb, in addition to the increased forward coverage of the ATLAS and CMS detectors, will be crucial to achieving the most precise measurement of \sineff possible at the HL-LHC. 

The uncertainties on the parton distribution functions translate into sizeable variations in the observed $\mathrm{A_{FB}}$ values, which have limited the precision of current measurements of \sineff at the LHC. However, the changes in PDFs affect the $\mathrm{A_{FB}}(M_{\llll},Y_{\llll})$ distribution in a different way from changes in \sineff. 
Because of this behaviour, the distribution of $\mathrm{A_{FB}}$ can itself be used to constrain the PDF uncertainties on the extraction of \sineff using either a Bayesian $\chi^2$ reweighting method~\cite{Giele:1998gw,Sato:2013ika,Bodek:2016olg} (in the case of PDFs with Monte Carlo replicas) or through a profiling procedure~\cite{Paukkunen:2014zia} (in the case of PDFs with Hessian error sets).

\begin{figure}[tb]
  \begin{center}
    \includegraphics[width=0.6\linewidth]{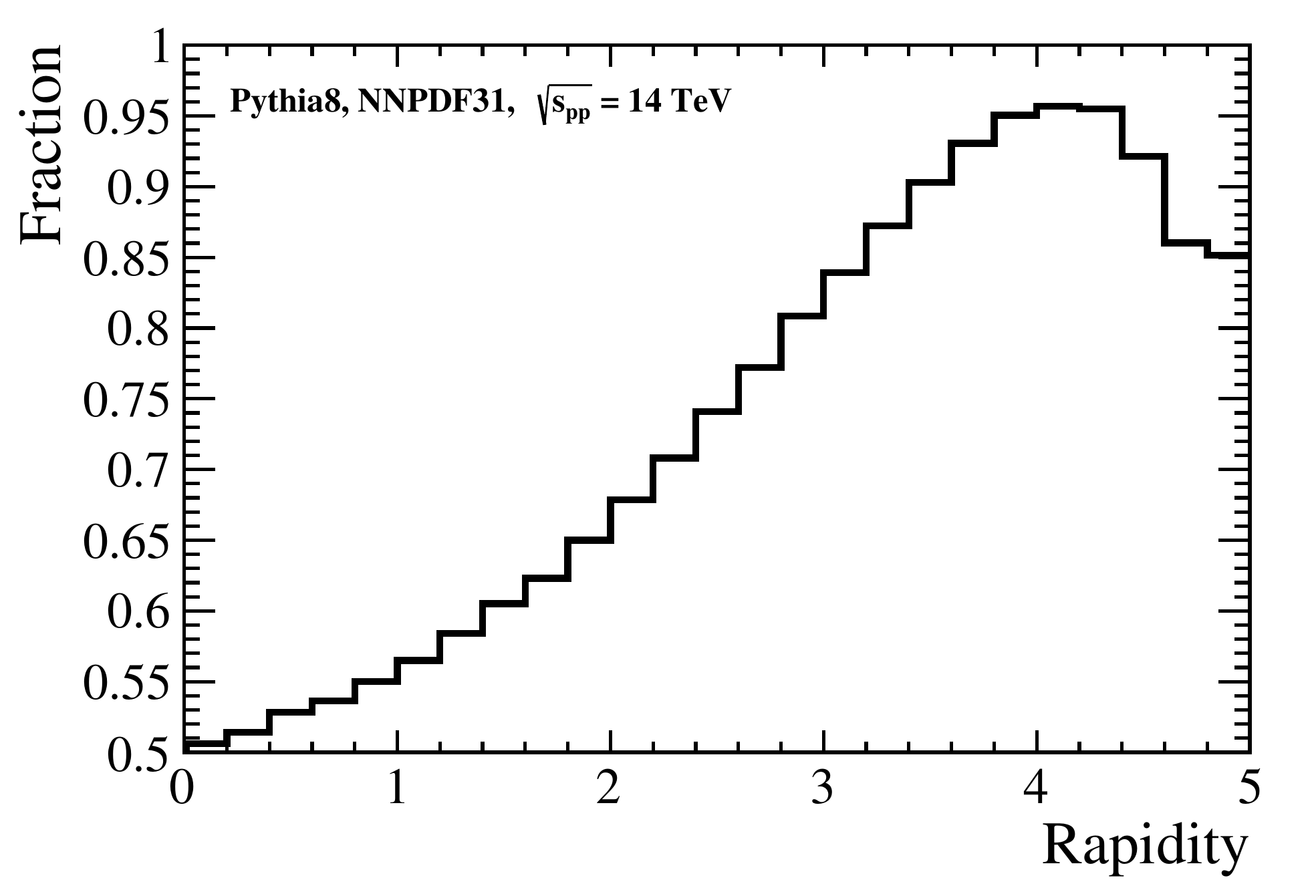}
   
    \vspace*{-0.5cm}
  \end{center}
  \caption{
    The fraction of events where the $Z$ boson travels in the same direction along the $z$-axis as the colliding quark, in proton-proton collisions with $\sqrt{s} = 14$ TeV. This increases as the event becomes more forward, reaching a maximum in the region probed by LHCb. The decrease once the rapidity is greater than 4 is because the fraction of collisions involving valence quarks decreases (the Bjorken-$x$ value of the high momentum quark in these collisions is typically greater than 0.3). No detector effects are simulated for this figure.
  }
  \label{fig:s2twlhcb1}
\end{figure}

Prospects for the measurement of the effective weak mixing angle using the forward-backward asymmetry, \afb, in Drell-Yan di-lepton events at the HL-LHC at ATLAS~\cite{ATL-PHYS-PUB-2018-037}, CMS~\cite{CMS-PAS-FTR-17-001}
and LHCb~\cite{Barter:2647836} have been performed and are reported here. The leptonic effective weak mixing angle  is extracted from measurements of \afb in dilepton events by minimising the $\chi^2$ value between the simulated data and template \afb distributions representing different \sineff values and PDF variations. The LHCb and CMS analyses consider the dimuon final state, while the ATLAS analysis considers the dielectron final state. For CMS and LHCb the samples and different \sineff templates are generated at next-to-leading order using the {\mbox{\textsc{POWHEG}}\xspace} event generator~\cite{POWHEG0,POWHEG1,POWHEG2,POWHEG3}, where the NNPDF3.0~\cite{NNPDF30} PDF set is used in the case of the CMS analysis, and the NNPDF3.1 PDF set~\cite{Ball:2017nwa} for LHCb. For CMS, the analysis is performed at generator level without the effect of smearing due to detector effects\footnote{A comparison of 8~TeV predictions and measured values suggests the effect is not significant.} while for LHCb, a smearing is performed where the momentum resolution and reconstruction efficiency is assumed to be similar to the performance of the current detector~\cite{Aaij:2014jba}. In the case of ATLAS, events are generated with {\mbox{\textsc{POWHEG}}\xspace} and overlaid with additional inelastic $pp$ collisions per bunch-crossing simulated with {\mbox{\textsc{PYTHIA}}\xspace}. Parameterisations of the expected ATLAS detector performances during the HL-LHC runs~\cite{CERN-LHCC-2015-020} are then applied on particle-level objects to emulate the detector response. Lepton trigger and identification efficiencies are derived as a function of $\eta$ and $p_{\rm T}$ and used to estimate the likelihood of a given lepton to fulfil either the trigger or identification requirements, which have been optimised for the level of pile-up expected at the HL-LHC~\cite{ATL-PHYS-PUB-2016-026}. The \afb distributions are generated, at leading order (LO) in QCD, with {\mbox{\textsc{DYTURBO}}\xspace}, an optimised version of {\mbox{\textsc{DYRES/DYNNLO}}\xspace}~\cite{PhysRevLett.103.082001} with NNLO CT14 PDF and the world average value for $\sineff = 0.23153$.





The HL-LHC CMS detector will extend the pseudorapidity, $\eta$, coverage of the muon reconstruction from the current configuration of 2.4 to 2.8. In the CMS analysis an event is selected if there are at least two muons with ${|\eta|} < 2.8$ and  with the leading \pt muon $\pt> 25\UGeV$ and the second leading muon $\pt>15\UGeV$.  Figure~\ref{figure:afb} shows the $\mathrm{A_{FB}}$ distributions in bins of dimuon mass and rapidity for different energies and pseudorapidity acceptances. As expected, at higher centre-of-mass energies the observed $\mathrm{A_{FB}}$ is smaller because the interacting partons have smaller $x$-values which results in a smaller fraction of dimuon events produced by the valence quarks, which also means more dilution. The samples are normalised to the integrated luminosities of 19\fbinv for $\sqrt{s} = 8$~TeV and to 10 -- 3000\fbinv for $\sqrt{s} = 14$~TeV samples and the simulated data are shown for $\sqrt{s} = 8$~TeV and $\sqrt{s}=14$~TeV for two different selection requirements, $|\eta| < 2.4$ and 2.8.  Extending the pseudorapidity acceptance significantly increases the coverage for larger $x$-values in the production and reduces both the statistical and PDF uncertainties, as shown below.

\begin{figure}[!htbp]
\centering
    \includegraphics[scale=0.8]{\main/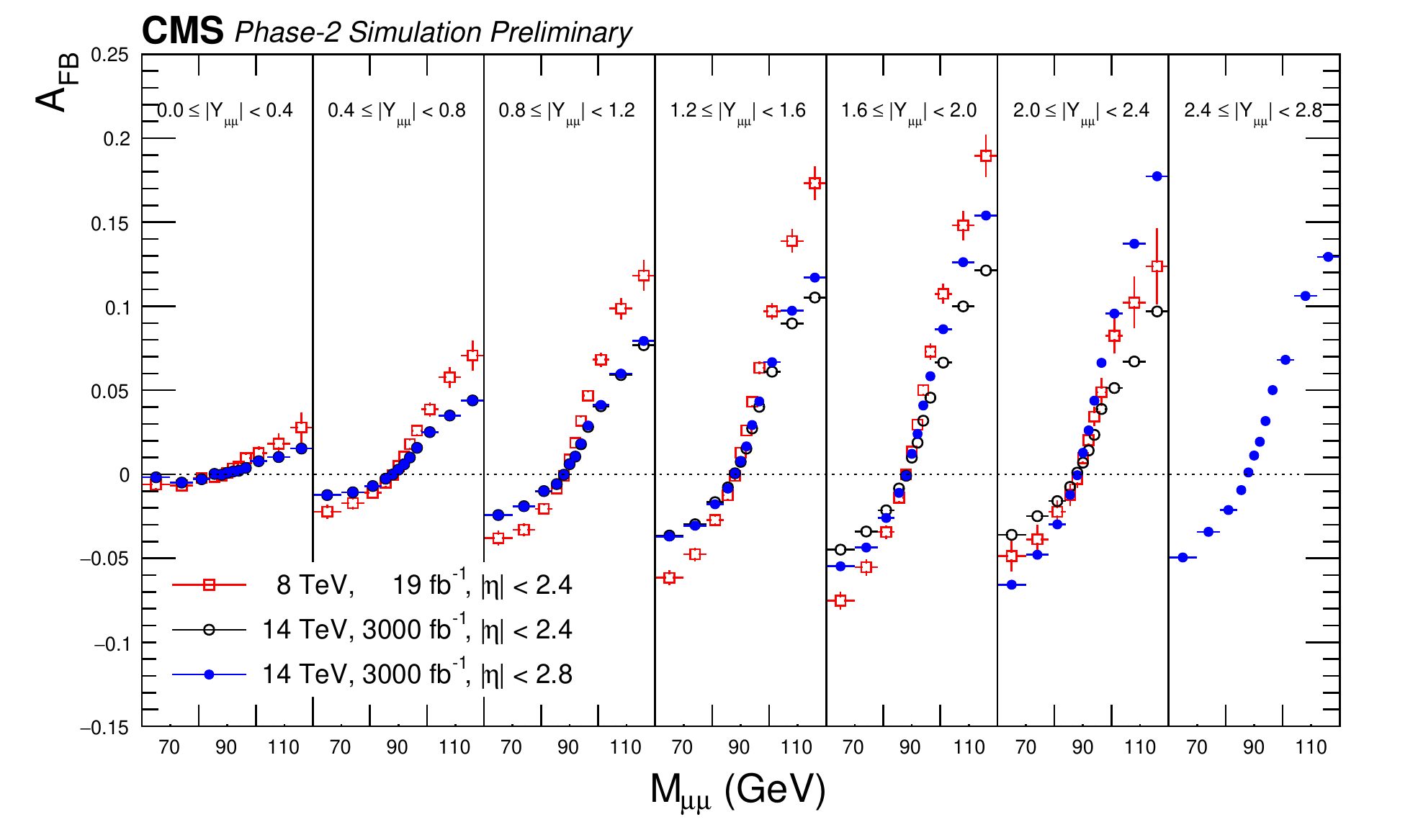}
    \caption{
	Forward-backward asymmetry distribution, 
    $\mathrm{A_{FB}}(M_{\mu\mu},Y_{\mu\mu})$, in dimuon events at $\sqrt{s}=8$~TeV and 14~TeV. 
	The distributions are made with {\mbox{\textsc{POWHEG}}\xspace} event generator using NNPDF3.0 PDFs and interfaced with {\mbox{\textsc{PYTHIA}}\xspace} v8 for parton-showering, 
	QED final-state radiation (FSR) and hadronization. 
	Following acceptance selections are applied to the generated muons after FSR: $|\eta|<2.4$ (or $|\eta|<2.8$), $\pt^\mathrm{lead}>25\UGeV$, $\pt^\mathrm{trail}>15$ GeV. 
	The error bars represent the statistical uncertainties for the integrated luminosities corresponding to 19\fbinv at $\sqrt{s}=8$~TeV and 3000\fbinv at $\sqrt{s}=14$~TeV.}
 \label{figure:afb} 
    
\end{figure}

In the case of the 14~TeV\; analysis with a large number of events ($>200$~\fbinv), the pseudo-data are too precise to estimate the PDF uncertainties with the Bayesian reweighting approach because the replica distributions are too sparse compared to the statistical uncertainties. Therefore, the PDF uncertainties after the Bayesian reweighting are estimated by extrapolating from the lower values of integrated luminosities.

The corresponding values for various luminosities at CMS are summarized in Table~\ref{table:extrapolation}. One can see from the table that with the extended pseudorapidity coverage of $|\eta|<2.8$, the statistical uncertainties are reduced by about 30\% and the PDF uncertainties are reduced by about 20\%, compared to $|\eta|<2.4$ regardless of the target integrated luminosity and for both nominal and constrained PDF uncertainties. 


\begin{table*}[h]
\centering
\caption{ Statistical, nominal NNPDF3.0, and constrained NNPDF3.0 uncertainties of the extracted \sineff value at CMS at 14~TeV for muon acceptances of $|\eta|<2.4$ and $|\eta|<2.8$ and for different values of integrated luminosity. For comparison, results of the 8~TeV estimate of this analysis are compared to the results obtained from 8 ~TeV measurement~\cite{CMS-PAS-SMP-16-007}. \label{table:extrapolation} }
\begin{tabular}{ | r | c  c | c  c |  c c |} \hline
    $L_{int}$ & \multicolumn{2}{c|}{$\delta_{\mathrm{stat}} [10^{-5}]$} & \multicolumn{2}{c|}{$\delta_{\mathrm{nnpdf3.0}}^{\mathrm{nominal}} [10^{-5}]$} & \multicolumn{2}{c|}{$\delta_{\mathrm{nnpdf3.0}}^{\mathrm{constrained}} [10^{-5}]$} \\ \hline
(\fbinv)  & $|\eta|<2.4$  & $|\eta|<2.8$  &  $|\eta|<2.4$ & $|\eta|<2.8$ & $|\eta|<2.4$ & $|\eta|<2.8$  \\ \hline\hline
  10  & 76  & 51 & 75  & 57 & 39 & 29\\
 100  & 24  & 16 & 75  & 57 & 27 & 20\\
 500  & 11  &  7 & 75  & 57 & 20 & 16\\
1000  &  8  &  5 & 75  & 57 & 18 & 14\\
3000  &  4  &  3 & 75  & 57 & 15 & 12\\
\hline
19 & 43  &  & 49  &  & 27 & \\
19 (from \cite{CMS-PAS-SMP-16-007}) & 44  &  & 54  &  & 32 & \\
\hline
\end{tabular}
\end{table*}



The LHCb detector has coverage in the pseudorapidity range $2 < \eta < 5$ and expects to install its `Upgrade II' in Long Shutdown 4. Following this upgrade, LHCb will collect at least 300 fb$^{-1}$ of data, allowing high precision measurements. The forward acceptance of LHCb brings a number of benefits in measurements of \sineff at the LHC. The lower level of dilution in the forward region results in a larger sensitivity to \sineff and the PDF effects are (in relative terms) smaller, providing both statistical precision in measurements of the weak mixing angle and a reduction in PDF uncertainties.  In addition, LHCb does not simply probe forward rapidities of the $Z$ boson: the leptons themselves are located over a significant range of rapidities, allowing extremal values of $\cos\theta^\ast$ to be probed, increasing sensitivity to the weak mixing angle. Finally, LHCb has the ability to select events at low momentum using a flexible full software trigger and real time analysis scheme (from Run-3 onwards). It is therefore foreseen that the LHCb Upgrade II will be able to select $Z$ boson decays where one lepton has transverse momentum above 20~GeV, while the other lepton has a transverse momentum above 5~GeV. Such low thresholds again increase the sensitivity to asymmetric events at high $|\cos\theta^\ast|$. In addition to the advantages of the extended forward acceptance for such measurements, as part of Upgrade II LHCb is expected to undergo a significant calorimeter upgrade\footnote{This upgrade will offer an extended dynamic range within the ECAL, offering improved electron momentum resolution.} allowing similar precision to be achieved in both the dielectron and dimuon final states.

LHCb has performed a study of projected sensitivities, considering the dimuon final state. The experiment is assumed to have coverage in the region $2.0<\eta<5$.  Toy measurements of the forward-backward asymmetry are used to determine the sensitivity of measurements at LHCb Upgrade II to the weak mixing angle. Only statistical uncertainties are considered alongside the effects of knowledge of PDFs. The statistical uncertainty on \sineff is expected to be below $5\times10^{-5}$ with 300~fb$^{-1}$ of data. The expected PDF uncertainty from current PDF knowledge is $\sim 20\times10^{-5}$, but with Bayesian reweighting this can be reduced to the level of about $10\times10^{-5}$ (with analysis of a dataset corresponding to an integrated luminosity of 300~fb$^{-1}$). This reduction assumes systematic effects are negligible in comparison to statistical uncertainties, though the current knowledge of PDFs means that any measurement in the forward region is expected to offer a smaller PDF uncertainty than the total uncertainties in the previous best measurements of the weak mixing angle. The main challenge of such measurements at LHCb Upgrade II will therefore be to control systematic uncertainties in order to ensure the overall measurement also achieves high precision; however, the large dataset of $J/\psi$ and $\Upsilon$ mesons to be recorded is expected to aid the understanding of effects such as the momentum scale (which introduced the largest systematic uncertainty in the Run-1 analysis at LHCb). This should enable a measurement at LHCb Upgrade II with a precision similar to or better than that achieved in the combination of measurements at LEP and SLD.


In the ATLAS analysis di-electron candidates are selected where each electron has $p_{\rm T}$ in excess of 25~GeV and the combined invariant mass is in the region of the $Z$ pole. A new inner tracking system (ITk) will extend the tracking coverage of the ATLAS detector from $|\eta|\leq 2.5$ up to $|\eta|\leq 4.0$ at the HL-LHC, providing the ability to reconstruct forward charged particle tracks, which can be matched to calorimeter clusters for forward electron reconstruction. The selected data sample is split into three channels, where both electrons are in the central region, satisfying $|\eta| < 2.47$ (the CC channel), where one electron is central and the other is forward, satisfying $2.5 <\eta < 4.2$ (the CF channel), and finally where both electrons are forward (the FF channel). Events are selected by requiring at least one electron firing the single electron trigger, except in the FF channel, where a dielectron trigger is required. 

\begin{figure}[!h]
\centering
\includegraphics[width=0.45\linewidth]{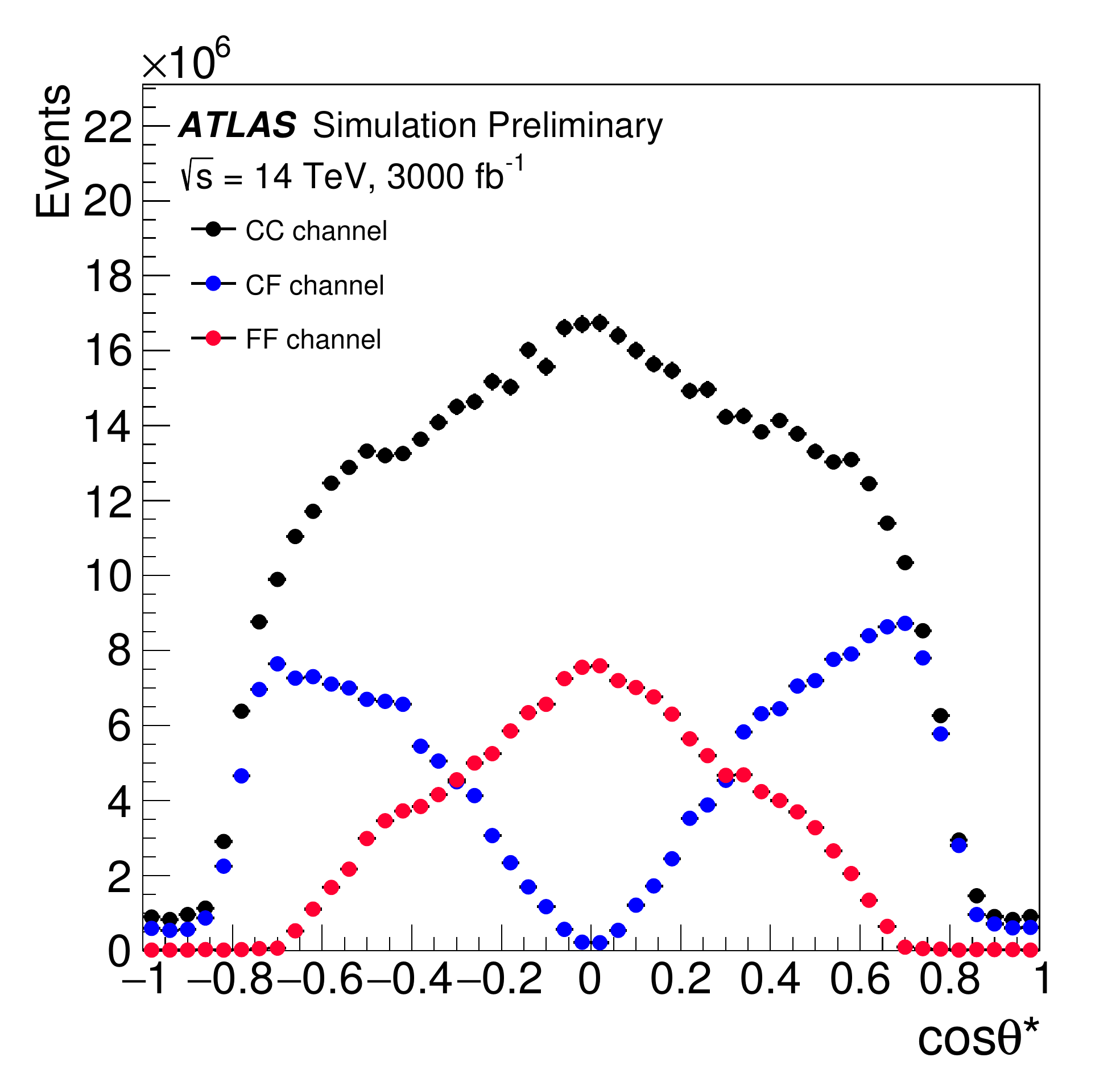}
\caption{
The $\cos\theta^\ast$  distribution for CC, CF and FF channels for selected Drell-Yan di-electron events expected for $3000\,\ifb$ of data at $\sqrt{s} = 14$ TeV.
\label{fig:AFB_Y} 
}
\end{figure}

As Fig.~\ref{fig:AFB_Y} shows, the CF channel selects events at high $\cos\theta^\ast$ values where the forward-backward asymmetry is more pronounced, and consequently the sensitivity to \sineff is higher in this channel. While the LHCb and CMS analyses consider only uncertainties due to statistics and PDFs, the ATLAS analysis considers also various sources of experimental uncertainty which affect the precision of the extraction of \afb. The main contributions arise from the limited knowledge of the momentum scale and resolution of the electrons, and the background contributions, which are mostly relevant in the CF and FF channels.

The extraction of \sineff is performed by minimising the $\chi^2$ value comparing particle-level \afb distributions with different weak mixing angle hypotheses in invariant mass and rapidity bins combining the CC, CF and FF channels. A global fit is performed where $\sineff$ is extracted while constraining the PDF uncertainties using a profiling procedure following that used in a previous ATLAS publication~\cite{Aaboud:2017ffb} and implemented in the xFitter package~\cite{Alekhin:2014irh}.

With this analysis, a significant reduction of the light quark uncertainties at low $x$ is seen and combining the three channels together, the measurement reaches a precision of $18\cdot10^{-5}$ ( $\pm16 \cdot 10^{-5}$ (PDF) $\pm 9 \cdot 10^{-5}$ (exp.) ). The uncertainty of the results remains dominated by the limited knowledge of the PDFs. 

In the context of the Yellow Report for the HL-LHC, prospect PDF fits including HL-LHC pseudo-data of future PDF-sensitive measurements from ATLAS, CMS and LHCb were performed (see Sec.~\ref{sec:ultimatepdfs}). Three prospect PDF scenarios were considered and compared with the reference PDF set PDF4LHC15~\cite{Butterworth:2015oua}. The expected sensitivity of the \sineff measurements with 3000 fb$^{-1}$ at $\sqrt{s} = 14$ TeV is improved by 10-25\% depending on the prospect PDFs scenario considered.  In Table~\ref{tab:ATLAS_WMA_Result}  the precision on \sineff obtained with the "ultimate" HL-LHC PDF set is compared with the with the one obtained with CT14NNLO PDF set.

The sensitivity of the analysis to the \sineff extraction is also estimated with a prospect PDF set including expected data from the LHeC collider~\cite{Klein:1564929}. In this case the PDF uncertainty is reduced by an additional factor of 5 with respect to the one obtained with the HL-LHC prospect PDFs.

\begin{table}[h]
\caption{
The  value of \sineff with the breakdown of uncertainties from the ATLAS preliminary results at $\sqrt{s} = 8$ TeV with $20\,\ifb$~\cite{ATLAS-CONF-2018-037} is compared to the projected \sineff measurements with $3000\,\ifb$ of data at $\sqrt{s} = 14$ TeV for two PDF sets considered in this note.   All the numbers values are given in units of  $10^{-5} $.  Note that other sources of systematic uncertainties, such as the impact of the MC statistical uncertainty, evaluated in Ref.~\cite{ATLAS-CONF-2018-037} are not considered in this prospect analysis.  
For the HL-LHC prospect PDFs the  "ultimate"  scenario is chosen.  }
   \label{tab:ATLAS_WMA_Result}
\small
  \begin{center}
    \resizebox*{\textwidth}{!}{
    \begin{tabular}{|lp{3.49cm}p{3.49cm}p{3.49cm}|}
    \hline
& ATLAS $\sqrt{s} =8$ TeV & ATLAS $\sqrt{s} =14$ TeV   & ATLAS $\sqrt{s} =14$ TeV   \\ 
 \hline
$\mathcal{L} ~ [\text{fb}^{-1}]$ & 20  & 3000   & 3000   \\ 
          
PDF set & MMHT14 & CT14   & PDF4LHC15$_{HL-LHC}$   \\ 
      \hline\hline
  \sineff\  {\footnotesize$[\times 10^{-5}]$} &  23140 &  23153 &  23153  \\ 
      \midrule
     Stat. & $\pm~21$ & $\pm~4$ &  $\pm~4$ \\  
      PDFs &$ \pm~24 $ &$ \pm~16 $& $ \pm~13 $  \\ 
  Experimental Syst. & $\pm~9$ & $\pm~8$ &  $\pm~6$ \\ 
    Other Syst. & $\pm~13$ & - & - \\  
      \hline
  Total &  $\pm~36 $ & $ \pm~18 $ & $ \pm~15 $  \\ 
      \hline
    \end{tabular}
  }
  \end{center}
\end{table}


To conclude, the accuracy of measurements of the weak mixing angle obtained with an analysis of $\mathrm{A_{FB}}$ in $Z$ events at $\sqrt{s} = 14$ TeV with $3000\,\ifb$ at ATLAS and CMS and $300\,\ifb$ at LHCb at the HL-LHC exceed the precision achieved in all previous single-experiment results  to date and the measurements are dominated by PDF uncertainties.
To explore the full potential of the HL-LHC data it will be therefore essential to reduce PDF uncertainties.  
A significant improvement of the sensitivity of the measurement is observed in the ATLAS analysis when using prospect PDF sets including ancillary Drell-Yan measurements performed with the data collected during the high luminosity phase of the LHC and at the LHeC collider.

%
%


\subsubsection[The global EW fit]{The global EW fit\footnote{Contribution by J. de Blas, M. Ciuchini, E. Franco, S. Mishima, M. Pierini, L. Reina, and L. Silvestrini.}}
\newcommand{\HEPfit}{{\mbox{\textsc{HEPfit}}\xspace}}
\newcommand{\cmrule}{\midrule[0.25mm]}
\newcommand{\lrD}{~\!\overset{\leftrightarrow}{\hspace{-0.1cm}D}\!}

The measurement of the Higgs Boson mass ($M_H$) at the Large Hadron
Collider (LHC) has provided the last input to the global fit of
electroweak (EW) precision observables (EWPO), which can now be used
to effectively constrain new physics.  Moreover, the measurement of
Higgs-boson production and decay rates that is at the core of the
physics program of the LHC Run-2 will further constrain those
interactions that directly affect Higgs-boson physics.

The HL-LHC will have the potential to provide
more constraining bounds on new physics via the global fit to EWPO and
Higgs data, thanks to the higher precision it will reach both in the
measurement of some of the crucial input parameters of global EW fits
(e.g. $M_W$, $m_t$, $M_H$, and
$\sin^2\theta_{\mathrm{eff}}^{\mathrm{lept}}$), and in the measurement
of Higgs-boson total and differential rates.  In this study the reach of the HL-LHC in constraining new physics is explored
via a global fit to EWPO. Earlier studies on the prospects for the LHC were performed in \cite{Baak:2014ora,deBlas:2016ojx}.

In the following, details are provided first on the parameters and procedure of the global EW fit. Next the results are interpreted within the Standard Model (SM). Finally, the EW fit is used to constrain new physics beyond the SM. The results are presented for both the current data and the projections in the HL-LHC scenario.



The global fit of EWPO is performed using the
\HEPfit~package~\cite{hepfitsite}, a general tool to combine direct
and indirect constraints on the SM and its extensions in any
statistical framework.  The default fit procedure, used here,
follows a Bayesian statistical approach and uses {\mbox{\textsc{BAT}}\xspace} (Bayesian Analysis
Toolkit)~\cite{Caldwell:2008fw}. Flat priors are used for all input
parameters, and the likelihoods are built assuming Gaussian distributions
for all experimental measurements. The output of the fit is therefore
given as the posterior distributions for each input parameters and
observables, calculated using a Markov Chain Monte Carlo method.

All EWPO are calculated as a SM core plus corrections. The SM core
includes all available higher-order corrections, including the latest
theoretical developments in the calculation of radiative corrections
to the EWPO of \cite{Dubovyk:2016aqv,Dubovyk:2018rlg}.\footnote{The
  uncertainties associated to missing higher-order corrections to the
  SM predictions for the EWPO are also taken into account in the
  fits, via nuisance parameters with Gaussian priors.}  New physics corrections are computed at the leading order.
The \HEPfit~code allows for the implementation of different models of
new physics. In particular, as explained below,
the study is specialised in the 
general framework of the so called SM effective field theory (SMEFT),
where the SM Lagrangian is extended by the addition of operators of
canonical mass dimension higher than four (limited to the basis of
operators of canonical dimension six in this study).

As far as EWPO are concerned, this study updates the EWPO fit of
Refs.~\cite{deBlas:2016ojx,deBlas:2016nqo,deBlas:2017wmn}, including
recent updates on the theory calculations~\cite{Dubovyk:2018rlg} and
experimental
measurements~\cite{Keshavarzi:2018mgv,Sirunyan:2017huu,ATLAS:2017lqh,Sirunyan:2018gqx,Sirunyan:2017exp,Aaboud:2018wps,Aaltonen:2018dxj,Sirunyan:2018swq,ATLAS-CONF-2018-037}. The uncertainties
on some input parameters that have
been obtained by including hadron collider data are further reduced, in order to account
for the level of accuracy expected for the HL-LHC. In all these
projections it is assumed that the central values for the HL-LHC measurements
will not change with respect to current data.  In particular the following assumptions are made:

\begin{enumerate}
\item{The $W$ mass, whose uncertainty obtained by combining ATLAS and
    Tevatron+LEP2 measurements is currently around
    12~MeV
    ~\cite{Schael:2013ita,Aaltonen:2013vwa,Aaboud:2017svj,Tanabashi:2018oca}
    could be measured at the HL-LHC with
    a precision of 7~MeV. This number is derived from the current
    estimate of the statistical plus PDF uncertainty using $1$
    fb$^{-1}$ of data reported in Sec.~\ref{sec:wmassprospects}, and
    assuming systematic errors to be of similar size to the
    statistical ones.  In this fit a measurement of $M_W=80.379\pm 0.007$~GeV is therefore added to the current
    combination.}
\item{An aggressive estimate of the current uncertainty on the
    top-quark mass, obtained by combining current Tevatron and LHC
    measurements, puts the uncertainty on $m_t$ at the level of 0.4 GeV.  It
    will be difficult to further reduce this number at the HL-LHC,
    since the remaining uncertainty is mainly of systematic and
    theoretical origin. In the current fit 
    $m_t=172.8\pm 0.4$~GeV is used. }
\item{The measurements of the effective angle
    $\sin^2\theta_{\mathrm{eff}}^{\mathrm{lept}}$ can also be improved
    at the HL-LHC.  Currently, a combination of the latest LHC and
    Tevatron results returns a precision for this observable of
    $\sim 0.00022-0.00027$, depending on the assumptions made in
    combining common uncertainties.  For the HL-LHC fit, the combination is repeated using the ATLAS projections outlined in Sec.~\ref{sec:s2tw} where the HL-LHC PDF set is used, corresponding to the value
    $\sin^2\theta_{\mathrm{eff}}^{\mathrm{lept}}= 0.23143 \pm 0.00015$.}
\item{ The error on the Higgs-boson mass, currently around 0.20 GeV,
    can be reduced to 0.05 GeV~\cite{Dawson:2013bba, Cepeda:2019klc}.}
\item{The HL-LHC should also be able to improve the current knowledge
    on the $W$ width, whose precision of 42~MeV is currently given by
    the combination of LEP2 and Tevatron measurements. This
    uncertainty is dominated by the hadron collider measurement.
    While there is no available information about a possible
    determination of this quantity at the (HL-)LHC, the
    conservative assumption that the HL-LHC can achieve a precision on
    $\Gamma_W$ at least as good as the one on the current average is used. An independent HL-LHC measurement of
    $\Gamma_W=2.085\pm 0.042$~GeV is therefore added. This gives a 30~MeV uncertainty
    when combined with the current average.}
\end{enumerate}
Finally, apart from the improved precision of the HL-LHC measurements,
the assumption is made that, by the end of the HL-LHC run, better measurements
of some of the SM input parameters are possible from other
experiments. In particular, following
Ref.~\cite{deBlas:2016ojx,deBlas:2016nqo}, it is assumed that: 1) the
uncertainty on $\Delta \alpha_{\mathrm{had}}^{(5)}(M_Z)$ can be
reduced to $\pm 5 \times 10^{-5}$ by using data from currently ongoing
and future experiments that measure the cross section for $e^+ e^-\to$
hadrons, and 2) future lattice QCD measurements will provide a
determination of the strong coupling constant with accuracy
$\delta \alpha_S(M_Z)=\pm 0.0002$. The measurements of all other EWPO
and input parameters have been kept to their currently available
values. The current values of all EWPO measurements, as well as the
corresponding HL-LHC projected uncertainties, are listed in the second
and third columns of Table \ref{tab:SMfit}, respectively.
\begin{table}[h]
{\footnotesize
\begin{center}
\caption{
  Current experimental measurement, HL-LHC projected uncertainty, posterior, and pull for the five input
  parameters ($\alpha_s(M_Z)$, $\Delta \alpha^{(5)}_{\mathrm{had}}(M_Z)$, $M_Z$,
  $m_t$, $M_H$), and for the main EWPO considered in the SM fit.
  The pulls in the last column are obtained comparing the experimental measurements with the predictions from a fit {\em removing} the corresponding observable(s) (See for e.g. Ref.~\cite{deBlas:2016ojx} for details.).
} 
\label{tab:SMfit}
\begin{tabular}{|lccccc|}
\hline
& Measurement & HL-LHC &\multicolumn{2}{c}{Posterior} &Pull\\
&  & uncertainty &Current& HL-LHC&Current/HL-LHC\\
\hline\hline
$\alpha_{s}(M_Z)$ & $ 0.1180 \pm 0.0010 $ & 
$   \pm 0.0002 $  & 
$  0.1180 \pm  0.0009 $ & 
$  0.1180 \pm  0.0002   $ & 
$ 0/0.5 $\\ 
$\Delta\alpha^{(5)}_{\rm had}(M_Z)$ & $ 0.027611 \pm 0.000111 $ & 
$   \pm 0.00005 $  & 
$  0.02758 \pm 0.00011$ & 
$  0.02759 \pm 0.00005  $ & 
$ 1.1/2.1  $\\ 
$M_Z$ [GeV] & $ 91.1875 \pm 0.0021 $ & 
$ $  & 
$ 91.1880  \pm 0.0020  $ & 
$ 91.1890  \pm 0.0020  $ & 
$ -1.3/\!\!-\!2.6 $\\ 
$m_t$ [GeV] & $ 172.8 \pm 0.7 $ & 
$  \pm 0.4 $  & 
$  173.2 \pm 0.66  $ & 
$  173.1 \pm 0.38  $ & 
$-1.7/\!\!-\!2.9  $\\ 
$M_H$ [GeV] & $ 125.13 \pm 0.17 $ & 
$   \pm  0.05 $  & 
$ 125.13  \pm 0.17  $ & 
$ 125.13  \pm 0.05  $ & 
$ 1.4/3 $\\ 
\hline
$M_W$ [GeV] & $ 80.379 \pm 0.012 $ & 
$ \pm 0.007  $  & 
$ 80.362  \pm 0.006   $ & 
$ 80.367  \pm 0.004   $ & 
$ 1.6 / 2.7 $\\ 
$\Gamma_{W}$ [GeV] & $ 2.085 \pm 0.042 $ & 
$   \pm 0.042  $  & 
$ 2.0885  \pm 0.0006  $ & 
$ 2.0889  \pm 0.0003   $ & 
$ -0.1 $\\ 
$\mathrm{BR}_{W\to\ell\nu}$ & $ 0.1086 \pm 0.0009 $ & 
$  $  & 
$ 0.10838  \pm 0.00002  $ & 
$ 0.10838  \pm 0.000005   $ & 
$ 0.2 $\\ 
$\mathrm{BR}_{W\to\mathrm{had}}$ & $ 0.6741 \pm 0.0027 $ & 
$  $  & 
$ 0.67486  \pm 0.00007  $ & 
$ 0.67486  \pm 0.00001   $ & 
$ -0.3 $\\ 
$\sin^2\theta_{\mathrm{eff}}^{\mathrm{lept}}(Q_{\rm FB}^{\rm had})$ & $ 0.2324 \pm 0.0012 $ & 
$  $  & 
$ 0.23151  \pm 0.00006  $ & 
$ 0.23150  \pm 0.00005   $ & 
$ 0.7 $\\ 
$P_{\tau}^{\rm pol}={A}_\ell$ & $ 0.1465 \pm 0.0033 $ & 
$  $  & 
$ 0.14711  \pm 0.0005  $ & 
$ 0.14713  \pm 0.0004  $ & 
$  -0.2$\\ 
$\Gamma_{Z}$ [GeV] & $ 2.4952 \pm 0.0023 $ & 
$  $  & 
$ 2.4946  \pm 0.0007  $ & 
$ 2.4947  \pm 0.0005   $ & 
$ 0.3 $\\ 
$\sigma_{h}^{0}$ [nb] & $ 41.540 \pm 0.037 $ & 
$  $  & 
$ 41.492 \pm 0.008  $ & 
$ 41.491 \pm 0.006   $ & 
$  1.3$\\ 
$R^{0}_{\ell}$ & $ 20.767 \pm 0.025 $ & 
$  $  & 
$ 20.749  \pm 0.008  $ & 
$ 20.749  \pm 0.006   $ & 
$ 0.7 $\\ 
$A_{\rm FB}^{0, \ell}$ & $ 0.0171 \pm 0.0010 $ & 
$  $  & 
$ 0.01623  \pm 0.0001  $ & 
$ 0.016247  \pm 0.00008   $ & 
$ 0.9 $\\ 
${A}_{\ell}$ (SLD) & $ 0.1513 \pm 0.0021 $ & 
$  $  & 
$ 0.14711  \pm 0.0005  $ & 
$ 0.14718  \pm 0.0004   $ & 
$  1.9$\\ 
$R^{0}_{b}$ & $ 0.21629 \pm 0.00066 $ & 
$  $  & 
$ 0.21586  \pm 0.0001  $ & 
$ 0.21586 \pm 0.0001   $ & 
$ 0.7 / 0.6 $\\ 
$R^{0}_{c}$ & $ 0.1721 \pm 0.0030 $ & 
$  $  & 
$ 0.17221  \pm 0.00005  $ & 
$ 0.17221  \pm 0.00005   $ & 
$ 0 $\\ 
$A_{\rm FB}^{0, b}$ & $ 0.0992 \pm 0.0016 $& 
$  $  & 
$ 0.10313  \pm 0.00032  $ & 
$0.10319  \pm 0.00026   $ & 
$ -2.4/\!\!-\!2.5 $\\ 
$A_{\rm FB}^{0, c}$ & $ 0.0707 \pm 0.0035 $ & 
$  $  & 
$ 0.07369  \pm 0.00024  $ & 
$  0.07373  \pm 0.0002   $ & 
$-0.9  $\\ 
${A}_b$ & $ 0.923 \pm 0.020 $ & 
$  $  & 
$ 0.93475  \pm 0.00004 $ & 
$ 0.93476  \pm 0.00004   $ & 
$-0.6  $\\ 
${A}_c$ & $ 0.670 \pm 0.027 $ & 
$  $  & 
$ 0.66792 \pm 0.0002  $ & 
$ 0.66794 \pm 0.0002   $ & 
$ 0.1 $\\ 
$\sin^2\theta_{\mathrm{eff (Had. coll.)}}^{\mathrm{lept}}$ & $ 0.23143 \pm 0.00027 $ & 
$   \pm 0.00015  $  & 
$ 0.23151  \pm 0.00006 $ & 
$  0.23150 \pm 0.00005 $ & 
$-0.5/\!\!-\!0.9  $\\ 
\hline
\end{tabular}
\end{center}
}
\end{table}
%


The results of the SM global fit to EWPO for both the present (LHC) and
future (HL-LHC) scenarios are collected in Table
\ref{tab:SMfit}. These are given in the form of the mean and standard
deviation for each of the observables, as derived from the posterior
of the fits. For each EWPO the ``pull'' is also computed, defined as
the difference between the experimental value and the SM prediction
computed by removing each observable from the fit (not shown in the
table), normalized to the total uncertainty.  As it is apparent, the
differences in the posteriors between both fits are quite small.
However, looking at the pulls one can see that, should the central
values of the SM input parameters remain the same, the expected
improvements in their experimental uncertainties, combined with the
more precise measurements of some EWPO at the HL-LHC, would
significantly increase the tension between the indirect determinations
of $M_Z$, $m_t$, and $M_H$ from the EW fit and the corresponding
experimental measurements, pushing them to the 3$\sigma$ level.  The
improvement in the precision on $m_t$ would also reduce the parametric
uncertainty on some observables, e.g. the $W$ mass, bringing the total
residual error very close to the intrinsic uncertainty associated to
missing higher-order corrections in the calculation of $M_W$.  As in
the case of some of the SM inputs, the expected improvement on the
experimental precision of $M_W$, without a significant deviation on
the central value, would add some tension between theory and
experiment, pushing the pull for this observable well beyond the
2$\sigma$ level. The impact of the HL-LHC measurements on the EW fit
is well illustrated in Fig.~\ref{fig:MW_X} where one can see the
comparison between direct (i.e. experimental) and indirect constraints
on the fit input parameters given for both the current and HL-LHC
scenarios in the $M_W$ vs. $m_t$ and the $M_W$
vs. $\sin^2{\theta_{\mathrm{eff}}^{\mathrm{lept}}}$ planes
respectively.

\begin{figure}[h]
\centering
  \includegraphics[width=.49\textwidth]{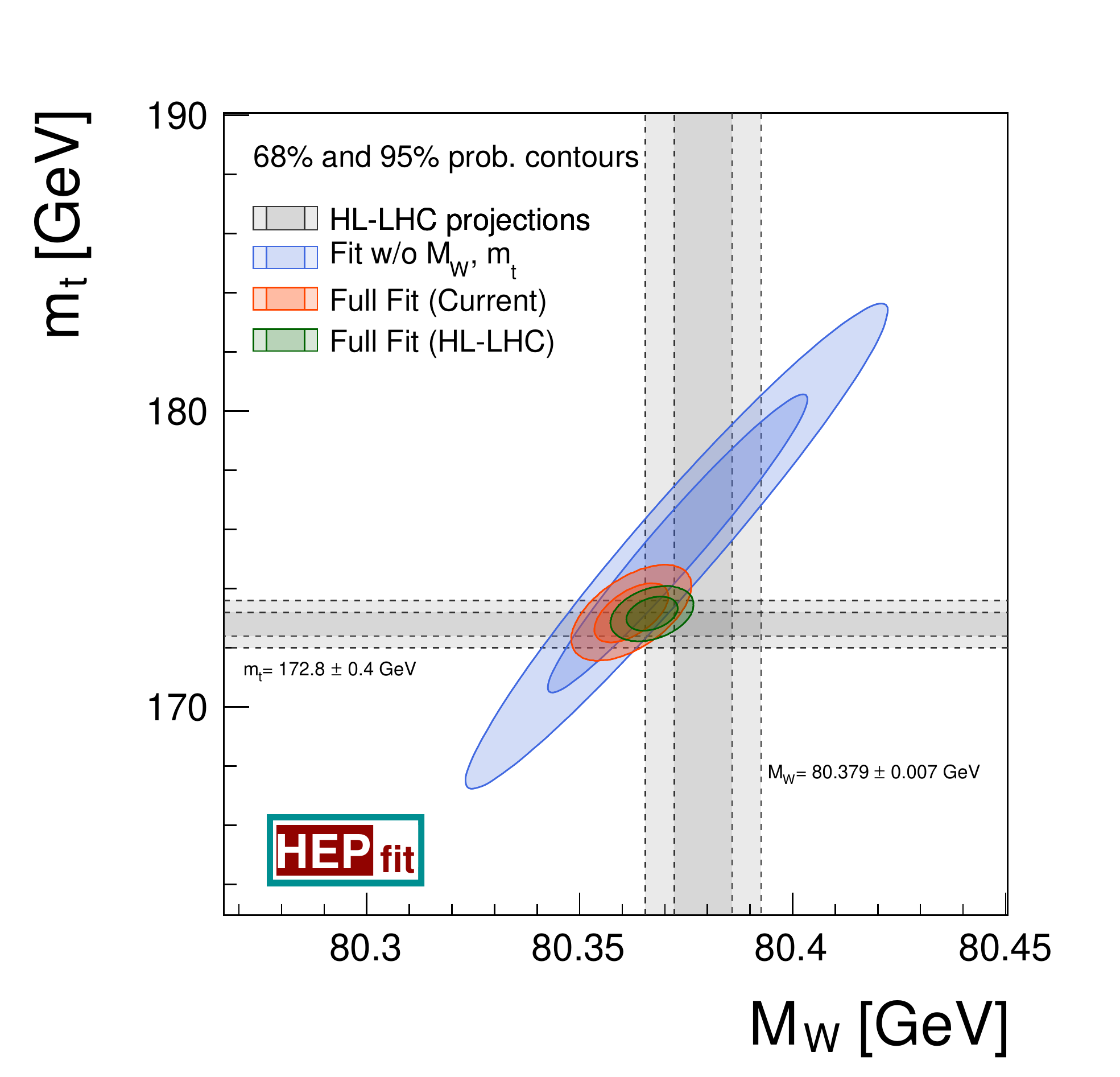} 
  \hspace{-3mm}
   \hspace{-0.1cm}\includegraphics[width=.49\textwidth]{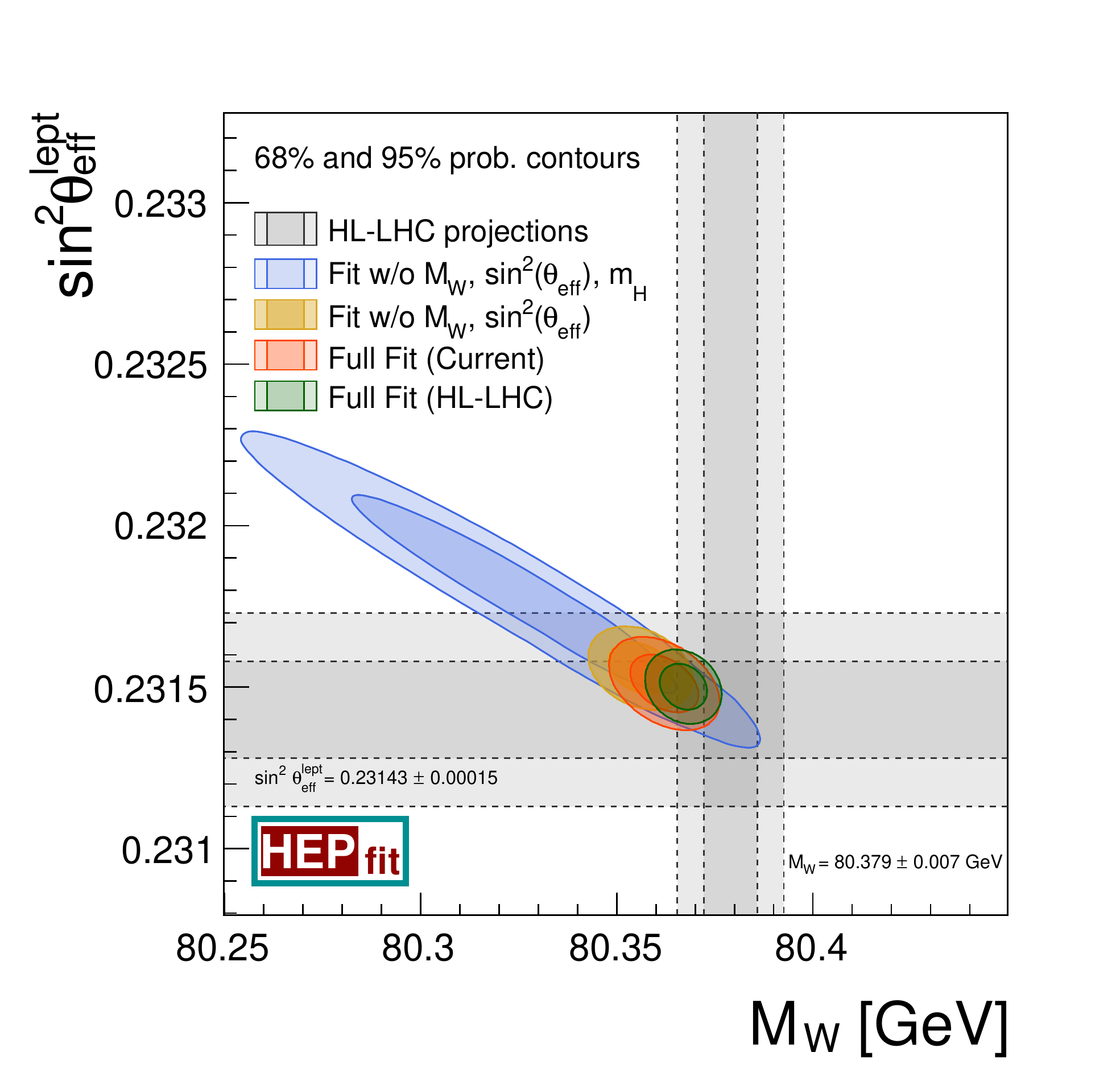}
  \vspace{-0.3cm} 
  \caption{Comparison of the indirect constraints on $M_W$ and
    $m_t$ with the current experimental measurements and the expected
    improvements at the HL-LHC (left). The same in the $M_W$-$\sin^2{\theta_{\mathrm{eff}}^\mathrm{lept}}$ plane (right).~\label{fig:MW_X}}
\end{figure}


The EWPO, being measured in processes mediated by the exchange of a
$Z$ or $W$ boson, are extremely sensitive to any new physics that
modifies the propagation of such particles. This results in a
universal modification of the interactions between the EW
gauge bosons and the SM fermions, which, from the point of view of
EWPO, can be described in terms of only three parameters: the
well-known $S$, $T$, and $U$ oblique
parameters~\cite{Peskin:1991sw}. The study of the constraints on the
$S$, $T$, and $U$ parameters is one of the classical benchmarks in the
study of EW precision constraints on new physics, and it is
well motivated from a theory point of view, within the context of
universal theories. The results of the fit to the $S$, $T$, and $U$
parameters are given in Table~\ref{tab:STUfit}. The results are presents in
terms of the full ($S$,$T$,$U$) fit and also assuming $U=0$, which is
motivated in theories where EW symmetry breaking is realised
linearly, since in that case $U\ll S,T$. In both cases the current
constraints are compared with the expected precision at the HL-LHC,
which, in some cases, could improve the sensitivity to such new
physics effects by up to $\sim 30\%$.  The results for the $ST$ fit
($U=0$) are shown in Fig.~\ref{fig:ST}, illustrating also the
constraints imposed by the different EWPO.

\begin{table}[h]
{\small
\begin{center}
 \caption{Results of the fit for the oblique parameters $S$, $T$, $U$;
   and $S$, $T$ $(U=0)$. Projections for the uncertainties at the
   HL-LHC are given in the last column.\label{tab:STUfit}}
\begin{tabular}{|c c rrr c |}
 \hline
 & ~~Result & \multicolumn{3}{c}{Correlation Matrix} & Precision at HL-LHC \\ 
 \hline\hline
$S$ & $0.04 \pm 0.10 $ & $\phantom{+}1.00$ &&&0.09 \\ 
$T$ & $0.08 \pm 0.12 $ & $\phantom{+}0.90$ & $1.00$&&0.12\\ 
$U$ & $0.00 \pm 0.09 $ & $-0.62$ & $-0.84$ & $1.00$&0.08 \\
 \hline
$S$ & $0.04 \pm 0.08 $ & $1.00$ &&&0.06\\ 
$T$ & $0.08 \pm 0.06 $ & $0.90$ & $1.00$ &&0.05\\ 
$(U=0)$ &  &  &  &  & \\
\hline
 \end{tabular}
 \end{center}
 }
\end{table}

\begin{figure}[h]
\centering
   \hspace{-0.1cm}\includegraphics[width=.49\textwidth]{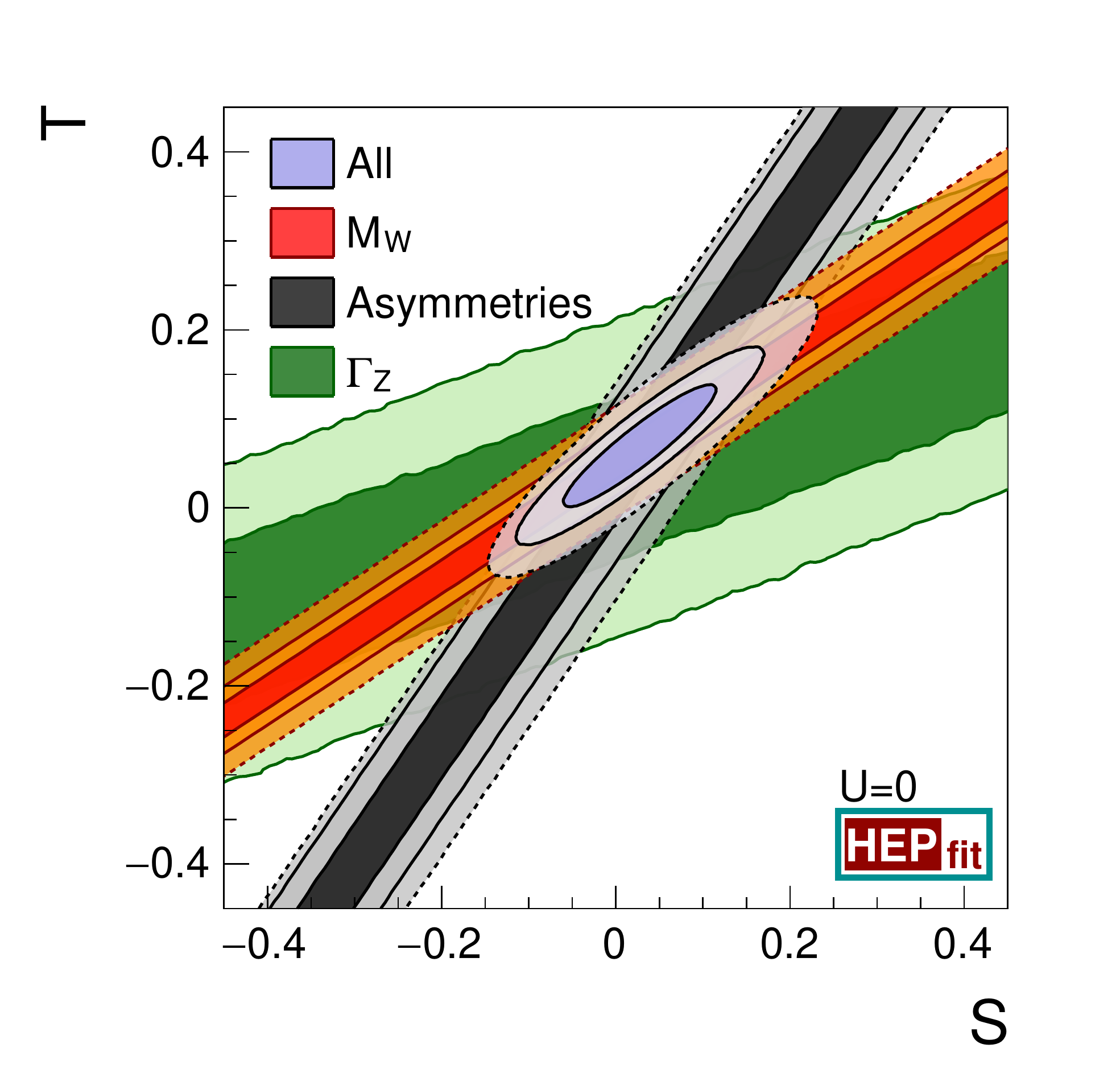}\vspace{-0.4cm}
  \caption{Comparison of the currently
    allowed 68\% and 95\% probability regions in the $S$, $T$ fit
    ($U=0$) (dashed contours) with the HL-LHC projections (solid contours). The different bands
    illustrate the bounds from the different EWPO included in the fit and the projected improvements
    at the HL-LHC.~\label{fig:ST}}
\end{figure}

As stressed above, the $STU$ parameterisation only describes universal
deformations with respect to the SM predictions.  In order to
systematically explore the impact of global EW precision fits on new
physics, the framework of the SMEFT is adopted in what follows.  In
this formalism, the SM Lagrangian is extended via operators of
dimension five and higher, i.e.
\begin{equation} 
\label{eq:smeft-lagrangian} 
{\cal L}_{\mathrm{eff}}={\cal
  L}_{\mathrm{SM}}+\sum_{d>4}\frac{1}{\Lambda^{d-4}}{\cal L}_d, 
~~\mbox{with}~~
{\cal L}_d=\sum_i C_i {\cal O}^{(d)}_i,~~\left[{\cal O}^{(d)}_i\right]=d\,, 
\end{equation}
where $\Lambda$ denotes the cut-off scale of the SMEFT.  This new
physics scale introduces a first hierarchical ordering between
contributions of operators of lower versus higher dimension, where
higher-dimension operators are suppressed by inverse powers of
$\Lambda$. Each term in ${\cal L}_d$ is a linear combination of
$d$-dimensional operators $O_i^{(d)}$ built in terms of SM fields,
with Wilson coefficients $C_i$ that can depend on both SM masses and
couplings, as well as new physics parameters.  For the analysis of
EWPO the leading new physics corrections come from dimension-six
operators (${\cal L}_6$). The study is limited to this order in the
effective theory expansion.  Using the complete basis of dimension-six
interactions presented in Ref. \cite{Grzadkowski:2010es}, the $Z$-pole
and $W$ observables in Table~\ref{tab:SMfit} are corrected at the
leading order by 10 different operators. The bosonic operators
$${\cal O}_{\phi D}=|\phi^\dagger D^\mu \phi|^2~~\mbox{ and }~~{\cal O}_{\phi WB}=(\phi^\dagger \sigma_a \phi) W_{\mu\nu}^a B^{\mu\nu},$$
modify the gauge-boson propagators in a way similar to the $T$ and $S$
parameters, respectively.  Among the remaining operators,
$${\cal O}_{\phi \psi}^{(1)}=(\phi^\dagger \lrD^\mu \phi) (\overline{\psi}\gamma_\mu \psi )~~\mbox{ and }~~{\cal O}_{\phi F}^{(3)}=(\phi^\dagger \sigma_a \lrD^\mu \phi) (\overline{F}\gamma_\mu \sigma_a F),$$
with $\psi=l,~e,~q,~u~,d$ and $F=l,q$ (where $l$ and $q$ denote the SM
left-handed fermion doublets, $e,u,d$ the SM right-handed fermion
singlets, and flavour universality is assumed), correct, upon
EW symmetry breaking, the EW couplings of the $Z$ and $W$
bosons to quarks and leptons. Finally, the four-lepton operator
${\cal O}_{ll}=\left(\overline{l}\gamma_\mu l\right) \left(\overline{l}\gamma^\mu l\right)$
modifies the muon decay amplitude and, by affecting the extraction of
the Fermi constant, propagates its effect to all the different
observables considered in the EW global fit.

The aim of a global fit to EWPO data is to constrain the corresponding
Wilson coefficients.  Of the ten operators considered, only eight
combinations can be constrained using EW precision data in the case of
flavour universal couplings.  This means that in the basis of
\cite{Grzadkowski:2010es} there are two flat directions which, for
simplicity are lifted by performing a field redefinition to exchange
$C_{\phi D}$ and $C_{\phi WB}$ with two interactions that do not enter
in EWPO.  The results of the fit to EWPO using the projected HL-LHC
data are shown in Fig.~\ref{fig:dim6}, both for the case in which
the eight remaining coefficients are active and fitted simultaneously
and for the case in which only one coefficient at a time is active and
independently fitted.  The results of both fits are also summarised in
Table~\ref{tab:dim6fit} where the HL-LHC bounds are additionally
compared to current bounds. It can be seen that the HL-LHC could improve
the current bounds on some of the considered Wilson coefficients by up
to a 10-30\%, although for most coefficients the effect is much milder
both when different effective interactions are fitted simultaneously
and individually.

\begin{figure}[h]
\centering
  \hspace{-0.1cm}\includegraphics[width=.8\textwidth]{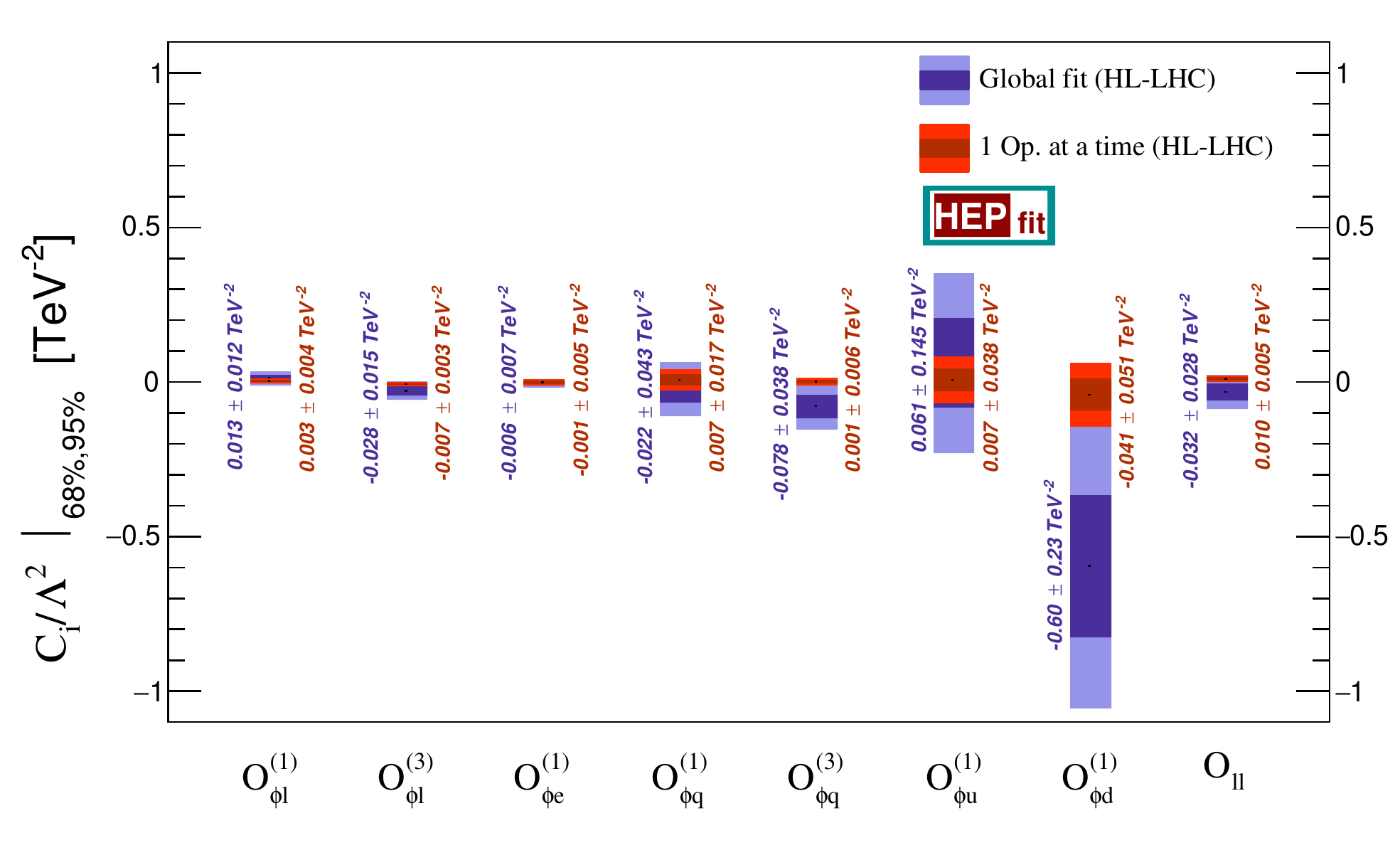}
  \vspace{-0.4cm}
  \caption{$68\%$ and $95\%$ probability limits on the dimension-six operator coefficients $C_i/\Lambda^2$ [TeV$^{-2}$] from the global fit to EWPO at HL-LHC including all operators (in blue), compared with the limits obtained assuming only one operator at a time (in red). See Table~\ref{tab:dim6fit} for the comparison with current uncertainties.~\label{fig:dim6}}
\end{figure}

\begin{table}[h]
{\small
\begin{center}
\caption{Results of the fit to the coefficients of the SMEFT
   dimension-six Lagrangian. The uncertainties shown refer to the fit
   performed assuming the presence of only one effective operator at a
   time and to the case when all (eight) operators are active at the
   same time (global fit). Projections for the uncertainties at the HL-LHC are given in the last two columns. Result shown for the ratios $\overline{C}_i \equiv C_i/\Lambda^2$. See text for details.
   \label{tab:dim6fit}}
\begin{tabular}{|l c c c c| }
 \hline
               & \multicolumn{2}{c}{Current uncertainty} & \multicolumn{2}{c|}{Precision at HL-LHC} \\ 
               & \multicolumn{2}{c}{[TeV$^{-2}$]} & \multicolumn{2}{c|}{ [TeV$^{-2}$]} \\ 
 Operator& 1 op. at& Global & 1 op. at  & Global  \\ 
 Coefficient &a time &fit  &a time & fit  \\ 
 \hline\hline
~~~$\overline{C}_{\phi l}^{(1)}$ & 
$0.004 $ & 
$0.012 $ & 
$0.004 $ & 
$0.012$  \\ 
~~~$\overline{C}_{\phi q}^{(1)}$ & 
$0.018 $ & 
$0.044 $ & 
$0.017 $ & 
$0.043 $  \\ 
~~~$\overline{C}_{\phi e}$ & 
$0.005 $ & 
$0.009 $ & 
$0.005 $ & 
$0.007 $  \\ 
~~~$\overline{C}_{\phi u}$ & 
$0.040 $ & 
$0.146 $ & 
$0.038 $ & 
$0.145 $  \\ 
~~~$\overline{C}_{\phi d}$ & 
$0.054 $ & 
$0.237 $ & 
$0.051 $ & 
$0.230$  \\ 
~~~$\overline{C}_{\phi l}^{(3)}$ & 
$0.004 $ & 
$0.017 $ & 
$0.003 $ & 
$0.015 $  \\ 
~~~$\overline{C}_{\phi q}^{(3)}$ & 
$0.007 $ & 
$0.040$ & 
$0.006 $ & 
$0.038$  \\ 
~~~$\overline{C}_{ll}$ & 
$0.007 $ & 
$0.028 $ & 
$0.005 $ & 
$0.028 $  \\ 
\cmrule
~~~$\overline{C}_{\phi WB}$ & 
$0.003 $ & 
$-$ & 
$0.002 $ & 
$-$  \\ 
~~~$\overline{C}_{\phi D}$ &
$0.007 $ & 
$-$ & 
$0.005 $ & 
$-$  \\ 
\hline
 \end{tabular}
 \end{center}
 }
\end{table}




\newpage

\section{Strong interactions}

This section presents studies at the HL-LHC and HE-LHC conditions for jet and photon production, parton density functions, underlying event and multi/double-parton interactions. Thanks to the larger integrated luminosity at the HL-LHC and HE-LHC and the jump in centre-of-mass energy at the HE-LHC, an increase in the kinematic reach is expected for light- and heavy-flavour jet production as well as photon production. An improvement is also expected in the experimental systematic uncertainty on the jet calibration. The measurements of jet and photon production cross sections in addition to other processes, e.g. Drell-Yan and top quark, at the HL-LHC will help improve the understanding of the parton density functions. The level of the underlying event activity is not expected to change significantly at the HL-LHC given the small increase in centre-of-mass energy from $\sqrt{s}=13$ TeV at the LHC Run-2 to $\sqrt{s}=14$ TeV at the HL-LHC, however a significant increase is expected at the HE-LHC energy of $\sqrt{s}=27$ TeV.
Multi-parton interactions are expected to play a more significant role at higher energies and, thanks to the large statistics available at both the HL- and HE-LHC, new measurements can be carried out to test more precisely the current theoretical models. 


\subsection[Jet and photon production]{Jet and photon production
\footnote{Contributed by the experimental collaborations, and by L. Cieri, G. Ferrera, A. Huss, and J. Pires.}}

This section presents phenomenological studies of inclusive jet, dijet, heavy-flavour jet production as well as inclusive photon, diphoton, and associated photon and jet production at future upgrades at the HL and HE stages of the LHC. In particular the reach in yields for these processes is investigated.  A comparison between the results expected at the future design centre-of-mass energies of $\sqrt{s}=14$~TeV and $\sqrt{s}=27$~TeV is presented.


\subsubsection{Inclusive jet production}
Jets are reconstructed using the anti-$k_{\rm T}$ algorithm \cite{Cacciari:2008gp} with distance parameter $R$=0.4 as implemented in the FastJet software package \cite{Cacciari:2011ma}, and calibrated following the procedure described in \cite{PERF-2016-04}.  The total jet energy scale (JES) uncertainty in ATLAS Run-2 measurements comprises of 88 sources, and all need to be propagated  through the analysis in order to  correctly account for uncertainty correlations in the jet calibration in the final result. 
Here follows a summary of the analysis detailed in Ref.~\cite{ATL-PHYS-PUB-2018-051}.

A reduced set of uncertainty components (nuisance parameters) is derived from eigenvectors and eigenvalues of the diagonalised total JES covariance matrix on the jet level. The globally reduced configuration  with 19  nuisance parameters (NPs) is used in this study.  Eight  NPs coming from the in~situ techniques are related to the detector description, physics modelling and measurements of the $Z/\gamma$  energies in the ATLAS calorimeters. Three  describe the physics  modelling, the statistics of the dijet MC sample and the non-closure of the method used to derive the $\eta$-intercalibration \cite{PERF-2016-04}.  Single-hadron response studies \cite{PERF-2015-05} are used to describe the JES uncertainty in the high-\pT\ jet regions, where the  in~situ studies have limited statistics.
Four NPs  are due to the pile-up corrections of the jet kinematics that take into account mis-modelling of $N_\textrm{PV}$ and $\langle\mu\rangle$  distributions, the average energy density $\rho$, and the residual \pT\ dependence. Finally, two uncertainty components take into account  the difference in the calorimeter response to the quark- and  gluon-initiated jets (flavour response) and the jet flavour composition, and one  uncertainty estimates the correction for the energy leakage beyond the calorimeter, the ``punch-through'' effect. 

In order to estimate the precision in the jet cross section measurements at the HL-LHC, three scenarios of possible uncertainties in the jet energy scale calibration are defined. 

In all three scenarios, the high-\pT\ uncertainty, the punch-through uncertainty and the flavour composition uncertainty are considered to be negligible. 
The JES uncertainty in the high-\pT\ range will be accessed using the multi-jet balance (MJB) method, rather than single hadron response measurements, since the high statistics at the HL-LHC will allow precision JES measurements in the high-\pT\ region. Flavour composition and flavour response uncertainties are derived from the MC generators. With the advances in the MC modelling and development of tunes, these uncertainties could be significantly reduced. The flavour composition  uncertainties are set to zero to highlight the maximal impact of possible future improvements in the understanding of parton shower and hadronisation modeling on the precision of the jet energy measurements. The flavour response uncertainties are kept the same as in Run-2 or reduced by a factor of two in conservative and optimistic scenarios, respectively.

The pile-up uncertainties, except the $\rho$ topology uncertainty, are considered to be negligible. Current small uncertainties in the JES due to mis-modelling of $N_\textrm{PV}$ and $\langle\mu\rangle$  distributions and the residual \pT\ dependence   lead to a very small uncertainties   at the HL-LHC conditions. With the advances of new pile-up rejection techniques, the $\rho$ topology uncertainty could be  maintained at a level comparable to the one in Run-2 or reduced by a factor of two. This is addressed in conservative  and optimistic scenarios.
\begin{figure}[t]
\centering
\includegraphics[width=0.65\linewidth]{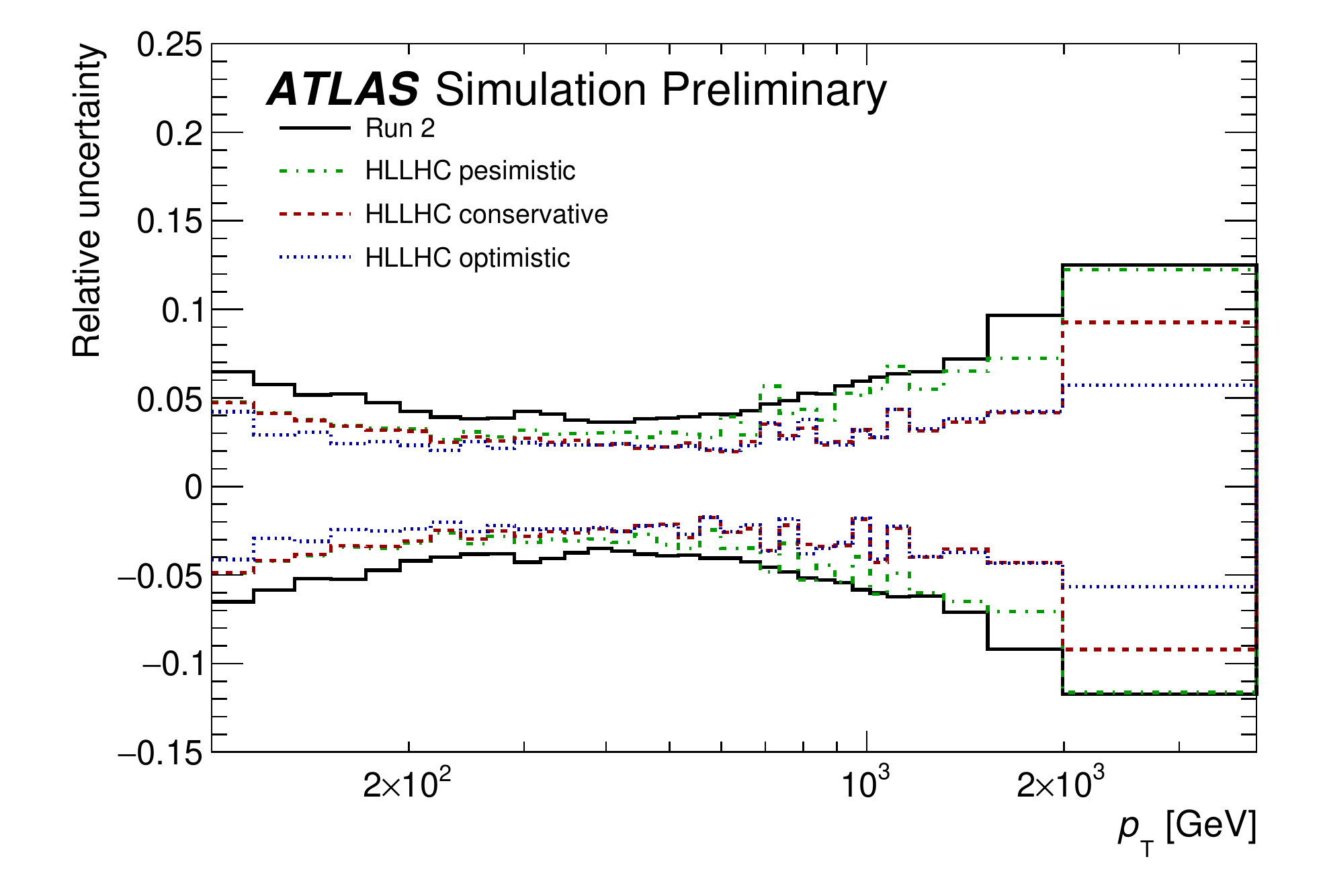}
\caption{\label{fig:uncertFinal} Relative uncertainties in the inclusive jet cross section measurements at the HL-LHC due the JES uncertainties. Three HL-LHC scenarios are compared to the Run-2 performance. Black line corresponds to the Run-2 performance. Green, red and blue lines represent pessimistic, conservative and optimistic scenarios, respectively. }
\end{figure}

Since the  Run-2 jet energy resolution (JER) uncertainty estimation is conservative, the final Run-2 JER uncertainty is expected (based on Run-1 experience) to be about twice as small as the current one. Therefore, the JER uncertainty is estimated to be half of that in Run-2. 
 
The remaining uncertainty sources are fixed in different scenarios as follows:

\begin{itemize}
\item Conservative scenario: 
\begin{itemize}
\item  All in~situ components are kept the same as in Run-2, except the uncertainties related to  the photon energy measurement in the  high-$E_{\mathrm{T}}$ range and the MJB method uncertainties whose uncertainties are reduced by a factor of two, since those are expected to be improved at the HL-LHC;
\item The MC modelling uncertainty in the $\eta$-intercalibration is reduced by a factor of two while the other two are neglected. Currently, the MC modelling uncertainty is derived through a comparison of leading-order (LO)  pQCD generators. With future advances in next-to-leading-order MC generators this uncertainty is expected to improve;
\item The flavour response uncertainty is set to the Run-2 value;
\item The $\rho$-topology uncertainty   is unchanged compared to Run-2 results;
\end{itemize}
\item Optimistic scenario:
\begin{itemize}
\item  All in~situ components  are treated identically to the conservative scenario;
\item All three  uncertainty sources in the $\eta$-intercalibration method  are set to zero;
\item The flavour response uncertainty is reduced by a factor of two compared to Run-2 results;
\item $\rho$-topology uncertainty   is two times smaller as in Run-2;
\end{itemize}
\item Pessimistic scenario:
\begin{itemize}
\item  Same as  the optimistic scenario, but all uncertainty sources of  in~situ methods  are retained from Run-2.
\end{itemize}
\end{itemize}

All components of the JES uncertainty  are propagated from the jet-level to the cross section level as follows. The jet \pT\  is scaled up and down by one standard deviation of each source of uncertainty.   The difference between the nominal detector-level  spectrum and the systematically shifted one is taken as a systematic uncertainty.  All JES uncertainties are treated as bin-to-bin correlated and independent from each other in this procedure. The unfolding of the detector-level distributions to the particle-level spectrum is not performed is this study.  A possible modification of the shapes of uncertainty components during the unfolding procedure is expected to be small and neglected in this study.

The inclusive jet cross-sections are studied as a function of the jet transverse momentum for jets with \pT\ $ >1 00$~GeV and within $|y|<3$.
The total JES uncertainty in the inclusive jet cross section measurement for the three HL-LHC  scenarios is depicted in Fig.~\ref{fig:uncertFinal} and is compared to the total JES uncertainty estimate for the  Run-2 jet cross section measurements.  The total JES uncertainty in the low \pT\ range is the same as in Run-2 and is about 2\% better in the high-\pT\ region. In the conservative and pessimistic scenarios the JES uncertainties in the cross section  are very similar in the intermediate and high-\pT\ range, while the JES uncertainty is about 1\% better in the low-\pT\ range for the optimistic scenario. 

The predicted number of events estimated using the program \mbox{\textsc{NNLOJET}}\xspace~\cite{Gehrmann:2018szu}, which includes next-to-next-to-leading order QCD calculations 
for both single jet inclusive~\cite{Currie:2016bfm} and dijet inclusive~\cite{Currie:2017eqf} production, is shown in Fig.~\ref{fig:jetevt} (left and right respectively). In the dijet analysis, 
a second jet with \pT~$>$ 75 GeV is required in the event. 
The lower panels show the ratios of events yields at 27~TeV and 14~TeV. This plot shows an
enhancement of the cross section growing with the jet \pT\ (left) and dijet mass (right). In summary, assuming ${\cal L}_{int}=3 ~{\rm ab}^{-1}$ of $pp$ collision data at $\sqrt{s} =$ 14 TeV the \pT\ reach 
of the measurement is 5~TeV with the observation of dijet events of mass up to 9~TeV.  At the HE-LHC upgrade, an increase in cross section by a factor between $10^3$ and $10^6$ in the tails
of the distributions extends the \pT\ range of the measurement by a factor of 2 up to 9~TeV, allowing the observation of dijet events of mass up to 16~TeV.
\begin{figure}[t!]
\begin{center}
\subfloat[]{\includegraphics[width=0.45\textwidth]{\main/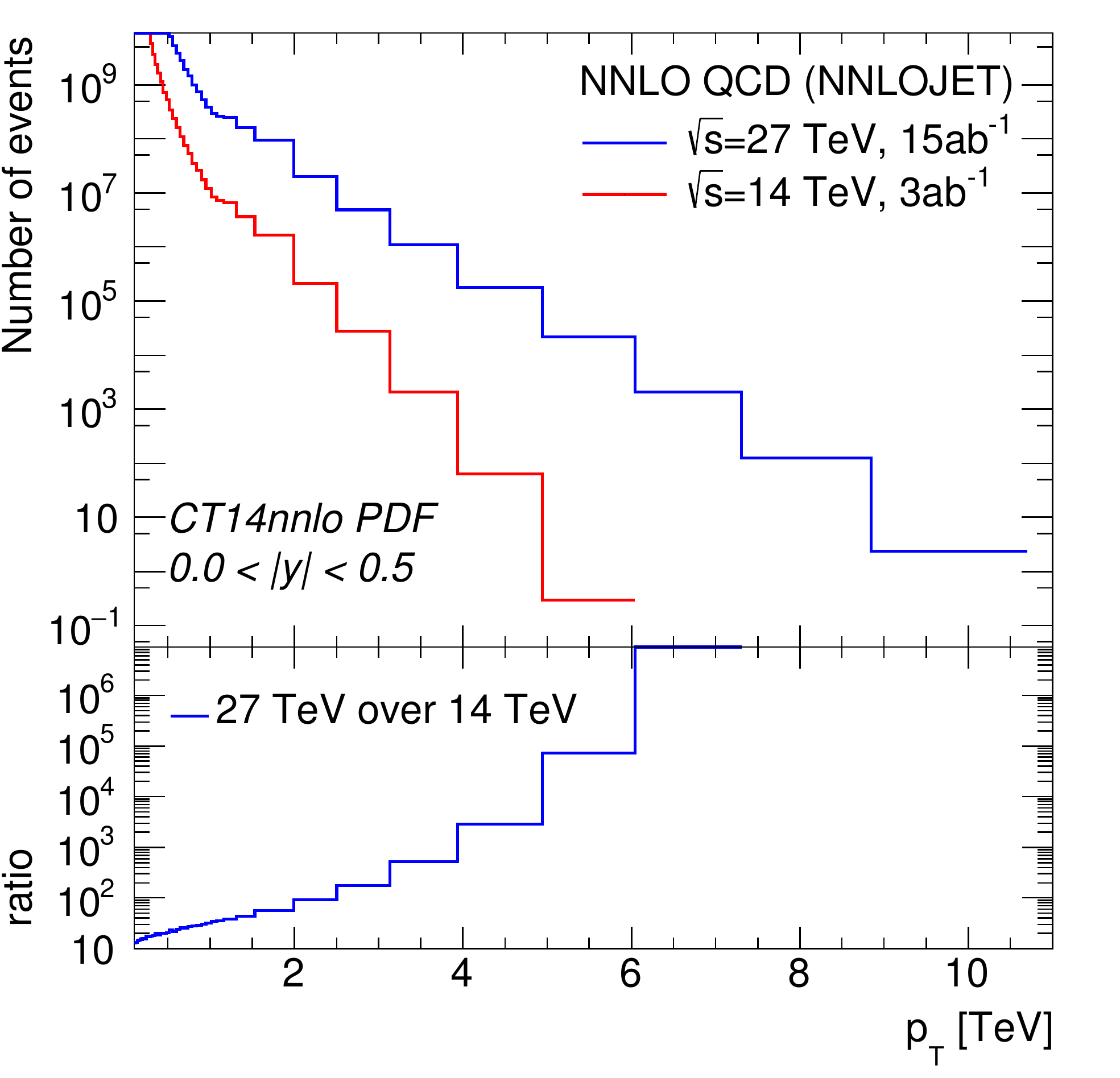}}
\subfloat[]{\includegraphics[width=0.45\textwidth]{\main/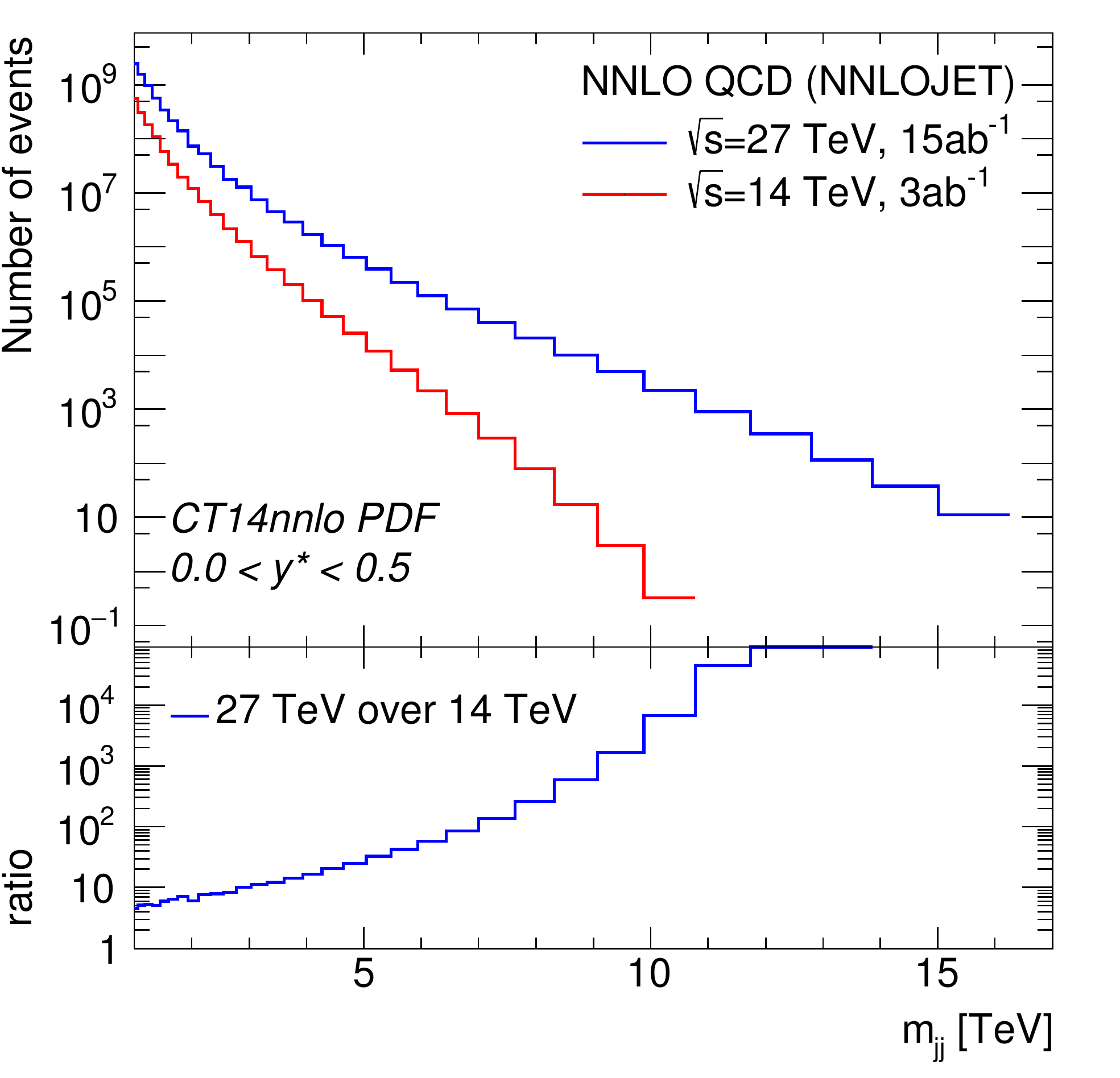}}
\end{center}
\caption{Predicted number of inclusive jet events as a function of the jet \pT\ (left) and dijet events as a function of dijet mass $m_{jj}$ (right) at NNLO,
assuming an integrated luminosity of 3 ${\rm ab}^{-1}$ (15 ${\rm ab}^{-1}$) of $pp$ collision data at $\sqrt{s}$=14 TeV ($\sqrt{s}$=27 TeV).}
\label{fig:jetevt}
\end{figure}

The increase in cross section in these scenarios will allow for a very precise multi-differential measurement of  inclusive jet production. Working at a fixed centre-of-mass energy, the high-\pT\, the high-$x$ and the large $Q^2$ region are probed and the sensitivity
to higher order QCD/EW effects and BSM signals is increased. On the other hand, at fixed-\pT, an increase in the collider energy 
and the inclusion of the forward detector regions increase the coverage to the low-$x$ regime, which is highly sensitive to small-$x$
resummation effects. For these reasons, it will be necessary to have accurate jet predictions covering both regions. 

To this end Fig.~\ref{fig:kfact} presents the double-differential $k$-factors at $\sqrt{s}=14$~TeV (left) and
$\sqrt{s}=27$~TeV (right) for the inclusive jet \pT\ (top), differentially in \pT\ and rapidity $|y|$ and dijet mass (bottom) differentially 
in dijet mass $m_{jj}$ and rapidity difference $y^*=1/2~|y_{j1}-y_{j2}|$. The shaded bands assess the scale uncertainty at different perturbative orders, LO,
NLO and NNLO. As for the value of the renormalization ($\mu_R$) and factorization ($\mu_F$) scales $\mu=\hat{H}_T$ is used, i.e.
the scalar sum of the \pT\ of all partons in the event, as recommended in~\cite{Currie:2018xkj} for the inclusive jet \pT, and 
the dijet mass $\mu=m_{jj}$ for the dijet mass distribution, as recommended in~\cite{Currie:2017eqf}.

For the inclusive jet \pT\ large NLO effects at high-\pT\ and central rapidity of approximately 
90\% (14~TeV) and 50\% (27~TeV) are observed with large NLO scale uncertainties of ${\cal O}(20-30\%)$. At NNLO moderate
corrections across the entire \pT\ and rapidity range are observed, except at high-\pT\ in the central rapidity slices where the
NNLO effects can reach between 10 to 30\%. An excellent convergence of the perturbative result is observed as well as a significant reduction in the scale uncertainty of the cross section when going from NLO to NNLO. 
The NNLO scale uncertainties are estimated at the $< 5\%$ level. 
\begin{figure}[t!]
  \begin{center}
    \subfloat[]{\includegraphics[width=0.45\textwidth]{\main/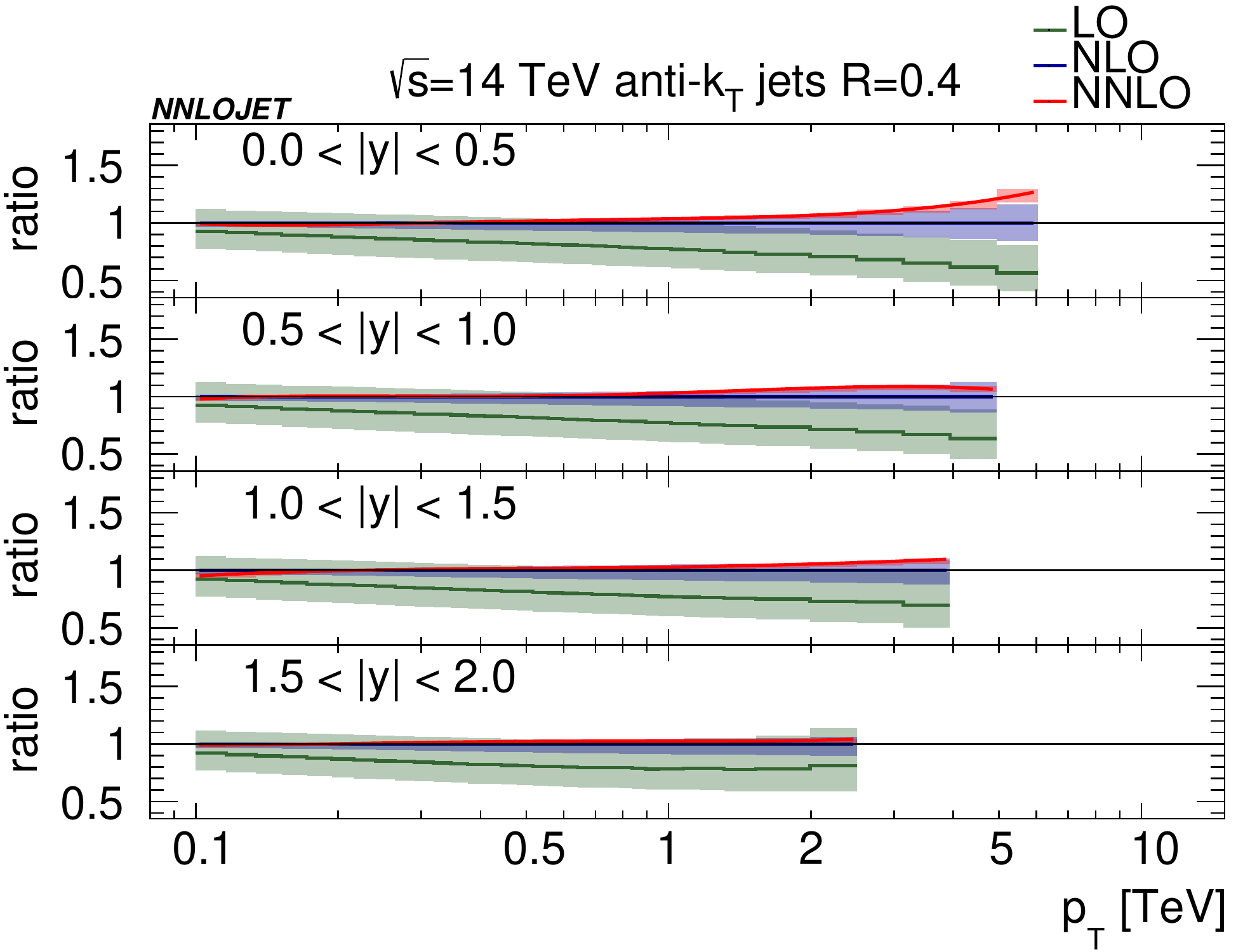}}
    \subfloat[]{\includegraphics[width=0.45\textwidth]{\main/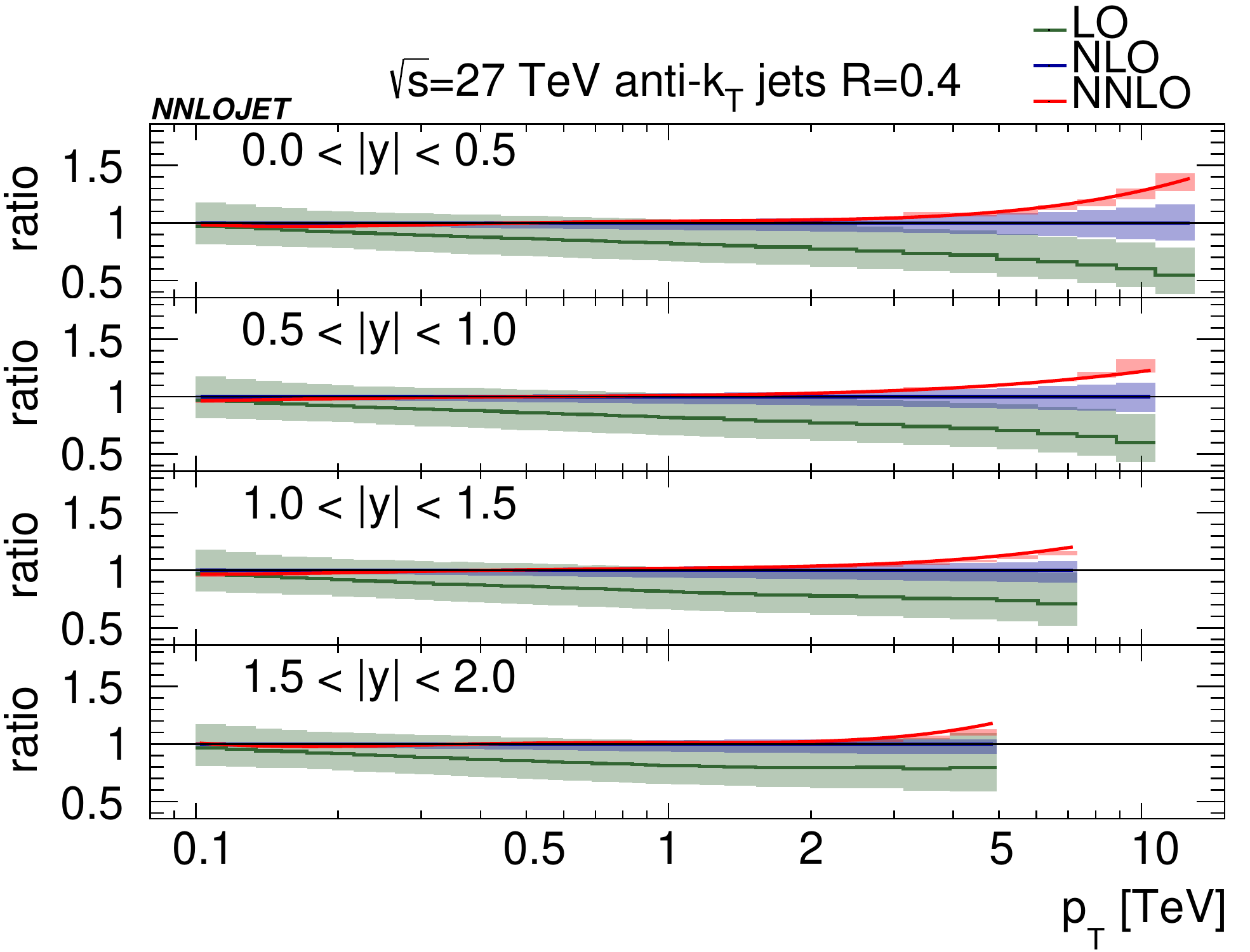}}\\
    \subfloat[]{\includegraphics[width=0.45\textwidth]{\main/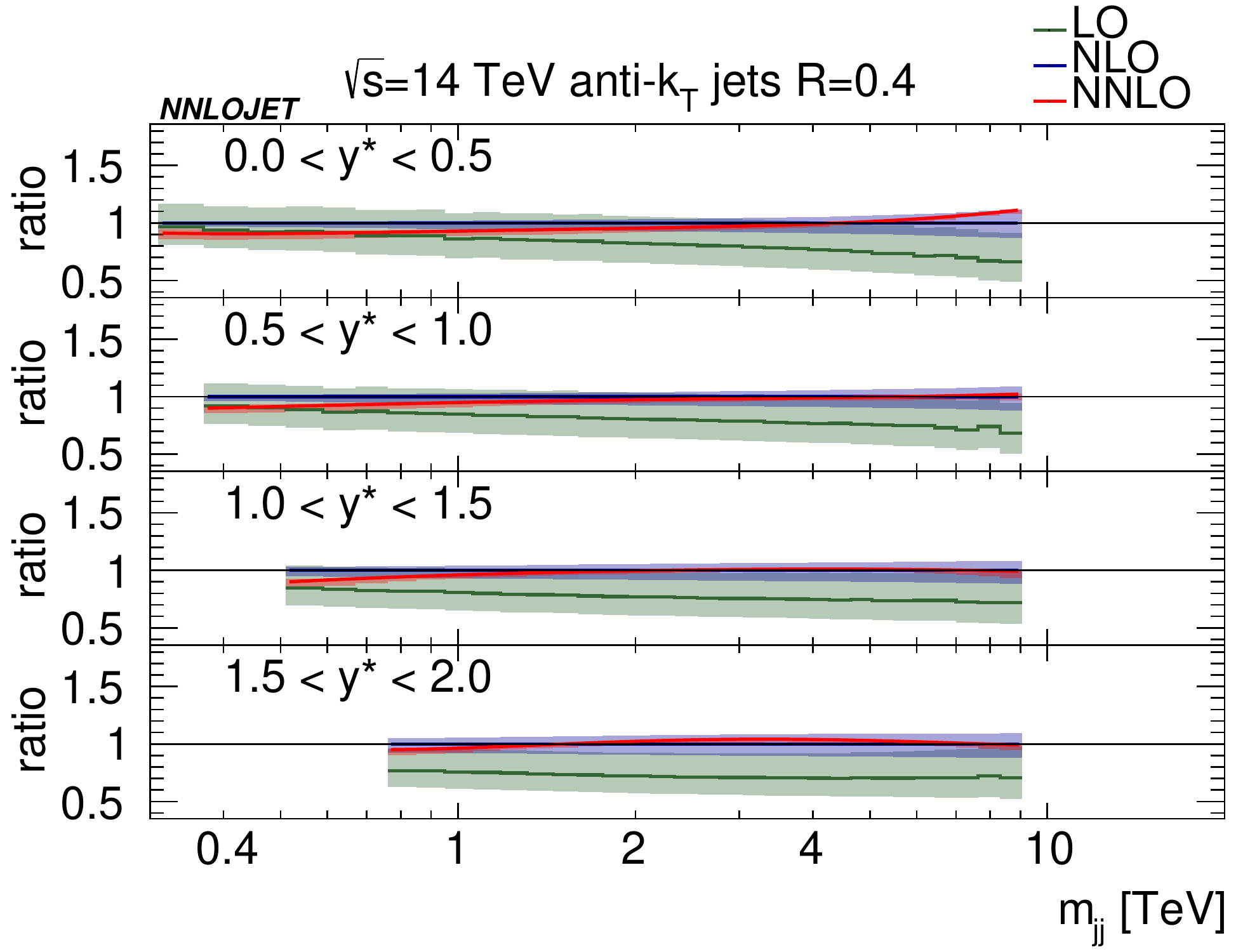}}
    \subfloat[]{\includegraphics[width=0.45\textwidth]{\main/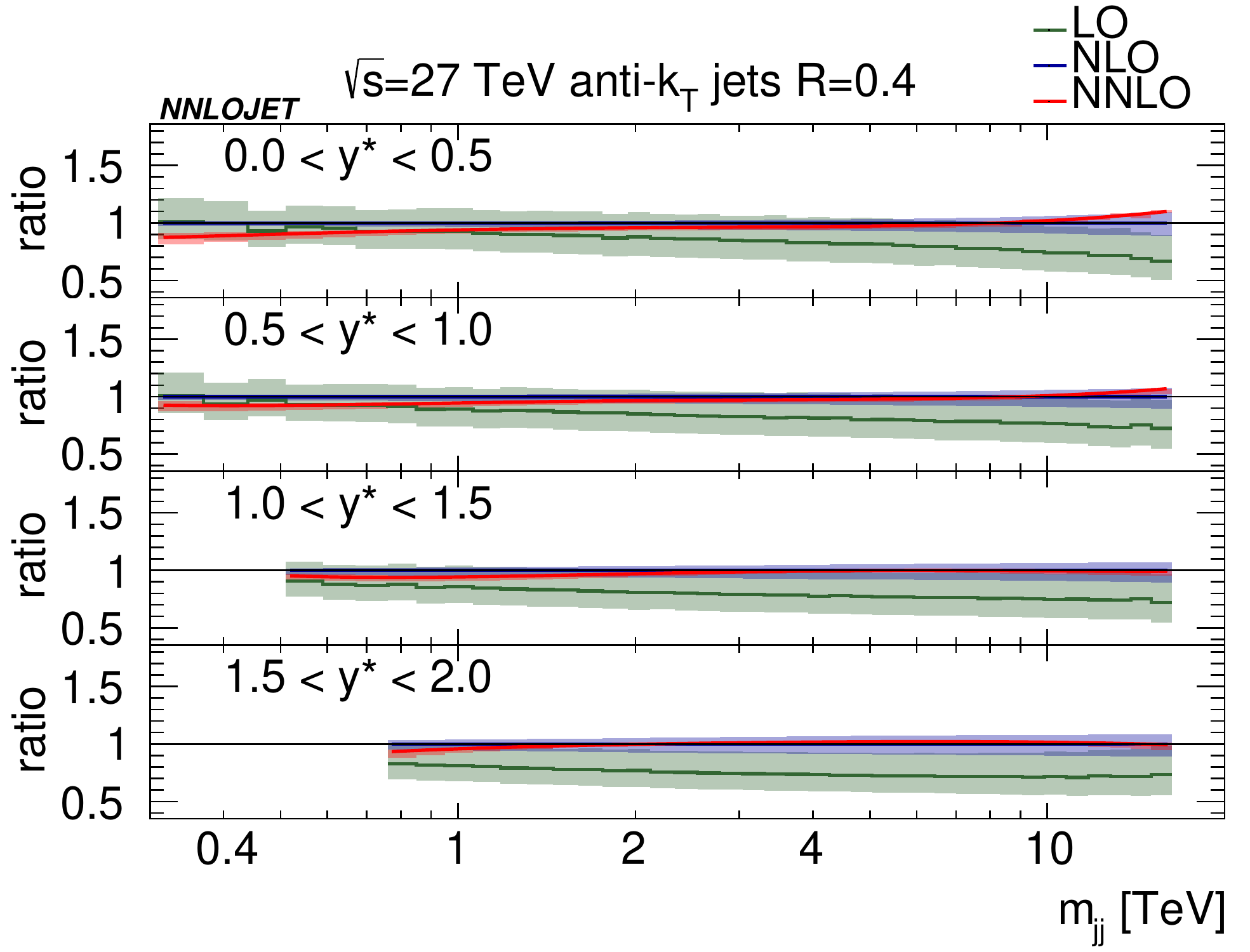}}
  \end{center}
\caption{Predictions for the inclusive jet \pT\ (top) and dijet mass $m_{jj}$ (bottom) at LO (green), NLO (blue) and NNLO (red)
at (a and c) $\sqrt{s}=14$~TeV and (b and d) $\sqrt{s}=27$~TeV normalised to the NLO result.}
\label{fig:kfact}
\end{figure}
Similarly to the inclusive jet \pT\ case, an excellent convergence of the perturbative result for the dijet mass is observed. The NNLO/NLO $k$-factors
are typically $< 10\%$ and alter the shape of the prediction at low $m_{jj}$ and low $y^*$. A large reduction is observed in the scale variation and NNLO 
scale uncertainties are estimated to be below the $5\%$ level, even at large $m_{jj}$. Scale uncertainties at this level are well below the PDF uncertainty, 
highlighting the huge potential to constrain PDFs with inclusive jet data.




Measurements of weak bosons \cite{Aad:2011dm}, top quarks \cite{ATLAS:2012aa}, photon and jet production \cite{Aad:2011fc} (and many others) performed by the LHC Collaborations have been already used by the global PDF groups \cite{Ball:2017nwa,Dulat:2015mca,Alekhin:2017kpj,Harland-Lang:2014zoa} in the determination of the proton structure.  Comparisons of inclusive jet and dijet production cross sections using different PDF sets at $\sqrt{s}=14$ and 27~TeV, show 5--10\% differences respectively between central values in the low and intermediate \pT\ and $m_{jj}$ regions, consistent with current PDF uncertainties. Larger differences between 
the predictions of the various PDF sets in the high-\pT\ and $m_{jj}$ range highlight the expected constraining power of future measurements at the HL-LHC and HE-LHC.

A study to estimate the impact of future PDF-sensitive measurements at the HL-LHC on PDFs determination was performed in \cite{Khalek:2018mdn} and reported in Sec.~\ref{sec:ultimatepdfs}. Three possible scenarios for the  experimental systematic uncertainties were considered. This study concluded that HL-LHC measurements will further reduce the PDF uncertainties, and published dedicated PDF sets, PDF4LHC~HL-LHC, with the inclusion of HL-LHC pseudo-data in the fits.   
Figure \ref{fig:ultPDF} ~depicts the comparison of PDF uncertainties in the inclusive jet and dijet production cross sections for CT14 and PDF4LHC~HL-LHC~(conservative scenario) in $pp$ collisions at $\sqrt{s}=14$ and 27~TeV. A significant reduction in the PDF uncertainty is expected with the inclusion of PDF-sensitive measurements in HL-LHC PDF fits.
\begin{figure}
  \begin{center}
    \subfloat[]{\includegraphics[width=0.45\textwidth]{\main/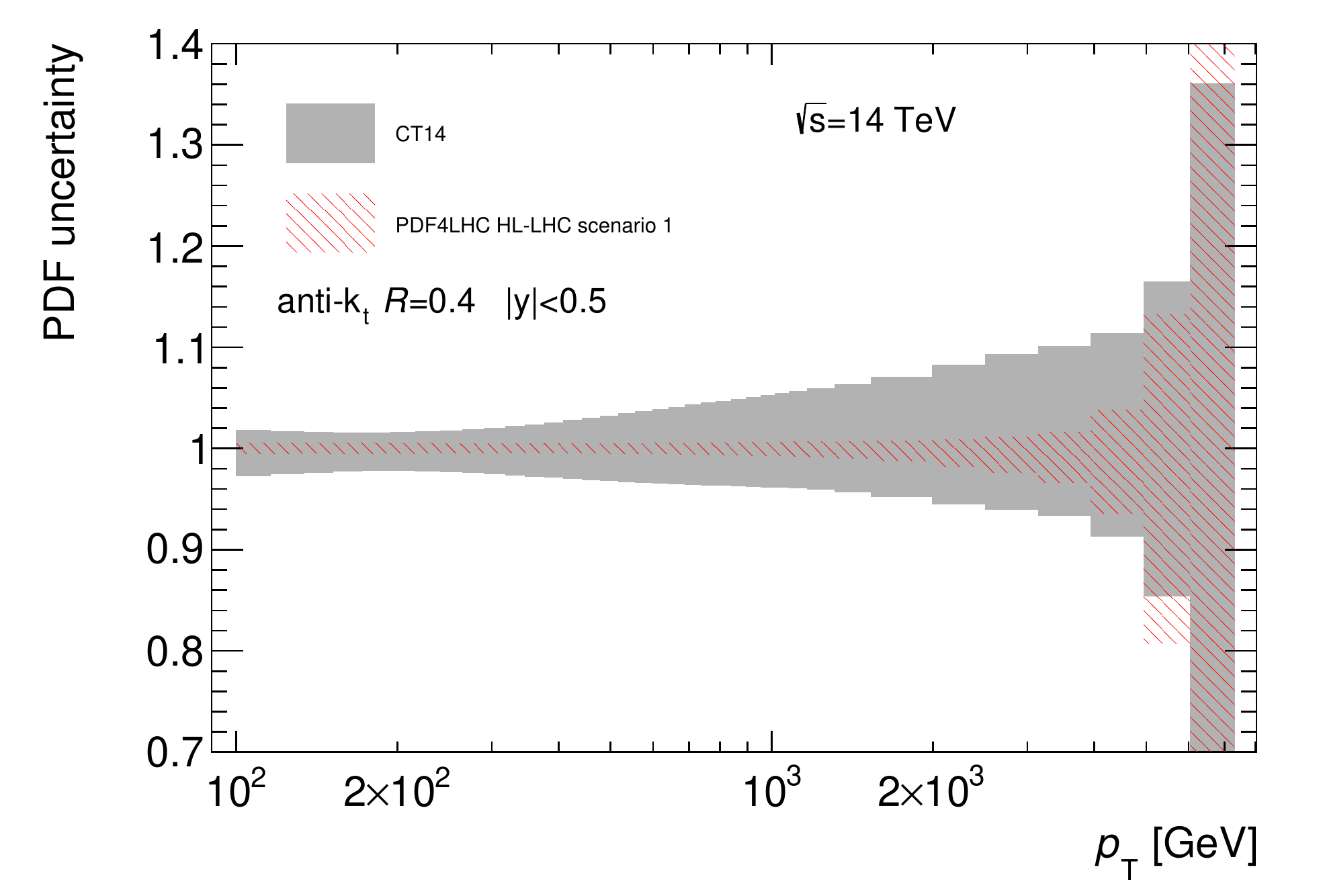}}
    \subfloat[]{\includegraphics[width=0.45\textwidth]{\main/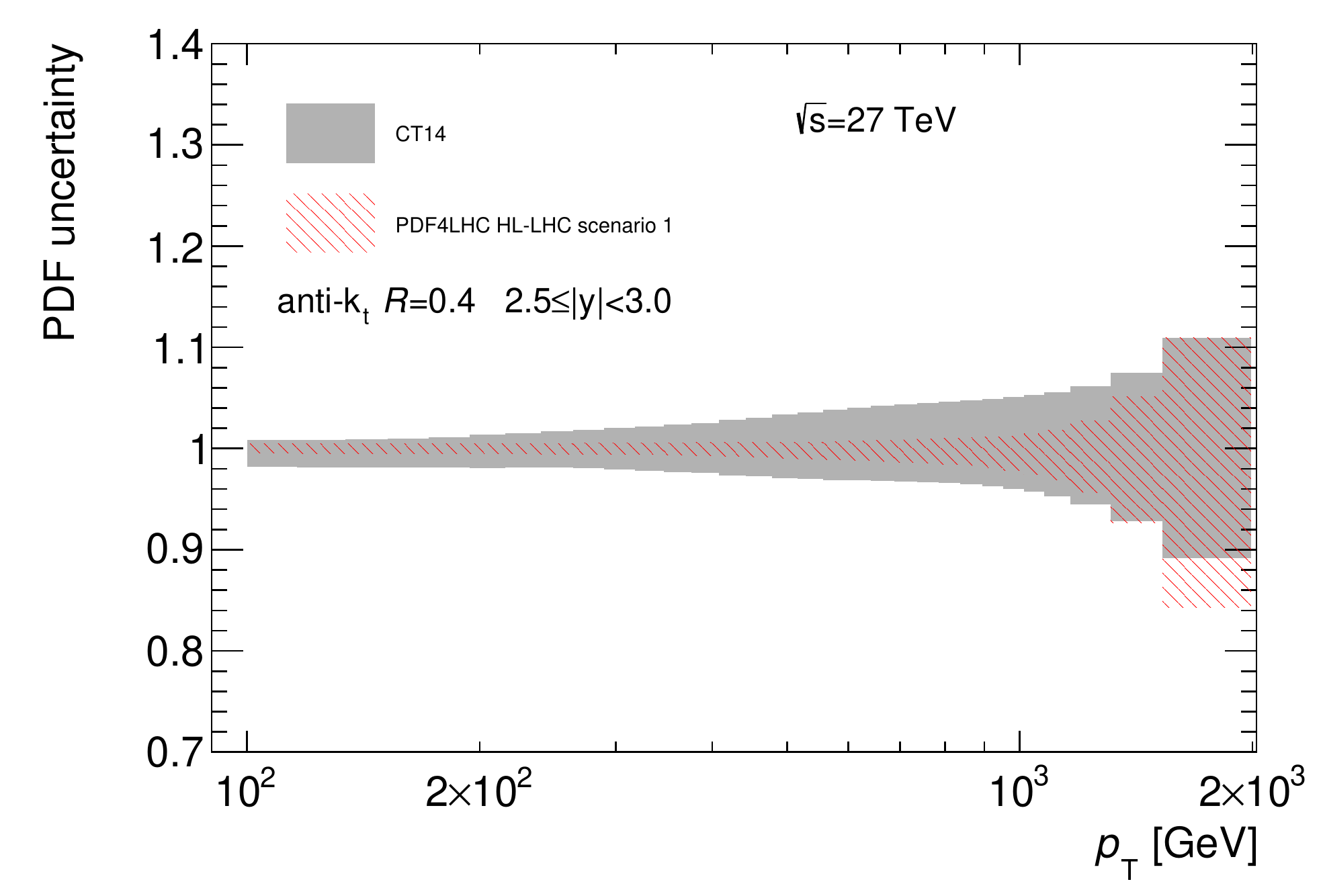}}\\
    \subfloat[]{\includegraphics[width=0.45\textwidth]{\main/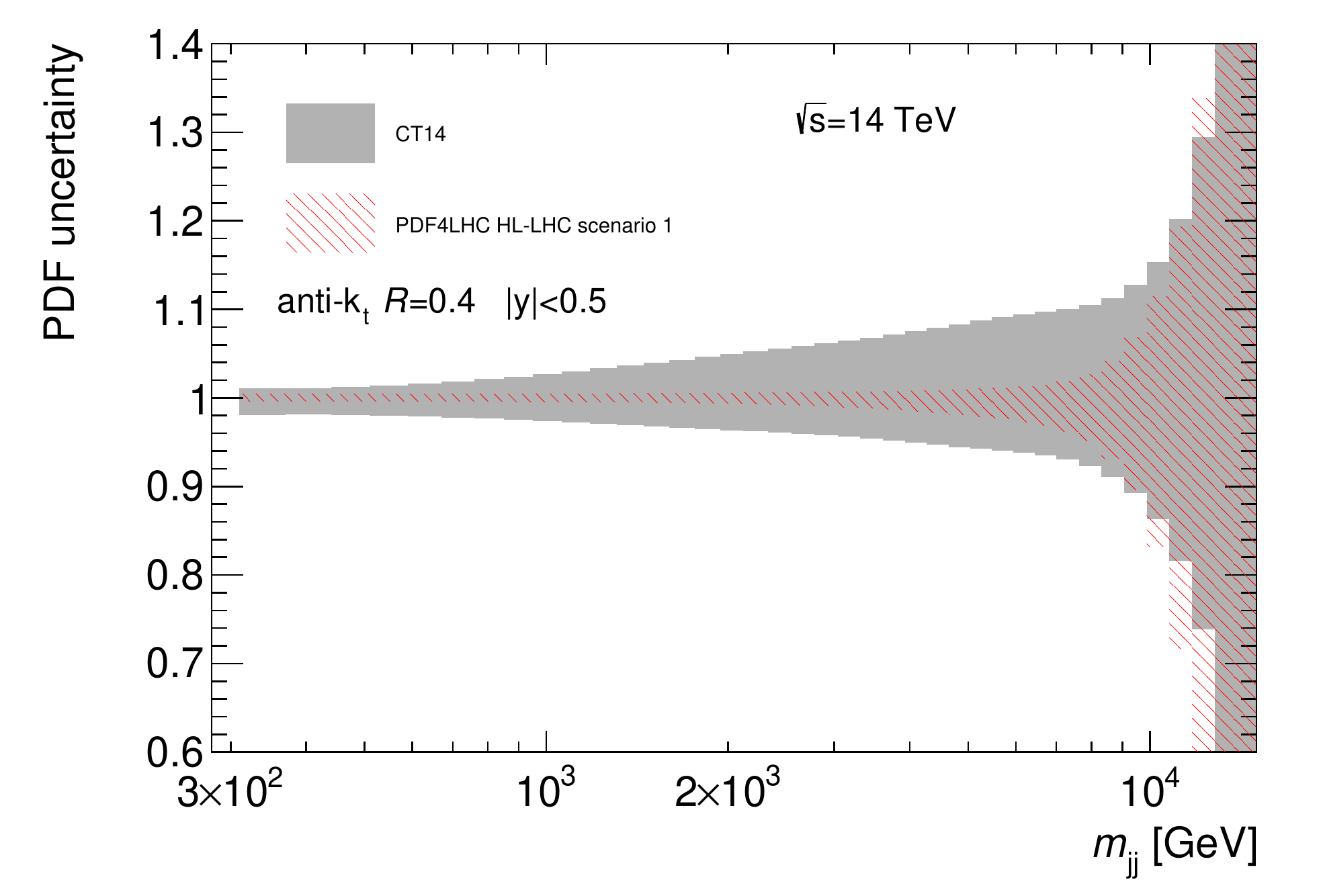}}
    \subfloat[]{\includegraphics[width=0.45\textwidth]{\main/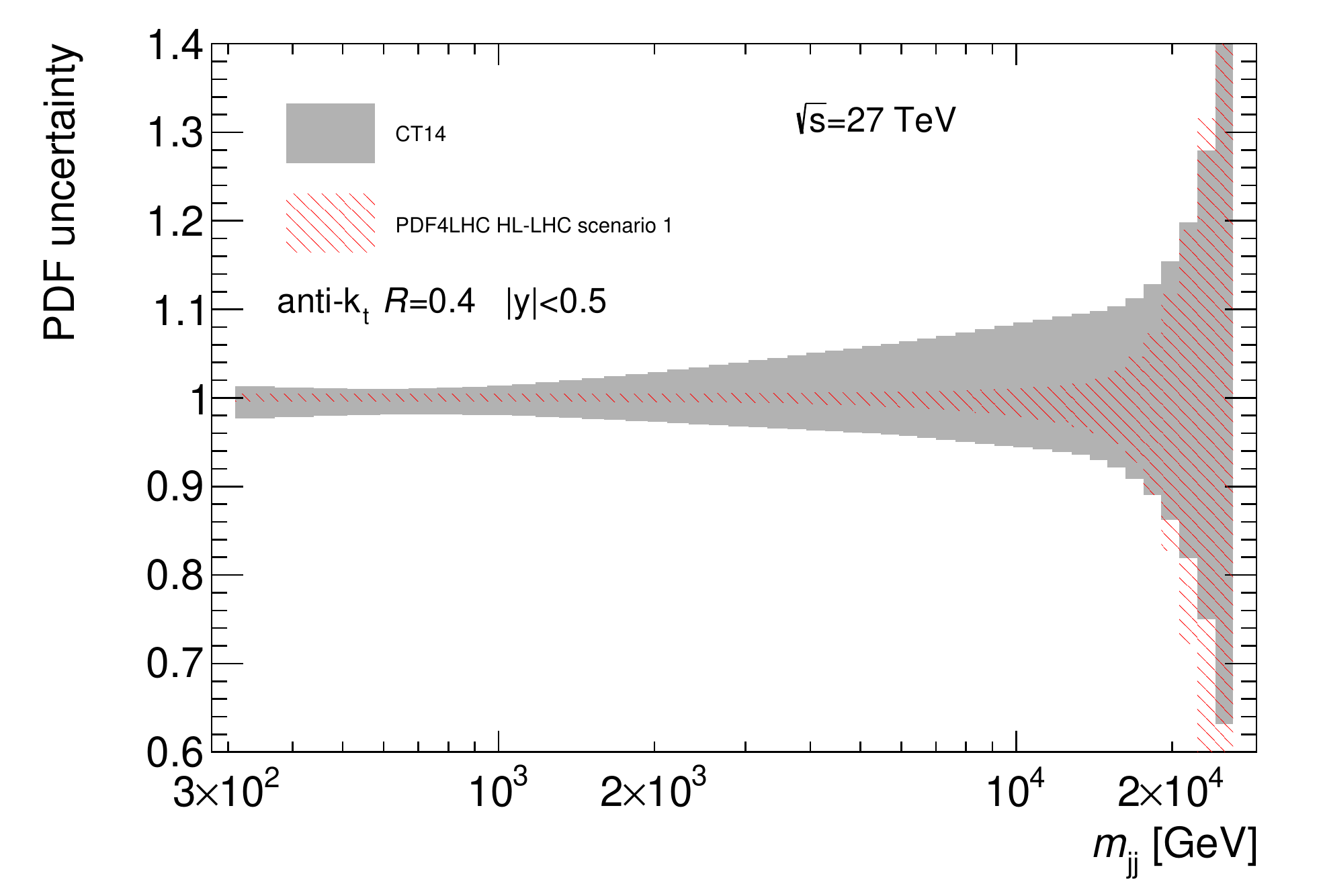}}
  \end{center}
\caption{\label{fig:ultpdf_14} Comparison of  PDF uncertainty in the inclusive jet (a,b) and dijet (c,d) cross sections calculated using the CT14 PDF set and the conservative PDF4LHC~HL-LHC scenario 1 (i.e. scenario C in Sec.~\ref{sec:ultimatepdfs}) \cite{Khalek:2018mdn} set 
at $\sqrt{s}=14$~TeV (left) and $\sqrt{s}=27$~TeV (right).}
\label{fig:ultPDF}
\end{figure}



\subsubsection{High--$p_{\rm T}$ light-- and heavy--flavour jet measurements at the HL--LHC}
The program of jet physics will substantially profit from the HL-LHC data since higher scales can be reached and the region of very low partonic momentum fractions $x$ can be accessed, where the parton density becomes large. 
Measurements of jets originating from $b$ quarks are important to investigate the heavy-flavor contribution to the total jet cross section and to study the agreement of the measurement with available theoretical predictions. In particular, inclusive $b$-jet production is very sensitive to higher-order corrections and to parton showers.
In top quark production processes, top jets can be defined when the top quark decays hadronically and all decay products can be clustered into a single jet.
The production of $W$ bosons is studied in the high-$p_{\rm T}$\ region, where the $W$ bosons decay hadronically and are reconstructed as  jets. Jet substructure techniques are applied to discriminate the jets originating from top quarks and $W$ bosons from  the QCD background. 

Higher order QCD radiation affects the distribution of the angular correlation, and the region where the jets are back-to-back in the transverse plane  is particularly sensitive to multiple ``soft'' gluon contributions, treated by all-order resummation and parton showers. This region is of particular interest since soft-gluon interference effects between the initial and final state can be significant \cite{Sun:2015doa,Collins:2007nk}. The azimuthal correlations $\Delta\phi = |\phi_2 -\phi_1|$  between the two leading $p_{\rm T}$ jets  and their dependency on the production process  is  of particular interest because of color interference effects \cite{Catani:2018mei,Catani:2014qha}.

Compared to Run-2 measurements at $\sqrt{s}=13 $ TeV the increase of the centre-of-mass energy leads to about twice larger cross section at highest $p_{\rm T}$.
Taking into account the much higher luminosity and the higher cross section, the statistical uncertainty is expected to be around six times smaller, compared to the analysis of the Run-2 data~\cite{CMS-PAS-FTR-18-032}.


Measurements of high-$p_{\mathrm{T}}$ jets originated from $b$-quarks are sensitive to the higher-order corrections, parton shower modeling and the parton densities of the proton. In
Fig.~\ref{bjetsY1} (left), the inclusive $b$-jet cross section differential in $p_{\mathrm{T}}$ is shown for centre-of-mass energy of $13$ and $14$~TeV and rapidity $|y| < 0.5$.
\begin{figure}[h]
    \centering
    \includegraphics[width=2.5in]{\main/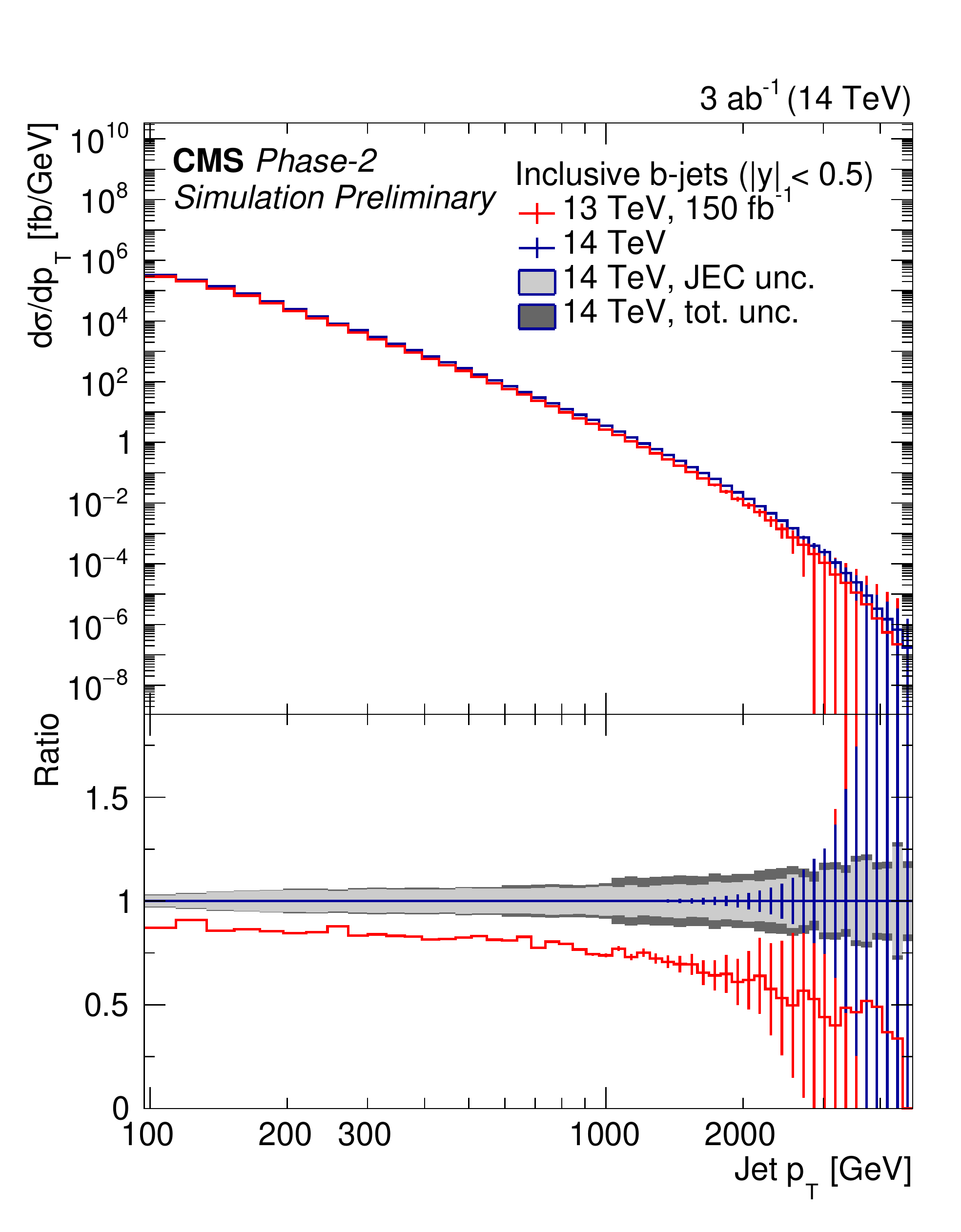}
    \centering
    \includegraphics[width=3.0in]{\main/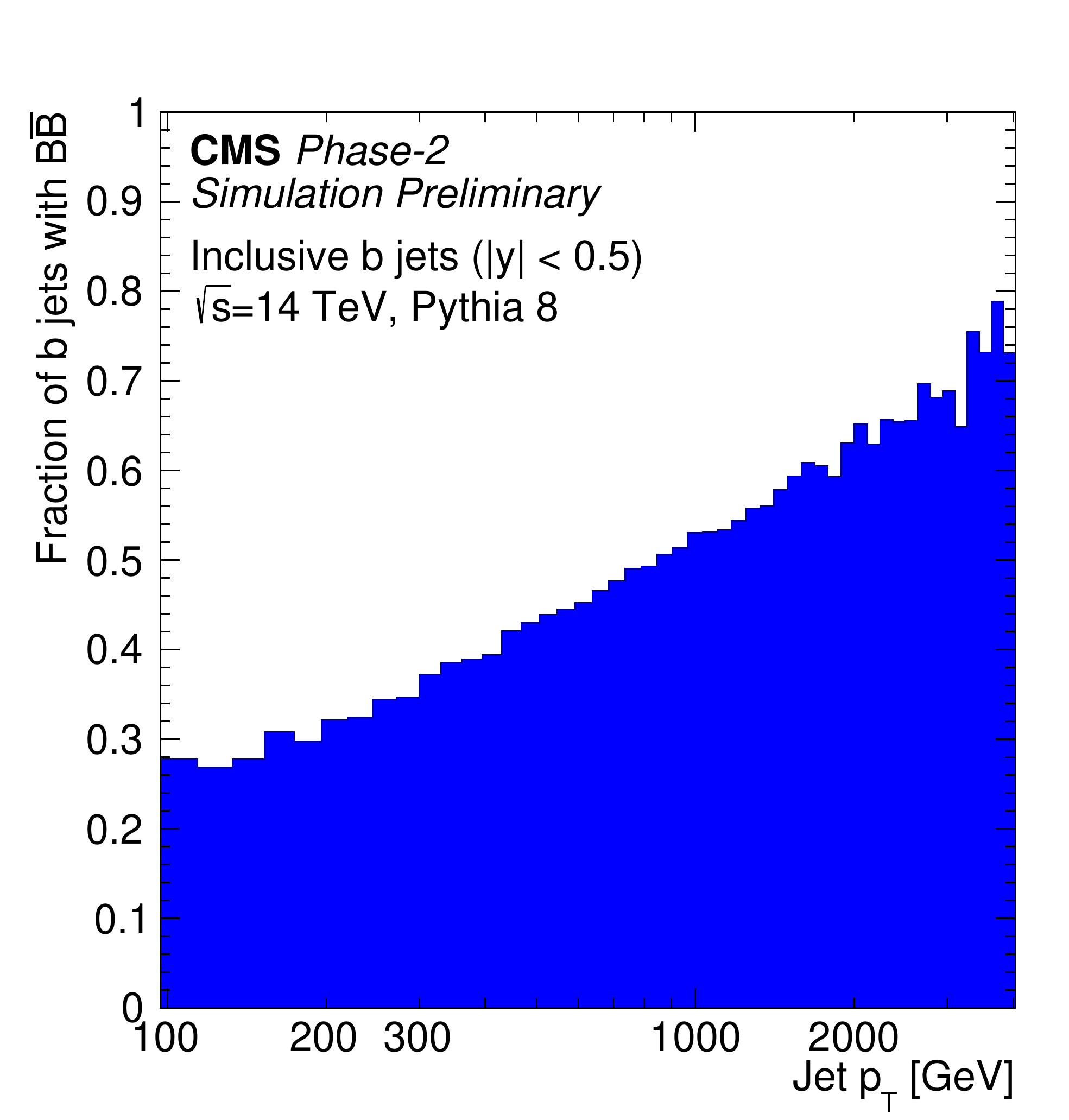}    
    \caption{The inclusive $b$-jet cross section differential in the $p_{\mathrm{T}}$ (left). The error bars show the statistical uncertainty corresponding to the given luminosity, while the gray band represent the systematic uncertainty from the jet-energy-scale and the total systematic uncertainty. The fraction of $b$-jets containing both $B$ and $\bar{B}$ hadrons as a function of $p_{\mathrm{T}}$ (right). }
    \label{bjetsY1}
\end{figure}
The depicted statistical uncertainties correspond to the luminosity $300\,\text{fb}^{-1}$ (13~TeV) and $3\,\text{ab}^{-1}$ (14~TeV).
The systematic uncertainty of the measurement is dominated by the jet energy scale uncertainty, which is of similar size as for inclusive jets, and the $b$-tagging uncertainty, which is expected to play a role mainly at higher $p_{\mathrm{T}}$ where it is about $10\%$.
It can be seen that the $p_{\mathrm{T}}$ reach at HL-LHC for the inclusive $b$-jets is about $3$~TeV, where about 30 events with $p_{\mathrm{T}} > 3\,\text{TeV}$ are expected.

It is worth noticing that at high-$p_{\mathrm{T}}$ the mass of the $b$-quark is nearly negligible with respect to the jet momentum.
This leads to the high probability that the $b$-quark is not produced in the hard sub-process, but in the parton shower.
As the mass of the $b$-quark becomes negligible, the probability of gluon splitting into $b\bar{b}$-pairs is similar to any other flavour (excluding top).
In this case, the pair of the $B$-hadrons is expected to be found inside the $b$-jet, where one consists of a $b$-quark, the second a $\bar{b}$-quark.
The fraction of such jets as a function of $p_{\mathrm{T}}$ as predicted by {\mbox{\textsc{Pythia}}\xspace}~v8 MC is shown on Fig.~\ref{bjetsY1} (right). In the future, it will be crucial to disentangle 
between $b$-jets with $b$-quarks produced in the shower, and $b$-jets with $b$-quarks produced in the hard sub-process. 


Figure~\ref{XsecAllGen} shows a comparison of the jet cross sections as a function of $p_{\rm T}$ and as a function of $\Delta\phi$ for the different processes applying the anti-$k_{\rm T}$ clustering algorithm~\cite{bib:antikt} with $R=0.8$. In Fig.~\ref{XsecAllGen} (left) the inclusive  $b$-jet cross section is shown (for comparison with the inclusive jet cross section), while in Fig.~\ref{XsecAllGen} (right) the two-$b$-jet  cross section is shown. Except for the cross section for $W$ production, the statistical uncertainties shown correspond to an integrated luminosity of 3 ab$^{-1}$ including efficiencies due to $b$-tagging and selection at the detector level, estimated using the  {\mbox{\textsc{Delphes}}\xspace} simulation. Details of the studies can be found in Ref.~\cite{CMS-PAS-FTR-18-032}.

\begin{figure}[htb]
    \begin{minipage}[b]{0.49\linewidth}
    \centering
    \includegraphics[width=\textwidth]{\main/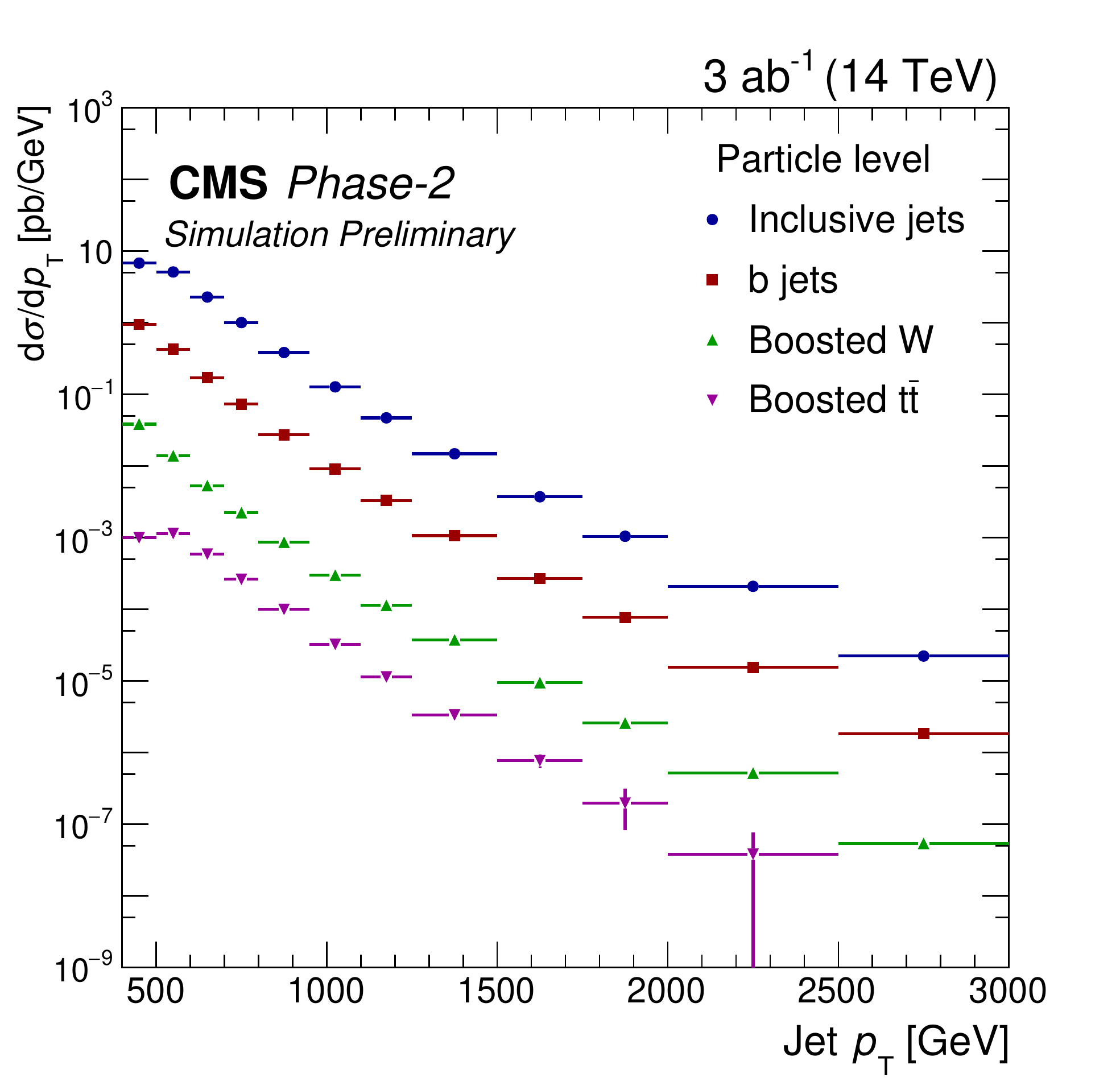}
    \end{minipage}
    \begin{minipage}[b]{0.49\linewidth}
    \centering
    \includegraphics[width=\textwidth]{\main/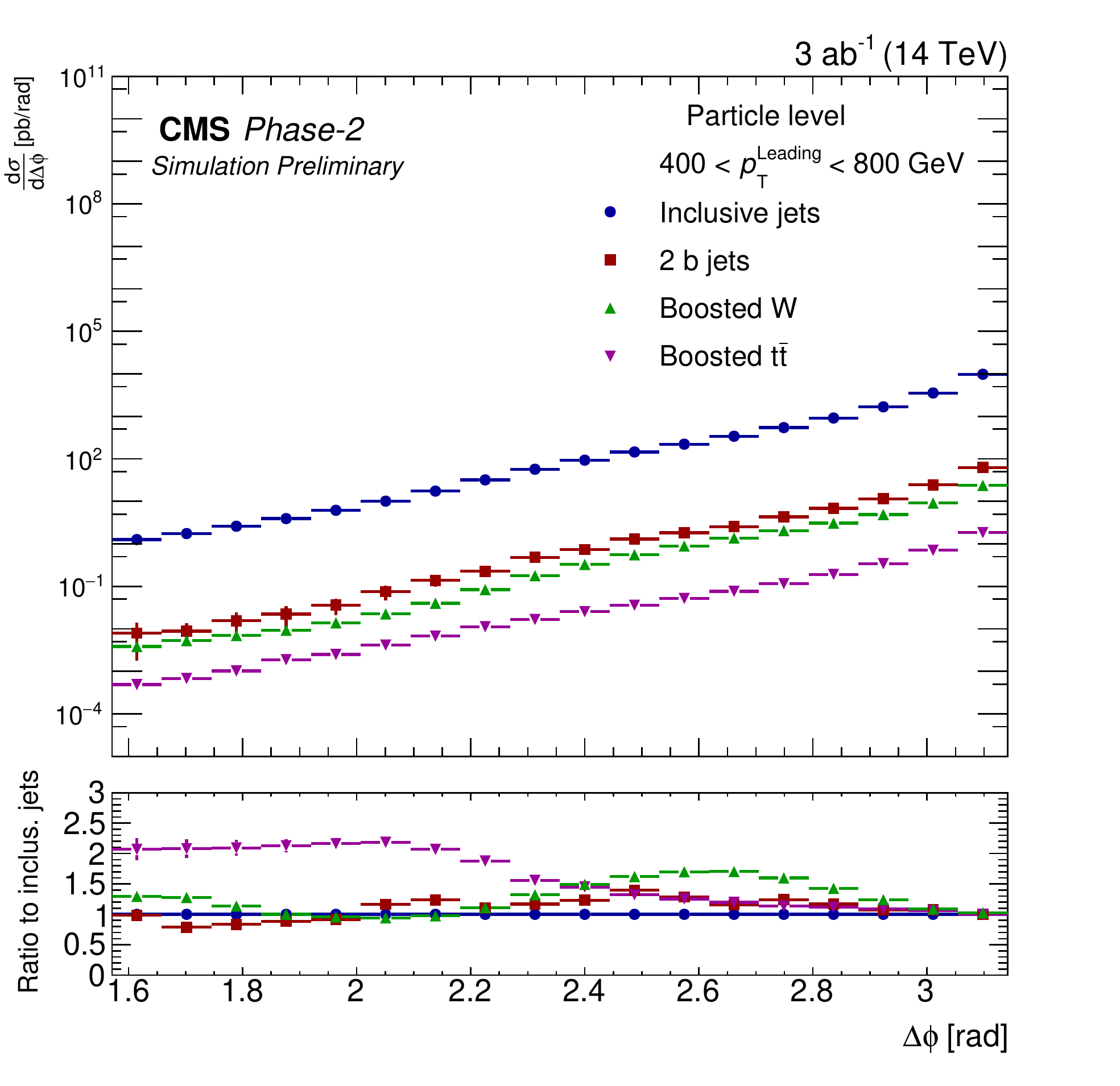}
    \end{minipage}
    \caption{
    The particle-level differential jet cross sections (with anti-$k_{\rm T}$ $R=0.8$) as a function of the jet $p_{\rm T}$ (left) and dijet $\Delta\phi$ (right) for various processes. In the left plot the inclusive  $b$ jet cross section is shown (for comparison with the inclusive jet cross section), while for  $\Delta\phi$ the two-$b$-jet  cross section is shown. For the ratio the normalization is fixed arbitrarily at $\Delta\phi=\pi$. 
   The cross section of W production does not include statistical uncertainties corrected for efficiencies and background subtraction.
    \label{XsecAllGen}}
\end{figure}

It can be seen that the shapes of the $p_{\rm T}$ spectra are comparable but the top-jet cross section is about ten thousand times smaller than the inclusive jet cross section.
The ratio to the inclusive dijet cross section as a function of $\Delta\phi$ illustrates the differences in shape of the $\Delta\phi$ distribution of the different processes (all processes are normalized at $\Delta\phi=\pi$), which depend on the partonic configuration of the initial state.



\subsubsection{Inclusive photon production}
\label{sec:photonresultsyr}
\begin{figure}[t!]
  \begin{center}
    \subfloat[]{\includegraphics[width=0.45\textwidth]{\main/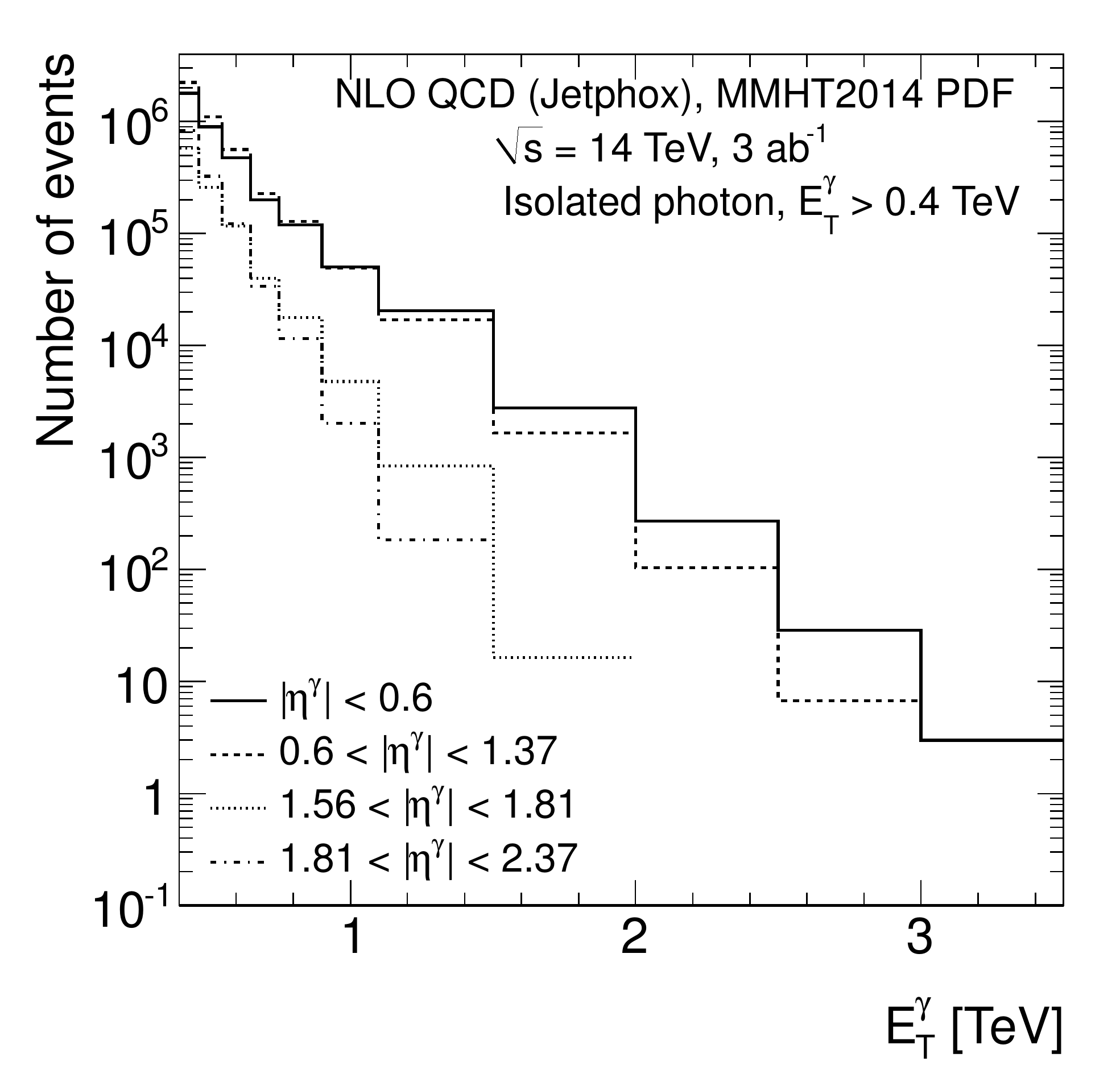}}
    \subfloat[]{\includegraphics[width=0.45\textwidth]{\main/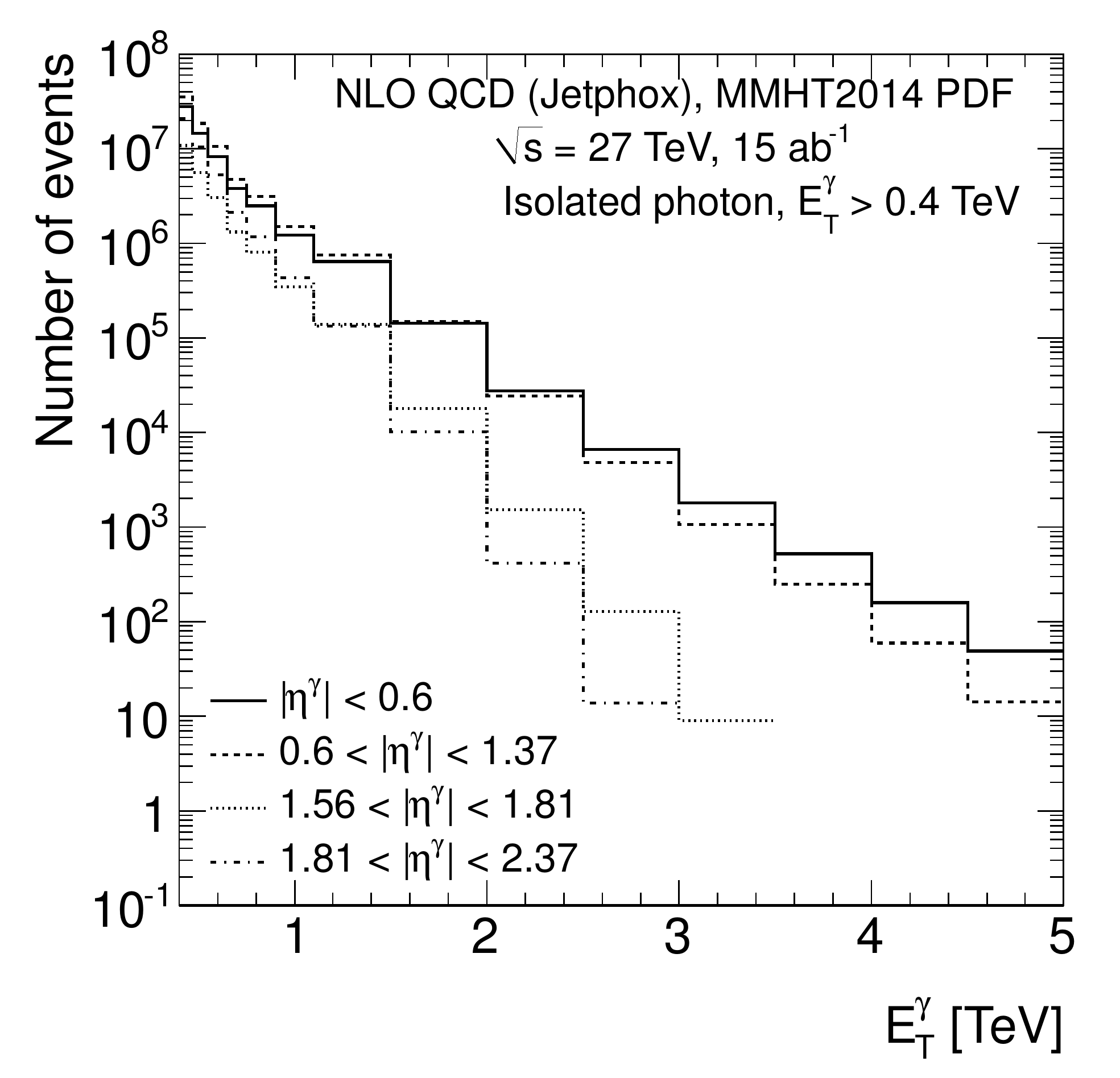}}\\
    \subfloat[]{\includegraphics[width=0.45\textwidth]{\main/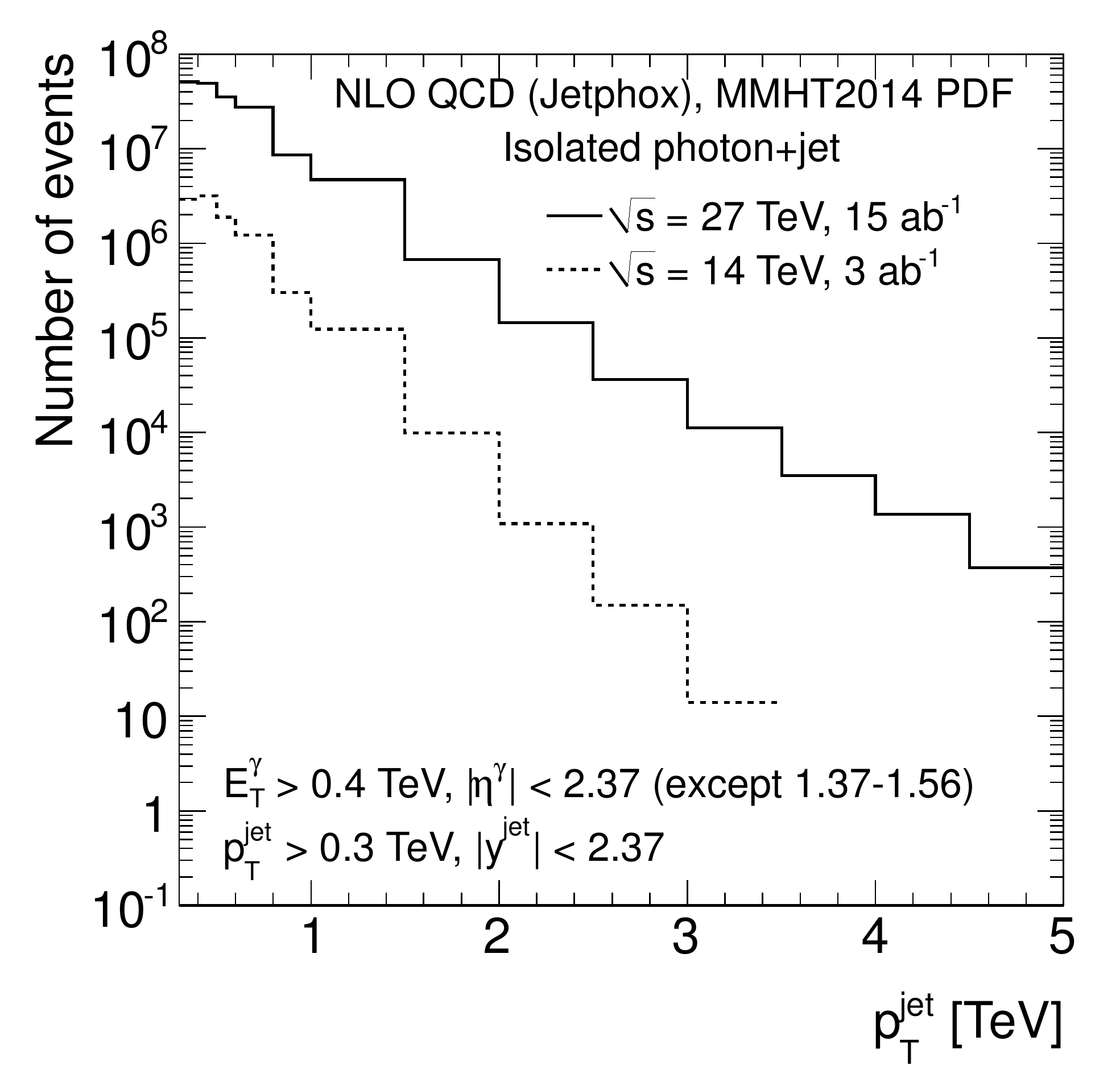}}
    \subfloat[]{\includegraphics[width=0.45\textwidth]{\main/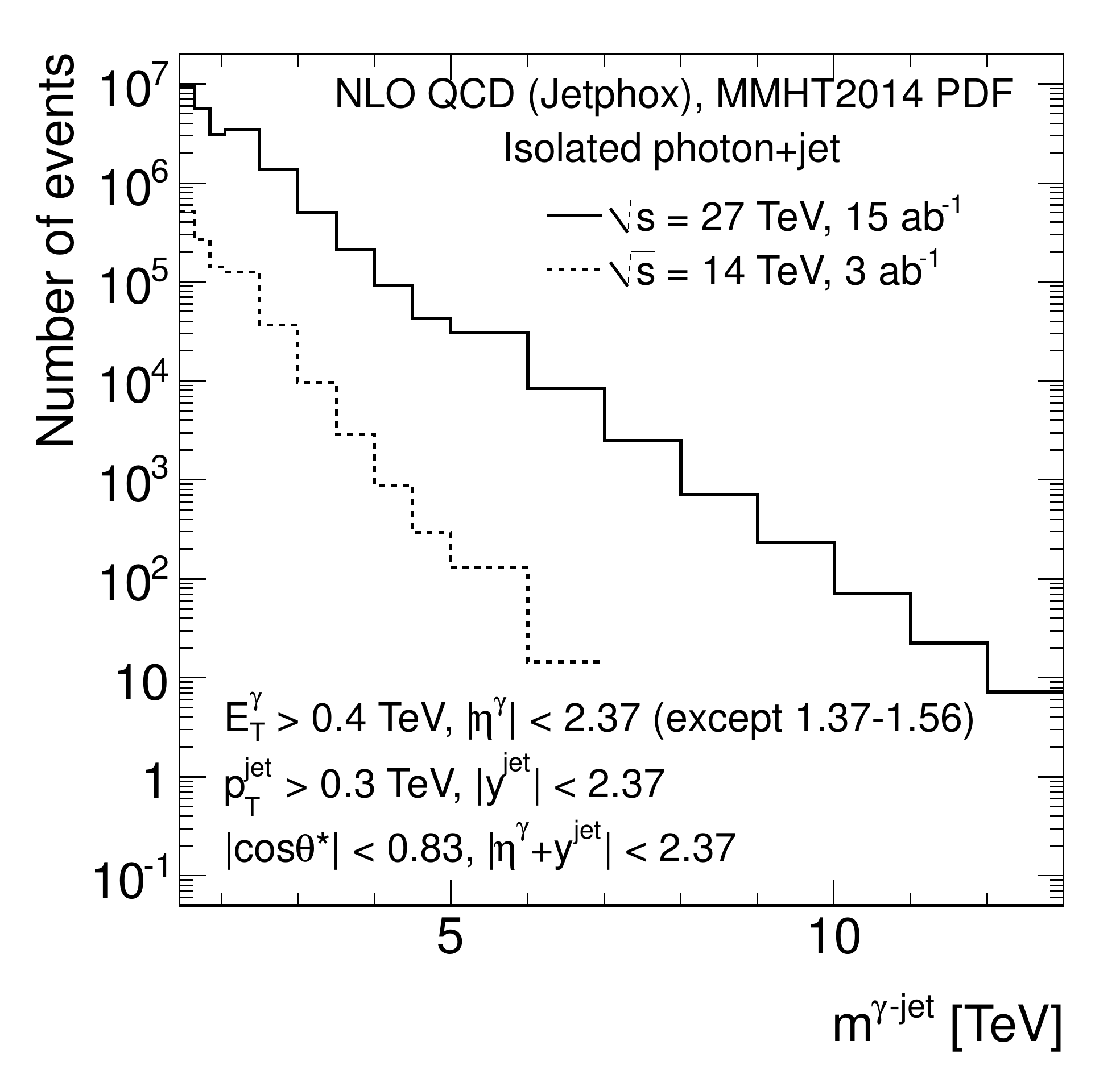}}
  \end{center}
\caption{ (a,b) Predicted number of inclusive isolated photon events as a function of $E_{\mathrm{T}}^{\gamma}$
 assuming an integrated luminosity of $3$~ab$^{-1}$ ($15$~ab$^{-1}$) of $pp$ collision data at
 $\sqrt s = 14$~TeV\ ($27$~TeV) in different ranges of photon pseudorapidity: $|\eta^{\gamma}|<0.6$
 (solid histogram),  $0.6<|\eta^{\gamma}|<1.37$ (dashed histogram), $1.56<|\eta^{\gamma}|<1.81$
 (dotted histogram) and $1.81<|\eta^{\gamma}|<2.37$ (dot-dashed histogram). (c,d)  Predicted number of
 photon$+$jet events assuming an integrated luminosity of $3$~ab$^{-1}$ ($15$~ab$^{-1}$) of $pp$
 collision data at $\sqrt s = 14$~TeV\ ($27$~TeV) as a function of (c) $p_{\mathrm{T}}^{\mathrm{jet}}$
 and (d) $m^{\gamma-{\mathrm{jet}}}$.}
\label{figurephotonyr}
\end{figure}
Here follows a summary of the studies detailed in Ref.~\cite{ATL-PHYS-PUB-2018-051} of inclusive isolated photon production and photon production in association with at least one jet. In both analyses the photon
is required to have a transverse energy in excess of $400$~GeV\ and the pseudorapidity
to lie in the range $|\eta^{\gamma}|<2.37$ excluding the region $1.37<|\eta^{\gamma}|<1.56$.
The photon is required to be isolated by imposing an upper limit on the amount of transverse
energy inside a cone of size $\Delta R  = 0.4$ in the $\eta$--$\phi$ plane around the photon,
excluding the photon itself: $E^{\mathrm{iso}}_{\mathrm{T}} < E^{\mathrm{iso}}_{\mathrm{T,max}}$.

In the inclusive photon analysis, the goal is the measurement of the differential cross section
as a function of $E_{\mathrm{T}}^{\gamma}$ in four regions of the photon pseudorapidity:
$|\eta^{\gamma}|<0.6$,  $0.6<|\eta^{\gamma}|<1.37$, $1.56<|\eta^{\gamma}|<1.81$ and
$1.81<|\eta^{\gamma}|<2.37$. Photon isolation is enforced by requiring
$E^{\mathrm{iso}}_{\mathrm{T}} < 4.2 \cdot 10^{-3} \cdot
E_{\mathrm{T}}^{\gamma} + 4.8$~GeV.

In the photon$+$jet analysis, jets are reconstructed using the anti-$k_{\rm T}$
algorithm~\cite{Cacciari:2008gp} with a
radius parameter $R=0.4$. Jets overlapping with the photon are not considered if the jet
axis lies within a cone of size $\Delta R  = 0.8$. The leading jet is required to have transverse
momentum above $300$~GeV\ and rapidity in the range $|y^{\mathrm{jet}}|<2.37$. No
additional condition is used for the differential cross section as a function of
$p_{\mathrm{T}}^{\mathrm{jet}}$. For the differential cross section as a function of the
invariant mass of the photon$+$jet system additional constraints are imposed:
$m^{\gamma-{\mathrm{jet}}}>1.45$~TeV, $|\cos\theta^*|<0.83$ and 
$|\eta^{\gamma}\pm y^{\mathrm{jet}}|<2.37$. These additional
constraints are imposed to remove the bias due to the rapidity and transverse-momentum
requirements on the photon and the leading jet~\cite{np:b875:483,np:b918:257}.
Photon isolation is enforced by requiring $E^{\mathrm{iso}}_{\mathrm{T}} < 4.2 \cdot 10^{-3} \cdot E_{\mathrm{T}}^{\gamma} + 10$~GeV.

The yields of inclusive isolated photons and of photon$+$jet events are estimated using the
program {\textsc{Jetphox}}~1.3.1\_2~\cite{jhep:0205:028,pr:d73:094007}. This
program includes a full next-to-leading-order QCD calculation of both the direct-photon and
fragmentation contributions to the cross sections for the
$pp\rightarrow\gamma+{\mathrm{X}}$ and $pp\rightarrow\gamma+{\mathrm{jet}}+{\mathrm{X}}$ reactions.
The number of massless quark flavours is set to five. The renormalisation ($\mu_{\mathrm{R}}$), factorisation
($\mu_{\mathrm{F}}$) and fragmentation ($\mu_{\mathrm{f}}$) scales are chosen to be
$\mu_{\mathrm{R}}=\mu_{\mathrm{F}}=\mu_{\mathrm{f}}=E_{\mathrm{T}}^{\gamma}$. The
calculations are performed using the MMHT2014~\cite{epj:c75:204} parameterisations of the proton
parton distribution functions (PDFs) and the BGF set II of parton-to-photon fragmentation functions
at NLO~\cite{epj:c2:529}. The strong coupling constant
$\alpha_{\mathrm{s}}(m_{\mathrm{Z}})$ is set to the value assumed in the fit to determine the PDFs.
The reliability of the estimated yields using the program {\textsc{Jetphox}} is supported by the
high purity of the signal photons, the mild unfolding corrections and the fact that the NLO QCD
predictions describe adequately the measurements of these processes using $pp$ collisions
at $\sqrt s = 13$~TeV~\cite{pl:b770:473,pl:b780:578}.

The predicted number of inclusive isolated photon events as a function of $E_{\mathrm{T}}^{\gamma}$ in the
different ranges of $|\eta^{\gamma}|$ assuming an integrated luminosity of $3$~ab$^{-1}$ 
($15$~ab$^{-1}$) of $pp$ collision data at $\sqrt s = 14$~TeV\ ($27$~TeV)
is shown in Figure ~\ref{figurephotonyr}(a) and \ref{figurephotonyr}(b). For the HL-LHC (HE-LHC), the reach in
$E_{\mathrm{T}}^{\gamma}$ is (a) $3$--$3.5$~($5$)~TeV\ for $|\eta^{\gamma}|<0.6$, 
(b) $2.5$--$3$~($5$)~TeV\ for $0.6<|\eta^{\gamma}|<1.37$, 
(c) $1.5$--$2$~($3$--$3.5$)~TeV\ for $1.56<|\eta^{\gamma}|<1.81$ and
(d) $1$--$1.5$~($2.5$--$3$)~TeV\ for $1.81<|\eta^{\gamma}|<2.37$.
This represents a significant extension of the region measured so far with $pp$ collisions at
$\sqrt s = 13$~TeV~\cite{pl:b770:473}; as an example, at the HL-LHC (HE-LHC) the $E_{\mathrm{T}}^{\gamma}$
reach is extended from $1.5$~TeV\ to $3$--$3.5$~($5$)~TeV\ for $|\eta^{\gamma}|<0.6$. 

The predicted number of photon$+$jet events as a function of $p_{\mathrm{T}}^{\mathrm{jet}}$
and $m^{\gamma-{\mathrm{jet}}}$ assuming an integrated luminosity of $3$~ab$^{-1}$
($15$~ab$^{-1}$) of $pp$ collision data at $\sqrt s = 14$~TeV\ ($27$~TeV) is shown in
Figs.~\ref{figurephotonyr}(c) and \ref{figurephotonyr}(d). 
In comparison with the latest measurements at $\sqrt s = 13$~TeV~\cite{pl:b780:578},
the expectations obtained at the HL-LHC (HE-LHC) extend significantly the reach in
$p_{\mathrm{T}}^{\mathrm{jet}}$ from $1.5$~TeV\ to $3.5$~($5$)~TeV\ and
$m^{\gamma-{\mathrm{jet}}}$ from $3.3$~TeV\ to $7$~($12$)~TeV.

\subsubsection{Diphoton production}
The production of photon pairs (diphotons) with high invariant mass  is a very important process for physics studies 
at high-energy hadron colliders.
Photons are very clean final states and photon energies and momenta can be measured with high precision
in modern electromagnetic calorimeters. 
Therefore \emph{prompt} photons represent ideal probes to test the properties of the Standard Model
(SM)~\cite{Aaltonen:2011vk}--\cite{Aaboud:2017vol}
and they are also important in searches for new-physics signals (see, e.g., Refs.~\cite{Aad:2012zza}--\cite{Khachatryan:2015qba}).
Owing to the above reasons, it is important to provide accurate theoretical predictions for diphoton production  at LHC energies.
This task requires in particular, the calculation of QCD and EW radiative corrections at high perturbative orders.

This contribution considers diphoton production in $pp$ collisions at the ${\sqrt s}=14$~GeV and  ${\sqrt s}=27$~GeV energies, and presents perturbative QCD
results up to the NNLO by using the smooth cone isolation criterion\,\footnote{The NNLO QCD calculation within the standard cone isolation criterion
  has not been performed yet.}. 
Within the smooth cone isolation criterion \cite{Frixione:1998jh} (see also Refs.~\cite{Frixione:1999gr,Catani:2000jh}) photons are selected by
fixing the size $R$ of the isolation cone and imposing a maximal amount of hadronic energy ($E_{\mathrm{T}}^{had}(r)$) allowed inside the cone
\begin{eqnarray}
\label{Eq:Isol_frixcriterion} 
 E_{\mathrm{T}}^{had}(r) \leq \,E_{\mathrm{T}\,{\rm max}} ~\chi(r;R) \;, \quad
 \mbox{ in {\it all} cones with} \;r \leq R 
\;\;,    
\end{eqnarray}
with a suitable choice of the $r$ dependence of the isolation function $\chi(r;R)$. The smooth isolation function $\chi(r;R)$ used
is\,\footnote{The same form of the isolation function is used in the
  NNLO predictions reported in Refs.~\cite{Aad:2012tba,Chatrchyan:2014fsa,Aaboud:2017vol}.}
\begin{equation}
\label{Eq:Isol_chinormal}
\chi(r;R) = \left( \frac{1-\cos (r)}{1-\cos (R)} \right)^{n}\;,
\end{equation}
 and the
 value of the power $n$ is set to the $n=1$. This value of $n$ avoids the sensitivity of the cross section to soft (collinear) photons
for large
(small) value of $n$~\cite{Catani:2018krb}. The radius of the photon isolation cone is set at the value $R=0.4$ and $E_{\mathrm{T}\,{\rm max}} =10$~GeV.  
Detailed comparisons between standard and smooth cone isolation criteria have been presented in
Refs.\cite{Catani:2018krb,Badger:2016bpw,Cieri:2015wwa,Andersen:2014efa}.

The following kinematic cuts are applied:
\begin{equation}
p_{\mathrm{T}}^{\gamma;~hard} > 40~GeV,\, \qquad p_{\mathrm{T}}^{\gamma;~soft} > 30~GeV,\qquad|y_\gamma|<2.8\,,
\end{equation}
where $p_{\mathrm{T}}^{\gamma;~hard}$ and $p_{\mathrm{T}}^{\gamma;~soft}$ are respectively  the transverse momenta of the harder and softer photon and  
$|y_\gamma|$ is the photon rapidity. The minimum angular distance between the two photons is $R_{\gamma \gamma}^{\rm min}=0.4$.

A lower limit $r_{\rm cut}$ is implemented on the ratio $p_{\mathrm{T}\gamma\gamma}/M_{\gamma\gamma}$ ($p_{\mathrm{T} \gamma \gamma} > r_{\rm cut} M_{\gamma\gamma}$) \cite{Grazzini:2017mhc},
and values in the range $r_{\rm cut}=0.08\%$--$0.15\%$ are used. The perturbative uncertainty is computed as the envelope of three-point scale variation by considering the two asymmetric scale configurations with $\{ \mu_R=\mu_0/2$, $\mu_F=2\mu_0 \}$ and  $\{ \mu_R=2\mu_0$ , $\mu_F=\mu_0/2 \}$ and the central scale $\{ \mu_R=\mu_F=\mu_0 \}$.

\begin{figure}[tb]
\centering
\includegraphics[width=.49\linewidth]{\main/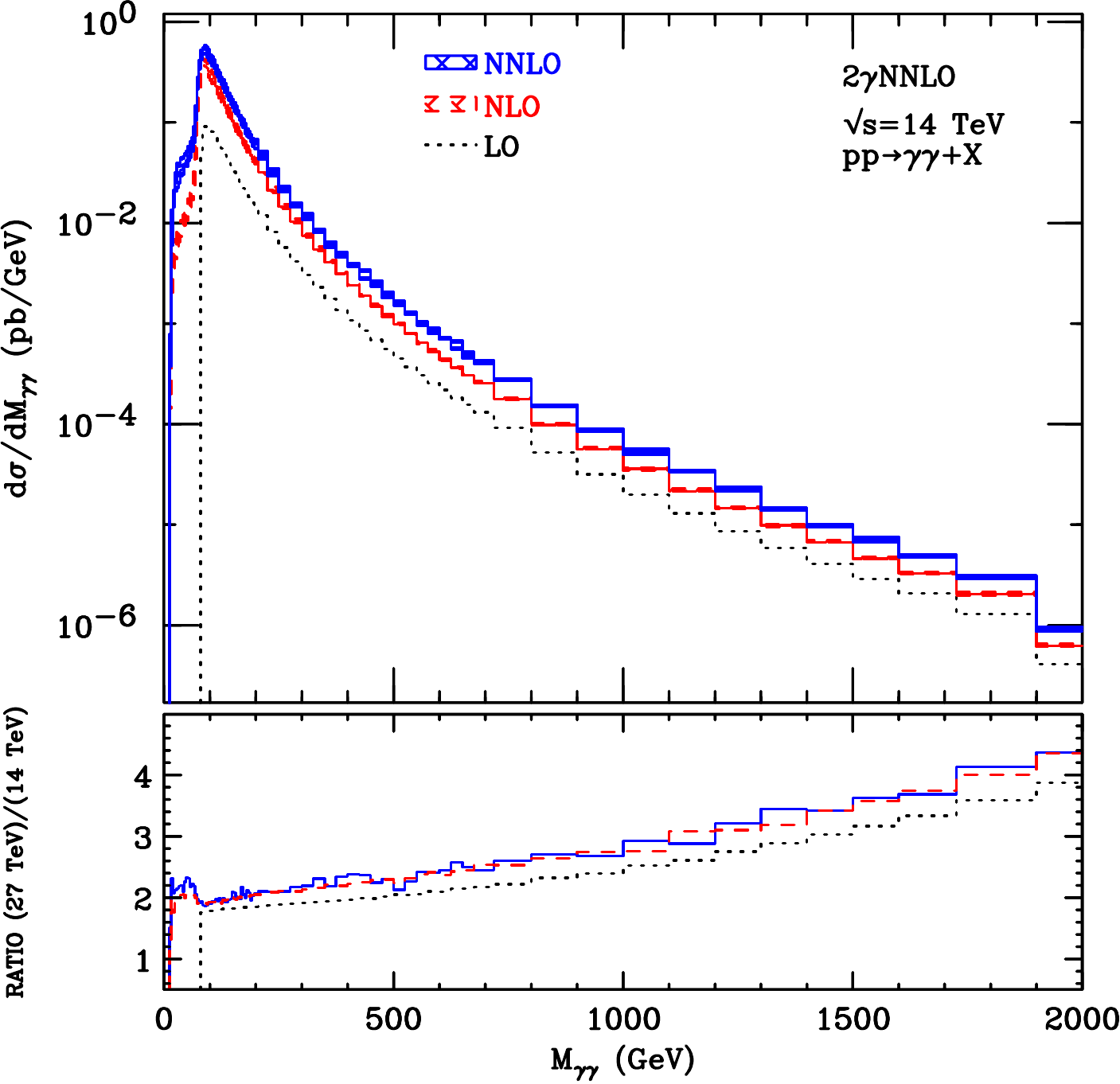}
\includegraphics[width=.475\linewidth]{\main/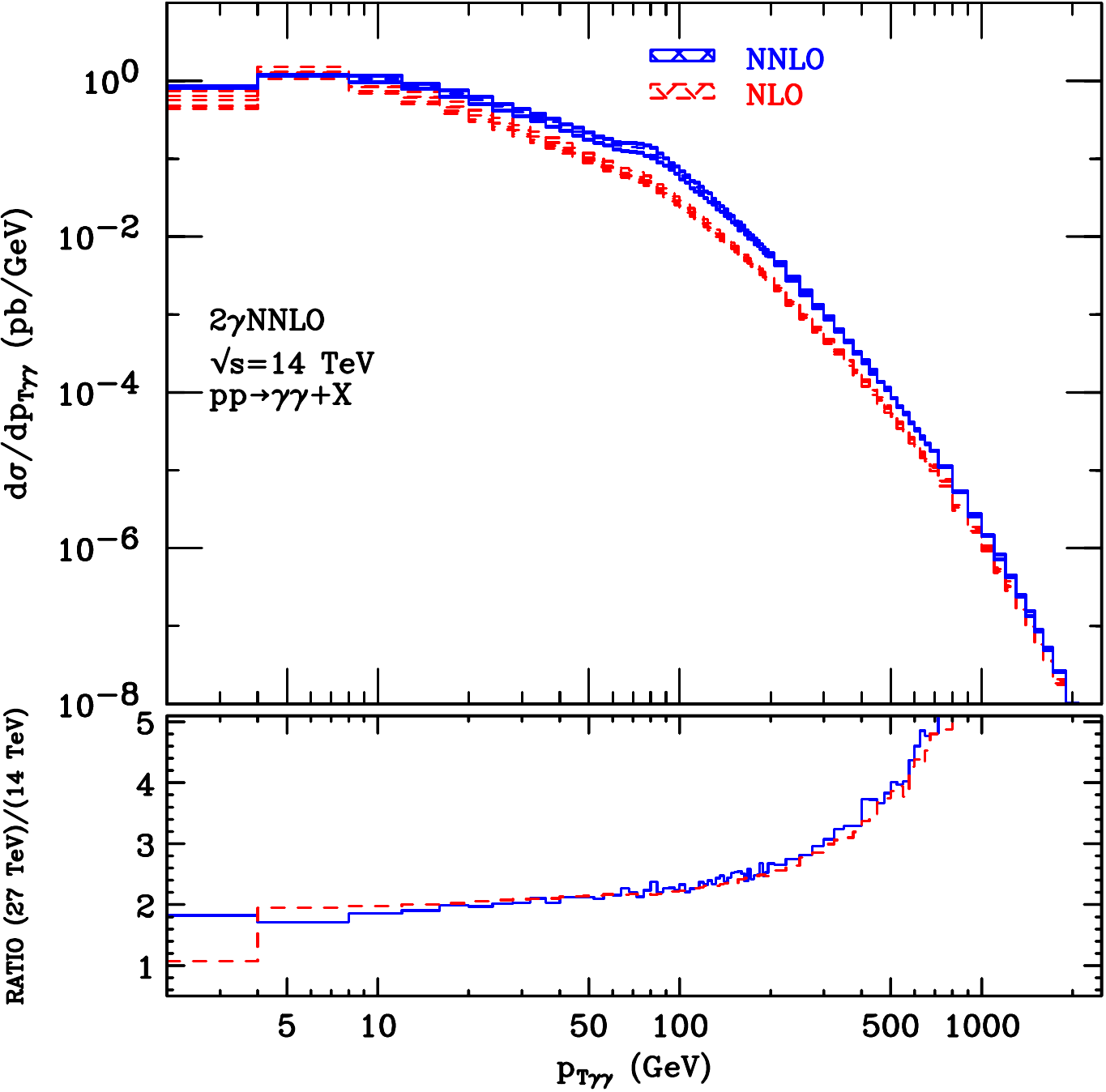}
\caption{\label{fig-gg}{ The differential cross sections $d\sigma/dM_{\gamma \gamma}$ (left)
    and $d\sigma/dp_{\mathrm{T} \gamma \gamma}$ (right)
    at $\sqrt{ s}=14$~TeV are shown in the upper panel at LO (black dotted), NLO (red dashed) and NNLO (blue solid). The NLO and NNLO scale variation bands are
    obtained as detailed in the text. In the lower subpanels the ratio between cross sections at two different  centre--of--mass energies ($\sqrt{ s}=27$~TeV and $\sqrt{ s}=14$~TeV) is also shown. The selection cuts are described in the text. }}
\end{figure}

This study begins by considering the invariant mass ($M_{\gamma\gamma}$) distribution up to value of 2\,TeV.
The LO, NLO and NNLO QCD results for a centre--of--mass energy of $\sqrt{ s}=14$~TeV
are presented in Fig.\ref{fig-gg} (left).
It is first observed the presence of a LO threshold at an invariant mass $M^{LO}=2p_{\mathrm{T}}^{\gamma;~hard}$.
The bulk of the cross section is concentrated in the region around $M^{LO}$ while for large values of $M_{\gamma\gamma}$  the distribution
rapidly decreases. At high invariant mass, $M_{\gamma\gamma}>1$~TeV, the cross section is dominated by the quark annihilation ($q\bar{q}$) partonic subprocess
(the other partonic subprocesses are suppressed by one order of magnitude or more). The NNLO $K$ factor,
$K^{NNLO}= \sigma^{NNLO}/\sigma^{NLO}$, is flat at large values of $M_{\gamma\gamma}$ and it is roughly equal to the NNLO $K$ factor of the
$q\bar{q}$ channel. 
The lower subpanel of Figure \ref{fig-gg} (left) presents results for the ratio ($R$) between the invariant mass distribution at $\sqrt{ s}=27$~TeV and
$\sqrt{ s}=14$~TeV. At LO the dynamic enhancement of the ratio can be described roughly as
${\cal G}(M^{2}_{\gamma\gamma}/27^{2}~{\rm TeV}^{2})/ {\cal G}(M^{2}_{\gamma\gamma}/(14^{2}~{\rm TeV}^{2})$, where
${\cal G}(\tau) = \log(\tau) \times  {\cal L}_{q\bar{q}}(\tau,\mu_{f})$ and ${\cal L}$ are the integrated parton luminosities. The
ratio at NLO and NNLO is numerically similar to the corresponding LO one. The enhancement of the ratio $R$ at
large values of invariant mass is directly related to the increasing the centre--of--mass energy and it reaches the value $R\sim 4$ at
$M_{\gamma\gamma}\simeq 1$~TeV.

Finally theoretical results are presented for the transverse momentum ($p_{\mathrm{T} \gamma \gamma}$) distribution.
The NLO and NNLO predictions with a centre--of--mass energy of $\sqrt{ s}=14$~TeV are shown in the upper panel of
Figure \ref{fig-gg} (right). 
Given the LO kinematical constraint $p_{\mathrm{T} \gamma \gamma}= 0$,
the (N)NLO  correction represent \textit{effectively} an (N)LO prediction. Moreover, in the small $p_{\mathrm{T} \gamma \gamma}$ region, the convergence of the
fixed order expansion is spoiled by the presence of large logarithmic corrections. Reliable perturbative results require an all order resummation
of these enhanced logarithmic contributions.

The lower subpanel of Figure \ref{fig-gg} (right) presents results for the ratio ($R$) between the 
transverse momentum distribution at $\sqrt{ s}=27$~TeV and $\sqrt{ s}=14$~TeV.
The ratio increases at large value of  $p_{\mathrm{T} \gamma \gamma}$, reaching $R\sim 4$ for $p_{\mathrm{T} \gamma \gamma}\simeq 1$\,TeV.

It is observed that the uncertainty bands for the NLO and NNLO results fail
to overlap in most of the kinematical regions.
This suggests that the computed scale dependence at NNLO cannot be considered
a reliable estimate of the true perturbative uncertainty.
As an alternative approach the perturbative uncertainty of the
NNLO result can be estimated by considering 
 half of the difference between the NNLO and NLO results at central values
of the scales~\cite{Catani:2018krb}.

It is finally observed that the photon fragmentation component (which is absent in the case of smooth cone isolation) mainly affects the
the low invariant mass region, where the cross section is strongly suppressed.
Conversely, the intermediate and high invariant mass region,  the transverse momentum distribution and the value of total cross section,
are less sensitive to photon fragmentation effects.
In particular, for isolation parameters commonly used in the experimental
analysis at the LHC, the quantitative differences between smooth and standard isolation predictions are much smaller than the corresponding perturbative
uncertainties. This observation justifies the use of the smooth cone criterion in the theoretical calculations.

\subsection[Ultimate Parton Densities]{Ultimate Parton Densities\footnote{Contribution by R. Abdul Khalek, S. Bailey, J. Gao, L. Harland-Lang and J. Rojo.}}

\label{sec:ultimatepdfs}
\newcommand{\be}{\begin{equation}}
\newcommand{\ee}{\end{equation}}

\newcommand{\bea}{\begin{eqnarray}}
\newcommand{\eea}{\end{eqnarray}}
\newcommand{\bi}{\begin{itemize}}
\newcommand{\ei}{\end{itemize}}
\newcommand{\ben}{\begin{enumerate}}
\newcommand{\een}{\end{enumerate}}
\newcommand{\la}{\left\langle}
\newcommand{\ra}{\right\rangle}
\newcommand{\lc}{\left[}
\newcommand{\rc}{\right]}
\newcommand{\lp}{\left(}
\newcommand{\rp}{\right)}
\newcommand{\as}{\alpha_s}
\newcommand{\aq}{\alpha_s\left( Q^2 \right)}
\newcommand{\amz}{\alpha_s\left( M_Z^2 \right)}
\newcommand{\aqq}{\alpha_s \left( Q^2_0 \right)}
\newcommand{\aqz}{\alpha_s \left( Q^2_0 \right)}
\newcommand{\Ord}{\mathcal{O}}
\newcommand{\MSbar}{\overline{\text{MS}}}
\def\toinf#1{\mathrel{\mathop{\sim}\limits_{\scriptscriptstyle
{#1\rightarrow\infty }}}}
\def\tozero#1{\mathrel{\mathop{\sim}\limits_{\scriptscriptstyle
{#1\rightarrow0 }}}}
\def\toone#1{\mathrel{\mathop{\sim}\limits_{\scriptscriptstyle
{#1\rightarrow1 }}}}
\def\frac#1#2{{{#1}\over {#2}}}
\def\gsim{\gtrsim}
\def\lsim{\lesssim}    
\newcommand{\mrexp}{\mathrm{exp}}
\newcommand{\dat}{\mathrm{dat}}
\newcommand{\one}{\mathrm{(1)}}
\newcommand{\two}{\mathrm{(2)}}
\newcommand{\art}{\mathrm{art}} 
\newcommand{\rep}{\mathrm{rep}}
\newcommand{\net}{\mathrm{net}}
\newcommand{\stopp}{\mathrm{stop}}
\newcommand{\sys}{\mathrm{sys}}
\newcommand{\stat}{\mathrm{stat}}
\newcommand{\diag}{\mathrm{diag}}
\newcommand{\pdf}{\mathrm{pdf}}
\newcommand{\tot}{\mathrm{tot}}
\newcommand{\minn}{\mathrm{min}}
\newcommand{\mut}{\mathrm{mut}}
\newcommand{\partt}{\mathrm{part}}
\newcommand{\dof}{\mathrm{dof}}
\newcommand{\NS}{\mathrm{NS}}
\newcommand{\cov}{\mathrm{cov}}
\newcommand{\gen}{\mathrm{gen}}
\newcommand{\cut}{\mathrm{cut}}
\newcommand{\parr}{\mathrm{par}}
\newcommand{\val}{\mathrm{val}}
\newcommand{\tr}{\mathrm{tr}}
\newcommand{\checkk}{\mathrm{check}}
\newcommand{\reff}{\mathrm{ref}}
\newcommand{\Mll}{M_{ll}}
\newcommand{\extra}{\mathrm{extra}}
\newcommand{\draft}[1]{}
\newcommand{\muf}{\mu_\text{F}}
\newcommand{\mur}{\mu_\text{R}}

\def\beq{\begin{equation}}  
\def\eeq{\end{equation}}  

\newcommand{\eq}[1]{eq.~\eqref{#1}}


%
The goal of this study is to quantify the precision that
can be expected in the determination of the parton distribution functions (PDFs)
of the proton in the HL-LHC era.
Such ``ultimate PDFs'' will provide an important ingredient
for the physics projections at the HL-LHC with a robust estimate
of theoretical uncertainties, including some of those presented in other chapters
of this Yellow Report.
With this motivation, HL-LHC pseudo-data have been generated for a number
of PDF-sensitive measurements such as top-quark, Drell-Yan, isolated photon,
and $W$+charm production, and then studied the constraints
that these pseudo-data impose on the global PDF analysis by means
of the Hessian profiling method.
While such
studies have been performed in the context of future lepton-hadron colliders,
see e.g.~\cite{AbelleiraFernandez:2012cc,AbelleiraFernandez:2012ty} for
the LHeC,
 this is the first time that such a
 systematic effort has been directed to the projections for a future hadron collider. The study below is described in further detail in~\cite{Harland-Lang:2018iur}.

\subsubsection{HL-LHC measurements for PDF studies}
The PDF-sensitive processes
that will be considered in this study are listed here first.
In all cases, pseudo-data is generated for a centre-of-mass
energy of $\sqrt{s}=14$ TeV assuming a total integrated
luminosity of $\mathcal{L}=3$ ab$^{-1}$ for the CMS and ATLAS experiments, and of
$\mathcal{L}=0.3$ ab$^{-1}$ for the LHCb experiment.
With these settings, HL-LHC pseudo-data has been generated for the following processes:
\begin{itemize}

\item High-mass Drell-Yan, specifically the dilepton
  invariant mass differential distributions $d\sigma(pp\to ll)/dm_{ll}$
  for $m_{ll}\gtrsim 110$ GeV for a central rapidity acceptance, $|\eta^{l}|\le 2.4$.
  This process is particularly useful for quark flavour separation, in particular of the  poorly known large-$x$ sea quarks.

\item   Differential distributions in top-quark pair production,
  providing direct information on the large $x$ gluon~\cite{Czakon:2016olj}.
  Specifically, pseudo-data has been generated for the top-quark
  transverse momentum $p_{\mathrm{T}}^t$ and rapidity $y_t$ as well as for
  the top-quark pair rapidity $y_{t\bar{t}}$ and invariant mass $m_{t\bar{t}}$.

\item The transverse momentum distribution of the $Z$ bosons in the
  large $p_{\mathrm{T}}^Z$ region for central rapidity $|y_{Z}|\le 2.4$
  and different bins of the dilepton invariant mass $m_{ll}$.
  This process is relevant to constrain the gluon and the antiquarks
  at intermediate values of $x$~\cite{Boughezal:2017nla}.

\item The production of $W$ bosons in association with charm quarks
 (both in the central and forward region).
  This process provides a sensitive handle to the strangeness content
  of the proton~\cite{Stirling:2012vh,Chatrchyan:2013mza}.
  The pseudo-data for this process has been generated as function of the 
  pseudorapidity $\eta_l$ of the charged lepton from the $W$ boson decay.

\item Prompt isolated photon production, which represents
  a complementary probe of the gluon PDF at intermediate
  values of $x$~\cite{dEnterria:2012yj,Campbell:2018wfu}.
  Here the pseudo-data have been generated as differential distributions
  in the photon transverse momentum $p_{\mathrm{T}}^\gamma$ for different bins
  in the photon pseudorapidity $\eta^\gamma$.

\item Differential distributions for on-peak $W$ and $Z$ boson
  production in the forward region, $2.0 \le \eta_{l} \le 4.5$, covered
  by detectors with large acceptance, including forward rapidity.
  These measurements constrain quark flavour separation,
  including the strange and charm content of the proton,
  in the large and small $x$ region~\cite{Rojo:2017xpe}.

\item The inclusive production of jets in different bins
  of rapidity (both in the central and forward region)
  as a function of $p_{\mathrm{T}}^{\rm jet}$.
  Jets have been reconstructed using the
  anti--$k_t$ algorithm~\cite{Cacciari:2008gp}
  with $R=0.4$, and provide information on the large-$x$ gluon
  and valence quarks~\cite{Rojo:2014kta}.
    
\end{itemize}

In all cases, the binning and kinematic cuts
from the most recent $\sqrt{s}=13$ TeV analyses or the corresponding
8 TeV analyses if the former are not available, are taken as the baseline.
The binning has been suitably extended to account for the extended
kinematic coverage achieved with $\mathcal{L}=3~(0.3)$ ab$^{-1}$.
The statistical uncertainties are computed from the number of events per bin,
while systematic errors are rescaled as compared to the 13~(or 8) TeV baseline
analysis, see below.
Various scenarios for the reduction of systematic errors are considered,
from a more conservative one to a more optimistic one.
The overall acceptance of the selection cuts (which affects the final event yield
per bin) is estimated globally again based on the reference experimental analysis.

As mentioned above, this list of processes is not exhaustive: several other important
processes will provide useful information on the parton distributions
in the HL-LHC era, from
inclusive dijet production~\cite{Currie:2017eqf} to
single top quark~\cite{Berger:2017zof} and $D$ meson
production~\cite{Gauld:2016kpd}, see also~\cite{Rojo:2015acz}.
In addition, progress may be expected from both the experimental and theory
sides leading to novel processes, not considered so far, being added to the PDF fitting toolbox.
Even with these caveats, the list above
is extensive enough to provide a reasonable snapshot of the PDF-constraining
potential of the HL-LHC.

It is worth emphasising that the projections are based on pseudo-data which have been generated
specifically
for this study.
They are thus not endorsed by
the LHC experiments, although the feedback received from the
ATLAS, CMS, and LHCb contact persons have been taken into account.

\subsubsection*{Generation of HL-LHC pseudo-data and fitting procedure}
For each of the HL-LHC processes listed above, theoretical predictions have been generated
at next-to-leading order (NLO) using {\sc MCFM}~\cite{Boughezal:2016wmq} interfaced to
{\sc APPLgrid}~\cite{Carli:2010rw} to produce
the corresponding fast grids.
The central value of the pseudo-data is first produced according
the central prediction of the  PDF4LHC15 NNLO set~\cite{Butterworth:2015oua}, and then  fluctuations as expected
by the corresponding experimental uncertainties are included.
Since the present study is based on pseudo-data, it does not account for higher-order QCD effects
or electroweak corrections.
As in the case of PDF closure tests~\cite{Ball:2014uwa}, here only the relative
reduction of PDF uncertainties once the HL-LHC data are added are of interest, while by construction
the central value will be mostly unaffected.

To be more specific, if $\sigma_i^{\rm th}$ is the theoretical cross-section
for  bin $i$ of a given process, computed with PDF4LHC15 NNLO, then the central value of
the HL-LHC pseudo-data $\sigma_i^{\rm exp}$ is constructed by means of
\be
\label{eq:pseudodataGen}
\sigma_i^{\rm exp} = \sigma_i^{\rm th} \times \lp 1 + r_i\cdot \delta^{\rm exp}_{{\rm tot},i}
 + \lambda\cdot \delta^{\rm exp}_{\mathcal L}\rp \, ,
\ee
where $r_i$, $\lambda$ are univariate Gaussian random numbers, $\delta^{\rm exp}_{{\rm tot},i}$ is the total
(relative) experimental uncertainty corresponding to this specific bin, and $\delta^{\rm exp}_{\mathcal L}$
is the luminosity uncertainty related to the experiment.
The latter are taken to be 1.5\% for each of the CMS, ATLAS, and LHCb experiments.
The motivation for adding the fluctuations on top of the central theoretical predictions
is to simulate the statistical and systematic uncertainties
of an actual experimental measurement.
In \eq{eq:pseudodataGen} the total experimental error is defined as
\be
\label{eq:totalExpError}
\delta^{\rm exp}_{{\rm tot},i} \equiv \lp \lp \delta^{\rm exp}_{{\rm stat},i}\rp ^2 +
\lp f_{\rm corr}\times f_{\rm red}\times
\delta^{\rm exp}_{{\rm sys},i}\rp^2 \rp^{1/2} \, .
\ee
In this expression, the relative statistical error $\delta^{\rm exp}_{{\rm stat},i}$ is
computed as
\be
\delta^{\rm exp}_{{\rm stat},i} = \lp f_{\rm acc} \times N_{{\rm ev},i}\rp^{-1/2} \, ,
  \ee
  where $N_{{\rm ev},i}=\sigma_i^{\rm th} \times \mathcal{L}$ is the expected number
  of events in bin $i$ at the HL-LHC with $\mathcal{L}=3~(0.3)$ ab$^{-1}$, and $f_{\rm acc}\le 1$
  is an acceptance correction which accounts for the fact that, for some of the processes
  considered, such as top quark pair production, there is a finite experimental acceptance
  and/or one needs to include the effects of branching fractions.
  The value of
  $f_{\rm acc}$ is then determined by extrapolation using the reference dataset. The one exception to this is the case of forward $W$+charm production, 
  for which no  baseline measurement has so far been performed by LHCb; here 
  the acceptance is set to $f_{\rm acc}=0.3$ to account for the anticipated $c$--jet tagging efficiency. 
  In \eq{eq:totalExpError},   $\delta^{\rm exp}_{{\rm sys},i}$ indicates the total
  systematic error of bin $i$ taken from the reference LHC measurement at either 8 TeV
  or 13 TeV. The  correction factor  $f_{\rm red}\le 1$ accounts for the expected improvement in the average systematic
  uncertainties  at the HL-LHC in comparison to Run-2, due to both detector
  improvements and the enlarged dataset for calibration. 
  
With the exception of the luminosity in \eq{eq:totalExpError} the systematic uncertainties have simply been added in quadrature with the statistical ones. That is, correlations between systematic errors are not taken into account. 
The full inclusion of such correlations goes beyond the scope of the closure tests being pursued in this exercise, which aim simply to provide a reasonable extrapolation of the expected PDF reach at the HL-LHC. In particular, the expected improvements in the overall size of the systematic uncertainties can only be based on the estimates and expectations provided by the LHC collaborations, and cannot be predicted with absolute certainty. The situation is certainly even more challenging in the case of the specific mutual correlations of the systematic uncertainties, which will be sensitive to the precise experimental setup in the future. 
  However, simply excluding the effects of correlations would artificially reduce the
  impact of the pseudo-data into the fit.

  For this reason, an effective correction factor $f_{\rm corr}$ is introduced to accounts
  for the fact that data with correlated systematic uncertainties is more constraining than the same
 data where all errors are added in quadrature.
  The value of $f_{\rm corr}$ has been checked against the available $\sqrt{s}=8$ TeV top quark ~\cite{Khachatryan:2015oqa,Aad:2015mbv} and the $13$ TeV $W$+charm ~\cite{CMS-PAS-SMP-17-014} differential
    distributions, that is $f_{\rm corr}$ is varied until the PDF impact is in line with the result including full experimental correlations.
  This turns out to have a value of between $f_{\rm corr}\simeq 1.0$ and $0.3$ depending on the data set and observable. A factor of $f_{\rm corr}=0.5$ is taken in what follows.

  In Table~\ref{tab:PseudoData} 
  a summary of the features of the HL-LHC pseudo-data generated for the present study is collected.
    For each process, the kinematic coverage, the number of pseudo-data
    points used  $N_{\rm dat}$, the values of the correction factors $f_{\rm acc}$,
    $f_{\rm corr}$, and $f_{\rm red}$; and finally the reference from the 8 TeV or
    13 TeV measurement used as baseline to define the binning and the systematic
    uncertainties of the HL-LHC pseudo-data are indicated.
    A total of $N_{\rm dat}=768$ pseudo-data points are then used in the PDF profiling.
    The values of the reduction factor for the systematic errors
    $f_{\rm red}$ are varied between 1 (0.5) and 0.4 (0.2) in the conservative
    and optimistic scenarios for a 8 TeV (13 TeV) baseline measurement.
    This choice is motivated because available 13 TeV measurements
    are based on a relatively small $\mathcal{L}$ and therefore
    cannot be taken as representative of the systematic errors expected at the
    HL-LHC, even in the most conservative scenario.
\begin{table}[t]
  \centering
   \caption{\small \label{tab:PseudoData}
    Summary of the features of the HL-LHC pseudo-data generated for the present
    study.
    For each process the kinematic coverage, the number of pseudo-data
    points used  $N_{\rm dat}$ across all detectors, the values of the correction factors
    $f_{\rm corr}$ and $f_{\rm red}$; and finally the reference from the 8 TeV or
    13 TeV measurement used as baseline to define the binning and the systematic
    uncertainties of the HL-LHC pseudo-data, as discussed in the text, are indicated.
  }
  \renewcommand{\arraystretch}{1.50}
  \small
  \begin{tabular}{|c|c|c|c|c|c|}
  \hline
    Process    &   Kinematics  &   $N_{\rm dat}$  &  
      $f_{\rm corr}$  &  $f_{\rm red}$ &  Baseline  \\
\hline\hline
\multirow{3}{*}{$Z$ $p_{\mathrm{T}}$}  &    $20\,{\rm GeV}\le p_{\mathrm{T}}^{ll} \le 3.5$ TeV           &
\multirow{3}{*}{338} &   \multirow{3}{*}{0.5}  & \multirow{3}{*}{$\lp 0.4, 1\rp$}
& \multirow{3}{*}{\cite{Aad:2015auj} (8 TeV)}  \\
  &    $12\,{\rm GeV}\le m_{ll} \le 150$ GeV           &              & & &\\
   &    $|y_{ll}|\le 2.4$            &                    &   &     &\\
\midrule
\multirow{2}{*}{high-mass Drell-Yan}  &   $p_{\mathrm{T}}^{l1(2)}\ge 40(30)\,{\rm GeV}$           &  \multirow{2}{*}{32}         &        \multirow{2}{*}{0.5}      
&       \multirow{2}{*}{$\lp 0.4, 1\rp$}        &    \multirow{2}{*}{\cite{Aad:2016zzw} (8 TeV)}     \\
&  $|\eta^l|\le 2.5$, $m_{ll}\ge 116\,{\rm GeV}$   & & & &  \\
\midrule
top quark pair  &  $|y_t|\le 2.4$       &       110                   & 0.5
&         $\lp 0.4, 1\rp$      &   \cite{Aad:2015mbv} (8 TeV)    \\
\midrule
\multirow{2}{*}{$W$+charm (central)}  &        $p_{\mathrm{T}}^\mu \ge26\,{\rm GeV}$, $p_{\mathrm{T}}^c \ge5\,{\rm GeV}$     &  \multirow{2}{*}{12}         &        \multirow{2}{*}{0.5}       
&       \multirow{2}{*}{$\lp 0.2, 0.5\rp$}       &    \multirow{2}{*}{\cite{CMS-PAS-SMP-17-014} (13 TeV)}     \\
&  $|\eta^\mu|\le 2.4$    & & & &\\
\midrule
\multirow{3}{*}{$W$+charm (forward)}  &      $p_{\mathrm{T}}^\mu \ge20\,{\rm GeV}$, $p_{\mathrm{T}}^c \ge20\,{\rm GeV}$           &          \multirow{3}{*}{10}     &        \multirow{3}{*}{0.5}   
&      \multirow{3}{*}{$\lp 0.4, 1\rp$}            &   \multirow{3}{*}{LHCb projection}         \\
&   $p_{\mathrm{T}}^{\mu+c} \ge20\,{\rm GeV}$    & & & & \\
&  $2\le \eta^\mu \le 5$, $2.2\le \eta^c \le 4.2$     & & & & \\
\midrule
Direct photon  &     $E_{\rm T}^\gamma \lsim 3$ TeV, $|\eta_{\gamma}|\le 2.5$          & 118              &      0.5        
&    \multirow{1}{*}{$\lp 0.2, 0.5\rp$}           &   \cite{Aaboud:2017cbm} (13 TeV)      \\
\midrule
\multirow{3}{*}{Forward $W,Z$}  &  $p_{\mathrm{T}}^{l}\ge 20\,{\rm GeV}$, $2.0\le \eta^l\le 4.5$           &  \multirow{3}{*}{90}         &        \multirow{3}{*}{0.5}     
&       \multirow{3}{*}{$\lp 0.4, 1\rp$}        &    \multirow{3}{*}{\cite{Aaij:2015zlq} (8 TeV)}     \\
&    $2.0\le y_{ll}\le 4.5$  & & & & \\
&    $60\le m_{ll}\le 120\,{\rm GeV}$  & & & & \\
\midrule
Inclusive jets  &       $|y| \le 3$, $R = 0.4$      &       58        &      0.5        &
\multirow{1}{*}{$\lp 0.2, 0.5\rp$} 
&   \cite{Khachatryan:2016wdh}    (13 TeV)               \\
\bottomrule
Total   &    &   712 &   &   &   \\
\bottomrule
  \end{tabular}
\end{table}
\subsubsection*{Hessian profiling}
There exist a number of techniques that can be used to quantify
the impact on PDFs of the pseudo-data listed  in Table~\ref{tab:PseudoData}.
In the case of Monte Carlo sets such as NNPDF, the Bayesian
reweighting method~\cite{Ball:2011gg,Ball:2010gb}
reproduces the result of a direct fit, but
it is restricted by the fact that information loss limits its
reliability when the measurements provide
significant new information.
For Hessian sets such as PDF4LHC15\_100 instead, the 
profiling technique~\cite{Paukkunen:2014zia} is more suitable to achieve
the same purpose.
This Hessian profiling is based on
the minimization of
\bea 
\nonumber
\chi^2\lp {\rm \beta_{exp}},{\rm \beta_{th}}\rp
&=&\sum_{i=1}^{N_{\rm dat}}\frac{1}{\lp\delta^{\rm exp}_{{\rm tot},i}\sigma_i^{\rm th} \rp^2}\lp \sigma_i^{\rm exp}
+\sum_j\Gamma_{ij}^{\rm exp}\beta_{j,\rm exp}
 -\sigma_i^{\rm th}
 +\sum_k\Gamma_{ik}^{\rm th}\,\beta_{k,\rm th} \rp^2 \\\label{eq:hessianchi2} &&
 +\sum_j \beta_{j,\rm exp}^2+T^2\sum_k \beta_{k,\rm th}^2 \; ,
 \eea
 with $\sigma_i^{\rm exp}~(\sigma_i^{\rm th})$ are the central values
 of a given experimental measurement (theory prediction),
 $\beta_{j,\rm exp}$ are the nuisance parameters corresponding
 to the set of fully correlated experimental systematic
 uncertainties, $\beta_{k,\rm th}$ are the nuisance parameters
 corresponding to the PDF Hessian eigenvectors, $N_{\rm dat}$ is the number of data
 points and $T$ is the tolerance factor.
 The matrices $\Gamma_{ij}^{\rm exp}$ and
 $\Gamma_{ik}^{\rm th}$ encode the effects of the corresponding
 nuisance parameters on the experimental data and on the
 theory predictions, respectively.

 As mentioned above, in this study the statistical
 and experimental uncertainties are added in quadrature excluding the luminosity, and then the effects of the missing
 correlations are accounted for by means of the factor $f_{\rm corr}$.
 For this reason there are only nuisance parameters for the luminosity
 errors, and for an overall normalization uncertainty of 5\% in forward
 $W$+charm production due to charm-jet tagging.
 If \eq{eq:hessianchi2} is minimised with respect to these nuisance parameters, this gives:
  \be
\label{eq:hessianchi2rev}
\chi^2\lp {\rm \beta_{th}}\rp
=\sum_{i,j=1}^{N_{\rm dat}}\lp \sigma_i^{\rm exp}
 -\sigma_i^{\rm th}
 +\sum_k\Gamma_{ik}^{\rm th}\,\beta_{k,\rm th}\rp \lp \text{cov} \rp_{ij}^{-1}\lp \sigma_j^{\rm exp}
  -\sigma_j^{\rm th}
  +\sum_m\Gamma_{jm}^{\rm th}\,\beta_{m,\rm th}\rp  +T^2\sum_k \beta_{k,\rm th}^2 \; ,
 \ee
 where:
 \be
 \label{eq:covariancematrix}
	 \lp \text{cov} \rp_{ij} = \delta_{ij}\lp \delta^{\rm exp}_{{\rm tot},i}\sigma_i^{\rm th}\rp^2 + 
	 \sum \Gamma_{i,\text{lumi/norm}}^{\rm exp}\Gamma_{j,\text{lumi/norm}}^{\rm exp}.
	 \ee
         \eq{eq:hessianchi2rev} is then minimised
         with respect to the Hessian
 PDF nuisance parameters $\beta_{k,\rm th}$, which can
  be interpreted as leading
 to PDFs that have been optimised 
 to describe this new dataset.
The resulting Hessian matrix on $\beta_{k,\rm th}$ at the minimum can be
diagonalised to construct the new eigenvector directions.
Finally, the PDF uncertainties are determined from the $\Delta\chi^2=T^2$
criteria.
In the studies presented here, a global $T=3$ is used which approximately corresponds to the average
tolerance determined dynamically in the CT14 and MMHT14 analyses.

 \subsubsection*{Results for individual processes}
 The results of the Hessian profiling of
 PDF4LHC15 from individual processes are now presented, and subsequently the corresponding results from the combination of all the HL-LHC
 processes are considered in different scenarios.
 First, the top-quark pair production case listed
 in Table~\ref{tab:PseudoData} is considered.
 In Fig.~\ref{fig:mtt_output_F_0_1} the
 comparison of the predictions for the $m_{t\bar{t}}$
  distribution in top-quark pair production at the HL-LHC using
  PDF4LHC15 is shown with the associated pseudo-data for ATLAS and CMS experiments, and with the profiled
  results with $F\equiv f_{\rm corr}\cdot f_{\rm red}=0.2$.
  The corresponding impact at the level of the gluon
  PDF at $Q=100$ GeV is also presented before and after profiling with all $t\bar t$
  data in Table~\ref{tab:PseudoData}.
  It is clear that the HL-LHC pseudo-data in this scenario will have much
  smaller uncertainties than the PDF uncertainties, so there is a marked
  reduction on the PDF errors on the gluon at large-$x$.
  Note that the two points in each of the bins in Fig.~\ref{fig:mtt_output_F_0_1} (left)
  correspond to the ATLAS and CMS pseudo-data.

\begin{figure}[t]
  \begin{center}
\includegraphics[width=0.49\linewidth]{\main/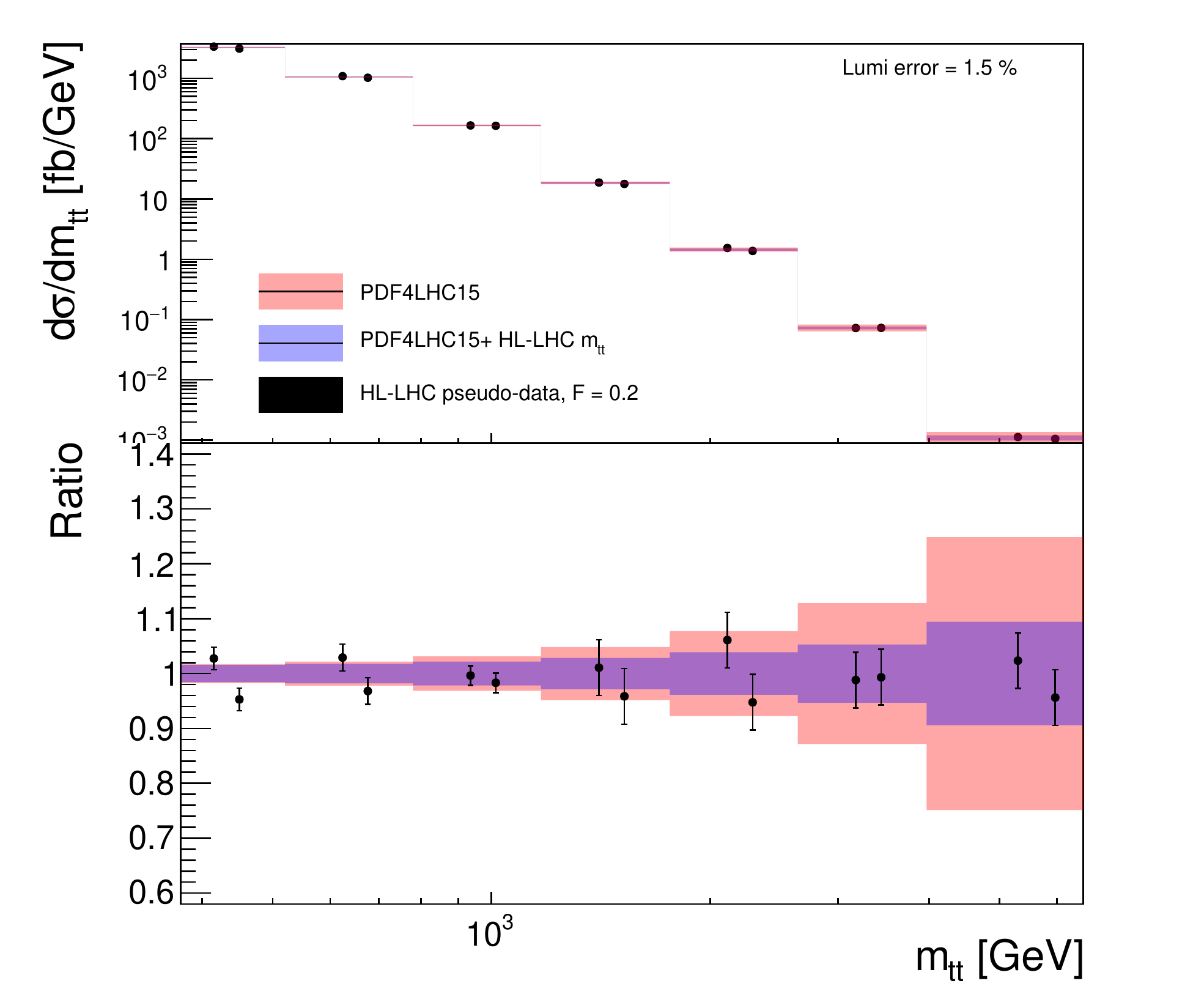}
\includegraphics[width=0.49\linewidth]{\main/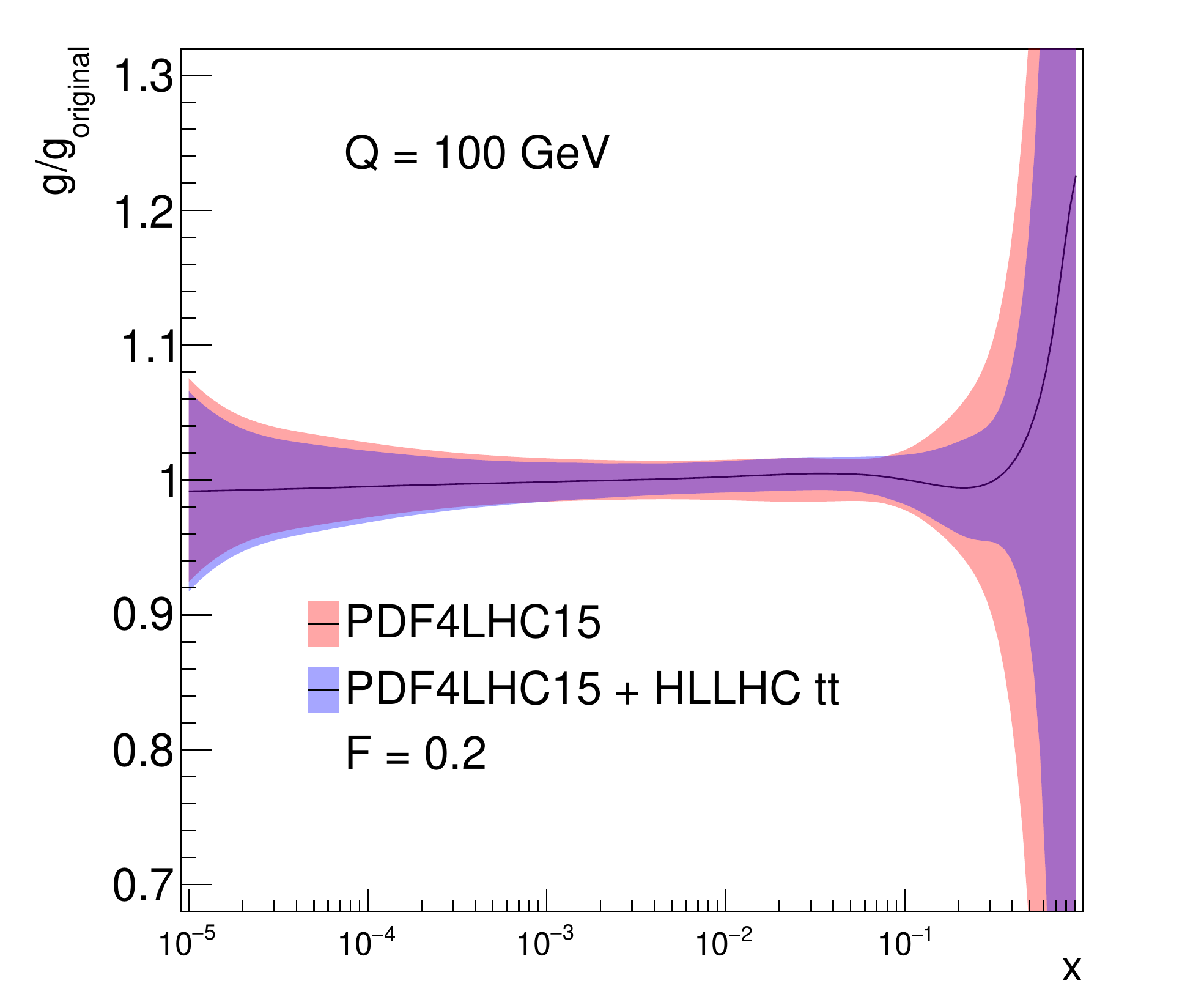}
\caption{\small Comparison of the predictions for the $m_{t\bar{t}}$
  distribution in top-quark pair production at the HL-LHC using
  PDF4LHC15 with the associated pseudo-data and with the profiled
  results with $F\equiv f_{\rm corr}\cdot f_{\rm red}=0.2$ (left).
  The corresponding differences at the level of the gluon
  PDF at $Q=100$ GeV before and after profiling all top-quark pair production observables (right).
     \label{fig:mtt_output_F_0_1} }
  \end{center}
\end{figure}

Two other  representative
processes are considered next: $W$+charm quark production in central rapidity region
and the high-mass Drell-Yan process.
In Fig.~\ref{fig:DY_WC_output_F_0_1} the same comparison is shown as in
Fig.~\ref{fig:mtt_output_F_0_1} for these two processes.
In the case of the $W$+charm quark production, a clear reduction of PDF
errors is observed in the strangeness, $s+\overline{s}$, at intermediate values of $x$, highlighting the
sensitivity of this measurement to the strange content
of the proton.
For the case of high-mass Drell-Yan, the uncertainties
on the $\bar u$ quark PDF are reduced at large $x$ region.
Here the impact is rather moderate, as experimental and PDF errors
are comparable even in the high $m_{ll}$ region.

\begin{figure}[t]
  \begin{center}
\includegraphics[width=0.49\linewidth]{\main/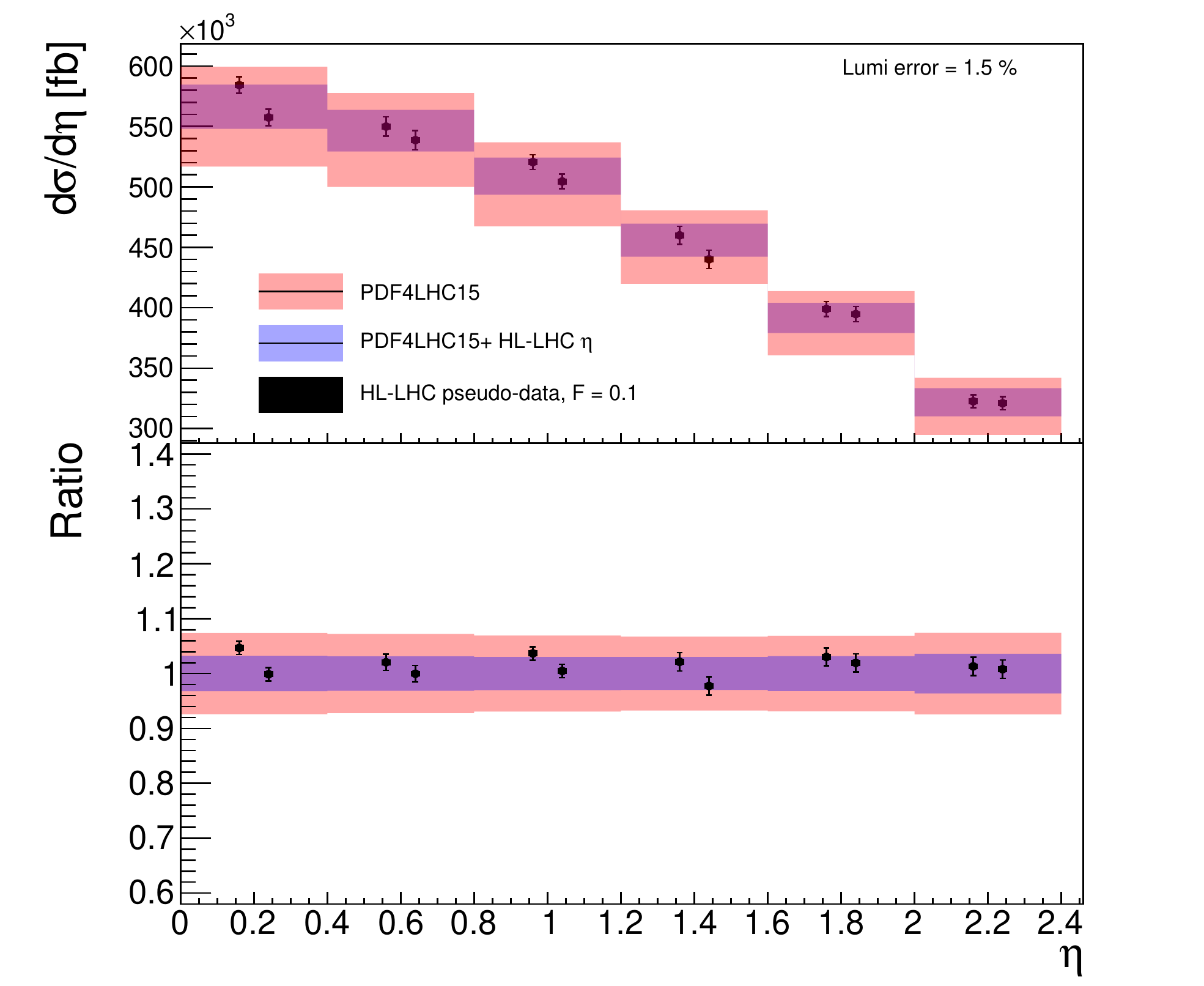}
\includegraphics[width=0.49\linewidth]{\main/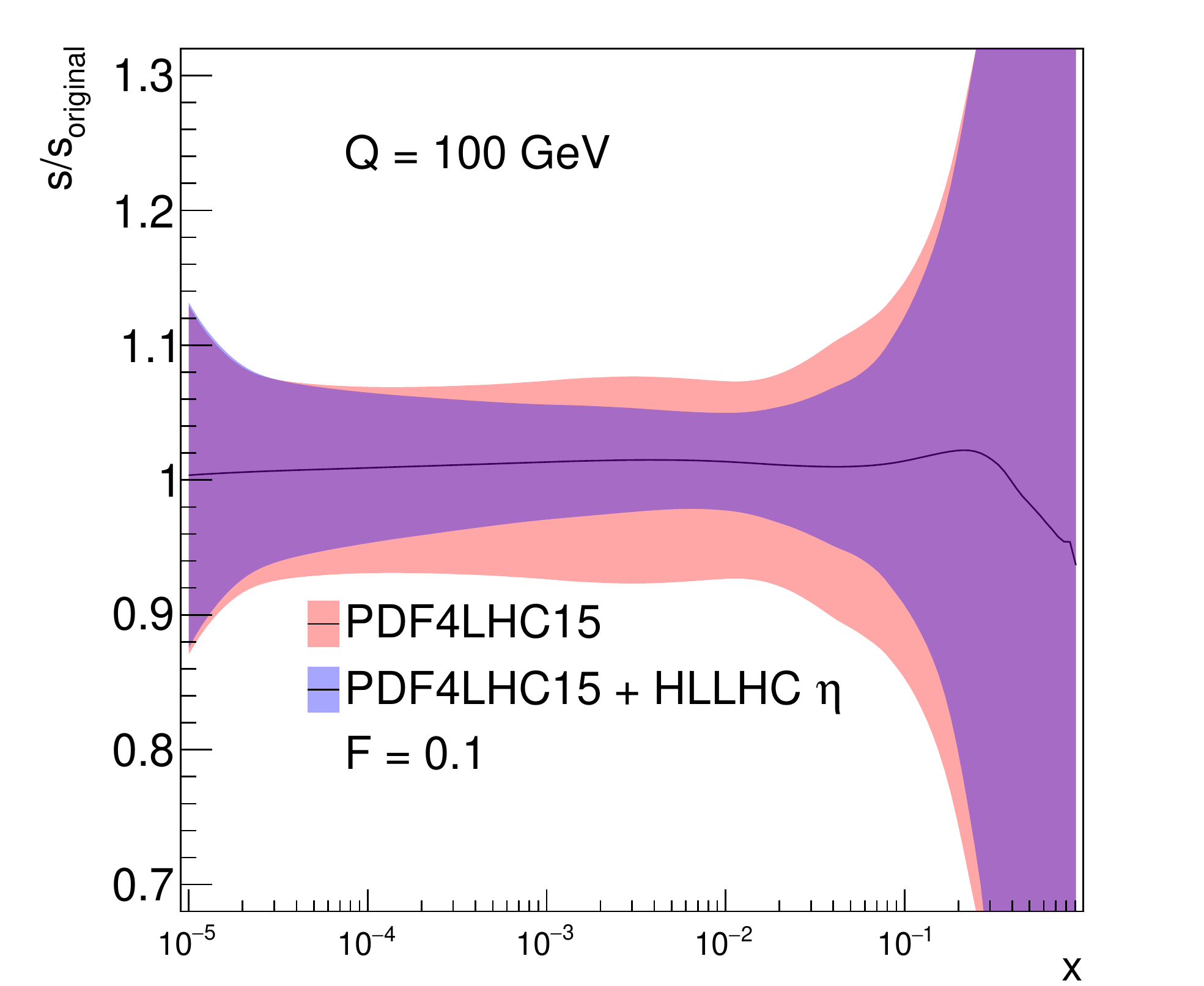}
\includegraphics[width=0.49\linewidth]{\main/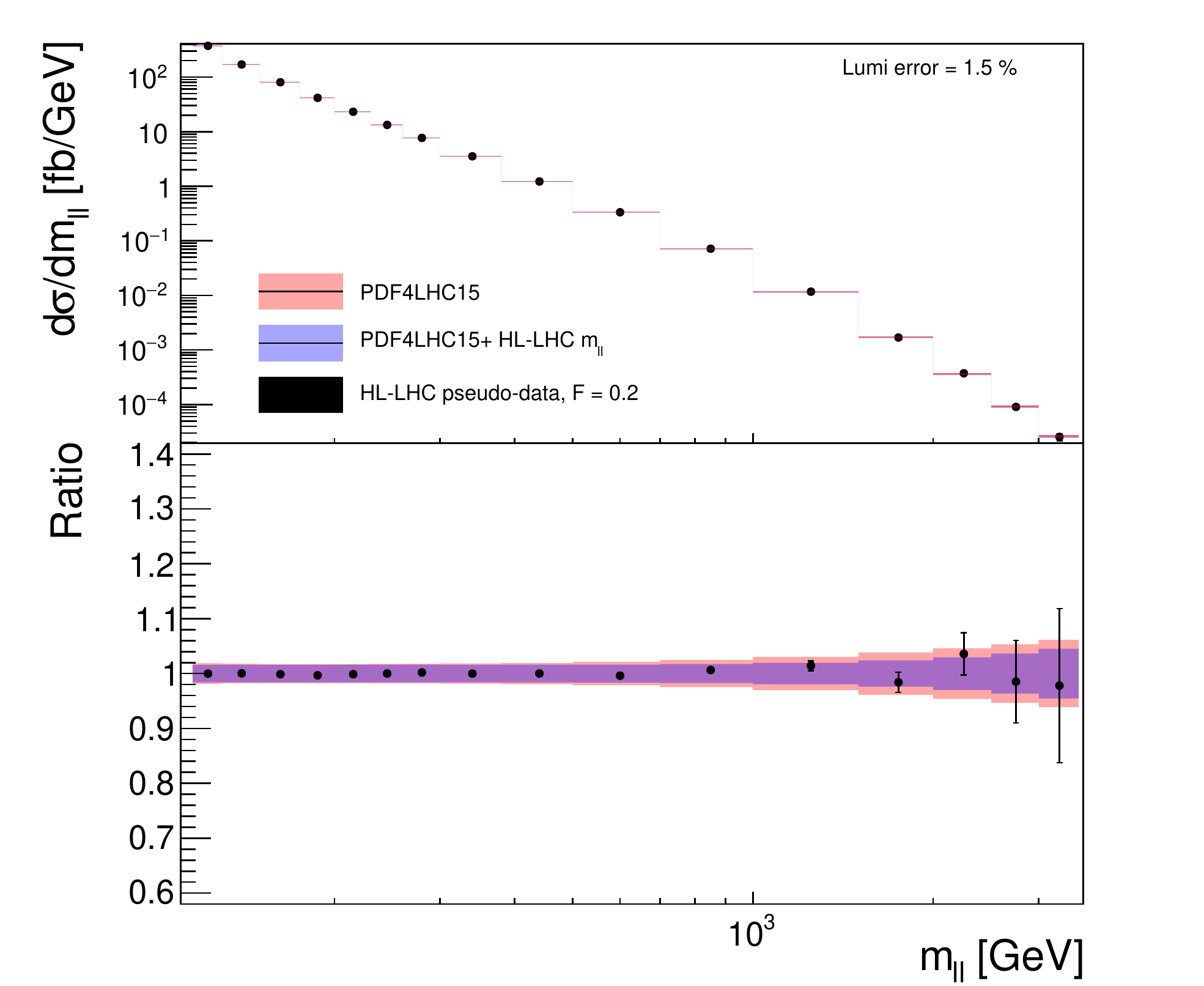}
\includegraphics[width=0.49\linewidth]{\main/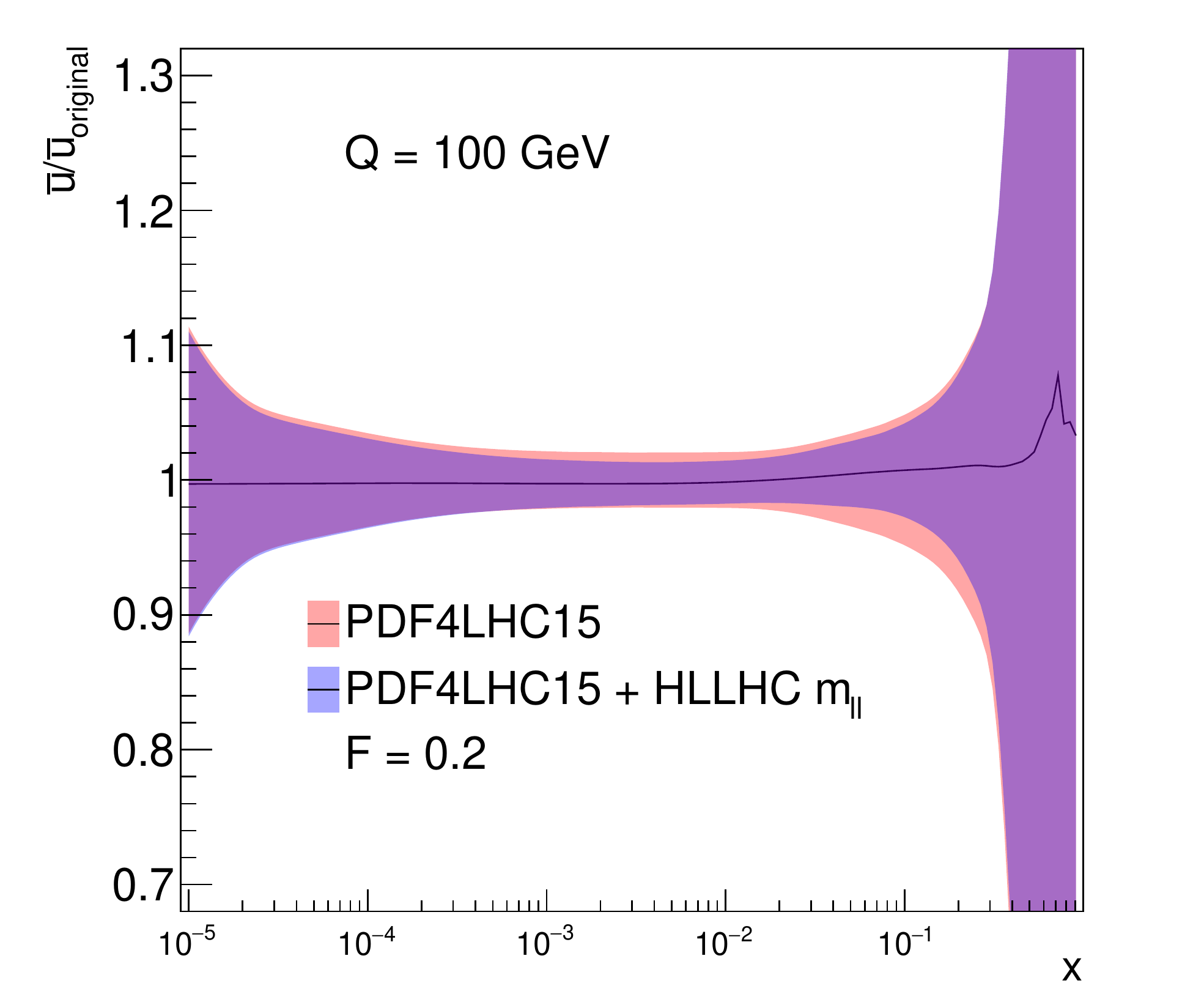}
\vspace{0.2cm}
\caption{\small Same as Fig.~\ref{fig:mtt_output_F_0_1}
  for $W$+charm quark production with impact on strange quark PDF (upper)
  and the high-mass Drell-Yan process with impact on $\bar u$ PDF (lower).
     \label{fig:DY_WC_output_F_0_1} }
  \end{center}
\end{figure}

\subsubsection{Ultimate PDFs from HL-LHC data}
The final profiled PDF sets are based on the combined datasets listed in Table~\ref{tab:PseudoData}; these provide an estimate of the impact of future HL-LHC measurements
into our knowledge of the quark and gluon structure of the proton.
In Table~\ref{tab:Scenarios} the
three scenarios for the systematic uncertainties of the HL-LHC pseudo-data assumed in the present exercise are listed.
    These scenarios, ranging from more conservative to more optimistic, differ among them in
    the reduction factor $f_{\rm red}$, \eq{eq:totalExpError},
    applied to the systematic errors of the reference
    8 TeV or 13 TeV measurements.
    In particular, in the optimistic scenario a reduction
    of the systematic errors by a factor 2.5 compared to the
    reference 8 TeV measurements is assumed. A large factor of 5 for the 13 TeV measurements is assumed, correcting for the fact that these are based in the initial datasets which generally have larger systematic errors in comparison to the 8 TeV case.
    The name of the corresponding {\tt LHAPDF} grid is also indicated in each case.

\begin{figure}[t]
  \begin{center}
\includegraphics[width=0.49\linewidth]{\main/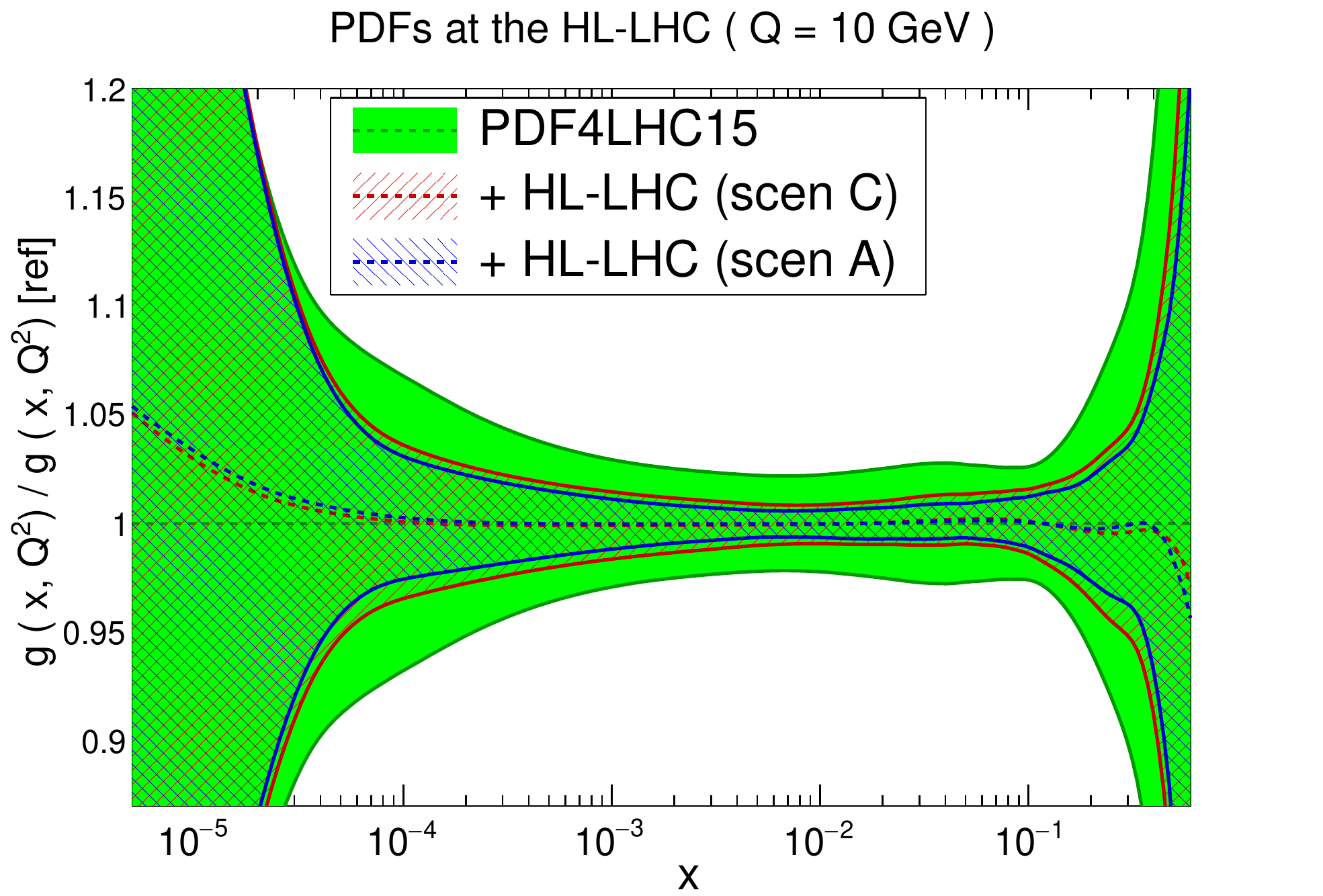}
\includegraphics[width=0.49\linewidth]{\main/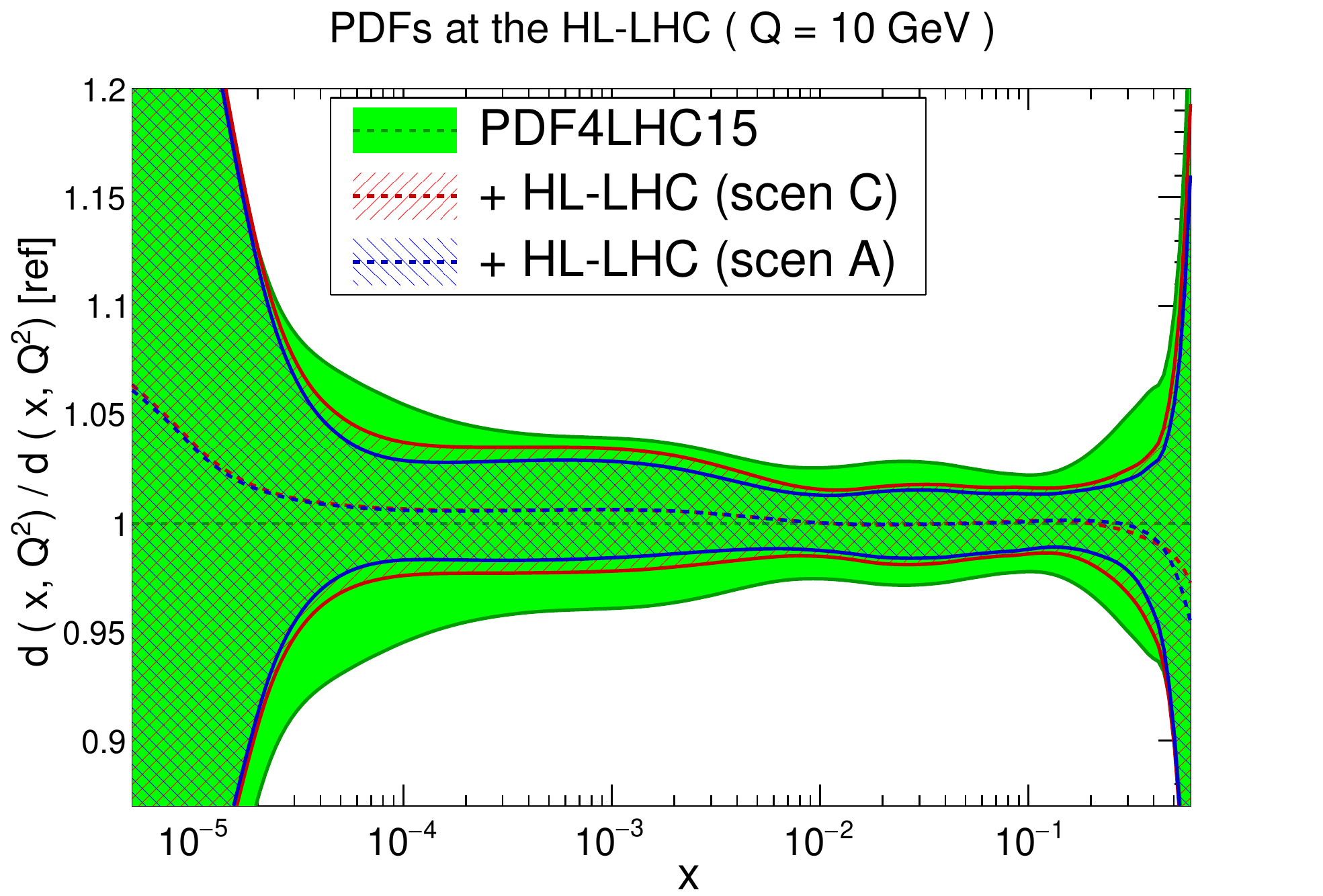}
\includegraphics[width=0.49\linewidth]{\main/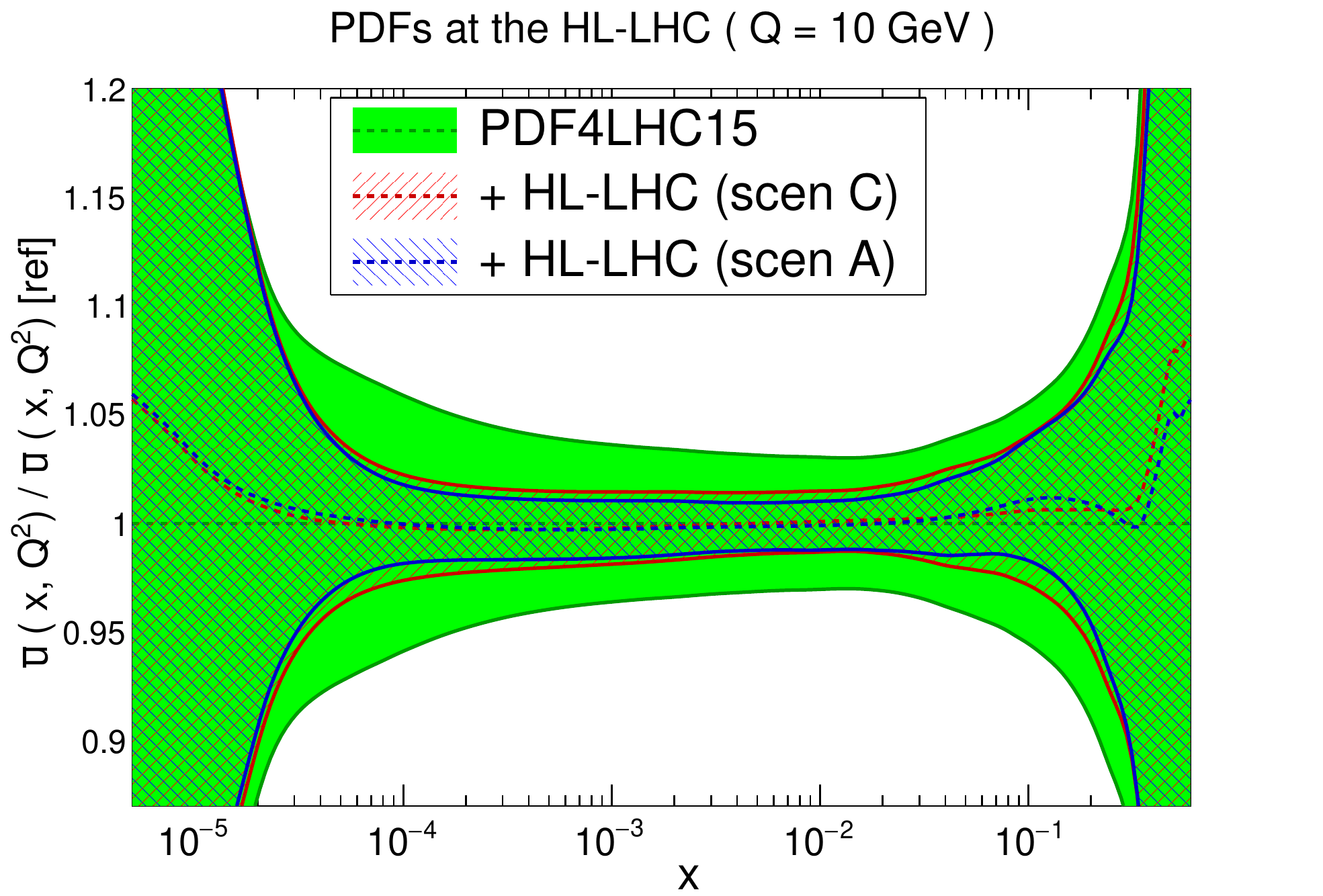}
\includegraphics[width=0.49\linewidth]{\main/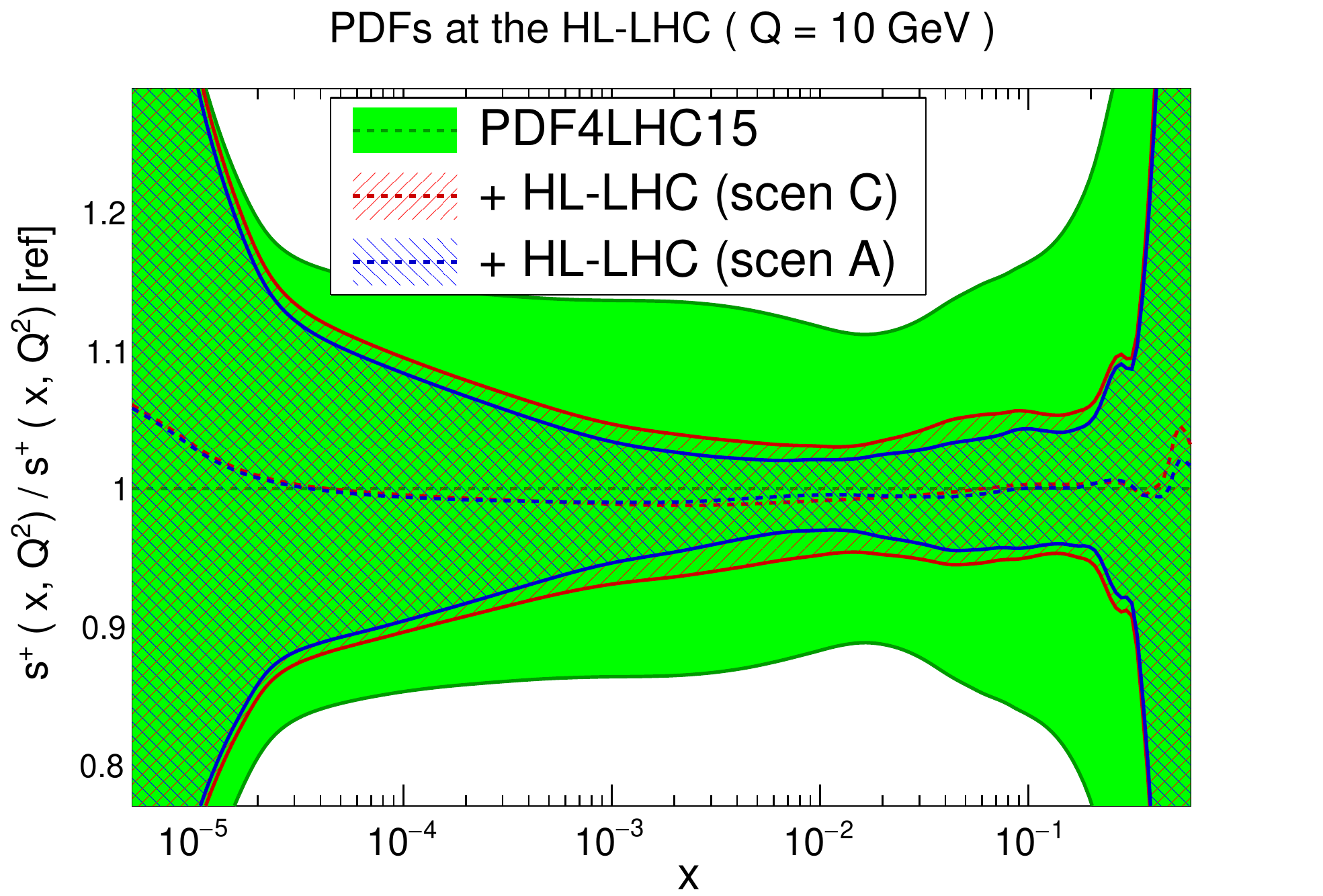}
\caption{\small Comparison of PDF4LHC15 with the profiled sets
  with HL-LHC data in scenarios A and C (see text).
  The gluon, down quark, up anti-quark, and total
  strangeness at $Q=10$ GeV are shown, normalized to the central value of
  the baseline.
     \label{fig:PDFratios} }
  \end{center}
\end{figure}

\begin{table}[h!]
  \centering
  \caption{\small \label{tab:Scenarios}
    The three scenarios for the systematic uncertainties of the HL-LHC pseudo-data
    assumed in the present exercise.
    These scenarios, ranging from conservative to optimistic, differ among them in
    the reduction factor $f_{\rm red}$, \eq{eq:totalExpError},
    applied to the systematic errors of the reference
    8 TeV or 13 TeV measurements.
    The name of the corresponding {\tt LHAPDF} grid is also indicated in each case.
  }
  \renewcommand{\arraystretch}{1.20}
  \begin{tabular}{|c|c|c|c|c|}
  \hline
    Scenario    &   $f_{\rm red}$ (8 TeV)  & $f_{\rm red}$ (13 TeV) &   {\tt LHAPDF} set  &
    Comments \\
    \hline\hline
    A          &   0.4   &  0.2  &  {\tt PDF4LHC\_nnlo\_hllhc\_scen3}  & Optimistic \\
    \midrule
    B         &   0.7   &  0.36  & {\tt PDF4LHC\_nnlo\_hllhc\_scen2}  & Intermediate \\
     \midrule
     C          &   1   &  0.5   & {\tt PDF4LHC\_nnlo\_hllhc\_scen1}  & Conservative \\
 \bottomrule
  \end{tabular}
  \end{table}

Then in Fig.~\ref{fig:PDFratios} a comparison of
the baseline PDF4LHC15 set is presented with the profiled sets
  based on HL-LHC pseudo-data from scenarios A and C in Table~\ref{tab:Scenarios}.
  Specifically, the gluon, down quark, up anti-quark, and total
  strangeness at $Q=10$ GeV are shown, normalized to the central value of
  the baseline.
  The predictions of scenarios A and C
  (optimistic and conservative respectively) are observed to be reasonably similar.
  This demonstrates that the results are relatively robust against the projections
  of how experimental errors will be reduced in HL-LHC measurements.
  A marked reduction of PDF uncertainties is visible in all cases, and
  is particularly significant for the gluon and the sea quarks,
  which are worse known than the valence quarks.

Next, the partonic luminosities are investigated, in particular
by quantifying the improvement in the PDF uncertainties in different
initial-state partonic combinations from the HL-LHC pseudo-data.
In Fig.~\ref{fig:PDFluminosities} the reduction of PDF uncertainties are shown
in the $gg$, $qg$, $q\bar{q}$,
  and $qq$ luminosities at $\sqrt{s}=14$ TeV due to the HL-LHC pseudo-data
  (in scenarios A and C)  with respect
  to the PDF4HC15 baseline.
  The average values of this PDF error reduction for three different invariant mass bins
  (low, medium, and high values of $M_X$)
  is shown in the table in Fig.~\ref{fig:PDFs-HL-LHC-summaryTable}.\footnote{The average is
  computed from 10 points per mass bin, log-spaced in $M_X$.}
  The value outside (inside) brackets correspond to scenario C (A).
  Note that in this table the $us$ luminosity is also listed,
  which contributes to processes such as inclusive $W^+$ production.

\begin{figure}[t]
  \begin{center}
\includegraphics[width=0.49\linewidth]{\main/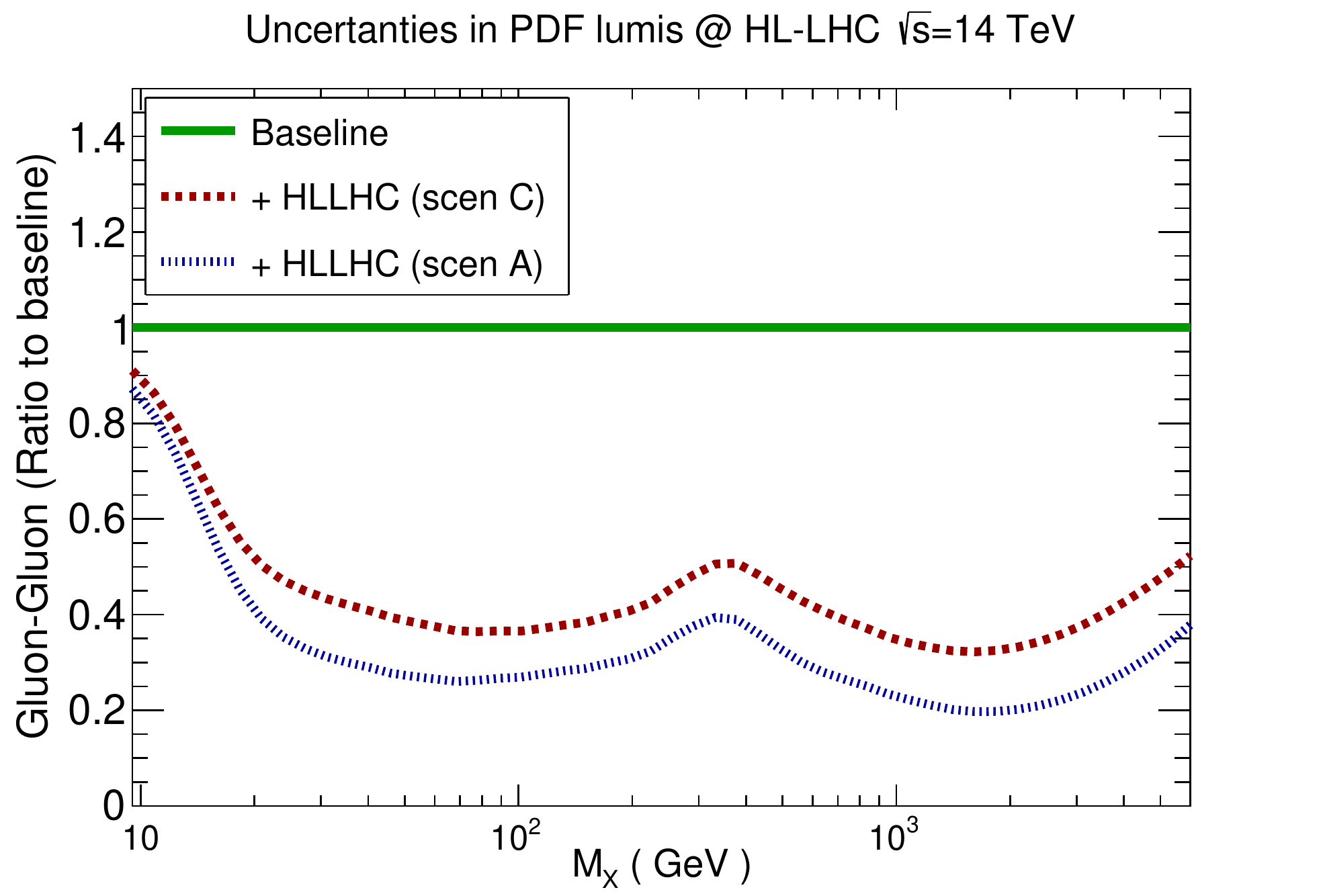}
\includegraphics[width=0.49\linewidth]{\main/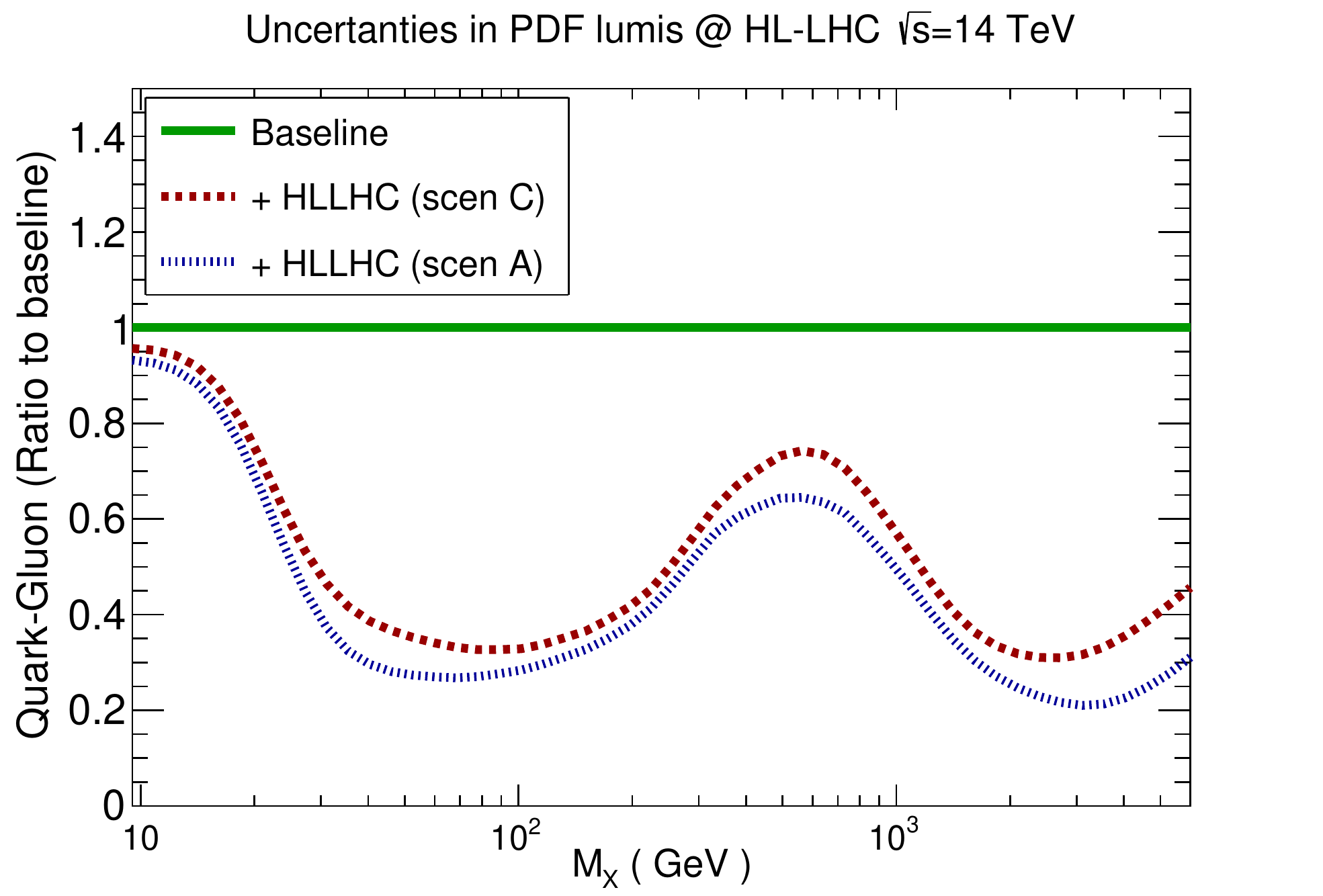}
\includegraphics[width=0.49\linewidth]{\main/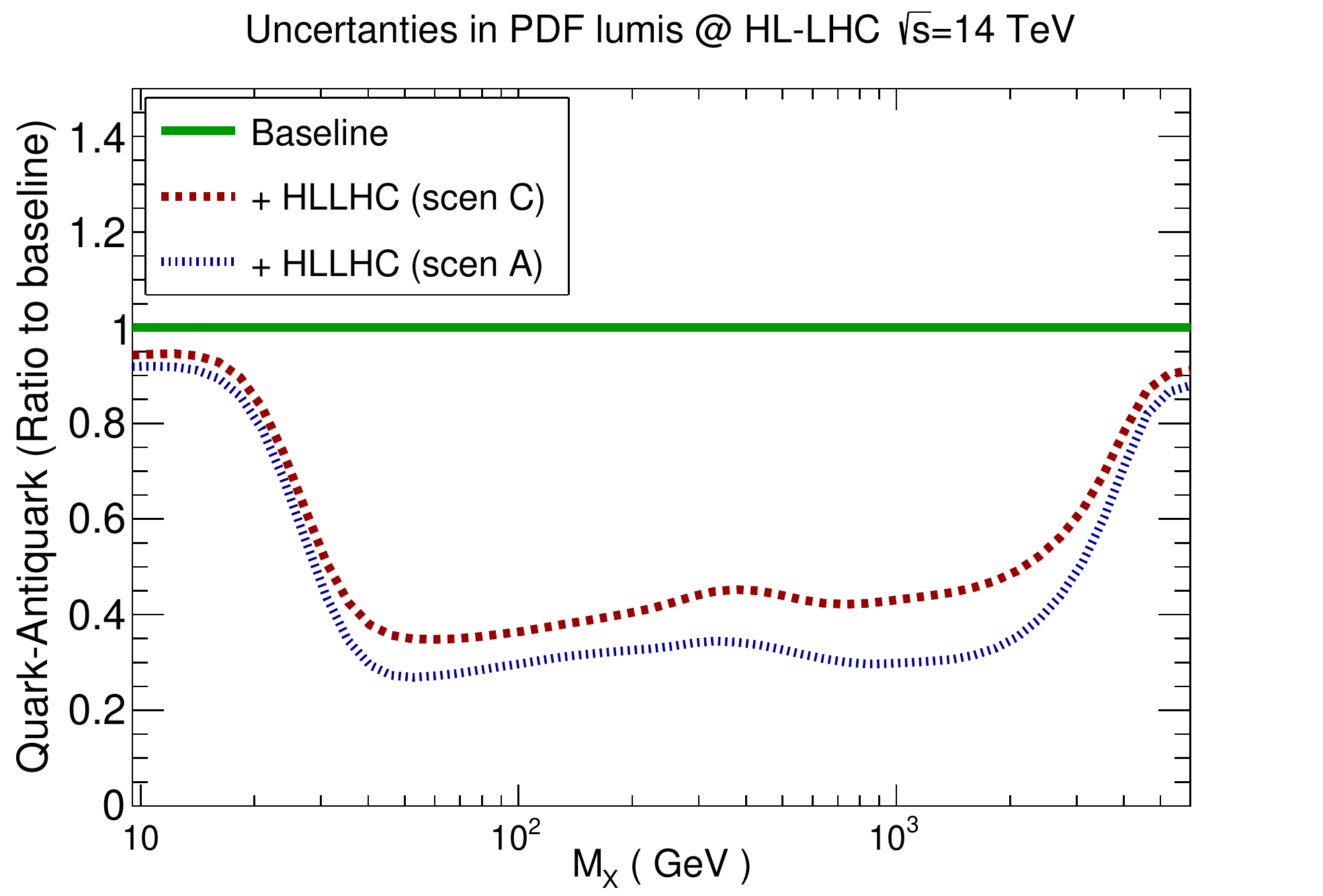}
\includegraphics[width=0.49\linewidth]{\main/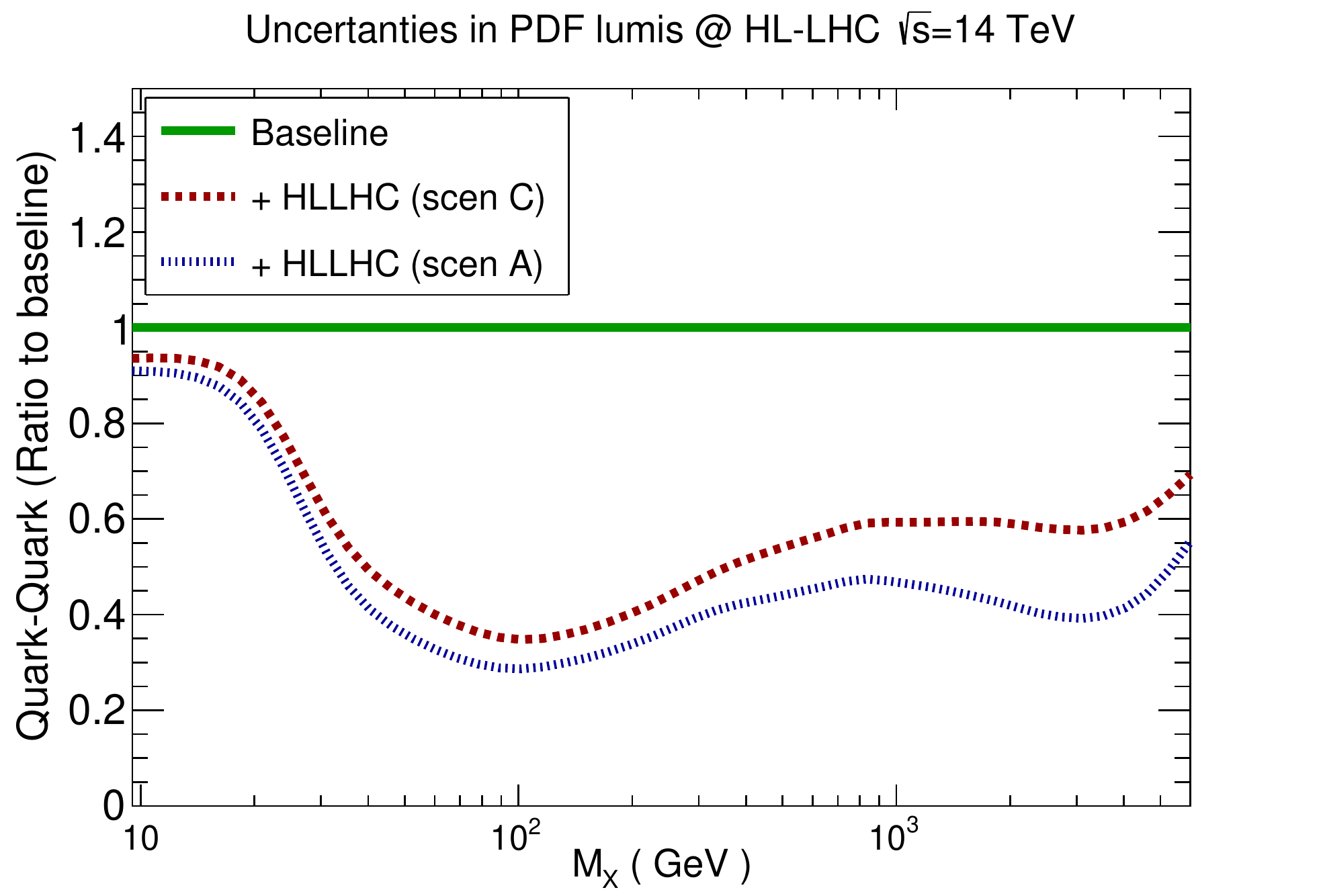}
\caption{\small The reduction of PDF uncertainties in the $gg$, $qg$, $q\bar{q}$,
  and $qq$ luminosities at $\sqrt{s}=14$ TeV due to the HL-LHC pseudo-data
  (in scenarios A and C)  with respect
  to the PDF4HC15 baseline.
     \label{fig:PDFluminosities} }
  \end{center}
\end{figure}

\begin{figure}[t]
  \begin{center}
\includegraphics[width=0.80\linewidth]{\main/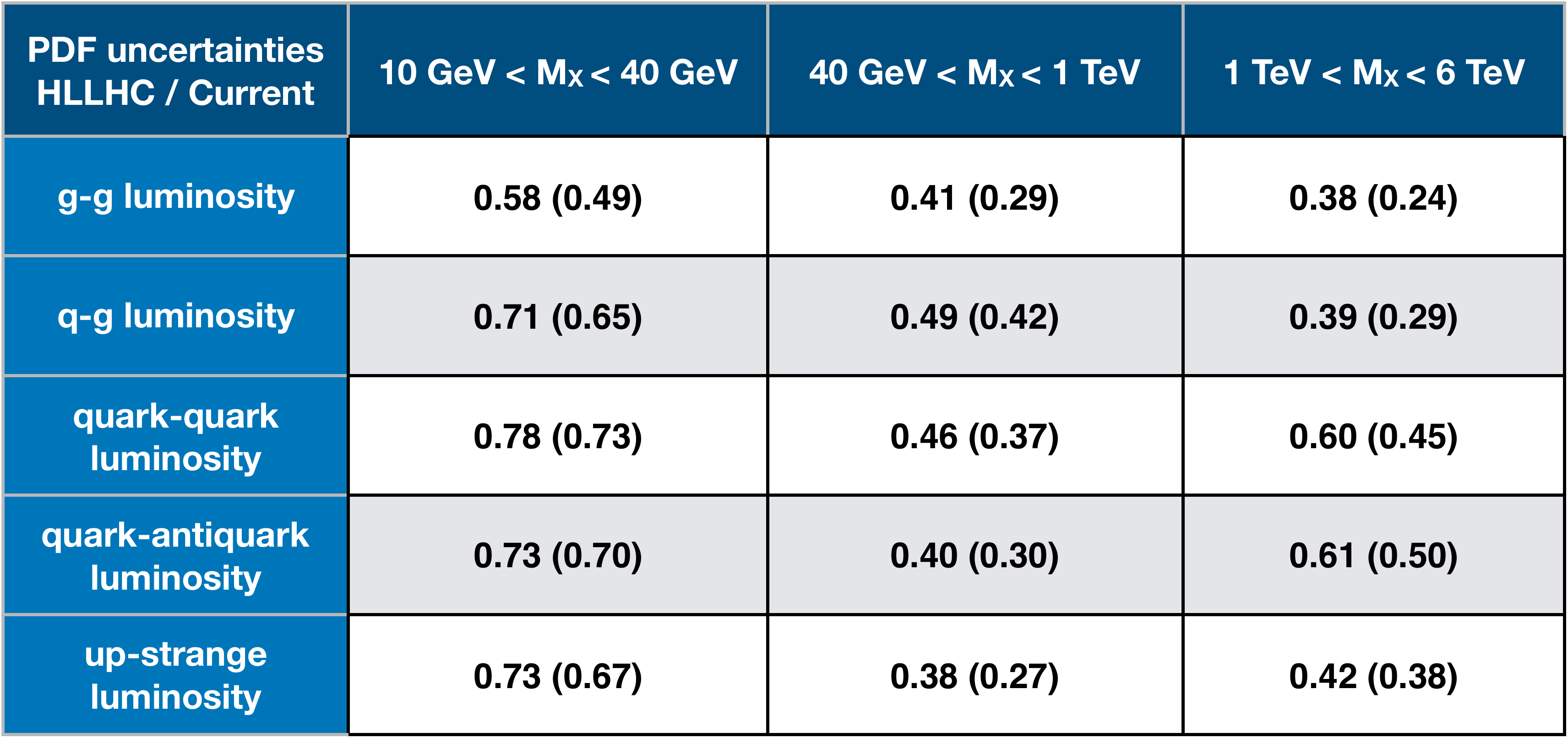}
\vspace{+0.1cm}
\caption{\small The uncertainties associated to different PDF luminosities,
  normalised to the uncertainties of the current baseline (PDF4LHC15).
  The average for three different invariant mass $M_X$ bins is computed.
  The numbers outside (inside) brackets correspond to the conservative
  (optimistic) scenario.
     \label{fig:PDFs-HL-LHC-summaryTable} }
  \end{center}
\end{figure}

From the comparisons in Fig.~\ref{fig:PDFluminosities} and in
Fig.~\ref{fig:PDFs-HL-LHC-summaryTable} it is observed
the overall error reduction is not too sensitive to the specific projections assumed
for the experimental systematic uncertainties.
In the intermediate mass bin, $40~{\rm GeV}\le M_X\le 1~{\rm TeV}$,
the reduction of PDF uncertainties ranges roughly between a factor of 2-4,
depending on the partonic channel and the scenario for the systematic errors.
For example, for the $gg$ luminosity in the range relevant for Higgs production,
a reduction by a factor $\simeq 3$ in scenario A is found.
A similar improvement is found in the high mass region, $M_X\ge 1$ TeV, directly
relevant for beyond-SM (BSM) searches.
In the optimistic scenario, the PDF error reduction at high masses
ranges between a factor 4
for the $gg$ luminosity to around a factor 2 for the $qq$ and $q\bar{q}$ ones.
On the other hand, the PDF error reduction is more moderate in the low
mass region, $ M_X\lesssim 20~{\rm GeV}$, since none of the processes
in Table~\ref{tab:PseudoData} is directly sensitive to it.

\subsubsection*{Implications for LHC phenomenology}
Now some selected phenomenological implications of these ``ultimate'' PDFs
at the HL-LHC are presented for a variety of processes, both within the SM and beyond.
First high-mass supersymmetric (SUSY) particle production at the
HL-LHC is considered, where sparticles masses up to $\simeq 3$ TeV can be searched for. While this SUSY scenario is considered for concreteness, similar results will hold for the production
of new BSM states within other models.
In Fig.~\ref{fig:susyxsects} the comparison between
the baseline PDF4LHC15 predictions
with the corresponding HL-LHC results is shown corresponding to scenarios
C and A (conservative and optimistic respectively),
normalised to the central value of the former.
Specifically, the cross-sections
  for gluino-gluino and squark-gluino are shown at $\sqrt{s}=14$ TeV.
  Theoretical predictions have been computed at leading order (LO) using
  {\sc Pythia8.235}~\cite{Sjostrand:2007gs}
  with the SLHA2 benchmark point~\cite{Allanach:2008qq} for a range
  of sparticle masses.
  For simplicity, underlying event and multiple interactions have been ignored.

\begin{figure}[t]
  \begin{center}
    \includegraphics[width=0.49\linewidth]{\main/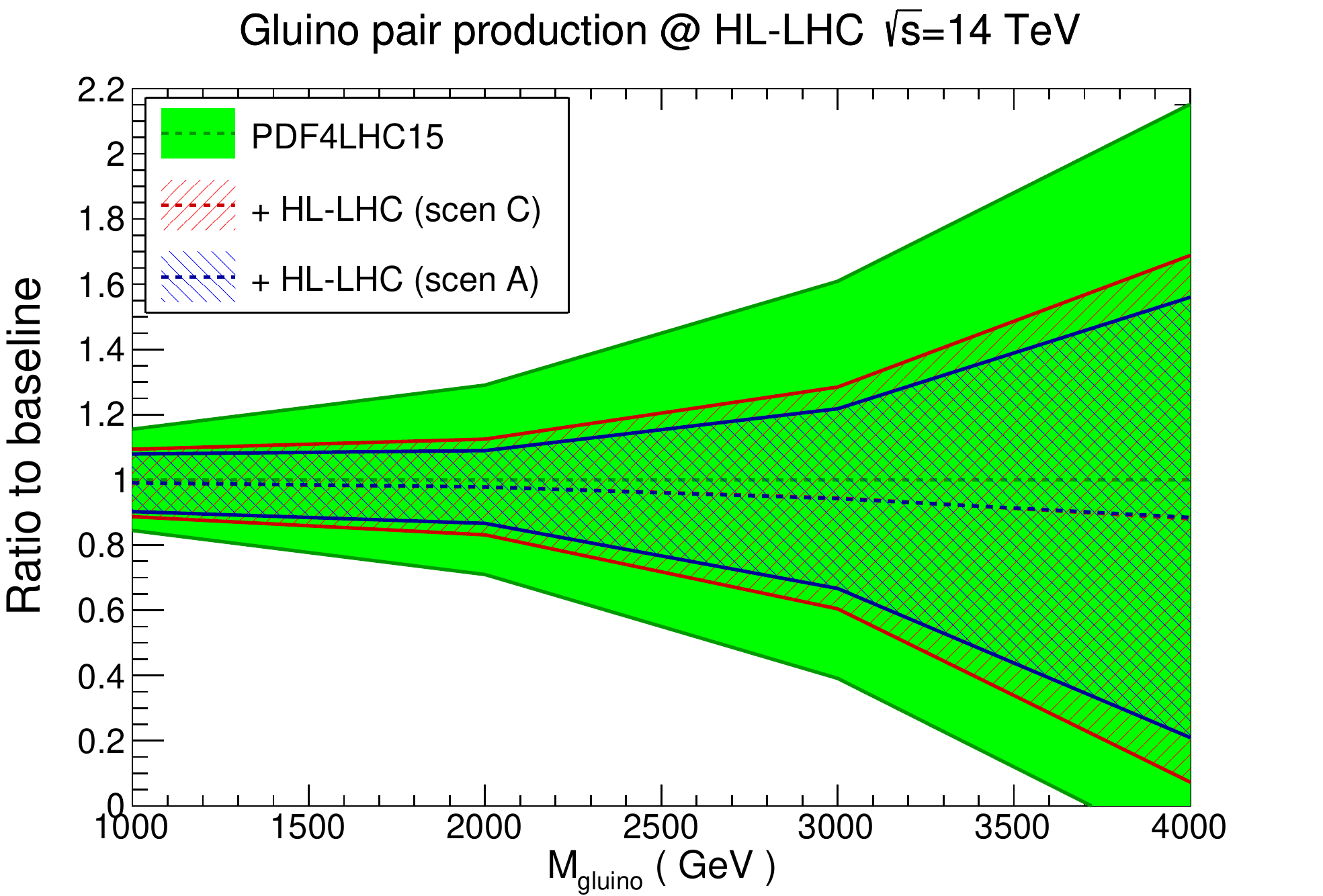}
    \includegraphics[width=0.49\linewidth]{\main/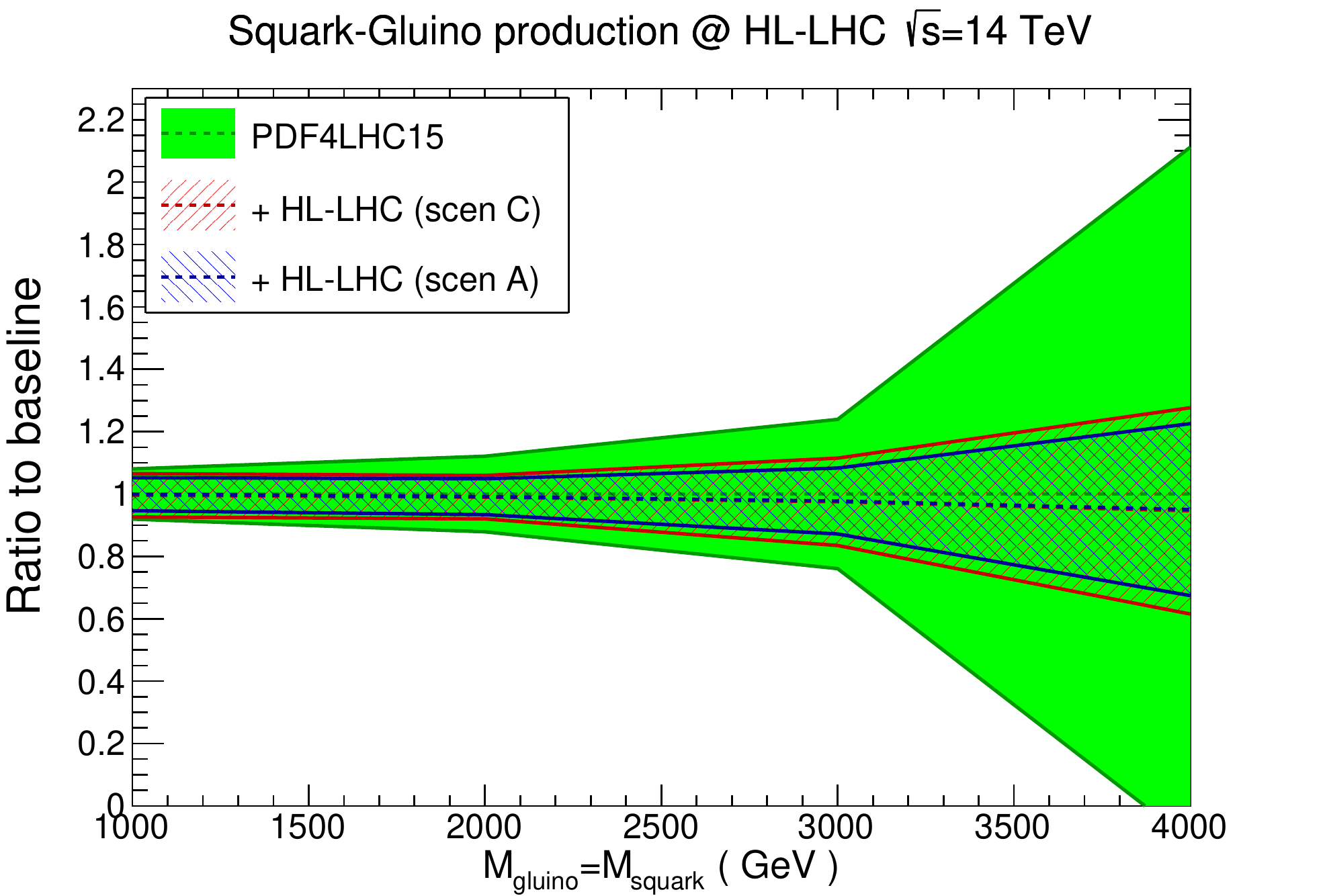}
\caption{\small Comparison between the baseline PDF4LHC15 predictions
  for  high-mass supersymmetric particle production at the
  HL-LHC
  with the corresponding HL-LHC projections corresponding to
  scenarios C and A, normalised 
 to the central value of the baseline.
  The results for gluino-gluino and squark-gluino
  production cross-sections are shown at $\sqrt{s}=14$ TeV.
     \label{fig:susyxsects} }
  \end{center}
\end{figure}

From the comparisons in Fig.~\ref{fig:susyxsects}, the constraints
on the PDFs from the HL-LHC pseudo-data lead to a marked reduction
to the uncertainties in the high-mass SUSY cross-sections,
consistent with the corresponding reduction at the level
of luminosities reported in Fig.~\ref{fig:PDFluminosities}.
For instance, for gluino pair-production with $M_{\widetilde{g}}=3$ TeV,
the PDF uncertainties are reduced from $\simeq 60\%$ to $\simeq 25\%$ in
the optimistic scenario.
An even more marked reduction is found for the squark-gluino cross-section,
specially at large sparticle masses.
More moderate improvements are found in the case of squark-antisquark
production, due to the limited constraints that the HL-LHC provides
on the large-$x$ antiquarks, at least for the processes considered here.
In this case, an error reduction of a factor of $\simeq 25\%$
is found for $M_{\widetilde{q}}=3$ TeV. 

Next, in Fig.~\ref{fig:MCFMxsects}~ a similar comparison is presented as that
of Fig.~\ref{fig:susyxsects}, now for various SM processes.
      The upper plots display diphoton (dijet) production as a function of
      the minimum invariant mass $M_{\gamma\gamma}^{\rm min}$
      ($M_{jj}^{\rm min}$).
      The bottom plots show Higgs boson production in gluon fusion, first inclusive
      and decaying into $b\bar{b}$ as a function of $p_b^{\mathrm{T},\rm min}$, and
      then in association with a hard jet as a function of $p_{\rm jet}^{\mathrm{T},\rm min}$.
These cross-sections have been computed at LO with {\tt MCFMv8.2}~\cite{Boughezal:2016wmq}
with the basic ATLAS and CMS acceptance cuts.
The use of leading-order theory is justified as only
the relative impact of the PDF error reduction is of interest, rather
than providing state-of-the-art predictions for the rates.

\begin{figure}[t]
  \begin{center}
    \includegraphics[width=0.49\linewidth]{\main/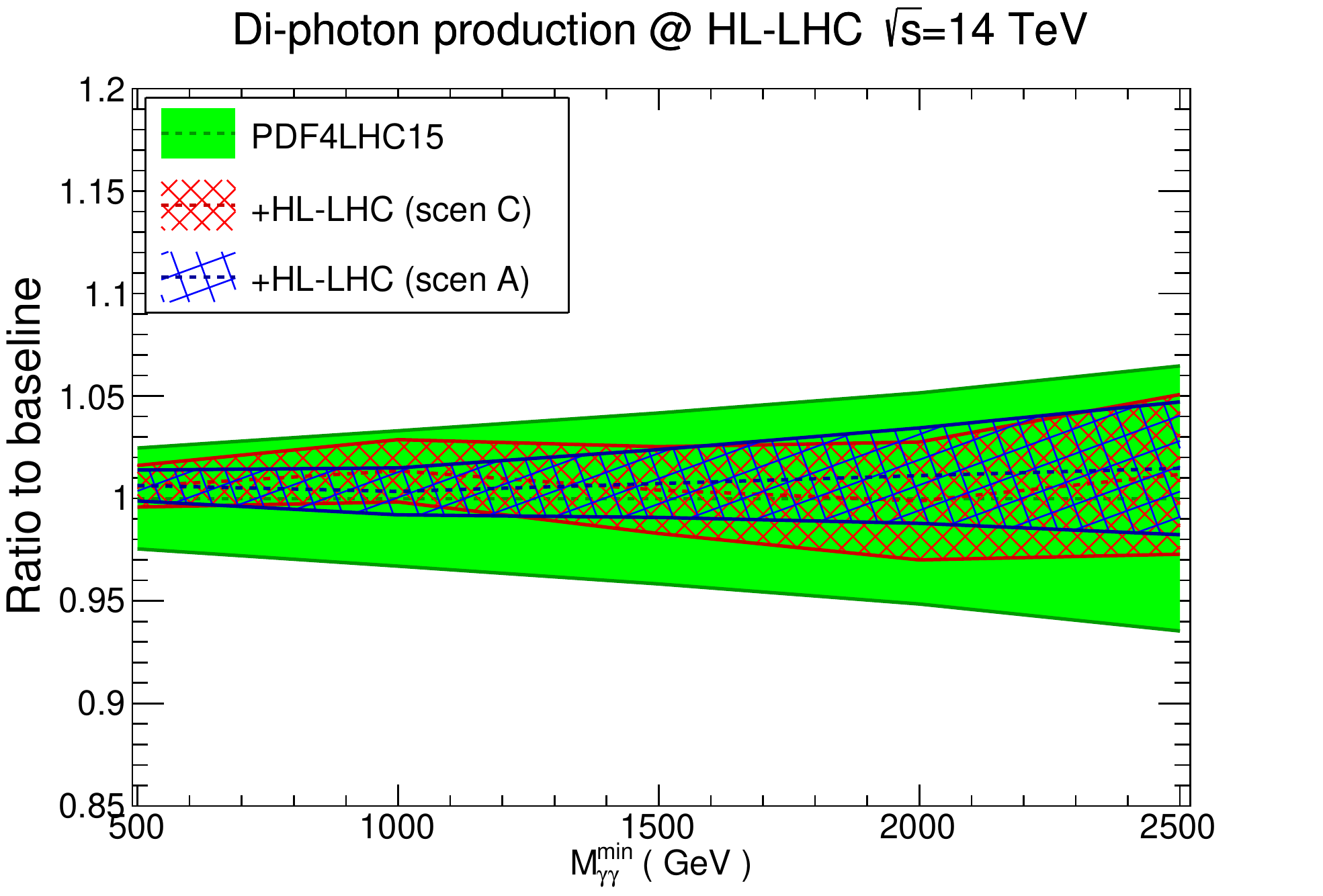}
    \includegraphics[width=0.49\linewidth]{\main/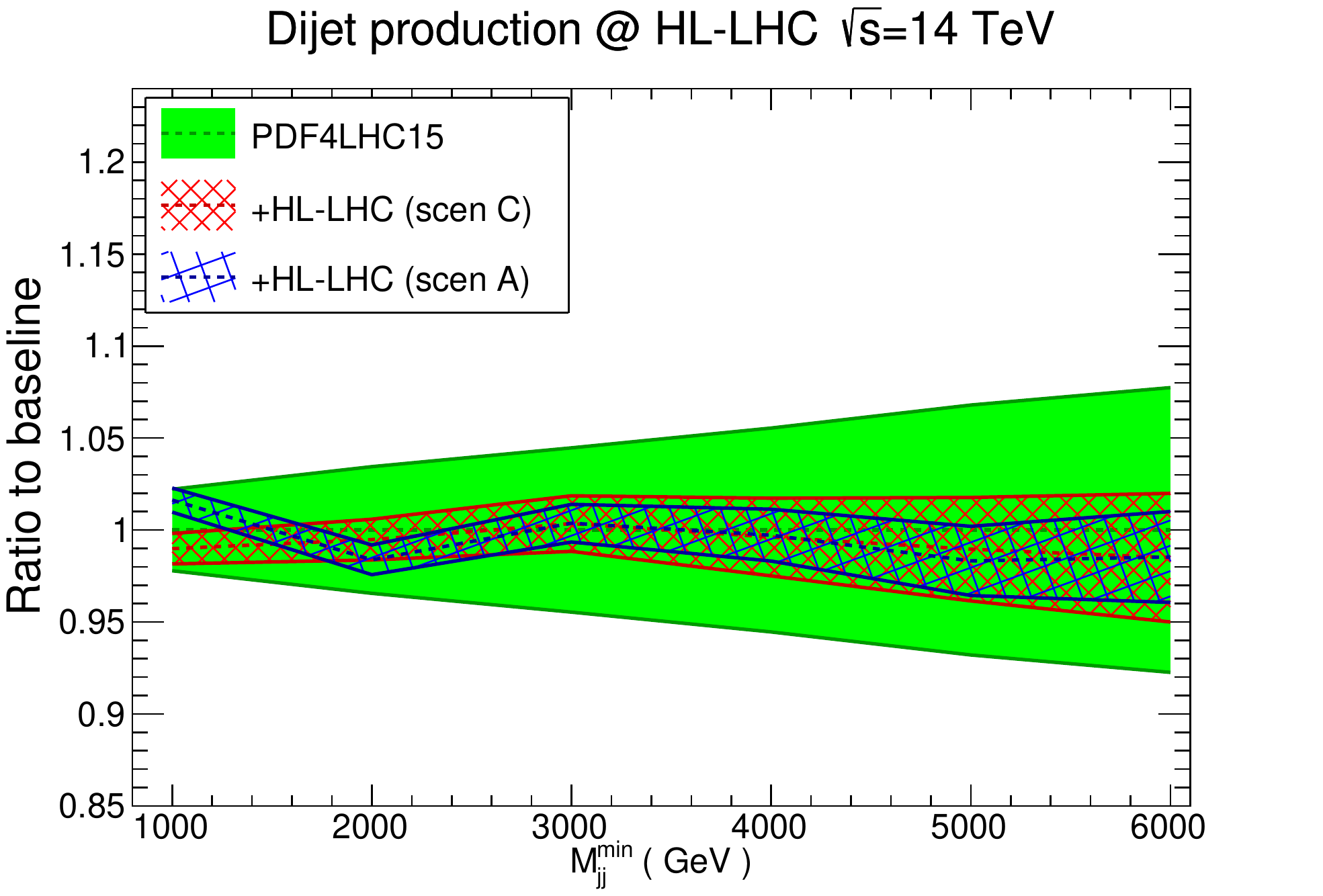}
    \includegraphics[width=0.49\linewidth]{\main/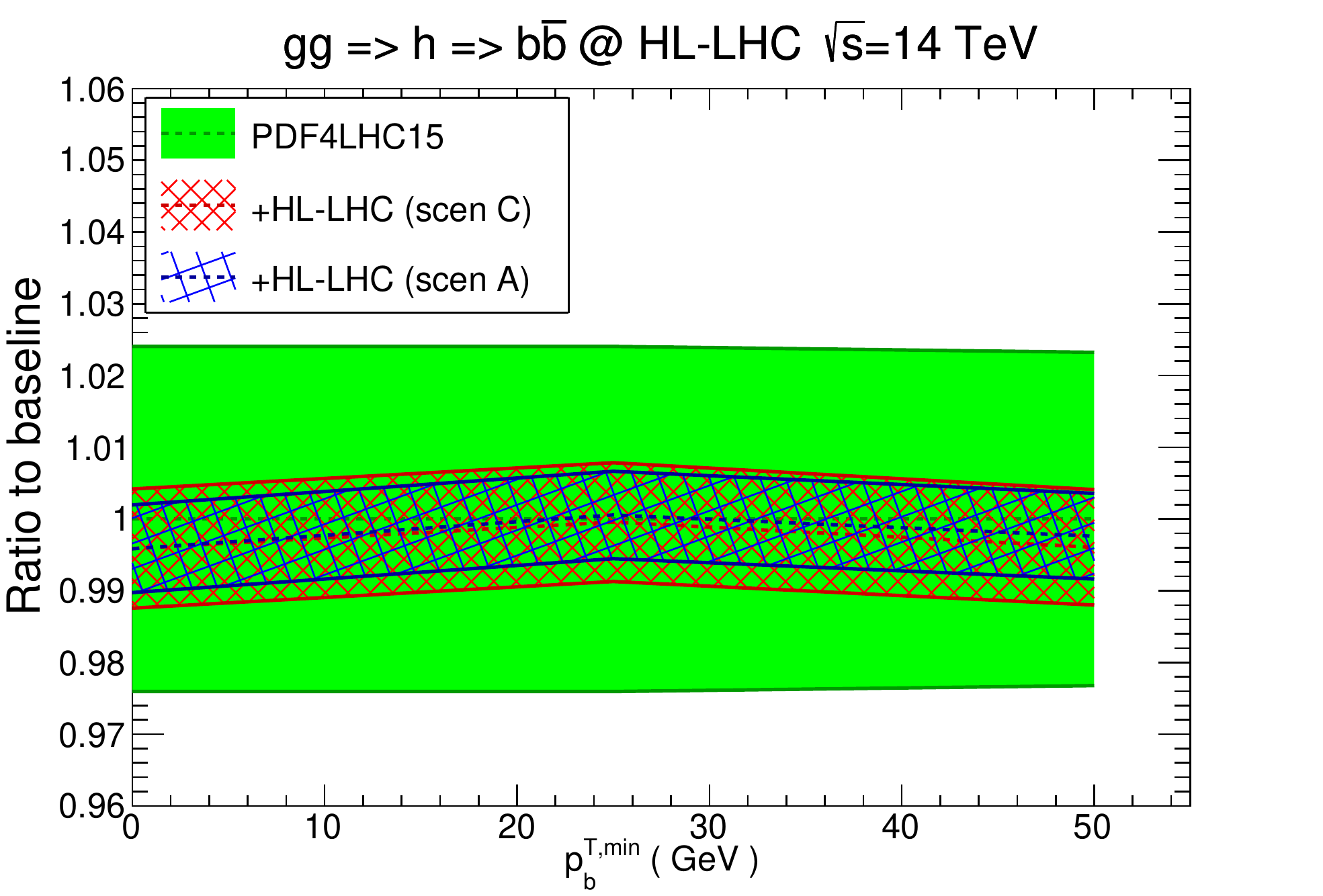}
    \includegraphics[width=0.49\linewidth]{\main/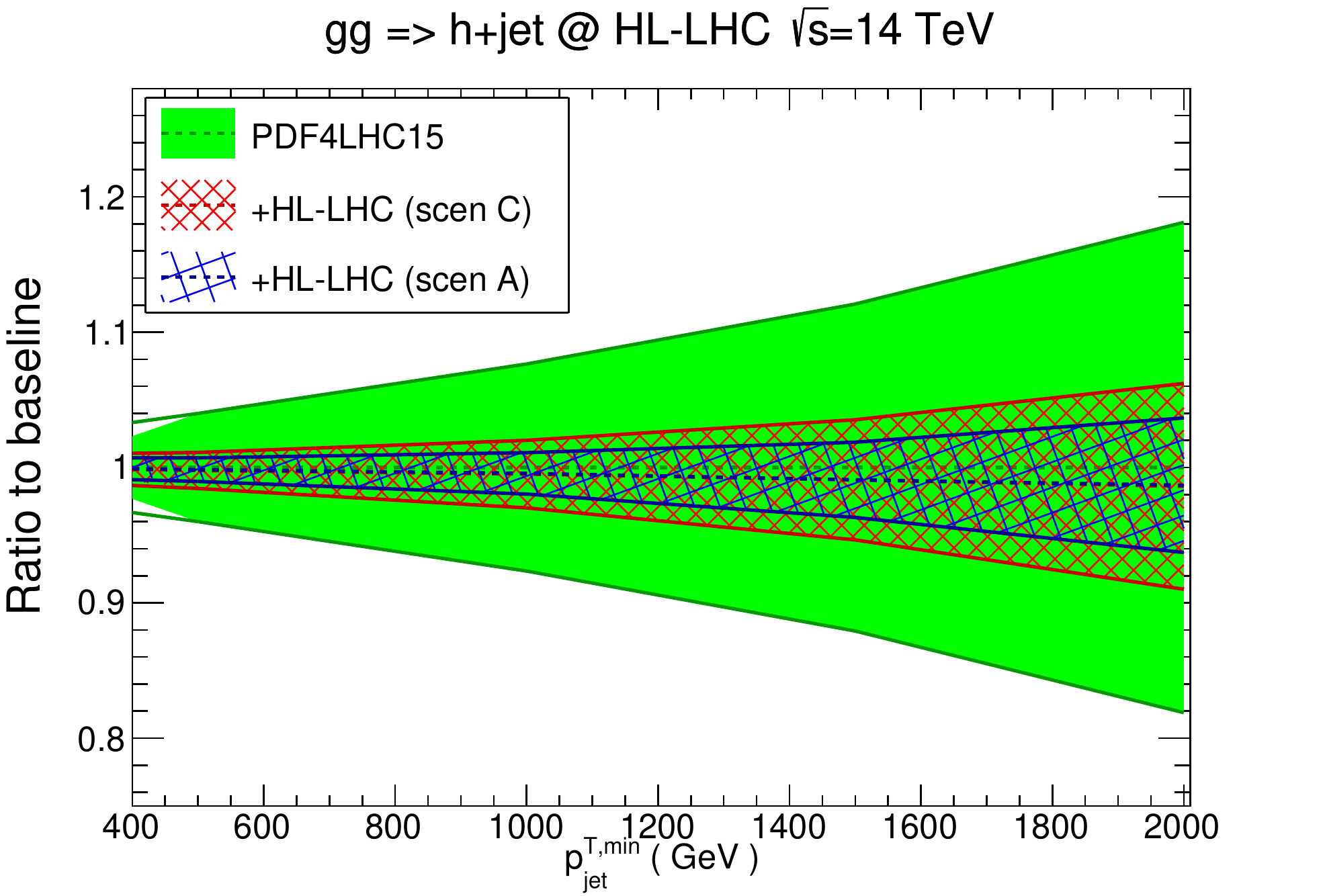}
    \caption{\small Same as Fig.~\ref{fig:susyxsects} for
      Standard Model processes.
      The upper plots show diphoton (dijet) production as a function of
      the minimum invariant mass $M_{\gamma\gamma}^{\rm min}$
      ($M_{jj}^{\rm min}$).
      The bottom plots show Higgs boson production in gluon fusion, first inclusive
      and decaying into $b\bar{b}$ as a function of $p_b^{\mathrm{T},\rm min}$, and
      then in association with a hard jet as a function of $p_{\rm jet}^{\mathrm{T},\rm min}$.
     \label{fig:MCFMxsects} }
  \end{center}
\end{figure}

From the comparisons in Fig.~\ref{fig:MCFMxsects}, the two scenarios, A and C, give similar results.
In the case of dijet production, which at large masses is dominated
by the $qq$ and $qg$ luminosities, PDF errors are expected to reduce down to $\simeq$2$\%$ even for invariant masses as large
as $M_{jj}=6$ TeV.
A similar conclusion can be drawn for diphoton production, also sensitive
to the $qq$ partonic initial state.
Concerning Higgs boson production in gluon fusion, in the inclusive
case the HL-LHC constraints should lead to PDF errors below the percent
level.
For Higgs boson production in association with a hard jet, 
a marked error reduction is found, suggesting that PDF uncertainties
in the $p_{\mathrm{T}}^h$ distribution should be down to at most the $\simeq$2\% level
at the HL-LHC in the entire relevant kinematical range.

 \subsubsection*{Summary and outlook}
 In this study, the constraints that
 HL-LHC measurements are expected to impose on the quark and gluon structure of the proton have been quantified.
 The impact of a range of physical processes have been assessed, from weak gauge
 boson and jet production to top quark and photon production, and the robustness
 of the results has been studied with respect to different projections for the experimental
 systematic uncertainties.
 It is found that, in the invariant mass region $M_X \gsim 100$ GeV, the HL-LHC
 measurements can be expected to reduce the PDF uncertainties in processes such
 as Higgs boson or SUSY particle production by a factor between 2 and 4, depending on the dominant
 partonic luminosity and on the scenario for the systematic errors.
 Therefore, the exploitation of the HL-LHC constraints
 on PDFs will feed into improved theoretical predictions for a range of phenomenologically
 relevant processes both within and beyond the SM.

 Two caveats are relevant at this point.
 First, only a non--exhaustive subset of all possible measurements of relevance for PDF fits has been considered.
 Other processes not considered here, due to currently anticipated measurements and those not foreseen but which may well added to the PDF toolbox in the future, will certainly increase the PDF impact in some regions.
 Second, any possible issues such as data incompatibility,
 theoretical limitations, or issues with the data correlation models, which may limit the PDF impact in some cases have been ignored.
 All these issues can only be tackled once the actual measurements
 are presented.

 The results of this study are made
 publicly available in the {\tt LHAPDF6} format~\cite{Buckley:2014ana},
 with the grid names listed in Table~\ref{tab:Scenarios}.
 This way, the ``ultimate'' PDFs produced here
 can be straightforwardly applied to related physics projections
 of HL-LHC processes taking into account our improved knowledge
 of the partonic structure of the proton which is expected by then.

 %
 %
 %
 %
%
 %
\subsection{Underlying Event and Multiple Parton Interactions}\label{sec:UEMPI}

Underlying event (UE), defined as a accompanying activity to hard proton-proton scattering process, is 
an unavoidable background to collider observables for most measurements and searches. The UE activity
is not constant on an event-by-event basis, so the contribution from UE cannot be subtracted. However by using
measurements sensitive to UE activity, the modelling of it in Monte Carlo (MC) event generators is \textit{tuned}.

Multiple parton interactions (MPI) are one of the most important contributors to UE. The dependence of MPI
on the centre-of-mass energy ($\sqrt{s}$) cannot be derived from first principles, rather modelled by looking at data at different
centrer-of-mass energies, from Tevatron to LHC. At the start of the LHC, it was found that the this energy extrapolation of MPI
based in Tevatron Run-1 and -2 data (at $\sqrt{s}=1.8$~TeV and $\sqrt{s}=1.96$~TeV) did not describe the LHC data at 
$\sqrt{s}=900$~GeV and at $\sqrt{s}=7$~TeV~\cite{Aad:2010fh}, and predictions of different MC generators varied significantly.
These generators were then tuned using LHC Run-1 and Run-2 ($\sqrt{s}=13$~TeV) data.

\subsubsection[Underlying Event at 27 TeV]{Underlying Event at 27 TeV\footnote{Contribution by D. Kar.}}


The level of UE activity at the HL-LHC centre-of-mass energy of $\sqrt{s}=14$~TeV is expected to be very similar to the one measured at $\sqrt{s}=13$~TeV in Run-2. Given such a small increment in centre-of-mass energy, it is expected that the current MC tunes will be generally valid at HL-LHC too.
On the other hand, to get a sense of the UE activity at HE-LHC, two state-of-the-art MC generators, \textsc{Pythia8}~\cite{Sjostrand:2007gs} (v235) 
with Monash tune~\cite{Skands:2014pea} and
\textsc{Herwig7}~\cite{Bahr:2008pv,Bellm:2015jjp} (v713) with default tune were used. As the first measurements at a new centre-of-mass energy data are easiest
to perform in inclusive (i.e minimum-bias) events, 5 million such events were generated in each case. The UE activity is measured
using the leading charged particle as the reference object, and defining the usual UE regions with respect to it, as shown in
Fig.~\ref{fig:ueregions}.

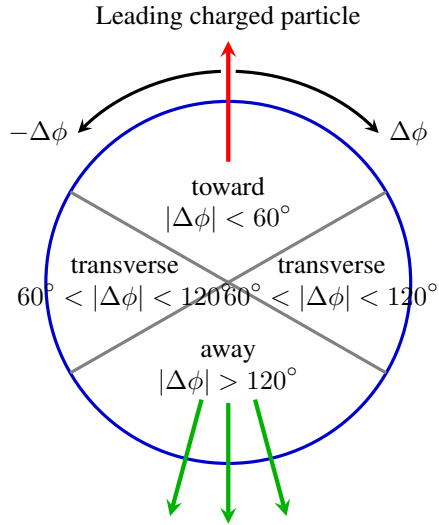
\begin{figure}[tbp]
  \begin{center}
    \begin{tikzpicture}[>=stealth, very thick, scale=0.8]
      \small
      \draw[color=blue!80!black] (0, 0) circle (3.0);
      \draw[rotate= 30, color=gray] (-3.0, 0) -- (3.0, 0);
      \draw[rotate=-30, color=gray] (-3.0, 0) -- (3.0, 0);

      \draw[->, color=black, rotate=-2] (0, 3.5) arc (90:47:3.5) node[right] {$\Delta{\phi}$};
      \draw[->, color=black, rotate=2] (0, 3.5) arc (90:133:3.5) node[left] {$-\Delta{\phi}$};

      \draw[->, color=red, ultra thick] (0, 2) -- (0, 4) node[above] {\textcolor{black}{Leading charged particle}};
      \draw[->, color=green!70!black, ultra thick] (0, -2) -- ( 0.0, -4);
      \draw[->, color=green!70!black, ultra thick, rotate around={-15:(0,-0.3)}] (0, -2) -- ( 0.0, -4);
      \draw[->, color=green!70!black, ultra thick, rotate around={ 15:(0,-0.3)}] (0, -2) -- ( 0.0, -4);

      \draw (0,  1.4) node[text width=2cm] {\begin{center} toward \goodbreak $|\Delta\phi| < 60^\circ$ \end{center}};
      \draw (0, -1.1) node[text width=2cm] {\begin{center} away \goodbreak $|\Delta\phi| > 120^\circ$ \end{center}};
      \draw ( 1.7, 0.3) node[text width=3cm] {\begin{center} transverse \goodbreak $60^\circ < |\Delta\phi| < 120^\circ$ \end{center}};
      \draw (-1.7, 0.3) node[text width=3cm] {\begin{center} transverse \goodbreak $60^\circ < |\Delta\phi| < 120^\circ$ \end{center}};
    \end{tikzpicture}
    \caption{Definition of UE regions in the azimuthal angle with respect to the leading charged particle}
    \label{fig:ueregions}
  \end{center}
\end{figure}

In Fig.~\ref{fig:ue}, the scalar sum (density in per unit $\eta$-$\phi$ area) of charged particles and 
charged particle multiplicity (density) as a function of leading charged particle $p_{\mathrm{T}}$ are shown.
The data is from the ATLAS measurement at $\sqrt{s}=13$~TeV~\cite{Aaboud:2017fwp}, while MC predictions both
at  $\sqrt{s}=13$~TeV and $\sqrt{s}=27$~TeV are shown. A few conclusions can be drawn.
The activity increases by about $25$ - $30\%$ by roughly doubling the centre-of-mass energy,
and the predictions by both generators are extremely consistent. The typical plateau-like behaviour
of the activity with increasing leading charged particle $p_{\mathrm{T}}$ can be seen at $\sqrt{s}=27$~TeV as well.

The similarity in predictions by two different generators is a welcoming sign, and perhaps indicates
that the modelling of MPI evolution with centre-of-mass energy is mature enough. Of course at 
$\sqrt{s}=27$~TeV, the events will be very active, and disentangling the effect of MPI in even typical
UE observables will be a challenge, and innovative topologies and observables will be have to be devised
in order perform UE measurements.

\begin{figure}[bt]
  \centering
  \includegraphics[width=0.45\textwidth]{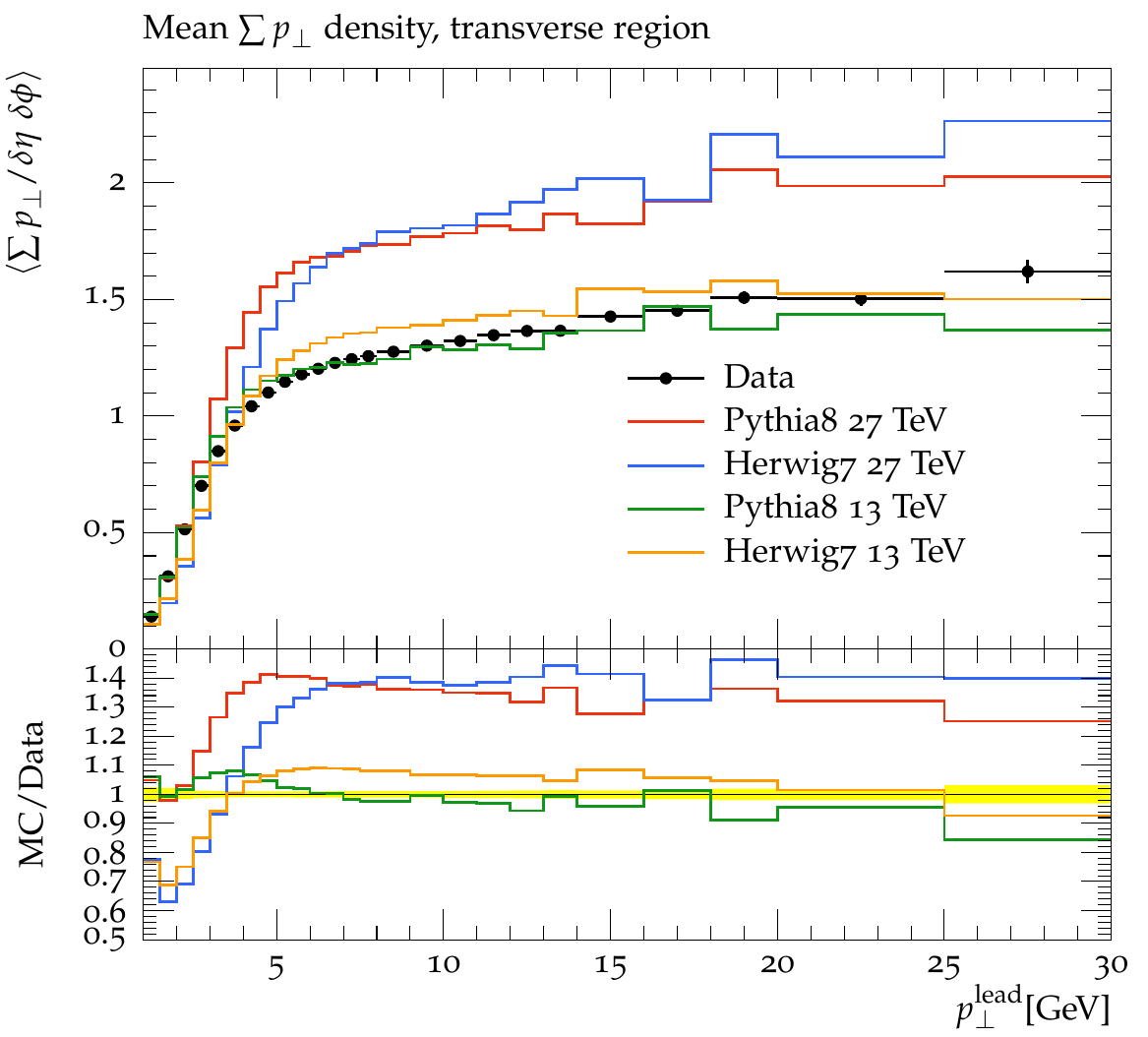}\qquad
  \includegraphics[width=0.45\textwidth]{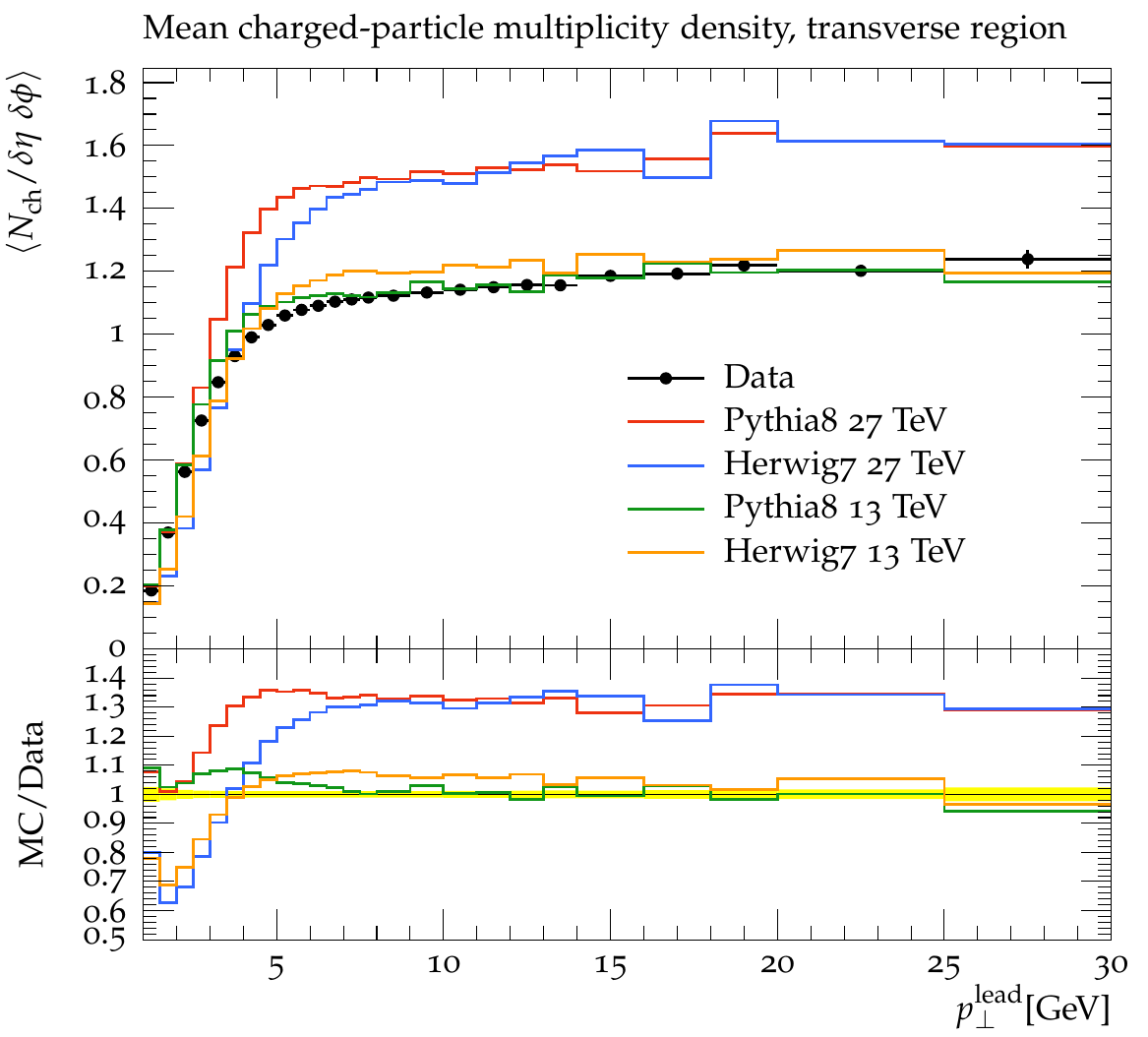}\\[1em]
  \caption{Comparison of the UE activities in different centre-of-mass energies.}
  \label{fig:ue}
\end{figure}

The analysis and plots are done using the Rivet~\cite{Buckley:2010ar} analysis framework.



\subsubsection[Double Parton Scattering]{Double Parton Scattering\footnote{Contribution by S. Cotogno, M. Dunser, J. R. Gaunt, T. Kasemets, and M. Myska.}}%
\label{sec:DPS}

\newcommand{\kbd}[1]{{\smaller{\texttt{#1}}}}
\def\pT{\ensuremath{p_\textrm{T}}\xspace}
\def\pTmin{\ensuremath{p_\textrm{Tmin}}}
\def\ZpT{\ensuremath{p_{\textrm{T}}^{\textrm{Z}}}}
\def\Nchg{\ensuremath{N_{\textrm{chg}}}\xspace}
\def\rivet{Rivet\xspace}
\newcommand{\Z}{{$Z$-boson}\xspace}
\def\ss{\ensuremath{\sqrt{s}}\xspace}
\def\pythiasix{\pythia\!6\xspace}
\def\pythiaeight{Pythia\,8\xspace}
\def\ptsum{\ensuremath{p_{\perp}^\text{sum}}\xspace}
\def\kt{\ensuremath{k_\perp}\xspace}

\newcommand{\EqRef}[1]{eq.~\eqref{#1}\xspace}
\newcommand{\SecRef}[1]{Section~\ref{#1}\xspace}
\newcommand{\TabRef}[1]{Table~\ref{#1}\xspace}
\newcommand{\FigRef}[1]{Figure~\ref{#1}\xspace}
\newcommand{\TabsRef}[1]{Tables~\ref{#1}\xspace}
\newcommand{\FigsRef}[1]{Figures~\ref{#1}\xspace}
\newcommand{\vect}[1]{\mathbf{#1}}
\newcommand{\eq}[1]{eq.~\eqref{#1}}
\renewcommand{\Eq}[1]{Eq.~\eqref{#1}}

An instance of MPI is the double parton scattering (DPS) that occurs when one has two distinct hard parton-parton collisions in a single proton-proton interaction. In terms of the total cross section to produce a final state $AB$ that may be divided into two subsets $A$ and $B$, DPS is formally power suppressed by $\sim \Lambda_{QCD}^2/\text{min}(Q_A^2,Q_B^2)$ compared to the more-familiar single parton scattering (SPS) mechanism. However, in practice there are various processes and kinematic regions where DPS contributes at a similar (or greater) level than SPS. Processes include those in which the SPS is suppressed by small/multiple coupling constants, such as same-sign $WW$ production, and processes where at least one part of the final state can be produced via a comparatively low scale scattering -- e.g. those involving a charm/bottom quark pair. 

The full theoretical description of DPS in QCD is rather complex, and many of the steps towards its formulation were achieved only recently \cite{Diehl:2011yj, Diehl:2015bca, Diehl:2017kgu, Buffing:2017mqm, Blok:2011bu, Manohar:2012jr}. As a result, many past studies of DPS have taken a strongly simplified approach in which it is assumed that the two colliding partons from each proton are entirely uncorrelated with one another, and that the (single) parton density in momentum fraction $x$ and impact parameter $\vect{b}$ may be factorised into the PDF and a transverse profile depending only on $\vect{b}$. In this case the DPS cross section simplifies into the so-called `pocket formula':
\begin{equation} \label{DPSeq:pocket}
\sigma^{AB}_{\text{DPS}} \simeq \dfrac{\sigma^{A}_{\text{SPS}}\sigma^{B}_{\text{SPS}}}{\sigma_{\text{eff}}}
\end{equation}
The quantity $\sigma_{\text{eff}}$ is a geometrical factor of order of the proton radius squared. The modelling of more general multiple parton interactions (MPI) in Monte Carlo event generators such as \mbox{\textsc{Herwig}}\xspace  and \mbox{\textsc{Pythia}}\xspace is based on similar approximations.

The \eq{DPSeq:pocket} does not take into account the possibility that the two partons from either or both protons may have arisen as the result of a perturbative $1\to 2$ splitting of a single parton into two. It also does not take into account a multitude of possible correlations between two partons in a proton, in spin, colour, and momentum fraction $x_i$, correlations between $x_i$ and the transverse separation between partons $\vect{y}$, as well as potential interference contributions in parton type. These correlations and QCD effects can result in a DPS cross section differing from the prediction of \eq{DPSeq:pocket}, both in terms of overall rate and also, crucially, in distributions.

Studies of DPS at the LHC and earlier colliders have essentially been restricted to extractions of a single number, the DPS rate, for several processes. From these early studies, in which the error bars are large and multiple factors change between measurements ($x$ values, parton channels, scales...), nothing conclusive can be determined thus far concerning correlations. However, the increased luminosity of the HL-LHC will provide the statistics needed to study differential distributions with sufficiently small uncertainties that it will be possible to probe quantum correlations between partons in the proton and the dynamics of the $1\to 2$ splitting for the first time. The results of these studies can be fed back and used to improve the theoretical modelling of DPS (and more general MPI), yielding improved DPS signal or background predictions.

As can be inferred from \eq{DPSeq:pocket}, DPS roughly scales as the fourth power of a parton distribution, whilst SPS only scales as the second power. This means that for given hard scales $Q_A, Q_B$, the DPS cross section grows faster than the SPS one as the collider energy increases (and decrease $x$), meaning that at a HE-LHC DPS will be more prominent and easily measurable than at the LHC. At the same time, at the lower $x$ values involved the effects of the correlations and $1\to 2$ splittings will be different - a combination of measurements of different processes at both the HL-LHC and HE-LHC should help us to separate out the effects of the different correlations.

Let us illustrate the general points above using a concrete process -- namely same-sign $WW$ production, where both $W$s decay leptonically into $e$ or $\mu$. A simple correlation-sensitive observable for this process is the asymmetry $a_{\eta_l}$:
\begin{equation}
a_{\eta_l}=\frac{\sigma(\eta_{1}\cdot\eta_{2}<0)-\sigma(\eta_{1}\cdot\eta_{2}>0)}{\sigma(\eta_{1}
  \cdot\eta_{2}<0)+\sigma(\eta_{1}\cdot\eta_{2}>0)}, 
\end{equation}
where $\eta_{1,2}$ are the rapidities of the two leptons. This quantity measures the discrepancy between the number of times the produced leptons emerge into opposite hemispheres of the detector and the number of times they emerge into the same hemisphere, normalised by the total number of lepton pairs produced. In the absence of parton correlations, it is found that $a_{\eta_l}=0$; any departure from this value indicates the presence of correlations. A more differential version of this asymmetry is the cross section differential in the product $\eta_1 \cdot \eta_2$. Here an absence of correlations yields a symmetric distribution under $\eta_1 \cdot \eta_2 \leftrightarrow -\eta_1 \cdot \eta_2$, and an asymmetric distribution indicates correlations. In the below studies a cut of $|\eta_i| < 2.4$ is always applied.

One type of correlation that will clearly affect these observables are correlations in momentum fraction $x$ between the partons. This type of effect was investigated in \cite{Ceccopieri:2017oqe}. Here, the double parton distributions (DPDs) were calculated at an input scale of $Q_0^2 = 0.26 \text{ GeV}^2$ from a constituent quark model calculation where the proton is taken as being composed only from the three quarks $uud$. At this scale there are necessarily strong correlations in $x$ space from the fact that there are only three quarks and due to the constraint $\sum_i x_i = 1$. These inputs were then evolved up to the $W$ mass scale via the double DGLAP equations, with effects of $1 \to 2$ splittings being ignored. In Fig. \ref{DPSfig:CMSTDR}, the green band represents their result at $\sqrt{s} = 14 \text{ TeV}$ for a quantity equal to $\sigma(\eta_{1}\cdot\eta_{2}<0)/\sigma(\eta_{1}\cdot\eta_{2}>0)$ -- their result corresponds to $a_{\eta_l} \sim 0.05$. On the same plot is given the anticipated sensitivity of the CMS experiment at the HL-LHC ($3 ~\text{ab}^{-1}$) \cite{CMSTDR} and the lowest values of this ratio that would allow one to reject the hypothesis of \eq{DPSeq:pocket} at the 95\% confidence level. These results indicate good prospects of the HL-LHC measuring $a_{\eta_l}$ values on the few per cent level for this process.

\begin{figure}[t]
\begin{center}
\includegraphics[scale=0.5]{\main/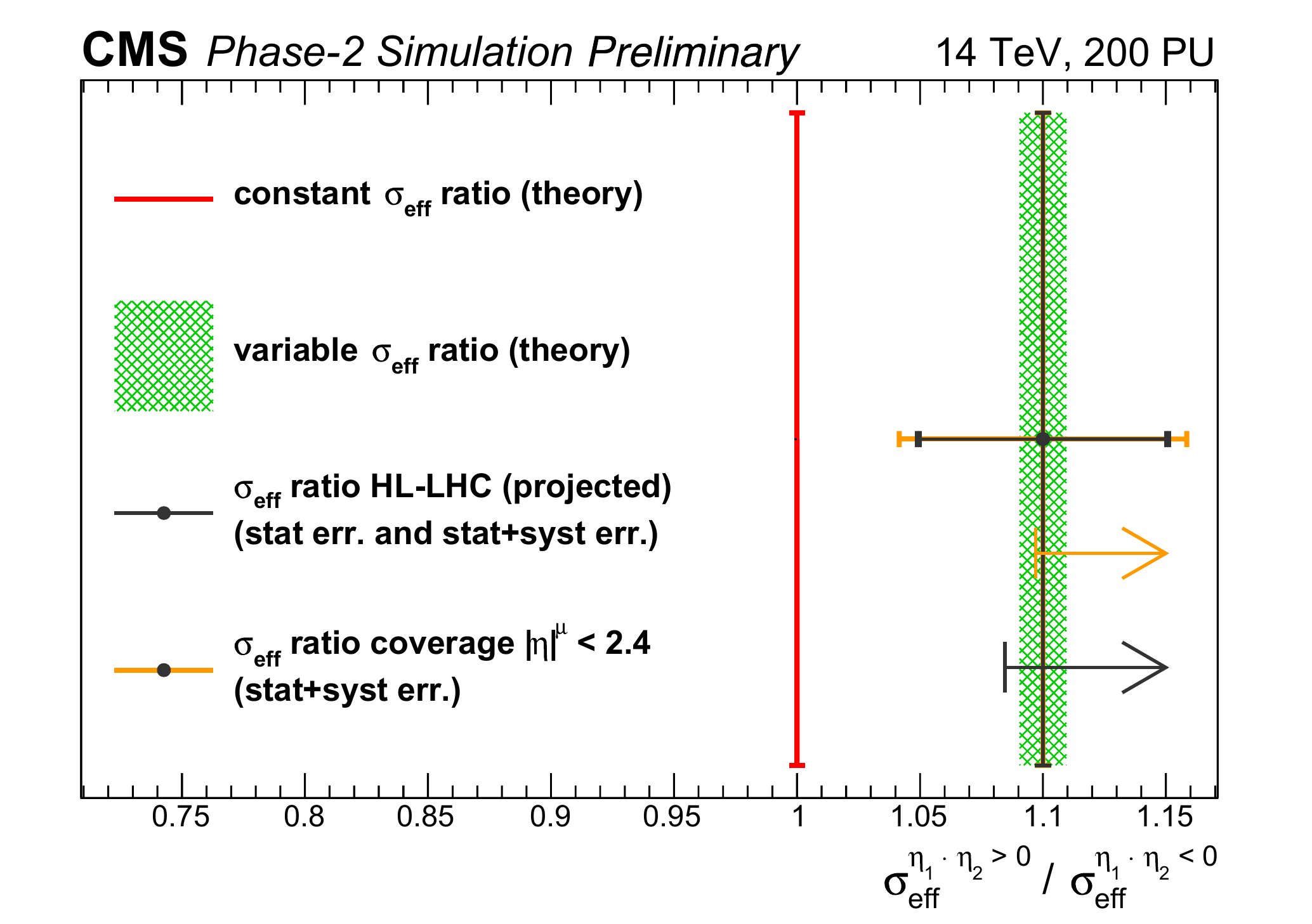}
\caption{Ratio of $\sigma_{eff}$ for $\eta_1 \cdot \eta_2>0$ and $\eta_1 \cdot \eta_2<0$, which is equal to the inverse ratio for $\sigma_{DPS}$. The value of this in the absence of parton correlations is $1$ (red line), whilst the prediction of \protect\cite{Ceccopieri:2017oqe} is given by the green band. The black error bars indicate systematic uncertainty attainable by the CMS experiment at $3$ ab$^{-1}$, the orange bars include systematic uncertainties assuming a conservative correlation of 0.8 between them for $\eta_1 \cdot \eta_2>0$ and $\eta_1 \cdot \eta_2<0$. The vertical line on the arrows indicates the lowest measured value of the ratio that would allow the exclusion of the uncorrelated parton hypothesis (i.e. \EqRef{DPSeq:pocket} with constant $\sigma_{eff}$) at 95\% CL. The black arrow corresponds to muon rapidity coverage $|\eta| < 2.8$, and the orange arrow $|\eta| < 2.4$.}
\label{DPSfig:CMSTDR}
\end{center}
\end{figure}
 
One simple feature that must necessarily be present in the true DPDs, and is taken into account by the DPDs of \cite{Ceccopieri:2017oqe} but not by \eq{DPSeq:pocket}, is the fact that removing one valence $u$ quark from the proton halves the probability to find another, and there is no chance to find two valence $d$ quarks (this requirement is formally expressed in the number sum rules of \cite{Gaunt:2009re}). This effect is highly relevant to $a_{\eta_l}$ as it results in a reduction of cross section for large $\eta_1\cdot\eta_2$ (which probes the `double valence' region in one DPD) whilst leaving the cross section elsewhere unchanged. To investigate the size of $a_{\eta_l}$ resulting from this effect only, DPD inputs are constructed at $Q_0 = 1$ GeV based on a factorised ansatz of a product of MSTW2008LO PDFs times a transverse factor, except that in the $uu$ and $dd$ cases the PDF part is given by $D^u(x_1)D^u(x_2) - \tfrac{1}{2}D^{u_v}(x_1)D^{u_v}(x_2)$ and $D^d(x_1)D^d(x_2) - D^{d_v}(x_1)D^{d_v}(x_2)$ respectively. Evolving these inputs and using them to calculate the $W^+W^+$ cross section at $\sqrt{s} = 13 \text{ TeV}$, an asymmetry of $\sim 0.017$ is observed, indicating that these simple `valence number effects' are at least one important driving force in the asymmetry of \cite{Ceccopieri:2017oqe}.

Correlations in (longitudinal) spin can affect the rapidity distributions of the produced leptons \cite{Kasemets:2012pr} and result in a nonzero $a_{\eta_l}$. The potential size of effects from spin correlations was investigated recently in \cite{Cotogno:2018mfv}. In this study the unpolarised double parton distributions were constructed according to an uncorrelated ansatz at an initial scale of $1$ GeV. The polarised double parton distributions, encoding parton spin correlations, were chosen at the initial scale to correspond to the maximal possible spin correlations (technically, saturate the positivity bounds \cite{Diehl:2013mla}), in such a way that the effects on the cross section would be maximal. These distributions were evolved to the $W$ mass and used to compute polarised and unpolarised $W^+W^+$ cross sections at $\sqrt{s} = 13 \text{ TeV}$. The resulting $\eta_1 \cdot \eta_2$ distribution is shown in Fig. \ref{DPSfig:polresults}(a) -- the corresponding value of $a_{\eta_l}$ is $0.07$, which is even larger than that resulting from $x$ correlations. One should, however, bear in mind that this is a maximal value, and that there are possibilities for the polarised distributions at the input scale, compatible with the positivity bounds, that also ultimately yield negative values for $a_{\eta_l}$ \cite{Sabrinathesis}. Figure \ref{DPSfig:polresults}(b) shows the expected significance of a measured non-zero asymmetry as a function of luminosity $L$, using a rapidity cut $|\eta_i| > 0.6 $ imposed such that the asymmetry $a_{\eta_l}$ rises to $0.11$ (but overall $W^+W^+$ cross section reduces from 0.51 fb to 0.29 fb). The blue band shows the sensitivity achievable using the $\mu^+\mu^+$ channel only, whilst the red band shows the sensitivity attainable using $\mu^+\mu^+$, $\mu^+e^+$, and $e^+e^+$ assuming a similar sensitivity can be achieved for electrons as for muons. This plot reinforces the notion that a few per cent level asymmetry can be measured at the HL-LHC.

\begin{figure} [t]
\begin{center}
\includegraphics[height=0.4\textwidth]{\main/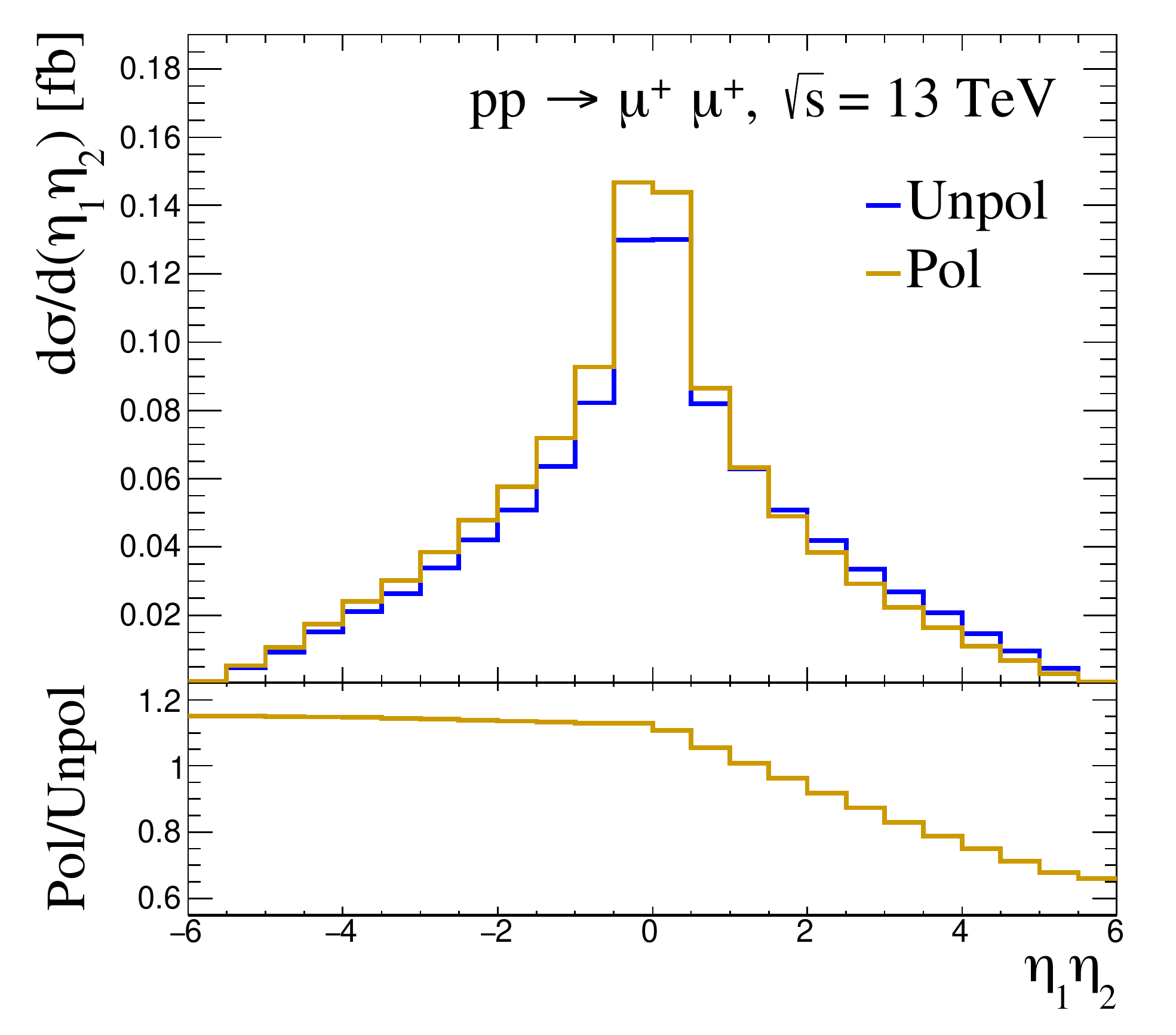}
\includegraphics[height=0.4\textwidth]{\main/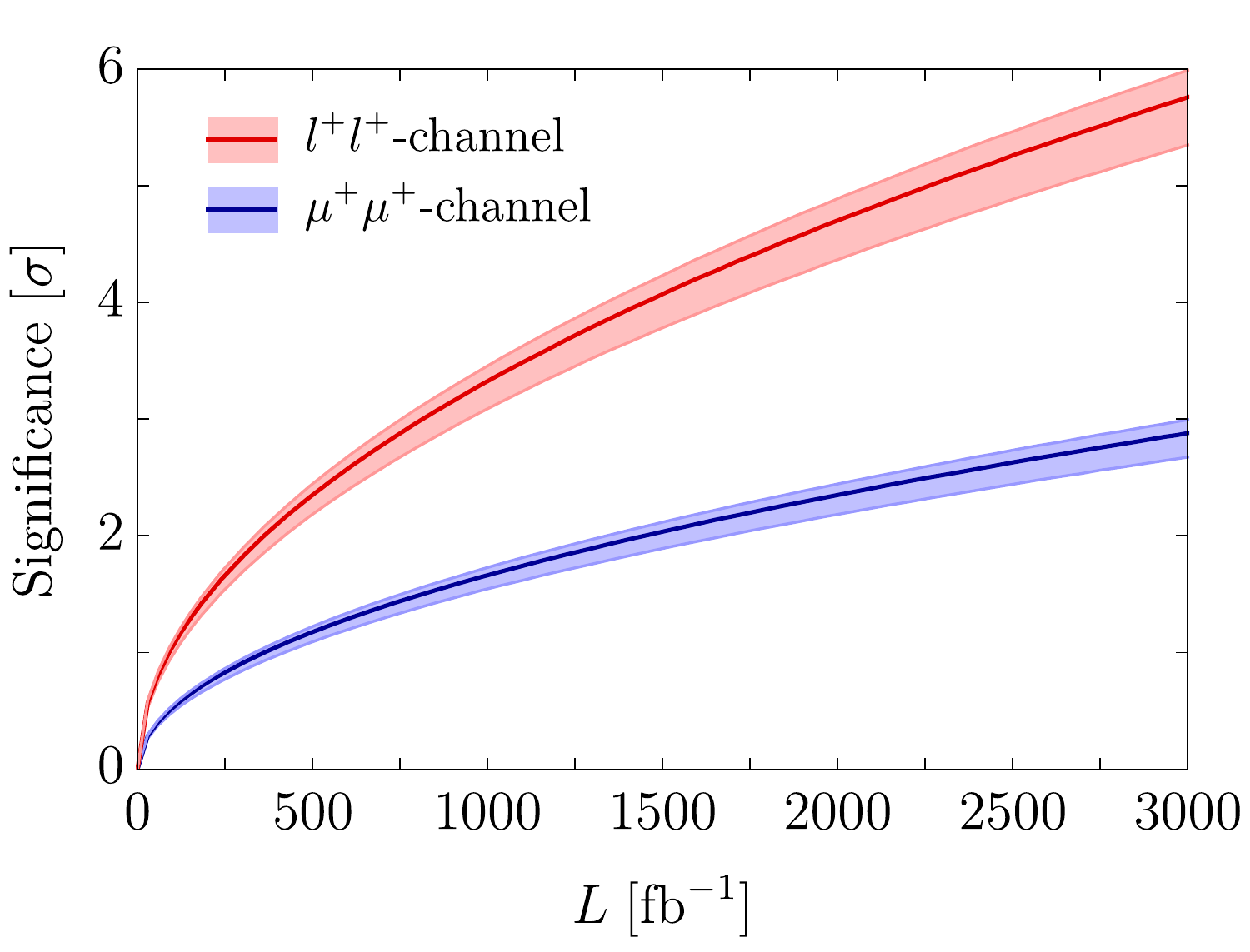}
\caption{Distribution in product of rapidities for two positively charged muons arising from $W^+W^+$ DPS. The blue plot includes only the unpolarised contribution, whilst the yellow also includes longitudinally polarised contributions (left). Estimated significance of a nonzero asymmetry as the distance in standard deviations of a measured asymmetry from zero, when the $W^+W^+$ cross section is 0.29 fb and asymmetry is 0.11 (right). This corresponds to the calculation of \protect\cite{Cotogno:2018mfv} with polarised contributions, and a cut on muon $|\eta| > 0.6$. The uncertainty bands indicate dependence of the sensitivity on assumptions regarding the subtraction of SPS backgrounds. More details regarding the set-up for both panels may be found in \protect\cite{Cotogno:2018mfv}.}
\label{DPSfig:polresults}
\end{center}
\end{figure}

To investigate how $1\to 2$ splittings may affect the asymmetry $a_{\eta_l}$, the code discussed in section 9 of \cite{Diehl:2017kgu} was upgraded to include charm and bottom quarks above the appropriate mass thresholds (chosen here to be equal to the MSTW 2008 values of $1.40$ GeV and $4.75$ GeV respectively). The `intrinsic' and `splitting' part of the DPDs were initialised as in \cite{Diehl:2017kgu} -- in particular, the intrinsic part was initialised according to an uncorrelated ansatz, up to a suppression factor near the phase space boundary $x_1+x_2=1$, that does not have a strong impact on $a_{\eta_l}$. Then, any nonzero value of $a_{\eta_l}$ will be almost entirely due to $1\to 2$ splitting effects. Computing $W^+W^+$ cross sections at $\sqrt{s} = 13$ TeV it is found that $a_{\eta_l} \sim 0.028$, which is of similar size to the asymmetry arising from other sources.

Note that the asymmetries from $x$ correlations, valence number effects and $1 \to 2$ splitting are in the same direction (favouring $\eta_1 \cdot \eta_2<0$ over $\eta_1 \cdot \eta_2 > 0$), whilst polarisation effects can potentially either favour a positive or negative asymmetry. 

At the HE-LHC, the asymmetry should be smaller for the same cuts on $|\eta_i|$ -- as $x$ is lowered, we move away from the `double valence' region where valence number effects are important, and the ratio of polarised to unpolarised quark distributions reduces (see Fig. 6 of \cite{Diehl:2014vaa}). Repeating the study above where a minimal modification of the uncorrelated ansatz at the input scale is made to take account of number effects, but at $\sqrt{s} = 27$ TeV, it is found that $a_{\eta_l} \sim 0.008$. Including instead the effects of the $1\to 2$ splittings yields $a_{\eta_l} \sim 0.013$ at $\sqrt{s} = 27$ TeV. At the HE-LHC (and the HL-LHC) it could be interesting to compare same-sign $WW$, which is comparatively weakly affected by $1 \to 2$ parton splitting (due to the fact there is no direct LO splitting yielding, for example $uu$), with processes that should receive stronger contributions from parton splitting, such as low mass Drell-Yan or $b\bar{b}b\bar{b}$ production, to probe in detail the effects of the $1 \to 2$ parton splitting and compare to theoretical predictions. More detailed studies in this direction are needed. 

In conclusion, the HL-LHC offers the opportunity to measure the effects of correlations between partons, via measurements of DPS processes, for the first time. In same-sign $WW$ production a good observable to probe correlations is the lepton pseudorapidity asymmetry $a_{\eta_l}$, which can only be nonzero in the presence of correlations -- theoretical calculations indicate values of $a_{\eta_l}$ at LHC energies on the order of a few per cent, which should be measurable at the HL-LHC. By combining measurements of various processes sensitive to DPS at the HL-LHC, and later and the HE-LHC, it will be ultimately possible to build up a picture of the various correlatons existing between partons in the proton.

\FloatBarrier

\newpage

\section{Top quark physics}

Precision measurements of top quark properties present an important test of the SM. As the heaviest particle in the SM, the top quark plays an important role for the electroweak symmetry breaking and becomes a sensitive probe for physics beyond the SM. 

\subsection{Top quark cross section}


\providecommand{\DELPHES}{\textsc{Delphes}\xspace}
\providecommand{\PUPPI}{\textsc{PUPPI}\xspace}
\providecommand{\PYTHIA}{\textsc{Pythia}\xspace}
\providecommand{\POWHEG}{\textsc{Powheg}\xspace}

\providecommand{\mg}{\textsc{MG5\_aMC@NLO}\xspace} 
\providecommand{\abmp} {ABMP16\xspace}
\providecommand{\ct} {CT14\xspace}
\providecommand{\nnpdf} {NNPDF3.1\xspace}
\providecommand{\chisq}{\ensuremath{\chi^2}\xspace}
\providecommand{\ndf}{dof\xspace}
\providecommand{\chisqndf}{\ensuremath{\chi^2}/\ndf\xspace}
\providecommand{\xfitter} {\textsc{xFitter}\xspace}
\providecommand{\applgrid} {\textsc{ApplGrid}\xspace}
\providecommand{\qcdnum} {\textsc{qcdnum}\xspace}
\providecommand{\difftop} {{\textsc{DiffTop}}\xspace}
\providecommand{\fastnlo} {{\textsc{fastNLO}}\xspace}
\providecommand{\amcfast} {{\textsc{aMCfast}}\xspace}
\providecommand{\lhapdf} {{\textsc{lhapdf}}\xspace}
\providecommand{\Mtop}{\ensuremath{m_{t}}\xspace}
\providecommand{\MW}{\ensuremath{m_{\mathrm{W}}}\xspace}
\providecommand{\tqh}{\ensuremath{\PQt_\mathrm{h}}\xspace}
\providecommand{\pt}{\ensuremath{p_\mathrm{T}}\xspace}
\providecommand{\ptvecmiss}{\ensuremath{\vec{p}_\mathrm{T}^\mathrm{miss}}\xspace}
\providecommand{\ttbar}{\ensuremath{{t\bar{t}}}\xspace}
\newcommand{\eq}[1]{eq.~\eqref{#1}}
\newcommand*{\ETMISS}{\ensuremath{E_{\rm{T}}^{miss}}\xspace}

\subsubsection[The $t\bar t$ production cross section: theoretical results]%
{The $t\bar t$ production cross section: theoretical
  results\footnote{Contributed by  M. Czakon, A. Mitov, and A. Papanastasiou.}}

\let\vaccent=\v 

\providecommand{\v}[1]{\ensuremath{\mathbf{#1}}} 

\providecommand{\perc}{\%}

\providecommand{\gv}[1]{\ensuremath{\mbox{\boldmath$ #1 $}}}
\providecommand{\uv}[1]{\ensuremath{\mathbf{\hat{#1}}}} 
\providecommand{\abs}[1]{\left| #1 \right|} 
\providecommand{\avg}[1]{\left< #1 \right>} 
\let\underdot=\d 
\providecommand{\d}[2]{\frac{d #1}{d #2}} 
\providecommand{\dd}[2]{\frac{d^2 #1}{d #2^2}} 
\providecommand{\pd}[2]{\frac{\partial #1}{\partial #2}}
\providecommand{\pdd}[2]{\frac{\partial^2 #1}{\partial #2^2}}
\providecommand{\pdc}[3]{\left( \frac{\partial #1}{\partial #2}
 \right)_{#3}} 
\providecommand{\ket}[1]{\left| #1 \right>} 
\providecommand{\bra}[1]{\left< #1 \right|} 
\providecommand{\braket}[2]{\left< #1 \vphantom{#2} \right|
 \left. #2 \vphantom{#1} \right>} 
\providecommand{\matrixel}[3]{\left< #1 \vphantom{#2#3} \right|
 #2 \left| #3 \vphantom{#1#2} \right>} 
\providecommand{\grad}[1]{\gv{\nabla} #1} 
\let\divsymb=\div 
\providecommand{\div}[1]{\gv{\nabla} \cdot #1} 
\providecommand{\curl}[1]{\gv{\nabla} \times #1} 
\let\baraccent=\= 
\providecommand{\=}[1]{\stackrel{#1}{=}} 
\theoremstyle{definition}
\theoremstyle{remark}

\def\PTavt{p_{\rm{T},{\rm avt}}}
\def\Yavt{y_{\rm avt}}


This sub-section provides a quick reference for the kinematic reach of the main $t\bar t$ differential distributions for both HL and HE-LHC. Figures~\ref{fig:14} and ~\ref{fig:27} are given in terms of expected events for the proposed ultimate luminosities for both colliders: 3\,ab$^{-1}$ for the HL-LHC running at 14 TeV and 15\,ab$^{-1}$ for the 27 TeV HE-LHC. The results are presented as plots of cumulative differential distributions and should be interpreted as follows: the histograms show the numbers of expected events (for the luminosities given above) above a given cut in any one of the four kinematic variables: $m_{t\bar t},\PTavt, \Yavt$ and $y_{t\bar t}$. Note that the cut corresponds to the left edge of a bin. The predictions are based on the CT14 parton distributions \cite{Dulat:2015mca} with value of the top quark mass $m_t=173.3$ GeV which is close to the current world average. The calculation is based on Ref.~\cite{Czakon:2015owf} and uses the dynamical scales of Ref.~\cite{Czakon:2016dgf}. 

\graphicspath{{top/img/}{../../}}
\begin{figure}[t]
\includegraphics[width = 0.5\textwidth]{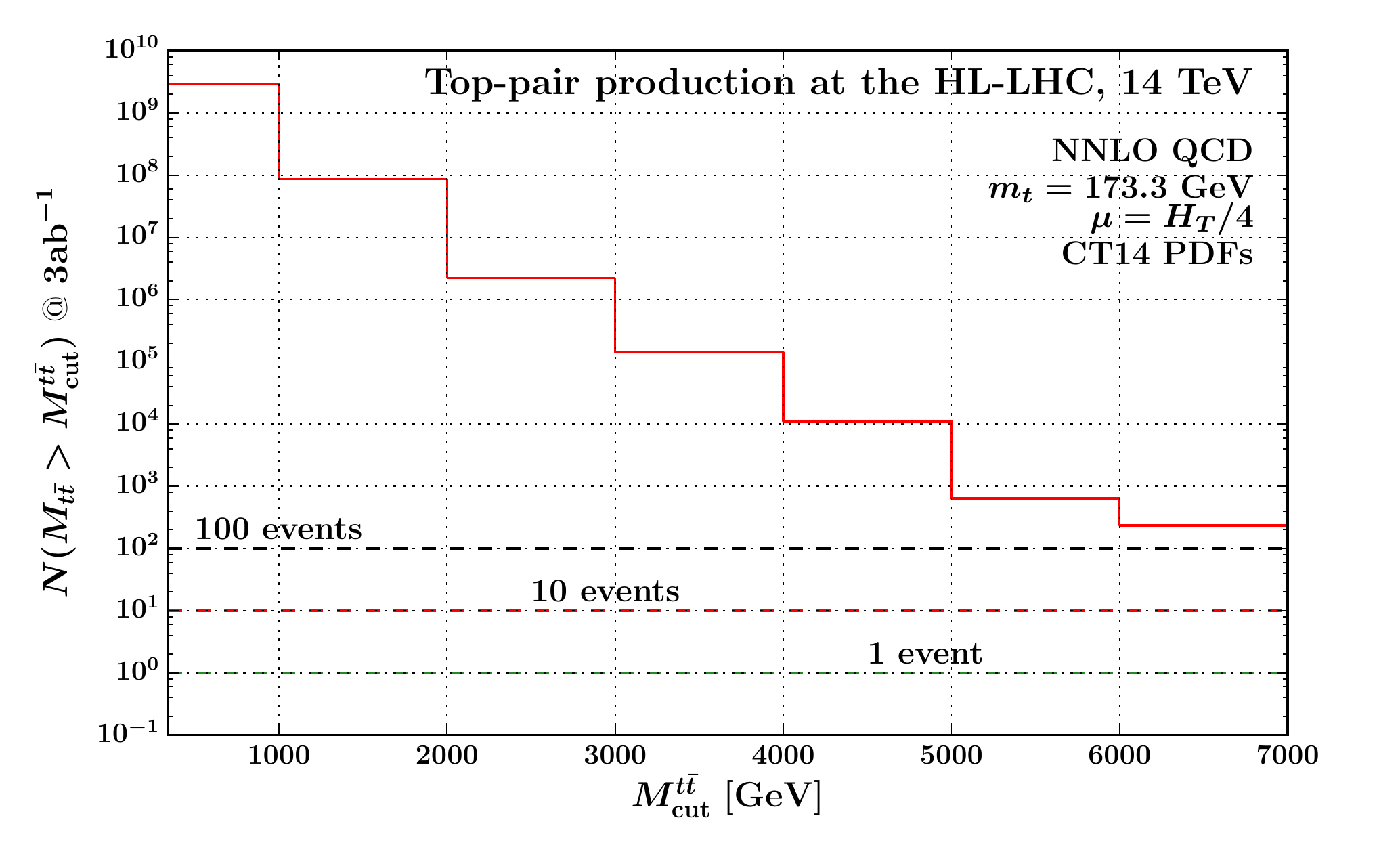}
\includegraphics[width = 0.5\textwidth]{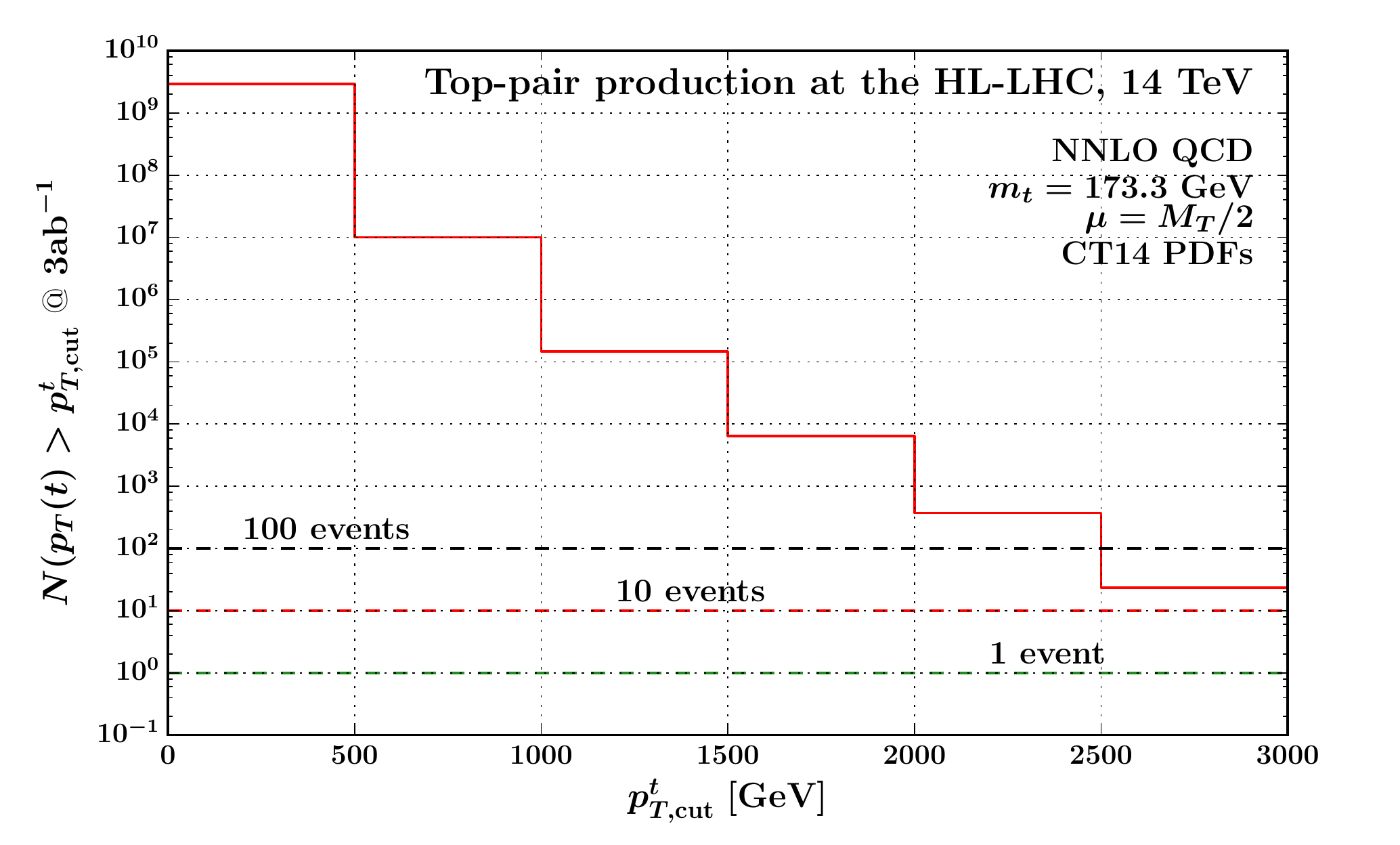}
\includegraphics[width = 0.5\textwidth]{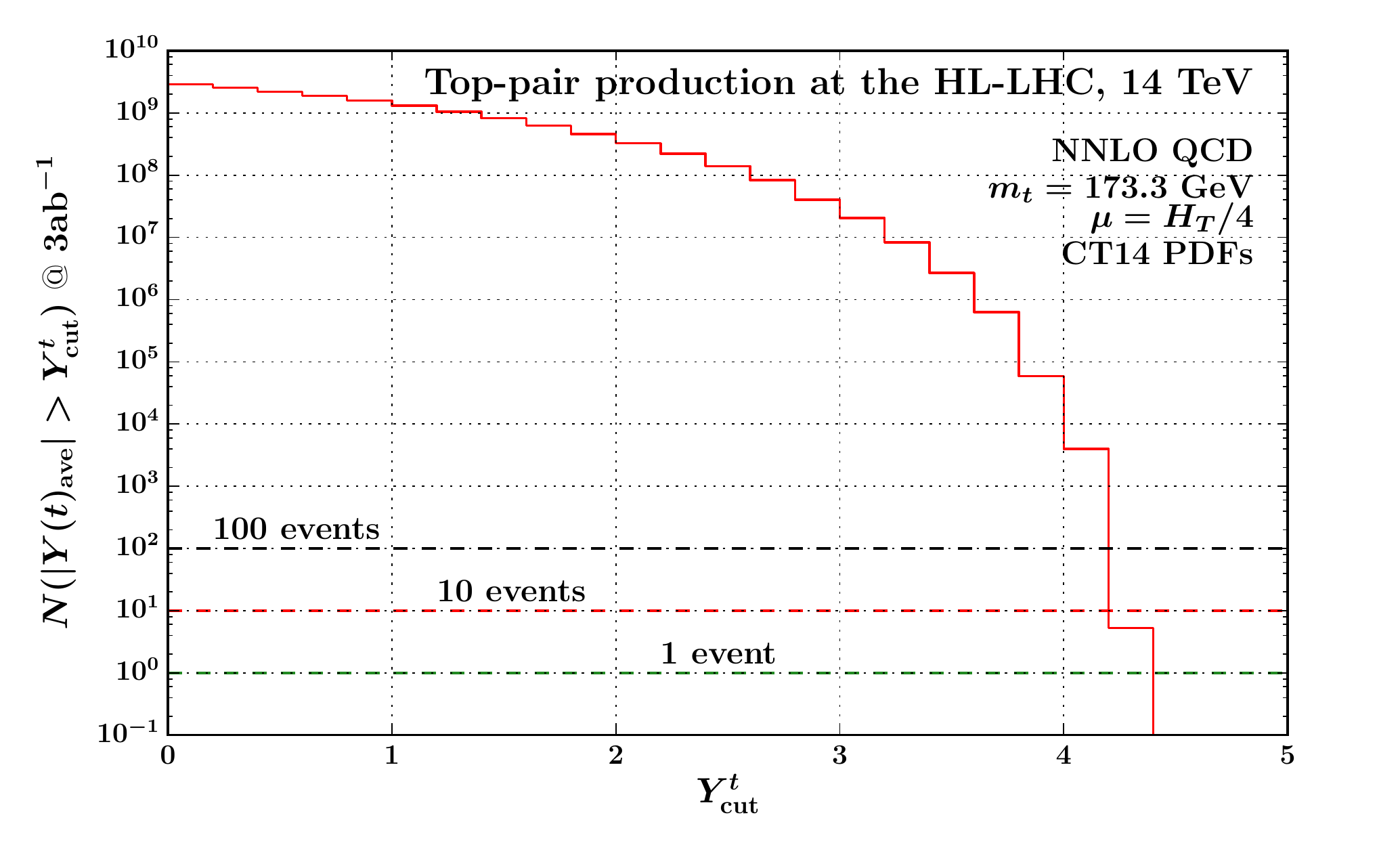}
\includegraphics[width = 0.5\textwidth]{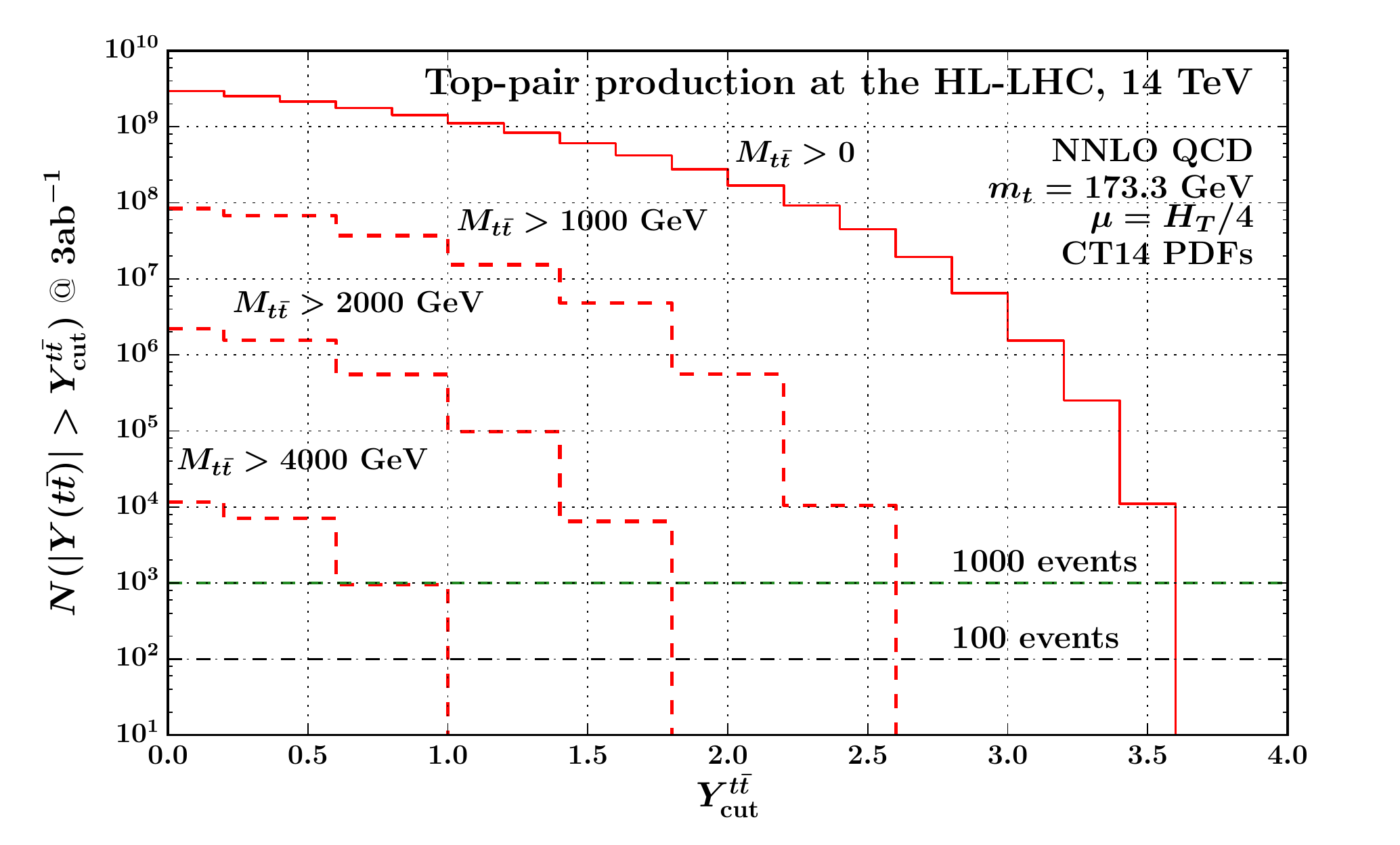}
\caption{Cumulative differential distributions for HL-LHC at 14 TeV.}
\label{fig:14}
\end{figure}

Figure~\ref{fig:14} presents predictions for the four cumulative distributions specified above in the case of the $t\bar t$ production at the HL-LHC (14 TeV), computed in NNLO QCD. In conclusion the HL-LHC allows detailed studies of top quark pair production with $m_{t\bar t}$ of up to about 7 TeV. Events with even larger values of $m_{t\bar t}$ are kinematically accessible and one expects about 10 events with $m_{t\bar t} > 7$ TeV. Therefore, the region $m_{t\bar t} > 7$ TeV provides a low SM background for, for example, searches for decays of BSM heavy particles to $t\bar t$ pairs. A detailed understanding of the SM background  - at the level of one expected event - will require a dedicated future effort due to the significant MC error in that region. 

The top quark $p_{\rm{T}}$ distribution can probe $p_{\rm{T}}$ values as high as 2.5 TeV, with a total of about 30 events expected beyond that value. 

The HL-LHC offers the possibility to access top production at high rapidity which might provide a link between top measurements at LHCb on one hand and ATLAS and CMS on the other. Indeed, in Fig.~\ref{fig:14} it can be observed that top quarks with rapidity $\Yavt$ as large as 4 will be copiously produced. The cross-section is a steeply falling function at large rapidity with a maximum attainable value of around 4.2 or so. Similarly, the rapidity of top quark pairs can be measured in detail up to values exceeding 3.4 with the maximum reach at about $y_{t\bar t} \sim  3.6$. In Fig.~\ref{fig:14} it is shown the $y_{t\bar t}$ distribution for a set of cuts on the top pair invariant mass. One should bear in mind that the NNLO $y_{t\bar t}$ calculation has significant MC error in the bins with 10 events or less.

%
\begin{figure}[t]
\includegraphics[width = 0.5\textwidth]{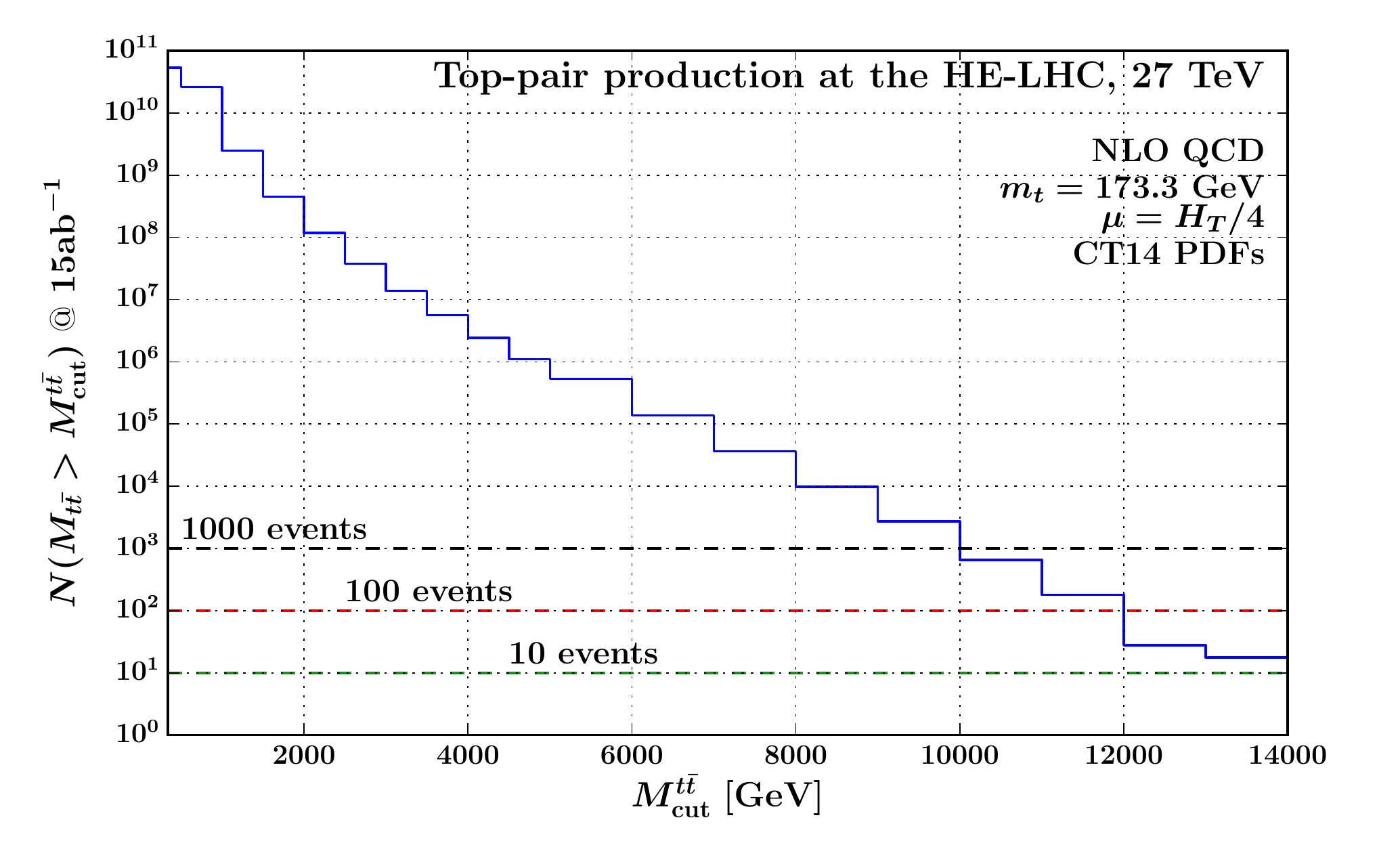}
\includegraphics[width = 0.5\textwidth]{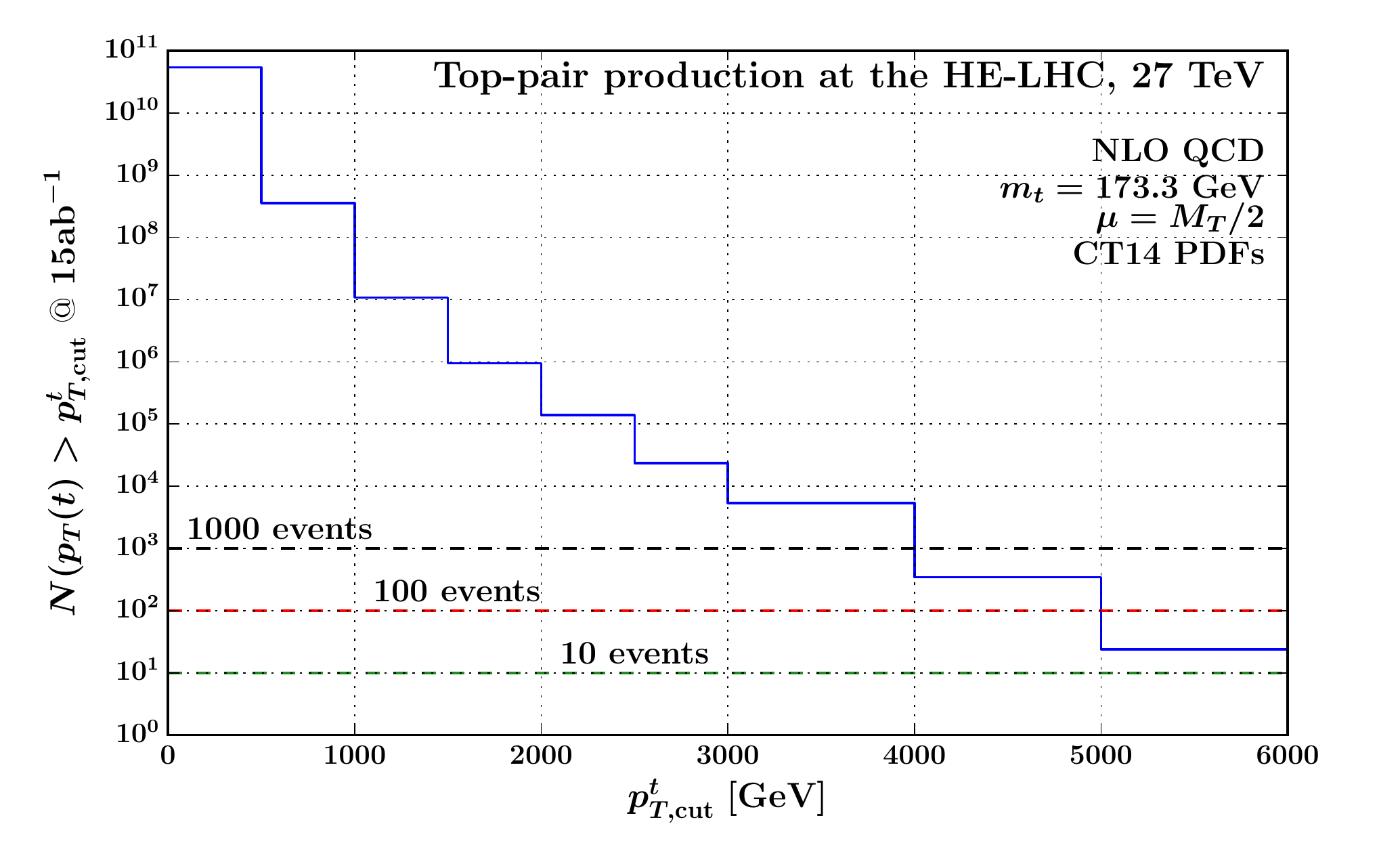}
\includegraphics[width = 0.5\textwidth]{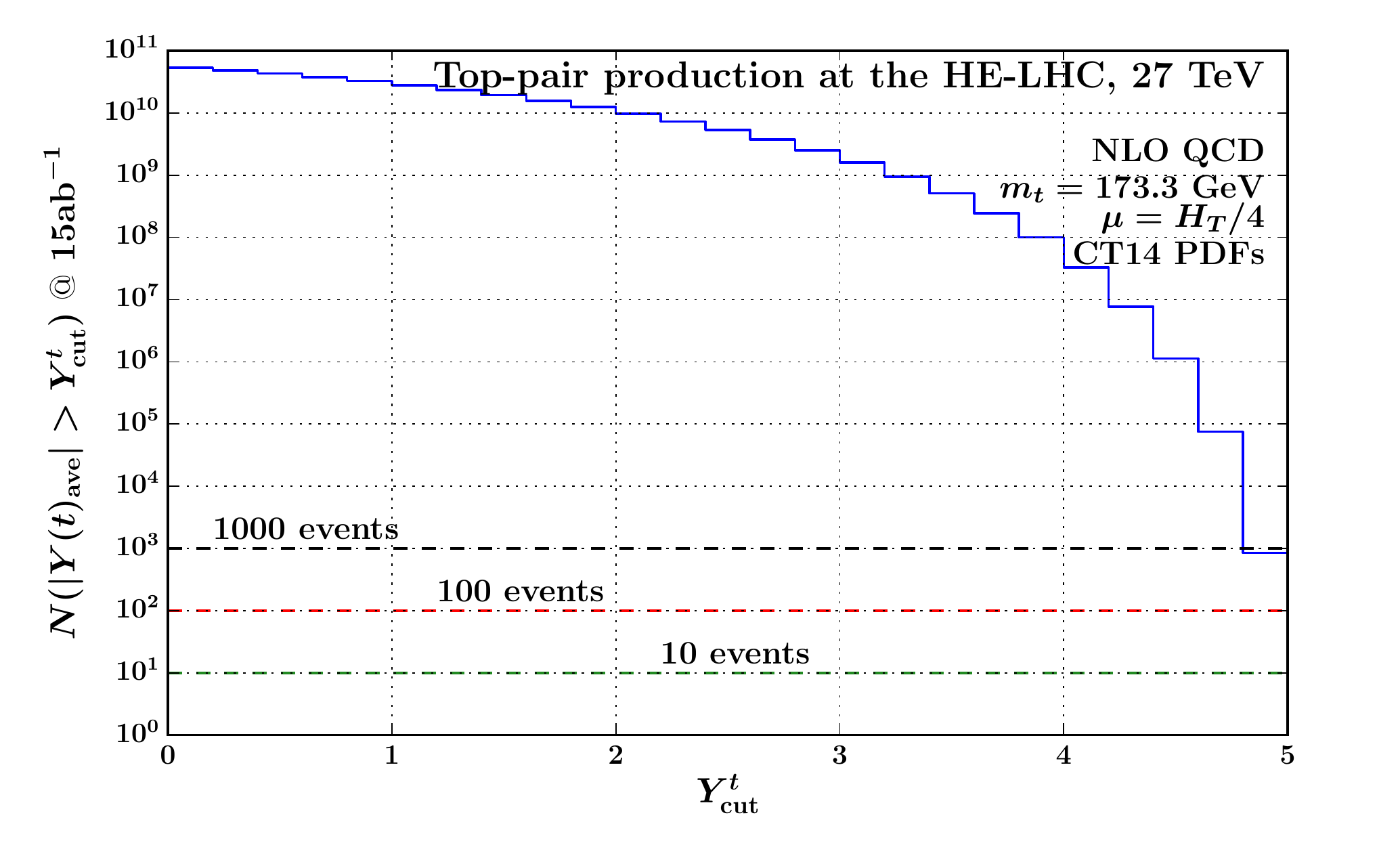}
\includegraphics[width = 0.5\textwidth]{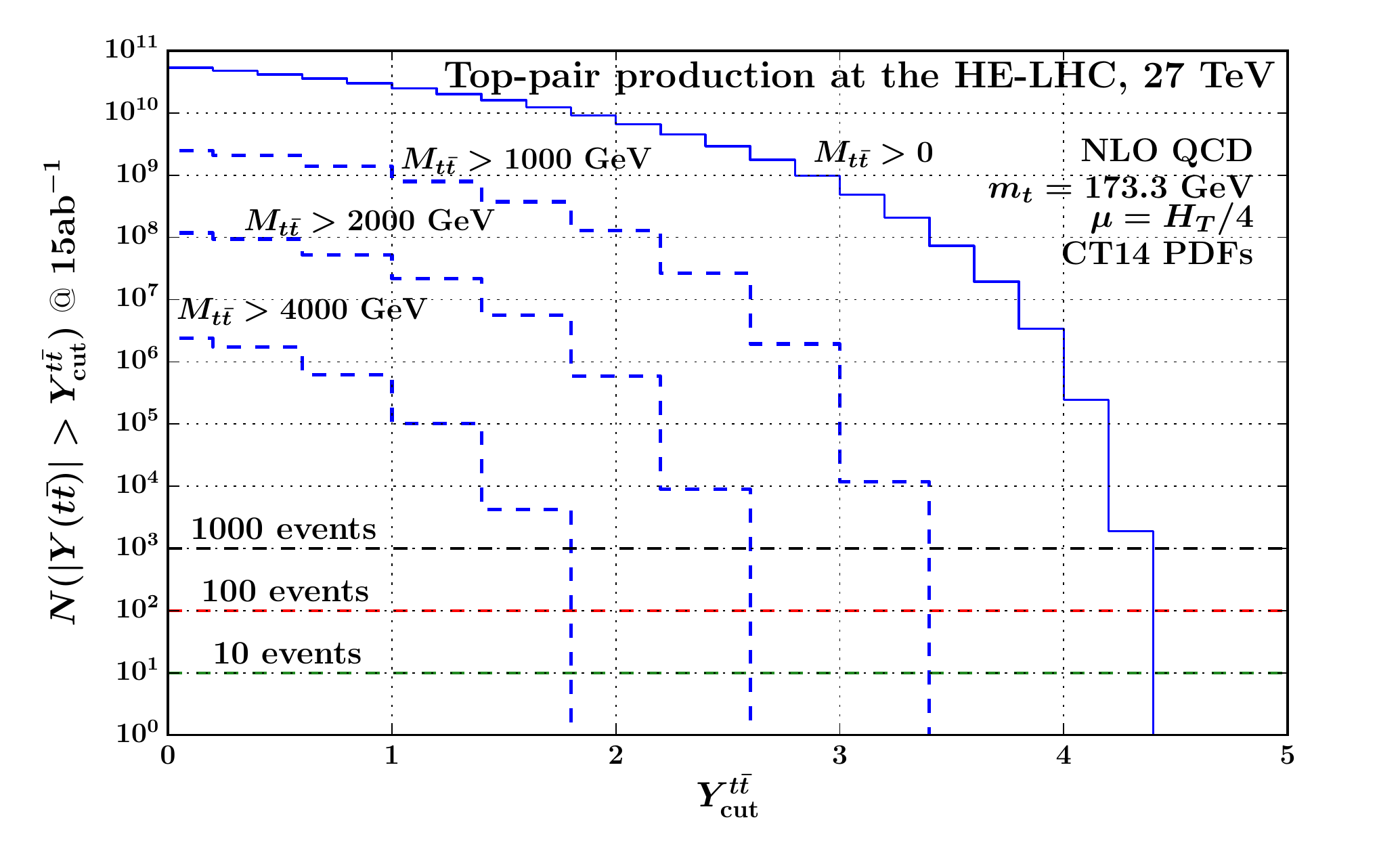}
\caption{Cumulative differential distributions for HE-LHC at 27 TeV.}
\label{fig:27}
\end{figure}

Figure~\ref{fig:27} presents the predictions for the same four cumulative distributions but in NLO QCD for the case of $t\bar t$ production at the HE-LHC (27 TeV). From this figure one can easily conclude that the increase in the kinematic reach over the HL-LHC is very substantial. There will be few hundred events with $m_{t\bar t}$ above 11 TeV and a similar number of events can be measured with $p_{\rm{T}}$ above 4 TeV. For the reliable description of such kinematics the inclusion of EW corrections as well as yet higher order soft and or collinear radiation will be essential; see Ref.~\cite{Czakon:2017wor,Czakon:2018nun}.

Very large rapidities can be attained at the HE-LHC. In particular, the top quark rapidity $\Yavt$ distribution can be measured to values as high as 4.8 with excellent statistics. Indeed, about 1000 events are expected above $\Yavt= 4.8$. The top pair rapidity can reach values as high as 4.4 and, if no additional cuts are applied, few thousand events will be produced with $y_{t\bar t} > 4.2$. As for the case of 14 TeV it is also show in Fig.~\ref{fig:27} the expected number of events as a function of $y_{t\bar t}$ for several cuts in $m_{t\bar t}$.


\subsubsection{Prospects in the measurement of differential \ttbar cross sections}
\label{subsec:ttbardiffxsec}

A study is presented for the resolved reconstruction of top quark pairs in the e/$\mu$+jets channels and a projection of differential \ttbar cross sections measurements with an integrated luminosity of 3\,ab$^{-1}$ at 14\,TeV \cite{CMS-PAS-FTR-18-015}. The analysis techniques are based on previous measurements of differential \ttbar cross sections at 13\,TeV\,\cite{TOP-16-008, TOP-17-002}. It is shown that such a measurement is feasible at the HL-LHC despite the expected large number of pileup interactions. The precision of the differential cross section can profit from the enormous amount of data and the extended $\eta$-range of the HL-LHC CMS detector. The results are used to estimate the improvement of measurements of parton distribution functions.

This study is based on a \DELPHES simulation of the HL-LHC CMS detector\,\cite{tracker, barrel, endcap, muon} using the Monte Carlo program \POWHEG~\cite{Nason:2004rx,Frixione:2007vw,Alioli:2010xd,Campbell:2014kua} (v2,hvq) in combination with \PYTHIA~\cite{Sjostrand:2006za,Sjostrand:2007gs} (v8.219) for the generation of \ttbar events at NLO accuracy. Events with a single isolated electron or muon with $\pt > 30$\,GeV and $|\eta| < 2.8$ are selected. Events with additional isolated electrons or muons with $\pt > 15$\,GeV and $|\eta| < 2.8$ are rejected. At least 4 jets with $\pt > 30$\,GeV and $|\eta| < 4.0$ are required, where at least 2 of the jets have to be identified as b jets. It is essential that the \PUPPI algorithm\,\cite{PUPPI} is used for the mitigation of pileup contribution when the jets are clustered and the \ptvecmiss is calculated.

A detailed description of the \ttbar reconstruction is presented in\,\cite{TOP-16-008, TOP-17-002}. For the reconstruction all possible permutations of assigning detector-level jets to the corresponding \ttbar decay products are tested and a likelihood that a certain permutation is correct is evaluated. In each event, the permutation with the highest likelihood is selected. The likelihood is constructed from the 2 dimensional \Mtop--\MW distribution of correctly assigned jets for the hadronically decaying top quark and the distribution of $D_{\nu,\mathrm{min}}$ obtained when calculating the neutrino momentum~\cite{Betchart:2013nba} for the leptonically decaying top quark. A comparison of the expected event yields and the migration matrices together with their properties are shown in Fig.\,\ref{fig:APPYR1} for the 
HL-LHC expectation. Despite the high pileup a performance of the \ttbar reconstruction similar to the one in 2016\cite{TOP-17-002} can be reached, while the portion of the direct measurable phase space is increased due to the extended $\eta$-range.

\begin{figure}[ht]
\centering
\includegraphics[width=0.40\textwidth]{\main/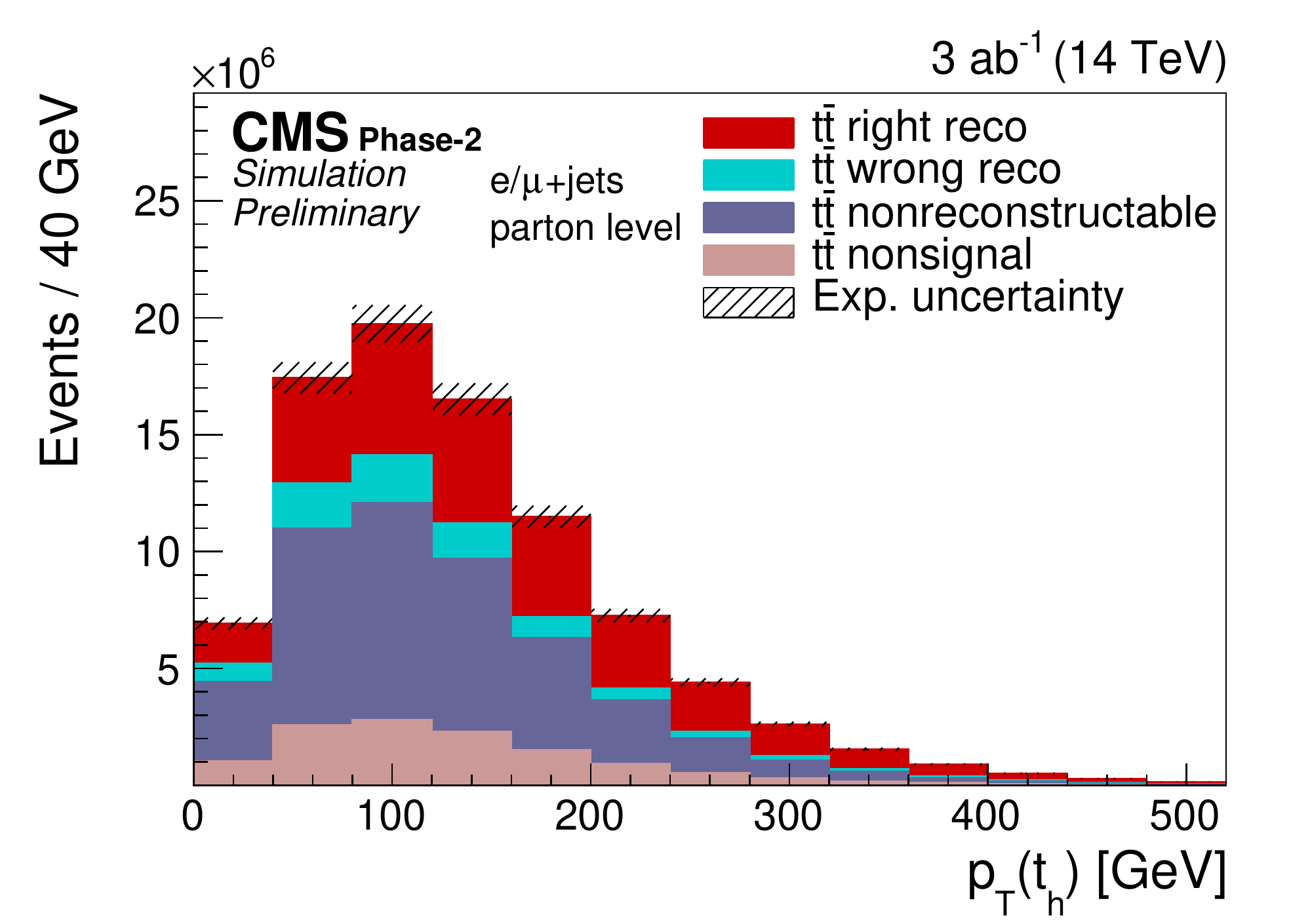}
\includegraphics[width=0.40\textwidth]{\main/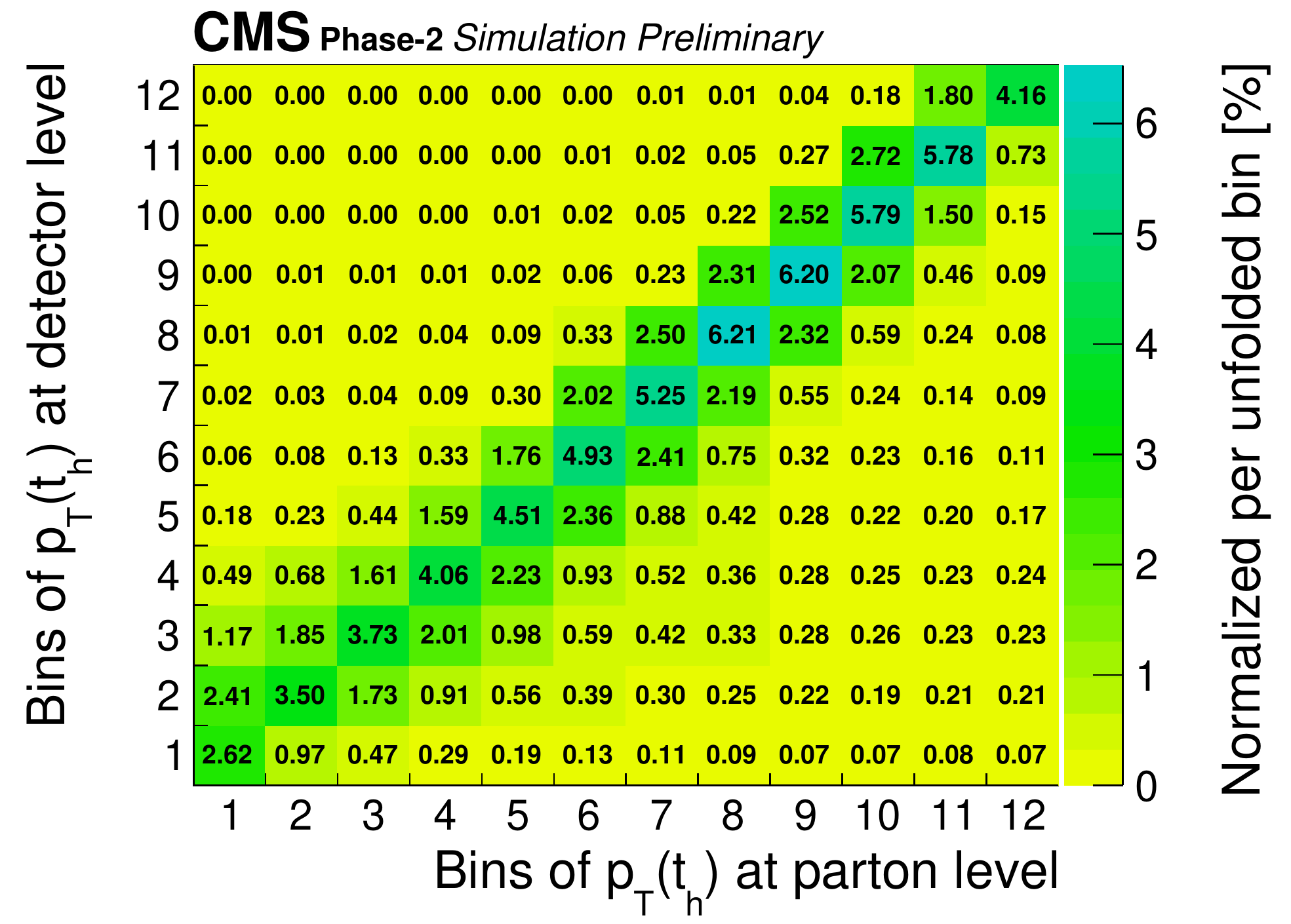}
\includegraphics[width=0.40\textwidth]{\main/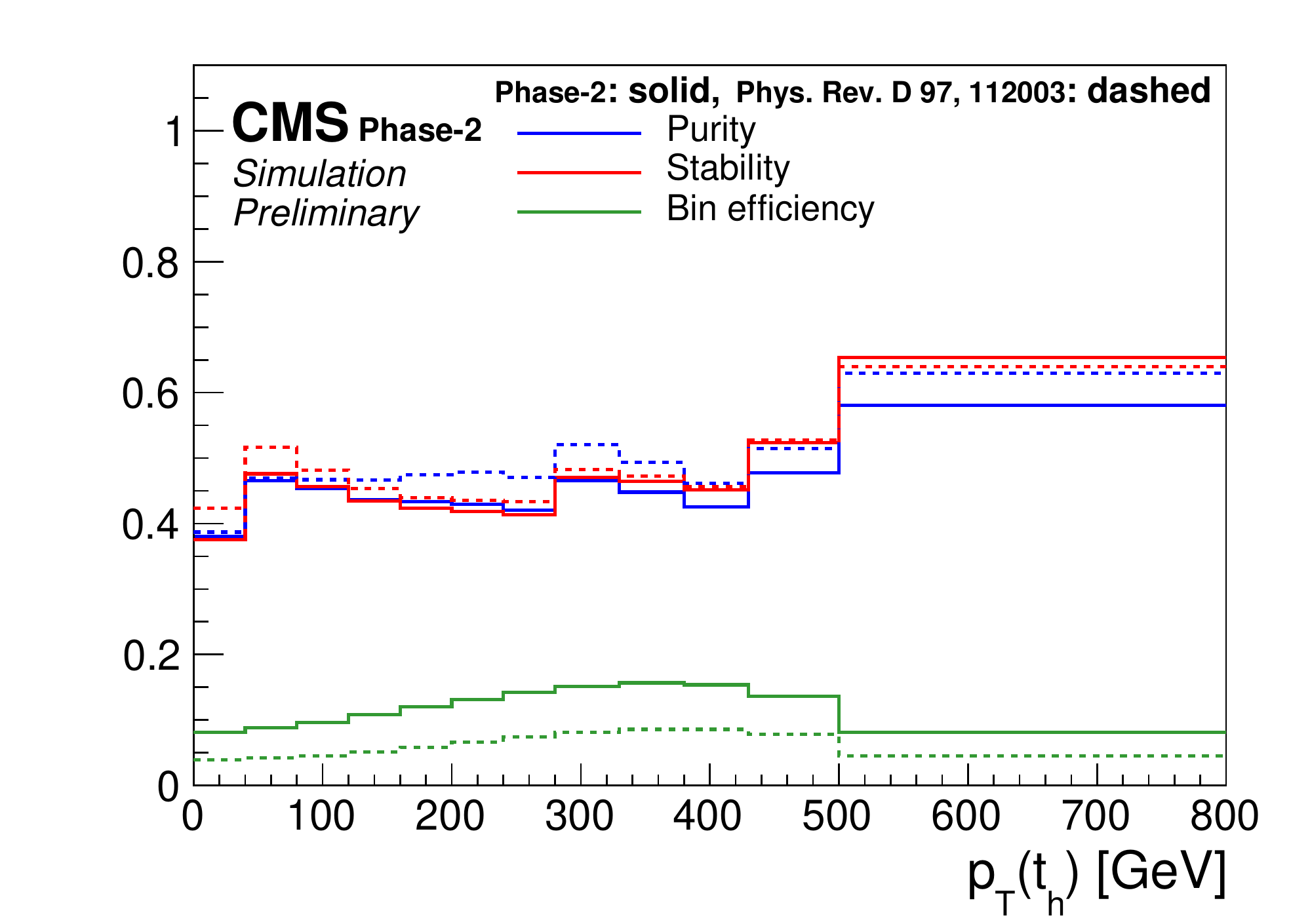}\\
 \caption{Expected signal yields (top-left), migration matrices (top-right), and its properties (bottom) for measurements of $\pt(\tqh)$ for the HL-LHC (Phase-2) simulation. 
 The purity is defined as the fraction of parton-level top quarks in the same bin at the detector level, the stability as the fraction of detector-level top quarks in the same bin at the parton level, and the bin efficiency as the ratio of the number of events found in a certain bin at detector level and the number of events found at parton-level in the same bin.}
 \label{fig:APPYR1}
\end{figure}

The following experimental uncertainties are estimated based on the expected performance of the HL-LHC CMS detector\cite{Collaboration:2650976}: electron and muon identification, $b$-tagging efficiencies, jet energy and \ptvecmiss calibration, and luminosity. All theoretical and modelling uncertainties have been reduced by a factor two. 


The unfolded results of the differential \ttbar cross section measurements as a function of \pt and rapidity $y$ of the hadronically decaying top quark (\tqh) are shown in Fig.\,\ref{fig:APPYR2}. In Fig.\,\ref{fig:APPYR3} the normalized double-differential cross section as a function of $M(\ttbar)$ vs $|y(\ttbar)|$ is shown. The strong impact of these measurement on PDF constraints is studied in Section\,\ref{subsec:pdfconst}. The high amount of data and the extended $\eta$-range of the HL-LHC detector allow for fine-binned measurements in phase-space regions --- especially at high rapidity --- that are not accessible in current measurements. The most significant reduction of uncertainty is expected due to an improved jet energy calibration.

\begin{figure}[ht]
\centering
\includegraphics[width=0.4\textwidth]{\main/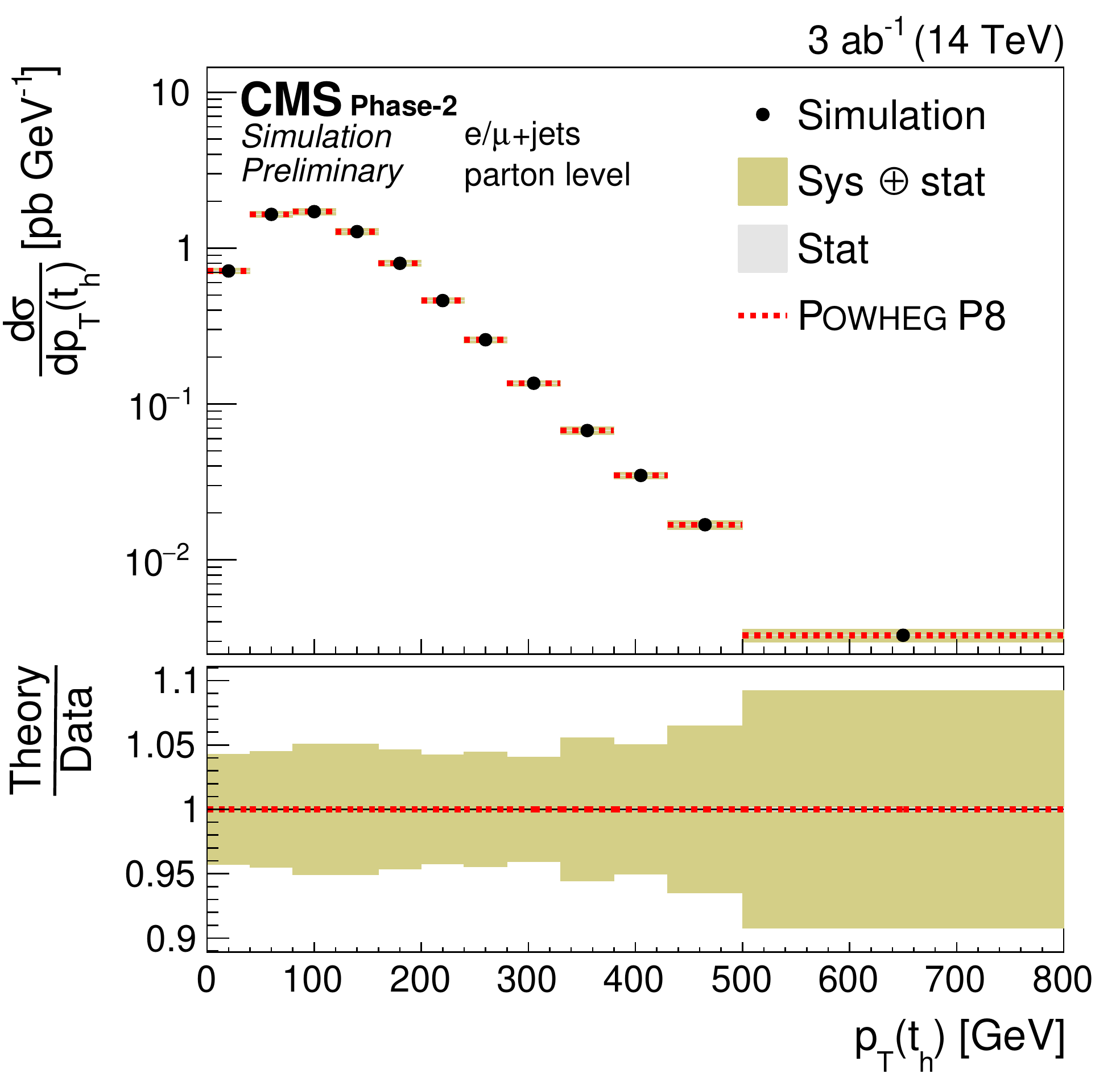}
\includegraphics[width=0.4\textwidth]{\main/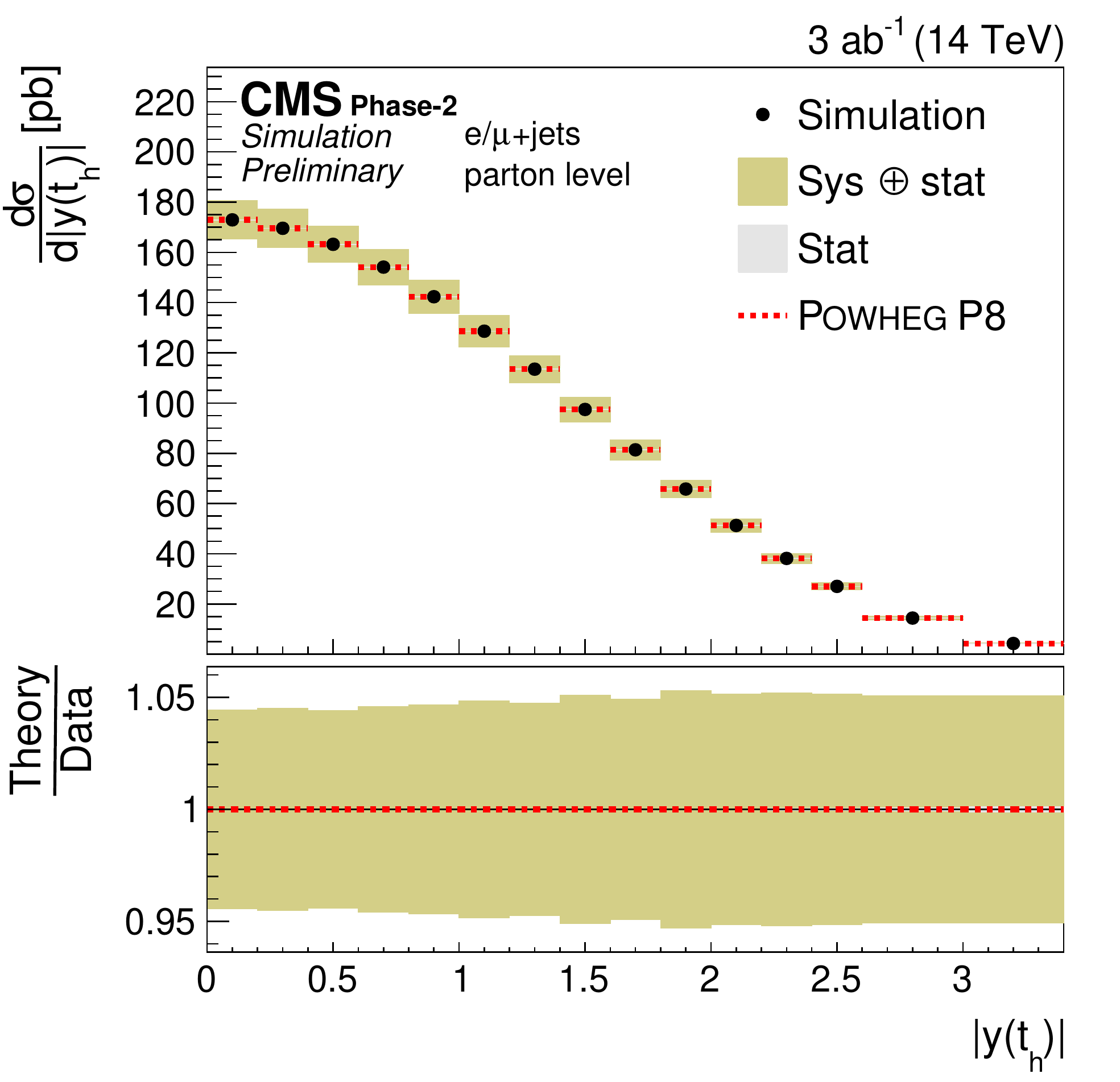}
 \caption{Projections of the differential cross sections as a function of $\pt(\tqh)$ (right) and $|y(\tqh)|$ (left).}
 \label{fig:APPYR2}
\end{figure}

\begin{figure}[ht]
\centering
\includegraphics[width=0.35\textwidth]{\main/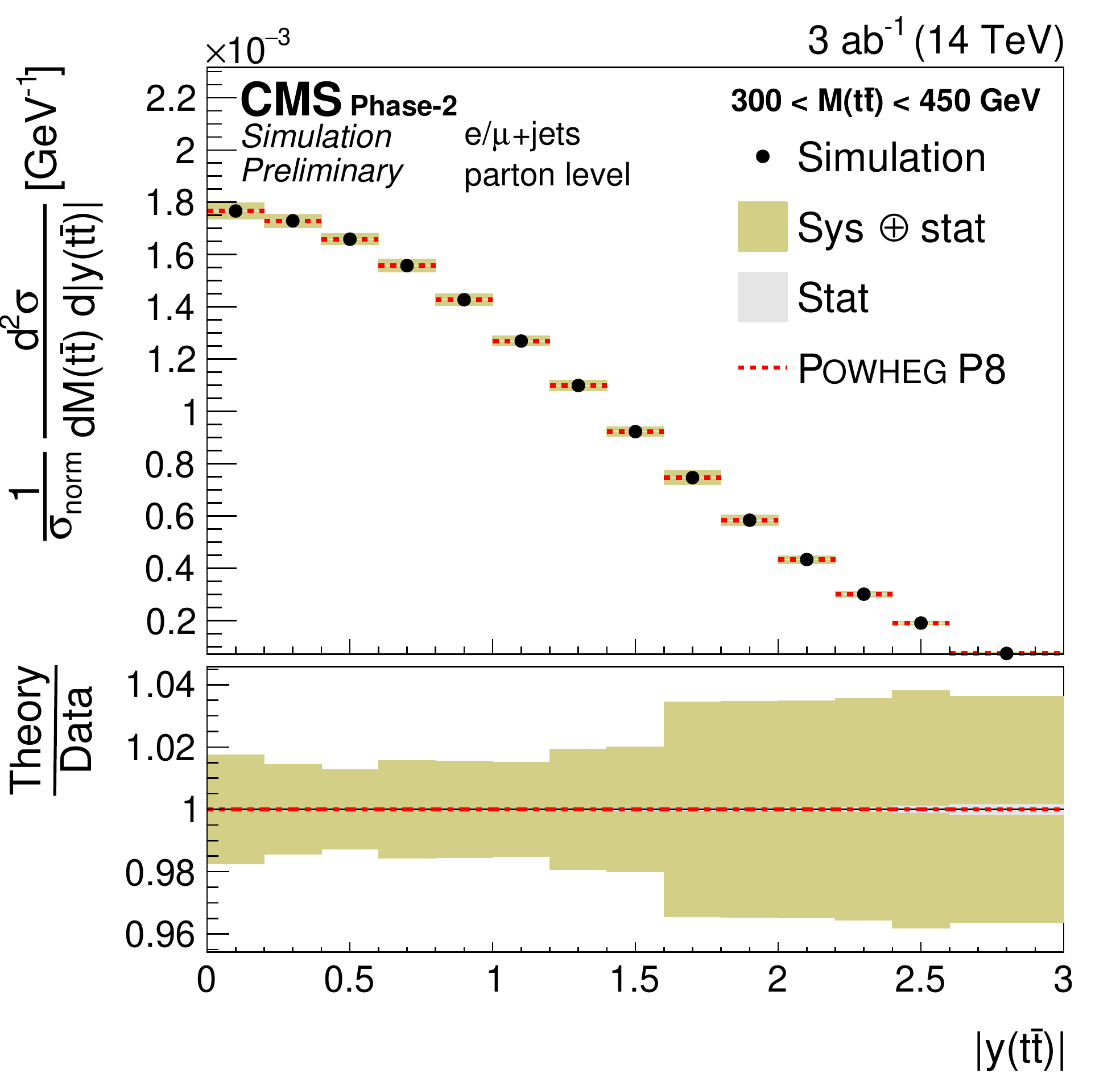}
\includegraphics[width=0.35\textwidth]{\main/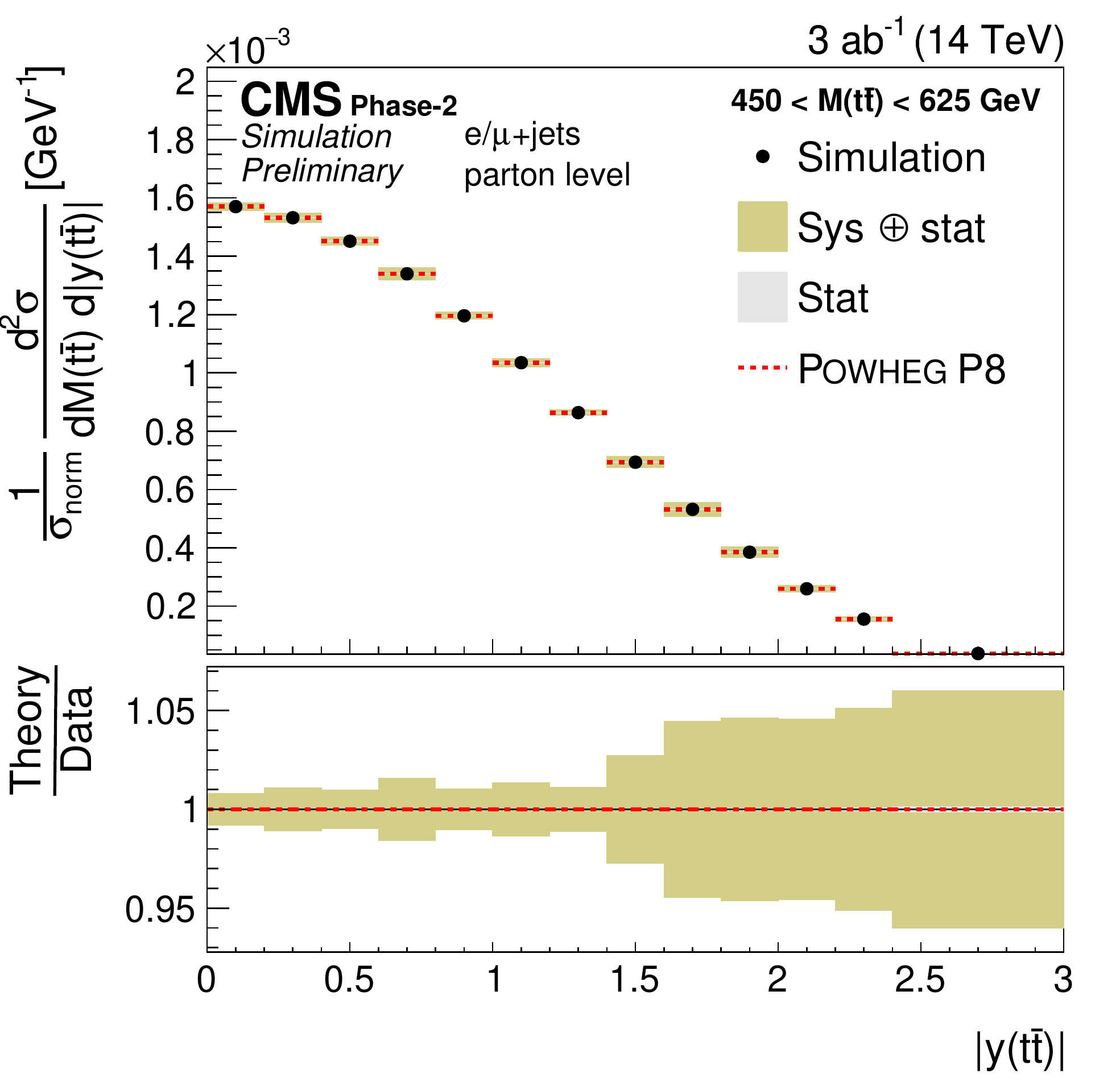}\\
\includegraphics[width=0.35\textwidth]{\main/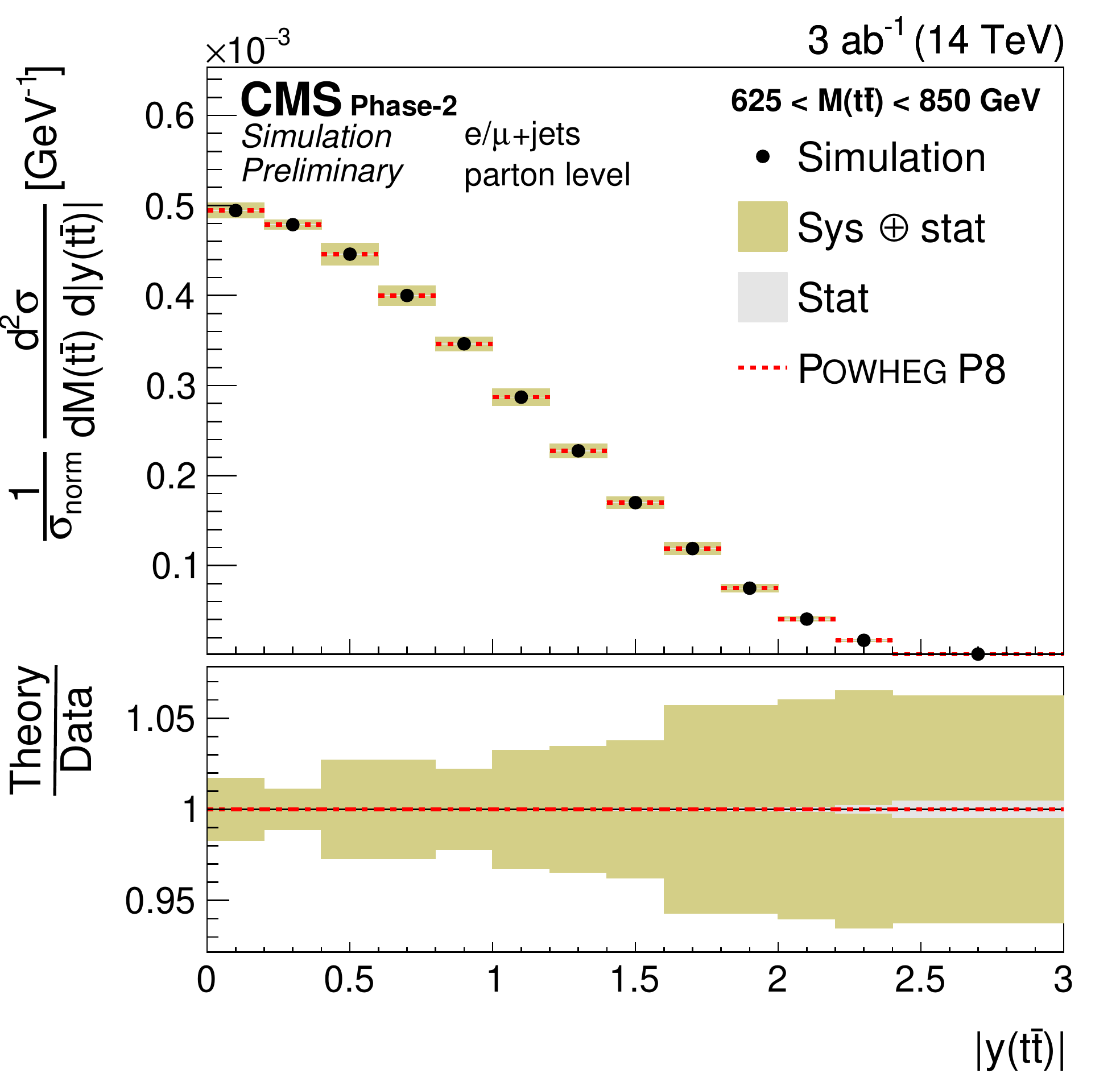}
\includegraphics[width=0.35\textwidth]{\main/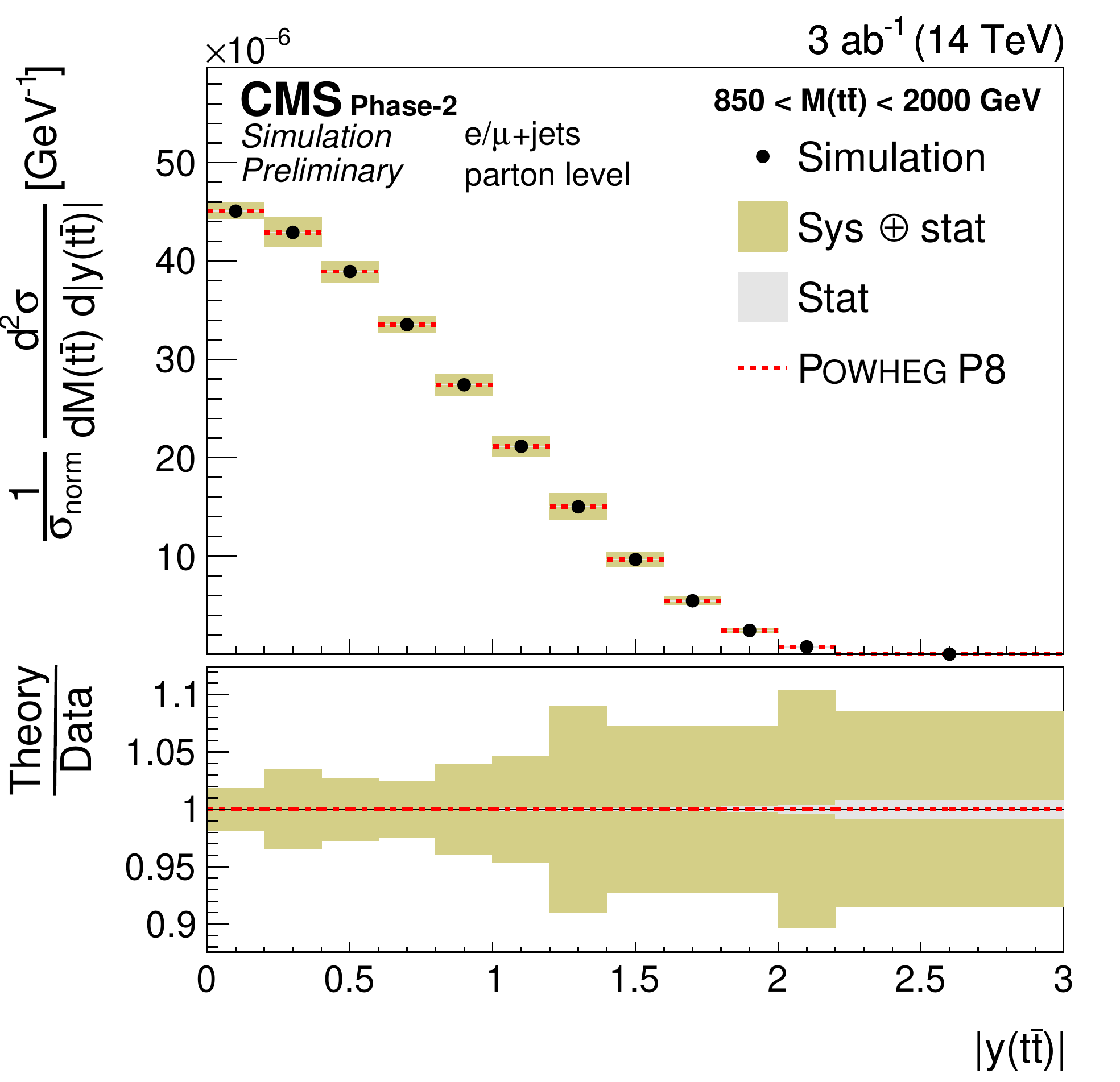}
 \caption{Projections of the double-differential cross section as a function of $|y(\ttbar)|$.}
 \label{fig:APPYR3}
\end{figure}

\subsubsection{PDF constraints from double-differential \ttbar cross sections}
\label{subsec:pdfconst}
The impact of differential \ttbar cross section measurements at the HL-LHC on the proton PDFs is quantitatively estimated using a profiling technique~\cite{Paukkunen:2014zia}, which is based on minimizing \chisq function between data and theoretical predictions taking into account both experimental and theoretical uncertainties arising from PDF variations. The analysis is performed using the \xfitter program~\cite{Alekhin:2014irh}, with the theoretical predictions for the \ttbar cross sections calculated at NLO QCD using the \mg~\cite{Alwall:2014hca} framework, interfaced with the \amcfast~\cite{Bertone:2014zva} and \applgrid~\cite{Carli:2010rw} programs. Three NLO PDF sets were chosen for this study: \abmp~\cite{Alekhin:2018pai}, \ct~\cite{Dulat:2015mca}, and \nnpdf~\cite{Ball:2017nwa}. The normalized double-differential \ttbar production cross sections as a function of $M(\ttbar)$ vs $|y(\ttbar)|$ are used which are expected to impose stringent constraints on the gluon distribution~\cite{Sirunyan:2017azo}. The \chisq value is calculated using the full covariance matrix representing the statistical and systematic uncertainties of the data, while the PDF uncertainties are treated through nuisance parameters. The values of these nuisance parameters at the minimum are interpreted as optimized or profiled PDFs, while their uncertainties determined using the tolerance criterion of $\Delta\chi^2 = 1$ correspond to the new PDF uncertainties. The profiling approach assumes that the new data are compatible with theoretical predictions using the existing PDFs, such that no modification of the PDF fitting procedure is needed. Under this assumption, the central values of the measured cross sections are set to the central values of the theoretical predictions. The original and profiled \abmp, \ct, and \nnpdf uncertainties of the gluon distribution at the scale $\mu_\mathrm{f}^2=30\,000\,GeV^2 \simeq m_{\PQt}^2$ are shown in Fig.~\ref{fig:pdf}. A consistent impact of the \ttbar data on the PDFs is observed for the three PDF sets. The uncertainties of the gluon distribution are drastically reduced once the \ttbar data are included in the fit.

\begin{figure}
\centering
\includegraphics[width=0.32\textwidth]{\main/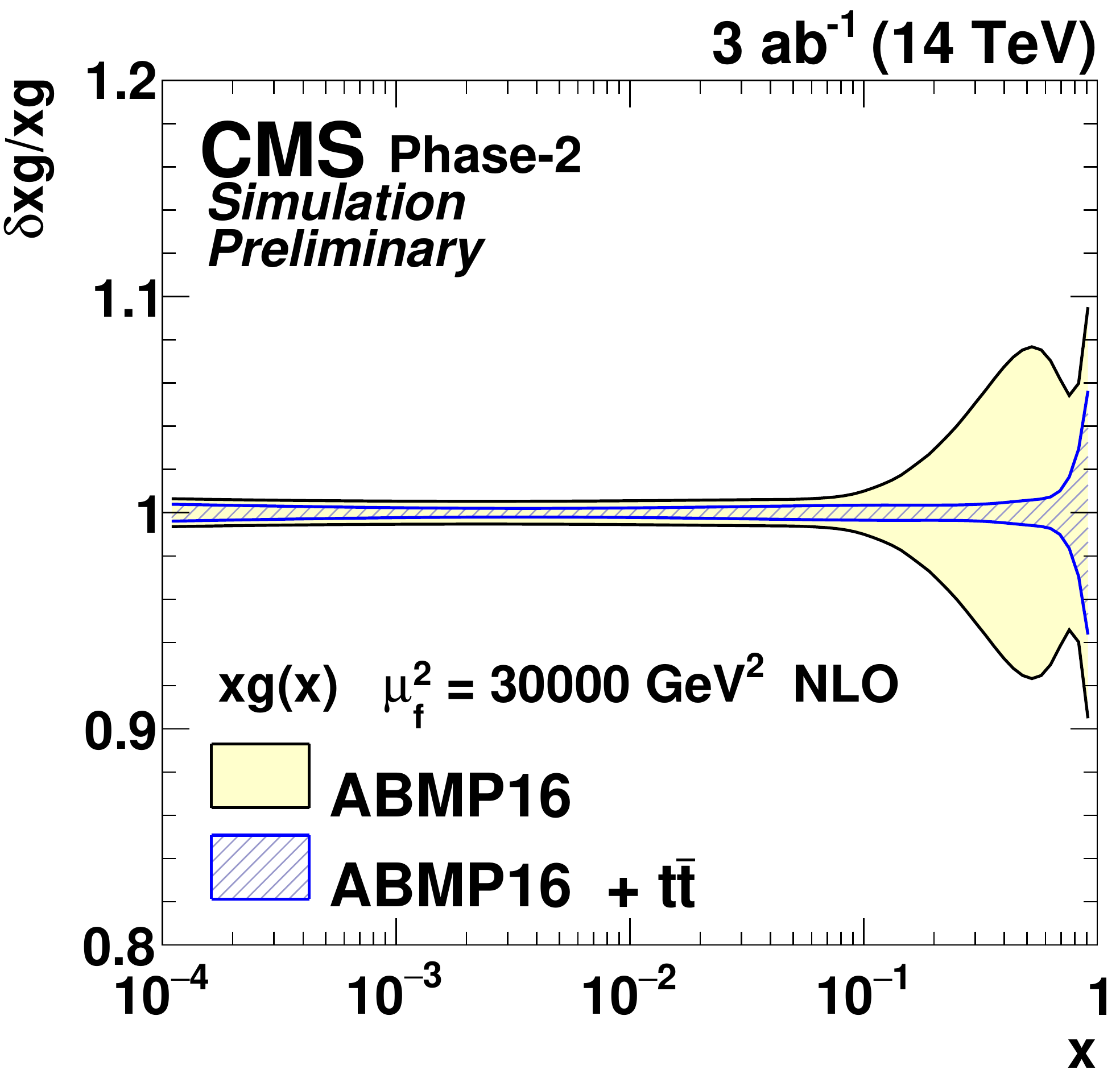}
\includegraphics[width=0.32\textwidth]{\main/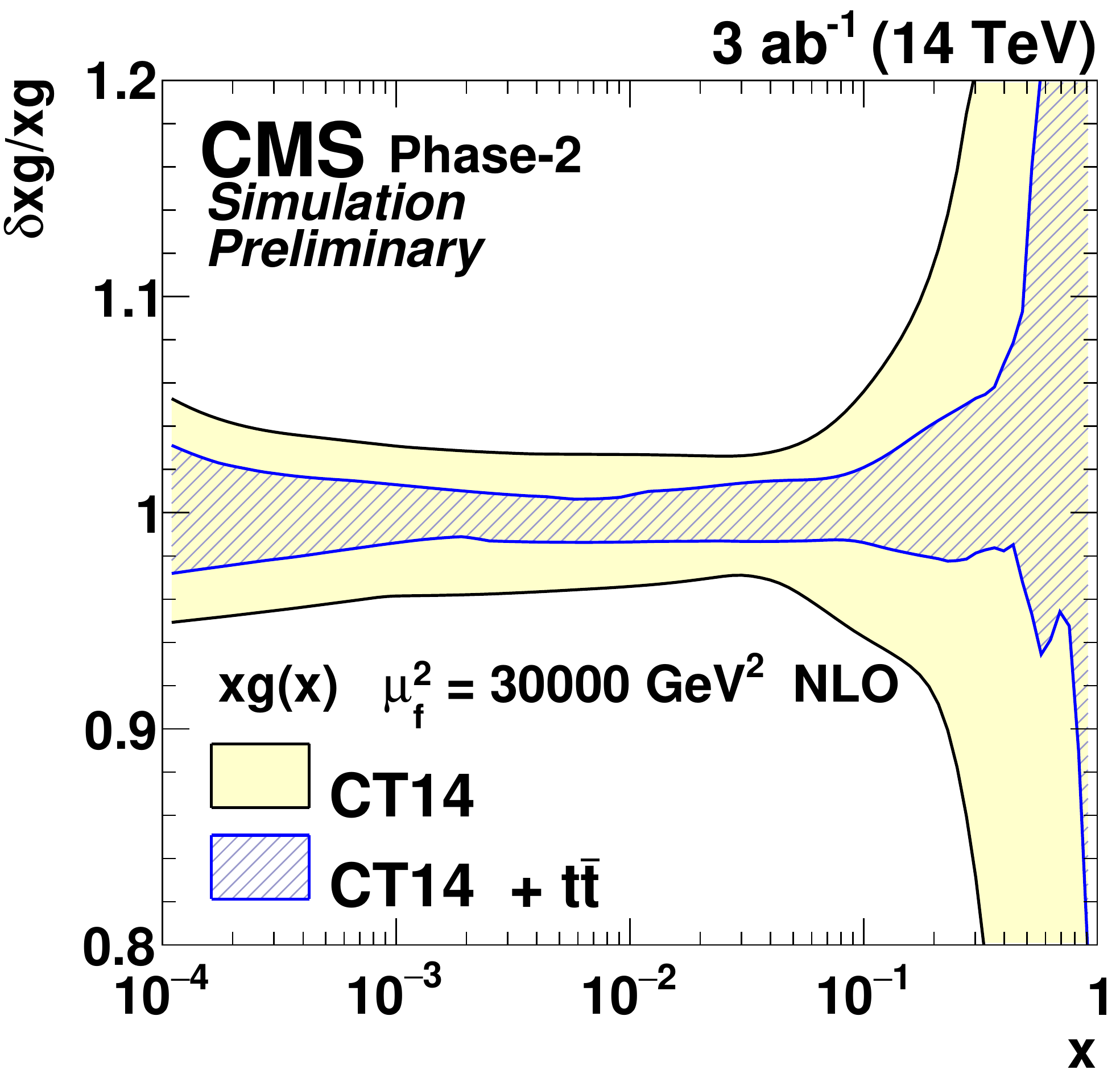}
\includegraphics[width=0.32\textwidth]{\main/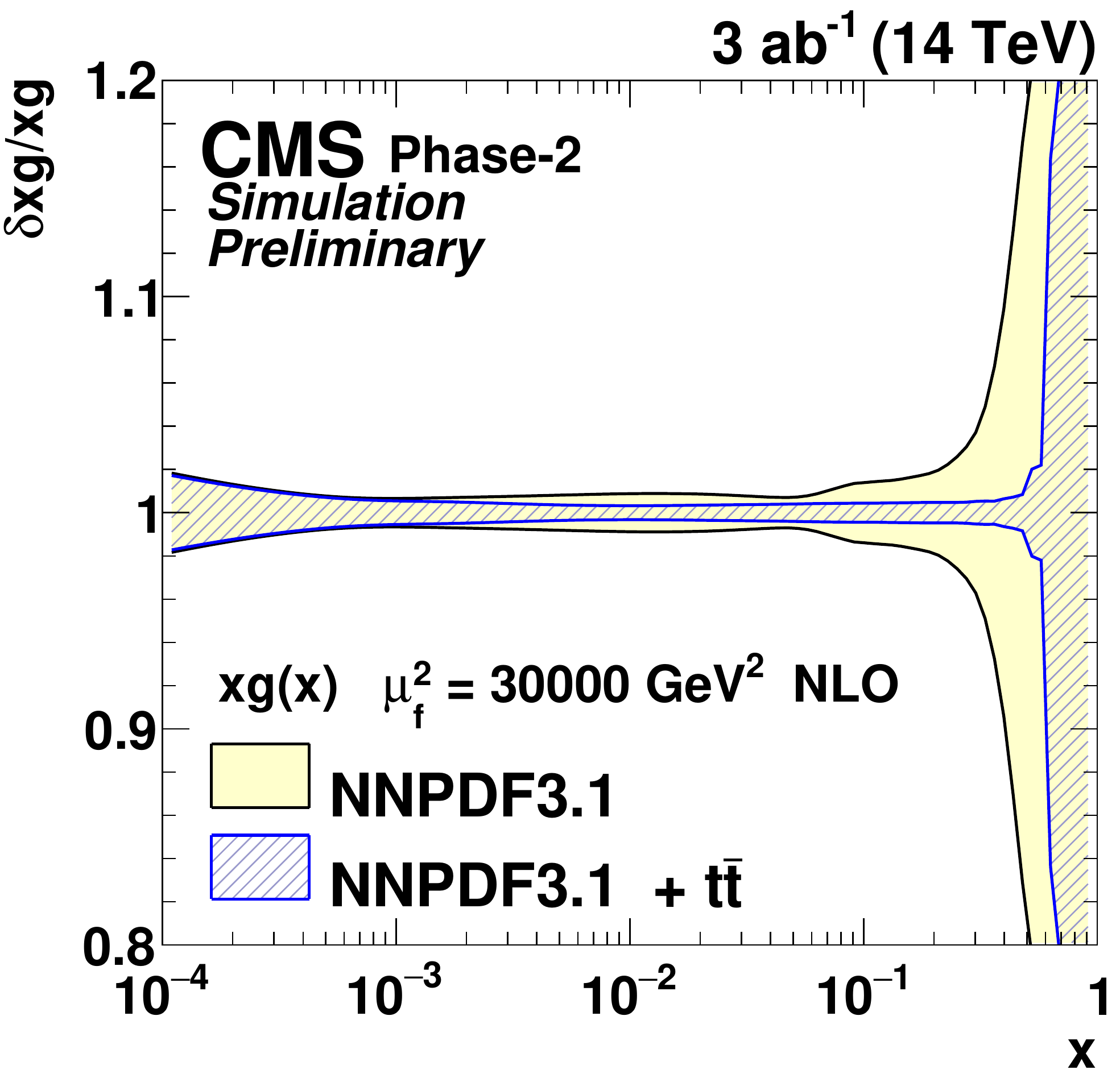}
\caption{The relative gluon PDF uncertainties of the original and profiled \abmp (left), \ct (middle) and \nnpdf (right) sets.}
\label{fig:pdf}
\end{figure}

\subsubsection{Forward top quark physics}
\label{sec:forwardtoplhcb}
Three measurements of top production have been performed by LHCb during Run-1 and -2 of the LHC with a precision of (20-40)\%, limited by the available data samples. As LHCb collects data at a lower rate than ATLAS and CMS, and has a limited acceptance, the measurements have focused on a partial reconstruction of the \ttbar final state in order to make optimal use of statistics. Additionally, as no estimate of missing energy is available, the measurements are performed at the level of the lepton and jets only, with no full top quark reconstruction performed. The first observation in the forward region was made in the $\mu b$ final state, where the top quark is identified by the presence of a muon and a $b$-jet~\cite{LHCb-PAPER-2015-022}. This final state has the highest signal yield, but suffers from the largest backgrounds, in particular from $W$ boson production in association with a $b$-jet. It also cannot separate single top and top pair production, which both contribute to the final state. Measurements were also performed in the $\ell b b$ final state\cite{LHCb-PAPER-2016-038} and $\mu e b$ final state~\cite{LHCb-PAPER-2017-050}, which suffer from lower statistics but select the signal with a higher purity.

While current measurements in the top sector at LHCb have been statistically limited, the available dataset at the HL-LHC, where LHCb is expected to collect 300 fb$^{-1}$, will permit precision measurements of the top quark pair production cross-section in the forward region, providing complementary information to ATLAS and CMS. The expected number of top pair events to be reconstructed at LHCb are given in Table~\ref{tab:topproj}, where the yields are obtained using next-to-leading predictions from the  {\mbox{\textsc{aMC@NLO}}\xspace} generator interfaced with  {\mbox{\textsc{Pythia}}\xspace} v8, with electroweak corrections approximated as described in Ref.~\cite{Gauld:2014pxa}. Leptons are required to satisfy $2.0<\eta<4.5$ and $p_{\rm T} > 20$ GeV, while jets are required to satisfy $2.2<\eta<4.2$ and $p_{\rm T} > 20$ GeV in all final states except the $\ell b$ final state, where the $p_{\rm T}$ threshold is raised to 60 GeV to combat the increased background. The detector efficiency is extrapolated from current measurements, where increases of between 10 and 50\% are expected due to  to improvements in the $b$-tagging algorithm and analysis techniques. Both muons and electrons are assumed to be employed for all analyses with similar efficiencies due to anticipated improvements in electron performance at LHCb during the HL-LHC. Measurements are expected to be made at sub-percent statistical precision in the $\ell b$ final state, and at the percent level in the $\mu eb$ and $\mu ebb$ final states. The dominant systematic uncertainties are expected to arise from the purity determination, particularly for the single lepton final states, and the knowledge of the $b$-tagging efficiency, which are both expected to be at the level of a few percent.

As \ttbar production in the LHCb acceptance probes very large values of Bjorken-$x$, it has the potential to provide significant constraints on the gluon PDF in this region. The potential of the $\mu e b$ final state was evaluated in Ref.~\cite{Gauld:2013aja}, where reductions of 20\% were found for a cross-section measurement with a precision of 4\%. Even more stringent constraints can be obtained through precise differential cross-section measurements, and measurements in the $\mu e bb$ final state, both of which will only be possible with the data available at the HL-LHC.

\begin{table}
\caption{
The  number of $t\bar{t}$ events expected to be reconstructed at LHCb per final state using a dataset corresponding to an integrated luminosity of 300 fb$^{-1}$. The mean value of  Bjorken-$x$ of the most energetic initiating parton is also shown for each final state.}

\begin{center}
\label{tab:topproj}
\begin{tabular}{|c c c| }
\hline
Final state &  300 fb$^{-1}$ & $<x>$ \\
\hline\hline
$\ell b$ & 830k & 0.295 \\
$\ell b \bar{b}$ & 130k & 0.368 \\
$\mu e b$ & 12k & 0.348 \\
$\mu e b \bar{b}$ & 1.5k & 0.415 \\
\hline
\end{tabular}
\end{center}
\end{table}

\def\figbase{\main/top/img/}

\subsubsection[Single top cross section: theoretical results]%
{Single top cross section: theoretical
  results\footnote{Contributed by F. Caola and E. Re.}}

\providecommand{\optW}{\mathcal{O}_{tW}}
\providecommand{\oppQ}{\mathcal{O}^{(3)}_{\phi Q}}
\providecommand{\opffff}{\mathcal{O}^{(3)}_{q Q, r s}}
\providecommand{\ctW}{C_{tW}}
\providecommand{\cpQ}{C^{(3)}_{\phi Q}}
\providecommand{\cffff}{C^{(3)}_{q Q, r s}}
\newcommand{\sss}{\scriptscriptstyle}
\newcommand{\OO}{\ensuremath{\mathcal{O}}}
\newcommand{\sst}{\scriptstyle}
\newcommand{\Op}[1]{\OO_{\sss #1}}
\def\lra#1{\overset{\text{\scriptsize$\leftrightarrow$}}{#1}}
\DeclareGraphicsRule{.tif}{png}{.png}{`convert #1 `dirname #1`/`basename #1 .tif`.png}
\graphicspath{{\main/top/img/}}
Although top quarks are predominantly produced in $t\bar{t}$ pairs
through strong interactions, a substantial fraction of them is also
produced through the exchange of electroweak bosons.  In the latter
case, only a single (anti-)top is produced per collision, hence one
refers to these processes as ``single-top'' production. Despite their
smaller rates with respect to pair production, single-top processes offer unique
opportunities to study the electroweak structure of top interactions.


The purpose of this section is to summarize the state-of-the-art for
the computation of single-top production cross sections, and highlight
what type of studies could be performed with an HL/HE-LHC upgrade.

It is customary to categorize single-top production in the SM
according to the virtuality of the $W$-boson involved in the
leading-order $2\to 2$ partonic process: the $s$-channel processes
($q\bar q' \to t \bar b$) involve the exchange of a time-like $W$ boson, the
$t$-channel processes $b q \to t q'$ involve the exchange of a
space-like $W$, while associated $Wt$-production ($bg \to t W^-$)
involves the production of a top quark in association with a $W$
boson.

Although convenient, the above characterization suffers two theoretical issues:
\begin{itemize}
\item 
a classification in terms of underlying $2\to2$ processes implicitly assumes
that the $b$-quark is treated as massless,  i.e.
the computations are performed in the so-called five-flavour number
scheme (5FNS). This framework effectively resums large logarithms of the form
$\ln m_b/Q$, where $Q$ is a typical transverse scale of the process and as such
it is particularly appropriate for observables that are only sensitive
to large $p_{\rm{T}}\gg m_b$ scales, like for example total cross sections. However,
especially in the $t$-channel case, there are important observables which are
sensitive to small transverse scales $p_{\rm{T}}\sim m_b$ (e.g. the kinematics of 
the ``spectator'' $b$-jet which
originates from initial state $g\to b\bar b$ splitting, particularly
at small $p_{\rm{T}}$). In this case, the 5FNS is not appropriate and 
it is important to treat the $b$-quark as massive,
i.e. to work in four-flavour mass scheme (4FNS). In this scheme, the $t$-channel LO process
becomes $2\to 3$: $g q\to t \bar{b} q'$. The 4FNS and 5FNS are
formally equivalent, but differences can arise when the perturbative
expansion is truncated, and in practice these effects might
be relevant for some observables~\cite{Campbell:2009gj,Campbell:2009ss,Frederix:2012dh}.
%
Within this context, the advantages of a HL/HE upgrade is twofold. On the one hand,
the larger dataset and increased energy would allow for more harsh selection
cuts that would effectively remove regions of the phase space sensitive to small
transverse scales. This would allow for a clean theoretical description using the 5FNS,
which does not suffer from large logarithmic contaminations. On the other hand, it would
allow one to explore with high accuracy the transition region between the range of validity
of the 4FNS and 5FNS, thus providing important information on their interplay. 

\item once higher-order corrections are included, the distinction between
$s$ and $t$ channels does not hold, due to interference effects. These
interference effects first appear at order
$\mathcal{O}(\alpha_s^2 \alpha^2)$, i.e. at NNLO in the 5FNS, or at
NLO in the 4FNS, and are color and (typically) kinematic suppressed. 
Given the large hierarchy
and small kinematic overlap between $t$- and $s$- channels, interference effects are
typically very small in $pp$ collisions, but may in principle play a role if very
high accuracy is required for specific observables. 
Moreover, once the $W$ and top decay products are
included, interferences arise also between $t\bar{t}$,
single-top (with $Wt$-production, as well as $t$-channel in the 4FNS) and 
$WWb\bar b$ production, 
unless the narrow-width limit $\Gamma_t\to 0$ is taken. These effects
can play a role for high precision studies, see e.g.~\cite{Cascioli:2014gda,Jezo:2016ujg}.
\end{itemize}

In spite of the above issues, as long as only NLO QCD corrections are
considered, it is possible to compute well-defined cross-sections for
$s$ and $t$-channel in the 5FNS, and, by imposing a jet-veto on
$b$-jets, to suppress the contamination of $t\bar t$ to the $Wt$
process, thereby allowing for a sensible definition of the cross section
for the latter channel as well. In Table~\ref{tab:xsections} the NLO cross sections are reported for the 3 channels at the LHC, for
centre-of-mass energies of 14 and 27 TeV. Scale and PDF uncertainties
are also reported. At both energies, the $t$-channel is the dominant
production mechanism. The relative importance of the $s$-channel decreases
with the collider energy, while it increases for $Wt$ associated production. 

\begin{table}[!h]
\begin{center}
\providecommand{\baselinestretch}{1.0}
\caption{\label{tab:xsections} Single-top inclusive cross sections at NLO for the LHC at 14 and 27 TeV, in the 5FNS. All results were obtained using PDF4LHC15\_nlo\_mc, the central value for the renormalization and factorizations scales ($\mu_R,\mu_F$) have been set equal to $m_t=173.2$ GeV and varied by a factor of two, with the constraint $1/2\le \mu_R/\mu_F\le 2$. For these predictions, $V_{tb}$ has been set to one. For $Wt$-channel only, a jet-veto on $b$-jets has been used ($p_{\rm{T},b_j}<50$ GeV), and the central value for $\mu_R$ and $\mu_F$ has been set to 50 GeV too.}
\begin{tabular}{| c | c c c | c c c | }
\hline
~ & ~ & 14 TeV & ~ & ~ & 27 TeV & ~ \\
\cline{2-7}
~ & $\sigma$ [pb] & $\Delta_{\mu_R,\mu_F}$ & $\Delta_{\mbox{\scriptsize PDF}}$
  & $\sigma$ [pb] & $\Delta_{\mu_R,\mu_F}$ & $\Delta_{\mbox{\scriptsize PDF}}$  \\
\hline
\hline
$t$-channel ($t$) & 156 & $~^{+3\%}_{-2.2\%}$ & $\pm$ 2.3 \% 
              & 447 & $~^{+3\%}_{-2.6\%}$ & $\pm$ 2\% \\
$t$-channel ($\bar{t}$) & 94 & $~^{+3.1\%}_{-2.1\%}$ & $\pm$ 3.1\% 
              & 299 & $~^{+3.1\%}_{-2.5\%}$ & $\pm$ 2.6\% \\
\hline
\hline
$s$-channel ($t$) & 6.8 & $~^{+2.7\%}_{-2.2\%}$ & $\pm$ 1.7\% 
              & 14.8 & $~^{+2.7\%}_{-3.2\%}$ & $\pm$ 1.8\% \\
$s$-channel ($\bar{t}$) & 4.3 & $~^{+2.7\%}_{-2.2\%}$ & $\pm$ 1.8\% 
              & 10.4 & $~^{+2.7\%}_{-3.3\%}$ & $\pm$ 1.8\% \\
\hline
\hline
$Wt$-channel ($t$ or $\bar t$) & 36 & $~^{+2.9\%}_{-4.4\%}$ & $\pm$ 5\% 
              & 137 & $~^{+3.8\%}_{-6.1\%}$ & $\pm$ 4\% \\
\hline              
\end{tabular}
\end{center}
\end{table}

Figure~\ref{fig:tchannel_cumulative} also shows, for the
$t$-channel case, the cumulative cross section with a minimum
$p_{\rm{T},\rm{min}}$ cut on the top, or antitop, transverse momentum, obtained
at NLO in the 5FNS.
\begin{figure}[!h]
  \caption{\label{fig:tchannel_cumulative} Cumulative cross section
    for $t$-channel single-(anti)top production in the 5FNS at 14 and
    27 TeV as a function of $p_{\rm{T},\rm{min}}$. The same settings used
    to obtain results in Table \ref{tab:xsections} were used here.}
  \begin{center}  
  \includegraphics[width=0.45\textwidth]{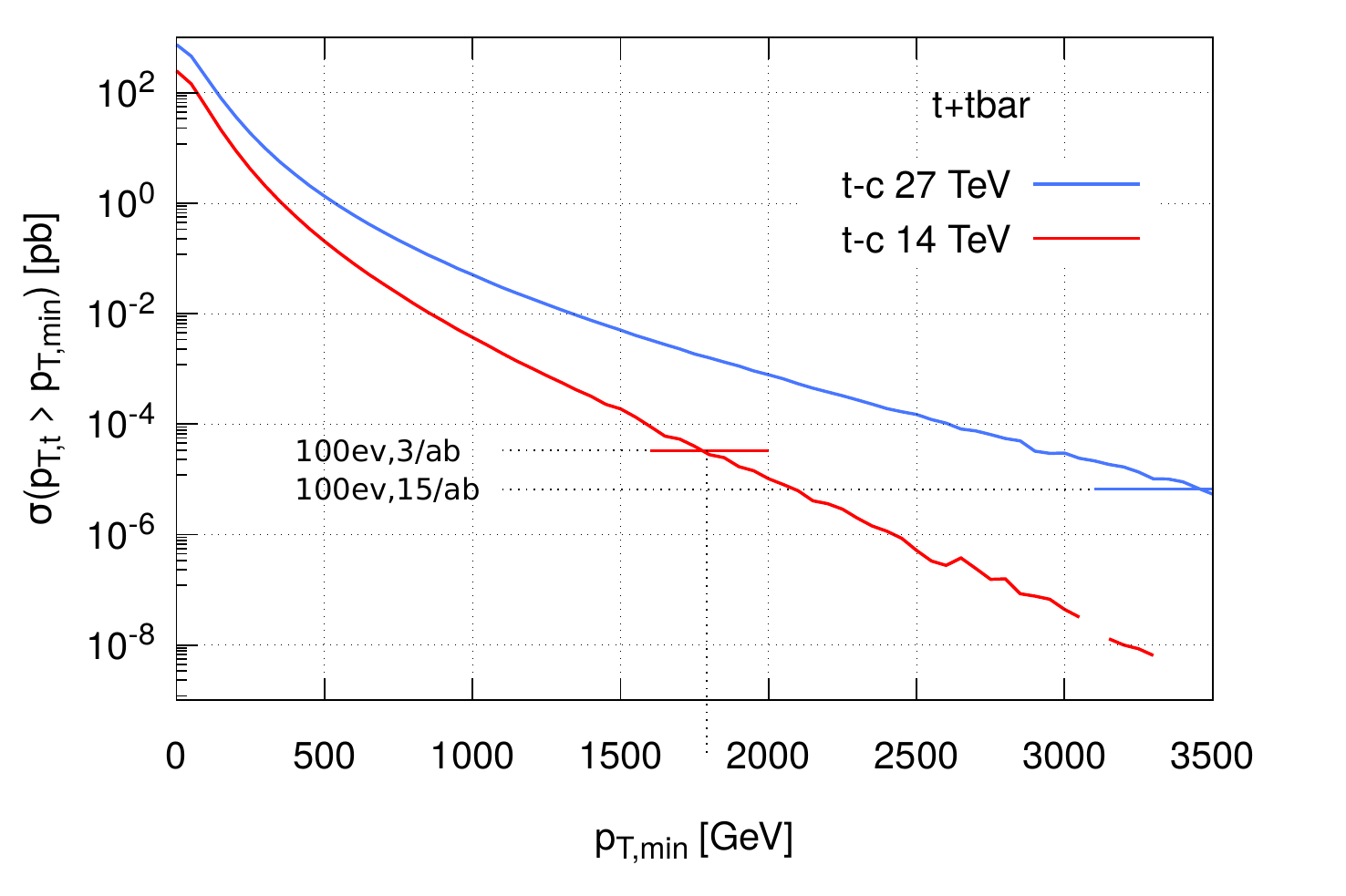}
  \end{center}
\end{figure}
The two horizontal bars in the plot correspond to the cross sections
for which one has 100 events, by assuming an integrated luminosity of
3 $\text{ab}^{-1}$ at 14 TeV (red) and of 15 $\text{ab}^{-1}$ at 27
TeV (blue).

For $t$-channel production, NNLO QCD corrections have also been
computed in
Refs.\cite{Brucherseifer:2014ama,Berger:2016oht,Berger:2017zof}.\footnote{NNLO
QCD results were also obtained for $s$-channel, see
Ref.~\cite{Liu:2018gxa}.} These corrections have been obtained in the
structure function approximations, where higher-order corrections to
the light and heavy-quark lines ($q\to q' W$ and $b\to t W$,
respectively) are computed separately. Within this approximation, the
terms which are not included at NNLO are color suppressed ($1/N_c^2$),
and hence estimated to be negligible for phenomenology, given the
moderate size of NNLO effects. Moreover, when working in these
approximations, interference effects between $s$ and $t$-channel are
also absent. The results obtained in
Refs\cite{Brucherseifer:2014ama,Berger:2016oht} indicate that NNLO
QCD corrections are small: the total cross sections at NNLO increase
by at most 2\% with respect to the NLO result (when the latter is
obtained with NLO PDFs), whereas the relative scale uncertainty is
reduced by at least $\sim 50\%$. Moreover, the NNLO result is
contained within the NLO uncertainty band, showing extremely good
convergence for the perturbative expansion.\footnote{When NLO
corrections are computed with NNLO PDFs, the NNLO/NLO ratio is instead
slightly smaller than one, but the conclusions remain the same.}
Despite the fact that the total cross section shows excellent
perturbative stability, more sizeable effects can be noticed in some
differential distributions, where NNLO/NLO corrections can reach
$\mathcal{O}(10\%)$ in certain regions of the transverse momentum
distributions of the top (anti-)quark and the pseudo-rapidity
distributions of the leading jet. In these cases, scale variation
may underestimate the actual theoretical uncertainty. 

NNLO corrections to the top quark decay are also
known\cite{Gao:2012ja,Brucherseifer:2013iv}, and they can be combined
with the NNLO corrections to production using the ``on-shell top-quark
approximation'' where the top width $\Gamma_t$ is kept finite, but
tree-level interference effects between the single top production and
decay stage are neglected, as well as loop diagrams with a virtual
gluon connecting the production and decay stages.  This is an
excellent approximation for inclusive-enough quantities, since omitted
corrections are suppressed by a factor $\Gamma_t/m_t< 1\%$~\footnote{
This is not the case for exclusive observables, which are sensitive to off-shell effects in 
the reconstructed top mass $M_{Wb}$, and beyond kinematic edges, see
Ref.~\cite{Papanastasiou:2013dta} for a thorough analysis.}.
More
details can be found in Ref.~\cite{Berger:2017zof}.

In presence of fiducial cuts, it is important to stress that QCD
corrections are more pronounced, with NNLO effects amounting about
$5\%$ on total rates as well as differential distributions. In this
case, corrections from pure decay are typically half of those from
pure production. Finally, it should be noted that NLO EW corrections to on-shell
single top production are small, $\sim$ few permille, see e.g.~\cite{Frederix:2018nkq}.
The EW effect can become more relevant in tails of distributions, or for
observables highly sensitive to off-shell effects. 

Single-top can also be produced in association with a $Z$ boson
($tZq$). Although the cross section is smaller than in the
aforementioned channels, a HL/HE upgrade at the LHC will allow one to
measure well this production process too. QCD NLO corrections to
$tZq$-production are known\cite{Campbell:2013yla}. Table~\ref{tab:tzq} reports the total cross sections at NLO in the
SM, for centre-of-mass energies of 14 and 27 TeV.
\begin{table}[!ht]
\begin{center}
\providecommand{\baselinestretch}{1.0}
\caption{\label{tab:tzq} Single-top production cross section in association with a $Z$ boson, at NLO for the LHC at 14 and 27 TeV, in the 5FNS. All results were obtained using PDF4LHC15\_nlo\_mc, the renormalization and factorizations scales have been set equal to $m_t=173.2$ GeV.}
\begin{tabular}{| c | c | c | }
\hline
~ & $\sigma$ [fb] @14 TeV & $\sigma$ [fb] @27 TeV\\
\hline
\hline
$tZq$-channel ($t$) & 639
                    & 2536  \\
$tZq$-channel ($\bar{t}$) & 350
                    &  1543    \\
\hline
\end{tabular}
%
\end{center}
\end{table}

As far as phenomenology is concerned, single-top offers the
possibility to perform several studies within and beyond the SM. Within
the ``SM only'' hypothesis, one can use it to extract information
about the SM $V_{tb}$ matrix element, as discussed for instance in
Ref.\cite{Alwall:2006bx}.
\begin{figure}[!t]
  \begin{center}  
  \includegraphics[width=0.45\textwidth]{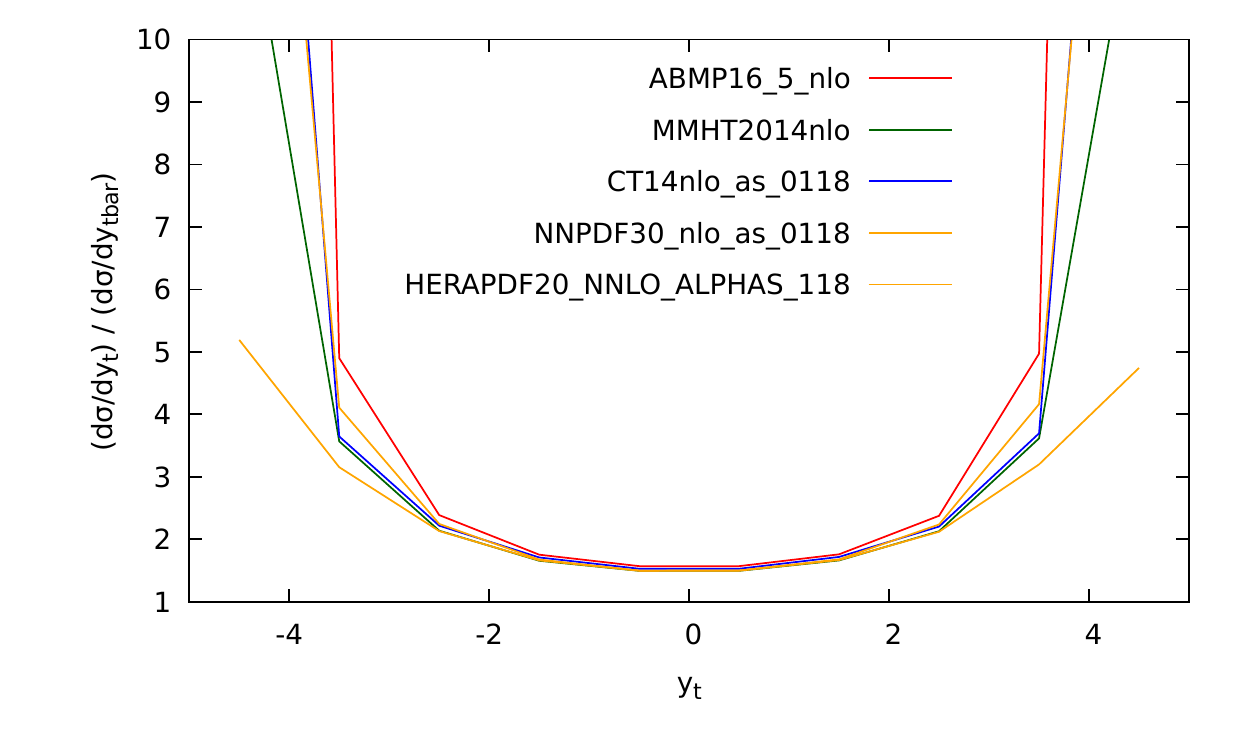}
  \includegraphics[width=0.45\textwidth]{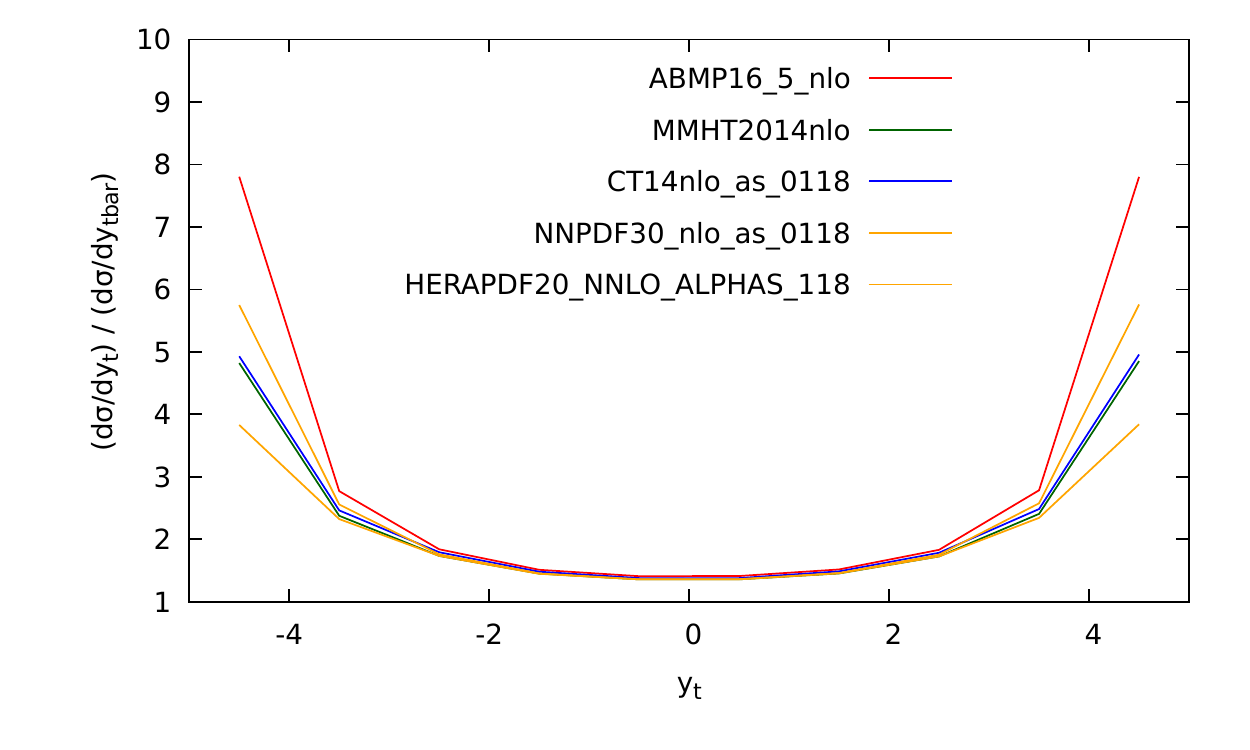}\\
  \includegraphics[width=0.45\textwidth]{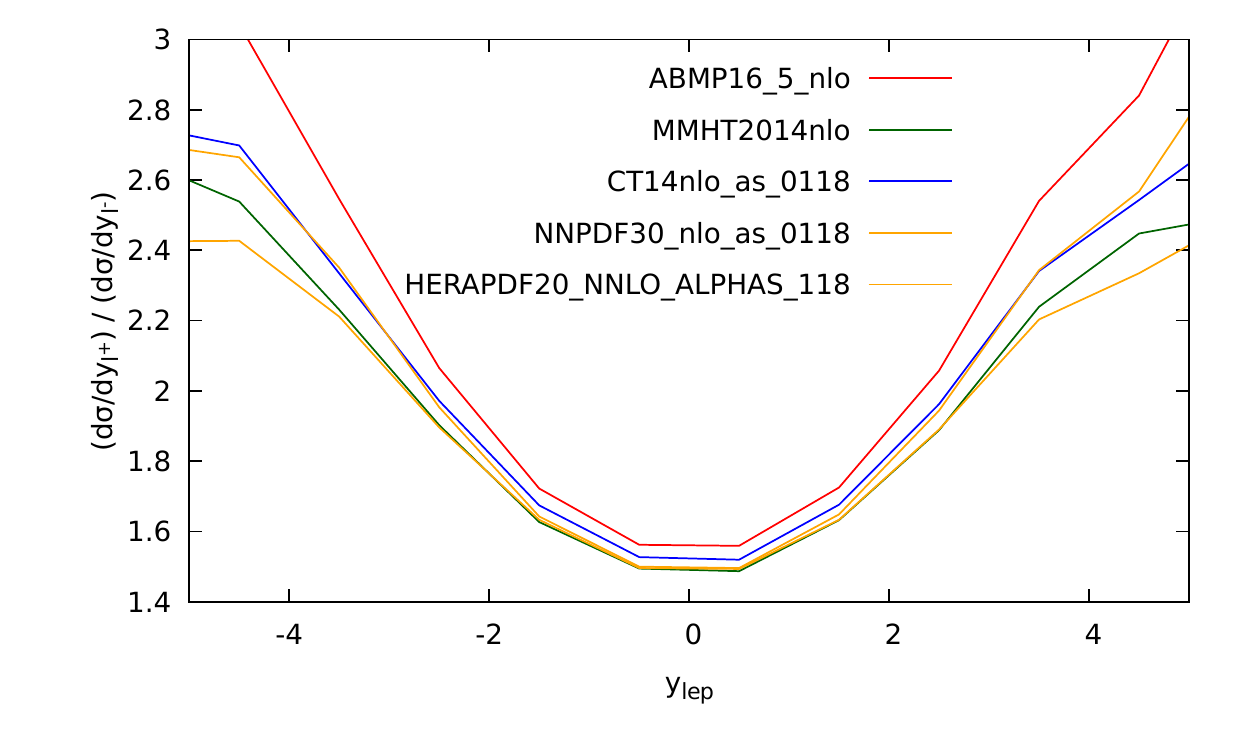}
  \includegraphics[width=0.45\textwidth]{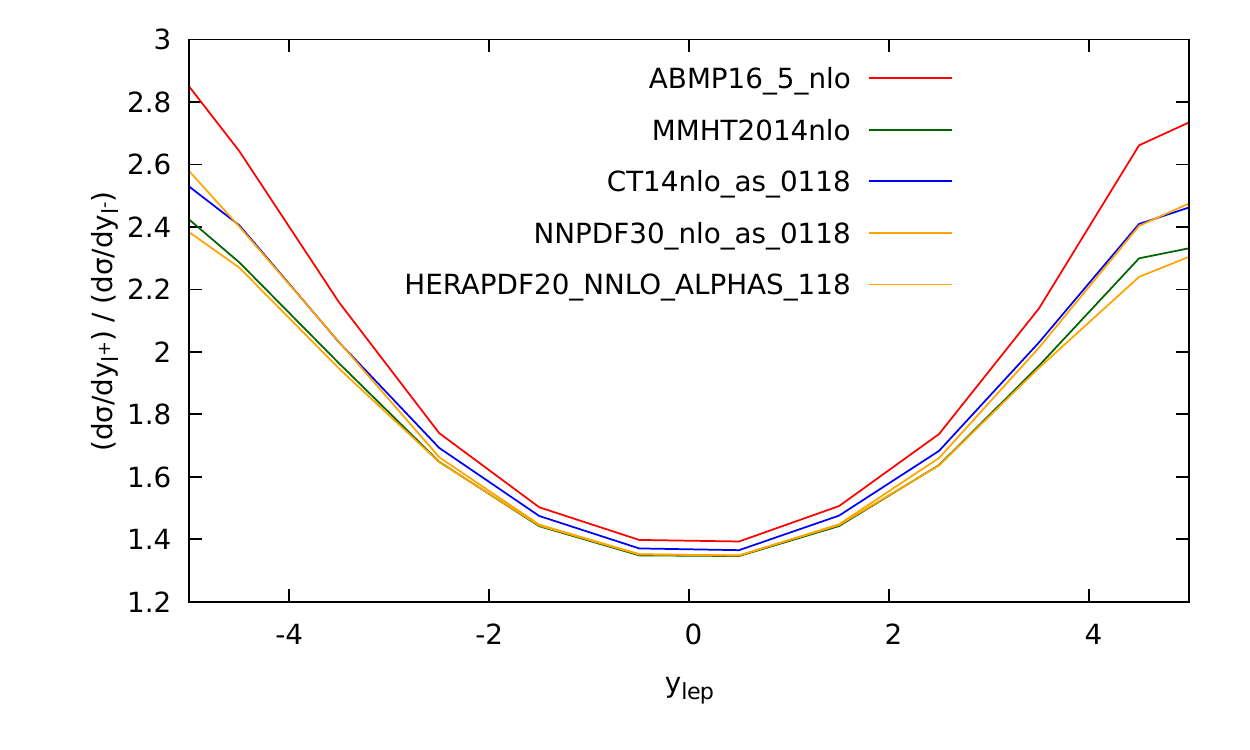}
  \end{center}
  \caption{\label{fig:charge_ratios} Differential charge ratios
  $\mathcal{O}_t/\mathcal{O}_{\bar t}$ at 14 (left panels) and 27
  (right panels) TeV for the top quark and charged lepton rapidities,
  in $t$-channel single-top production.}
\end{figure}
Setting constraints on the $b$-quark PDF might also be possible, by
looking at charge ratios, i.e. ratios of $t/\bar t$ cross
sections. These ratios depend in general upon the PDFs used, and
notably, in the $t$-channel case, on the $b$-quark PDF. Moreover, they
can be predicted quite accurately, as most of the theoretical
uncertainties cancel out in the ratio, leaving a residual theoretical
uncertainty from scale variation (at NNLO) of few percent for each PDF
set, as shown for instance in Fig.~29 of
Ref.~\cite{Berger:2017zof}. Although the charge ratio for total cross
sections $\sigma_t/\sigma_{\bar t}$ exhibits a dependence upon the PDF
set\cite{Aaboud:2016ymp,Sirunyan:2016cdg}, slightly more pronounced
sensitivity might be obtained by looking at differential
distributions, such as $(d\sigma/dy_t)/(d\sigma/dy_{\bar t})$ and
$(d\sigma/dy_{\ell^+})/(d\sigma/dy_{\ell^-})$, which also allow one to
constrain the $u/d$ ratio in the proton. In
Fig.~\ref{fig:charge_ratios} such a comparison among different PDF
sets is shown, for LHC collisions at 14 and 27 TeV: differences among
different PDF sets can be observed, especially at large rapidities. It
is clear that a HL upgrade will allow one to reduce the statistical
uncertainty at large rapidities, giving the chance to discriminate
among different PDF sets. As the available phase space opens up,
further sensitivity might be expected at 27 TeV.

Single-top processes offer also several opportunities to probe some
new-physics scenarios\footnote{In the following the
discussion is limited to the $t$-channel case, and the production in association
with a $Z$ boson.}.
In order to systematically interpret potential
deviations from the SM, it is particularly convenient to work in the
SM Effective Field Theory
(SMEFT)\cite{Weinberg:1978kz,Buchmuller:1985jz}, where the SM is augmented by a set of higher-dimension operators. If the discussion is limited
to dimension-6 operators, the SMEFT Lagrangian has the form
\begin{equation}
\mathcal{L}_{\rm SMEFT} = \mathcal{L}_{\rm SM} + \sum_i\frac{C_i}{\Lambda^2} \mathcal{O}_i + \mathcal{O}(\Lambda^{-4}).
\label{eq:smeft}
\end{equation}
where the sum runs over all the dimension-6 operators that maintain
the SM symmetries. The remarkable virtue of $t$-channel single-top
production is that its cross section only depends upon a limited
number of dimension-6 operators, thereby allowing to set bounds on
them relatively easily. At LO and in the 5FNS only three operators
contribute:
\begin{eqnarray}
\Op{tW}&=&
     i\big(\bar{Q}\sigma^{\mu\nu}\,\tau_{\sss I}\,t\big)\,
     \tilde{\phi}\,W^I_{\mu\nu}
     + \text{h.c.}\,,\label{eq:O1}\\
\Op{\phi q}^{\sss(3)}&=&
     i\big(\phi^\dagger\lra{D}_\mu\,\tau_{\sss I}\phi\big)
     \big(\bar{q}_i\,\gamma^\mu\,\tau^{\sss I}q_i\big)
     + \text{h.c.}\,,\label{eq:O2}\\
\Op{Qq}^{\sss (3,1)}&=&
    \big(\bar{q}_i\,\gamma_\mu\,\tau_{\sss I}q_i\big)
    \big(\bar{Q}\,\gamma^\mu\,\tau^{\sss I}Q\big)\,,\label{eq:O3}
\end{eqnarray}
in agreement with the notation of \cite{AguilarSaavedra:2018nen}.
The operators of \eq{eq:O1}-\eq{eq:O2} modify the $Wtb$
interaction in the following way
\begin{eqnarray}
  {\mathcal L}^{\mathrm{dim-6}}_{Wtb} &=& -\frac{g}{\sqrt{2}}  \bar{b}(x) \gamma^\mu 
                                      P_L t(x)\,W_{\mu}(x) \left(1+  \frac{C^{(3)}_{\varphi Q} v^2}{\Lambda^2}  \right) \nonumber \\
                                  &+&  \frac{2  \,v  \, C_{tW}}{\Lambda^2}  \bar{b}(x) \sigma^{\mu\nu} 
                                      P_R t(x)\,\partial_{\nu} W_{\mu}(x) +  {\mathrm{h.\; c.}} \,,\label{eq:st-interaction}
\end{eqnarray} 
where $v = 246$ GeV is the Higgs doublet vacuum expectation value, and
$y_t$ the top quark Yukawa coupling. Here and below it is assumed
$V_{tb}=1$.  Note that the four-fermion operator of eq.~\eqref{eq:O3}
introduces a contact $udtb$ interaction. From eq.~\eqref{eq:st-interaction}
it is clear that setting bounds on the SMEFT using single-top
measurements allows to probe in detail the structure of the $Wtb$
coupling. A comprehensive discussion can be found
in Ref.~\cite{deBeurs:2018pvs}, where a NLO study of the effect of these
operators on total and differential distributions in single top
production and decay is performed.

In the SMEFT, the single top cross section can be parameterised as
\begin{flalign}
	\sigma=\sigma_{SM}+\sum_i\frac{1{\rm TeV}^2}{\Lambda^2}C_i\sigma_i
	+\sum_{i\leq j}
	\frac{1{\rm TeV}^4}{\Lambda^4}C_iC_j\sigma_{ij}.
	\label{eq:xsecpara}
\end{flalign}
To establish the impact of the operators on single top production at
HL/HE-LHC, Table~\ref{tab:sensitivity} shows the ratio
$r_i=\sigma_i/\sigma_{SM}$ for 14 TeV and 27 TeV both for the
inclusive cross section and the high transverse momentum region.
Results are obtained in the 5FNS with NNPDF3.0 LO
PDFs \cite{Ball:2014uwa}.
Central scales for $\mu_R,\mu_F$ are chosen as $m_t$. It is found that the
impact of the operator in eq.~\eqref{eq:O2} remains unchanged when going
from 14 to 27 TeV, as its effect is to only rescale the SM coupling.
The impact of the dipole operator in eq.~\eqref{eq:O1} is only mildly affected
by going to the HE-LHC, whereas the sensitivity to the four-fermion
operator is the one which benefits most by probing the high $p_{\rm{T}}$ tail
and by the HE-LHC.
\begin{table}[!ht]
\begin{center}
\caption{\label{tab:sensitivity}
Comparison among the LO sensitivities of $t-$ channel single-top to
the three operators described in eq.~\eqref{eq:O1}-\eqref{eq:O3}, for
the inclusive cross-section and with a cut $p_{\rm{T}}^{t}>350$ GeV, at 14
and 27 TeV. Results are obtained in the 5FNS with NNPDF3.0 LO
PDFs~\cite{Ball:2014uwa}, the renormalization and factorizations
scales have been set equal to $m_t=173.2$ GeV. The interference term
$r_i=\sigma_i/\sigma_{SM}$ (when non-zero) and the square
$r_{i,i}=\sigma_{i,i}/\sigma_{SM}$ are given for each
operator. $\sigma_i$ and $\sigma_{i,i}$ are defined in
eq.~\eqref{eq:xsecpara}.}
   \begin{tabular}{| l | c c | c c |}
   \hline
   ~ &  \multicolumn{2}{c|}{$t$-channel 14 TeV} & \multicolumn{2}{c|}{$t$-channel 27 TeV ~}

       \tabularnewline
        & 
        & ($p_{\rm{T}}^{t}>350$ GeV) 
        & 
        & ($p_{\rm{T}}^{t}>350$ GeV) 
       \tabularnewline
        \hline
        \hline
        $\sigma_{SM}$
        & 225 pb 
        & 0.746 pb
        & 640 pb
        &  3.40 pb

        \tabularnewline\hline
        $r_{ tW}$
        & 0.025
        & 0.052
        & 0.022
        & 0.040

        \tabularnewline
        $r_{ tW,tW}$
        & 0.014
        & 0.31
        & 0.016
        & 0.34

        \tabularnewline\hline
        $r_{ \phi Q^{(3)}}$
        & 0.12
        & 0.12
        & 0.12
        & 0.12

        \tabularnewline
        $r_{ \phi Q^{(3)},\phi Q^{(3)}}$
        & 0.0037
        & 0.0037
        & 0.0037
        & 0.0037

        \tabularnewline\hline
        $r_{ Qq^{(3,1)}}$
        &  -0.36
        & -6.45
        & -0.39
        & -6.79

        \tabularnewline
        $r_{ Qq^{(3,1)},Qq^{(3,1)}}$
        & 0.135
        & 18.8
        & 0.222
        & 26.8

        \tabularnewline\hline
   \end{tabular}
   \end{center}
   \renewcommand{\baselinestretch}{1.0}
\end{table}

Production in association with a $Z$ boson is also important in the
BSM context. A complete study of its sensitivity to BSM effects was
performed in Ref.~\cite{Degrande:2018fog}, where the interplay with
$t$-channel single-top, as well as single-top production in
association with a Higgs boson, is discussed thoroughly, and at
NLO. Table 6 of~\cite{Degrande:2018fog} reports a comparison among the
sensitivity of these processes to various operators. Current limits
from other processes, as well as current and future projections for
bounds that can be achieved looking into $tZj$ production are also
discussed (e.g. in Fig. 6 of Ref.~\cite{Degrande:2018fog}). For some operators, notably $\Op{tW}$ and
$\Op{\phi q}^{\sss(3)}$, the improvement due to considering $tZj$
measurements at HL are remarkable, especially when tails of
distributions are considered.  It is likely that even more promising
results could be obtained at HE.

Another goal of a HL/HE upgrade is to extract bounds on (or find
evidence of) $WWZ$ anomalous gauge couplings, or FCNC. In this
context, $tZq$ is quite important both because it is sensitive to
these effects, as well as because it's an irreducible background, as
its production rate is competitive with $t\bar{t}Z$ production, where
these effects are typically looked for.


\let\vaccent=\v 
\providecommand{\v}[1]{\ensuremath{\mathbf{#1}}} 
\providecommand{\perc}{\%}
\providecommand{\gv}[1]{\ensuremath{\mbox{\boldmath$ #1 $}}}
\providecommand{\uv}[1]{\ensuremath{\mathbf{\hat{#1}}}} 
\providecommand{\abs}[1]{\left| #1 \right|} 
\providecommand{\avg}[1]{\left< #1 \right>} 
\let\underdot=\d 
\providecommand{\d}[2]{\frac{d #1}{d #2}} 
\providecommand{\dd}[2]{\frac{d^2 #1}{d #2^2}} 
\providecommand{\pd}[2]{\frac{\partial #1}{\partial #2}}
\providecommand{\pdd}[2]{\frac{\partial^2 #1}{\partial #2^2}}
\providecommand{\pdc}[3]{\left( \frac{\partial #1}{\partial #2}
 \right)_{#3}} 
\providecommand{\ket}[1]{\left| #1 \right>} 
\providecommand{\bra}[1]{\left< #1 \right|} 
\providecommand{\braket}[2]{\left< #1 \vphantom{#2} \right|
 \left. #2 \vphantom{#1} \right>} 
\providecommand{\matrixel}[3]{\left< #1 \vphantom{#2#3} \right|
 #2 \left| #3 \vphantom{#1#2} \right>} 
\providecommand{\grad}[1]{\gv{\nabla} #1} 
\let\divsymb=\div 
\providecommand{\div}[1]{\gv{\nabla} \cdot #1} 
\providecommand{\curl}[1]{\gv{\nabla} \times #1} 
\let\baraccent=\= 
\providecommand{\=}[1]{\stackrel{#1}{=}} 
\newtheorem{prop}{Proposition}
\newtheorem{thm}{Theorem}[section]
\newtheorem{lem}[thm]{Lemma}
\theoremstyle{definition}
\newtheorem{dfn}{Definition}
\newtheorem*{rst}{Result}
\theoremstyle{remark}
\newtheorem*{rmk}{Remark}

\providecommand{\gev}[1]{$\unit{#1}{\giga\electronvolt}$}
\providecommand{\tev}[1]{$\unit{#1}{\tera\electronvolt}$}
\providecommand{\gevm}[1]{\unit{#1}{\giga\electronvolt}}
\providecommand{\tevm}[1]{\unit{#1}{\tera\electronvolt}}
\providecommand{\half}{$\frac{1}{2}$}
\providecommand{\ddk}[1]{\frac{d^d k_{#1}}{(4\pi)^d}}
\providecommand{\sidenote}[1]{\todo[noline]{#1}}
\providecommand\calo[1]{{\cal O}\hspace{-0.2em}\left(#1\right)}
\providecommand{\cala}{{\cal A}}
\providecommand{\bbH}{\ensuremath{b\bar{b}H}}
\providecommand{\ttH}{\ensuremath{t\bar{t}H}}
\providecommand{\bbphi}{\ensuremath{b\bar{b}\phi}}
\providecommand{\yt}{\ensuremath{y_t}}
\providecommand{\ytsq}{\ensuremath{y_t^2}}
\providecommand{\yb}{\ensuremath{y_b}}
\providecommand{\ybsq}{\ensuremath{y_b^2}}
\providecommand{\ybyt}{\ensuremath{y_b\, y_t}}

\providecommand{\pt}{\ensuremath{p^T}}
\providecommand{\pth}{\ensuremath{p^T_H}}
\providecommand{\ptb}{\ensuremath{p^T_b}}
\providecommand{\ptj}{\ensuremath{p^T_j}}
\providecommand{\ptbone}{\ensuremath{p^T_{b_1}}}
\providecommand{\ptjone}{\ensuremath{p^T_{j_1}}}
\providecommand{\ptbtwo}{\ensuremath{p^T_{b_2}}}
\providecommand{\ptjtwo}{\ensuremath{p^T_{j_2}}}
\providecommand{\Mbb}{\ensuremath{M(bb)}}
\providecommand{\Rbb}{\ensuremath{\Delta R(bb)}}
\providecommand{\MBB}{\ensuremath{M(BB)}}
\providecommand{\RBB}{\ensuremath{\Delta R(BB)}}

\providecommand{\citere}[1]{ref.\,\cite{#1}}
\providecommand{\citeres}[1]{refs.\,\cite{#1}}

\providecommand{\eqn}[1]{eq.~(\ref{#1})}
\providecommand{\neqn}[1]{eqs.~(\ref{#1})}
\providecommand{\fig}[1]{figure~\ref{#1}}
\providecommand{\Fig}[1]{Figure~\ref{#1}}
\providecommand{\figs}[1]{figures~\ref{#1}}
\providecommand{\tab}[1]{table~\ref{#1}}
\providecommand{\sct}[1]{section~\ref{#1}}
\providecommand{\scts}[1]{sections~\ref{#1}}
\providecommand{\app}[1]{appendix~\ref{#1}}

\def\ft{t \bar t t \bar t}
\def\ttbar{t \bar t}
\def\fbinv{fb$^{-1}$}

\subsection{Four top production at the HL/HE-LHC}

The production of four top quarks is one of the rare processes in top quark physics that has large sensitivity to variety of
new physics effects (including effective field theory sensitivity and sensitivity to anomalous top-Higgs couplings), while at the same time
it is interesting in the Standard Model context as a complex QCD process. The cross section at 13 TeV is about fifty times smaller 
than $\ttbar\mathrm{H}$ production, with multiple precision calculations predicting values of $\sigma_{\ttbar\ttbar} = 9.2^{+2.9}_{-2.4} \rm{fb}$ (NLO) 
and  $\sigma_{\ttbar\ttbar} = 11.97^{+2.15}_{-2.51}\rm{fb}$ (NLO+EW)\cite{Alwall:2014hca,Frederix:2017wme,Bevilacqua:2012em}. 



ATLAS and CMS have published multiple papers where limits on $\ttbar\ttbar$ production were presented as 
SM-oriented searches \cite{Sirunyan:2017roi,Sirunyan:2017tep,Khachatryan:2014sca} and/or derived as a side product of 
searches for new physics, typically coming from searches for vector-like quarks or MSSM SUSY signatures \cite{Khachatryan:2016kod,Aad:2014pda,Aad:2015kqa,Aaboud:2018xuw,Aaboud:2018xpj}.

The production of $\ttbar\ttbar$  is a rare SM process that is expected to be discovered by future LHC runs, including HL-LHC and HE-LHC. The increase in collision energy is important for $\ttbar\ttbar$ production because the cross section is largely induced by gluons in the initial state, leading to a substantial improvement in the signal-to-background ratio when the collision energy of the LHC is increased.  Analyses looking for the production of $\ttbar\ttbar$ also are well-suited for interpretation in SMEFT\ \cite{AguilarSaavedra:2018nen}.

The $\ttbar\ttbar$ process has not yet been observed at the LHC. Once closer to observation, and considering the sensitivity
of $\ttbar\ttbar$ production to new physics scenarios in the top quark and scalar section, it is prudent to instead consider 
how accurately the cross section can be measured. Of course in the future analysis techniques are also expected
to improve, and dedicated analyses will surely improve this sensitivity, but this is beyond the scope of this study. It is however
important to keep in mind that such a study is less sensitive to systematic uncertainties on the background determination,
while being more sensitive to the signal modelling uncertainties and overall branching fraction and acceptance of the selection. 


\subsubsection[The complete NLO corrections to four-top production]%
{The complete NLO corrections to four-top
  production\footnote{Contributed by R. Frederix, D. Pagani and M. Zaro.}}
\label{sec:4top-smNLO}
\def\LO{{\rm LO}}
\def\NLO{{\rm NLO}}
\def\LNLO{{\rm (N)LO}}
\def\NLOQCD{{\rm NLO}_{\rm QCD}}
\def\NLOEW{{\rm NLO}_{\rm EW}}
\def\LOQCD{{\rm LO}_{\rm QCD}}
\def\ft{t \bar t t \bar t}

In this section the so-called ``complete''-NLO corrections to four-top production at the HE and HL-LHC is computed. Four-top 
production can proceed through different terms of order $\alpha_s^p \alpha^q$ with $p+q=4,\;5$ at LO and at NLO respectively. The term complete-NLO 
refers to computation of all terms with $p+q\le5$, which has been performed for the first time in Ref.~\cite{Frederix:2017wme}
by employing the newly-released version of \mg~\cite{Alwall:2014hca} capable of computing mixed QCD and electroweak corrections~\cite{Frederix:2018nkq}. 
Among the various contributions, the NLO QCD corrections ($p=5$, $q=0$) are also included, which have 
been known for some years~\cite{Bevilacqua:2012em, Maltoni:2015ena}. Despite that power-counting arguments suggest that the 
larger $q$ the more suppressed a contribution is, it has been shown in Ref.~\cite{Frederix:2017wme} that this is not the case for $\ft$ production. In fact, 
terms with up to two powers of $\alpha$ still contribute to several 10\%s with respect to the $\mathcal O(\alpha_s^4)$ LO contribution. One of the reasons why this happens is
because of the large Higgs-top Yukawa coupling; furthermore,
important cancellations appear among these terms, which may be spoiled by non-SM effects.\\
This short paragraph reports inclusive predictions for the HL and HE-LHC, with a centre-of-mass energy of respectively
14 TeV and 27 TeV. For differential distributions, the qualitative and quantitative behaviour is very similar to the predictions at 13 TeV reported 
in Ref.~\cite{Frederix:2017wme}. 
The same setup and notation of Ref.~\cite{Frederix:2017wme}, is used, where the interested reader can find more details
as well as predictions for 13 and 100 TeV.
\begin{table}[h]
\small
\providecommand{\arraystretch}{1.5}
\caption{Cross section for four-top production at the HL and HE-LHC, in various approximations, for $\mu=H_{\rm{T}}/4$. See Ref.~\cite{Frederix:2017wme} for details.}
\label{table:4t-xsect}
\begin{center}
\begin{tabular}{|c c c c c c|}
\hline
$\sigma[\textrm{fb}]$ & LO${}_{\textrm{QCD}}$ & $\LOQCD+\NLOQCD$ & LO & $\LO+\NLO$ & $\frac{\LO(+\NLO)}{\LOQCD(+\NLOQCD)}$\\
\hline\hline
 14 TeV &  $ 9.04 ^{+69\perc}_{-38\perc} $  &  $ 14.72 ^{+19\perc}_{-23\perc} $  &  $ 10.04 ^{+63\perc}_{-35\perc} $  &  $ 15.83 ^{+18\perc}_{-21\perc} $ & $ 1.11 $  ($ 1.08 $)\\
 27 TeV &  $ 81.87 ^{+62\perc}_{-36\perc} $  &  $ 135.19 ^{+19\perc}_{-21\perc} $  &  $ 91.10 ^{+56\perc}_{-33\perc} $  &  $ 143.93 ^{+17\perc}_{-20\perc} $  &  $ 1.11 $  ($ 1.06 $)\\
\hline
\end{tabular}
\end{center}
\end{table}
 \begin{table}[t]
\small

\caption{$\ft$: $\sigma_{\LNLO_i}/\sigma_{\LOQCD}$ ratios at 14 and 27 TeV, for different values of $\mu=\mu_R=\mu_F$. See Ref.~\cite{Frederix:2017wme} for details.}
\label{table:4t-orders}
\begin{center}
\begin{tabular}{|c | c c c | c c c| }
\hline
\multirow{2}{*}{$\delta[\%]$}
 & \multicolumn{3}{c|}{14 TeV}
 & \multicolumn{3}{c|}{27 TeV}\\
& $\mu= H_{\rm{T}}/8$ & $\mu=H_{\rm{T}}/4$ & $\mu = H_{\rm{T}}/2$     & $\mu= H_{\rm{T}}/8$ & $\mu=H_{\rm{T}}/4$ & $\mu = H_{\rm{T}}/2$ \\
\hline\hline 
 $ \rm{LO_2} $  &  $ -25.8 $  &  $ -28.1 $  &  $ -30.4 $      &  $ -23.6 $  &  $ -25.9 $  &  $ -28.2 $ \\ 
 $ \rm{LO_3} $  &  $ 32.5 $  &  $ 38.9 $  &  $ 45.8 $         &  $ 30.7 $  &  $ 37.0 $  &  $ 43.8 $  \\
 $ \rm{LO_4} $  &  $  0.2 $  &  $  0.3 $  &  $  0.4 $         &  $  0.1 $  &  $  0.2 $  &  $  0.2 $  \\
 $ \rm{LO_5} $  &  $  0.0 $  &  $  0.0 $  &  $  0.1 $         &  $  0.0 $  &  $  0.0 $  &  $  0.1 $  \\
\midrule 
 $ \rm{NLO_1} $  &  $ 14.7 $  &  $ 62.9 $  &  $ 103.3 $       &  $ 21.7 $  &  $ 65.1 $  &  $ 101.9 $  \\
 $ \rm{NLO_2} $  &  $  8.1 $  &  $ -3.5 $  &  $ -15.1 $       &  $  5.0 $  &  $ -4.4 $  &  $ -13.9 $  \\
 $ \rm{NLO_3} $  &  $ -10.0 $  &  $  1.8 $  &  $ 15.8 $       &  $ -7.8 $  &  $  1.6 $  &  $ 13.2 $  \\
 $ \rm{NLO_4} $  &  $  2.2 $  &  $  2.7 $  &  $  3.4 $        &  $  1.6 $  &  $  2.0 $  &  $  2.4 $  \\
 $ \rm{NLO_5} $  &  $  0.1 $  &  $  0.2 $  &  $  0.2 $        &  $  0.1 $  &  $  0.2 $  &  $  0.2 $  \\
 $ \rm{NLO_6} $  &  $ <0.1 $  &  $ <0.1 $  &  $ <0.1 $        &  $ <0.1 $  &  $ <0.1 $  &  $ <0.1 $  \\
\midrule 
$\NLO_2+\NLO_3$ & $-1.9$ & $-1.7$ & $0.7$                     & $-2.8$ & $-2.8$ & $-0.7$ \\
\bottomrule
\end{tabular}
\end{center}
\end{table}

Table~\ref{table:4t-xsect} reports the total-cross section for $\ft$ production in different approximations, and Table~\ref{table:4t-orders} the breakdown of the
different orders contributing at LO and NLO, as fraction of the $\mathcal O(\alpha_s^4)$ LO contribution, LO$_1$.
It is observed that the pattern of relative corrections is rather similar between 14 and 27 TeV. In particular, besides NLO$_1$ which is entirely of QCD origin, and thus displays a strong
dependence on the renormalisation and factorisation scales, such a feature is present also for NLO$_2$ and NLO$_3$, which witnesses the fact that they receive an important contribution through
QCD corrections from LO$_2$ and LO$_3$ respectively, on top of the electroweak corrections from LO$_1$ and LO$_2$. Furthermore, NLO$_2$ and NLO$_3$ tend to cancel each other almost exactly,
leading to a complete-NLO prediction well within the uncertainty band of the one at NLO QCD accuracy. Such a feature may be spoiled by effects beyond the Standard Model,
such as anomalous Higgs-top couplings. Thus, NLO corrections cannot be neglected when similar studies are performed, such
as those presented in Sec.~\ref{sec:4top-yt}.

\subsubsection{Prospect for experimental measurements}
\label{sec:4top-exp}

ATLAS has studied the potential to measure the Standard Model $\ttbar\ttbar$ cross section using 
3000~${\rm{fb}}^{-1}$ of HL-LHC data in the channel with several leptons~\cite{ATL-PHYS-PUB-2018-047}.
%
%
Events are selected if they contain at least two isolated leptons with the same charge or at least three isolated leptons.
At least six jets among which at least three are $b$-tagged are required.
In addition the scalar sum of the $p_{\rm{T}}$ of all selected jets and leptons ($H_{\rm{T}}$)
is requested to be $H_{\rm{T}} > 500$~GeV and the missing transverse momentum  \ETMISS $> 40$~GeV.
In order to extract the measured $\ttbar\ttbar$ cross section a fit is performed to the $H_{\rm{T}}$ distributions in several
signal regions according to the jets and $b$-jets multiplicities: at least 6 jets and exactly 
3 $b$-jets, or at least 6 jets and at least 4 $b$ jets.
These regions are further split in events with two same-charge leptons or with at least three leptons leading to 4 signal regions.

The background arises from $\ttbar V$ process, multiboson and $\ttbar H$ events as well as events with fake, non prompt or charge mis-identified leptons.
The rate of this difficult instrumental background is computed from the ATLAS 36~$\rm{fb}^{-1}$ analysis~\cite{Aaboud:2018xpj}
in the relevant regions with different lepton and $b$-tagged jet multiplicities.
The number of events selected in the different signal regions are shown in Fig.~\ref{fig:altas4top}.

The main sources of systematic uncertainties taken into account come from uncertainties on the fake lepton background and on the 
SM background and signal normalisations. 
A maximum-likelihood fit is performed simultaneously in the four signal regions to extract the $\ttbar \ttbar$ signal cross section
normalised to the prediction from the SM.
The impact of systematic uncertainties on the background expectations is described by nuisance parameters.
As a result of the fit, the expected uncertainty on the measured $\ttbar \ttbar$ cross section is found to be 11\%.
The systematic uncertainty that impacts the precision the most is uncertainty in the normalisation of the $\ttbar V$ and instrumental background in the region with at least 6 jets and exactly 
3 $b$-jets.
Overall the impact of the systematic uncertainties remain limited as a fit without systematic uncertainties
leads to a precision of 9\% on the extracted $\ttbar \ttbar$ cross section.

\begin{figure}[h!] \centering
 $\vcenter{\hbox{\includegraphics[width=0.5\textwidth]{\main/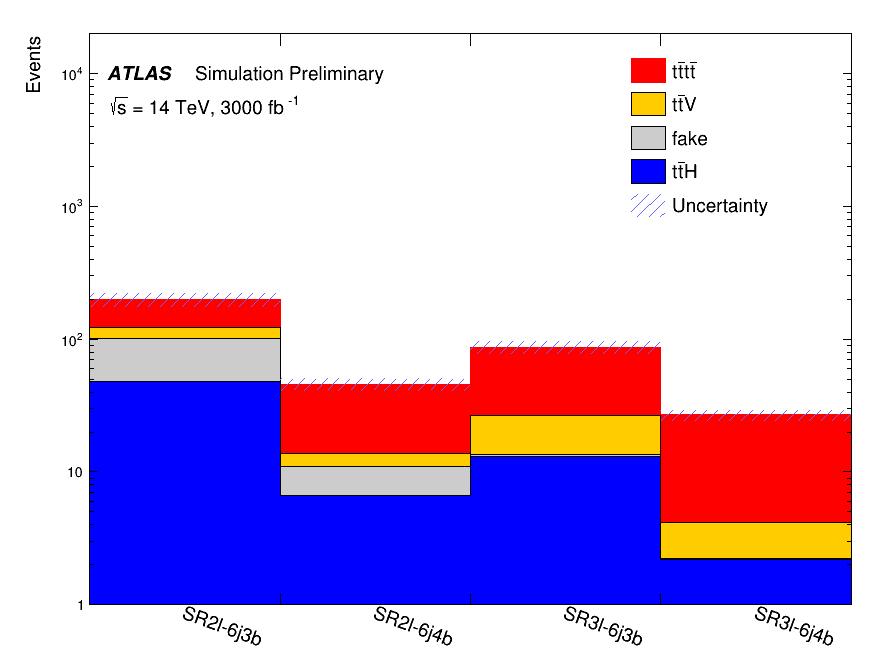}}}$
        \caption{\label{fig:altas4top}
Event yields of signal and background processes in the different signal regions used to extract the $\ttbar \ttbar$ cross section for an integrated luminosity of 3000~$fb^{-1}$~\cite{ATL-PHYS-PUB-2018-047}.}
\end{figure}

\par

The most sensitive result of the CMS collaboration on the Standard Model $\ttbar\ttbar$ process~\cite{Sirunyan:2017roi}
is based on an integrated luminosity of 35.9~\fbinv and a centre-of-mass energy of 13~TeV,
and relies on events with 2 same-sign leptons or 3 or more leptons. This Run-2 analysis sets an 
expected 95\% CL upper limit on the $\ttbar\ttbar$ production cross section of $20.8^{+11.2}_{-6.9}$~fb,
and an expected significance (based on a cross section of 9.2~fb) of 1.0 standard deviations above the background-only hypothesis.

The result of Ref.~\cite{Sirunyan:2017roi} is used to derive extrapolations for HL and HE-LHC, which are described in Ref.~\cite{CMS-PAS-FTR-18-031} and summarized below.
The extrapolations rely on a simple rescaling of the signal and background cross sections,
and make different assumptions on the systematic uncertainties. First, the statistical uncertainties are considered, 
then the same systematic uncertainties as the Run-2 published result are used,
and finally these systematics are progressively reduced as a function of the integrated luminosity. 

The expected sensitivity on the $\ttbar \ttbar$ cross section for different HL and HE-LHC scenarios is listed in Table~\ref{tab:hllhcsensitivity}.
Based on these results, evidence for $\ttbar \ttbar$ production will become possible with around 300 fb$^{-1}$ of HL-LHC data at $\sqrt{s}=14$~TeV, 
at which point the statistical uncertainty on the measured cross section will be of the order of 30\% and  the measurement will have a total uncertainty of around 33-43\%, 
depending on the systematic uncertainty scenario considered. For larger datasets at HL-LHC, all scenarios considered become dominated by systematic uncertainties. 
With 3 ${\rm ab^{-1}}$ the cross section can be constrained to 9\% statistical uncertainty, and the total uncertainty of a measurement ranges between 18\% and 28\% 
depending on the considered systematic uncertainties. At HE-LHC the $\ttbar \ttbar$ cross section is expected to be constrained to within a 1-2\% statistical uncertainty,
and the systematic uncertainties also decrease due to the improved signal to background ratio at $\sqrt{s}=27$~TeV.
Future changes to the analysis strategy might allow improvements based on optimizing the interplay between statistical and systematic uncertainties.

The $\ttbar\ttbar$ cross section measurements can also be used to constrain the Wilson coefficients of the $\mathcal{O}_{R}$, $\mathcal{O}_{L}^{(1)}$, $\mathcal{O}_{B}^{(1)}$ 
and $\mathcal{O}_{B}^{(8)}$ dimension-6 operators of the Effective-Field-Theory (EFT) Lagrangian. These constraints are included in Ref.~\cite{CMS-PAS-FTR-18-031} 
for both HL-LHC and HE-LHC scenarios.

\begin{table*}[htb]
\centering
\caption{Expected sensitivity for the production cross section of $\ttbar \ttbar$ production, in percent, at 68\% confidence level. The fractional uncertainty on the cross section signal strength is given for various LHC upgrade scenarios. Cross sections are corrected for the changes expected by $\sqrt{s}$. For the $15~{\rm ab^{-1}}$ 27~TeV scenario, the systematic uncertainty extrapolation is no longer valid, so only the statistical uncertainty is provided.
\label{tab:hllhcsensitivity} }
\begin{tabular}{|cccccc|}
\hline
 Int. Luminosity & $\sqrt{s}$ & {\em Stat. only} (\%) & {\em Run-2} (\%) &  {\em YR18} (\%) & {\em YR18+} (\%) \\
 \hline\hline
  $300~{\rm fb^{-1}}$ & 14 TeV   &   ${+30},{-28}$  &  ${+43},{-39}$ &  ${+36},{-34}$ &  ${+36},{-33}$  \\ 
 $3~{\rm ab^{-1}}$ & 14 TeV  &   $\pm9$  &  ${+28},{-24}$ &  ${+20},{-19}$ &  $\pm18$   \\ 
 $3~{\rm ab^{-1}}$ & 27 TeV  &   $\pm2$ &  ${+15},{-12}$ &  ${+9},{-8}$ &  ${+8},{-7}$  \\ 
   $15~{\rm ab^{-1}}$ & 27 TeV  &   $\pm 1$ &   & & \\ 
   \hline
\end{tabular}
\end{table*}

\subsection{Four top quarks as a probe of new physics}

Heavy coloured resonances decaying into a pair of top quarks are present in many
new physics theories~\cite{Fox:2002bu,Burdman:2006gy,Kribs:2007ac,%
Benakli:2008pg,Kilic:2009mi}. Such particles are typically pair-produced at
large rate and their decay then leads to a substantial enhancement of four-top
production. Current bounds on such a setup are driven by a recent CMS
analysis of four-top events~\cite{Sirunyan:2017roi}, using 35.9~fb$^{-1}$ of LHC
collisions at a centre-of-mass energy of 13~TeV. Those bounds however are expected to
strongly improve in the upcoming years, as illustrated in following contributions,
with the example of a scalar colour-octet field $O$, traditionally
dubbed a sgluon.

\subsubsection[Limits on pseudoscalar colour-octets]%
{Limits on pseudoscalar colour-octets\footnote{Contributed by
    B. Fuks, L. Darm\'e and M.D.~Goodsell.}}
{

\providecommand{\scr}[1]{\ensuremath{\mathcal{#1}}}
\providecommand{\amc}{{\sc MadGraph5}\_a{\sc MC@NLO}}
\providecommand{\fr}{{\sc Feyn\-Rules}}
\providecommand{\fa}{{\sc Feyn\-Arts}}
\providecommand{\nloct}{{\sc NloCT}}
\providecommand{\mloop}{{\sc MadLoop}}
\providecommand{\mspin}{{\sc MadSpin}}
\providecommand{\mw}{{\sc MadWidth}}
\providecommand{\ma}{{\sc MadAnalysis~5}}
\providecommand{\py}{{\sc Pythia~8}}
\providecommand{\del}{{\sc Delphes~3}}
\providecommand{\fj}{{\sc FastJet}}

\def\so{O_R}
\def\pso{O_I}
\def\mpso{M_{O_I}}
\def\mso{M_{O_R}}
\def\ggo{g_8}
\def\gqL{{y_{8}^L}}
\def\gqR{{y_{8}^R}}

The effective Lagrangian describing the couplings of such a sgluon to the
Standard Model is given by~\cite{Calvet:2012rk}
\begin{align}
  \scr{L} ~\supset~ \! \ggo d_{abc} O^a G_{\mu \nu}^b G^{\mu \nu c}\! +
    \tilde{g}_8 d_{abc} O^a G_{\mu \nu}^b \tilde{G}^{\mu \nu c} \!+\! \Big\{
    \bar{q} \Big[ {\bf \gqL} P_L + {\bf \gqR} P_R \Big] O^a T_a q + {\rm  h.c.}
    \Big\} \ ,
\end{align}
where $T^a$ and $d_{abc}$ are respectively the fundamental representation
matrices and symmetric structure constants of $SU(3)$. Moreover, flavour and
fundamental colour indices are understood for simplicity and the gluon
field strength (dual field strength) tensor is denoted by $G_{\mu \nu}^a$
($\tilde{G}_{\mu \nu}^{a}$). The focus here is on the case of a pseudoscalar sgluon with
$g_8 = 0$ and purely imaginary ${\bf y_8}$ matrices, and it is additionally enforced
$\tilde g_8=0$ as in Dirac gaugino supersymmetric scenarios. A non-vanishing
$\tilde g_8$ coupling would however weaken the bounds by reducing the sgluon
branching ratio into top quarks. In order to assess the impact of future search
on the potential discovery of a sgluon, recasting strategy is followed here, as
detailed in Ref.~\cite{Darme:2018dvz}. An NLO UFO
module~\cite{Degrande:2011ua} is generated through \fr~\cite{Alloul:2013bka},
\nloct~\cite{Degrande:2014vpa} and \fa~\cite{Hahn:2000kx} and it is used to generate
events within the \mg framework~\cite{Alwall:2014hca}, the hard-scattering
matrix elements being convolved with the NNPDF3.0 NLO set of parton
densities~\cite{Ball:2014uwa} and the sgluon decays being achieved with
\mspin~\cite{Artoisenet:2012st} and \mw~\cite{Alwall:2014bza}. Parton showering
and hadronisation are performed by \py~\cite{Sjostrand:2014zea} and the response of the CMS detector is simulated with \del~\cite{deFavereau:2013fsa} and
\fj~\cite{Cacciari:2011ma}. Finally, the four-top selection strategy
of CMS~\cite{Sirunyan:2017roi} is mimicked by using the \ma~\cite{Conte:2012fm,%
Conte:2014zja,Dumont:2014tja} framework.

The best signal region (SR6) from Ref.~\cite{Sirunyan:2017roi}, in terms of
constraints, focuses on a topology featuring one pair of same-sign leptons, at
least 4 $b$-jets and at least 5 hard jets. It is shown the observed and expected
limits on the pseudoscalar octet cross section times the corresponding
branching ration into four top quarks in Fig.~\ref{fig:CS} (left). While the
analysis of Ref.~\cite{Sirunyan:2017roi} targeted a Standard Model four-top
signal, future studies adopting a new physics signal selection strategy relying
on the large differences in the final-state kinematics could be more adapted and
lead to sizeable improvement in the reach~\cite{Darme:2018dvz}.

\begin{figure}[h] \centering
 $\vcenter{\hbox{\includegraphics[width=0.5\textwidth]{\main/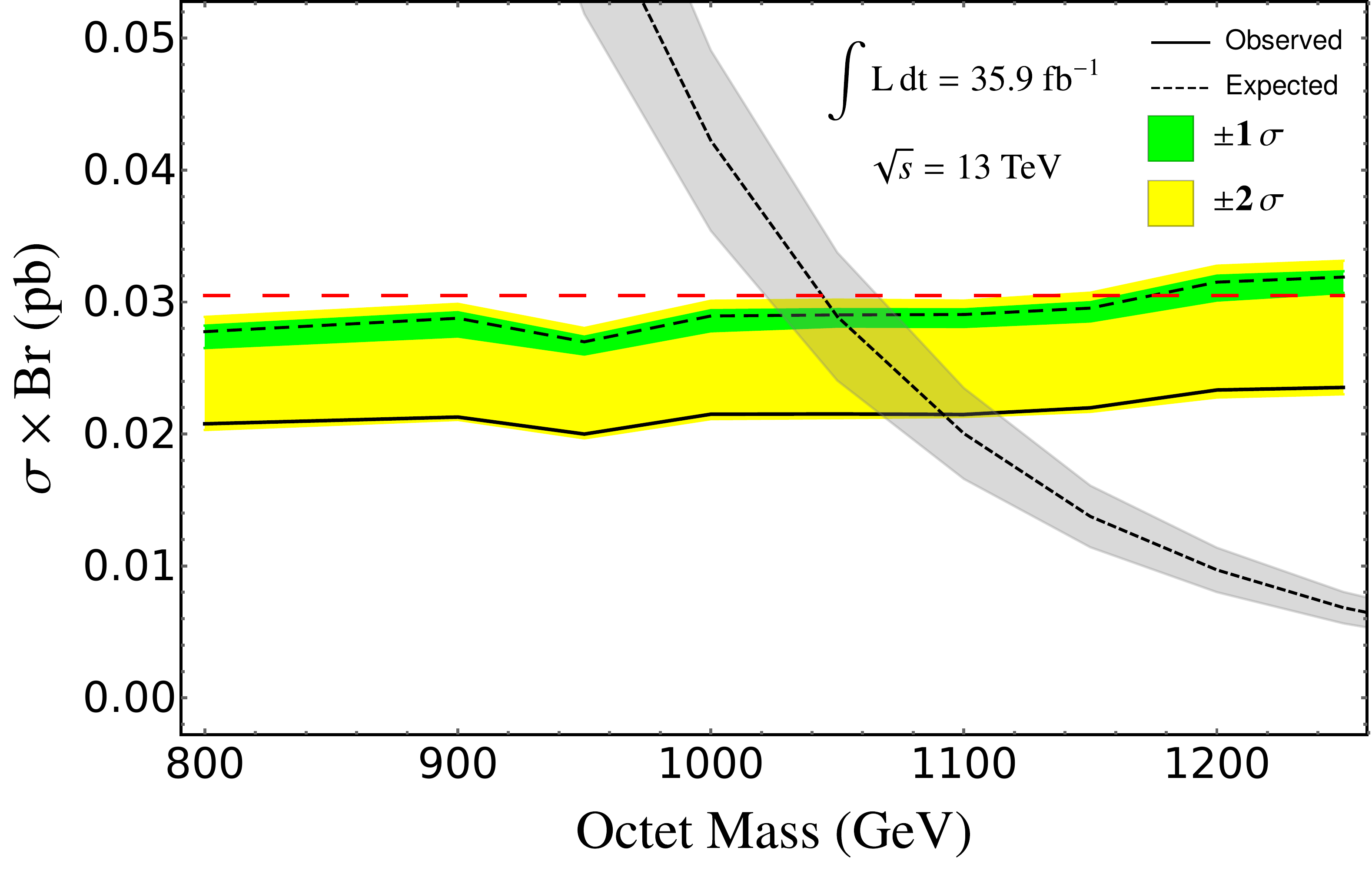}}}$ \begin{tabular}{c}\includegraphics[width=0.35\textwidth]{\main/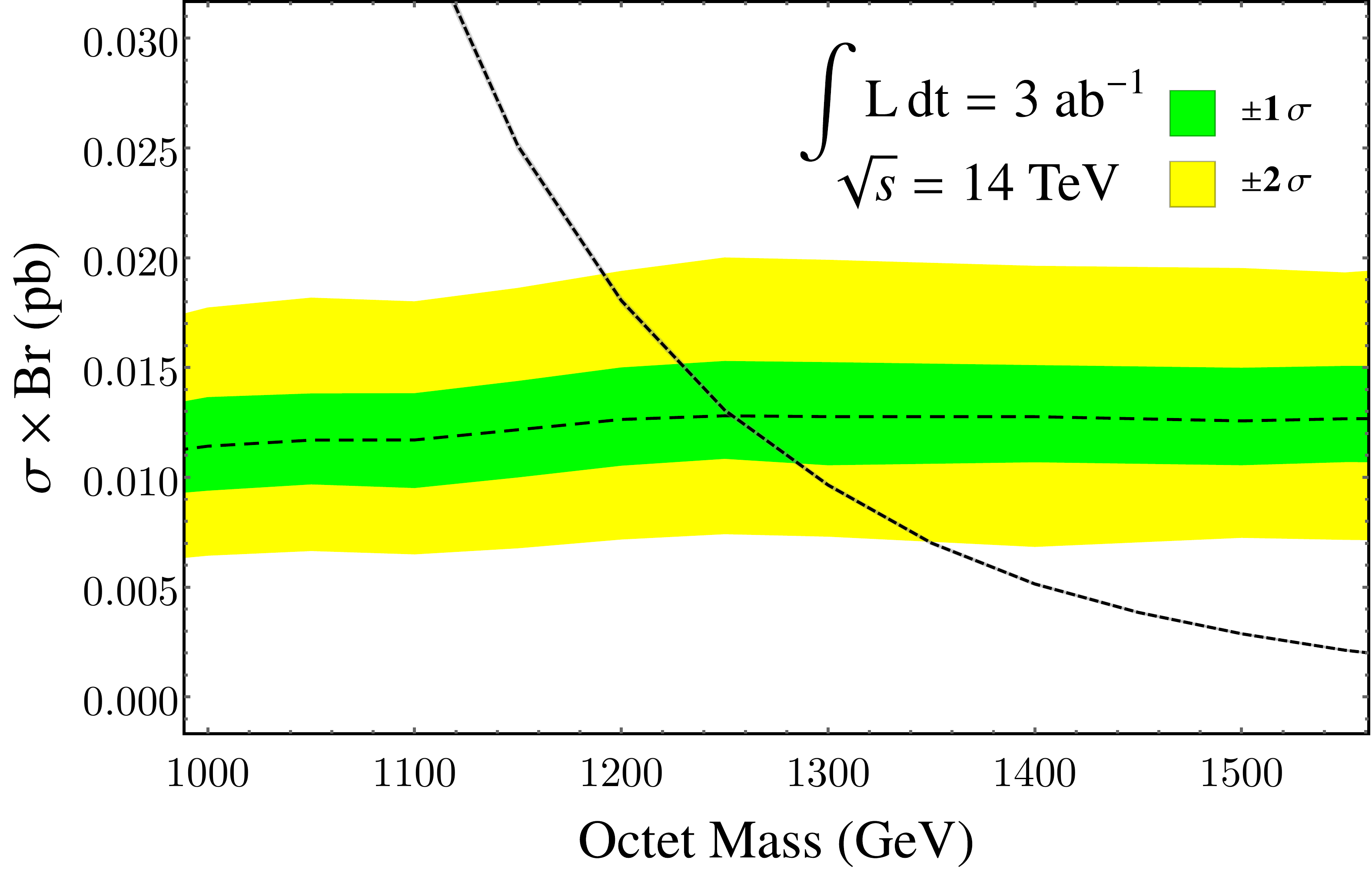} \\\includegraphics[width=0.35\textwidth]{\main/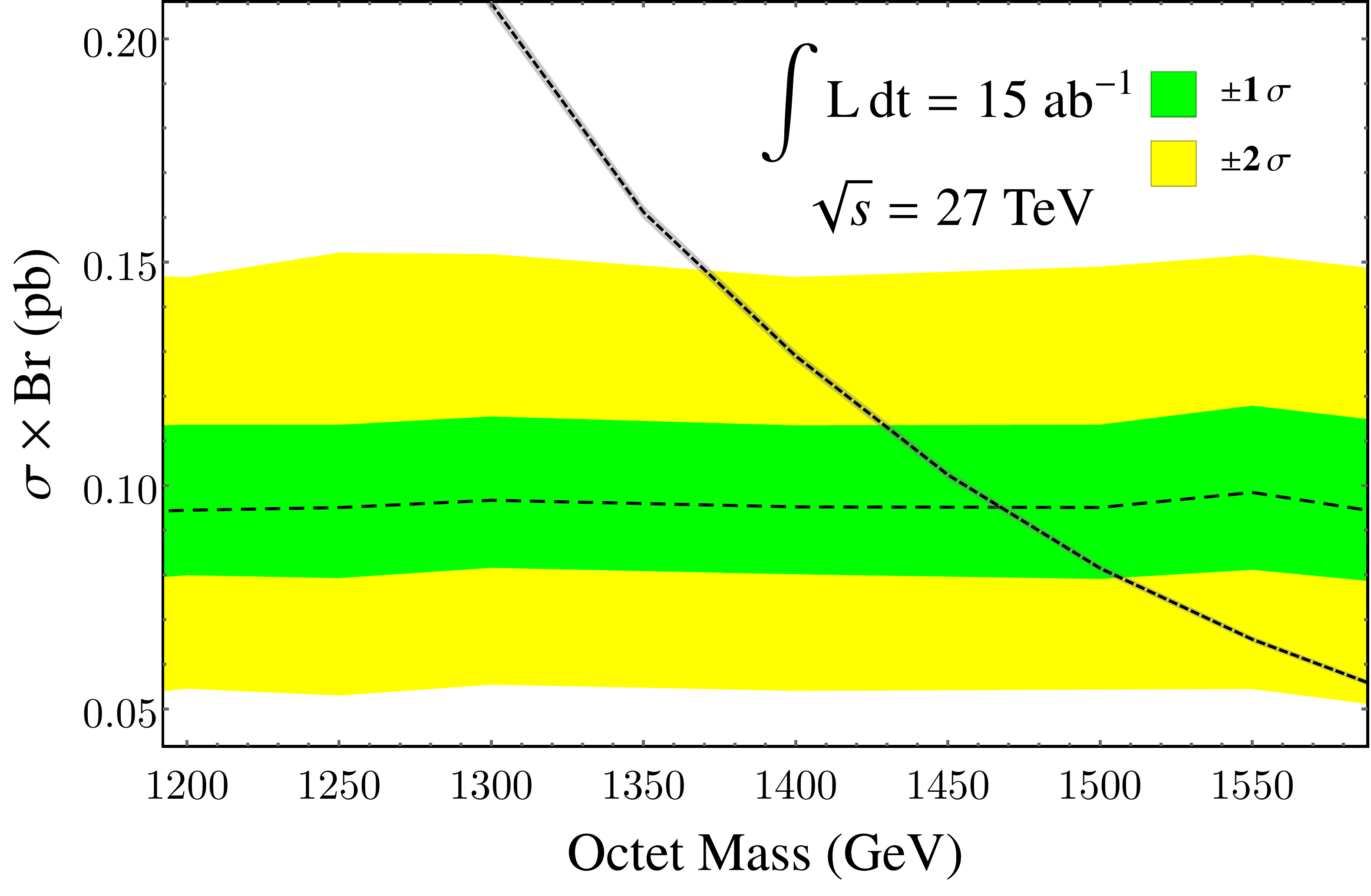} \end{tabular}
 \caption{Left: Expected (dashed) and observed (solid) pseudoscalar sgluon
   pair-production cross section excluded at the 95\%
   confidence level when making use of the results associated with the SR6
   region of the four-top CMS analysis of Ref.~\cite{Sirunyan:2017roi}.
   Theoretical predictions for the signal rate are indicated by the grey band. Right: expected limits for proton-proton collisions at centre-of-mass energies of
  $14$ (top) and $27$ (bottom) TeV, with the sgluon cross-section as the fine dotted line. }
 \label{fig:CS}
\end{figure}

To calculate the projected sensitivity of the HL/HE-LHC, it is assumed that the
current selection efficiencies at $13$ TeV are similar to the future ones, and
moreover rescale the four-top and other SM backgrounds by the appropriate
partonic luminosities relative to those at 13~TeV. The rescaling factor for the
non-four-top SM background is taken to be the largest ratio of the $ttZ$ and
$ttW$ background component, using the projected cross-sections reported in Sec.~\ref{sec:eft}. Factors of 1.3 and 12 are obtained for the 14 and 27 TeV cases,
respectively. According to Sec.~\ref{sec:4top-smNLO}, the four-top cross section is then set to 15.83~fb and 144~fb at
14 and 27~TeV, respectively, recalling that the 13~TeV cross section is of
11.97~fb. The results for the projected mass limits are then given in the
following Table~\ref{tab:octet},  together with the $13$ TeV value for reference.

\begin{table}[!h]\centering
\caption{Results for the projected mass limits on pseudo-scalar color octets. \label{tab:octet}}

\begin{tabular}{|cccc|} 
\hline
& $35.9\, \mathrm{fb}^{-1}$, $13$ TeV & $3\, \mathrm{ab}^{-1}$, $14$ TeV  & $15\, \mathrm{ab}^{-1}$, $27$ TeV  \\
\hline\hline
Octet mass (GeV)  & 1060 & 1260   & 1470  \\
\hline 
\end{tabular}
\end{table}
}

\subsubsection[Limits on top-Higgs interaction from multi-top final state]%
{Limits on top-Higgs interaction from multi-top final
  state\footnote{Contributed by Qing-Hong Cao, Shao-Long Chen and Yandong Liu.}}
\providecommand{\met}{\not{\!\!{\rm E}}_{\rm T}}

\label{sec:4top-yt}
Four top-quark ($t\bar{t}t\bar{t}$) production provides a powerful tool to probe the Top-quark Yukawa coupling ($y_t$)~\cite{Cao:2016wib}. In the SM the $t\bar{t}t\bar{t}$ production can be induced either by the pure gauge interaction (involving the gluon, Z-boson or photon in the intermediate state)~\cite{Barger:1991vn} or by the Higgs boson mediation~\cite{Cao:2016wib}. Defining the general top-Higgs coupling as $y_t\equiv \kappa_t y_t^{\rm SM}$ with $y_t^{\rm SM}$ the top-Yukawa coupling in the SM, the leading-order cross section of $t\bar{t}t\bar{t}$ production can be parameterised as
\begin{equation}
\sigma(t\bar{t}t\bar{t}) = \sigma(t\bar{t}t\bar{t})^{\rm SM}_{g/Z/\gamma} + \kappa_t^2 \sigma(t\bar{t}t\bar{t})^{\rm SM}_{\rm int} + \kappa_t^4 \sigma(t\bar{t}t\bar{t})^{\rm SM}_H,
\end{equation}
where $\sigma(t\bar{t}t\bar{t})^{\rm SM}_{g/Z,\gamma,~H,~{\rm int}}$ denotes the cross section induced by the pure gauge interaction, Higgs-boson mediation and the interfere effect, respectively. Note that $\sigma_{H, \rm{int}}^{\rm SM}$ is comparable to $\sigma(t\bar{t}t\bar{t})^{\rm SM}_{g/Z,\gamma}$ as $y_t^{\rm SM}\sim 1$ in the SM. For example, the leading order calculation with the renormalization/factorization scale ($\mu$) fixed to the dynamics scale~\cite{Alwall:2014hca} yields
\begin{eqnarray}
{\rm HL-LHC}~(\sqrt{s}=14~{\rm TeV}) &~:~&  \sigma(t\bar{t}t\bar{t})= ~13.14 - ~~2.01\kappa_t^2 + ~1.52\kappa_t^4 ~~[{\rm fb}]\nonumber \\
{\rm HE-LHC}~(\sqrt{s}=27~{\rm TeV}) &~:~& \sigma(t\bar{t}t\bar{t})= 115.10 - 15.57\kappa_t^2 + 11.73\kappa_t^4 ~~[{\rm fb}]
\end{eqnarray}
Clearly, $\sigma(t\bar{t}t\bar{t})$ depends only on $\kappa_t$ such that it directly probes $y_t$ without any assumption on Higgs boson. The above values suffer from a large $\mu$ dependence; when varying the scale by a factor 2, the cross section varies by about $50\%$. It is crucial to take the full next-to-leading order corrections~\cite{Bevilacqua:2012em,Frederix:2017wme} into account to get a realistic simulation. Here, the tree level events are generated and the cross section rescaled to the NLO.

A special signature of the $t\bar{t}t\bar{t}$ events is the same-sign charged leptons (SSL) from the two same-sign top quarks.  The other two top quarks are demanded to decay hadronically to maximize the event rate. Therefore, the topology of the signal event consists of two same-sign charged leptons, four $b$-quarks, four light-flavor quarks, and two invisible neutrinos. In practice it is challenging to identify four $b$-jets. Instead, it is required for at least 5 (6) jets to be tagged and three of them to be identified as $b$-jets at the HL(HE)-LHC, respectively. The two invisible neutrinos appear as a missing transverse momentum \ETMISS in the detector. The SM backgrounds contain $t\bar{t}+X$, $W^\pm W^\pm jj$ and $W^\pm W^\pm jj$ processes. See Ref.~\cite{Cao:2016wib} for the details of those kinematic cuts used to disentangle the $t\bar{t}t\bar{t}$ signal from the huge backgrounds. It is demanded that  \ETMISS $>100~{\rm GeV}$ at the HL-LHC and \ETMISS $>150~{\rm GeV}$ at the HE-LHC. 
Table~\ref{tbl:tttt} displays the numbers of signal and background events after applying the kinematics cuts listed in each row sequentially. In Table~\ref{tbl:tttt},
at the HL-LHC the $t\bar{t}t\bar{t}$ production cross section is multiplied by a constant $K$-factor of 1.27 with uncertainty $27\%$ (see Ref. \cite{Bevilacqua:2012em}), while at the HE-LHC
the cross section is rescaled to NLO order of $143.93^{+17\%}_{-20\%}$ fb (see Table~\ref{table:4t-xsect} in Sec.~\ref{sec:4top-smNLO}).

The MC simulation shows that the $t\bar{t}t\bar{t}$ production ($\kappa_t=1$) can be discovery at a 5$\sigma$ confidence level with an integrated luminosity of $2075 
~{\rm fb}^{-1}$ at the HL-LHC and  $146
~{\rm fb}^{-1} $ at the HE-LHC, respectively. The event rate is not enough for measuring $y_t$ precisely at the HL/HE-LHC but it is good for bounding $y_t$; for example, a direct bound $\kappa_t \le 1.41~ [1.37, 1.47]$ is obtained at the HL-LHC and $\kappa_t \le 1.15 ~[1.12, 1.17]~(1.12 ~[1.10, 1.13], ~1.10 ~[1.08, 1.12])$ with an luminosity of $10~(20, ~30)~{\rm ab}^{-1}$ at the HE-LHC, respectively.

A few words of care on the interpretation of 
results from this study are however necessary:
as it has been discussed in Sec.~\ref{sec:4top-smNLO}, the complete-NLO corrections to $t\bar{t}t\bar{t}$ are large and can involve terms 
proportional to $y_t^3$, $y_t^5$ and $y_t^6$ (on top of $y_t^2$ and $y_t^4$ already present at LO). However, since in such corrections $y_t$ is
renormalised, an extension of our study will not be immediately possible at NLO.

\begin{table}[h!]
\caption{The numbers of signal and background events at the HL-LHC with an integrated luminosity of $300~{\rm fb}^{-1}$ (left) and at the HE-LHC with an integrated luminosity of $1~{\rm ab}^{-1}$. The cuts listed in the row are applied sequentially~\cite{Cao:2016wib}.}
\label{tbl:tttt}
\centering
\begin{tabular}{|l|r|r|r|r|r|r|} \hline
HL-LHC&Basic &SSL&Jets&$\met$ &$m_{\rm{T}}$&$H_{\rm{T}}$ 
 \\ \hline \hline
 $\bar{t}t\bar{t}t_H$ &577.22&9.82& 4.68&2.43&1.33& 1.21 
  \\ \hline
 $\bar{t}t\bar{t}t_{g/Z/\gamma}$&5006.34  &78.15&37.02&19.25 &11.09&10.16 
  \\ \hline
 $\bar{t}t\bar{t}t_{\rm int}$&-764.67&-12.79&-6.19&-3.23&-1.93&-1.77 
  \\ \hline\hline
 $\bar{t}t$&$2.5\times 10^8$ & 28802.4 & 44.1 & 18.9 &0 &0 
  \\ \hline
 $\bar{t}tW^+$& 32670& 2359.5&36.9& 17.7& 12.3& 8.7 
  \\ \hline
 $\bar{t}tW^-$& 16758 & 1397.1&49.5 & 9.9& 4.5 &  4.5 
  \\ \hline
 $\bar{t}tZ$ &24516& 2309.4& 20.1 & 10.8&10.8& 9.3 
  \\ \hline
 $W^\pm W^\pm j j$ &4187.7& 1147.5& 0.11 &0&0&0 
  \\ \hline
\end{tabular}
\\[5mm]
%
\begin{tabular}{|l|r|r|r|r|r|r|} \hline
HE-LHC &Basic &SSL&Jets&$\met$ &$m_{\rm{T}}$&$H_{\rm{T}}$ \\ \hline \hline
 $\bar{t}t\bar{t}t_H$ &15174.4& 260.09&  84.61& 27.92& 15.42& 15.17 \\ \hline
   $\bar{t}t\bar{t}t_{g/Z/\gamma}$&148898.& 2421.08& 814.77& 268.02& 168.55& 166.77 \\ \hline
   $\bar{t}t\bar{t}t_{\rm int}$&-20141.9& -347.81& -117.95& -36.17& -20.14& -19.66 \\ \hline\hline
   $\bar{t}t$&$3.3\times 10^7$ & 130207 & 291.9 & 0 &0 &0 \\ \hline
   $\bar{t}tW^+$& $1.3\times 10^6$& 11488.5&171.0& 39.6& 27.1& 27.1 \\ \hline
   $\bar{t}tW^-$& $7.6\times 10^5$ & 7387.1&99.5 & 19.9& 9.9 &  9.9 \\ \hline
   $\bar{t}tZ$ &$3.9\times 10^6$& 20748.7& 507.2 & 129.7&70.8& 70.8 \\ \hline
   $W^\pm W^\pm j j$ &888700& 7947.0& 4.7 &3.5&0&0 \\ \hline
\end{tabular}
\end{table}

\subsubsection[Constraining four-fermion operators in the EFT]%
{Constraining four-fermion operators in the EFT\footnote{Contributed by Cen Zhang.}}

The four-top total cross section measurement can be interpreted within the SMEFT framework~\cite{Zhang:2017mls}\footnote{This interpretation is also present in Ref.\cite{CMS-PAS-FTR-18-031}.}. 
Following the notation in
Refs.~\cite{Zhang:2017mls} and~\cite{AguilarSaavedra:2018nen}, the relevant operators
consist of four independent four-top-quark operator coefficients, $\tilde{C}_{\text{tt}} ,\
\tilde{C}_{\text{QQ}}^{(+)} ,\  \tilde{C}_{\text{Qt}}^{(1)} ,\
\tilde{C}_{\text{Qt}}^{(8)}$, 
and fourteen independent two-light-two-top-quark ($qqtt$) operator
coefficients, $\tilde{C}_{\text{td}}^{(8)}$, $  \tilde{C}_{\text{td}}^{(1)}$,
$\tilde{C}_{\text{Qd}}^{(8)}$, $ \tilde{C}_{\text{Qd}}^{(1)}$, 
$\tilde{C}_{\text{tu}}^{(8)}$, $\tilde{C}_{\text{tu}}^{(1)}$, 
$\tilde{C}_{\text{Qu}}^{(8)}$, $ \tilde{C}_{\text{Qu}}^{(1)}$, 
$\tilde{C}_{\text{Qq}}^{(8,1)}$, $ \tilde{C}_{\text{Qq}}^{(1,1)}$, 
$\tilde{C}_{\text{Qq}}^{(8,3)}$, $\tilde{C}_{\text{Qq}}^{(1,3)}$, 
$\tilde{C}_{\text{tq}}^{(8)}$, $ \tilde{C}_{\text{tq}}^{(1)}$. Here $\tilde
C_i\equiv C_i/\Lambda^2$. $O_{tG}$ is relevant but better constrained by other processes.

To estimate the projected limits on these coefficients, a few simple
assumptions are made: 1) the effective operators do not significantly change the
distribution of events, so the sensitivity mainly comes from inclusive
measurements; 2) a kinematic cut $M_{cut}$ of a few TeV can be applied to the
total mass of the four tops to make sure the SMEFT can be matched to BSM models
with scales larger than this energy (i.e.~following Ref.~\cite{Contino:2016jqw}); and
3) $M_{cut}$ does not significantly change the projected sensitivity on cross
section measurements. By combining the expected experimental sensitivity discussed in Sec.~\ref{sec:4top-exp}
and the theoretical predictions presented in Sec.~\ref{sec:4top-smNLO}
it is estimated that the total cross section can be determined with an uncertainty of
$102\%$, $58\%$, and $40\%$, at 95\% CL level, for the 13, 14 and 27 TeV runs respectively. The corresponding
integrated luminosities are 300 fb$^{-1}$, 3 ab$^{-1}$ and 15 ab$^{-1}$.

For illustration, Fig.~\ref{fig:eftlimitsinfourtop} shows the signal
strength dependence on two operator coefficients: one four-top coefficient
(left) and one $qqtt$ coefficient (right), assuming a 3 TeV $M_{cut}$.  The
cross section becomes more sensitive to the four-top operator coefficient at
larger energies.  Together with smaller uncertainties, the limit on this
coefficient is significantly improved with the 27 TeV run.  On the other hand,
the cross section becomes less sensitive to the $qqtt$ operator coefficient as
the energy increases.  The limits are thus not very much affected by energy.
Table~\ref{tab:eftlimitsinfourtop} presents individual limits on all 18 operator
coefficients, assuming $M_{cut}=3$ TeV. 

\begin{figure}[h!]
	\hfill
	\includegraphics[width=.45\textwidth]{\main/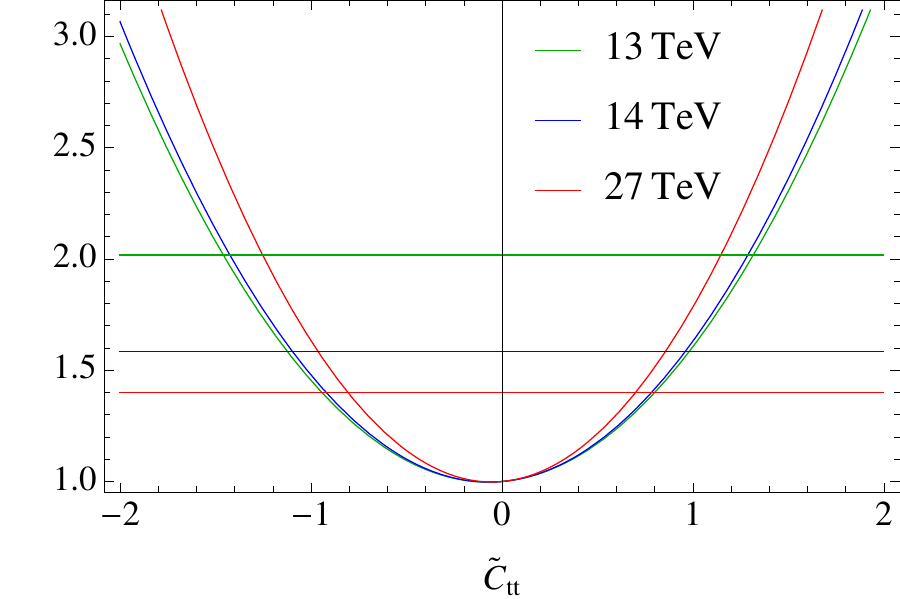}
	\hfill
	\includegraphics[width=.45\textwidth]{\main/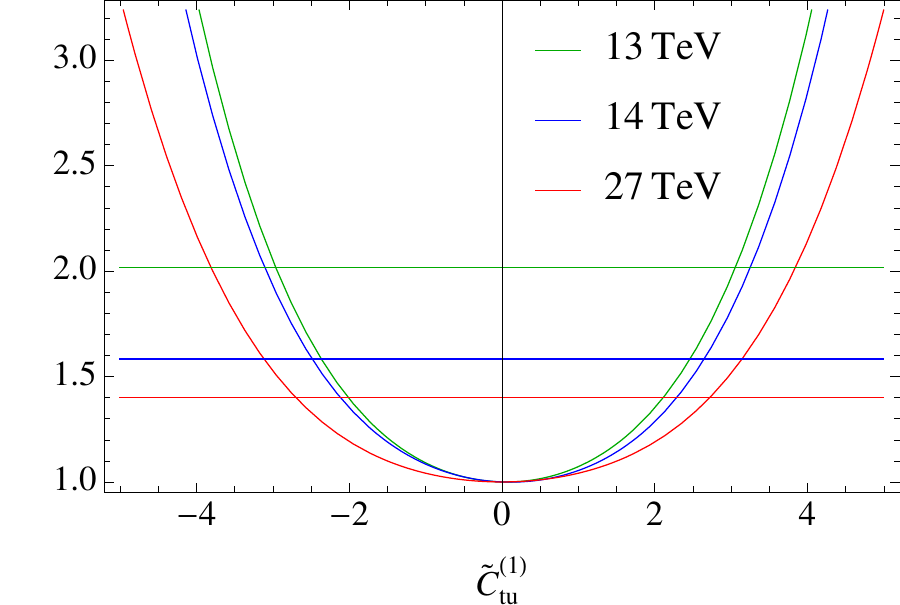}
	\hspace*{\fill}
	\caption{\label{fig:eftlimitsinfourtop}
	Four-top signal strength as a function of operator coefficients, $\tilde C_{tt}$ (left)
	and $\tilde C^{(1)}_{tu}$ (right).  Horizontal lines represent the expected measurements
	at each energy. $M_{cut}=3$ TeV is applied.}
\end{figure}

\begin{table}[h!]
\centering
\caption{\label{tab:eftlimitsinfourtop}
Limits on 14 $qqtt$ operator coefficients and 4 four-top operator coefficients,
expected at the 13, 14 and 27 TeV scenarios, at the 95\% CL level.}
\begin{tabular}{|c|ccc|}
	\hline
        $ \text{} $   &    13 \text{TeV} &  14 \text{TeV} &  27 \text{TeV}\\   
        \hline\hline
        $ \tilde{C}_{\text{td}}^{(8)} $   &  \text{[-9.8, 6.4]}  &  \text{[-8.8, 5.4]}  &  \text{[-6.6, 5.4]} \\   
        $ \tilde{C}_{\text{td}}^{(1)} $   &  \text{[-3.9, 4.1]}  &  \text{[-3.3, 3.4]}  &  \text{[-3.3, 3.3]} \\    
        $ \tilde{C}_{\text{Qd}}^{(8)} $   &  \text{[-9.6, 6.2]}  &  \text{[-8.8, 5.2]}  &  \text{[-7.6, 5.2]} \\   
        $ \tilde{C}_{\text{Qd}}^{(1)} $   &  \text{[-4., 4.]}  &  \text{[-3.3, 3.3]}  &  \text{[-3.4, 3.3]} \\   
        $ \tilde{C}_{\text{tu}}^{(8)} $   &    \text{[-8.2, 4.8]}  &    \text{[-6.4, 4.3]}  &    \text{[-9.6, 4.5]} \\    
        $ \tilde{C}_{\text{tu}}^{(1)} $   &  \text{[-3., 3.1]}  &  \text{[-2.5, 2.6]}  &  \text{[-2.7, 2.7]} \\   
        $ \tilde{C}_{\text{Qu}}^{(8)} $   &  \text{[-7.8, 4.6]}  &  \text{[-7.8, 4.]}  &  \text{[-5.8, 4.2]} \\   
        $ \tilde{C}_{\text{Qu}}^{(1)} $   &  \text{[-3., 3.]}  &  \text{[-2.6, 2.6]}  &  \text{[-2.7, 2.7]} \\    
        $ \tilde{C}_{\text{Qq}}^{(8,1)} $  &  \text{[-7.5,   4.2]}  &  \text{[-6.,   3.6]}  &  \text{[-6.5,   3.7]} \\ 
        $ \tilde{C}_{\text{Qq}}^{(1,1)} $   &  \text{[-2.5, 2.7]}  &  \text{[-2.1, 2.3]}  &  \text{[-2.2, 2.3]} \\  
        $ \tilde{C}_{\text{Qq}}^{(8,3)} $   &  \text{[-5.8, 4.8]}  &  \text{[-4.7, 4.2]}  &  \text{[-5.4, 4.]} \\    
        $ \tilde{C}_{\text{Qq}}^{(1,3)} $   &  \text{[-2.6, 2.6]}  &  \text{[-2.1, 2.2]}  &  \text{[-2.2, 2.2]} \\  
        $ \tilde{C}_{\text{tq}}^{(8)}   $   &  \text{[-7.1, 3.9]}  &  \text{[-6.9, 3.3]}  &  \text{[-5.1, 3.4]} \\  
        $ \tilde{C}_{\text{tq}}^{(1)}   $   &    \text{[-2.6, 2.6]}  &    \text{[-2.2, 2.2]}  &    \text{[-2.3, 2.2]} \\   
   \hline
        $ \tilde{C}_{\text{tt}}         $   &  \text{[-1.5, 1.3]}  &  \text{[-1.1, 0.96]}  &  \text{[-0.81, 0.7]} \\  
        $ \tilde{C}_{\text{QQ}}^{(+)}   $   &  \text{[-1.5, 1.3]}  &  \text{[-1.1, 0.96]}  &  \text{[-0.81, 0.7]} \\  
        $ \tilde{C}_{\text{Qt}}^{(1)}   $   &  \text{[-2.4, 2.4]}  &  \text{[-1.8, 1.8]}  &  \text{[-1.3, 1.3]} \\    
        $ \tilde{C}_{\text{Qt}}^{(8)}   $   &  \text{[-5.3,   4.4]}  &    \text{[-4.1, 3.1]}  &    \text{[-3., 2.3]} \\ 
   \hline
\end{tabular}
\end{table}

\subsubsection[Top quark dipole moment in multi-top production]%
{Top quark dipole moment in multi-top
  production\footnote{Contributed by
    J. Ebadi, H. Khanpour, S. Khatibi and M. Mohammadi Najafabadi.}}

This paragraph presents the study of the sensitivity of the four top quark production on the strong dipole moments of the top quark \cite{Malekhosseini:2018fgp}. 
Within the SM framework, the top quark dipole moments are zero at tree level, however, 
higher-order corrections could generate non-zero strong dipole moments for the top quark. 
The top quark strong dipole moments have very small values in the SM, so that they would not be observable at the  LHC experiments. 
However, there are extensions of the SM in which sizable contributions to these
dipole moments arise, making them accessible by the experiments at the LHC~\cite{Ibrahim:2010hv,Yang:2013ula}.
As a result,  observation of any significant deviation of dipole moments
from zero would point to beyond the SM physics. 
The most general effective Lagrangian describing the $gt\bar{t}$ coupling considering dimension-6 
operators can be parametrized as~\cite{AguilarSaavedra:2008zc}:
\begin{eqnarray}\label{Lag:gttbar}
\mathcal{L}_{gt\bar{t}} = -g_{s}\bar{t}\frac{\lambda^a}{2}\gamma^{\mu}tG^a_{\mu}
-g_s\bar{t}\lambda^a\frac{i\sigma^{\mu\nu}q_{\nu}}{m_t} (d_V^{g}+id_A^g\gamma_5)tG^a_{\mu}, \nonumber
\end{eqnarray}
where  the chromomagnetic and chromoelectric dipole moments of the top quark are denoted by $d^g_{V}$ and $d^g_{A}$ (both are zero in the SM at leading order).
Direct bounds on both $d^g_{V}$ and $d^g_{A}$ were obtained  from  the top quark pair cross section measurements  
at  the  LHC and the Tevatron. The bounds on the dipole moments using the $t\bar{t}$ cross section at the
LHC and Tevatron were found to be:
$-0.012\leq  d_V^g \leq 0.023~,~|d_A^g|\leq 0.087$~\cite{Aguilar-Saavedra:2014iga}.
Four-top quark production is also affected by the $gt\bar{t}$ effective coupling 
and  provides a powerful way to probe the chromomagnetic and chromoelectric dipole moments of the top quark.
The representative Feynman diagrams with the effective $gt\bar{t}$ coupling denoted by filled red circles are shown in Fig.~\ref{fig:FeynmanDM}.
\begin{figure}[tbh]
\vspace{5mm}
\centering
\includegraphics[width=0.50\textwidth]{\main/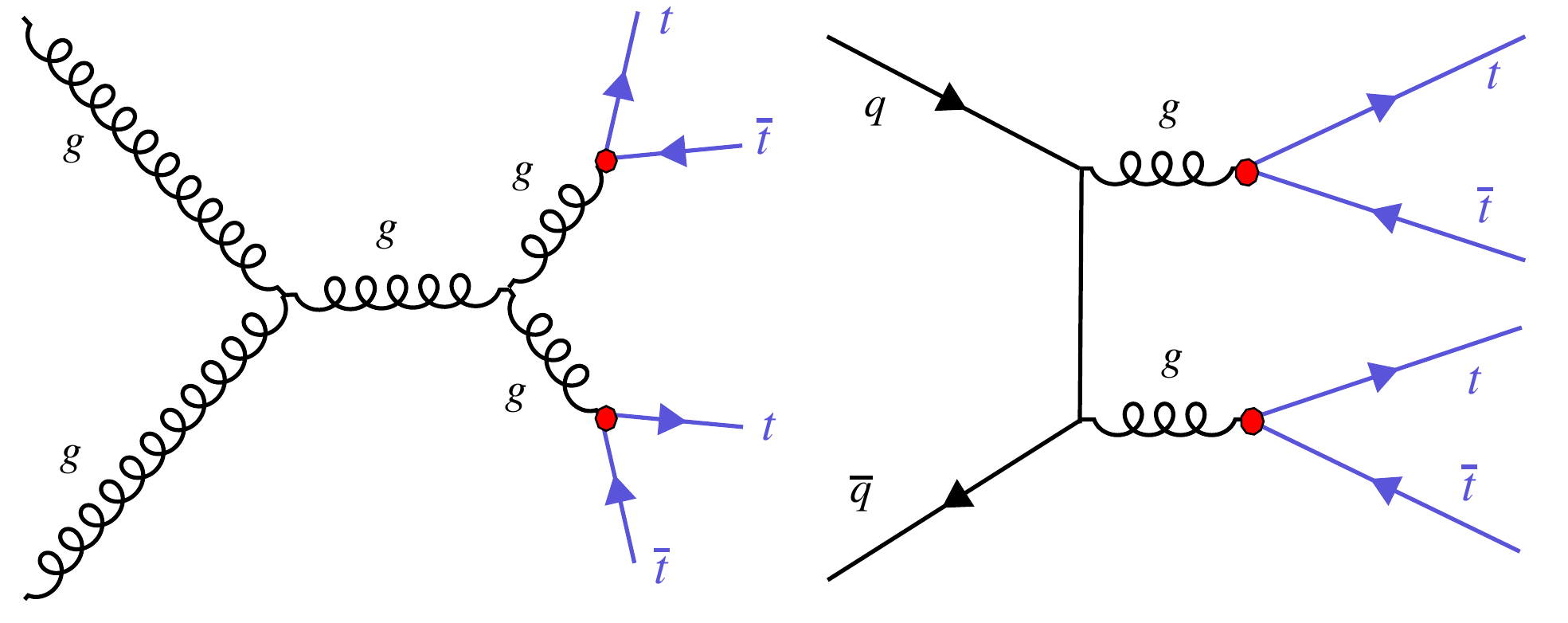}
\caption{\small{Representative Feynman diagrams for the $t\bar t t\bar t$ production where 
the effects of the strong dipole moments are shown as filled red circles.\label{fig:FeynmanDM}}}
\end{figure}
The contribution of the top quark dipole moments to the $t\bar{t}t\bar{t}$ production cross section is determined
with the \mg package~\cite{Alwall:2014hca}. 
By taking into account at most an effective vertex in each diagram,  the total four top cross section at $\sqrt{s} = 14$ TeV has the following form:
\begin{eqnarray}\label{csqq}
\sigma(pp\rightarrow t\bar{t}t\bar{t}) (\text{fb}) &=& \sigma_{\rm SM} + 154.8 \times  d_{V}^{g} + 3404.4  \times  (d_{V}^{g} )^{2},  \nonumber \\
\sigma(pp\rightarrow t\bar{t}t\bar{t}) (\text{fb}) &=& \sigma_{\rm SM} +2731.3  \times  (d_{A}^{g} )^{2},  
\end{eqnarray}
where the SM four top quark cross section is denoted by $\sigma_{\rm SM}$. The linear terms are due to the interference between the 
new physics and SM with the contribution of the order of $\Lambda^{-2}$.
The quadratic terms suppressed by $\Lambda^{-4}$ power are the pure contributions of the strong  dipole moments.
To estimate the sensitivity of the four top process to dipole moments, the same-sign dilepton channel is the focus here due to its clean signature and very low background contribution.
The main background contributions come from the $t\bar{t}W$ and $t\bar{t}Z$ processes.
Signal and the background processes are generated with the \mg package at leading order. 
\PYTHIA v6~\cite{Sjostrand:2003wg} is used for hadronization, showering and decay of unstable particles. 
Jets are reconstructed using the anti-$k_{\rm T}$ algorithm~\cite{Cacciari:2008gp}.
Signal events are selected by requiring  exactly two same-sign charged leptons 
with $p_{\rm{T}}^{\ell} > 25$ GeV and $|\eta^{\ell}| < 2.5$. The missing transverse energy has to be 
larger than 30 GeV. Each event is required to have at least eight jets with $p_{\rm{T}} > 30$ GeV and $|\eta|< 2.5$ 
from which at least three should be $b$-tagged jets. All objects in the final state are required to be well isolated objects 
by requiring $\Delta R(i,j) > 0.4$. Table~\ref{tab:dipole} presents limits at 95\% CL 
on the chromoelectric ($d^{g}_{A}$) and chromomagnetic ($d^{g}_{V}$) dipole moments for the HL-LHC and HE-LHC.
\begin{table}[]
\vspace{1mm}
\centering
\caption{\small{Limits at 95\% CL on the chromoelectric and chromomagnetic dipole moments $d^{g,Z}_{V}$ at $95\%$ CL for the HL-LHC and HE-LHC.}}\label{tab:dipole}
\begin{tabular}{llll} \hline\hline
Coupling    &     HL-LHC, 14 TeV, 3 ab$^{-1}$          &      HE-LHC, 27 TeV, 15 ab$^{-1}$  \\ \hline  
$d^{g}_{V}$ &     [-0.084, 0.009]                      &      [-0.063, 0.001]        \\
$d^{g}_{A}$ &     [-0.030, 0.030]                      &      [-0.011, 0.011]        \\  \hline
\end{tabular}
\end{table}
The HE-LHC improves the HL-LHC bound on $d^{g}_{A}$ by about a factor 
of three and the upper bound on $d^{g}_{V}$ by one order of magnitude. 
The four top-quark production at the HE-LHC would be able to tighten the 
upper limit on $d^{g}_{A}$ ($d^{g}_{V}$) by a factor of two (eight) with respect to the 
top pair production at the HL-LHC \cite{Aguilar-Saavedra:2014iga}.

\providecommand{\ttv}{t\bar t V}
\providecommand{\ttw}{t\bar t W}
\providecommand{\ttwp}{t\bar t W^+}
\providecommand{\ttwm}{t\bar t W^-}
\providecommand{\ttwpm}{t\bar t W^\pm}
\providecommand{\ttz}{t\bar t Z}


\subsection{The $\ttv$ production at the HL/HE-LHC}

\subsubsection[The $\ttz$ cross sections at NLO QCD and EW]%
{$\ttz$ cross sections at NLO QCD and EW\footnote{Contributed by R. Frederix, D. Pagani and M. Zaro.}}
\label{sec:ttznloew}

This section provides the cross section for $\ttz$ production at the HL and HE-LHC. The results are accurate up to NLO QCD and NLO EW accuracy~\cite{Frixione:2015zaa}. NLO 
QCD and EW corrections are computed 
simultaneously with \mg~\cite{Alwall:2014hca}, more specifically by using the recently-released version capable of mixed-coupling expansions~\cite{Frederix:2018nkq}. The same setup as in Ref.~\cite{deFlorian:2016spz} is used (see in particular Sec.~1.6.7.a), except for the PDF set, for which the 
PDF4LHC15\_nlo\_30\_pdfas set~\cite{Butterworth:2015oua} is employed.
In fact, at variance with the predictions in Ref.~\cite{deFlorian:2016spz}, photon-initiated contributions are not included, since recent studies on the photon distribution became 
available~\cite{Manohar:2016nzj,Manohar:2017eqh}, and the corresponding photon density gives negligible contributions for $\ttz$. The quoted EW corrections include the LO term
at $\mathcal O (\alpha^2 \alpha_s)$ and the NLO one at $\mathcal O (\alpha^2 \alpha_s^2)$. At variance with $\ttw$ production, for which other contributions, subleading in the couplings,
turn instead to be relevant (see Sec.~\ref{sec:ttwnlo}), it has been shown in Ref.~\cite{Frederix:2018nkq} that such contributions can be safely neglected for $\ttz$.\\

Cross-sections for $\ttz$ are quoted in Table~\ref{table:ttz-xsect}, together with the NLO/LO QCD $K$-factor, the relative impact of EW corrections, and the theory uncertainties. For
the latter, the uncertainty coming from scale variations, the PDF uncertainty and the $\alpha_s$ one are quoted separately. 

\begin{table}[h]
\small
\caption{Cross section, in pb, for $\ttz$ production at the HL and HE-LHC. Uncertainties on the cross sections are at the per-mil level.}
\label{table:ttz-xsect}
\providecommand{\arraystretch}{1.5}
\begin{center}
\begin{tabular}{|c c c c c c c c|}
\hline
$\sqrt s$  &  $\sigma_{\rm QCD}^{\rm NLO}$  &  $\sigma_{\rm QCD+EW}^{\rm NLO}$  &  $K_{\rm QCD}$  &  $\delta_{\rm EW}\; [\%]$  &  scale $[\%]$  &  PDF $[\%]$  &  $\alpha_s$ $[\%]$ \\
\hline\hline
14 TeV  &  1.018  &  1.015  &  1.40  &  -0.3  &  +9.6 \;\; -11.2  &  $\pm$2.7  &  $\pm$  2.8\\
27 TeV  &  4.90  &  4.81  &  1.45  &  -2.0  &  +9.9 \;\; -10.4  &  $\pm$2.0  &  $\pm$  2.0\\
\hline
\end{tabular}
\end{center}
\end{table}

\subsubsection[The complete-NLO corrections to $\ttw$]%
{The complete-NLO corrections to $\ttw$\footnote{Contributed by R. Frederix, D. Pagani and M. Zaro.}}
\label{sec:ttwnlo}
\def\LO{{\rm LO}}
\def\NLO{{\rm NLO}}
\def\LNLO{{\rm (N)LO}}
\def\NLOQCD{{\rm NLO}_{\rm QCD}}
\def\NLOEW{{\rm NLO}_{\rm EW}}
\def\LOQCD{{\rm LO}_{\rm QCD}}

This section presents the so-called ``complete''-NLO corrections to $\ttwpm$ production. This process 
can proceed through different terms of order $\alpha_s^p \alpha^{q+1}$ with $p+q=2,\;3$ at LO and at NLO respectively. The term complete-NLO 
refers to computation of all terms with $p+q\le3$, which has been performed for the first time in Ref.~\cite{Frederix:2017wme}
by employing the newly-released version of~\mg~\cite{Alwall:2014hca} capable of computing mixed QCD and electroweak corrections~\cite{Frederix:2018nkq}. 
Among the various contributions, the complete-NLO corrections include the NLO QCD ones 
($p=3$, $q=0$)~\cite{Hirschi:2011pa, Garzelli:2012bn, Campbell:2012dh, Maltoni:2014zpa}, and the NLO EW corrections 
($p=2$, $q=2$)~\cite{Frixione:2015zaa}. This short paragraph reports inclusive predictions 
for the HL and HE-LHC, with a centre-of-mass energy of respectively
14 TeV and 27 TeV.  The same setup and notation of Ref.~\cite{Frederix:2017wme} is used, where the interested reader can find more details
as well as predictions for 13 and 100 TeV. 
\begin{table}[h]
\small
\caption{Cross section for $\ttwpm$ production at the HL and HE-LHC, in various approximations, for $\mu=H_{\rm{T}}/2$. Number in parentheses are computed with a jet veto. See Ref.~\cite{Frederix:2017wme} for details.}
\label{table:ttw-xsect}
\providecommand{\arraystretch}{1.5}
\begin{center}
\begin{tabular}{|c c c c c c|}
\hline
$\sigma[\textrm{fb}]$ & LO${}_{\textrm{QCD}}$ & $\LOQCD+\NLOQCD$ & LO & $\LO+\NLO$ & $\frac{\LO(+\NLO)}{\LOQCD(+\NLOQCD)}$\\
\hline\hline
14 TeV  &  $  414 ^{+23\perc}_{-18\perc} $  &  $  628 ^{+11\perc}_{-11\perc} $  ($  521 ^{+5\perc}_{-7\perc} $)  &  $  418 ^{+23\perc}_{-17\perc} $  &  $  670 ^{+12\perc}_{-11\perc} $  ($  548 ^{+6\perc}_{-7\perc} $)  &  $ 1.07 $  ($ 1.05 $)\\
27 TeV  &  $ 1182 ^{+21\perc}_{-16\perc} $  &  $ 2066 ^{+14\perc}_{-11\perc} $  ($ 1561 ^{+7\perc}_{-7\perc} $)  &  $ 1194 ^{+21\perc}_{-16\perc} $  &  $ 2329 ^{+14\perc}_{-11\perc} $  ($ 1750 ^{+7\perc}_{-7\perc} $)  &  $ 1.13 $  ($ 1.12 $)\\
\hline
\end{tabular}
\end{center}
\end{table}
 \begin{table}[t]
\small
\caption{$\ttw$: $\sigma_{\LNLO_i}/\sigma_{\LOQCD}$ ratios at 14 and 27 TeV, for different values of $\mu=\mu_r=\mu_f$. LO$_2$ is identically zero and is not quoted in the table. 
 Number in parentheses are computed with a jet veto.
See Ref.~\cite{Frederix:2017wme} for details.}
\label{table:ttw-orders}
\begin{center}
\begin{tabular}{|c | r@{\,}l r@{\,}l r@{\,}l | r@{\,}l r@{\,}l r@{\,}l |}
\hline
\multirow{2}{*}{$\delta[\%]$}
 & \multicolumn{6}{c}{14 TeV}
 & \multicolumn{6}{c|}{27 TeV}\\
 & \multicolumn{2}{c}{$\mu= H_{\rm{T}}/4$} & \multicolumn{2}{c}{$\mu=H_{\rm{T}}/2$} & \multicolumn{2}{c}{$\mu = H_T$}     & 
     \multicolumn{2}{c}{$\mu= H_{\rm{T}}/4$} & \multicolumn{2}{c}{$\mu=H_{\rm{T}}/2$} & \multicolumn{2}{c|}{$\mu = H_{\rm{T}}$} \\
\hline\hline 
LO${}_3$ & \multicolumn{2}{c}{$0.8$} & \multicolumn{2}{c}{$1.0$} & \multicolumn{2}{c}{$1.1$}  & 
            \multicolumn{2}{c}{$0.9$} & \multicolumn{2}{c}{$1.0$} & \multicolumn{2}{c|}{$1.2$}\\
\hline
 $ \rm{NLO_1} $  &  $ 37.4 $ & ($  7.7 $)  &  $ 51.8 $ & ($ 25.9 $)  &  $ 64.7 $ & ($ 41.9 $)  &
            $ 67.4 $ & ($ 18.4 $)  &  $ 74.8 $ & ($ 32.0 $)  &  $ 82.0 $ & ($ 44.3 $)\\
 $ \rm{NLO_2} $  &  $ -4.5 $ & ($ -4.7 $)  &  $ -4.3 $ & ($ -4.5 $)  &  $ -4.1 $ & ($ -4.3 $)  &
            $ -5.1 $ & ($ -5.4 $)  &  $ -5.0 $ & ($ -5.2 $)  &  $ -4.8 $ & ($ -5.1 $)\\
 $ \rm{NLO_3} $  &  $ 13.0 $ & ($  9.7 $)  &  $ 13.3 $ & ($  9.9 $)  &  $ 13.6 $ & ($ 10.1 $)  &
            $ 25.5 $ & ($ 19.8 $)  &  $ 26.1 $ & ($ 20.2 $)  &  $ 26.6 $ & ($ 20.6 $)\\
 $ \rm{NLO_4} $  &  $ 0.02 $ & ($ -0.00 $)  &  $ 0.03 $ & ($ 0.00 $)  &  $ 0.05 $ & ($ 0.01 $)  &
            $ 0.06 $ & ($ 0.01 $)  &  $ 0.08 $ & ($ 0.02 $)  &  $ 0.10 $ & ($ 0.03 $)  \\
\hline
\end{tabular}
\end{center}
\end{table}

Table~\ref{table:ttw-xsect} reports the total-cross section for $\ttwpm$ production in different approximations, and Table~\ref{table:ttw-orders} 
the breakdown of the
different orders contributing at LO and NLO, as fraction of the $\mathcal O(\alpha_s^2 \alpha)$ LO contribution, LO$_1$.  Number in 
parentheses are computed by vetoing hard central jets, with $p_{\rm{T}} > 100\, \textrm{GeV}$ and $\eta < 2.5$. As it can be gathered from the tables, the
jet veto is beneficial in order
to reduce the NLO QCD corrections, in particular the large contribution coming from hard real emissions with a soft or collinear $W$ boson. It can be appreciated how the NLO$_3$ contribution is actually larger than the NLO$_2$ (the EW corrections) despite the extra power of $\alpha$, and how such a contribution
grows with the collider energy. As explained in 
Ref.~\cite{Frederix:2017wme}, this is due to the $t-W$ scattering process~\cite{Dror:2015nkp}. Since the size of NLO$_3$ is not much affected by the jet 
veto, a measurement of the $t-W$ scattering from the $\ttw$ cross section should be possible.

\subsection{Top mass}
\newcommand{\fbinv} {\mbox{\ensuremath{\,\text{fb}^\text{$-$1}}}\xspace}
\newcommand{\mt}{\ensuremath{m_{t}}\xspace}
\newcommand{\ttbar}{\ensuremath{{\mathrm{t}\overline{\mathrm{t}}}}\xspace} 
\newcommand{\TeV}{\ensuremath{\,\text{Te\hspace{-.08em}V}}\xspace}
\newcommand{\JPsi}{\ensuremath{J/\psi}\xspace}
\newcommand{\lambdaqcd}{\ensuremath{\Lambda_{\text{QCD}}}\xspace}
\providecommand{\PYTHIA}{\textsc{Pythia}\xspace}
\providecommand{\HERWIG}{\textsc{Herwig}\xspace}

\newcommand\MSB{\ensuremath{\overline{{\rm MS}}}}
\newcommand\MSR{\ensuremath{{\rm MSR}}}


\subsubsection[Theoretical issues]{Theoretical issues\footnote{Contribution by  G. Corcella, P. Nason, A. Hoang and H. Yokoya.}}
\label{sec:top_mass_th}
The currently most precise methods for top mass measurements at the LHC
are the so called ``direct measurements'' which are obtained exploiting
information from the kinematic reconstruction of the measured top
quark decay products, and their corresponding combinations.  The
typical errors currently quoted for the direct LHC top mass
measurements are of the order of 500-600 MeV, and with the prospect of
the high luminosity operations, as can be seen from Fig.~\ref{fig:topmassextr} of the
following section, the projected future experimental uncertainty is
around 200 MeV. Such a high precision entails also a high level of
scrutiny concerning the extracted top mass value. In direct
measurements, the measured top mass is the value of the top mass
parameter in the Monte Carlo generator that is used to fit top-mass
sensitive distributions,
because the complexity of the measurement
is such that the extraction of these distributions
corrected for detector effects, to be compared with analytic
calculations, is not feasible.
In this respect, the scrutiny must also regard
theoretical aspects dealing with how the Monte Carlo models the
relevant mass sensitive distributions, keeping in mind that all
effects that can lead to variations of the result in the 100 MeV range
should be considered.

The top mass parameter, as all coupling constants characterizing the
underlying field theory, requires renormalization, and its precise
value depends upon the adopted renormalization scheme.
The differences in the top mass in different renormalization prescriptions used in the theoretical community
are parametrically of order $R\,\alpha_s(R )$, with R between about 1 GeV and $m_t$,
and thus can amount from a few 
hundred MeV to several GeV.
 It is thus clear that an
experimental result, in order to be of any use, must specify to which
scheme the measured value corresponds to.

At present, the experimental collaborations have renounced to
qualify direct mass measurements by also specifying a renormalization
scheme.  This is a consequence of the fact that no full agreement has
been reached among theorists on this issue.  Some authors have argued
that, in view of the inherent leading-order nature of the Monte Carlo
generators, no scheme can be specified for the mass measured in direct
measurements, since at leading order all schemes are equivalent.  This
argument was also used as part of the motivation
in favour of alternative measurements where the
mass-sensitive observable is directly computed in perturbation theory
at NLO or NNLO accuracy, and is compared to experimental distributions
already corrected for detector
effects~\cite{Alioli:2013mxa,Bevilacqua:2017ipv}.  For example, the
total cross section for $t\bar{t}$ production is sensitive to the top
mass, it has been computed up to the NNLO order in
QCD~\cite{Czakon:2013goa}, and can be used to extract a top mass
value~\cite{Khachatryan:2016mqs, Aad:2014kva, Langenfeld:2009wd}.
Similarly, in
Ref.~\cite{Alioli:2013mxa,Bevilacqua:2017ipv}, shape observables
constructed out of the $t{\bar t}+{\rm jet}$ kinematics are used.

Several theoretical works have appeared proposing alternative
techniques to measure the top mass, partly to provide predictions
with at least NLO precision to allow for a mass determination in a well-defined
mass scheme,
and partly to circumvent other aspects of
direct measurements that may be considered problematic.
The authors of Ref.~\cite{Kawabataa:2014osa}
presented a method, based upon the charged-lepton energy spectrum, that is not
sensitive to top production kinematics, but only to top decay, and does not
make use of jets. Since top decays have been computed at NNLO accuracy~\cite{Gao:2012ja,
  Brucherseifer:2013iv}, they argue that a very accurate measurement may be achieved
in this way. Other authors have advocated using the invariant mass of boosted top jets
supplemented by light grooming (see Ref.~\cite{Hoang:2017kmk} and references therein).  In
Ref.~\cite{Agashe:2016bok}, the $b$-jet energy peak position
is proposed as mass-sensitive observable, that is
claimed to have a reduced sensitivity to production dynamics.  In
Ref.~\cite{Frixione:2014ala}, the use of lowest Mellin moments of lepton
kinematic distributions is discussed.  In the leptonic channel, it is also
possible to use distributions based on the ``transverse'' mass
variable~\cite{Sirunyan:2017idq}, which generalizes the concept of transverse
mass for a system with two identical decay branches~\cite{Lester:1999tx,
  Barr:2009jv}.
Some of these methods have been effectively exploited by the experimental
collaborations~\cite{CMS-PAS-TOP-13-006,
  CMS-PAS-TOP-15-002, Aad:2015waa, Sirunyan:2017idq, Aaboud:2017ujq} to yield
alternative determinations of $\mt$. They are consistent within errors with
direct measurements, and thus provide valuable checks.
It turns out, however, that at the moment their errors are not competitive with direct measurements,
mostly because the (less direct)
observables of the alternative methods do not have the top mass discriminating
power of the direct method.
Furthermore, in view of the larger errors, the assessment of
their eventual theoretical uncertainties is a less demanding task in comparison
to the case of direct measurements.

The notion that the Monte Carlo mass parameter cannot be qualified as
a field theoretical mass has extensively permeated the discussions
regarding the interpretation of top mass measurements.
This notion, however, oversimplifies the situation,
because more precise 
statements on the
Monte Carlo mass parameter can be made.
In reality, the accuracy of Shower Monte Carlo's depends upon the
observables one considers.
As a trivial example, the total cross section for the production of top quarks is predicted 
at leading order by standard Shower Monte Carlo’s,  so that the value of the top mass extracted 
by fitting it to the measured total production cross section would indeed carry a scheme ambiguity 
of order $m_t \alpha_s$, because the pole or 
the \MSB{} schemes can be used for computing the total
cross section at higher orders. 
Such measurement cannot be qualified by specifying any
particular scheme.\footnote{In fact, at the moment, Monte Carlo
  generators that achieve NLO accuracy for sufficiently inclusive
  cross section are routinely used in top mass studies.}  This is not
the case if one considers as an observable the mass of the top decay
products.  In Ref.~\cite{Nason:2017cxd}, for example, it is pointed
out that, in the narrow width limit, a perturbative calculation of the
mass of the top decay products performed in the pole mass scheme
yields the pole mass at any perturbative order.
Since Monte Carlo generators, when performing heavy particle decay,
strictly conserve the mass of the decaying particle, it can be inferred that
the Monte Carlo mass parameter should be identified with the pole mass
up to non-perturbative effects\footnote{In the narrow width limit the
  top can propagate a long time before decay, and long-distance
  non-perturbative effects can manifest themselves there, and affect
  the mass by a few hundred MeV.} as far as the mass of the decay
products is concerned.  From a different point of view, in
Ref.~\cite{Hoang:2008xm} it is argued that since the top-quark decay
is treated with a Breit-Wigner form in the Monte Carlo generators,
and due to the infrared shower cutoff $Q_0 \approx 1\,$GeV, the
top mass parameter
should be close to top mass schemes that are compatible with the
Breit-Wigner form. In turn, these schemes yield mass values that
differ from the pole mass by terms of order $\alpha_s(R) R$, with
$R\approx \Gamma_t\approx Q_0$. In
a subsequent work~\cite{Hoang:2018zrp}, it is argued that, in the narrow width limit, one
can relate the Monte Carlo mass parameter to a running mass (such as
the \MSR{} mass~\cite{Hoang:2017suc}) evaluated at the scale of the Monte Carlo shower
cutoff $Q_0$, as long as $Q_0\gtrsim 1\,$GeV. These arguments entail that the Monte
Carlo mass parameter differs from the top pole mass by several hundred MeV.
 It must also be noted that theoretical papers that make use of the direct
top mass  (noticeably those on electroweak precision
fits~\cite{Patrignani:2016xqp, Baak:2014ora}, and calculations
inherent to the issue of the SM vacuum stability~\cite{Degrassi:2012ry,
  Buttazzo:2013uya,Andreassen:2017rzq}) interpret the direct
measurement results as being close to the pole mass, up
to a theoretical error of few hundred MeV.

A problem that has received much attention is the presence of
an infrared renormalon in the pole mass definition.
The QCD perturbative series for the difference of the
pole mass and the \MSB{} mass has factorially divergent
coefficients~\cite{Bigi:1994em, Beneke:1994sw}. This is related to an
ambiguity of the order of a typical hadronic scale in the
 pole mass.
 Estimates of this inherent ambiguity
vary from 110 to 250~MeV~\cite{Beneke:2016cbu,
  Hoang:2017btd,Pineda:2001zq, Bali:2013pla}.  It should be stressed,
however, that the finite width of the top screens the effects of soft
radiation, so that this ambiguity \emph{does not affect the physics}
of top production and decay.
This mean that the pole mass ambiguity does not represent in principle
a limitation on the precision of top quark mass measurements, since
short-distance mass schemes that are free of the pole mass ambiguity
can be adopted.  So in view of the considerable time to the start of
the LHC HL program, the pole mass ambiguity, if it becomes a limiting
factor, can be easily avoided, and is thus not discussed further here.

Accepting the fact that the difference between the top mass in direct
measurements and the top pole mass is of the order of few hundred MeV,
and in view of the current and projected accuracy of the direct
measurements, several works have appeared in the literature to better
quantify the difference. In \cite{Butenschoen:2016lpz} 
numerical relations between the
Monte Carlo mass parameter and the pole mass as well as the MSR mass
\cite{Hoang:2017suc} were determined from comparing hadron level resummed analytic
NNLL calculations performed in SCET factorization and Monte Carlo
output (using \PYTHIA v8.2) for the 2-Jettiness distribution at the top
mass resonance for boosted top jets in $e^+e^-$ annihilation.\footnote{
  This procedure is often quoted as a form of calibration of the Monte
  Carlo top mass parameter.  It must be noted that the same terminology has
  also been used in a different context in Ref.~\cite{Kieseler:2015jzh},
  where it is suggested that the Monte Carlo mass parameter can be
  constrained by fitting it from kinematic normalized distributions
  predicted from the Monte Carlo generator, simultaneously with an
  inclusive cross section measurement, that is then compared to a
  fixed order calculation.}
In the
work of Ref.~\cite{Hoang:2018zrp},
exploiting the fact that soft emission effects both in
shower Monte Carlo and in full QCD can be computed as long as the
shower cut $Q_0$ is a perturbative scale, the analytic structure of
angular ordered shower algorithms was examined in detail and compared
to the one of resummed calculations in SCET factorization for
hemisphere masses for boosted top jets in $e^+e^-$ annihilation. From the
analysis an analytic relation at ${\cal O}(\alpha_s)$ between the shower mass
parameter and the pole mass was calculated which is proportional to
$Q_0\,\alpha_s(Q_0)$.

The results of Ref.\cite{Butenschoen:2016lpz,Hoang:2018zrp} are obtained in
the context of global event-shape-type top jets observables in $e^+e^-$
annihilation, which are different from observables involving jets of
the top decay product that enter the direct measurements. Furthermore,
the findings of Ref.\cite{Hoang:2018zrp} represent parton level results and refer
exclusively to angular ordered parton showers. Future work should be
aimed to lift these limitations and to extend studies of this sort to
observables that enter the direct measurements at the LHC. Such
studies are also valuable to expose effects that should be included to
eventually match the experimental accuracy.

Direct measurements are not the only context where theoretical effects
in the top mass
that are linear in the strong interaction scale, i.e. of the order of
few hundred MeV, do arise. In Ref.~\cite{FerrarioRavasio:2018ubr}, the
production and decay of a top quark is considered in a very simplified
context, and in a particular approximation, such that non-perturbative
corrections can be examined in relation to the factorial growth of the
coefficients of the perturbative expansions. Linear power corrections
are found to affect all observables that make use of jets.  But it was
also found that typical leptonic observables are also affected by
linear power corrections.  Notice that this implies that the total
cross section is also affected by linear power corrections, as soon as
selection cuts are imposed.  These kind of studies can also be
extended to more complex measurement procedures, eventually making use
of jet calibration, in order to understand to what extent these
theoretical limitations to the precision can be removed.

The discussion carried out so far has highlighted theoretical issues
that should be studied in more depth in order to advance our
understanding of the theoretical precision of the measurements. In
essence these issues are related to the physics of different stages of soft emission,
where a deeper insight would allow to draw conclusions motivated by perturbation
theory, that may be extrapolated to low scales.
There are also aspects of the event simulations that on the one hand only
have to
do with relatively hard scales, and can be reliably computed,
and on the other hand are more related to
the modeling of hadronization effects that currently cannot be computed
from first principles. There is a current research effort,
aimed at improving the simulation of top production and decay, in both
these directions.  It includes both the improvement of perturbative
accuracy, and the improvement in the overall
shower-hadronization aspects.
Regarding the perturbative accuracy, recent progress has been achieved in the
Monte Carlo implementation of finite width and off-resonance
effects~\cite{Jezo:2016ujg}, whose impact has also
been investigated in Ref.~\cite{Heinrich:2017bqp}. Regarding
the hadronization aspects, the importance of the colour reconnection
models has been recognized and
investigated in Ref.~\cite{Argyropoulos:2014zoa,Christiansen:2015yqa}.
Furthermore, studies of the sensitivity of top-mass sensitive
observables to the perturbative accuracy, to the shower implementation
and to the hadronization model, are being carried out.  In one such
study~\cite{Ravasio:2018lzi}, significant differences were found when
comparing \HERWIG v7 and \PYTHIA v8, where the former adopts an angular
ordered shower, and the latter has a dipole shower, in the description
of top-mass sensitive observables.  In general, there is a range of
equally plausible simulation models than can be used to describe heavy
quark production and decay, that will include different Monte Carlo
generators, different Monte Carlo tunes in a given generator, and
different implementations of some component of a generator, like for
example the colour reconnection model.  As more work is done by
exploring different options for simulation models, the range of models
may enlarge, and potentially also the error in mass measurement may
increase. This increase in the error should be contrasted by limiting
the range of models, typically by requiring that some key observables
are in reasonable agreement with data, or by scrutiny concerning the
models themselves.  An example of a study in this
direction is given in Ref.~\cite{Corcella:2017rpt}, where the
sensitivity of the top-mass error upon the uncertainties in key Monte
Carlo tuning parameters is studied, and a set of calibration
observables strongly sensitive to the Monte Carlo parameters, but with
very mild sensitivity to the top mass, is considered in order to
reduce the parametric uncertainties.

A complementary way of reducing the error is to find variants of
measurement methods that reduce the dependence of the
extracted mass from the range of models. In situ jet calibration is
routinely used by the experimental collaborations in top mass
measurement. This procedure not only reduces the experimental error
associated with the jet energy scale, but it may also reduces the
theoretical error, by reducing the sensitivity of the measurements
from features of jet simulations in the generators.  More specific
proposals in this direction have appeared in
Ref.~\cite{Andreassen:2017ugs}, where the impact of adopting jet grooming
techniques to the jets in direct top mass measurements is examined.

As mentioned earlier, alternative techniques for mass measurements are
currently explored, and will become more precise at the HL-LHC.
As shown in Fig.~\ref{fig:topmassextr} in the
following subsection, the mass measurement from single top production
will acquire a precision similar to the one available today from
direct measurements.  The end-point measurement using the $J/\Psi$
will also reach a precision near 600~MeV. Thus, at the HL-LHC there will
likely be one highly precise measurement technique, plus a
number of independent methods supporting its results. It should not be
forgotten however, that high luminosity and/or high energy may also
offer opportunities for new techniques.  In Ref.~\cite{Hoang:2017kmk},
the use of grooming techniques applied to boosted top jets is studied,
with the goal of directly extracting a short distance mass. To what
extent the high luminosity phase can make this technique feasible is a
matter for future studies.  Another example is given in the work of
Ref.~\cite{Kawabata:2016aya}, where it is argued that a glitch in the
dilepton spectrum should be visible for a dilepton invariant mass near
twice the top mass. This effect is due to the diphoton production
subprocess $gg\to \gamma\gamma$ mediated by a top loop. The projected
statistical error for the mass determination using this method is of
2-3 GeV for the High Luminosity LHC, and 0.3-0.6 GeV for the 27~TeV
High Energy option. A 1~GeV error systematic from the EM calorimeter
calibration should also accounted for. Furthermore, a complete study
of the projected theoretical error is not yet available.  It is
nevertheless interesting to remember that ``out of the box'' thinking may
lead to progress in this area.

In summary, from a theoretical point of view, much
work is still needed to put the top mass measurements at the HL-LHC on a solid
ground.  Such work should comprise more thorough experimental work
aimed at understanding and reduce the sources of errors; theoretical
work in the framework of Monte Carlo studies and simulation; and
formal theoretical work aimed at understanding conceptual aspects.
Such work is already under way, and it is expected that
much more will be understood by the time the High Luminosity program
starts.  Thus, in spite of the many challenges, one can expect that
a theoretical precision matching the foreseeable experimental errors
for top mass measurements at the HL-HLC can be achieved.


\providecommand{\GeV}{\ensuremath{\text{Ge\kern -0.1em V}}}
\providecommand{\TeV}{\ensuremath{\text{Te\kern -0.1em V}}}
\providecommand{\ifb}{\mbox{fb$^{-1}$}}
\providecommand{\Wboson}{$W$~boson}
\providecommand{\Zboson}{$Z$~boson}
\providecommand{\Jpsi}{$J/\psi$}
\providecommand{\mtop}{\ensuremath{m_t}}

\graphicspath{{\main/top/img/}}

\subsubsection{Experimental projections}

The input material for the experimental summary is collected
in Ref~\cite{CMS-PAS-FTR-16-006,ATL-PHYS-PUB-2018-042}.
The measurement of the top quark mass \mtop with high precision is a crucial task for 
the expected 3000~\ifb of $pp$ collision data expected in HL-LHC.
The top quark mass is one of the free parameters within the Standard Model and
its Yukawa coupling is predicted to be close to unity.
Therefore it may play a special role in the electroweak symmetry breaking.
The top quark mass dominantly contributes to the quantum corrections of the
Higgs field, which become important for any extrapolation of the Standard
Model to extremely high energies, from a few hundred GeV and above.
At these high energies some of the fundamental deficiencies of the Standard
Model can be further investigated, such as the stability of the electroweak
vacuum state in the Higgs potential.
Thus, precise measurements of the top quark mass allow for consistency tests
of the Standard Model and to look for signs of new physics beyond.

The top quark mass is measured using various techniques and in different
decays channels by the ATLAS and CMS experiments following two 
different approaches.
Firstly, direct \mtop measurements are obtained exploiting
information from the kinematic reconstruction of the measured top quark decay
products, and their corresponding combinations.
This information is obtained from Monte Carlo (MC) simulated events using
different assumed values for the top quark mass parameter in the program.
Therefore, such results relate to measurements of the input parameter of MC
event generators, and differences between different MC are covered by a
specific systematic uncertainty.
The relation between the measured Monte-Carlo top quark mass parameter
and theoretical mass schemes such as the pole mass is discussed in
detail in Section~\ref{sec:top_mass_th}. Secondly, indirect determinations of \mtop~are obtained
based on the comparison of inclusive or differential $t{\bar t}$ production
cross-section to the corresponding theory calculations, thus sensitive
to \ensuremath{m_t^\textrm{pole}}.

The methods exploited for the measurement of \ensuremath{m_t}
directly using the kinematic properties of the $t{\bar t}$
(or single-top quark) decay products are the template, the matrix element
and the ideogram methods.
In the template method, based on a full
($t{\bar t}\rightarrow$lepton+jets, $t{\bar t}\rightarrow$all-jets)   
or partial ($t{\bar t}\rightarrow$dilepton and single-top quark) reconstruction 
of the kinematics underlying the top-quark(s) decay, probability density
functions (templates) for observables sensitive to the underlying
\mtop, and to additional parameters, are constructed based on MC simulation.
These templates are fitted to functions interpolating between the different
input values of \mtop, fixing all other parameters of the functions.
Finally, an unbinned likelihood fit to the observed data distribution of the
observable is used to obtain the value of \mtop\; describing the data best.
Typically, for single top and dilepton events the \ensuremath{m(lb)} variable
is used, whereas for the lepton+jets events the
\ensuremath{m_t^\textrm{reco}} obtained from a kinematic fit is more
appropriate.
The ideogram method can be considered as a computational effective
approximation of a matrix element method.
After a kinematic fit of the decay products to a $t{\bar t}$ hypothesis,
MC-based likelihood functions are exploited for each event (ideograms)
that depend only on the parameters to be determined from the data.
The ideograms reflect the compatibility of the kinematics of the event with
a given decay hypothesis.
As in the case of the template method, ideograms can be generalised in
multiple dimensions depending on the number of input observables used.

The latest ATLAS combination of direct \mtop\; measurements
leads to of top quark mass value of 
\ensuremath{m_t=172.69\pm 0.48}~\TeV
with a total precision of $\sim$~0.28\%~\cite{TOPQ-2017-03}.
The latest CMS combination of direct \mtop\; measurements
leads to of top quark mass value of
\ensuremath{m_t=172.44\pm 0.48}~\TeV
with a total precision of $\sim$~0.28\%~[arXiv:1509.04044].
The precision in each of these analyses is primarily limited by
systematic effects, in particular by the modelling of top quark
production and decay and by the jet energy scale.
Analysis techniques have been developed to use in-situ constraints
from the data on a global jet energy scale factor or light jet and $b$-jet
energy scale (3D fits)~\cite{TOPQ-2017-03}, which still suffer from
statistical uncertainties, which will be reduced strongly at the HL-LHC.
The total amount of 3000~\ifb of 14~\TeV data would clearly decrease the
statistical uncertainty in these analyses.
Therefore, the statistical precision in each analysis should be traded in
various ways for a reduced total systematic uncertainty by cutting into phase
space regions where the systematic uncertainties are high.

A variety of alternative methods are exploited to supplement the top quark mass
measurements from direct mass reconstruction based on jet observables.
One source of alternative observables is the usage of the $b$-jet information
in the $t{\bar t}$ decay, e.g. via final states featuring \Jpsi\; produced in
the $b$-hadron decays or secondary vertices in $b$-jets.
With the alternative approaches, a large variety of other 
\ensuremath{m_t^\textrm{MC}} measurements can be done, which have
different sensitivities to the top quark production and decay mechanisms and
making therefore different contributions to the systematic uncertainties.
Compared to the template method with the standard final states, the
sensitivity to the light-jet and $b$-jet energy scale (respectively JES and
$b$-JES) is expected to be reduced.
One of the limiting factors of this approach is the small branching fraction,
$\mathcal{B}(t{\bar t}\rightarrow(W^+b)(W^-b)\rightarrow (\ell\nu_\ell J/\psi(\rightarrow\mu^+\mu^-) X)(qq'b)~\sim 4.1\times 10^{-4}$, where $\ell=e,\mu$.
On the other hand the modelling of $b$-fragmentation and $b$-decay are
expected to be among the dominating sources of systematic uncertainties
of these two analyses and need to be studied extensively in a dedicated study
to reduce the signal modelling uncertainties.
Both measurements can contribute in different ways to the final combination
to improve the precision measurement of \mtop.
Individual \mtop\; results resting on various techniques
and $t{\bar t}$ (or single-top quark) decay channels, have different
sensitivities to statistical and systematic effects, and to the details of
the MC simulation.
To exploit the full physics potential of the available measurements, and to
profit from their diversity and complementarity, they are combined, thereby
further increasing our knowledge on \mtop.



In some alternative techniques the top quark mass is extracted
by comparing cross sections or distributions that can be calculated directly in QCD at either NLO or NNLO, to corresponding distributions extracted from
data. The mass parameter used in the NLO or NNLO calculation (either
the \MSB{} or the Pole top mass) is obtained by fitting the theoretical 
cross-section or distribution to the measured one.
In this framework, mass measurements have been performed using
as observables the inclusive $t{\bar t}$
cross-section, the differential decay rate in $t{\bar t}$+1 jet events, 
lepton and dilepton differential cross-sections.

Due to the changes of the detector performance for the HL-LHC,
it is difficult to estimate precisely the effects of systematic uncertainties.
The sources of uncertainty are assumed to be the same as the current ones.
The estimated Run-2 uncertainties are scaled to align with HL-LHC
extrapolations developed by the ATLAS and CMS Collaborations and documented
in Ref.~\cite{HLLHCSystRecommendations}.
The impact of the experimental systematic uncertainties will likely be
reduced relative to their effect on the Run-2 analysis given the large datasets
available, allowing precise performance studies to be conducted. 
The jet reconstruction uncertainties on \mtop\; are
expected to be divided by a factor up to two, while uncertainties related to
the reconstruction of electrons and muons remain the same as in Run-2.
The theory modelling uncertainties are expected to be divided by a
factor two compared to existing values.
The larger HL-LHC dataset will allow for dedicated tuning and good
understanding of NLO MC generators matched to parton showers,
as already started with Run-2 data~\cite{ATL-PHYS-PUB-2018-009}.
Another large contribution to the uncertainties stems from the modelling of
QCD interactions, which can be investigated and constrained using differential
measurements of the mass parameter itself or other ancillary measurements
in parts of the phase space not yet accessible.
These measurements are partially already being performed~\cite{CMS-PAS-TOP-13-007, CMS-PAS-TOP-12-030, CMS-PAS-TOP-15-017}, but will benefit from more statistics, therefore strong constraints from the high statistics at the HL-LHC are expected.

For this report, ATLAS Collaboration presents projections for the top quark
mass measurement accuracy using $t{\bar t}\rightarrow$ lepton+jets events with $J/\psi \rightarrow\mu^+\mu^-$ in the final state~\cite{ATL-PHYS-PUB-2018-042}. 
Samples of simulated events for signal and background processes are produced
at 14~\TeV centre-of-mass energy.
They include the production of $t{\bar t}$ pairs, single-top quarks and
$W/Z$ bosons in association with jets.
After the event generation step, a fast simulation of the trigger and detector effects 
is added with the dedicated ATLAS software framework.
The event selection follows the analysis done
at 8~\TeV~\cite{ATLAS-CONF-2015-040}.
Events are required to have at least one charged isolated lepton with
$p_\textrm{T}>25$~\GeV\ and $|\eta|<4$ and at least 4 jets with
$p_\textrm{T}>25$~\GeV\ and $|\eta|<4.5$.
No requirement is applied on the number of $b$-tagged jets. 
\Jpsi\; candidates are reconstructed using all pairs of opposite charge
sign soft muons with $p_\textrm{T}>4$~\GeV and $|\eta|<4.5$
The top quark mass is obtained from a template method with unbinned likelihood maximisation approach.
A statistical uncertainty of 0.14~\GeV is expected, with a systematic
uncertainty of 0.48~\GeV.



This paragraph discusses the potential of selected top quark mass measurements
at the HL-LHC done by the CMS Collaboration,
as described in detail in Ref~\cite{CMS-PAS-FTR-16-006}.
The extrapolations are based on measurements performed at 7 and
8~\TeV centre-of-mass energy using 5~\ifb and 19.7~\ifb, respectively.
The numbers presented here do not include the possible ambiguity in the
interpretation of the measured value with respect to a well defined
renormalisation scheme.
However, also the measurement of the pole mass from the inclusive \ttbar
cross-section cross section is extrapolated to HL-LHC conditions.

Typically, the jet energy scale uncertainties play a dominant role for top
quark mass measurements.
The contribution from background processes, important only for the measurement
using single top events, is expected to be well under control.
For the extrapolation of the extraction of \mtop\; from the total cross-section, 
the cross-section measurement is assumed to be ultimately limited by the luminosity uncertainty, here assumed to be 1.5\%.
For the prediction, no predictions beyond NNLO are assumed, such that the
uncertainty due to scale variations is constant.

The resulting extrapolated uncertainties on the top quark mass measurements are
summarised in Fig.~\ref{fig:topmassextr}.
The measurement using \Jpsi\;  mesons and using in general
secondary vertices benefit the most from higher statistics.
But also the other measurements improve significantly, mostly from more
precise understanding of systematic uncertainties, as discussed above,
such that ultimately, the precision will range between 0.1\%
(which is of the order of \ensuremath{\Lambda_{\textrm{QCD}}}) and 0.7\%.

\begin{figure}[htbp]
\centering{
\includegraphics[width=0.80\textwidth]{\main/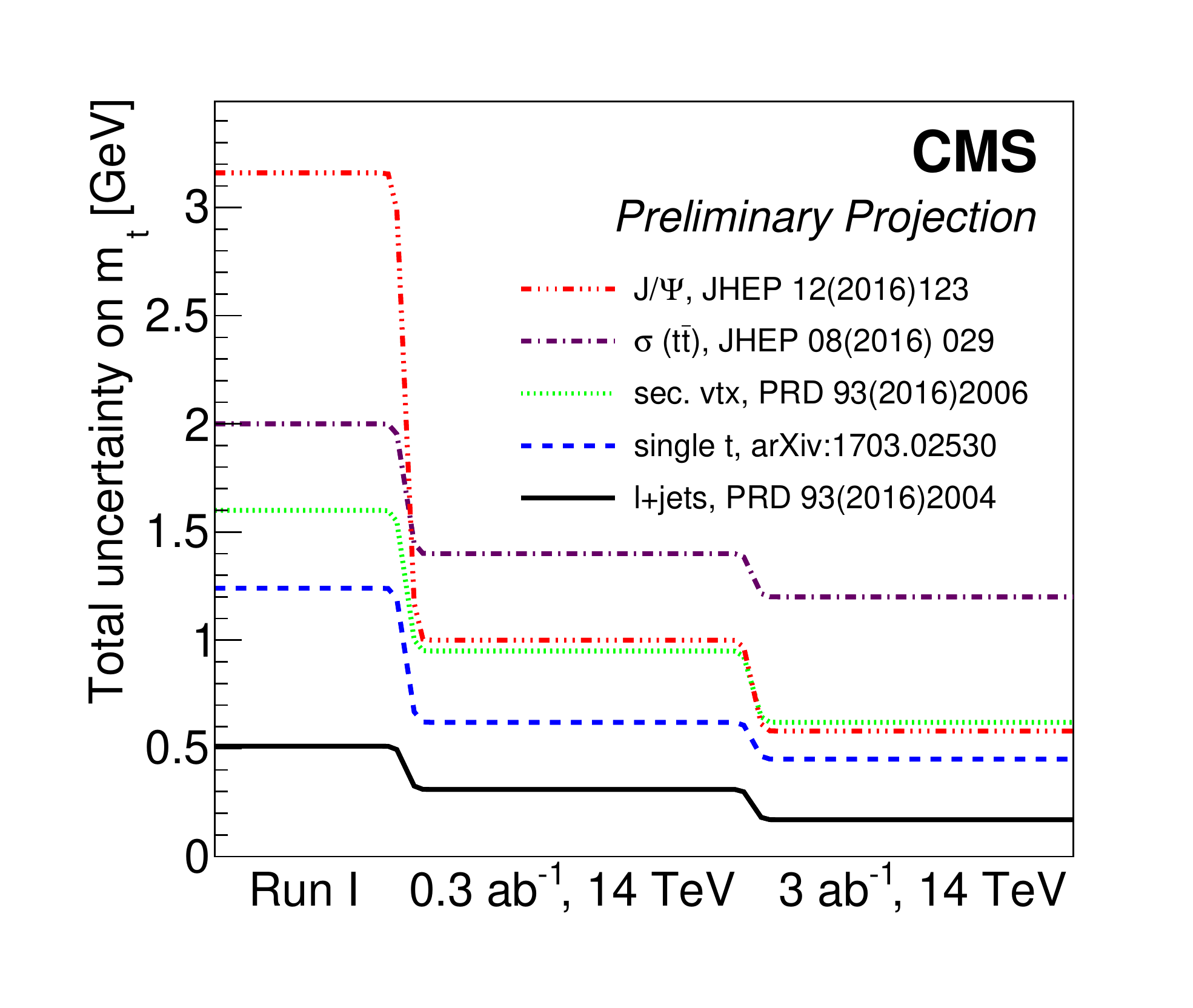}
\caption{ The top mass measurement uncertainty for different methods as a function of integrated luminosity as obtained by CMS.}
\label{fig:topmassextr}
}
\end{figure}

\begin{table}
\caption{Projected total uncertainties on the top quark mass for 3\,ab$^{-1}$ and $\sqrt{s}$=14~\TeV obtained with different methods as obtained by CMS.}
\centering{
\begin{tabular}{|lccc|}
\hline
Method & Statistical & Systematic & Total (\GeV) \\
\hline\hline
$t\bar t$ lepton+jets & $0.17$ & $0.02$ & $0.17$\\
single-$t$ t-channel & $0.45$ & $0.06$ & $0.45$ \\
$m_{sv\ell}$ & $0.62$ & $0.02$ & $0.62$ \\
\Jpsi & $0.24$ & $0.53$ & $0.58$ \\
$\sigma_{t\bar t}$ & $0.4\%$ (exp) & $0.4\%$ (theory) & $1.2$ \\
\hline\hline
\end{tabular}
}
\end{table}



\subsection{Top quark properties and couplings}


\subsubsection{Top quark charge asymmetries at LHCb}
The top quark charge asymmetry present in quark-initiated production is diluted by the presence of gluon-gluon fusion and the increased quark content in the proton at forward rapidities gives LHCb additional sensitivity to this observable. As LHCb takes data at a lower rate than ATLAS and CMS, and has a limited acceptance, a partial reconstruction of the $t\bar{t}$ final state is anticipated in order to make optimal use of statistics, as described in Sec.~\ref{sec:forwardtoplhcb}. The expected differential single lepton asymmetry at LHCb, inferred from the rate of $\ell^{+}b$ and $\ell^{-}b$ production as a function of lepton pseudorapidity, is shown in Fig.~\ref{fig:lhcb_top_asymm}~\cite{LHCb-PII-Physics}. The expected statistical precision of a dataset corresponding to 300 fb$^{-1}$ of integrated luminosity, the total expected at LHCb during the HL-LHC, is shown, along with the theoretical uncertainties due to scale, $\alpha_{s}$ and PDF uncertainties. The projection indicates that LHCb will have sufficient statistics to make a non-zero observation of the $t\bar{t}$ charge asymmetry at the HL-LHC. The dominant systematic uncertainty on the measurement is expected to come from the knowledge of the background contributions, particularly from $W$ production in association with $b$-jets. Other final states, where an additional $b$-jet or lepton are required to be present will provide additional information as, despite the lower statistical precision, they probe larger values of Bjorken-$x$ and select the data sample with a higher purity.
\begin{figure}
\begin{center}
\includegraphics[width=0.7\textwidth]{\main/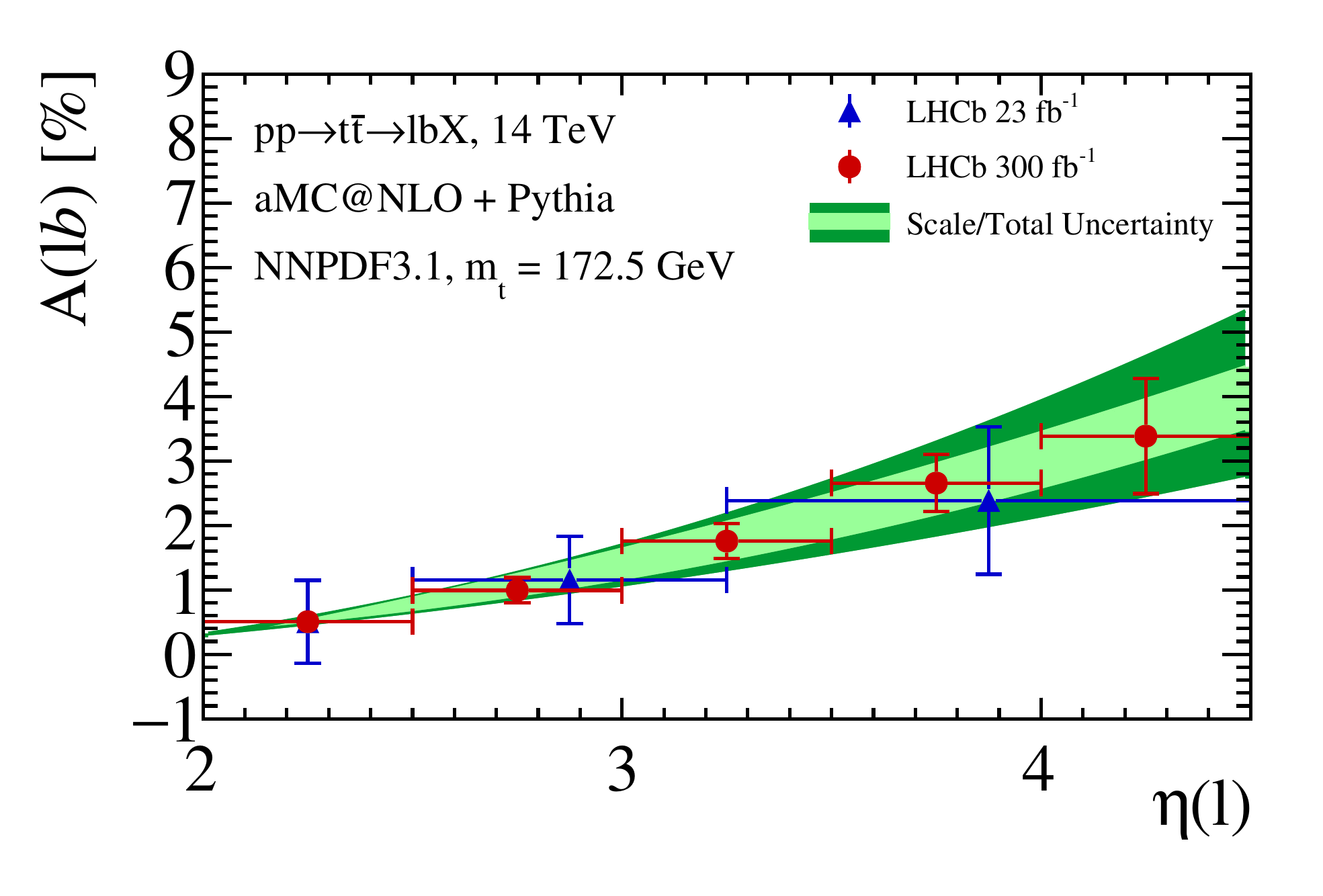}
\end{center}
\caption{
The predicted SM asymmetry at LHCb as a function of lepton pseudorapidity in the $\ell b$ final state at 14 TeV. The bands show the uncertainty on the theoretical predictions due to scale variations (light green) and due to combined scale, PDF and $\alpha_{s}$ variations (dark green). The expected statistical precision on measurements performed by LHCb using 23 and 300 fb$^{-1}$ of data is indicated by the error bars on the points.}
\label{fig:lhcb_top_asymm}
\end{figure}




\newcommand{\beq}{\begin{equation}}
\newcommand{\eeq}{\end{equation}}
\newcommand{\bea}{\begin{eqnarray}}
\newcommand{\eea}{\end{eqnarray}}
\newcommand{\ql}{\ensuremath{q_{\ell}}}
\newcommand{\sig}{{\cal S}}
\newcommand{\Dc}{{\cal D}_c}
\newcommand{\Dz}{{\cal D}_0}
\newcommand{\Nhad}{{\cal N}_{had}}
\newcommand{\Ncb}{{\cal N}_{cb}}
\newcommand{\Ncc}{{\cal N}_{cc}}
\newcommand{\Nbb}{{\cal N}_{bb}}
\newcommand{\Nlb}{{\cal N}_{\ell b}}
\newcommand{\Nlc}{{\cal N}_{\ell c}}
\newcommand{\bg}{{\cal B}}
\newcommand{\Ba}{\ensuremath{{\cal B}_1}}
\newcommand{\Bb}{\ensuremath{{\cal B}_2}}
\newcommand{\BF}{{\cal BF}}
\newcommand{\Rlcp}{{\cal R}_{\ell c}'}
\newcommand{\Rcbt}{\widetilde{\cal R}_{cb}}
\newcommand{\Rcct}{\widetilde{\cal R}_{cc}}
\newcommand{\Rlbt}{\widetilde{\cal R}_{\ell b}}
\newcommand{\Rlc}{{\cal R}_{lc}}
\newcommand{\Rcb}{{\cal R}_{cb}}
\newcommand{\Rcc}{{\cal R}_{cc}}
\newcommand{\Rlb}{{\cal R}_{\ell b}}
\newcommand{\Co}{\ensuremath{{\cal C}_0}}
\newcommand{\Mbcq}{\ensuremath{{\cal P}_{Bcl}}}
\newcommand{\Mbqc}{\ensuremath{{\cal P}_{Blc}}}
\newcommand{\Mcbq}{\ensuremath{{\cal P}_{cBl}}}
\newcommand{\Mcqb}{\ensuremath{{\cal P}_{clB}}}
\newcommand{\Mqbc}{\ensuremath{{\cal P}_{lBc}}}
\newcommand{\Mqcb}{\ensuremath{{\cal P}_{lcB}}}
\newcommand{\Mbqq}{\ensuremath{{\cal P}_{Bll}}}
\newcommand{\Mqbq}{\ensuremath{{\cal P}_{lBl}}}
\newcommand{\Mqqb}{\ensuremath{{\cal P}_{llB}}}
\newcommand{\MBcb}{\ensuremath{{\cal P}_{Bcb}}}
\newcommand{\MBbc}{\ensuremath{{\cal P}_{Bbc}}}
\newcommand{\McBb}{\ensuremath{{\cal P}_{cBb}}}
\newcommand{\McbB}{\ensuremath{{\cal P}_{cbB}}}
\newcommand{\MbcB}{\ensuremath{{\cal P}_{bcB}}}
\newcommand{\MbBc}{\ensuremath{{\cal P}_{bBc}}}
\newcommand{\Vcb}{\ensuremath{|V_{cb}|}}
\newcommand{\Vcbsq}{\ensuremath{|V_{cb}|^2}}
\newcommand{\DVcbsq}{\Delta(|V_{cb}|^2)}
\newcommand{\lum}{{\cal L}}
\newcommand{\epspre}{\ensuremath{\epsilon_0}}
\newcommand{\eflav}{\ensuremath{\epsilon_B^2}}
\newcommand{\epsflav}{\ensuremath{\epsilon_B^2}}
\newcommand{\fB}{\ensuremath{f_{{\cal B}}}}
\newcommand{\fflava}{\ensuremath{f_{{\cal B}1}}}
\newcommand{\fflavb}{\ensuremath{f_{{\cal B}2}}}
\newcommand{\epsB}{\ensuremath{\epsilon_{B}}}
\newcommand{\epsl}{\ensuremath{\epsilon_{\ell}}}
\newcommand{\epsb}{\ensuremath{\epsilon_{b}}}
\newcommand{\epsc}{\ensuremath{\epsilon_{c}}}
\newcommand{\epsDP}{\ensuremath{\epsilon_{DP}}}
\newcommand{\epsbB}{\ensuremath{\epsilon_{bB}}}
\newcommand{\epsbb}{\ensuremath{\epsilon_{bb}}}
\newcommand{\epscc}{\ensuremath{\epsilon_{cc}}}
\newcommand{\epsll}{\ensuremath{\epsilon_{ll}}}
\newcommand{\fDP}{\ensuremath{f_{DP}}}
\newcommand{\fia}{\ensuremath{f_{i\alpha}}}
\newcommand{\fcB}{\ensuremath{f_{cB}}}
\newcommand{\fcb}{\ensuremath{f_{cb}}}
\newcommand{\fll}{\ensuremath{f_{ll}}}
\newcommand{\fcc}{\ensuremath{f_{cc}}}
\newcommand{\fbb}{\ensuremath{f_{bb}}}
\newcommand{\fbc}{\ensuremath{f_{bc}}}
\newcommand{\flB}{\ensuremath{f_{lB}}}
\newcommand{\flb}{\ensuremath{f_{lb}}}
\newcommand{\flc}{\ensuremath{f_{lc}}}
\newcommand{\fcl}{\ensuremath{f_{cl}}}
\newcommand{\fbl}{\ensuremath{f_{bl}}}
\newcommand{\fcbt}{\ensuremath{\widetilde{f}_{cb}}}
\newcommand{\ttbar}{\ensuremath{t\overline{t}}}
\newcommand{\tbar}{\ensuremath{\overline{t}}}
\newcommand{\bbar}{\ensuremath{\overline{b}}}
\newcommand{\cbar}{\ensuremath{\overline{c}}}
\newcommand{\qbar}{\ensuremath{\overline{q}}}
\newcommand{\bi}{\begin{itemize}}
\newcommand{\ei}{\end{itemize}}
\newcommand{\delS}{\delta_{\epsb}}
\newcommand{\delB}{\delta_{\bg}}
\newcommand{\delBi}{\delta_{\bg i}}
\newcommand{\delBa}{\delta_{\Ba}}
\newcommand{\delBb}{\delta_{\Bb}}
\newcommand{\sigBeps}{\ensuremath{1.3\times 10^4}}
\newcommand{\sigFac}{\ensuremath{23}}
\renewcommand{\ni}{\noindent}
\newcommand{\wlc}{\ensuremath{W\rightarrow\ql\,c}}
\newcommand{\wcc}{\ensuremath{W\rightarrow "c\,c"}}
\newcommand{\wlb}{\ensuremath{W\rightarrow\ql\,b}}
\newcommand{\wbb}{\ensuremath{W\rightarrow "b\,b"}}
\newcommand{\wll}{\ensuremath{W\rightarrow \ql\,\ql}}
\newcommand{\simlt}  {\raisebox{-.6ex}{$\stackrel{\textstyle <}{\sim}$}}

\subsubsection{A method to determine $|V_{cb}|$ at the weak scale in top quark decays}


In a recent paper \cite{Harrison:2018bqi}, a new method was proposed to measure the $|V_{cb}|$ element of the Cabibbo Kobayashi Maskawa (CKM) quark mixing matrix at the scale $q\simeq m_W$, using top decays at the LHC. 
To date, \Vcb\ has always been measured in $B$ decays, i.e.~at an energy scale $q\simeq \frac{m_b}{2}$, far below the weak scale, and it is currently known to an uncertainty of about 2\% \cite{Tanabashi:2018oca}:
\beq
\Vcb =(42.2\pm0.8)\times 10^{-3}.
\eeq
In the proposed measurement at the LHC, $|V_{cb}|$ will be measured at the scale $q\simeq m_W$, more representative of the weak scale. The motivation for such a measurement is that the traditional extraction of \Vcb\ in $B$ decays relies heavily on the operator product expansion, and its sensitivity is significantly affected by theoretical uncertainties \cite{Tanabashi:2018oca}. In contrast, in dealing with decays of on-shell $W$s, as here, 
the theoretical situation is likely to be much cleaner and the systematic uncertainties will be very different. 
Moreover, there could be significant evolution of \Vcb\ between $q\simeq \frac{m_b}{2}$ and $q\simeq m_W$ due to radiative corrections: e.g.~the application (somewhat inappropriately) of the Standard Model (SM) six-quark evolution equations \cite{Cheng:1973nv} at two-loop order \cite{Machacek:1983fi} to the CKM matrix between $q\simeq\frac{m_b}{2}$ and $q\simeq m_W$ yields a fractional increase in \Vcb\ of $\simeq5\%$, see Fig.~\ref{evol}. 
\begin{figure}[h]
\centering
\includegraphics[height=7.cm]{\main/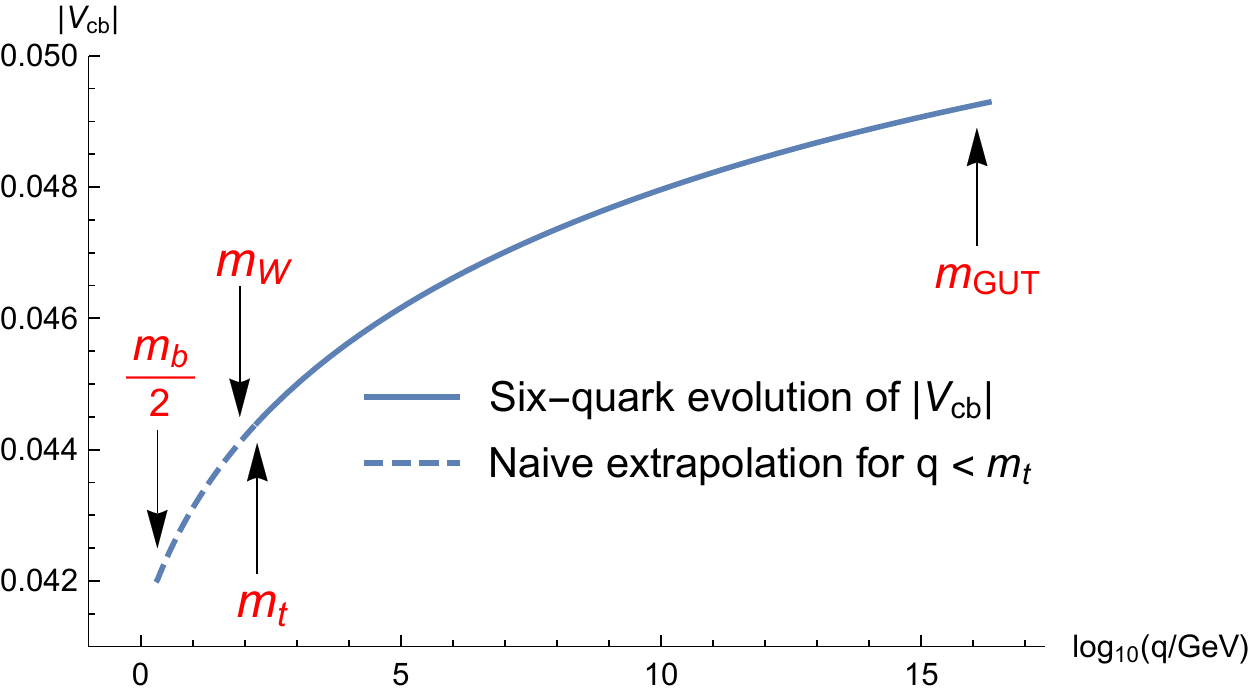}
\caption{Renomalisation Group evolution of \Vcb\ using the six-quark running scheme \cite{Cheng:1973nv, Machacek:1983fi, Balzereit:1998id} between $m_{GUT}$ and $\frac{m_b}{2}$. Previous publications stop at $m_t$, while a correct procedure would use a five-quark scheme for $q\,\simlt\, m_t$. This naive procedure at least suggests the possibility of significant low-energy evolution of \Vcb.}
\label{evol}
\end{figure}
While the correct treatment for SM evolution at such low energies is rather to use an effective field theory, integrating out the top quark below $q\sim m_t$ \cite{Balzereit:1998id}, such a calculation of the \Vcb\ running has not yet appeared in the literature. Thus the low-energy evolution of \Vcb\ is currently completely uncertain, while the naive calculation outlined above at least opens the possibility that its running might be observable, if \Vcb\ can be measured at or above the weak scale.

The proposed method uses the decays of tagged \ttbar\ pairs with one semileptonic top decay, (the tag), $\tbar\rightarrow \bbar W^- \rightarrow \bbar \ell^- \overline{\nu}_{\ell}$, and the other a hadronic decay, $t\rightarrow bW\rightarrow b\overline{q}c$, where $\overline{q}$ is a charge $\frac{1}{3}$ anti-quark (charge-conjugate decays will be assumed everywhere unless otherwise stated). The fraction of these in which $\overline{q}=\overline{b}$
is (up to negligible phase-space factors), exactly \Vcbsq.
Using this ratio, otherwise leading experimental uncertainties in most of the tagging efficiencies are cancelled. 
Thus the required signal will contain three tagged $b$-jets and a tagged $c$-jet, in addition to a charged lepton and missing transverse momentum.

Taking as a starting point, efficiencies from existing ATLAS and CMS \ttbar\ cross-section analyses, already-achieved experimental tagging performances \cite{ATLAS:2018ac006, Aaboud:2018xwy, ATLAS:2018bpl, ATLAS:2018xcf, Aaboud:2018fhh, Sirunyan:2017ezt}, and reasonable assumptions about backgrounds, it is estimated \cite{Harrison:2018bqi}, that the fractional uncertainty on \Vcb\ which can be obtained at a single experiment using the Run-2 dataset is:
\bea
\frac{\Delta\Vcb}{\Vcb}
&\sim&0.07,
\label{relErr2}
\eea
which is statistics-limited.  Averaging the two experiments would give a fractional error of $\sim5\%$.

Since the values of the systematic uncertainties on the tagging performances used to calculate  eq.~(\ref{relErr2}) were based roughly on their present determinations, the result is generalised in Fig.~\ref{sensitivity}, 
\begin{figure}[h]
\centering
\includegraphics[height=7.5cm]{\main/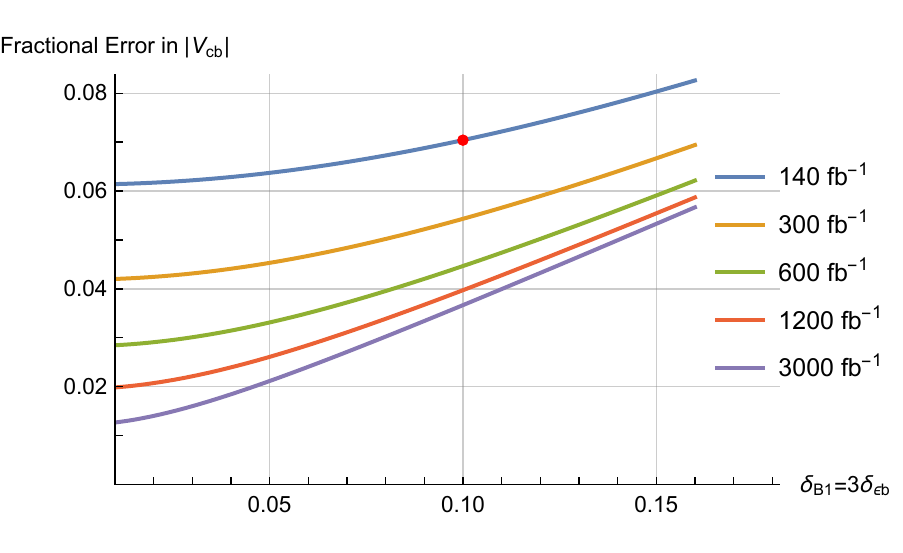}
\caption{Estimated fractional error in $\Vcbsq$ as a function of the systematic uncertainties $\delS$ in the $b$-jet tagging efficiency and $\delBa$ in the light-to-$b$ jet flavour mis-tag probability, and integrated luminosity. For ease of presentation, we assume $\delBa\simeq3\delS$ as it is at the time of writing. The top curve represents the Run-2 statistics and the red point on it indicates the illustrative values used to obtain eq.~(\ref{relErr2}). The second curve corresponds to luminosity projections for Run-3, while the bottom curve is for the projected integrated luminosity for HL-LHC. We have allowed for a 15\% increase in the $\ttbar$ cross section in the lower three curves, corresponding to an increase in beam collision energy to 14 TeV.}
\label{sensitivity}
\end{figure}
to show the dependence of the obtained fractional error on the systematic uncertainties as they vary. Also shown in Fig.~\ref{sensitivity} are the results using larger datasets, corresponding to various future LHC luminosity scenarios. The systematics-limited regime is represented by the linear-sloping region towards the bottom-right part of the figure, while the statistics-limited regime lies close to the $y$-axis, where the benefit of more statistics is most marked. The figure shows that making the measurement with future LHC data promises further improvements from both increased statistics and if tagging performance uncertainties can be reduced. E.g.~if $\delBa=3\delS$ can be reduced to $\simeq 0.05$, then at the end of Run-3, the uncertainty on \Vcb\ per experiment using this method could be as low as 4.5\%, giving a fractional uncertainty on the average of the two \Vcb\ measurements of $\sim3\%$. HL-LHC would then deliver a further reduction in the measurement uncertainty of better than a factor of 2. Either of these higher statistics measurements could give sensitivity for the first time to the renormalisation group running of \Vcb.

\providecommand{\pt}{\ensuremath{p_\mathrm{T}}\xspace}
\subsection{Flavour changing neutral current}


Processes with flavour-changing neutral currents (FCNC) are forbidden
at tree level and are strongly suppressed in higher orders
by the Glashow-Iliopoulos-Maiani (GIM) mechanism~\cite{PhysRevD.2.1285}. 
The SM predicts the branching fractions for top quark FCNC decays of $\mathcal{O}(10^{-12}$--$10^{-16})$~\cite{PhysRevD.44.1473, Mele:1998ag, AguilarSaavedra:2004wm}. However, various extensions of the SM allow a significant enhancement of the FCNC top quark decay rates arising from possible contributions of new particles~\cite{AguilarSaavedra:2004wm, Larios:2006pb, Agashe:2013hma}. Any deviations from heavily suppressed top FCNC rates would be a clear sign of new physics. The FCNC interactions of the top quark with the SM gauge and Higgs bosons can be described through the following anomalous coupling Lagrangian:
\begin{center}
\begin{eqnarray}
\mathcal{L} =
\sum_{q=u,c}\big[\sqrt{2}g_{s}\frac{\kappa_{gqt}}{\Lambda}\bar{t}\sigma^{\mu\nu}T_{a}
(f^{L}_{Gq}P_{L}+f^{R}_{Gq}P_{R})qG_{\mu\nu}^{a}+ \nonumber \\
+\frac{g}{\sqrt{2}c_{W}}\frac{\kappa_{zqt}}{\Lambda}\bar{t}\sigma^{\mu\nu}
(f^{L}_{Zq}P_{L}+f^{R}_{Zq}P_{R})qZ_{\mu\nu}+
\frac{g}{4c_{W}}\zeta_{zqt}\bar{t}\gamma^{\mu}
(f^{L}_{Zq}P_{L}+f^{R}_{Zq}P_{R})qZ_{\mu}- \nonumber \\
-e\frac{\kappa_{\gamma qt}}{\Lambda}\bar{t}\sigma^{\mu\nu}
(f^{L}_{\gamma q}P_{L}+f^{R}_{\gamma q}P_{R})qA_{\mu\nu}+ \nonumber \\
+\frac{g}{\sqrt{2}}\bar{t}\kappa_{Hqt}
(f^{L}_{Hq}P_{L}+f^{R}_{Hq}P_{R})qH
\big]+h.c.,
\end{eqnarray}
\end{center}

\noindent where $P_{L}$ and $P_{R}$ are chiral projection operators in spin space,
$\kappa_{Xqt}$ is the anomalous coupling
for $tXq$ vertex ($X=g,Z,\gamma,H$), $\zeta_{Zqt}$ is the additional
anomalous coupling for $tZq$ vertex, $f_{Xq}^L$ and $f_{Xq}^R$ are the left and right-handed complex chiral parameters with an unitarity constraint of $|f^L_{Xq}|^2 +
|f^R_{Xq}|^2 = 1$. Each of the anomalous couplings can be probed in events with the top quark pair production where one of the top quark decays via FCNC interaction, as well as in events with the associated production of the single top quark with a gluon, Z boson, $\gamma$, or Higgs boson.




\subsubsection*{Top-gluon}

\newcommand{\tGu}{\ensuremath{\rm t\rightarrow \rm ug}\xspace}
\newcommand{\tGc}{\ensuremath{\rm t\rightarrow \rm cg}\xspace}
\newcommand{\kLu}{\ensuremath{\lvert\kappa_{\rm tug}\rvert/\Lambda}\xspace}
\newcommand{\kLc}{\ensuremath{\lvert\kappa_{\rm tcg}\rvert/\Lambda}\xspace}
\newcommand{\fbinv} {\mbox{\ensuremath{\,\text{fb}^\text{$-$1}}}\xspace}

The $gqt$ FCNC process was studied by CMS~\cite{CMS-PAS-FTR-18-004} in single top quark events. The event signature includes the requirement of one isolated lepton and exactly one $b$ and one non-$b$ jet to be present in the final state with the dominant background arising from the $t\bar{t}$+jets and $W$+jets production. The signal events are simulated in the \textsc{SingleTop}\xspace  Monte-Carlo (MC) generator~\cite{Boos:2006af} based on the  \textsc{CompHEP}\xspace v4.5.2 package~\cite{Boos:2004kh}. The backgrounds processes are estimated 
with the \textsc{MG5\_a\scshape MC@NLO}\xspace~v2.5.2~\cite{Alwall:2011uj} package, showered and hadronized with \textsc{Pythia}\xspace  v8.230~\cite{Sjostrand:2014zea}. The full detector simulation has been performed for the signal and background events. A Bayesian neural network technique is used to separate signal from background events. The shape of the neural networks discriminants are used in the statistical analysis to estimate the expected sensitivity to the contribution from FCNC. Bayesian inference is used to obtain the posterior probabilities based on an Asimov data set of the background-only model. We assume the same systematic scenario as in Ref.~\cite{Collaboration:2293646}. To obtain the individual exclusion limits on \ensuremath{\lvert\kappa_{ tug}\rvert/\Lambda}\xspace and \ensuremath{\lvert\kappa_{ tcg}\rvert/\Lambda}\xspace we assume the presence of only one corresponding FCNC parameter in the FCNC signal Monte Carlo model.
These individual limits can be used to calculate the upper limits on the branching fractions $\mathcal{B}(\ensuremath{ t\rightarrow  ug}\xspace)$ and $\mathcal{B}(\ensuremath{ t\rightarrow  cg}\xspace)$~\cite{Zhang:2008yn}.
The expected exclusion limits at 95\% C.L. on the FCNC couplings and the corresponding branching fractions are given in Table~\ref{tab:1D_fcnc_limits}. In addition the two-dimensional contours that reflect the possible simultaneous presence of both FCNC parameters are shown in Fig.~\ref{tdr_fcnc:contour2d}. In this case both FCNC couplings are implemented in the FCNC signal Monte Carlo model. The expected limits can be compared with the recent CMS results~\cite{Khachatryan:2016sib} for the upper limits on the branching fractions of $2.0\times10^{-5}$ and $4.1\times10^{-4}$ for the decays $t\to ug$ and $t\to cg$, respectively.
\begin{table}[hbt]
\center{
\def\arraystretch{1.25}
\caption{The expected exclusion 1D limits at 95\% C.L. on the FCNC couplings and the corresponding branching fractions 
for an integrated luminosity of 300\fbinv and 3000\fbinv. In addition, a comparison with statistic-only uncertainties is shown.}
\label{tab:1D_fcnc_limits}
\begin{tabular}{|c|c|c|c|c|}
\hline
{Integrated luminosity} & {$\mathcal{B}( t~\to~ug)$}   & {$\lvert\kappa_{tug}\rvert/\Lambda$} & {$\mathcal{B}( t~\to~cg)$} & {$\lvert\kappa_{tcg}\rvert/\Lambda$} \\  
\hline\hline
300\fbinv  & $9.8 \cdot 10^{-6}$ & 0.0029 $\mathrm{TeV^{ -1}}$ & $99 \cdot 10^{-6}$ & 0.0091 $\mathrm{TeV^{ -1}}$ \\
3000\fbinv & $3.8 \cdot 10^{-6}$ & 0.0018 $\mathrm{TeV^{ -1}}$ & $32 \cdot 10^{-6}$ & 0.0052 $\mathrm{TeV^{ -1}}$ \\
3000\fbinv Stat. only & $1.0 \cdot 10^{-6}$ & 0.0009 $\mathrm{TeV^{ -1}}$ & $4.9 \cdot 10^{-6}$ & 0.0020 $\mathrm{TeV^{ -1}}$ \\
\hline
\end{tabular}
}
\end{table}

\begin{figure}[!h!]
	\begin{minipage}[t]{\linewidth}
		\begin{center}
			\includegraphics[width=0.49\textwidth]{\main/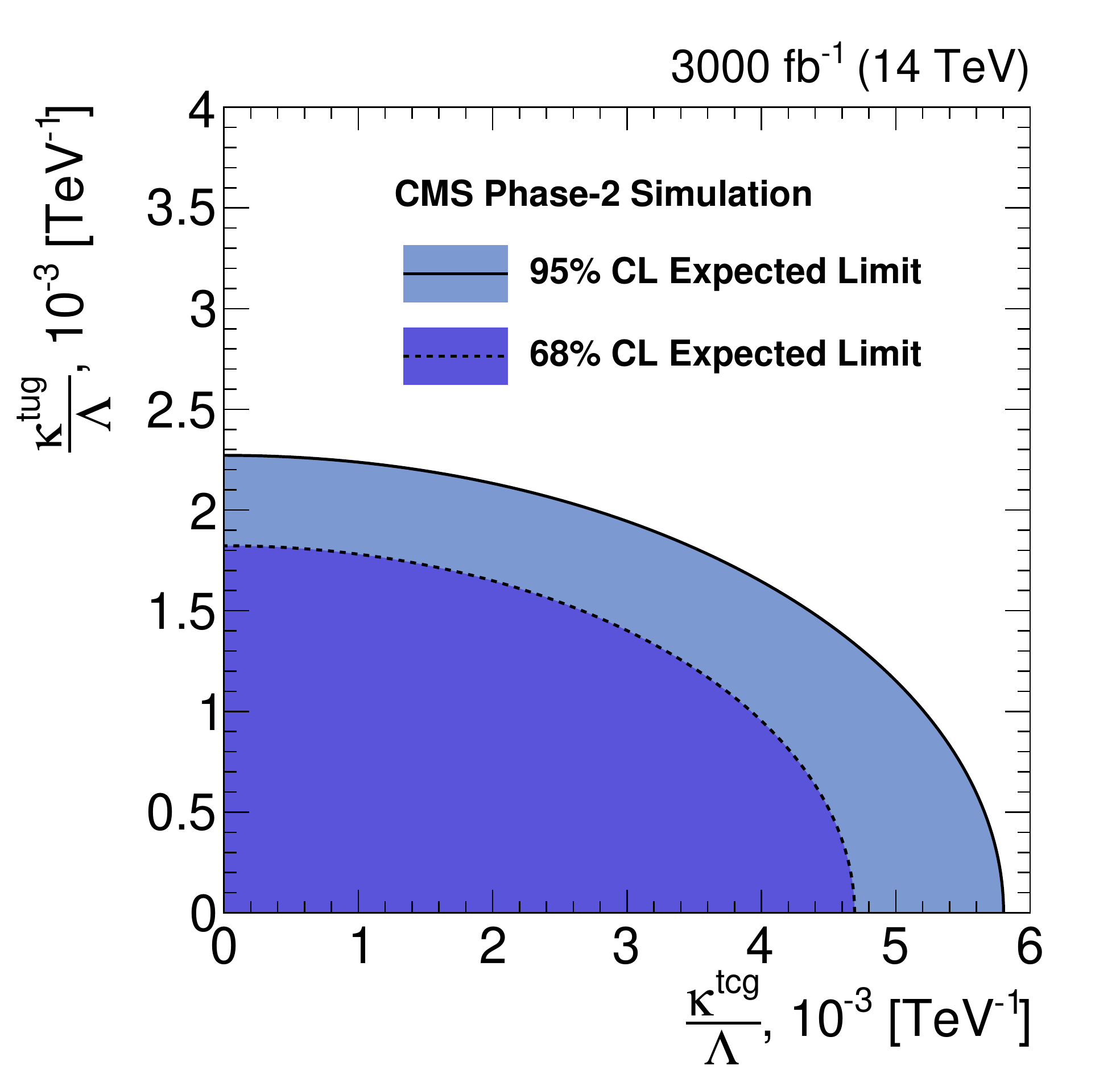}
			\includegraphics[width=0.49\textwidth]{\main/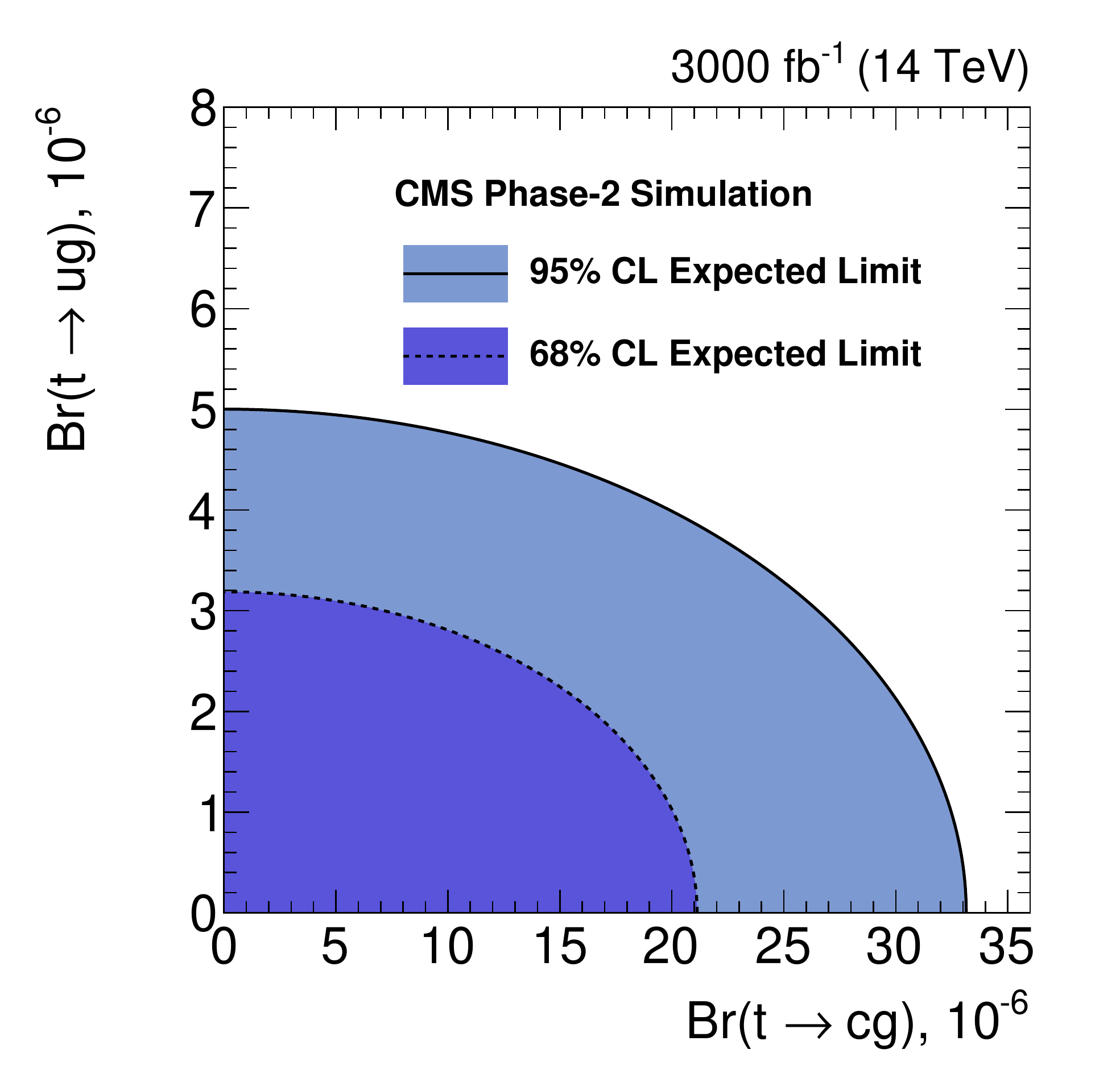}
		\end{center}
	\end{minipage}
	\caption{Two-dimensional expected limits on the FCNC couplings and the corresponding branching fractions at 68\% and 95\% C.L. for an integrated luminosity of 3000\fbinv .}
	\label{tdr_fcnc:contour2d}
\end{figure}

\subsubsection*{Top-Z}


The ATLAS Collaboration studied the sensitivity to the $tqZ$ interaction, by performing an analysis, detailed in Ref.~\cite{ATL-PHYS-PUB-2019-001}, based on simulated samples and following the strategy detailed in Ref~\cite{Aaboud:2018nyl} for the analysis of Run-2 data at 13~TeV centre-of-mass energy and the general recommendations for HL-LHC studies for this report. The study is performed in the three charged lepton final state of $t\bar t$ events, in which one of the top quarks decays to $qZ$, $(q=u,c)$ and the other one decays to $bW$ ($t\bar t\to bWqZ\to b\ell\nu q\ell\ell$). The kinematics of the events are reconstructed through a $\chi^2$ minimisation and dedicated control regions are used to normalize the main backgrounds and constrain systematic uncertainties. The main uncertainties, in both the background and signal estimations, are expected to come from theoretical normalization uncertainties and uncertainties in the modeling of background processes in the simulation. Different scenarios for the systematic uncertainties are considered, ranging from the conservative estimations obtained with the 13~TeV data analysis, to those that assume a factor two improvement due to expected advances in theoretical predictions. Figure~\ref{atlas_fcnc:tZq} shows the $\chi^2$ distribution for the events reconstructed in the signal region, after the combined
fit of signal and control regions under the background-only hypothesis. A binned likelihood function $L(\mu, \theta)$ is used to extract the signal normalisation. An improvement by a factor of five is expected with respect to the current 13~TeV data analysis results. The limits on the branching ratio are at the level of 4 to 5 $\times 10^{-5}$ depending on the considered scenarios assumed for the systematic uncertainties.

\begin{figure}[!h!]
\begin{center}
\includegraphics[width=0.49\textwidth]{\main/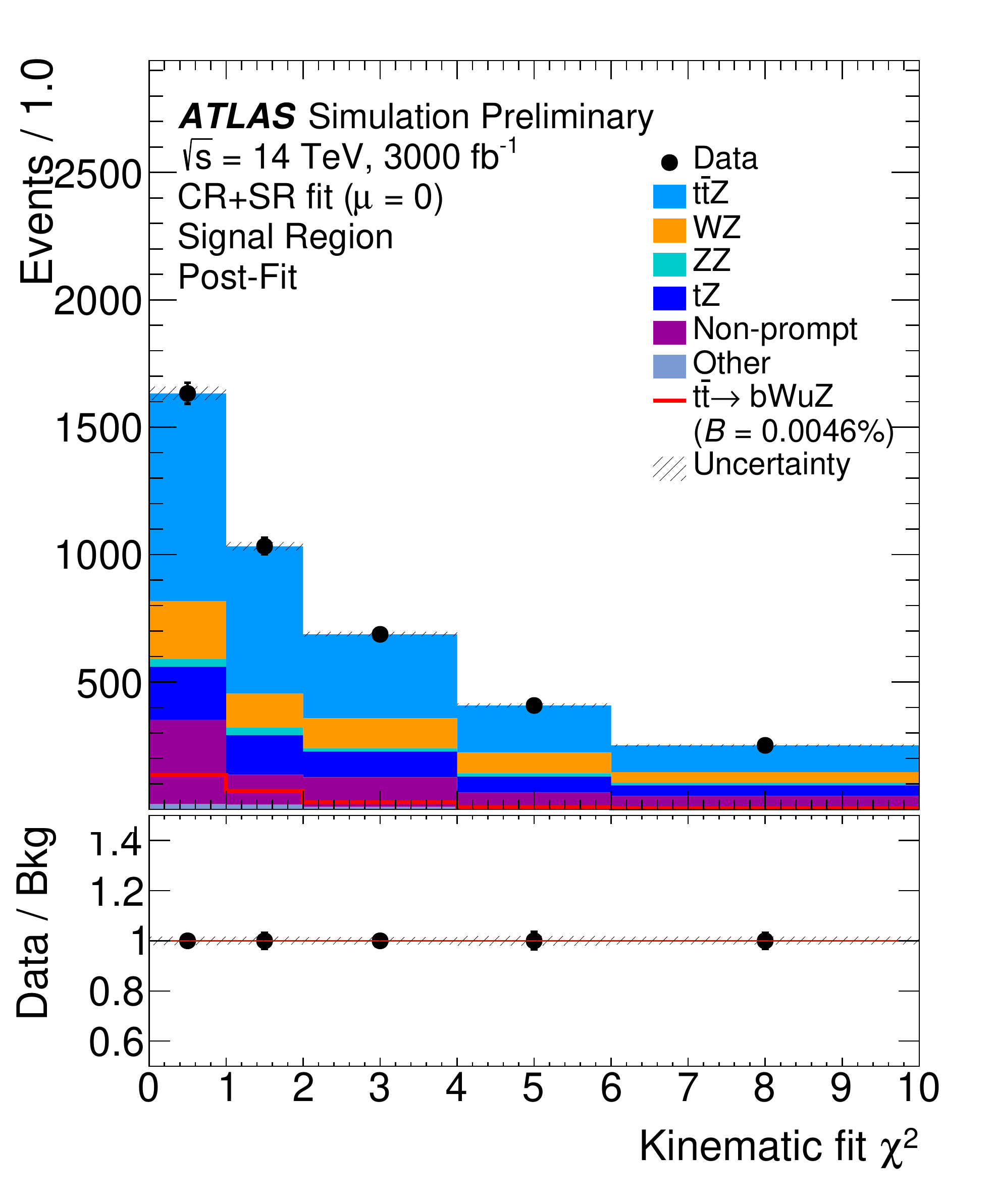}
\end{center}
\caption{The distributions for the $\chi^2$ for events reconstructed in the signal region, after the combined fit of signal and control regions under the background-only hypothesis. The data points are from the "Asimov dataset", defined as a total expected pre-fit background. The number of signal events is normalized to the expected branching ratio limit of $B(t\rightarrow uZ) = 4.6 \cdot 10^{-5}$. The dashed area represents the systematic uncertainty on the background prediction.}
\label{atlas_fcnc:tZq}
\end{figure}

\subsubsection*{Top-$\gamma$}
The $t\gamma q$ anomalous interactions have been probed by CMS at 8 TeV in
events with single top quarks produced in association with a photon~\cite{Khachatryan:2015att} and the resulting exclusion limits are $\mathcal{B}(t \to \gamma u) < 1.3~(1.9) \times 10^{-4}$ and $\mathcal{B}(t \to \gamma c) < 2.0~(1.7) \times 10^{-3}$. 

In this section, the sensitivity of the upgraded CMS detector to $tq\gamma$ FCNC transitions is estimated for integrated luminosities of 300 and 3000 fb$^{-1}$ using single top quark production via $q\rightarrow q\gamma$, with $q$ being a $u$ or a charm quark~\cite{Collaboration:2293646}. This analysis focuses on subsequent SM decays of the top quark in a $W$ boson and bottom quark, with the $W$ boson decays leptonically to a muon or electron and a neutrino. 
The finale state signature is the presence of a single muon or electron, large missing transverse momentum, a $b$-jet, and an isolated high energy photon, with a broad $\eta$ spectrum. 
The photon properties themselves provide good separation with respect to the dominant background processes from $W+$jets, and single top or top quark pair production in association with photons. 
For the discrimination of signal and background events, and to set the limits on the FCNC couplings, the events are split into two categories depending on the pseudo-rapidity of the photon (central region with $|\eta_{\gamma}| < 1.4$ and forward region with $1.6<|\eta_{\gamma}| < 2.8$). In the central (forward) region the photon \pt (energy) is used as a discriminating distribution: the low \pt(energy) is background dominated, while the high \pt(energy) region is populated by signal events. The distributions are shown in Fig.~\ref{fig:tqgamma}.

\begin{figure}[!h!]
	\begin{minipage}[t]{\linewidth}
		\begin{center}
        \includegraphics[width=0.49\textwidth]{\main/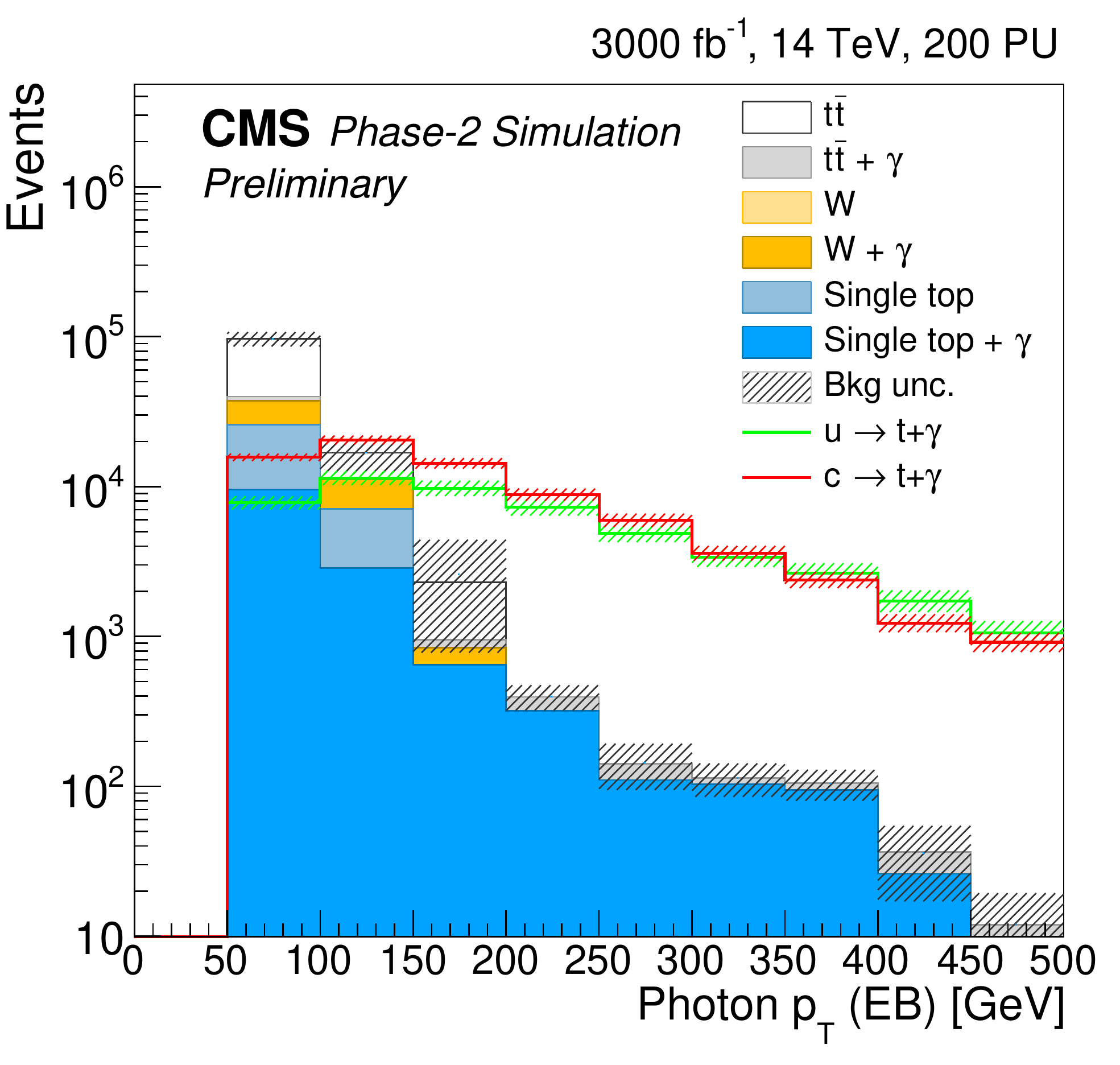}
         \includegraphics[width=0.49\textwidth]{\main/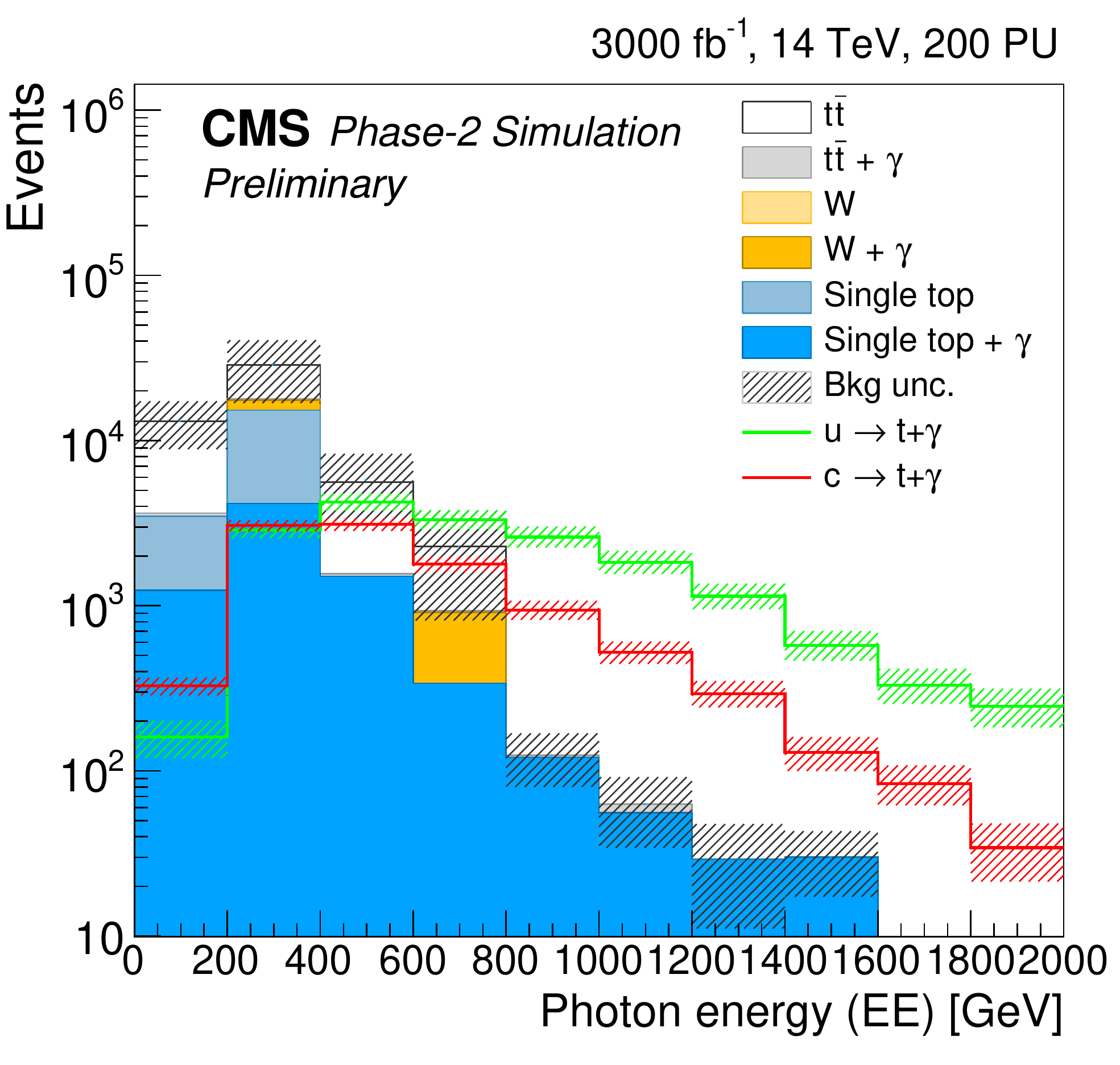}
		\end{center}
	\end{minipage}
	\caption{Transverse momentum of photon candidates for the central $\eta$ region (left) and
energy of photon candidates in the forward region (right).}
	\label{fig:tqgamma}
\end{figure}

The limits on the cross section for the single top quark production via $tq\gamma$ are obtained considering systematic uncertainties from variations of the renormalization and factorization scale, $b$-tagging and jet energy scale corrections and their effects as propagated to missing transverse energy, lepton efficiency and luminosity. 

These studies yield the following upper limite on the branching ratios at 95\%C.L.:  $\mathcal{B}(t \to \gamma u) < 8.6 \times 10^{-6}$, $\mathcal{B}(t \to \gamma c) < 7.4 \times 10^{-5}$.

\subsubsection*{Top-Higgs}

The $tHq$ interactions are studied by ATLAS in top quark
pair events with $t \rightarrow qH, H \rightarrow \gamma\gamma$~\cite{Aaboud:2017mfd} and $H \rightarrow WW$~\cite{Aaboud:2018pob} at 13
TeV. The former analysis explores the final state with two isolated
photons. For leptonic top quark decays the selection criteria includes
the requirement of one isolated lepton, exactly one $b$ jet, and at least
one non-$b$ jet. In case of hadronic top quark decays the analysis
selects events with no isolated leptons, at least one $b$ jet, and at
least three additional non-$b$ jets. The dominant background processes
are associated with the production of non-resonant $\gamma\gamma$+jets, $t\bar{t}$+jets and
$W$+$\gamma\gamma$ events. 
The resultant limits are $\mathcal{B}(t \to Hu) < 2.4~(1.7) \times
10^{-3}$ and $\mathcal{B}(t \to Hc) < 2.2~(1.6) \times 10^{-3}$.
The search for FCNC in $H \rightarrow WW$ includes the analysis of multilepton final states with either two same-sign or three leptons. The dominant backgrounds arising from the $ttW$, $ttZ$ and non-prompt lepton production are suppressed with a BDT.
The obtained limits are $\mathcal{B}(t \to Hu) < 1.9~(1.5) \times
10^{-3}$ and $\mathcal{B}(t \to Hc) < 1.6~(1.5) \times 10^{-3}$.
The $tHq$ anomalous couplings are probed by CMS in $H \rightarrow
b\bar{b}$ channel in top quark pair events, as well as in single top
associated production with a Higgs boson, at 13
TeV~\cite{Sirunyan:2017uae}. The event selection includes the
requirement of one isolated lepton, at least two $b$ jets, and at least
one additional non-$b$ jet. The dominant $t\bar{t}$ background is
suppressed with a BDT discriminant to set the exclusion limits of
$\mathcal{B}(t \to Hu) < 4.7~(3.4) \times 10^{-3}$ and $\mathcal{B}(t
\to Hc) < 4.7~(4.4) \times 10^{-3}$. Preliminary projections suggest $\mathcal{B}(t \to Hq) < \mathcal{O}(10^{-4})$~\cite{ATL-PHYS-PUB-2016-019, ATL-PHYS-PUB-2013-012}.

\begin{table}
\centering
\caption{Summary of the projected reach for the 95\% C.L. limits on the branching ratio for anomalous flavor changing top couplings.}
\begin{tabular}{lcccc}
\hline\hline
$\mathcal{B}$ limit at 95\%C.L. & 3 ab$^{-1}$, 14 TeV & 15ab$^{-1}$, 27 TeV & Ref. \\
\hline
$t\to gu $ & $3.8\times 10^{-6}$ & $5.6\times 10^{-7}$ & \cite{CMS-PAS-FTR-18-004}\\
$t\to gc $ & $32.1\times 10^{-6}$ &  $19.1\times 10^{-7}$ & \cite{CMS-PAS-FTR-18-004}\\
$t\to Zq$ & $2.4-5.8\times 10^{-5}$ &  & \cite{ATL-PHYS-PUB-2016-019}\\
$t\to \gamma u $ & $8.6\times 10^{-6}$ & & \cite{Collaboration:2293646} \\
$t\to \gamma c $ & $7.4\times 10^{-5}$ & & \cite{Collaboration:2293646} \\
$t\to Hq $ & 10$^{-4}$ & & \cite{ATL-PHYS-PUB-2016-019} \\
\hline\hline
\end{tabular}
\end{table}



\subsection[Effective coupling interpretations for top quark cross sections and properties]{Effective coupling interpretations for top quark cross sections and properties\footnote{Contributed by L. Lechner, D. Spitzbart, R. Sch\"ofbeck, D. Azevedo,F. D\'eliot, A. Ferroglia, M. C. N. Fiolhais, E. Gouveia, A. Onofre, E. Vryonidou, and M. Moreno Llacer.}}
\label{sec:eft}

\def\PZ{Z\xspace}
\def\PW{W\xspace}
\def\PQt{t\xspace}
\def\PAQt{\bar{t}\xspace}
\def\PQq{q\xspace}
\def\cPgn{\nu\xspace}
\def\Pl{l}
\def\jet{$j$}

\def\ttZ{$t\bar{{t}}Z$\xspace}
\def\ttG{$t\bar{{t}}\gamma$\xspace}
\def\ttW{$t\bar{{t}}W$\xspace}
\def\ttbar{$t\bar{{t}}$\xspace}
\def\WZ{$WZ$\xspace}
\def\tZq{$tZq$\xspace}
\def\tWZ{$tWZ$\xspace}

\def\pT{p$_\text{T}$}
\def\pTZ{p$_\text{T}(Z)$\xspace}
\def\cosThetaStar{$\cos\theta^*_{Z}$\xspace}
\def\nlep{N$_\text{lep}$\xspace}

\def\ctZ{C$_{tZ}$\xspace}
\def\ctZI{C$_{tZ}^\text{[Im]}$\xspace}
\def\cpt{C$_{\phi t}$\xspace}
\def\cpQM{C$_{\phi {Q}}$\xspace}

\def\delphes{\mbox{\textsc{Delphes}}\xspace}
\def\MGvATNLO{{MADGRAPH5aMC@NLO}\xspace}
\def\PYTHIA{\mbox{\textsc{Pythia}}\xspace}
\def\MadSpin{\mbox{\textsc{MadSpin}}\xspace}

\def\TeV{TeV\xspace}
\def\GeV{GeV\xspace}
\newcommand{\sss}{\scriptscriptstyle}
\newcommand{\OO}{\ensuremath{\mathcal{O}}}
\newcommand{\sst}{\scriptstyle}
\newcommand{\Op}[1]{\OO_{\sss #1}}
 \def\lra#1{\overset{\text{\scriptsize$\leftrightarrow$}}{#1}}

\newcommand{\Vl}{V_{\rm{L}}}
\newcommand{\Vr}{V_{\rm{R}}}
\newcommand{\gl}{g_{\rm {L}}}
\newcommand{\gr}{g_{\rm{R}}}





Effective Field Theory (SMEFT) \cite{Weinberg:1978kz,Buchmuller:1985jz}, where the SM is augmented by a set of higher-dimension operators
\begin{equation} 
	\mathcal{L}_\mathrm{SMEFT}=\mathcal{L}_\mathrm{SM}+
	\sum_i\frac{C_{i}}{\Lambda^2}\OO_{i}+\mathcal{O}(\Lambda^{-4}).
\label{eq:smeft1}
\end{equation}

As an example the relevant operators for the $tWb$ vertex  are: 
\begin{eqnarray}
\Op{tW}&=&
     i\big(\bar{Q}\sigma^{\mu\nu}\,\tau_{\sss I}\,t\big)\,
     \tilde{\phi}\,W^I_{\mu\nu}
     + \text{h.c.}\label{O1}\\
     \Op{\phi q}^{\sss(3)}&=&
     i\big(\phi^\dagger\lra{D}_\mu\,\tau_{\sss I}\phi\big)
     \big(\bar{q}_i\,\gamma^\mu\,\tau^{\sss I}q_i\big)
     + \text{h.c.}\label{O2}
\end{eqnarray}in agreement with the notation of \cite{AguilarSaavedra:2018nen}.

The operators of eq.~\eqref{O1}-\eqref{O2} modify the $Wtb$
interaction in the following way
\begin{eqnarray}
  {\mathcal L}^{\mathrm{dim-6}}_{Wtb} &=& -\frac{g}{\sqrt{2}}  \bar{b}(x) \gamma^\mu 
                                      P_L t(x)\,W_{\mu}(x) \left(1+  \frac{C^{(3)}_{\varphi Q} v^2}{\Lambda^2}  \right) \nonumber \\
                                  &+&  \frac{2  \,v  \, C_{tW}}{\Lambda^2}  \bar{b}(x) \sigma^{\mu\nu} 
                                      P_R t(x)\,\partial_{\nu} W_{\mu}(x) +  {\mathrm{h.\; c.}} \,,\label{interaction}
\end{eqnarray} 
where $v = 246$ GeV is the Higgs doublet vacuum expectation value, and
$y_t$ the top quark Yukawa coupling. Here and below it is assumed
$V_{tb}=1$.  It must be noted that a slightly different approach
  \cite{AguilarSaavedra:2006fy,AguilarSaavedra:2008gt,Aguilar-Saavedra:2014eqa,Prasath:2014mfa,Jueid:2018wnj},
  not using operators but anomalous couplings, has also been used in
  the literature. It is straightforward to connect the operator coefficients with the anomalous couplings description. The connection between the operator coefficients to
  the anomalous couplings is discussed in Ref.~\cite{AguilarSaavedra:2008zc}. 
The $Wtb$ vertex can be probed in single top production ($t-, Wt, s-$channel top production), $W$ helicity fractions and forward-backward asymmetries.

Similarly the coupling of the top to the Z and photon can be parameterised by the dimension-6 operators as discussed in Ref.~\cite{AguilarSaavedra:2018nen}, where the relevant degrees of freedom are discussed. The relevant degrees of freedom for the top-Z interaction $c^-_{\phi Q}, c^3_{\phi Q}, c_{\phi t}, c^{[I]}_{tZ}$ whilst the photon-top interaction depends on $c^{[I]}_{tA}$ as defined in Ref.~\cite{AguilarSaavedra:2018nen}. Phenomenological studies of top production in association with a vector boson or a photon exist in the literature  \cite{ Rontsch:2014cca,  Rontsch:2015una, Zhang:2016omx, Bylund:2016phk} including NLO QCD corrections.

This section examines the prospects of probing top charged and neutral couplings at the HL-LHC. 

\subsubsection{The top quark couplings to the $W$ boson}

The latest and most precise measurements on single top quark production cross sections ($t-, Wt-$ and $s-$channels)
~\cite{Aaltonen:2015cra, CDF:2014uma, Sirunyan:2016cdg, Aaboud:2017pdi, Aad:2015upn, Chatrchyan:2012ep, Chatrchyan:2012zca, Khachatryan:2016ewo},
$W$ boson helicity fractions ($F_{\rm{0}}$,$F_{\rm{L}}$ and $F_{\rm{R}}$) ~\cite{Aaltonen:2012rz, Aaboud:2016hsq} and forward-backward asymmetries ($A^{\ell}_{FB}$,$A^N_{FB}$,$A^T_{FB}$)~\cite{Aaboud:2017yqf}, measured at different centre-of-mass energies i.e., 2~TeV at Tevatron and 7, 8 and 13~TeV at the LHC, were used to set stringent 95\% CL limits on possible new physics that affect the $Wtb$ vertex structure. The results were extrapolated to the HL-LHC phase of the LHC, by assuming the full expected luminosity (3000~fb$^{-1}$) and scaling the uncertainties obtained at  the LHC for $\sqrt{s}=13$~TeV (the central value of the observables were assumed to be the Standard Model prediction at 14~TeV). 
The statistical and simulation related uncertainties were scaled according to the total integrated luminosity at the HL-LHC. All generator and signal modelling related systematic uncertainties of these observables were extrapolated to be half of their current value, in accordance with the recent ATLAS and CMS official recommendations for the High-Luminosity studies. All experimental performance related uncertainties (leptons and jets, efficiencies, energy resolutions, etc.) were considered to maintain the current value at 13~TeV, at the exception of the efficiency of tagging jets from the hadronization of $b-$quarks ($b$-tagging), which is expected to be reduced by half. These extrapolated measurements were included in the global fit, in combination with the current measurements, in order to estimate expected limits on the real and imaginary components of the top quark couplings. The allowed regions of the new couplings are presented in Figure~\ref{fig:observables} and Table~\ref{tab:ReCoup}.  Figure~\ref{fig:observables} allows also for a comparison between current LHC results and the HL-LHC projections. 

\begin{figure}
\begin{center}
\includegraphics[width=8cm]{\main/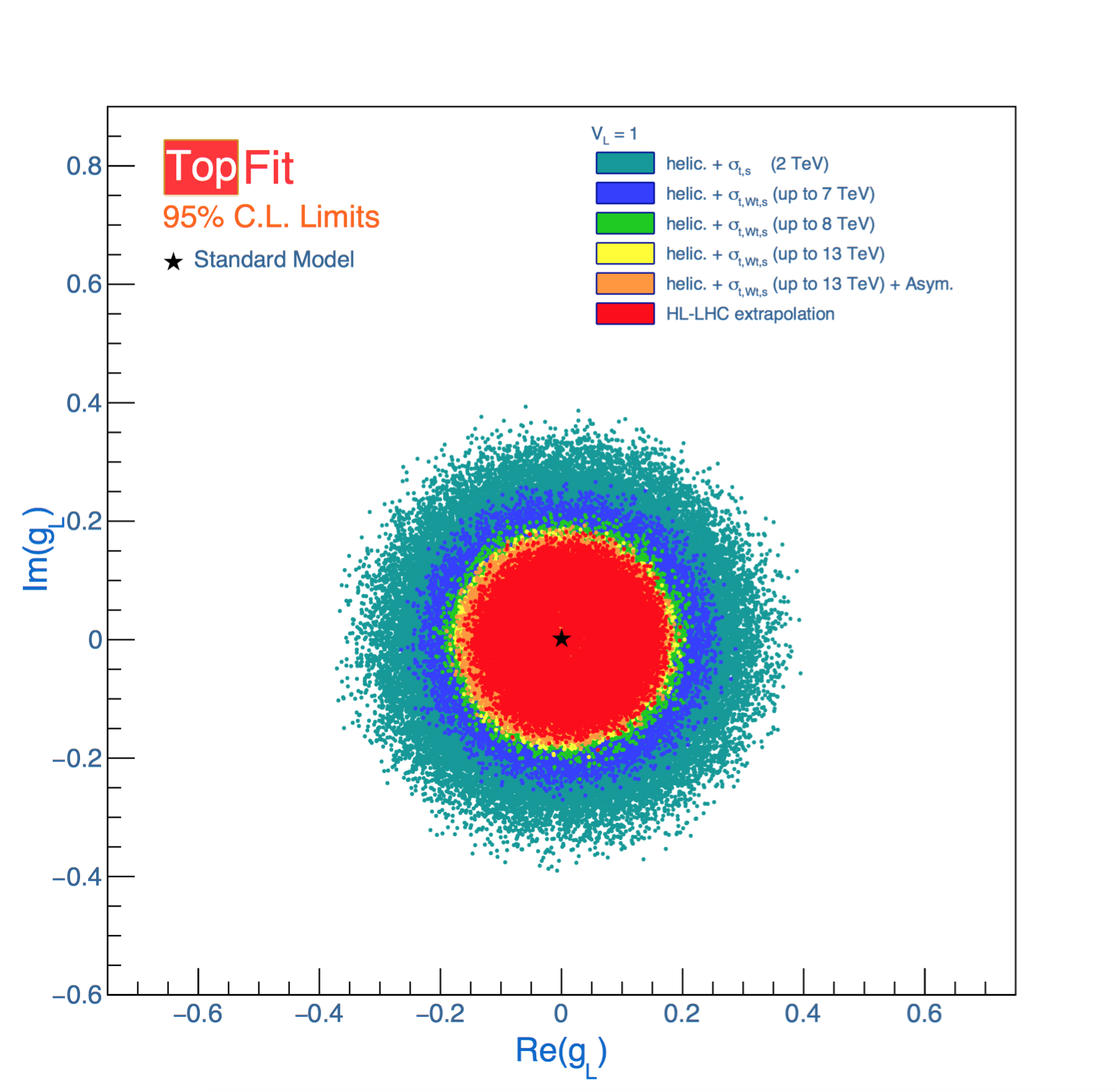}\includegraphics[width=8cm]{\main/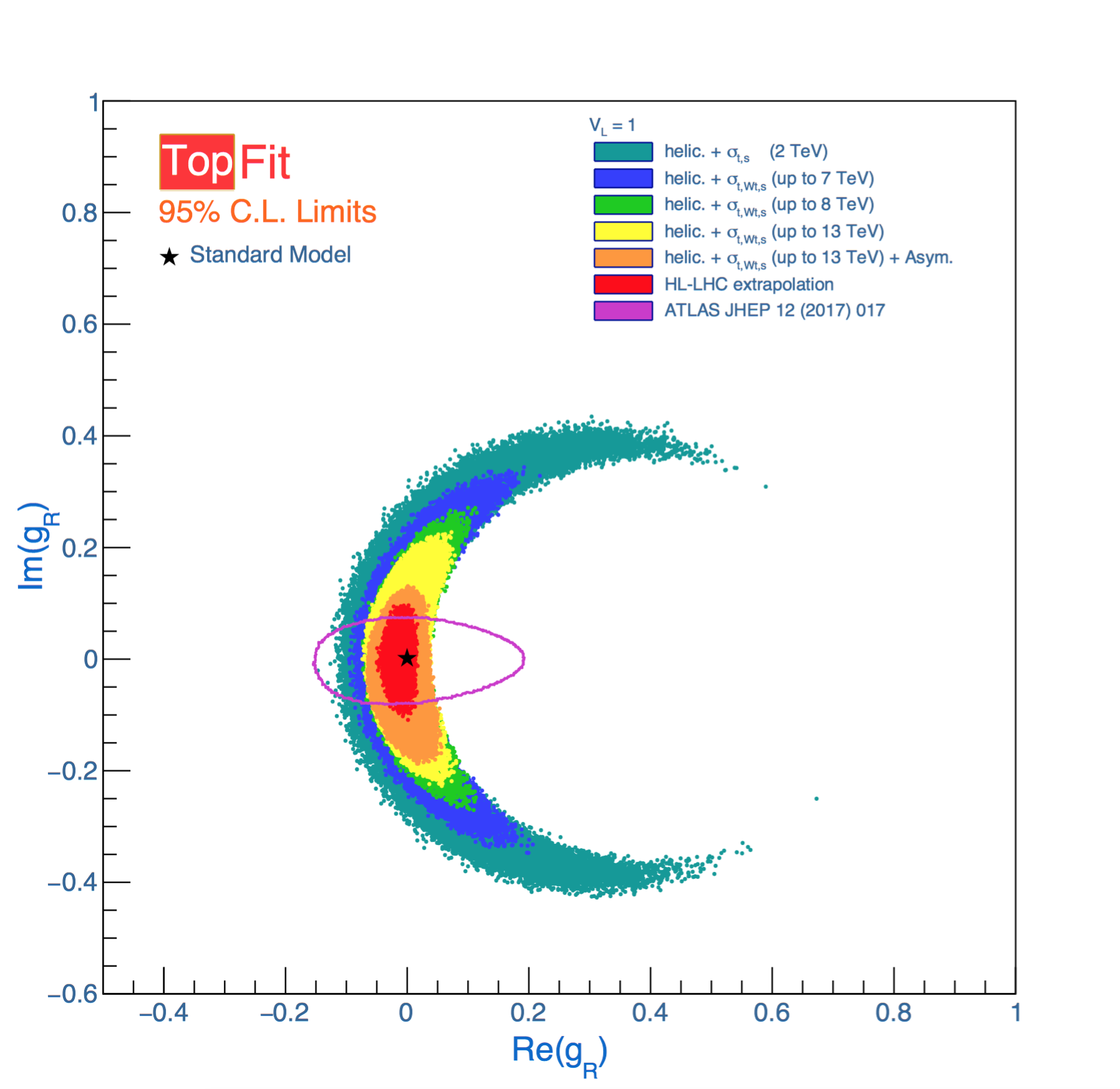} \\
\caption{Limits at 95\% CL on the allowed regions for anomalous couplings\cite{Deliot:2018jts}. The two-dimensional distributions of the $Re$ versus the $Im$ components of $g_{\rm{L}}$ (left) and $g_{\rm{R}}$ (right), are shown.  }
\label{fig:observables}
\end{center}
\end{figure}

\begin{table}[h]
\begin{center}
\caption{Allowed regions for anomalous couplings. \label{tab:ReCoup}}
\begin{tabular}{|c|c|c|c|}
\hline
      HL-LHC &      $\gr$      &       $\gl$        &     $\Vr$ \\
\hline\hline
     Allowed Region ($Re$)   	& [-0.05 , 0.02]  &    [-0.17 , 0.19]  & [-0.28 , 0.32] \\
\hline
     Allowed Region ($Im$)   	& [-0.11 , 0.10]  &    [-0.19 , 0.18]  & [-0.30 , 0.30] \\
\hline
\end{tabular}
\end{center}
\end{table}

\subsubsection{The \ttG~production}
Measurements of \ttG production at the HL-LHC are studied by ATLAS in terms of the expected precision 
for the measurements of fiducial and differential cross sections in leptonic final states
and the expected limits that can be imposed on the Wilson coefficients of operators relevant to \ttG production~\cite{ATL-PHYS-PUB-2018-049}.
These operators are the $\ensuremath{O}_{tB}$, $\ensuremath{O}_{tG}$, and $\ensuremath{O}_{tW}$ in Ref.~\cite{AguilarSaavedra:2018nen}.
The analysis is performed in the same way as the 13~TeV \ttG analysis~\cite{Aaboud:2018hip}, 
by selecting leptonic decay final states of the $t\bar{t}$ pair with an isolated high-$p_{\rm T}$ photon.
Compared to the 13~TeV analysis, data statistical uncertainty is scaled down according to the integrated luminosity at the HL-LHC.
Monte Carlo (MC) statistical uncertainty is ignored as it is expected to have enough MC events generated.
Theoretical uncertainties are reduced by a factor of two due to the expected improvement in the theoretical tools
and background estimation and experimental uncertainties are in general kept the same, with respect to the uncertainties in the 13~TeV analysis.
The fiducial cross-section measurement can reach an uncertainty as low as 3\% (8\%) in the channel with two (one) leptons and requiring a photon candidate with $p_{\rm T}$ larger than 20 (500)~GeV.
The expected uncertainties of differential cross-section measurements, normalised to unity, for several typical observables like the photon $p_{\rm T}$ and $\eta$, 
are found to be in general below 5\%.
The expected uncertainty of the absolute differential cross-section as a function of the photon $p_{\rm T}$ is interpreted as 95\% CL limits for the relevant EFT operators, as shown in Table~\ref{tab:ttGlimits} for single-lepton and dilepton final states.

\begin{table}[H]
\caption{Expected 95~\% CL intervals for the three Wilson coefficients relevant to \ttG production.}\label{tab:ttGlimits}
\begin{center}
\begin{tabular}{|c|c|c|c|}
\hline
Operator & $\ensuremath{\mathcal{O}}_{tB}$ & $\ensuremath{\mathcal{O}}_{tG}$ & $\ensuremath{\mathcal{O}}_{tW}$ \\
\hline
\hline
Single lepton & [-0.5,0.3] &[-0.1,0.1] &[-0.3,0.5] \\
Dilepton & [-0.6,0.4] &[-0.1,0.1] &[-0.4,0.3] \\
\hline
\end{tabular}
\end{center}
\end{table}

\subsubsection{The \ttZ~production}

Many beyond the Standard Model (BSM) predictions include anomalous couplings of the top quark to the electroweak gauge bosons~\cite{Hollik:1998vz,Agashe:2006wa,Kagan:2009bn,Ibrahim:2010hv,Ibrahim:2011im,Grojean:2013qca,Richard:2013pwa}.
While this study is restricted to the \ttZ channel and the CMS HL-LHC detector with a luminosity scenario of 3~ab${}^{-1}$, it goes beyond earlier work~\cite{Rontsch:2015una} and studies the sensitivity of the \ttZ process using differential cross section data\cite{CMS-PAS-FTR-18-036}.
The results are interpreted in terms of the SM effective field theory~\cite{AguilarSaavedra:2018nen} and limits are set on the relevant Wilson coefficients of the Warsaw basis~\cite{Grzadkowski:2010es} \ctZ, \ctZI, \cpt and \cpQM~\cite{Brehmer:2016nyr, Brehmer:2017fyp}.

Events are generated at the parton level at LO using {\mbox{\textsc{MG5\_aMC@NLO}}\xspace}~v2.3.3~\cite{Alwall:2014hca}, and decay them using \MadSpin~\cite{Artoisenet:2012st, Frixione:2007zp}.
Parton showering and hadronization are done using \PYTHIA v8.2~\cite{Sjostrand:2007gs,Sjostrand:2014zea}.
Fast detector simulation was performed using \delphes~\cite{deFavereau:2013fsa}, with the CMS reconstruction efficiency parameterisation for the HL-LHC upgrade.
The mean number of interactions per bunch crossing (pileup, PU) is varied from 0 to 200.
Jets are reconstructed with the \mbox{\textsc{FastJet}}\xspace  package~\cite{Cacciari:2011ma} and using the anti-$k_{\rm T}$ algorithm~\cite{Cacciari:2008gp} with a cone size $R=0.4$.
Besides the signals, the main backgrounds are also generated in the leptonic final states in order to achieve a realistic background prediction.
The \WZ, \tZq, \tWZ, \ttG and \ttZ processes are normalized to cross sections calculated up to next-to-leading order (NLO) in perturbative QCD.

The results on the inclusive \ttZ cross section from ATLAS~\cite{Aad:2015eua,ATLAS:2016ttVarticles} and CMS~\cite{PRL-110-172002,EPJC-C74-2014-9,JHEP-1601-2016-096,Sirunyan:2017uzs} show that the three lepton channel, where the $Z$ and one of the $W$~bosons originating from a top~quark decay leptonically is the most sensitive.
Thus, it is required to have three reconstructed leptons (e~or~$\mu$) with \pT($\Pl$) thresholds of 10, 20, and 40~\GeV, respectively, and $|\eta(\Pl)|<3.0$.
It is furthermore required that there is among them a pair of opposite-sign same-flavor leptons consistent with the \PZ~boson by requiring $|m(\Pl\Pl)-m_\text{\PZ}|<10$~\GeV.
Reconstructed leptons are removed within a cone of $\Delta R < 0.3$ to any reconstructed jet satisfying \pT(\jet) $>$ 30~\GeV.
Furthermore, at least 3 jets are required with \pT(\jet) $>$ 30~\GeV and $|\eta($\jet$)|<4.0$, where one of the jets has been identified as a b-tag jet according to the \delphes specification.

The distributions of the above-mentioned observables are considered in equally sized bins of the transverse $Z$~boson momenta \pTZ~\cite{Rontsch:2014cca} and \cosThetaStar, the relative angle of the negatively charged lepton to the $Z$~boson direction of flight in the rest frame of the boson.
The differential cross sections for the SM (black) and BSM (colored lines) interpretations in \ttZ with respect to \pTZ and \cosThetaStar are shown in Fig.~\ref{fig:ttZ_pt_cos} for \ctZ = 2~($\Lambda$/\TeV)$^2$ and \ctZI = 2~($\Lambda$/\TeV)$^2$.
The BSM distributions are normalized to the SM yield in the plots to visualize the discriminating features of the parameters.
The part of the signal which does not contain information on the Wilson coefficients is shown hatched, backgrounds are shown in solid colors.

\begin{figure}[tbp]
  \centering
    \includegraphics[trim={0.2cm 1.cm 0.2cm 0.cm},clip,width=0.40\textwidth]{\main/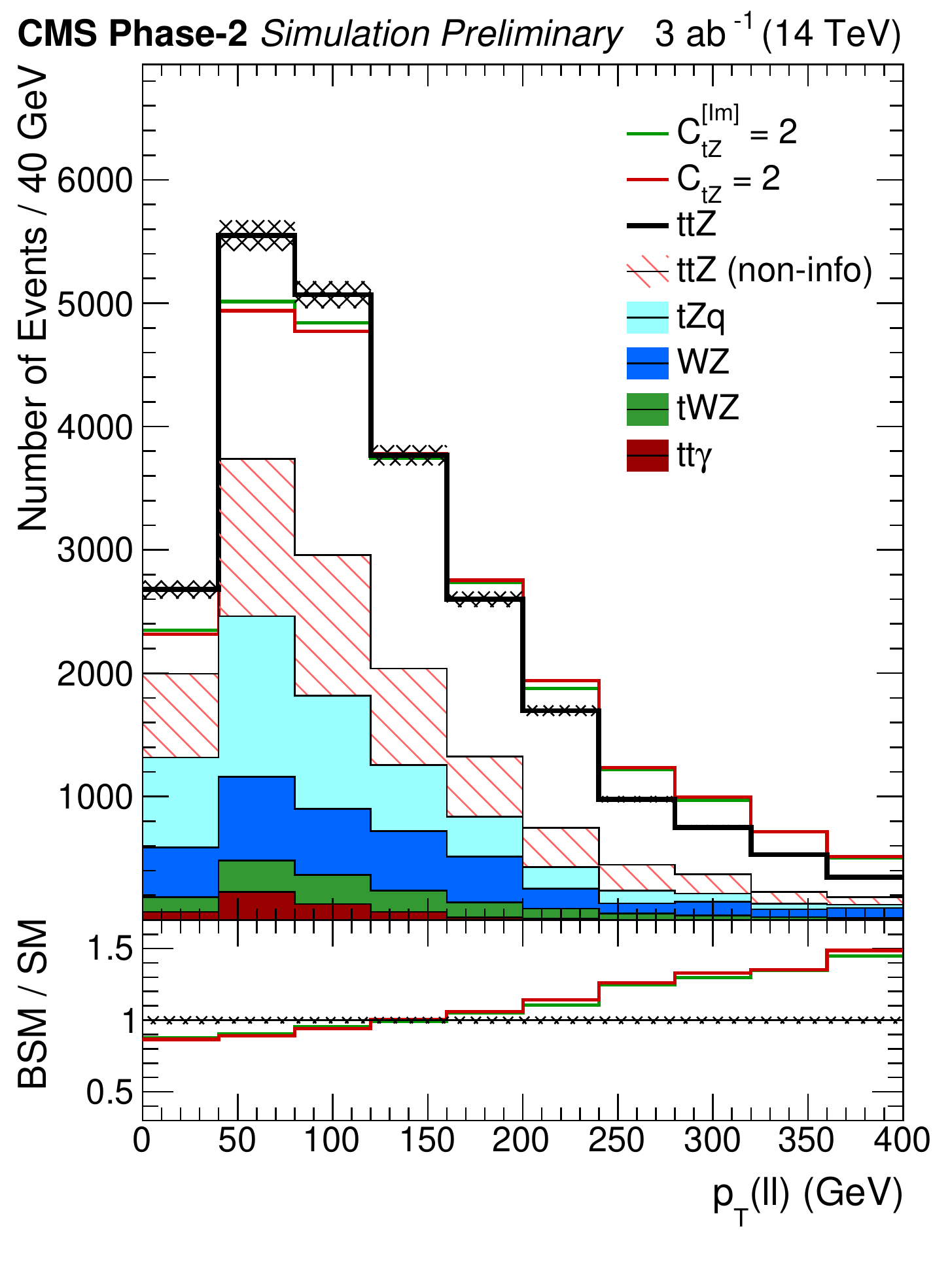}
    \hspace{1.7cm}
    \includegraphics[trim={0.2cm 1.cm 0.2cm 0.cm},clip,width=0.40\textwidth]{\main/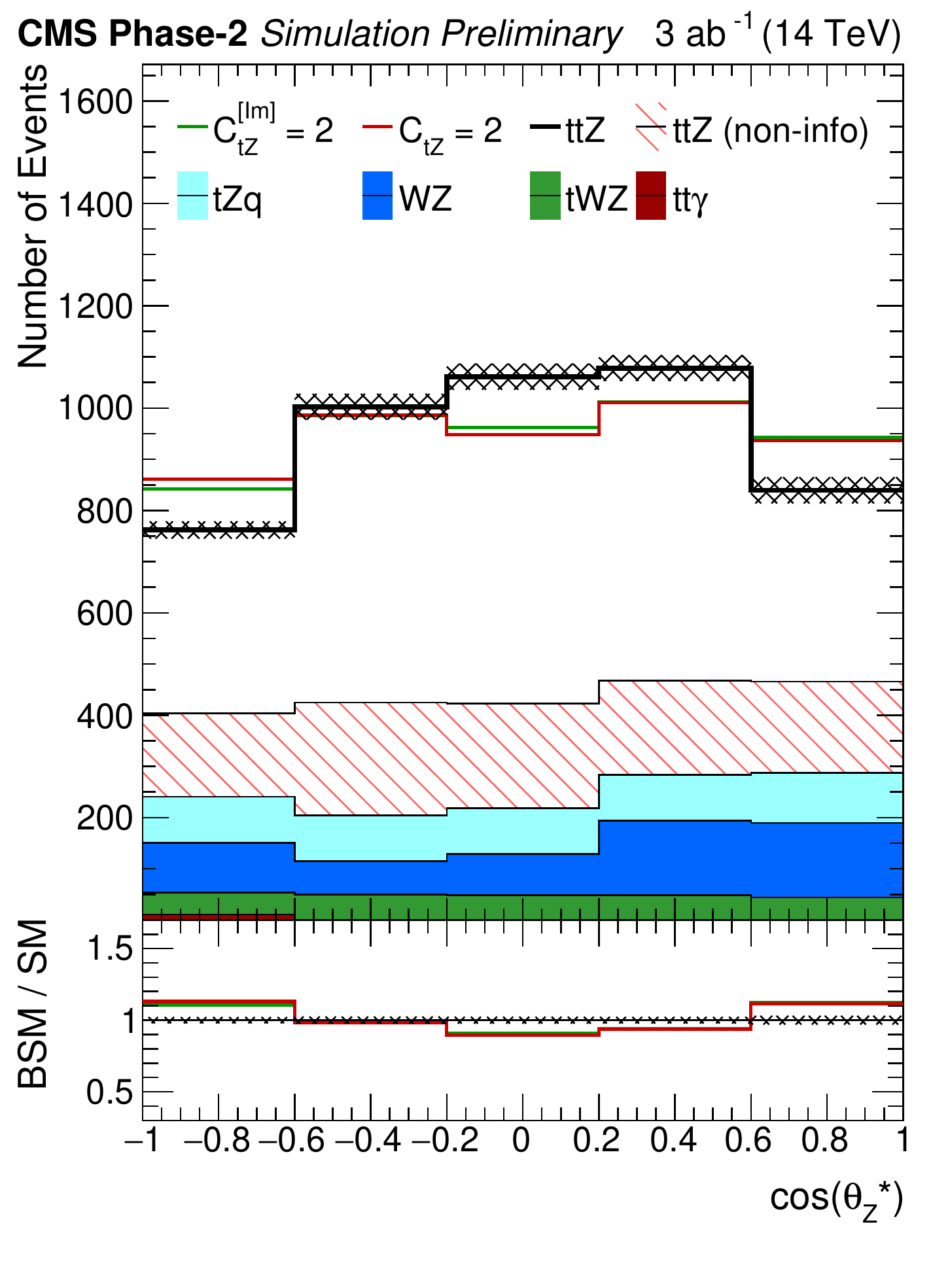}
    \hspace{.3cm}
\caption{
Differential cross sections of \pTZ (left) and \cosThetaStar (right) for the in the text mentioned selection and the HL-LHC scenario.
For \cosThetaStar, additionally \pTZ $>$ 200~\GeV is applied. 
}\label{fig:ttZ_pt_cos}
\end{figure}

The predicted yields are estimated for the 3~ab${}^{-1}$ HL-LHC scenario at $\sqrt{s}=13$~\TeV and scaled to 14~\TeV, where an additional small background from non-prompt leptons is taken from Ref.~\cite{Sirunyan:2017uzs} and scaled to 3~ab${}^{-1}$.
A profiled maximum likelihood fit of the binned likelihood function $L(\theta)$ is performed and it is considered $q(r)=-2\log(L(\hat{\theta})/L(\hat{\theta}_{\textrm{SM}}))$, where $\hat{\theta}$ and
$\hat{\theta}_\textrm{SM}$ are the set of nuisance parameters maximizing $L(\theta)$ at the BSM and SM point, respectively.
Experimental uncertainties are estimated based on the expected performance of the HL-LHC CMS detector.
In Table~\ref{tab:limits}, the 68\% and 95\% CL intervals of the likelihood scan for the \ttZ process are shown, where one non-zero Wilson coefficient is considered at a time, and all others are set to zero.

Table~\ref{tab:proflimits} shows the 68\% and 95\% CL intervals of the likelihood ratios for two pairs of Wilson coefficients corresponding to modified neutral current interactions (\cpt and \cpQM) and dipole moment interactions (\ctZ and \ctZI).
The corresponding second Wilson coefficient is included in the profiling of nuisance parameters.

In Fig.~\ref{fig:ttZ_nll}, the log-likelihood scan for the \ttZ process is shown in the \cpQM/\cpt parameter plane~(left) and the dipole moment parameter plane \ctZ/\ctZI~(right).
The green (red) lines show the 68\% (95\%) CL contour line and the SM parameter point corresponds to \cpt~=~\cpQM~=~0 and \ctZ~=~\ctZI~=~0.

\begin{table}[H]
\caption{Expected 68~\% and 95~\% CL intervals, where one Wilson coefficient at a time is considered non-zero.}\label{tab:limits}
\begin{center}
\begin{tabular}{|c|c|c|}
\hline
Wilson coefficient & 68~\% CL $(\Lambda/$TeV$)^2$ & 95~\% CL $(\Lambda/$TeV$)^2$ \\
\hline
\hline
\cpt   & [-0.47, 0.47]                 & [-0.89, 0.89]                 \\
\hline
\cpQM  & [-0.38, 0.38]                 & [-0.75, 0.73]                 \\
\hline
\ctZ   & [-0.37, 0.36]                 & [-0.52, 0.51]                 \\
\hline
\ctZI  & [-0.38, 0.36]                 & [-0.54, 0.51]                 \\
\hline
\end{tabular}
\end{center}
\end{table}

\begin{table}[H]
\caption{Expected 68~\% and 95~\% CL intervals for the selected Wilson coefficients in a profiled scan over the 2D parameter planes \cpQM/\cpt and \ctZ/\ctZI. The respective second parameter of the scan is left free.}\label{tab:proflimits}
\begin{center}
\begin{tabular}{|c|c|c|}
\hline
Wilson coefficient & 68~\% CL $(\Lambda/$TeV$)^2$ & 95~\% CL $(\Lambda/$TeV$)^2$ \\
\hline
\hline
\cpt   & [-1.65, 3.37]                 & [-2.89, 6.76]                 \\
\hline
\cpQM  & [-1.35, 2.92]                 & [-2.33, 6.69]                 \\
\hline
\ctZ   & [-0.37, 0.36]                 & [-0.52, 0.51]                 \\
\hline
\ctZI  & [-0.38, 0.36]                 & [-0.54, 0.51]                 \\
\hline
\end{tabular}
\end{center}
\end{table}

\begin{figure}[H]
  \centering
    \includegraphics[trim={0.4cm 0.cm 0.8cm 0.cm},clip,width=0.45\textwidth]{\main/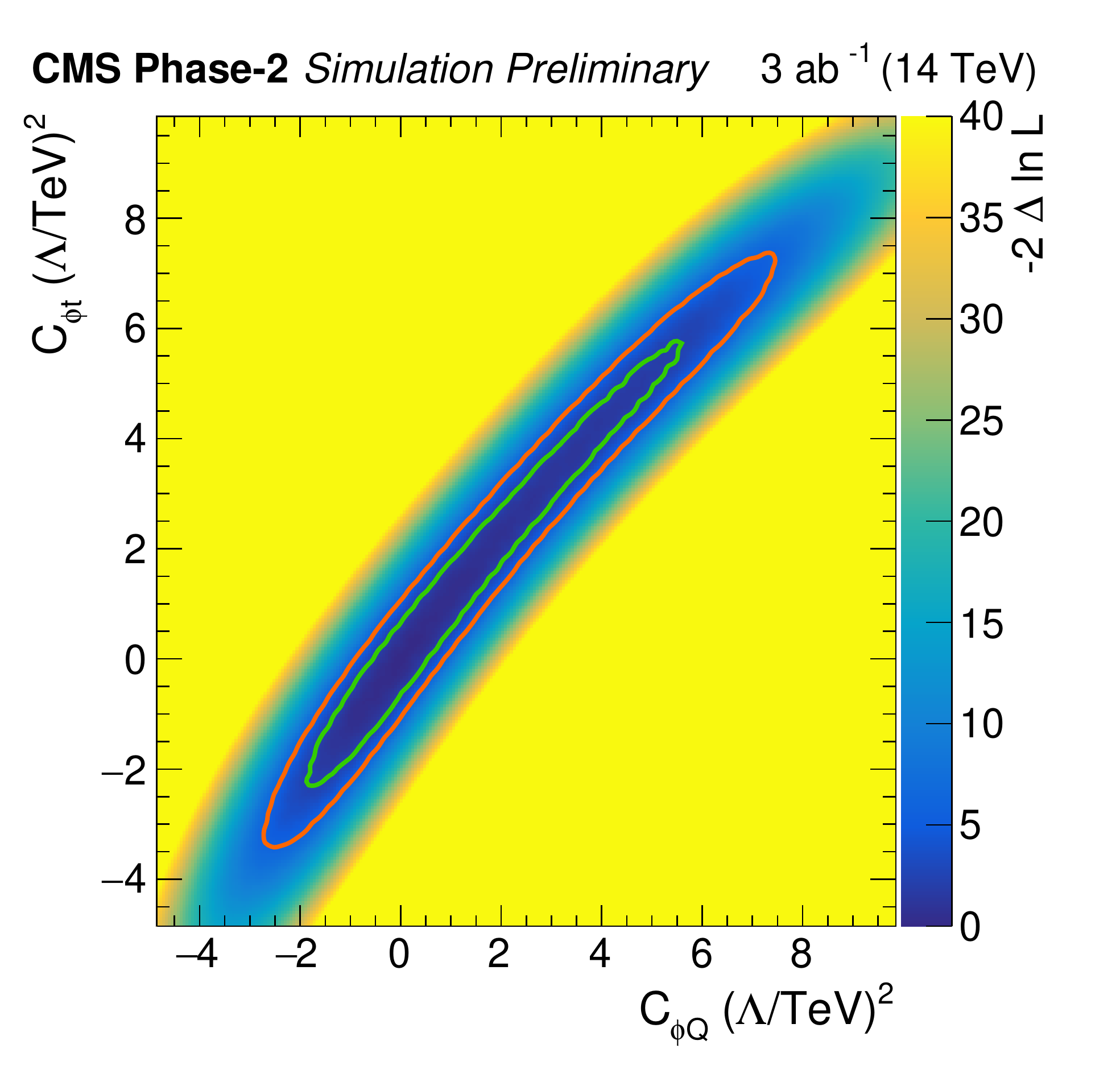}
    \includegraphics[trim={0.4cm 0.cm 0.8cm 0.cm},clip,width=0.45\textwidth]{\main/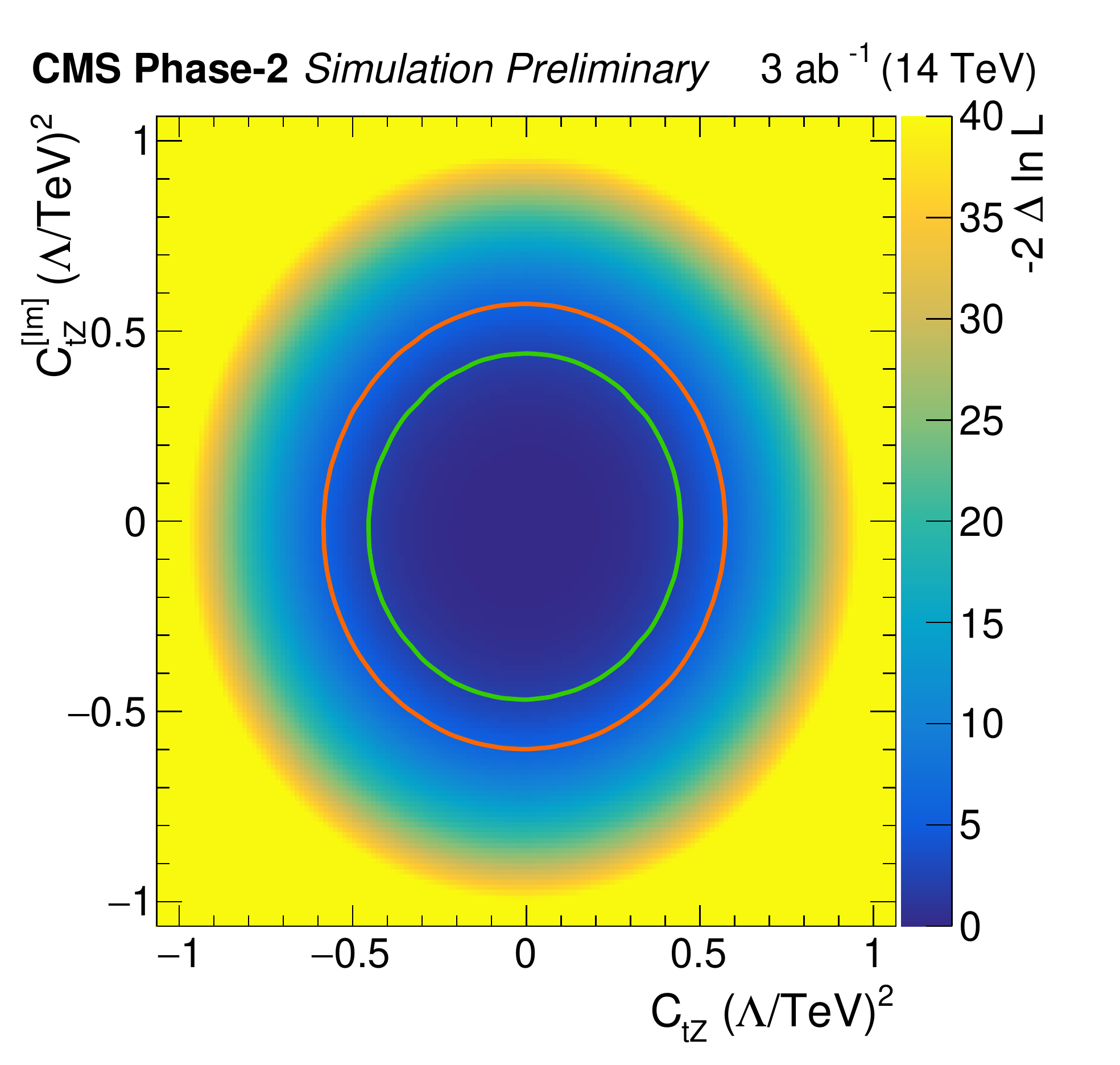}
  \caption{Scan of the negative likelihood in the \cpQM/\cpt (left) and \ctZ/\ctZI parameter planes (right) for the \ttZ process under the SM hypothesis.
           The 68\% (95\%) CL contour lines are given in green (red).
           }
  \label{fig:ttZ_nll}
\end{figure}

\newpage
\clearpage

\section{Forward physics}


\renewcommand{\bs}[1]{\boldsymbol{#1}}
\def\d{\mathrm{d}}
\def\be{\begin{equation}}
\def\ee{\end{equation}}

\newcommand*\widefbox[1]{\fbox{\hspace{1em}#1\hspace{em}}}
\newcommand{\eq}[1]{eq.~\eqref{#1}}

\setlength{\textheight}{8.35truein}
\setlength{\textwidth}{6.53truein}
\setlength{\topmargin}{-0.2truein}
\setlength{\oddsidemargin}{-0.truein}
\setlength{\evensidemargin}{\oddsidemargin}


\subsection[Photon-induced collisions at the HL--LHC]{Photon-induced collisions at the HL--LHC\footnote{Section edited by L. Harland-Lang.}}



\begin{figure}
\centering
\includegraphics[width=0.4\textwidth]{\main/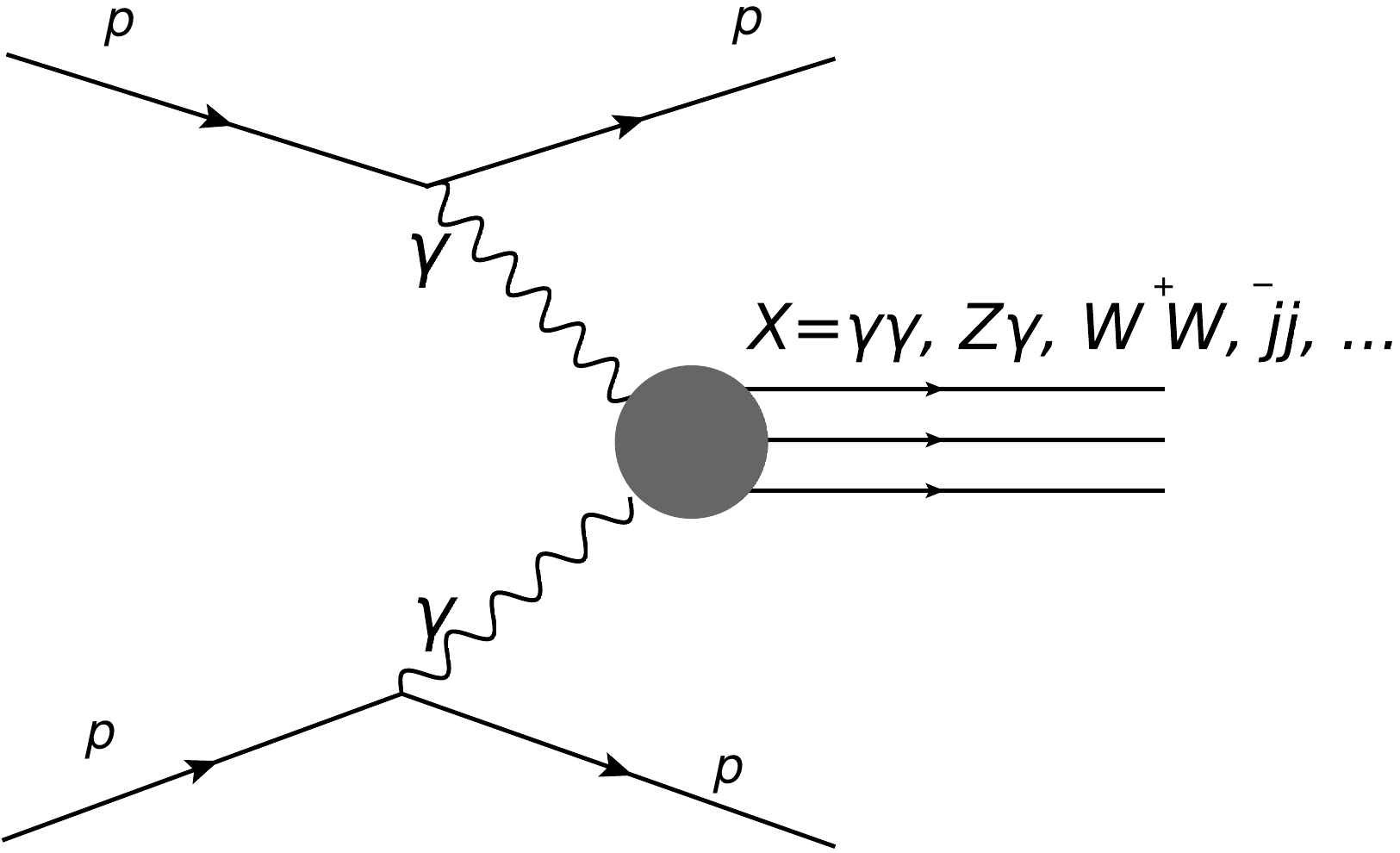}\quad
\includegraphics[width=0.48\textwidth]{\main/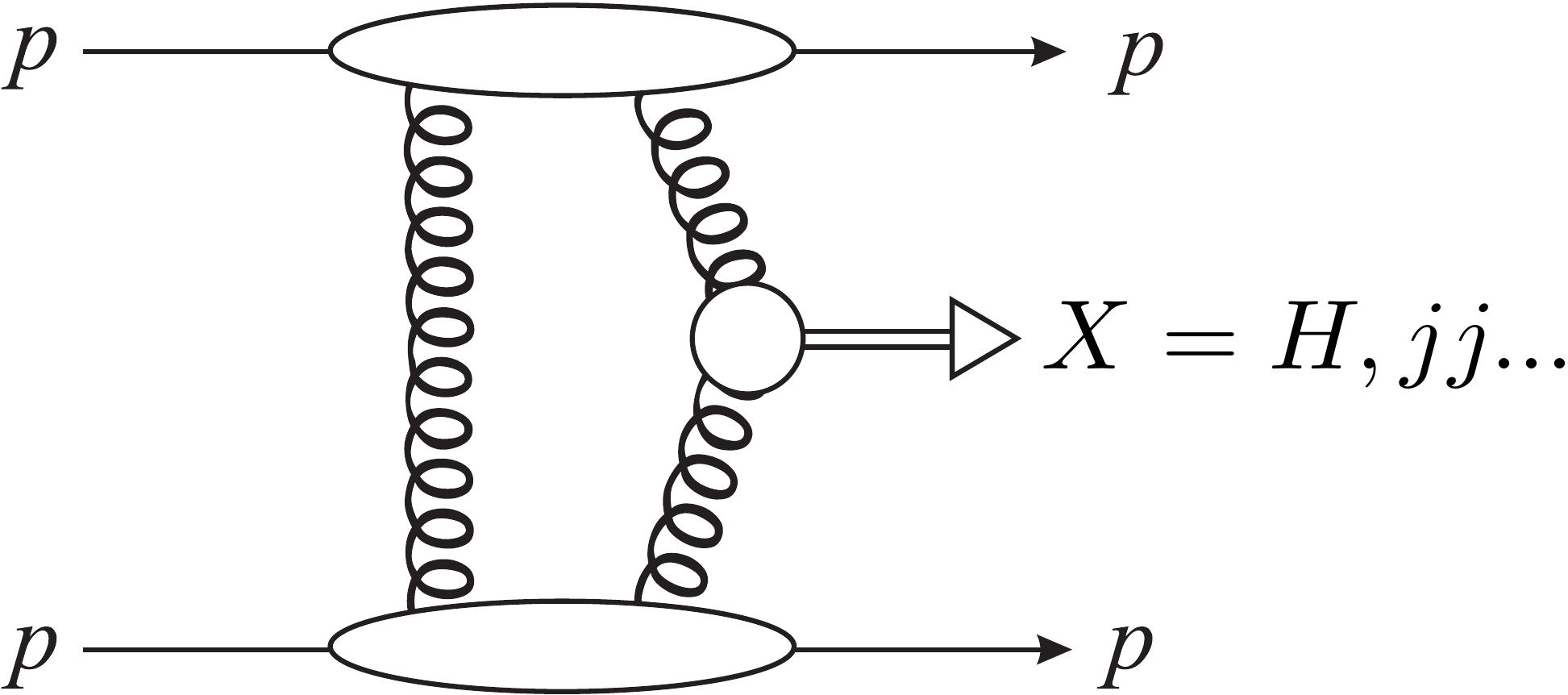}
\caption{\label{exclusive_production} Schematic diagram of the production of a system $X$ in (left) two--photon (right) QCD--initiated central exclusive production.}
\end{figure}

Central exclusive production (CEP) corresponds to the production of a central system $X$, and nothing else, with two outgoing intact protons:
\begin{equation}
\label{eqn:exclusive_production}
pp \to p \;+\; X \;+\;p\;.
\end{equation}
Such a process may be mediated by photon exchange, with the elastic photon emission vertex leaving the protons intact, see Fig.~\ref{exclusive_production} (left). A range of SM (e.g. $X = \gamma\gamma$, $Z\gamma$, $ZZ$, $\ell\bar{\ell}$) and BSM states (e.g. $X=$ axion--like particles,  monopoles, SUSY particles) may be produced in this way. These have the benefit of:

\begin{itemize}

\item The theoretical framework to model the underlying production mechanism, based on the equivalent photon approximation~\cite{Budnev:1974de}, is very well understood. Moreover, due to the peripheral nature of the interaction the possibility for additional inelastic proton--proton interactions (in other words of multiple--particle interactions) is very low. 

\item As the mass of the central system increases, the relative size of any contribution from QCD--initiated production, see section~\ref{sec:qcdcep}, becomes increasingly small~\cite{Khoze:2001xm}, due to the strong Sudakov suppression in vetoing on additional QCD radiation.

\end{itemize}

CEP therefore offers a unique opportunity at the LHC to observe the purely photon--initiated production of electromagnetically charged objects at the LHC in a clean and well understood environment; in this context the LHC is effectively used as a photon--photon collider. The cross sections for such processes can be relatively small, in particular at higher mass, and therefore to  select such events it is essential to run during nominal LHC running with tagged protons. The increased statistics available during the HL--LHC stage will allow these studies to push to higher masses and lower cross sections, increasing the discovery potential. A detailed study for the example case of anomalous quartic gauge couplings is discussed below. 

\subsubsection[Anomalous quartic gauge couplings with proton tagging at the HL--LHC]{Anomalous quartic gauge couplings with proton tagging at the HL--LHC\footnote{Contribution by C. Baldenegro and C. Royon.}}




	This section discusses the discovery potential of anomalous quartic gauge interactions at the LHC via the measurement of central exclusive production (see Refs.~\cite{Baldenegro:2018hng,Baldenegro:2017aen,Fichet:2016pvq,Fichet:2015vvy,Fichet:2014uka,Fichet:2013gsa,Chapon:2009hh,Kepka:2008yx}).
	The central system $X$ is reconstructed in the central detector (CMS, ATLAS) while the outgoing protons, which remain intact due to the coherent photon exchange, can be reconstructed with dedicated tracking detectors located in the very forward region at about $\pm$ 210 m (220 m) with respect to the interaction point of the CMS (ATLAS) experiment. The fractional momentum loss of the outgoing protons $\xi = \Delta p/p$ is reconstructed offline.
	 Central exclusive production processes satisfy $m_X = m_{X}^\text{fwd} = \sqrt{\xi_1\xi_2 s}$ and $y_X = y_{X}^\text{fwd} = \frac{1}{2}\log \big( \frac{\xi_1}{\xi_2} \big)$, where $m_X$ and $y_X$ are the mass and rapidity of the system $X$ reconstructed with the central detector, $m_{X}^\text{fwd}$ and $y_{X}^\text{fwd}$ are the mass and rapidity of the system $X$ reconstructed with the forward detectors and $\sqrt{s}$ is the proton-proton centre-of-mass energy. This relationship sets a powerful offline selection tool for background suppression, since non-exclusive events are not correlated to the forward protons.\\

 In these projections, it is assumed that a similar set-up as with the CT-PPS and AFP detectors is possible at the HL-LHC. An overview of the physics case for light-by-light scattering is given as the prototype example, and the quartic $\gamma\gamma\gamma Z$ coupling is given as an instance of other gauge couplings that could be studied at the HL-LHC. These projections consider also the impact of the difference of the measured time-of-flight for the intact protons with various timing precisions (on the order of 10 ps), which can be used to determine the longitudinal coordinate of the event vertex down to $\sim 2 $ mm. Time-of-flight measurements can help further reduce the background, especially at the HL-LHC where the number of interactions per bunch crossing will range from 140-200.\\


\subsubsubsection*{Scattering of light-by-light in p-p collisions}


Under the assumption that there exists a New Physics energy scale $\Lambda$ much heavier than the experimentally accessible energy $E$, new physics manifestations can be described using an effective Lagrangian valid for $\Lambda \gg E$. Among these operators, the pure photon dimension-eight operators $\mathcal{L}_{4\gamma}= 
\zeta_1^{4\gamma} F_{\mu\nu}F^{\mu\nu}F_{\rho\sigma}F^{\rho\sigma}
+\zeta_2^{4\gamma} F_{\mu\nu}F^{\nu\rho}F_{\rho\lambda}F^{\lambda\mu}$ induce the $\gamma\gamma\gamma\gamma$ interaction. This coupling can be probed in $pp\rightarrow p(\gamma\gamma\rightarrow\gamma\gamma)p$ reactions. This sub-process and the SM light-by-light scattering one are implemented in the Forward Physics Monte Carlo (FPMC)~\cite{Boonekamp:2011ky} event generator. The Equivalent Photon Approximation is used to calculate the emitted coherent photon flux off the protons.\\

With proton tagging, one can probe $\gamma\gamma\rightarrow\gamma\gamma$ collisions from about 300 GeV to 2 TeV. The mass acceptance on the photon pair is limited mainly by the acceptance of $\xi$ of the proton taggers ($0.015\leq \xi \leq 0.15$).
 The background is dominated by non-exclusive diphoton production events overlapped with uncorrelated events with intact protons coming from the secondary collisions occurring in the same bunch crossing. This background can be suppressed by looking at the central and forward systems kinematic correlations (the aforementioned mass and rapidity matching). The irreducible background coming from the SM exclusive diphoton production is negligible at large invariant masses. The background can be further suppressed if the time-of-flight difference of each of the scattered protons is measured. The precision of the event vertex longitudinal position determined with the time-of-flight measurement is given by $\delta z = c\,\delta t /\sqrt{2}$, where $c$ is the speed of light and $\delta t$ is the timing precision. In these projections, the average pileup of 200 collisions per bunch crossing was considered.\\

The expected bounds on the anomalous couplings $\zeta_{1,2}^{4\gamma}$ at 95\% CL are calculated based on the total expected background rate and can be seen in Fig. \ref{fig:sensitivity_4photon}. The reach on the quartic couplings $\zeta^{4\gamma}_{1,2}$ down to $5\cdot 10^{-14}$ GeV$^{-4}$ with 300 fb$^{-1}$ at 14 TeV, and down to $1\cdot 10^{-14}$ GeV$^{-4}$ at the HL-LHC with a luminosity of 3000 fb$^{-1}$ without using time-of-flight information. The last bound can be improved by a factor of $\sim$ 1.2 if the timing precision is of 10 ps.
\begin{figure}
\centering
\includegraphics[scale=0.15]{\main/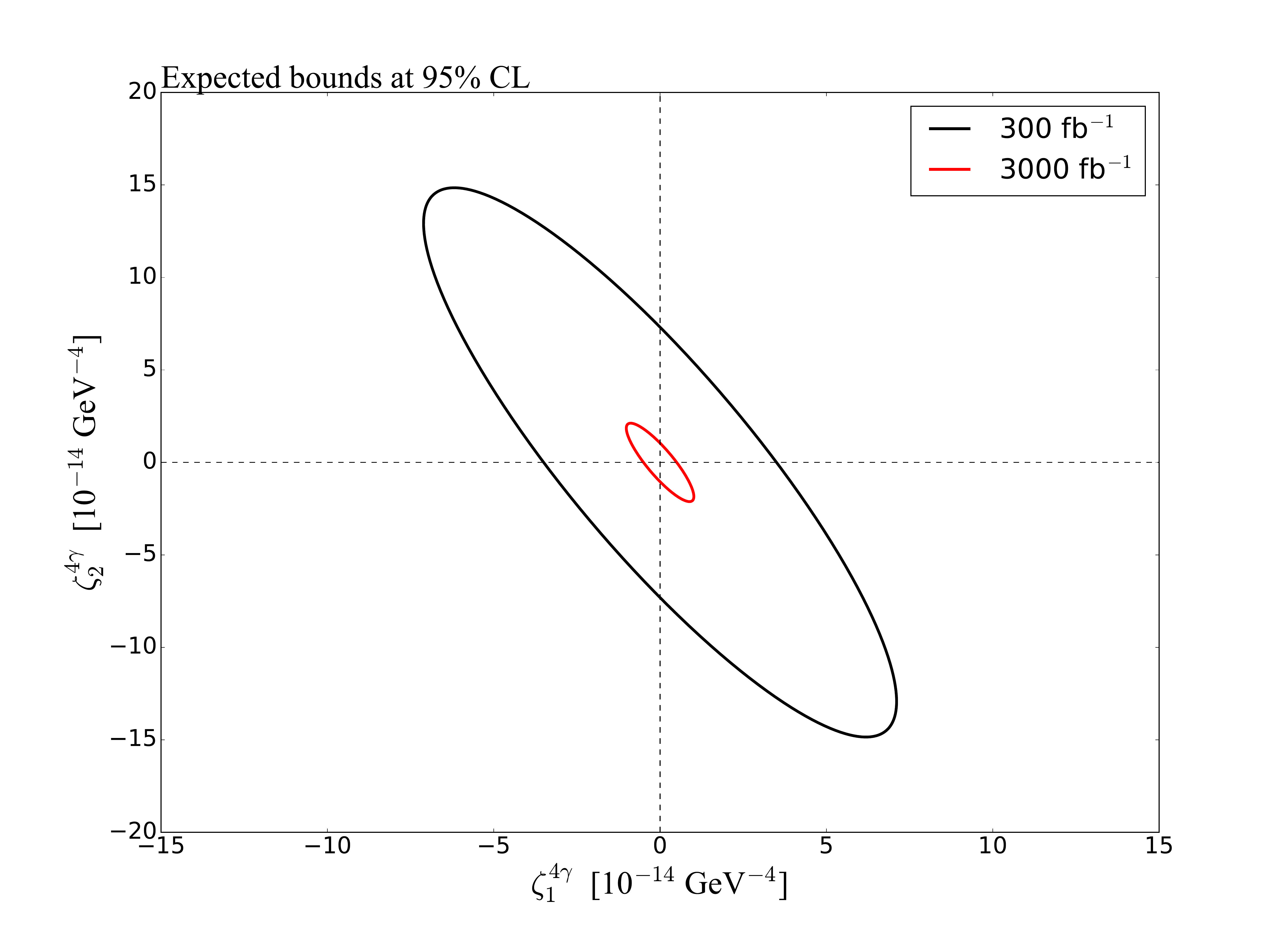}
\includegraphics[scale=0.135]{\main/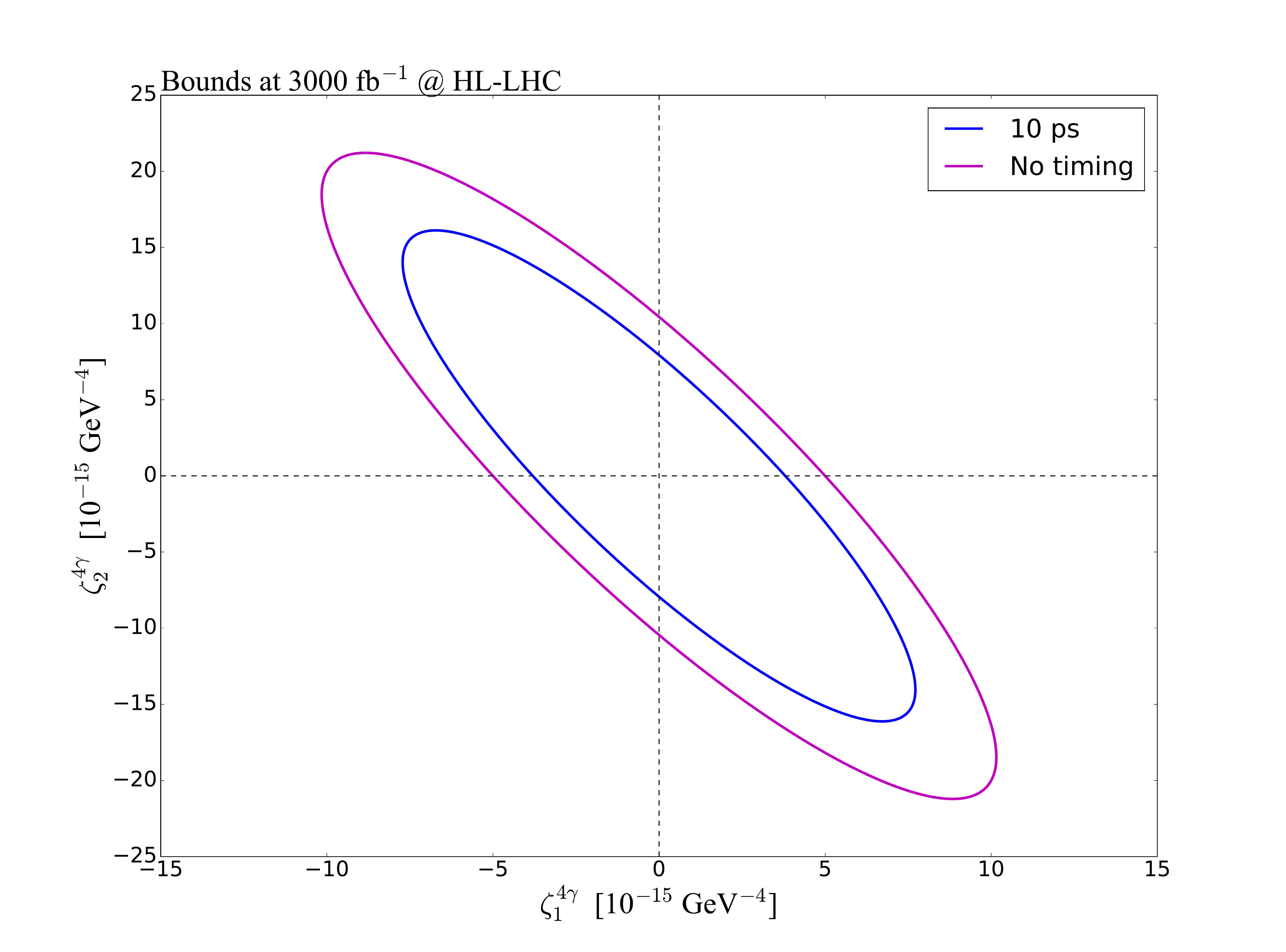}
\caption{\label{fig:sensitivity_4photon} Expected bounds at 95\% CL on the anomalous quartic coupling for 300 fb$^{-1}$ and at the HL-LHC with 3000 fb$^{-1}$ (no time-of-flight measurement) (left). Expected bounds at 95\% CL on the anomalous couplings at the HL-LHC with time-of-flight measurement with precision of 10 ps and without time-of-flight measurement (right).}
\end{figure}

\subsubsubsection*{Constraining $\gamma\gamma\gamma Z$ coupling via $pp\rightarrow p(\gamma \gamma\rightarrow \gamma Z)p$}

The $\gamma\gamma\gamma Z$ interaction is induced at one-loop level in the SM via loops of fermions and $W^\pm$ bosons. Loops of heavy particles charged under SU(2)$_L \times$ U(1)$_Y$ contribute to the $\gamma\gamma\gamma Z$ couplings. The dimension-eight effective operators are  $\mathcal{L}_{\gamma\gamma\gamma Z}=\zeta_1^{3\gamma Z} F^{\mu\nu}F_{\mu\nu}F^{\rho\sigma}Z_{\rho\sigma}+\zeta_2^{3\gamma Z} F^{\mu\nu} \tilde{F}_{\mu\nu}F^{\rho\sigma}\tilde{Z}_{\rho\sigma}$ , which induce the $\gamma\gamma\gamma Z$ interaction. This induces the anomalous $\gamma\gamma \rightarrow \gamma Z$ scattering and generates the rare SM decay $Z\rightarrow\gamma\gamma\gamma$. This coupling can be probed in $pp\rightarrow p(\gamma\gamma\rightarrow\gamma Z)p$ reactions. The sub-process was implemented in the FPMC event generator as well.\\


Since the exclusive channel is very clean, it allows the possibility of studying exclusive $Z\gamma$ production with the $Z$ boson decaying into a charged lepton pair or to hadrons (dijet or large radius jet signature). The signature $(Z\rightarrow\ell\bar{\ell})+\gamma$ is much cleaner, but has vastly fewer events than $(Z\rightarrow \text{hadrons})+\gamma$ final states. A similar event selection is applied on the exclusive $Z\gamma$ production as in the exclusive $\gamma\gamma$ case.
 The sensitivity on the anomalous coupling at 95\% CL combining both channels at 14 TeV with 300 fb$^{-1}$ of data is on the order of 1 $\cdot$ 10$^{-13}$ GeV $^{-4}$ (see Fig. \ref{fig:sensitivity_3gammaZ}). For the HL-LHC with 3000 fb$^{-1}$ it scales down to $1\cdot 10^{-14}$ GeV$^{-4}$ when combining both channels. The time-of-flight measurement can improve the expected bounds by a factor of $\sim 2$	.\\



\begin{figure}
\centering
\includegraphics[scale=0.15]{\main/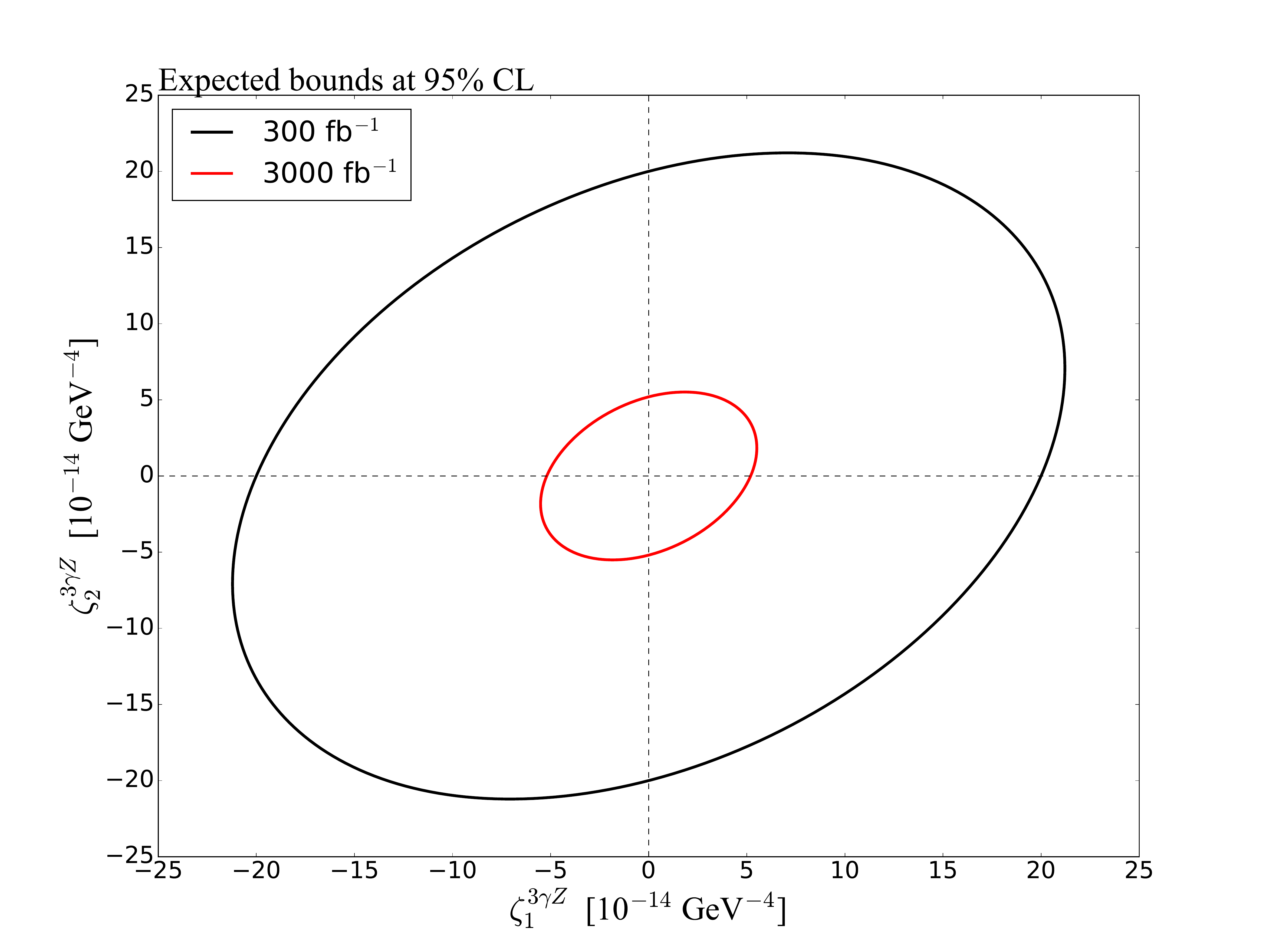}
\includegraphics[scale=0.15]{\main/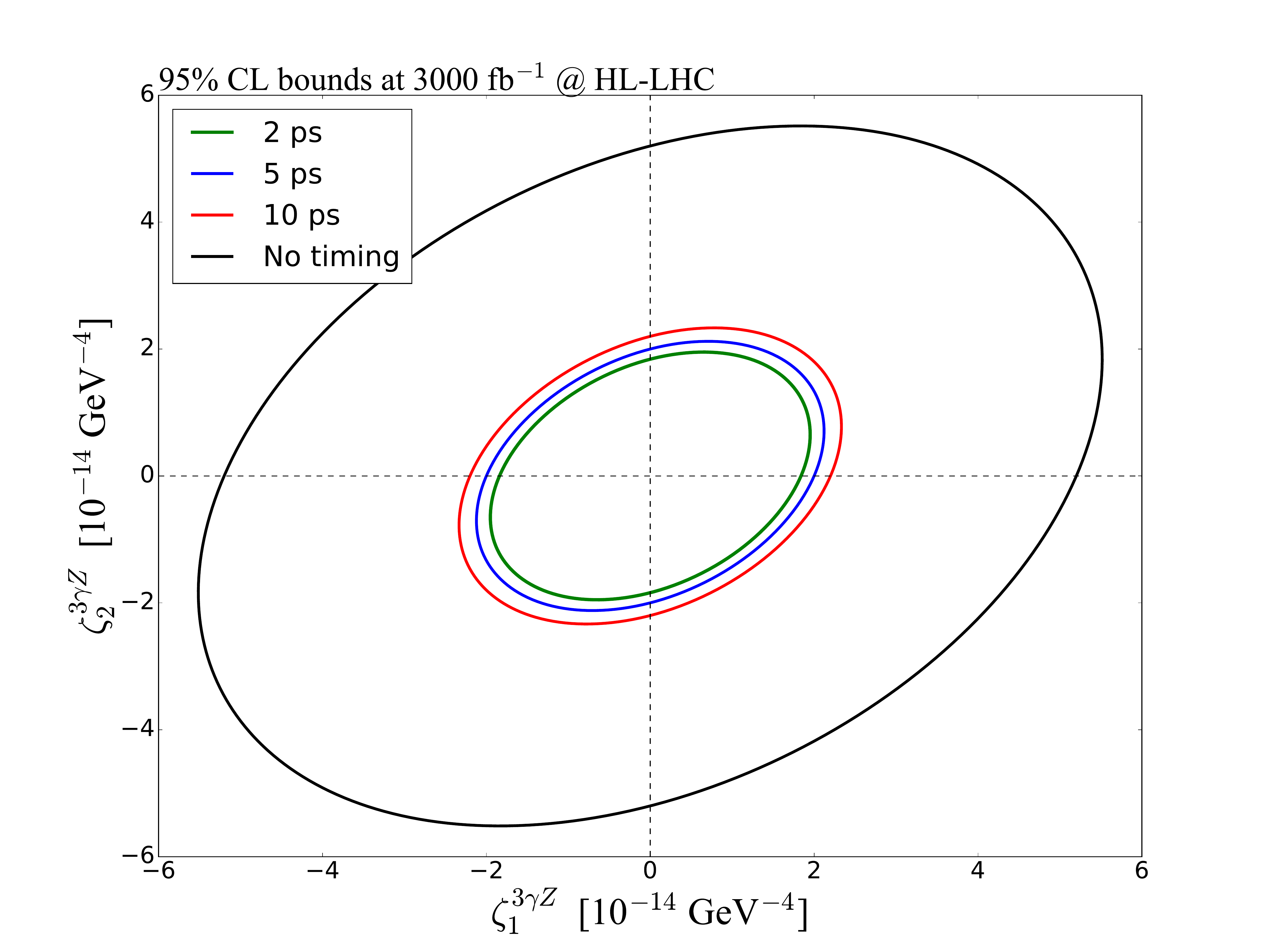}
\caption{\label{fig:sensitivity_3gammaZ} Expected bounds on the anomalous couplings at 95\% CL with 300 fb$^{-1}$ and 3000 fb$^{-1}$ at the HL-LHC (no time-of-flight measurement) (left). Expected bounds at 95\%CL for timing precisions of $\delta t = 2,\,5,\,10$ ps at the HL-LHC (right).}
\end{figure}

\subsection[Central exclusive production: QCD prospects]{Central exclusive production: QCD prospects\footnote{Contribution by L. Harland-Lang.}}
\label{sec:qcdcep}


The CEP process may be mediated purely by the strong interaction, and in such a case if the mass of the central system is large enough a perturbative approach may be applied, via the diagram shown in Fig.~\ref{exclusive_production} (right), see~\cite{Albrow:2010yb,Harland-Lang:2014lxa} for reviews. As well as probing QCD in a novel regime, the exclusive nature of this process has the benefit that the produced object obeys a quantum number selection rule. Namely the object must be $C$ even, while the production of $P$ even states with $J_z=0$ angular momentum projection on the beam axis is strongly dominant. From the point of view of the production of new BSM states or the understanding of existing QCD bound states (e.g. exotic quarkonia) this therefore has the benefit of identifying the produced object quantum numbers. The $J_z=0$ selection implies that only certain helicity configurations in the underlying $gg \to X$ production process contribute, which also leads to unique phenomenological consequences. A detailed discussion of this selection rule can be found in~\cite{Harland-Lang:2014lxa} and the references therein. Two example processes, namely exclusive jet and Higgs boson production, are discussed briefly below. These represent higher mass test cases relevant to HL--LHC running with tagged protons at ATLAS or CMS. The possibilities for the observation of lower mass objects with the ALICE detector will be addressed in section~\ref{sec:ALICECEP}.

The exclusive production of jets provides a new and unexplored area of QCD phenomenology. 
This process has been first observed at the Tevatron~\cite{Aaltonen:2007hs,Abazov:2010bk}.
The quantum number selection rule discussed above has a number of consequences that are quite distinct from the standard inclusive channels. In particular, the production of purely gluonic dijets is predicted to be strongly dominant, allowing a study of purely $gg$ jets from a colour--singlet initial state. In the three--jet case the presence of `radiation zeros'~\cite{Harland-Lang:2015faa}, that is a complete vanishing in the leading order amplitudes for certain kinematic configurations, is expected. This phenomena is well known in electroweak processes, but this is the only known example of a purely QCD process where this occurs. Some representative predictions for the HL--LHC are shown in Table~\ref{table:jets}. These are calculated using the \textsc{SuperChic 2.5} MC generator~\cite{Harland-Lang:2015cta}, which provides the most up to date predictions for CEP processes. The cross sections are suppressed relative to the inclusive case, but are nonetheless relatively large. On the other hand, in the three jet case, in particular in the invariant mass region that may be relevant for the acceptance of proton tagging detectors, the cross sections are lower and would clearly benefit from as large a data sample as possible for studies of novel features, such as radiation zeros and other jet shape variables.

\begin{table}
\begin{center}
\caption{Parton--level predictions for exclusive two and three jet production cross sections (in pb) at the LHC for different cuts on the minimum central system invariant mass $M_X$ at $\sqrt{s}=14$ TeV. The jets are required to have transverse momentum $p_{\mathrm{T}}>$ 20 GeV for $M_X({\rm min})=75,150$ GeV and $p_{\mathrm{T}}>$ 40 GeV for $M_X({\rm min})=250$ GeV and pseudorapidity $|\eta|<2.5$.  The anti-$k_{\rm T}$ algorithm with jet radius $R=0.6$  is used in the three jet case and the $q\overline{q}$ cross sections correspond to one massless quark flavour.}
\label{table:jets}
\begin{tabular}{|c|c|c|c|c|c|}
\hline
$M_X({\rm min})$ & $gg$ & $q\overline{q}$ & $b\overline{b}$ & $ggg$ & $gq\overline{q}$ \\ 
\hline\hline
75 & 130 & 0.032 & 0.082 &5.0 &0.11\\
150 &4.5  & $6.1 \times 10^{-4}$ &  $1.1 \times 10^{-3}$ &0.70 & 0.019\\
250 & 0.15 & $2.2 \times 10^{-5}$ &$2.7 \times 10^{-5}$   &0.016 &$4.3 \times 10^{-4}$ \\
\hline
\end{tabular}
\end{center}
\end{table}

The production of the Higgs boson through exclusive $gg$ fusion would represent a completely new observation channel. As discussed in more detail in~\cite{Harland-Lang:2014lxa}, this has the potential to shed light on the $CP$ properties of the state, as well as its coupling to $b$ quarks in a distinct way to inclusive channels. The cross section for a SM Higgs, as predicted by  \textsc{SuperChic 2.5}~\cite{Harland-Lang:2015cta}, is $\sigma(|y_H|<2.5) = (1\,{}^{\times}_{\div}\,2)\,{\rm fb}$, where the dominant uncertainties are due to PDFs and modelling of the soft gap survival probability. The predicted rate is therefore relatively small, and would again benefit both from the increased statistics available in HL running, and even more crucially from the potential installation of new tagging detectors at a larger distance from the ATLAS or CMS interaction points (IPs), see section~\ref{sec:tagpros}, which would extend the existing mass acceptance into the Higgs region.

\subsection{Tagged proton at the HL--LHC: experimental prospects}\label{sec:tagpros}



This section discusses possible locations for movable near-beam 
detectors along the outgoing beam lines near IP5, designed for detecting the 
leading protons from central production processes (Fig.~\ref{exclusive_production}, \eq{eqn:exclusive_production}). While the results which follow consider the possibilities for detectors in association with the CMS experiment, similar qualitative prospects are expected in the case of the ATLAS detector, although this is not discussed explicitly here. After identifying the 
best-suited positions, the proton detection acceptance and hence the 
central-mass tagging reach is calculated for each of these positions as a 
function of beam parameters and based on present-day assumptions on optics
, collimation 
scheme and near-beam-detector insertion rules from machine protection 
arguments. Given that at the time of this report the crossing-angle plane in 
IP5 (horizontal as until LS3, or vertical) has not yet been decided, both options have been investigated. It has to be pointed out that the crossing planes of IP1 and IP5 have to be different: one will be horizontal, the other vertical.

While in the CT-PPS (later PPS) project~\cite{ctpps}
in Run-2 the near-beam detectors were Roman Pots inherited from 
the TOTEM experiment~\cite{totem-tdr,totem-jinst} and upgraded for 
high-luminosity operation~\cite{totem-upgrade}, no 
technological assumptions are made at this early stage of preparation for 
HL-LHC. The highly demanding engineering and detector physics challenges are 
not addressed here. 

\subsubsection*{Possible Locations for Near-Beam Detectors}
The search for suitable detector locations around IP5 is driven by the goal to cover the widest possible range of central masses $M$ to be measured via the fractional momentum losses 
\begin{equation}
\xi_{1/2} = \frac{\Delta p_{1/2}}{p}
\end{equation}
of the two surviving protons using the relation
\begin{equation}
\label{eqn:mass-xi1-xi2}
M^{2} = \xi_{1}\, \xi_{2}\, s \: ,
\end{equation}
where $\sqrt{s} = 14\,$TeV is the centre-of-mass energy.

The minimum accessible $\xi$ of leading protons at a location $z$~\footnote{In this article the variable $z$ is used for the longitudinal coordinate instead of $s$ to avoid confusion with the Mandelstam $s$.} along the beam line is given by
\begin{equation}
\label{eqn:xi-min}
\xi_{\rm min}(\alpha, \beta^{*}, z) = \frac{[n_{\rm TCT}(\beta^{*}) + \Delta n] \sigma_{\rm XRP}(\beta^{*},z) + \Delta d + \delta}{D_{x, \rm XRP}(\alpha, \xi_{\rm min}, z)} \: ,
\end{equation}
where $\sigma_{x}$ is the horizontal beam width depending on the optics (characterised by $\beta^{*}$), $D_{x}$ is the horizontal dispersion depending on the crossing-angle $\alpha$, $n_{\rm TCT}$ is the half-gap of the tertiary collimators (TCT) as defined by the collimation scheme, $\Delta n = 3$ is the retraction of the near-beam detector housings (e.g. Roman Pots) relative to the TCT position in terms of $\sigma_{x}$, $\Delta d = 0.3\,$mm is an additional safety retraction to allow for beam orbit fluctuations, and the constant $\delta$, typically 0.5\,mm, accounts for any distance between the outer housing surface closest to the beam and the sensitive detector. The dependence of the dispersion on $\xi$ implies that \eq{eqn:xi-min} has to be resolved for $\xi_{\rm min}$ after parameterising $D_{x}(\xi)$.

The first step of the study is to plot the $z$-dependent quantities, $\sigma_{x}$ and $D_{x}$, along the outgoing beam line for one typical HL-LHC optics configuration (Fig.~\ref{fig:d-sigma-xi_vs_s}, left). The resulting $\xi_{\rm min}$ is shown in Fig.~\ref{fig:d-sigma-xi_vs_s} (right). Note that for vertical crossing smaller values are reached. The locations most suitable for the measurement of small $|\xi|$ values are marked in red. Closer layout inspection of the region around the minimum at 232\,m (inside the quadrupole Q6) indicated two promising locations: at 220\,m (just before the collimator TCL6) and at 234\,m (after the exit of Q6). Even smaller momentum losses can be reached at 420\,m (the ``missing magnet'' region already studied previously by the FP420 project~\cite{fp420}). 

\begin{figure}[h!]
\begin{center}
\includegraphics[width=0.35\textwidth]{\main/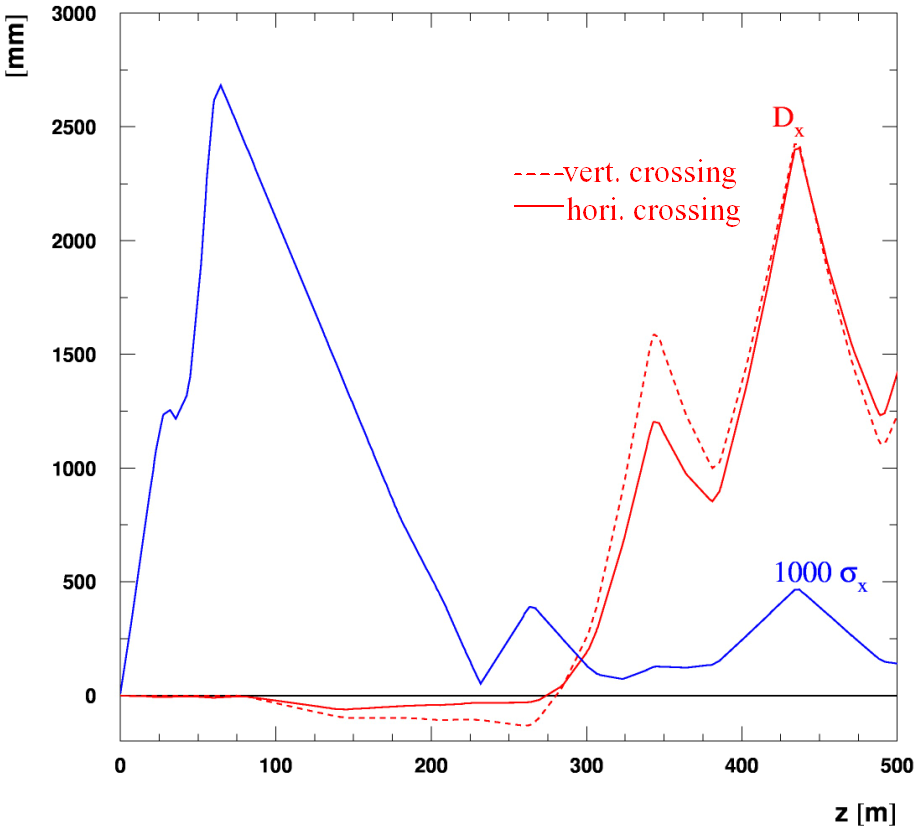}
\includegraphics[width=0.35\textwidth]{\main/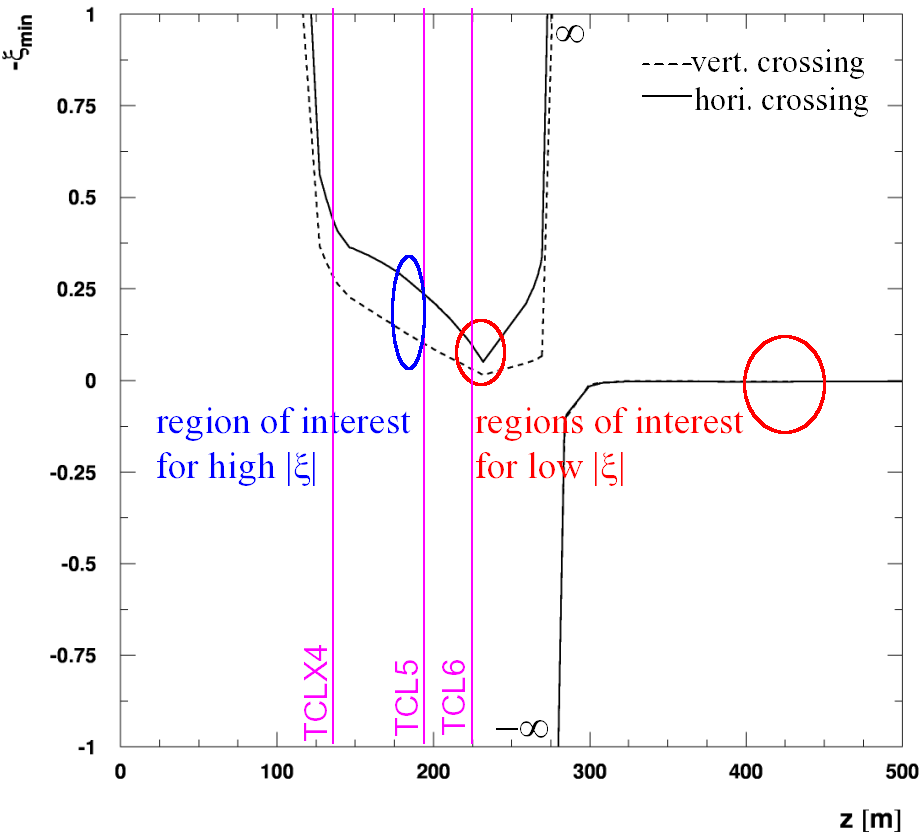}
\end{center}
\vspace{-0.5cm} 
\caption{Horizontal dispersion and beam width (scaled by 1000) as a function of the distance $s$ from IP5 for Beam 1, i.e. in LHC Sector 5-6 (left). Minimum accepted $\xi$ as a function of $z$ according to \eq{eqn:xi-min} for $(\alpha/2, \beta^{*}) = (250\,\rm \mu rad, 15\,cm)$ and $n_{\rm TCT} = 12.9$ (right). 
The TCL collimator positions are indicated. In both pictures the continuous and dashed lines represent horizontal and vertical crossing in IP5, respectively.} 
\label{fig:d-sigma-xi_vs_s}
\end{figure}

The apparent sign change of $\xi_{\rm min}$ at $z \approx 270\,$m reflects the sign change of the dispersion at that location (as seen in the left panel of the figure). It means that the diffractive proton trajectories transition from $x > 0$ to $x < 0$. The implication for the potential detector location at 420\,m is that detectors need to be placed in the confined space between the incoming and the outgoing beam pipes, excluding conventional Roman Pot technology. A further complication is that in this location the beam pipes are in a cryostat, necessitating more involved engineering changes.

A region of interest for the detection of higher masses lies at 196\,m just upstream of the collimator TCL5 that intercepts protons with large $|\xi|$ (section~\ref{sec:masslimits}). Locations even further upstream, before TCLX4, would give an even 
higher upper mass cut but are excluded due to the prohibitively high low-mass limit leaving no acceptance interval.

In summary, for the more detailed discussions in the following sections, four detector locations have been retained: 196\,m, 220\,m, 234\,m, 420\,m.

\subsubsection*{Crossing-Angle and Optics Dependence of the Mass Acceptance Limits}
\label{sec:masslimits}
In the previous section, only one specific combination of crossing-angle $\alpha$ and beam optics ($\beta^{*}$) was considered. However, at HL-LHC luminosity levelling will be performed in all fills by changing $\alpha$ and $\beta^{*}$ in a pre-defined sequence. For the present study the $(\alpha/2$, $\beta^{*})$ trajectories envisaged in~\cite{levellingtalk} were used.

\paragraph*{Minimum Mass \\}

The minimum mass accepted at a location $z$ for given $\alpha$ and $\beta^{*}$ can be calculated using \eq{eqn:mass-xi1-xi2} and~(\ref{eqn:xi-min}). For simplicity, symmetric optics in the two beams, i.e. equal $\xi_{\rm min}$, are assumed:
\begin{equation}
M_{\rm min} = |\xi_{\rm min}|\,\sqrt{s} \: .
\end{equation}
The $\alpha$ and $\xi$ dependencies of $D_{x}$ can be parameterised based on simulations with {\mbox{\textsc{MAD-X}}\xspace}~\cite{MADX}. The $\alpha$ dependence is linear, and the $\xi$-dependence can be linearly approximated within the $\xi$-ranges relevant in practice.

The $\beta^{*}$ dependence of $\sigma_{\rm XRP}$ was calculated analytically, profiting from invariance properties of the presently planned family of ATS optics. This is likely to change in the future and will need to be adapted.

\begin{figure}[h!]
\begin{center}
\includegraphics[width=0.45\textwidth]{\main/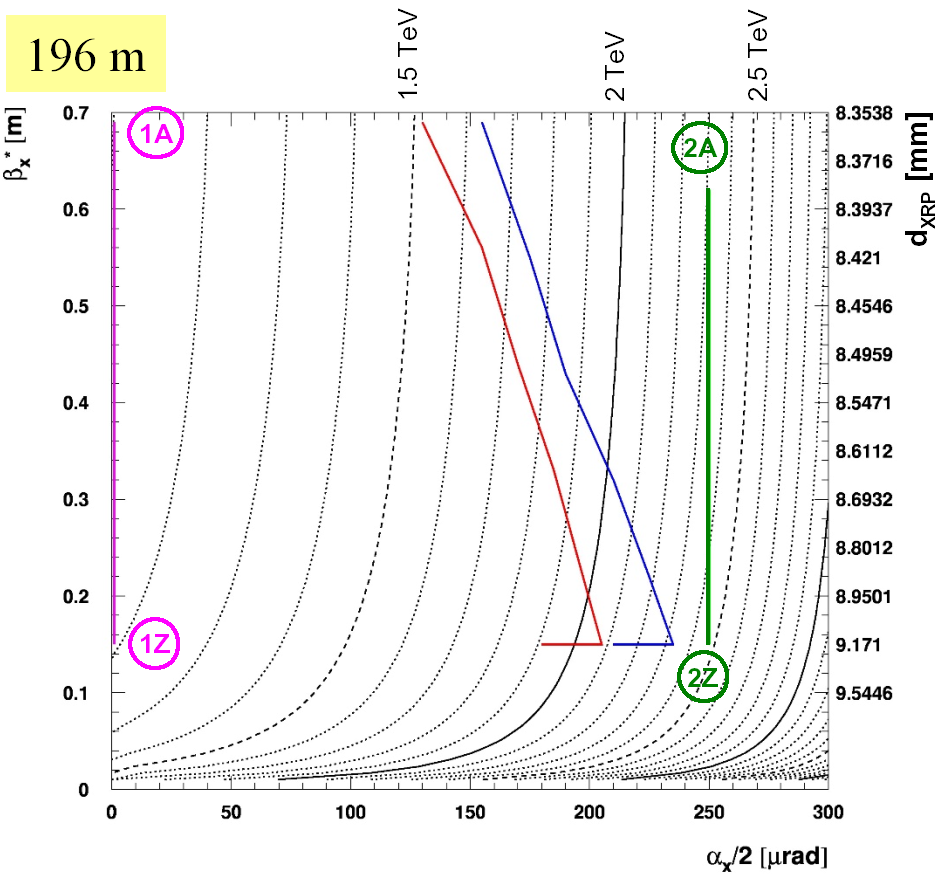}\hfill
\includegraphics[width=0.45\textwidth]{\main/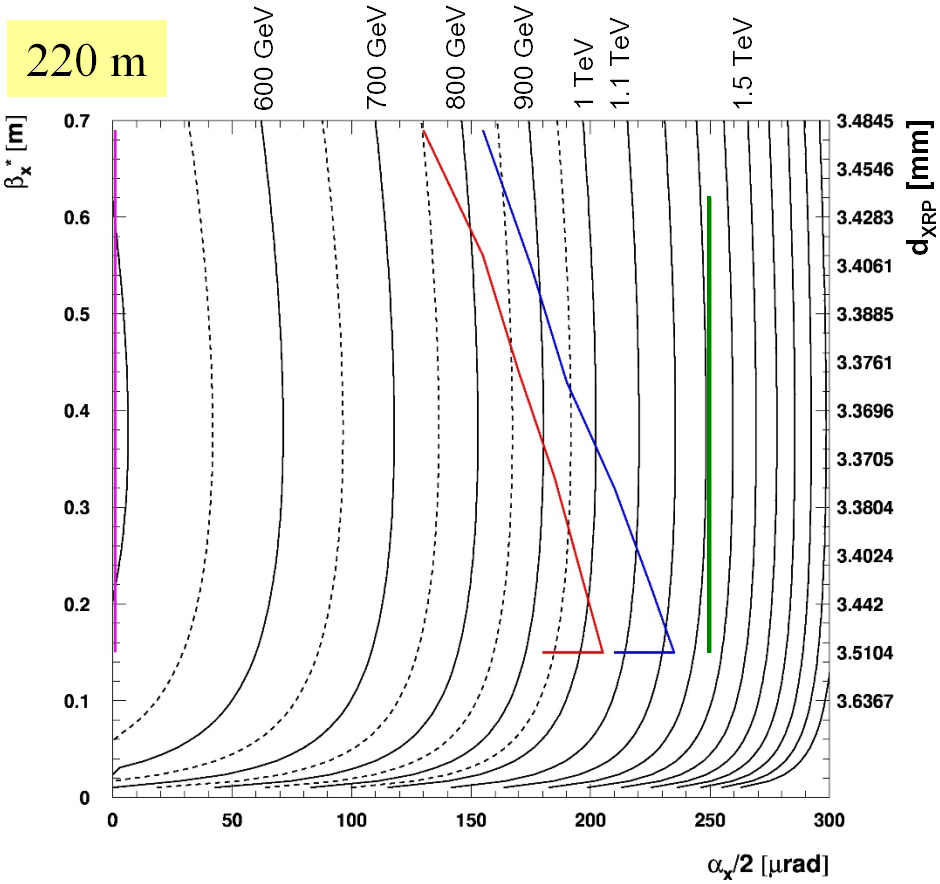}
\includegraphics[width=0.45\textwidth]{\main/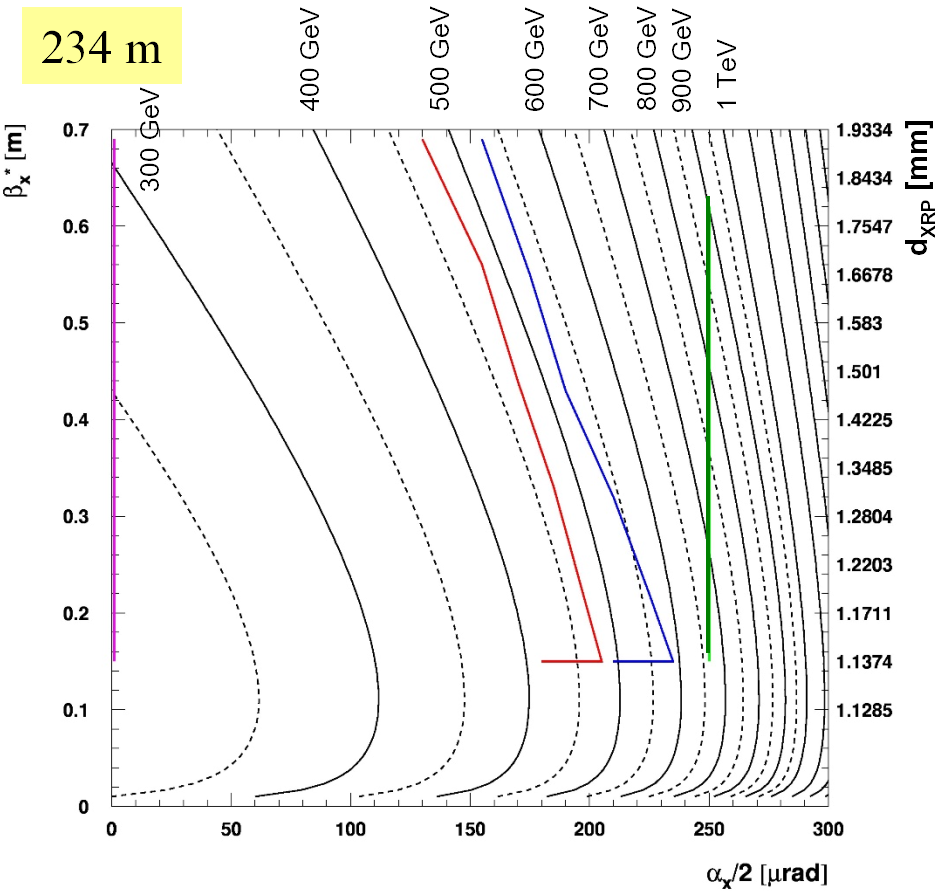}\hfill
\includegraphics[width=0.45\textwidth]{\main/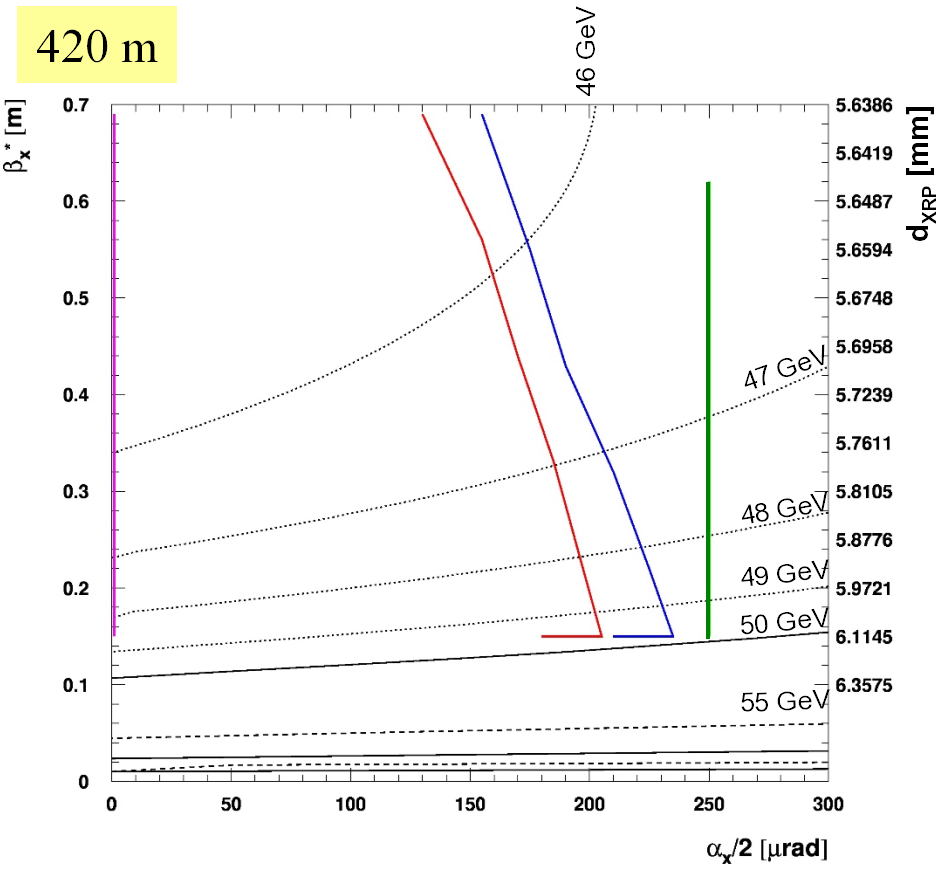}
\end{center}
\vspace{-0.5cm} 
\caption{Contour lines for the minimum accepted mass $M_{\rm min} = |\xi|_{\rm min} \sqrt{s}$ in the crossing-angle/optics parameter space $(\alpha/2, \beta^{*})$. On the right-hand ordinate the XRP approach distance is calculated from $\beta^{*}$. The coloured lines represent possible luminosity-levelling trajectories~\cite{levellingtalk}. For horizontal crossing: green corresponds to ``baseline'', blue to ``relaxed adaptive'', red to ``aggressive adaptive''; for vertical crossing: violet corresponds to any trajectory. The labels (1A) -- (2Z) in the first panel define the trajectory start and end points used in Figs.~\ref{fig:m-y} and~\ref{fig:m-acceptance}.}
\label{fig:m-min}
\end{figure}
The $\beta^{*}$ dependence of $n_{\rm TCT}$ follows the presently foreseen collimation strategy~\cite{collimation} of keeping the TCT gap constant at 
$d_{\rm TCT} = 12.9\, \sigma_{\rm TCT}(\beta^{*}=15\,\rm cm)$ (for nominal emittance $\varepsilon_{n} = 2.5\,\mu$m\,rad), implying \mbox{$n_{\rm TCT}(\beta^{*}) = \frac{d_{\rm TCT}}{\sigma_{\rm TCT}(\beta^{*})}$}, where an analytical expression for $\sigma_{\rm TCT}(\beta^{*})$ can be derived.

The result of this calculation, contour lines of $M_{\rm min}$ in the beam parameter space $(\alpha/2, \beta^{*})$, is shown in Fig.~\ref{fig:m-min} for the four detector locations chosen in the previous section. Some possible luminosity-levelling trajectories are drawn, too. The start point at the beginning of the fill is always at the maximum $\beta^{*}$ value.

From these graphs the following conclusions are drawn:
\begin{itemize}
\item The main driving factor for the minimum mass is the dispersion which in turn is fully determined by the crossing-angle. The optics (via $\beta^{*}$) plays
a minor role.
\item If the 420\,m location can be instrumented, the minimum mass is about 50\,GeV with only a very weak dependence on the optics, the crossing-angle and its plane (horizontal or vertical).
\item Without the 420\,m location, the vertical crossing gives a much better low-mass acceptance (210\,GeV) than the horizontal crossing (660\,GeV).
\end{itemize}

\paragraph*{Maximum Mass \\}
The maximum mass accepted at a location $z$ is determined by the tightest aperture restriction $d_{A}$ upstream of $z$ and the dispersion there:
\begin{equation}
M_{\rm max} = |\xi_{\rm max}|\,\sqrt{s} =  \frac{d_{A}}{D_{A}(\alpha, \xi_{\rm max})}\: .
\end{equation}

\begin{figure}[h!]
\begin{center}
\includegraphics[width=0.45\textwidth]{\main/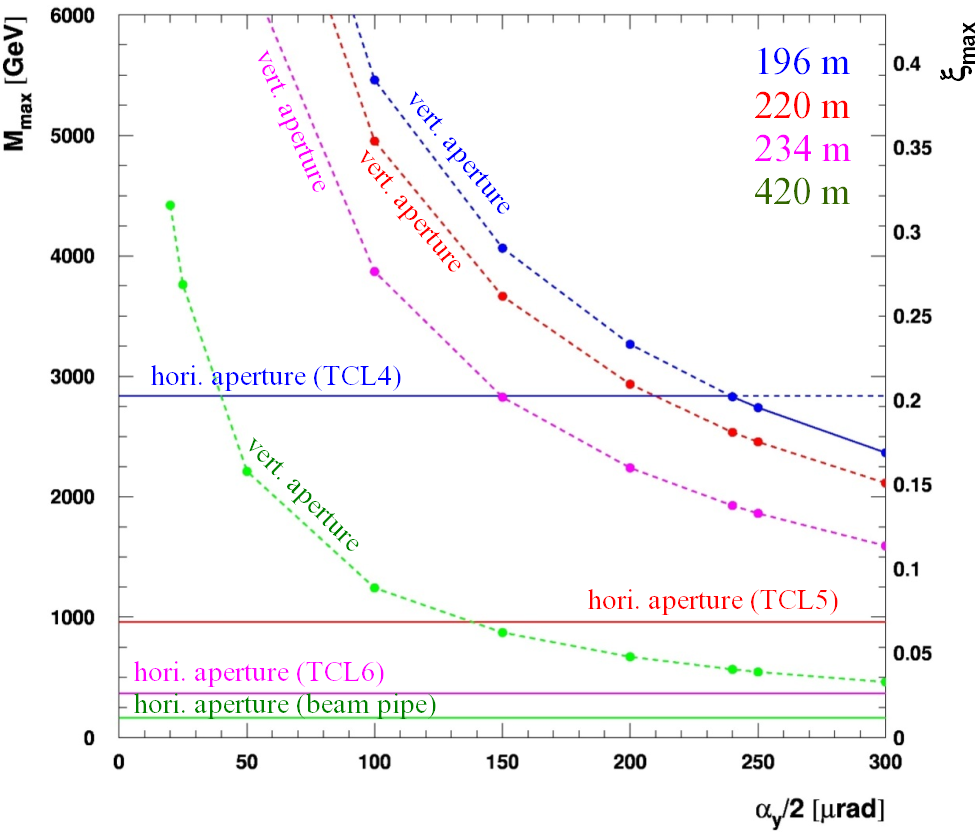}
\hspace{0.5cm}
\includegraphics[width=0.45\textwidth]{\main/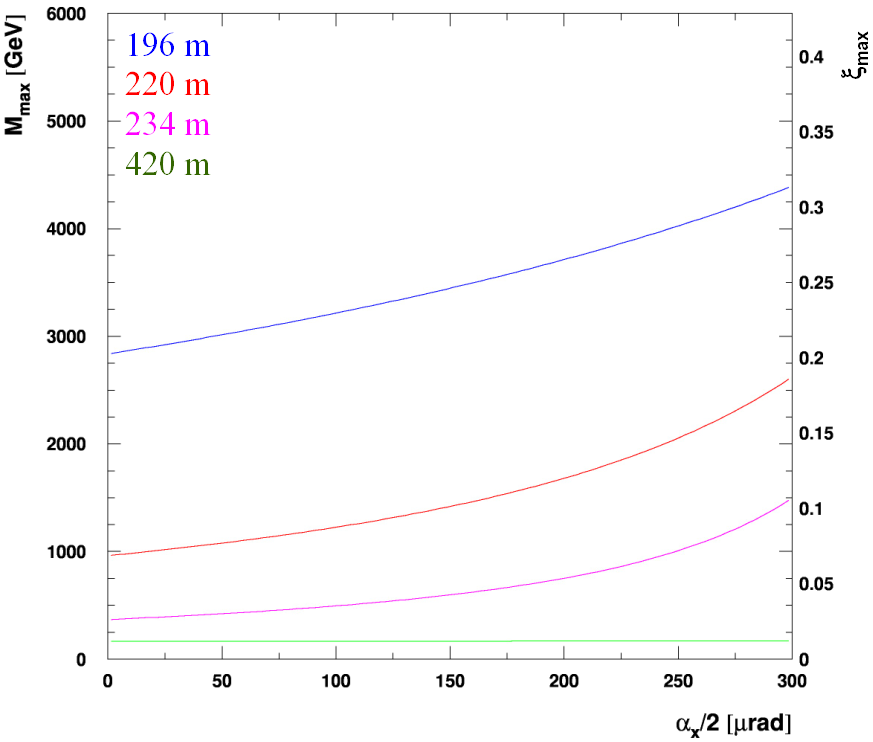}
\end{center}
\vspace{-0.5cm} 
\caption{Maximum accepted diffractive mass for each detector location as a 
function of the crossing-angle. Vertical crossing (left): both horizontal and
vertical apertures contribute to the mass limits. The continuous lines denote the most restrictive, i.e. dominant, limitations.
Horizontal crossing (right): only the horizontal apertures contribute.} 
\label{fig:m-max}
\end{figure}
In the case of the vertical beam crossing in IP5, both the horizontal and vertical apertures may impose limitations, whereas in the case of the horizontal crossing there is no substantial vertical dispersion and hence no acceptance loss from the vertical aperture. Figure~\ref{fig:m-max} shows the results of a complete aperture study. It was concluded that even for vertical crossing most limitations come from the horizontal aperture and that for all locations, except 420\,m, this horizontal aperture is limited by the TCL collimators. 
At 420\,m, on the other hand, the beam-pipe absorbs diffractive protons with $|\xi| > 0.012$.
The highest masses are accepted by the unit at 196\,m: up to 2.7\,TeV for vertical crossing and up to 4 TeV for horizontal crossing.

\subsubsection*{Mass-Rapidity Acceptance}
The CEP acceptance for a given point in the beam parameter space $(\alpha, \beta^{*})$ can be visualised by drawing for every instrumented detector location the $|\xi|$-acceptance bands -- whose limits are calculated according to the previous section -- in the mass-rapidity plane
\begin{equation}
\left(\ln \frac{M}{\sqrt{s}}, y\right) = \left(\frac{1}{2}(\ln \xi_{1} + \ln \xi_{2}), \frac{1}{2}(\ln \xi_{1} - \ln \xi_{2})\right) \: .
\end{equation}
Figure~\ref{fig:m-y} shows these $(M,y)$ contour plots for the start and end points of the two extreme levelling cases defined in Fig.~\ref{fig:m-min}: 
points (1A) and (1Z) for any trajectory with vertical crossing in IP5, points (2A) and (2Z) for the ``Baseline'' trajectory with horizontal crossing. The projections on the mass axis, under the approximation of flat rapidity distributions, are given in Fig.~\ref{fig:m-acceptance}.

\begin{figure}[h!]
\begin{center}
\includegraphics[width=\textwidth]{\main/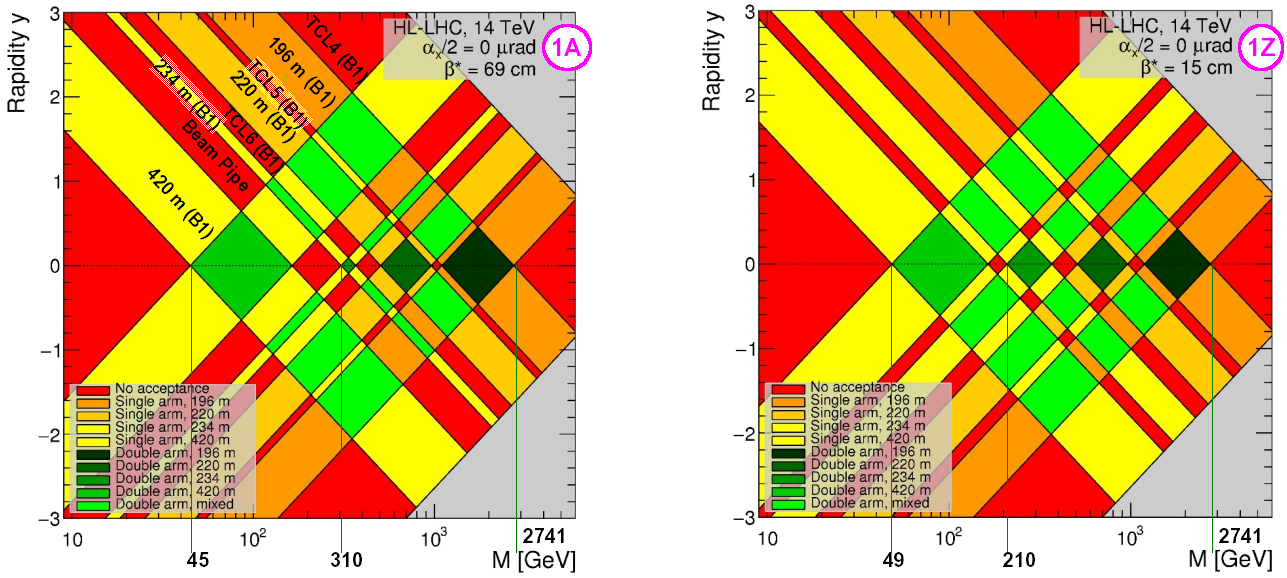}
\includegraphics[width=\textwidth]{\main/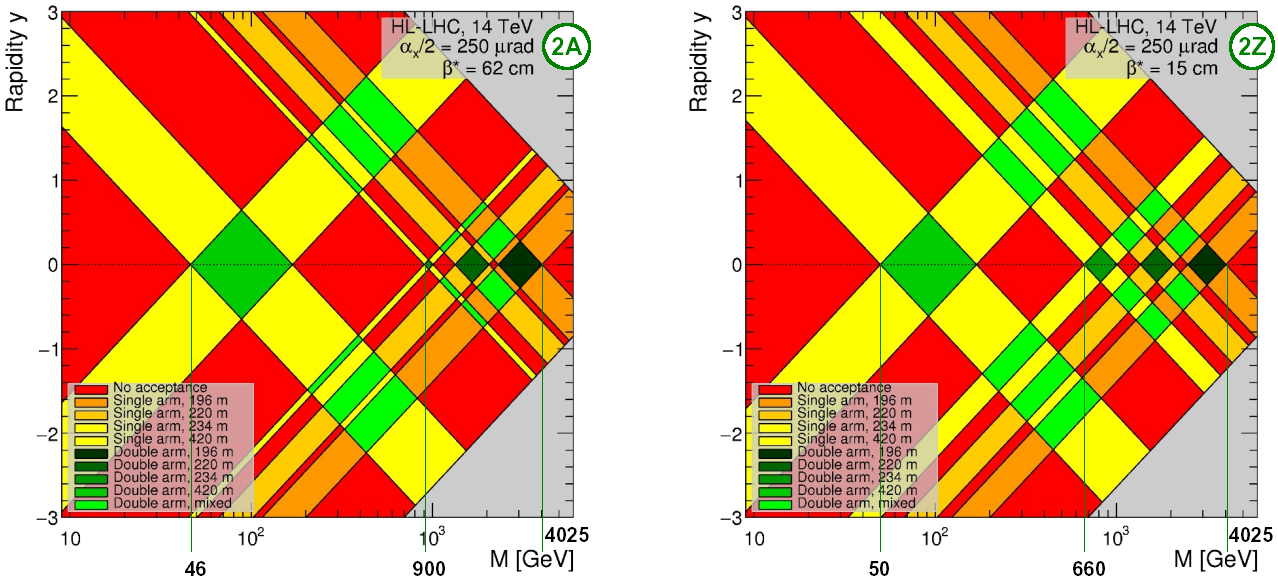}
\end{center}
\caption{Acceptance for the protons from central diffraction in the mass-rapidity plane. The yellow/orange colour tones mark single-arm proton acceptance, the green tones mark double-arm acceptance. Top: start and end point of any levelling trajectory for vertical crossing, bottom: start and end point of the baseline levelling trajectory for horizontal crossing.} 
\label{fig:m-y}
\end{figure}

\begin{figure}[htbp]
\begin{center}
\includegraphics[width=0.45\textwidth]{\main/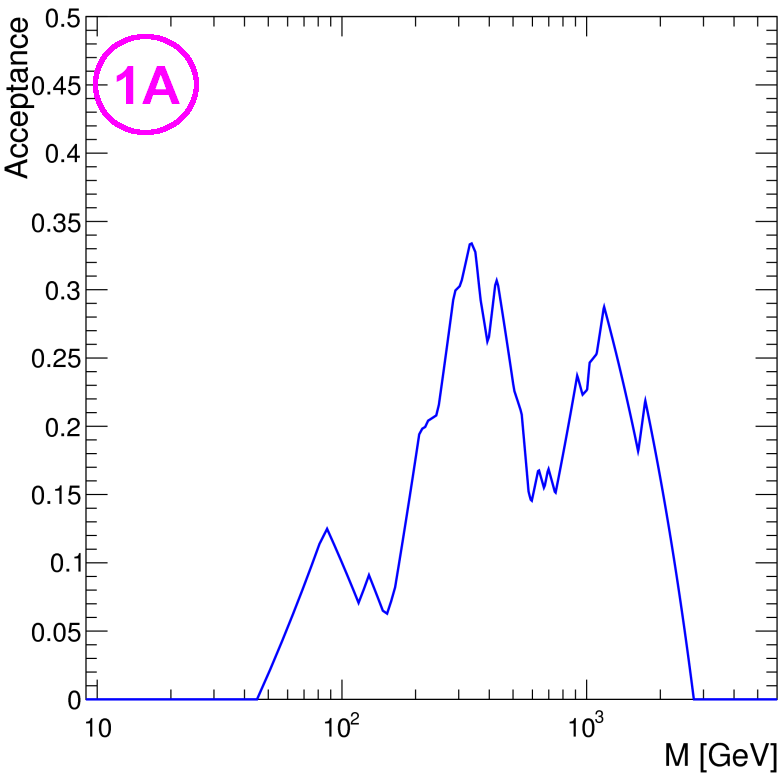}\hfill
\includegraphics[width=0.45\textwidth]{\main/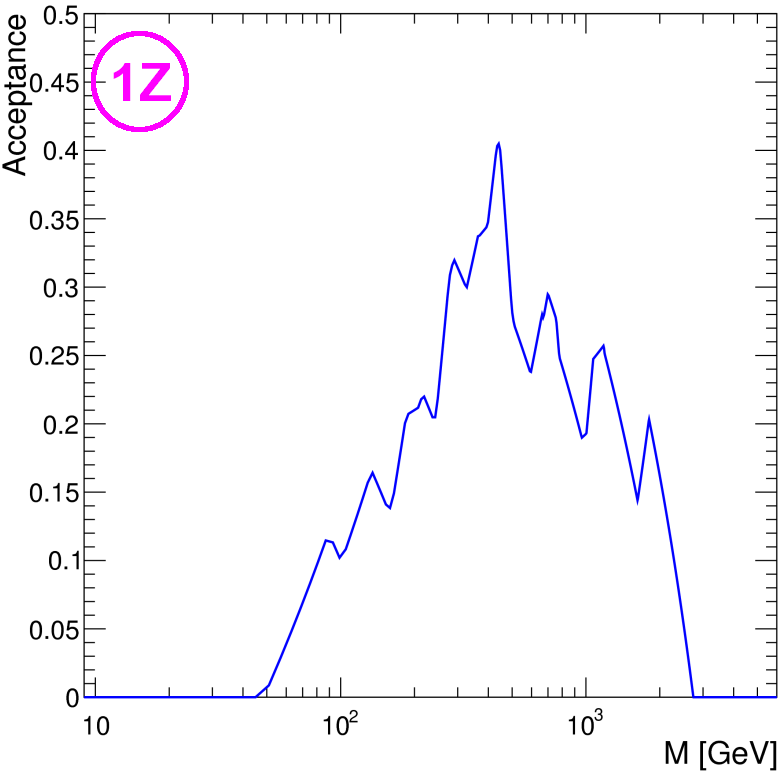}
\includegraphics[width=0.45\textwidth]{\main/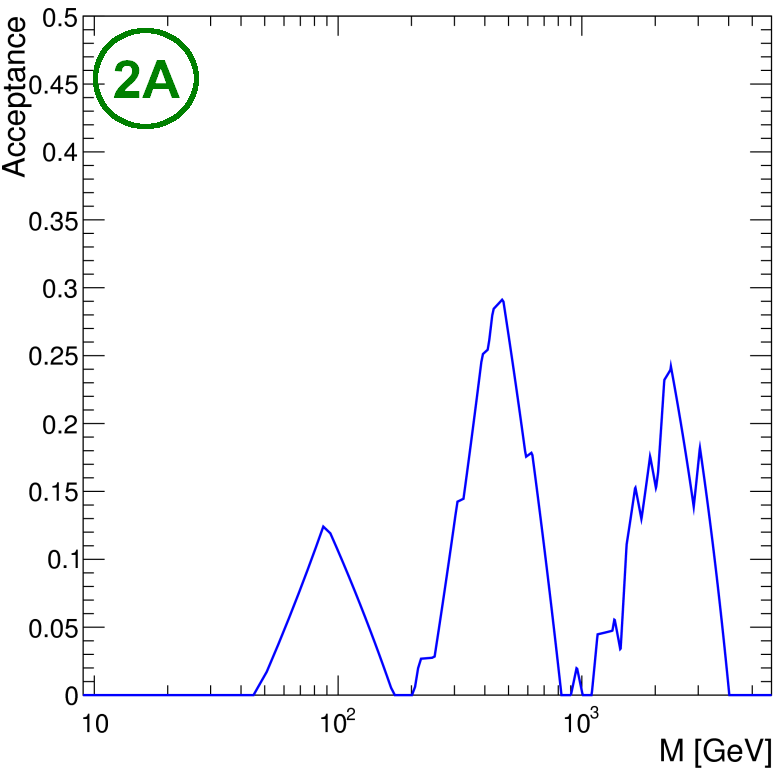}\hfill
\includegraphics[width=0.45\textwidth]{\main/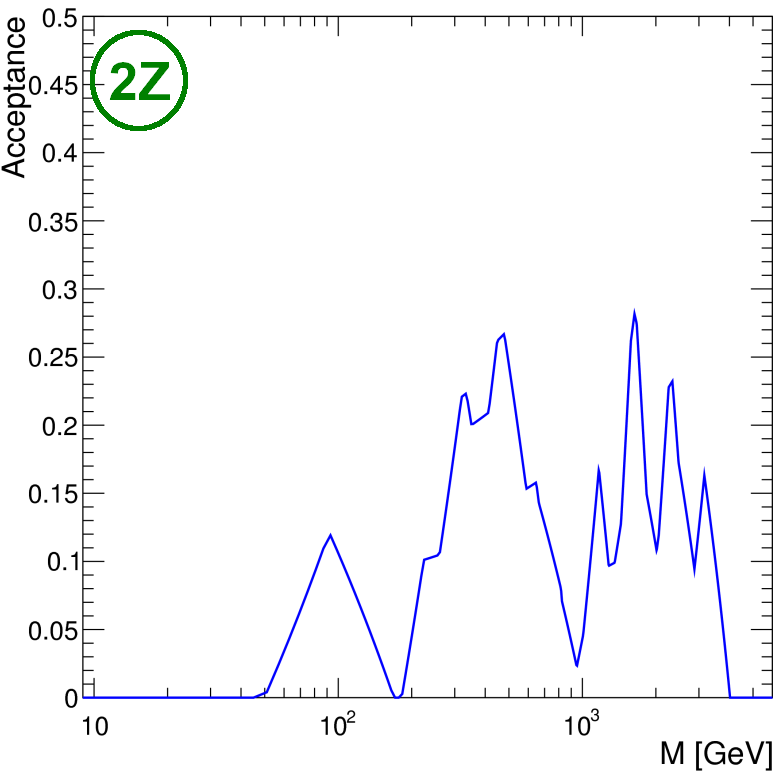}
\end{center}
\caption{Projection of the $(M, y)$ acceptance on the mass axis, adding up all the double-arm areas of Fig.~\ref{fig:m-y} for the same points in the $(\alpha, \beta^{*})$ beam parameter space.}
\label{fig:m-acceptance}
\end{figure}

The following observations are made:
\begin{itemize}
\item The acceptance zones of the four detector locations are non-overlapping and separated by gaps. For horizontal crossing the gaps are wider than for vertical crossing.
\item Although the double-arm acceptance has mass gaps at central rapidities, the mixed acceptance zones combining different detector units in the two arms of the experiment (e.g. 420\,m left + 234\,m right) fill some of these mass gaps by providing acceptance at forward rapidities.
\item The gaps between the acceptances of 196\,m, 220\,m and 234\,m can potentially be closed by opening TCL5 and TCL6 a little further if allowable from machine protection arguments. On the other hand, the gap between 234\,m and 420\,m is caused by the beam pipe at $z > 300\,$m limiting the aperture. It could only be closed by adding a detector unit near 300\,m. 
\end{itemize}


\subsection{Low-mass central exclusive production}\label{sec:ALICECEP}
Central exclusive production of low-mass diffractive states in $pp$ collisions at the LHC may serve as a valuable source of information on the non-perturbative aspects of strong interaction. At low masses, CEP is usually described in terms of a double pomeron exchange (DPE) mechanism. DPE is expected to be an ideal process for the investigation of meson resonances with $I^{G}(J^{PC})=0^{+} (0^{++}$, $2^{++}$, \dots)  quantum numbers and gluonic bound states. Glueball searches in CEP are of particular interest because lattice QCD calculations predict the lightest glueballs to have masses $M_G(0^{++}) = 1710$ MeV and $M_G(2^{++}) = 2390$ MeV~\cite{Morningstar:1999rf}. Pure glueballs are predicted to decay equally well into pair of pions, kaons or $\eta$ mesons with suppressed two photon decays. However this simple signature is spoiled by the fact that glueballs are expected to mix with nearby $q\bar q$ states.

Central-exclusive production of low-mass resonances in $\pi\pi$ and $KK$ channels has been extensively studied in fixed target experiments at CERN and Fermilab (see review in~\cite{Albrow:2010yb}) and recent collider experiments at RHIC~\cite{Sikora:2018cyk}, Tevatron~\cite{Aaltonen:2015uva} and the LHC~\cite{Khachatryan:2017xsi}. The partial-wave analysis (PWA) has been performed in several experiments to investigate the spin-parity nature of the centrally produced system~\cite{Reyes:1997ei,Barberis:1999am,Barberis:1999ap}. There is a clear evidence of supernumerous light scalar meson states, not fitting well into the conventional groundstate $q\bar q$ nonet and suggesting that some of these states have significant gluonic component. The $f_0(1370)$, $f_0(1500)$ and $f_0(1710)$ mesons are considered as most promising  glueball-meson mixing state candidates but the nature of all these states is still open for discussion~\cite{Crede:2008vw}. In the tensor sector, the lightest isoscalar $q\bar q$ states $f_2(1270)$ and $f_2'(1525)$ are well established however there are at least four additional reported tensor resonances requiring confirmation. 

CEP can be also used to investigate the spin structure of the Pomeron and its coupling to hadrons. Historically, the Pomeron was considered as effective spin 1 quasiparticle supported by successful fits of total and differential $pp$ cross sections~\cite{Close:1999bi}. Recently, an alternative approach based on the tensorial Pomeron has been 
developed~\cite{Lebiedowicz:2013ika} providing definitive predictions and restrictions of spin-parity, polarization and rapidity of the produced diffractive system in CEP at the LHC~\cite{Lebiedowicz:2016ioh,Lebiedowicz:2018eui,Lebiedowicz:2018sdt}. 

Multidifferential measurements and PWA of $\pi\pi$, $KK$ and $p\bar p$ final states in a wide range of invariant masses in CEP at the LHC would also allow one to constrain poorly known Pomeron-meson couplings and form-factors in various phenomenological models~\cite{Lebiedowicz:2016ioh,Harland-Lang:2013dia} and also build a transition to perturbative QCD models of hadron pair production in CEP~\cite{HarlandLang:2011qd} valid at high invariant masses and transverse momenta of the produced pair. Another important outcome of CEP measurements would be a determination of the absorptive corrections, the probability that the rapidity gaps would be filled with particles from accompanying initial- or final-state interactions. The central exclusive production of meson pairs therefore represents a process of much phenomenological interest, which can shed light on both perturbative and non-perturbative aspects of QCD.

Perturbative aspects of QCD can be also investigated in CEP of heavy quarkonium states~\cite{Harland-Lang:2014lxa}. Detailed studies of $\chi_c$ resonances in CEP at the LHC would provide a valuable input to test the ideas and methods of the QCD physics of bound states. Measurements of the outgoing proton momentum distributions, cross sections and relative abundances of $\chi_{c0}$, $\chi_{c1}$ and $\chi_{c2}$ states would be important for the test of the overall theoretical formalism.

Measurements of CEP processes rely on the selection of events with only few tracks in an otherwise empty detector, therefore large pseudorapidity coverage and low pileup conditions are essential to guarantee the event emptiness. The ALICE detector nicely matches these requirements. Low material budget, access to low transverse momenta and excellent particle identification capabilities in ALICE serve as additional advantages. First CEP measurements have been already performed by ALICE in the LHC Run-1 and -2. Figure~\ref{fig:alice_cep} illustrates raw invariant mass spectra of $\pi^+\pi^-$ and $K^+K^-$ pairs in CEP events collected by ALICE in proton-proton collisions at~$\sqrt{s}=13$ TeV, where one can easily identify several resonance structures. ALICE is going to collect a much larger sample of central exclusive events and significantly extend the scope of the CEP program in proton-proton collisions in LHC Run-3 with expected integrated luminosity of about $200$ pb$^{-1}$ at $\sqrt{s}=14$ TeV and 6 pb$^{-1}$ at $\sqrt{s}=5.5$ TeV profiting from much better efficiency in the continuous readout mode.
The CEP program includes glueball searches and precision hadron spectroscopy in $\pi^+\pi^-$, $K^+K^-$, $p\bar p$, $2\pi2K$, $4\pi$ and other channels. The expected high integrated luminosity will also allow ALICE to measure the spectrum of heavy quarkonium states in various decay channels, e.g. a yield of at least 50,000 $\chi_{c0} \to \pi^+\pi^-$ decays is expected in CEP events by the end of Run-3 based on cross section estimates from {\mbox{\textsc{SuperChic}}\xspace} generator~\cite{Harland-Lang:2015jpt}.

\begin{figure}
 \includegraphics[width=0.48\textwidth]{\main/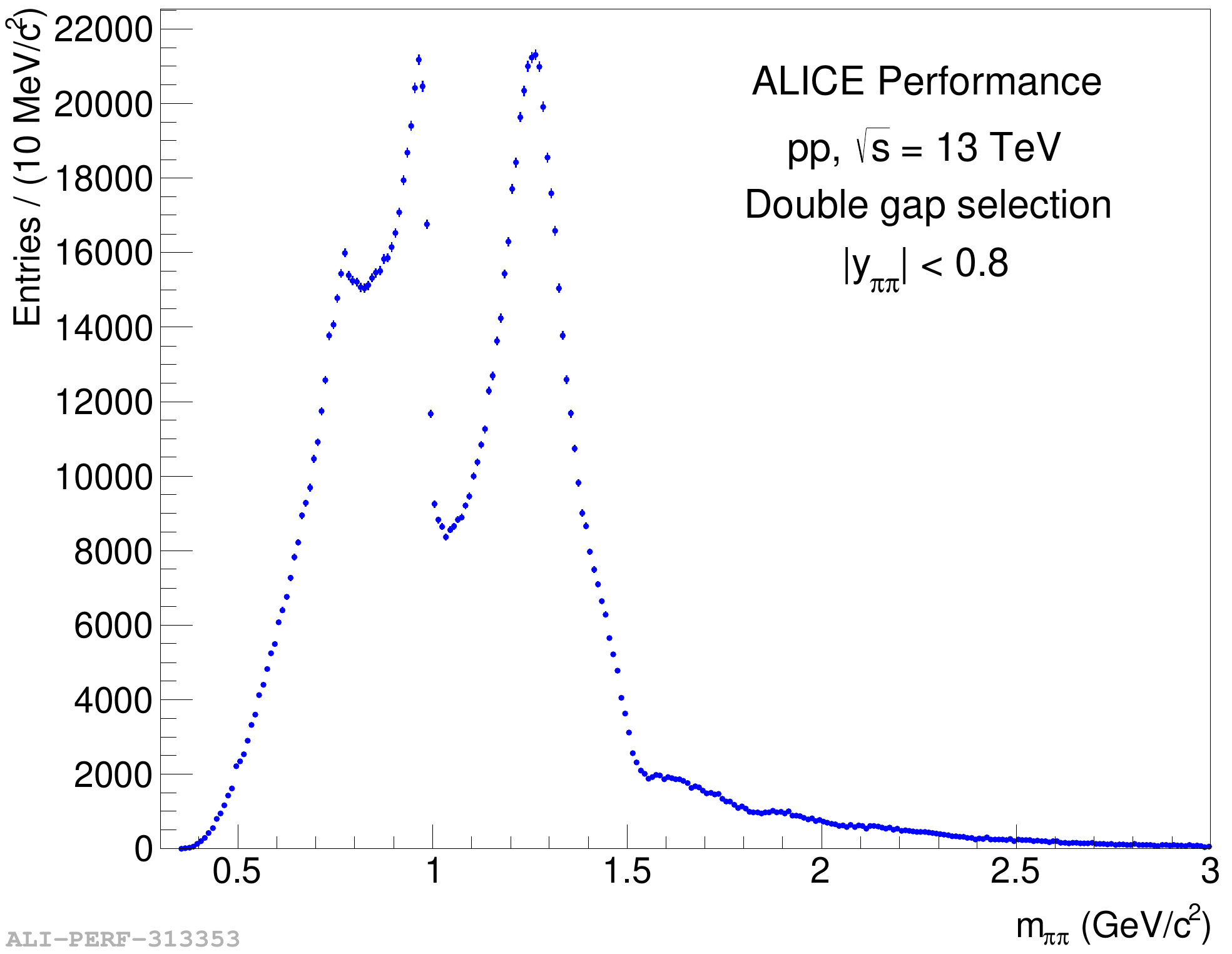}
 \includegraphics[width=0.48\textwidth]{\main/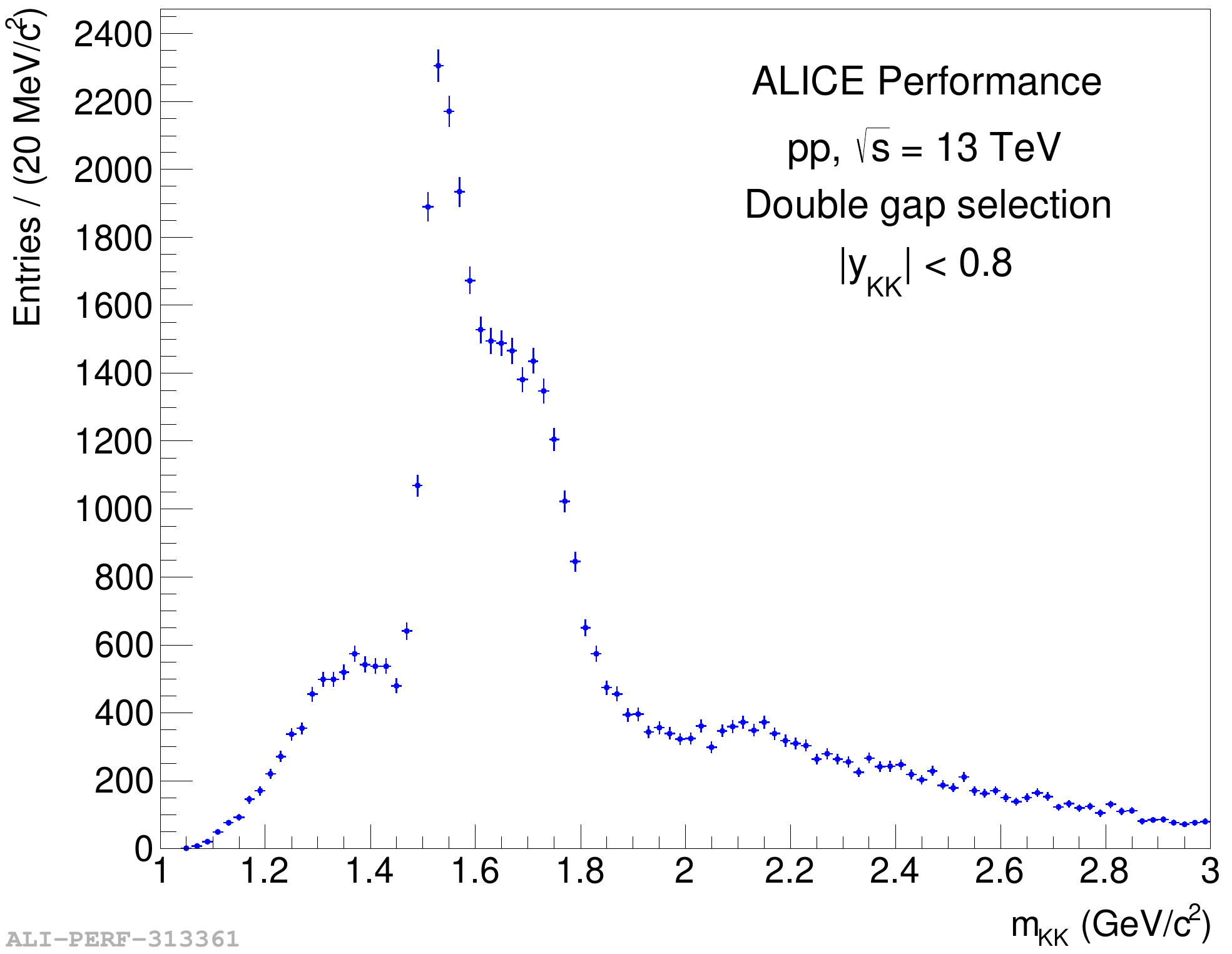}
 \caption{Raw invariant mass spectra of $\pi^+\pi^-$ (left) and $K^+K^-$ (right) pairs in CEP events collected by ALICE in proton-proton collisions at~$\sqrt{s}=13$ TeV.
 \label{fig:alice_cep}
 }
\end{figure}

The LHCb experiment can extend the CEP program to forward rapidities. High luminosity at moderate pileup and good hadron PID capabilities would be particularly useful for the studies of heavy quarkonium states in central exclusive events. Measurements of low-mass central exclusive production processes with proton tagging might be also possible with the ATLAS and CMS detectors during low pile-up runs at high $\beta^*$.

\newpage
\newpage
\section*{Acknowledgements}
We would like to thank the LHC experimental Collaborations and the WLCG for their essential support.
We are especially grateful for the efforts by the computing, generator and validation groups who were instrumental for the creation of large simulation samples. We thank the detector upgrade groups as well as the physics and performance groups for their input. Not least, we thank the many colleagues who have provided useful comments on the analyses.

\newpage

\bibliographystyle{report}
\bgroup
\interlinepenalty=10000
\parskip0pt plus\baselineskip
\bibliography{\bibfiles}
\egroup

\end{document}